\documentclass[11pt,twoside]{ckmArticle}
\usepackage{epsf}
\usepackage{rotating}
\usepackage{epsfig}
\usepackage{supertabular}
\usepackage{subfigure}
\usepackage{multirow}
\usepackage{eufrak}
\usepackage{refmerge} 
\input{myBabarsym}
\newcommand{\bei}{\begin{itemize}}
\newcommand{\eei}{\end{itemize}}
\newcommand{\beq}{\begin{equation}}
\newcommand{\eeq}{\end{equation}}
\newcommand{\beqn}{\begin{eqnarray}}
\newcommand{\eeqn}{\end{eqnarray}}
\newcommand{\beqns}{\begin{eqnarray*}}
\newcommand{\eeqns}{\end{eqnarray*}}
\newcommand{\vs}{\\[0.7\baselineskip]\indent}

\newcommand{\intl}{\int\limits}

\newcommand{\mc}{\multicolumn}

\newcommand{\hmm}{\hspace{-0.2cm}}
\newcommand{\hmmm}{\hspace{-0.3cm}}

\newcommand\rs{\raisebox{1.3ex}[-1.5ex]}

\newcommand\ph{\phantom}
\newcommand{\GF}{G_{\rm F}}

\newcommand{\MSbar}{\overline{\rm MS}}
\newcommand{\tmsix}{\times10^{-6}}

\newcommand{\HFAG}{{\it Heavy Flavor Averaging Group}}
\newcommand\NP{{ Nucl. Phys.}}
\newcommand\PL{{ Phys. Lett.}}

\newcommand\PRL{{ Phys. Rev. Lett.}}
\newcommand\NIM{{ Nucl. Inst. Meth.}}

\newcommand\ea{{\em et al.}}
\newcommand\EPJ{{ Eur. Phys. J.}}
\newcommand\PRD{{ Phys. Rev.}}
\newcommand\tho{{\rm theo}}
\newcommand\experi{{\rm exp}}
\newcommand\mod{{\rm mod}}
\newcommand\QCD{{\rm QCD}}
\newcommand\mini{{\rm min}}

\newcommand\opt{{\rm opt}}
\newcommand\iA{I}
\newcommand\iB{II}
\newcommand\iC{III}
\newcommand\iD{IV}

\newcommand\TKpi{T_{K\pi}^{+-}}
\newcommand\PKpi{P_{K\pi}^{+-}}
\newcommand\PKzpi{P_{K\pi}^{0+}}
\newcommand\PrhoKz{P_{\rho K}^{+0}}
\newcommand\PKstzpi{P_{K^*\pi}^{0+}}
\newcommand\Tpm{T^{+-}}
\newcommand\Tmp{T^{-+}}
\newcommand\Ppm{P^{+-}}
\newcommand\Pmp{P^{-+}}

\newcommand\Tzp{T^{0+}}

\newcommand\dmt{t}

\newcommand\Amptpbar{\kern 0.18em\overline{\kern -0.18em {\cal A}}_{3\pi}}

\newcommand\Amptpbarkappa{\kern 0.18em\overline{\kern -0.18em A}^{\kappa}{}}
\newcommand\Amptpbarsigma{\kern 0.18em\overline{\kern -0.18em A}^{\sigma}{}}

\newcommand\Aij{A^{ij}}
\newcommand\Abij{\Abar^{ij}}

\newcommand\Kzpiznn{{\cal B}(K^0_{\rm L}\to\pi^0\nu\nub)}
\newcommand\Kppipnn{{\cal B}(K^+\to\pi^+\nu\nub)}
\newcommand\Bpmun{{\cal B}(B^+\to\mu^+\nu_\mu)}
\newcommand\Bptaun{{\cal B}(B^+\to\tau^+\nu_\mu)}
\renewcommand\Re{{\rm Re}}
\renewcommand\Im{{\rm Im}}
\newcommand\rhoeta{(\rhobar,\etabar)}
\newcommand\CL{{\rm CL}}

\newcommand\dmd{\Delta m_d}

\newcommand\dms{\Delta m_s}
\newcommand\Dms{\Delta m_s}
\newcommand\CLcut{{\rm CL}_{\rm cut}}
\newcommand\epsk{\varepsilon_K}
\newcommand\epe{\varepsilon^\prime/\varepsilon}
\newcommand\fbdbd{\fbd\sqrt{B_d}}

\newcommand\RqcdP{R^{+-}}
\newcommand\RqcdM{R^{-+}}
\newcommand\SL{{\rm SL}}

\newcommand\delpmmp{\hat\delta}
\newcommand\rTpm{r_{T^{+-}}}

\newcommand\lampm{\lambda^{+-}}
\newcommand\lammp{\lambda^{-+}}
\newcommand\lampmmp{\lambda^{+-(-+)}}
\newcommand\kappm{\kappa^{+-}}
\newcommand\kapmp{\kappa^{-+}}
\newcommand\kappmmp{\kappa^{+-(-+)}}
\newcommand\lamCP{\lambda_{CP}}
\newcommand\lamFL{\lambda_{\rm tag}}
\newcommand\alphaeffpm{\alpha_{\rm eff}^{+-}}

\newcommand\alphaeffmppm{\alpha_{\rm eff}^{-+(+-)}}
\newcommand\alphaeffmp{\alpha_{\rm eff}^{-+}}
\newcommand\alphaeffave{\alpha_{\rm eff}^{\pm}}
\newcommand\Czz{C_{\rho\pi}^{00}}
\newcommand\Szz{S_{\rho\pi}^{00}}
\newcommand\Crhopi{C_{\rho\pi}}
\newcommand\dCrhopi{\Delta C_{\rho\pi}}
\newcommand\Srhopi{S_{\rho\pi}}
\newcommand\dSrhopi{\Delta S_{\rho\pi}}

\newcommand\Acp{{\cal A}_{\rho\pi}}
\newcommand\Acppz{{\cal A}_{\rho\pi}^{+0}}
\newcommand\Acpzp{{\cal A}_{\rho\pi}^{0+}}
\newcommand\Acppm{{\cal A}_{\rho\pi}^{+-}}
\newcommand\Acpmp{{\cal A}_{\rho\pi}^{-+}}

\newcommand\Acpave{{{\cal A}_{\rho\pi}^\pm}}
\newcommand\AcpKstrpi{{\cal A}_{K^*\pi}^{+-}}
\newcommand\AcprhoK{{\cal A}_{\rho K}^{-+}}

\newcommand\Nbpm{{\kern 0.18em\overline{\kern -0.18em N}}^{+-}}
\newcommand\Nbmp{{\kern 0.18em\overline{\kern -0.18em N}}^{-+}}
\newcommand\pdf{PDF}
\newcommand\pdfs{PDFs}
\newcommand\rB{r_B}

\newcommand\sta{\stwoa}
\newcommand\staeff{\stwoa_{\rm eff}}
\newcommand\stb{\stwob}
\newcommand\stbeff{\stwob_{\rm eff}}
\newcommand\ctb{\ctwob}

\newcommand\stbwa{\stwob_{[c\bar c]}}

\newcommand\stbpg{\sin(2\beta+\gamma)}
\newcommand\ctbpg{\cos(2\beta+\gamma)}

\newcommand\alphaeff{\alpha_{\rm eff}}

\newcommand\Cpizpiz{{C_{\pi\pi}^{00}}}
\newcommand\Spizpiz{{S_{\pi\pi}^{00}}}
\newcommand\Cpipi{{C_{\pi\pi}^{+-}}}
\newcommand\Spipi{{S_{\pi\pi}^{+-}}}
\newcommand\Ckspiz{{C_{K_{\scriptscriptstyle S}\pi}^{00}}}
\newcommand\Skspiz{{S_{K_{\scriptscriptstyle S}\pi}^{00}}}
\newcommand\BRpimpiz{{\cal B}^{+0}_{\pi\pi}}
\newcommand\BRKpi{{\cal B}_{K\pi}^{+-}}
\newcommand\BRpipi{{\cal B}^{+-}_{\pi\pi}}
\newcommand\BRpippiz{{\cal B}^{+0}_{\pi\pi}}
\newcommand\BRpizpiz{{\cal B}^{00}_{\pi\pi}}
\newcommand\AKpi{A_{K\pi}^{+-}}
\newcommand\Apipi{A^{+-}}
\newcommand\Abarpipi{\Abar^{+-}}

\newcommand\Apippiz{A^{+0}}
\newcommand\Abarpippiz{\Abar^{+0}}
\newcommand\Apimpiz{A^{-0}}

\newcommand\Apizpiz{A^{00}}
\newcommand\Abarpizpiz{\Abar^{00}}

\newcommand\CC{c}
\newcommand\CCbar{\overline\CC}

\newcommand\Tpipi{T^{+-}}
\newcommand\Ppipi{P^{+-}}
\newcommand\PoTpipi{|P^{+-}/T^{+-}|}
\newcommand\rpipi{r^{+-}}
\newcommand\deltapipi{\delta^{+-}}

\newcommand\ToTpipi{|T^{00}_\mathrm{C}/T^{+-}|}

\newcommand\Tpipiz{T^{+0}}

\newcommand\PpippizEW{P^{\rm EW}}

\newcommand\Tcpizpiz{T_\mathrm{C}^{00}}
\newcommand\Ppizpiz{P^{00}}

\newcommand\deltaAlpha{\Delta\alpha}

\newcommand\Tkk{T^s_{KK}}
\newcommand\Pkk{P^s_{KK}}
\newcommand\Ckk{C^s_{KK}}
\newcommand\Skk{S^s_{KK}}
\newcommand\BRkk{\BR^s_{KK}}

\def\U{{\mathfrak U}}
\def\C{{\mathfrak C}}
\def\T{{\mathfrak T}}

\newcommand\reBoundl{{\cal B}^{00}_{\alpha:{\rm -}}}
\newcommand\reBoundu{{\cal B}^{00}_{\alpha:{\rm +}}}
\newcommand\reBoundlu{{\cal B}^{00}_{\alpha:{\rm \pm}}}
\newcommand\piotwo{{\pi\over 2}}
\newcommand\BrooGLSSl{{\cal B}^{00}_{\rm GLSS ^-}}
\newcommand\BrooGLSSu{{\cal B}^{00}_{\rm GLSS ^+}}
\newcommand\BrooGLSSlu{{\cal B}^{00}_{\rm GLSS ^{\pm}}}

\newcommand\BRpm{{\cal B}_{\rho\pi}^{+-}}
\newcommand\BRmp{{\cal B}_{\rho\pi}^{-+}}

\newcommand\BRpmb{{\cal \kern 0.18em\overline{\kern -0.18em  B}}{}_{\rho\pi}^{+-}}
\newcommand\BRmpb{{\cal \kern 0.18em\overline{\kern -0.18em  B}}{}_{\rho\pi}^{-+}}
\newcommand\BRipm{{\cal B}_{\rho^+\pi^-}}
\newcommand\BRimp{{\cal B}_{\rho^-\pi^+}}
\newcommand\BRipmb{{\cal \kern 0.18em\overline{\kern -0.18em  B}}{}_{\rho^+\pi^-}}
\newcommand\BRimpb{{\cal \kern 0.18em\overline{\kern -0.18em  B}}{}_{\rho^-\pi^+}}
\newcommand\BRall{{\cal B}_{\rho\pi}^{\pm\mp}}
\newcommand\BRpmave{{\cal B}_{\rho\pi}^{\pm}}
\newcommand\BRpz{{\cal B}_{\rho\pi}^{+0}}
\newcommand\BRzp{{\cal B}_{\rho\pi}^{0+}}
\newcommand\BRzz{{\cal B}_{\rho\pi}^{00}}
\newcommand\BRrhoK{{\cal B}_{\rho K}^{-+}}
\newcommand\BRKstrpi{{\cal B}_{K^*\pi}^{+-}}
\newcommand\BRrhoKC{{\cal B}_{\rho K}^{+0}}
\newcommand\BRKstrpiC{{\cal B}_{K^*\pi}^{0+}}
\newcommand\Abar{\kern 0.18em\overline{\kern -0.18em A}{}}
\newcommand\abar{\overline a}
\newcommand\apm{a^{+-}}
\newcommand\amp{a^{-+}}
\newcommand\apmb{\abar^{+-}}
\newcommand\ampb{\abar^{-+}}
\newcommand\Apm{A^{+-}}
\newcommand\Amp{A^{-+}}
\newcommand\Apmb{\Abar^{+-}}
\newcommand\Ampb{\Abar^{-+}}
\newcommand\Atpm{A_\prime^{+-}}
\newcommand\Atmp{A_\prime^{-+}}
\newcommand\Atpmb{\Abar_\prime^{+-}}
\newcommand\Atmpb{\Abar_\prime^{-+}}
\newcommand\Apz{A^{+0}}

\newcommand\Apzb{\Abar^{+0}}
\newcommand\Azp{A^{0+}}
\newcommand\Azpb{\Abar^{0+}}
\newcommand\Azz{A^{00}}
\newcommand\Azzb{\Abar^{00}}

\newcommand\BRPhiKz{{\cal B}_{\Phi K}^{00}}
\newcommand\msRun{{\overline m}_s}
\newcommand\mcRun{{\overline m}_c}

\newcommand\mtRun{{\overline m}_t}
\newcommand\smallfig{10.5cm}
\newcommand\mediumfig{11.6cm}

\newcommand\rfit{{\em R}fit}
\newcommand\erfit{{\em ER}fit}
\newcommand\ckmfitter{{CKMfitter}}

\newcommand\ie{{i.e.}} 
\newcommand\cf{{cf.}} 
\newcommand\eg{{e.g.}} 
\newcommand\via{via} 

\newcommand\Ampli{{\cal A}}
\newcommand\xo{x_0}
\newcommand\xobar{\overline\xo}
\newcommand\xexp{x_{\experi}}
\newcommand\xthe{x_{\tho}}
\newcommand\ymod{y_{\mod}}
\newcommand\ymodopt{y_{\mod}^{\opt}}
\newcommand\ythe{y_{\tho}}
\newcommand\yQCD{y_{\QCD}}
\newcommand\yNP{y_{\rm NP}}

\newcommand\sigexp{\sigma_{\experi}}
\newcommand\sigxo{\sigma_o}
\newcommand\Lik{{\cal L}}
\newcommand\Hatsyst{{\cal L}_{\rm syst}}
\newcommand\expHatsyst{_{\rm exp}{\cal L}_{\rm syst}}
\newcommand\F{{\cal F}}
\newcommand\Ndof{N_{\rm dof}}
\newcommand\VCKM{{V}}
\newcommand\CKM{{\rm CKM}}
\newcommand\Likexp{{\cal L}_{\experi}}
\newcommand\Likthe{{\cal L}_{\tho}}
\newcommand\J{{J}}
\def\a{a}
\newcommand\Mu{\mu}
\newcommand\ChiMinGlob{\chi^2_{\mini ;\ymod}}
\newcommand\Prob{{\cal P}}
\newcommand\ProbCERN{{\rm Prob}}
\newcommand\w{{w}}
\newcommand\Nthe{N_{\tho}}
\newcommand\Nmod{N_{\mod}}
\newcommand\NQCD{N_{\QCD}}

\newcommand\Neff{N_{\experi}^{\rm eff}}
\newcommand\Na{N_{\a}}
\newcommand\Nmu{N_{\Mu}}
\newcommand\Nexp{N_{\experi}}

\newcommand\NNP{N_{\rm NP}}
\newcommand\arhoeta{\{\rhobar,\,\etabar\}}
\newcommand\astastb{\{\sta,\,\stb\}}
\newcommand\etaB{\eta_B}
\newcommand\etaBb{\overline\eta_B}
\newcommand\x{ x}
\newcommand\xt{\x}
\newcommand\xm{\x_{m}}
\newcommand\xmp{\x_{m}^\prime}
\newcommand\xmin{x_{\rm min}}
\newcommand\xmax{x_{\rm max}}
\newcommand\chitwo{\chi^2}
\newcommand\delchitwop{\Delta\chi^{2\prime}}
\newcommand\minchitwo{\chi^2_{\rm min}}
\newcommand\Erfc{{\rm erfc}}
\newcommand\OR{\vee}
\newcommand\AND{\wedge}
\def\Combiner{Combiner}

\def\xmes{x_{\rm exp}}
\def\xmesbar{\langle x_{\rm exp}\rangle}
\def\sigmes{\sigma_{x_{\rm exp}}}
\def\sigmesbar{\sigma_{\langle x_{\rm exp}\rangle}}
\def\Sigmesbar{\Sigma_{\langle x_{\rm exp}\rangle}}
\def\rmsC{\sigma[C]}

\def\rmsRWM{\sigma[{\rm RWM}]}
\def\xtrue{x}
\def\cfull{c_{\rm all}}
\def\DeltaxMesTwo{\Delta x^2_{\rm exp}}
\def\DeltaxMes{\Delta x_{\rm exp}}
\def\meanT{\langle x[C]\rangle}

\begin{document}

\begin{titlepage}

\setcounter{page}{1}

{\small
\begin{flushright} 
        CERN-PH-EP/2004-031\\
        CPT-2004/P.030\\
        LAL 04-21 \\
        LAPP-EXP-2004-01 \\
        LPNHE 2004-01 \\
        hep-ph/0406184\\[0.3cm]
        \today
\end{flushright} 
}

\begin{center}
\vspace{0.5cm}
{\Large\bf\boldmath
\CP Violation and the CKM Matrix: \\[0.2cm]
Assessing the Impact of the Asymmetric \B Factories
}\\[0.5cm] 
\vspace{0.5cm}
{\large The \ckmfitter\  Group} \\
\vspace{0.4cm}
{\large
J.~Charles$^{\,a}$,
A.~H\"ocker$^{\,b}$,  
H.~Lacker$^{\,c}$,
S.~Laplace$^{\,d}$,    \\[0.05cm]
F.R.~Le~Diberder$^{\,b}$,
J.~Malcl\`es$^{\,e}$, 
J.~Ocariz$^{\,e}$, 
M.~Pivk$^{\,f}$, 
L.~Roos$^{\,e}$
} \\
\vspace{0.0cm}
\end{center}

\vspace{0.4cm}

\centerline{\small{\bf Abstract}} 
\vspace{0.1cm}
\noindent
{\small
We present an up-to-date profile of the Cabibbo-Kobayashi-Maskawa
matrix with emphasis on the interpretation of recent 
\CP-violation results from the \B factories. 
For this purpose, we review all relevant experimental and theoretical
inputs  from the contributing domains of electroweak interaction. We
give the ``standard'' determination of the apex of the Unitarity Triangle, 
namely the Wolfenstein parameters $\rhobar$ and  $\etabar$,  by means of
a global CKM fit. The fit is dominated by the precision 
measurement of $\stwob$ by the \B factories. A detailed numerical and 
graphical study of the impact of the results is presented.
We propose to include $\sin2\alpha$ from the recent measurement  of
the time-dependent \CP-violating asymmetries in $\Bz\to\rho^+\rho^-$,
using isospin relations to discriminate the penguin contribution. 
The constraint from $\epe$ is discussed. We study the impact from
the branching fraction measurement of the rare kaon decay
$\Kp\to\pip\nu\nub$, and give an
outlook into the reach of a future measurement of $\KL\to\piz\nu\nub$.
The \B system is investigated in detail. We display the constraint 
on $2\beta+\gamma$ and $\gamma$ from $\Bz\to D^{(*)\pm}\pi^\mp$ and 
$\Bp\to D^{(*)0}\Kp$ decays, respectively. 
A significant part of this paper is dedicated to the understanding
of  the dynamics of \B decays into $\pi\pi$, $K\pi$,
$\rho\pi$,  $\rho\rho$ and modes related to these by flavor symmetry. 
Various phenomenological approaches and theoretical frameworks are 
discussed.
We find a remarkable agreement of the $\pi\pi$ and $K\pi$ data with the 
other constraints in the unitarity plane when the hadronic
matrix elements are calculated within QCD Factorization, where we
apply a  conservative treatment of the theoretical uncertainties. 
A global fit of QCD Factorization to all $\pi\pi$ and $K\pi$ data  
leads to precise predictions of the related observables. However sizable
phenomenological power corrections are preferred.
Using an isospin-based phenomenological parameterization, we analyze
separately the $\B\to K\pi$ decays, and the impact of electroweak  
penguins in response to recent discussions. We find that the present 
data are not sufficiently precise to constrain either electroweak parameters
or hadronic amplitude ratios.  We do not observe any unambiguous
sign of New Physics, whereas there is some evidence for potentially
large non-perturbative rescattering  effects.
Finally we use a model-independent description of a large class  of
New Physics effects in both $\Bz\Bzb$ mixing and $B$ decays, namely  in
the $b\to d$ and $b\to s$ gluonic penguin amplitudes, to perform a new
numerical analysis. Significant non-standard  corrections
cannot be excluded yet, however Standard Model solutions  are favored
in most cases.
In the appendix to this paper we propose a frequentist method to extract 
a confidence level on $\dms$ from the experimental information on $\Bs\Bsb$ 
oscillation. In addition we describe a novel approach to combine potentially
inconsistent measurements.
All results reported in this paper have been obtained with the numerical 
analysis package \ckmfitter,  featuring the frequentist statistical 
approach \rfit.

}
\vspace{\stretch{1}}

\hrule\vspace{0.1cm}
{\small\noindent http://ckmfitter.in2p3.fr
      \hfill
      http://www.slac.stanford.edu/xorg/ckmfitter}

\thispagestyle{empty}

\newpage

\begin{center}
\noindent
{\small \em $^{a}$Centre de Physique Th\'eorique, \\
                   Campus de Luminy, 
                   Case 907, F-13288 Marseille Cedex 9, France\\
                   (UMR 6207 du CNRS associ\'ee aux
                   Universit\'es d'Aix-Marseille I et II \\et
                   Universit\'e du Sud Toulon-Var; laboratoire
                   affili\'e \`a la FRUMAM-FR2291) \\
                {e-mail: charles@cpt.univ-mrs.fr}} \\[0.5cm]
{\small \em $^{b}$Laboratoire de l'Acc\'el\'erateur Lin\'eaire,\\
                   B\^at. 200 BP34 F-91898 Orsay, France\\
                   (UMR 8607 du CNRS-IN2P3 associ\'ee \`a
                   l'Universit\'e Paris XI) \\
                {e-mail: hoecker@lal.in2p3.fr, diberder@in2p3.fr}}\\[0.5cm]
{\small \em $^{c}$Technische Universit\"at Dresden, \\
                   Institut f\"ur Kern- und Teilchenphysik, 
                   D-01062 Dresden, Germany \\
                {e-mail: h.lacker@physik.tu-dresden.de} }\\[0.5cm]
{\small \em $^{d}$Laboratoire d'Annecy-Le-Vieux de Physique des
                   Particules \\
                   9 Chemin de Bellevue, BP 110, F-74941
                   Annecy-le-Vieux Cedex, France\\
                   (UMR 5814 du CNRS-IN2P3 associ\'ee \`a
                   l'Universit\'e de Savoie) \\
                {e-mail: laplace@lapp.in2p3.fr}}\\[0.5cm]
{\small \em $^{e}$Laboratoire de Physique Nucl\'eaire et de Hautes Energies \\
                   4 place Jussieu, F-75252 Paris Cedex 05, France\\
                   (UMR 7585 du CNRS-IN2P3 associ\'ee aux
                   Universit\'es Paris VI et VII) \\
                {e-mail: malcles@lpnhep.in2p3.fr, 
                         ocariz@in2p3.fr, lroos@lpnhep.in2p3.fr}} \\[0.5cm]
{\small \em $^{f}$CERN, PH Department \\ CH-1211 Geneva 23, Switzerland \\
                {e-mail: muriel.pivk@cern.ch}}
\end{center}

\thispagestyle{empty}

\end{titlepage}

\vfill
%
%
 \newpage
{\normalsize
 \tableofcontents 
}

 \clearpage
%
%
 \newpage\part{Introduction}\setcounter{section}{0}
\markboth{\textsc{Part I -- Introduction}}{\textsc{Part I -- Introduction}}
\label{sec:introduction}

Within the Standard Model (SM), \CP  violation (CPV) is generated by a
single non-vanishing phase in the unitary Cabibbo--Kobayashi--Maskawa
(CKM) quark mixing matrix $\VCKM$~\cite{cabibbo,kmmatrix}. A useful
parameterization of $\VCKM$ follows from the observation
that its elements exhibit a hierarchy in terms of the
parameter  $\lambda\simeq|V_{us}|$~\cite{wolfenstein,buras}.
Other parameters are $A$, $\rhobar$
and $\etabar$, where \CP  violation requires $\etabar\ne0$. The
parameters  $\lambda$ and $A$ are  obtained from
measurements of semileptonic decay  rates of $K$ mesons, and of $B$ meson
decays involving beauty-to-charm  transitions, respectively. The constraints on
$\rhobar$ and $\etabar$
are conveniently displayed in the complex plane where they
determine the apex of the Unitarity Triangle\footnote
{
        Throughout this paper, we adopt the $\alpha$, $\beta$, $\gamma$
        convention for the angles of the Unitarity Triangle. They are
        related to the $\phi_1$, $\phi_2$, $\phi_3$ ``historical''
        convention~\cite{bigietal} as
        $\alpha=\phi_2$, $\beta=\phi_1$ and $\gamma=\phi_3$. Angles are
        given in units of degrees.
} (UT),
which is a graphical representation of the unitarity relation
between the first and the third column of the CKM matrix.
For example, semileptonic $B$ decays yielding $|V_{ub}|$,
predictions of $\BzBzb$
oscillation and of indirect \CP  violation in the neutral
kaon sector depend on $\rhobar$, $\etabar$. However the
understanding of this dependence is limited by theoretical
uncertainties, which are mainly due to long distance QCD.
In the era of the \B factories, a large number of measurements
has appeared that are related to the CKM phase. The most famous
of them is the measurement of the \CP-violation parameter
$\stb$ in $b\to c\bar c s$ transitions, which is theoretically clean.
Other modes are
sensitive to the angles $\alpha$ and $\gamma$ of the UT, where in many
cases one has to deal with interfering amplitudes with different
\CP-violating phases, complicating the extraction of the
CKM-related parameters.
\vs
A focus of this work is the phenomenological interpretation of \B-physics
results. The spectacular performance of the first five
years of the asymmetric-energy \B factories, PEP-II and KEK$\,$B, and their
experiments \babar and Belle, with published
results on  up to $270\invfb$ integrated
luminosity (combined), has produced an avalanche of publications,
many of which are related to \CP violation. In spite of the difficulties
due to small branching fractions and/or hadronic uncertainties, the
goal of overconstraining the UT from tree-level-dominated \B decays seems
achievable even if the precision may turn out insufficient to reveal
a failure of the SM. Since tree decays are not expected to lead to large
inconsistencies with the SM, more and more experimental and theoretical
effort goes into the determination of UT angles and/or other parameters
from \B decays dominated by penguin-type diagrams, the most prominent of
which are $b\to s\gamma^{(*)}$ and $b\to s\bar s s$ (\eg,
$\Bz\to\phi\Kz$). A number of other decay modes with net strangeness in 
the final state, which are now being studied, may reveal specific signs 
of physics beyond the SM through unexpected \CP violation or enhanced 
branching fractions.
\vs
The CKM analysis performed in this paper is threefold:
the first goal of the global CKM fit is to probe the validity
of the SM, that is to quantify the agreement between the SM and the
experimental information; if this is confirmed, one secondly enters
the metrology phase where allowed ranges for the CKM matrix elements and
related quantities are determined, assuming explicitly the SM
to be correct; finally, within an extended theoretical framework, one
may search for signals of New Physics and constrain parameters of specific
New Physics scenarios.
\vs
Analyzing data in a well defined theoretical scheme ceases to be
a straightforward task when one moves away from Gaussian statistics.
This is the case for the theoretically limited precision on the SM
predictions of the neutral $K$ and $B$ mixing observables and, to a
lesser extent, for the semileptonic decay rates
of $B$ decays to charmed and charmless final states. Also the
interpretation of results on \CP  violation in terms of the UT
angles often invokes unknown phases occurring in absorptive parts
of non-leptonic transitions. The statistical approach \rfit\  developed
and described in detail in Ref.~\cite{CKMfitter} treats these uncertainties
in a frequentist framework, which allows one to determine
confidence levels. The ensemble of the statistical analyses
reported here is realized with the use of the program package
\ckmfitter\footnote
{
        \ckmfitter\  is a framework package that hosts several
        statistical approaches to a global CKM fit and the
        interpretation of \CP-violation results.
        It is available to the public~\cite{CKMweb}. Please
        contact the authors for more information.
}. \vs This paper is the second edition~\cite{CKMfitter} of our effort
to collect all significant information on the CKM matrix and to combine
it in a global CKM fit~\cite{Achille1,CernCkmWS,otherCkm}. All figures
given in this document as well as partially updated results can be
found on the \ckmfitter\  web site~\cite{CKMweb}. \vs The paper is
organized as follows. Part~\ref{sec:introduction} provides a brief
introduction of the CKM matrix and mainly serves to define the
conventions adopted in this paper. We review in
Part~\ref{sec:statistics} the statistical approach and the analysis
tools implemented in \ckmfitter, the understanding of which is
necessary for an adequate interpretation of the results derived in this
work. Part~\ref{sec:standardFit} first defines the observables that are
used as input in the so-called ``standard CKM fit'', which is defined
as the global CKM fit that includes only those observables, which
provide competitive constraints and of which the SM prediction can be
considered to be quantitatively under control. We put emphasis on the
discussion of the theoretical uncertainties. This introduction is
followed by a compendium of numerical and graphical results of the
standard CKM fit for all parameters and observables of the electroweak
sector that significantly depend on the CKM matrix. Beginning with
Part~\ref{sec:kaons} we perform rather detailed investigations of
specific subsystems, related to \CP violation in the quark sector and
to the Unitarity Triangle, with emphasis on the discussion of
observables not used in the standard CKM fit. We study direct \CP
violation in the kaon system and specifically derive constraints on the
non-perturbative bag parameters. We discuss the impact of the
measurement of rare kaon decays and give an outlook into the future
where we attempt to quantify the expected uncertainties.
Part~\ref{sec:gamma} displays the constraints related to the Unitarity
Triangle angle $\gamma$ from the time-dependent analysis of $\Bz\to
D^{(*)\pm}\pi^\mp$ decays and the Dalitz analysis of $\Bp\to\Dz\Kp$.
Part~\ref{sec:charmlessBDecays} describes in detail the analysis of
charmless \B decays to $\pi\pi$, $K\pi$, $\rho\pi$ and $\rho\rho$,
which, besides the global CKM fit, represents a central pillar of this
work. We discuss constraints on the \CP-violating CKM phase using
various phenomenological and theoretical approaches based on flavor
symmetries and factorization. In Part~\ref{sec:newPhysics} we use a
model-independent parameterization of a large class of New Physics
effects in both $\Bz\Bzb$ mixing and $B$ decays, namely in the $b\to d$
and $b\to s$ gluonic penguin amplitudes, to perform a tentative
numerical analysis.
In the appendix to this work we describe the frequentist treatment of
the measurement of $\BszBszb$ oscillation incorporated in our CKM
fit, and we propose a novel method to handle the problem
of (apparently) inconsistent measurements.
%
%
\section{The CKM Matrix}

Invariance under local gauge transformation prevents the bare
masses of leptons and quarks to appear in the
$SU(3)\times SU(2)\times U(1)$ Lagrange density of the SM.
Instead, the spontaneous breakdown of electroweak symmetry
dynamically generates masses for the fermions due to the Yukawa
coupling of the fermion fields to the Higgs
doublet. Since the latter has a non-vanishing vacuum
expectation value, the Yukawa couplings $g$ give rise to
the $3\times3$ mass matrices
\beq
\label{eq:mumd}
   M_i=\frac{v g_i}{\sqrt{2}}~,
\eeq
with $i=u(d)$ for up(down)-type quarks and $i=e$ for the massive leptons.
To move from the basis of the flavor (electroweak) eigenstates to the basis
of the mass eigenstates, one performs the transformation
\beq
        U_{u(d,e)} M_{u(d,e)} \tilde U_{u(d,e)}^\dag
        = {\rm diag}\left(m_{u(d,e)},
                          m_{c(s,\mu)},
                          m_{t(b,\tau)}
                     \right)~,
\eeq
where $U_i$ and $\tilde U_i$ are unitary complex rotation matrices
and the masses $m_i$ are real.
The neutral-current part of the Lagrange density in
the basis of the mass-eigenstates remains unchanged (\ie, there are
no flavor-changing neutral currents present at tree level),
whereas the charged current part of the quark sector is
modified by the product of the up-type and down-type quark
mass matrices,
\beq
        \VCKM = U_u U_d^\dag~,
\eeq
which is the CKM mixing matrix. By convention, $\VCKM$ operates on the
$-1/3$ charged down-type quark mass eigenstates
\beq
\label{eq:ckm1}
\VCKM = \left(
        \begin{array}{ccc}
        V_{ud} & V_{us} & V_{ub} \\
        V_{cd} & V_{cs} & V_{cb} \\
        V_{td} & V_{ts} & V_{tb} \\
        \end{array}
        \right)
\eeq
and, being the product of unitary matrices, $\VCKM$ itself is unitary
\beq
\label{eq:unitarity}
        \VCKM \VCKM^\dag={\rm I}~.
\eeq
There exists a hierarchy between the elements of $\VCKM$ both for
their value (the diagonal elements dominate) and their errors
(since they dominate, they are better known). The unitarity and
the phase arbitrariness of fields reduce the initial nine
complex elements of $\VCKM$ to three real numbers
and one phase, where the latter accounts for \CP  violation.
It is therefore interesting to over-constrain $\VCKM$
since deviations from unitarity would reveal the existence of new
generation(s) or new couplings.
\vs
The charged current couplings among left-handed quark fields
are proportional to the elements of $\VCKM$. For right-handed
quarks, there exist no $W$ boson interaction in the SM and
the $Z$, photon and gluon couplings are flavor diagonal.
For left-handed leptons the analysis proceeds similarly
to the quarks with the notable difference that, since
the neutrinos are (almost) massless, one can choose to
make the same unitary transformation on the left-handed
charged leptons and neutrinos so that the analog of $\VCKM$
in the lepton sector becomes the unit matrix.
\vs
There are many ways of parameterizing the CKM matrix in terms of four
parameters. The following section summarizes the
most popular representations.

\subsection{The Standard Parameterization}

The Standard Parameterization of $\VCKM$ was proposed by
Chau and Keung~\cite{chau} and is advocated by the Particle Data
Group (PDG)~\cite{PDG}. It is obtained by the product
of three (complex) rotation matrices, where the rotations are
characterized by the Euler angles $\theta_{12}$, $\theta_{13}$
and $\theta_{23}$, which are the mixing angles
between the generations, and one overall phase $\delta$~\footnote
{
        This phase $\delta$ is a \CP-violating phase; it should not be
        confused with the \CP-conserving hadronic phases that will
        be introduced later with the same symbol.
}
\beq
\label{eq:ckmPdg}
\VCKM = \left(
        \begin{array}{ccc}
        c_{12}c_{13}
                &    s_{12}c_{13}
                        &   s_{13}e^{-i\delta}  \\
        -s_{12}c_{23}-c_{12}s_{23}s_{13}e^{i\delta}
                &  c_{12}c_{23}-s_{12}s_{23}s_{13}e^{i\delta}
                        & s_{23}c_{13} \\
        s_{12}s_{23}-c_{12}c_{23}s_{13}e^{i\delta}
                &  -c_{12}s_{23}-s_{12}c_{23}s_{13}e^{i\delta}
                        & c_{23}c_{13}
        \end{array}
        \right)
\eeq
where $c_{ij}=\cos\theta_{ij}$, $s_{ij}=\sin\theta_{ij}$ for
$i<j=1,2,3$. This parameterization strictly satisfies the unitarity
relation~(\ref{eq:unitarity}).

\subsection{The Wolfenstein Parameterization}

Following the observation of a hierarchy between the different
matrix elements, Wolfenstein~\cite{wolfenstein} proposed an
expansion of the CKM matrix in terms of the four parameters $\lambda$,
$A$, $\rho$ and $\eta$ ($\lambda\simeq|V_{us}|\sim 0.22$ being the expansion
parameter), which is widely used in contemporary literature, and
which is the parameterization employed in this work and in \ckmfitter.
We use in the following as definitions to {\em all orders} in
$\lambda$~\cite{buras}
\beqn
\label{eq:burasdef}
        s_{12}             &\equiv& \lambda~,\nonumber \\
        s_{23}             &\equiv& A\lambda^2~, \\
        s_{13}e^{-i\delta} &\equiv& A\lambda^3(\rho -i\eta)~,\nonumber
\eeqn
inserted into the standard parameterization~(\ref{eq:ckmPdg}), so that
unitarity of the CKM matrix is achieved to all orders\footnote
{
        The Taylor expansion of Eqs.~(\ref{eq:burasdef}), inserted
        into (\ref{eq:ckmPdg}), up to order $\mathcal{O}(\lambda^9)$
        reads
        \beqns
        \label{eq:ckmWolf}
       V_{ud} &=&
               1-\frac{1}{2}\lambda^2-\frac{1}{8}\lambda^4
               -\frac{1}{16}\lambda^6\left(1+8A^2(\rho^2+\eta^2)\right)
               -\frac{1}{128}\lambda^8
                \left(5-32A^2(\rho^2+\eta^2\right)~, \nonumber\\
       V_{us} &=&
               \lambda - \frac{1}{2}A^2\lambda^7(\rho^2+\eta^2)~, \nonumber\\
       V_{ub} &=&
               A\lambda^3(\rho-i\eta)~, \nonumber\\
       V_{cd} &=&
               -\lambda + \frac{1}{2}A^2\lambda^5
               \left(1 - 2(\rho+i\eta)\right)
               + \frac{1}{2}A^2\lambda^7(\rho+i\eta)~, \nonumber\\
       V_{cs} &=&
               1-\frac{1}{2}\lambda^2-\frac{1}{8}\lambda^4(1+4A^2)
               - \frac{1}{16}\lambda^6
               \left(1 - 4A^2 + 16A^2(\rho+i\eta)\right)
               - \frac{1}{128}\lambda^8\left(5-8A^2+16A^4\right)~, \nonumber\\
       V_{cb} &=&
               A\lambda^2 - \frac{1}{2}A^3\lambda^8
                            \left(\rho^2+\eta^2\right)~,
               \nonumber\\
       V_{td} &=&
               A\lambda^3 \left(1 - \rho - i\eta\right)
               + \frac{1}{2}A\lambda^5(\rho+i\eta)
               + \frac{1}{8}A\lambda^7(1+4A^2)(\rho+i\eta)~, \nonumber\\
       V_{ts} &=&
               -A\lambda^2 + \frac{1}{2}A\lambda^4
               \left(1 - 2(\rho+i\eta)\right)
               + \frac{1}{8}A\lambda^6
               + \frac{1}{16}A\lambda^8\left(1 + 8A^2(\rho+i\eta)\right)~,
               \nonumber\\
       V_{tb} &=&
               1 - \frac{1}{2}A^2\lambda^4
               - \frac{1}{2}A^2\lambda^6
               \left(\rho^2+\eta^2\right) - \frac{1}{8}A^4\lambda^8~.
                \nonumber
        \eeqns
}.

\subsection{The Jarlskog Invariant}

It was shown by Jarlskog~\cite{jarlskog} that the
determinant of the commutator of the up-type and
down-type unitary mass matrices~(\ref{eq:mumd}) reads
\beq
\label{eq:commu}
        {\rm det}[M_u,M_d] = -2 i F_u F_d J~,
\eeq
with $F_{u(d)} = (m_{t(b)}-m_{c(s)})(m_{t(b)}-m_{u(d)})
(m_{c(s)}-m_{u(d)})/m_{t(b)}^3$.
The phase-convention independent measurement of \CP  violation,
$J$, is given by
\beq
\label{eq:jarlskog}
{\rm Im}\left[V_{ij}V_{kl}V_{il}^*V_{kj}^*\right]
        = J \!\sum_{m,n=1}^3\!\! \varepsilon_{ikm}\varepsilon_{jln}~,
\eeq
where $V_{ij}$ are the CKM matrix elements and $\varepsilon_{ikm}$
is the total antisymmetric tensor. One
representation of Eq.~(\ref{eq:jarlskog}) reads, for
instance, $J={\rm Im}[V_{ud}V_{cs}V_{us}^*V_{cd}^*]$.
A non-vanishing CKM phase and hence \CP  violation
necessarily requires $J\ne0$.
The Jarlskog parameter expressed in the Standard
Parameterization~(\ref{eq:ckmPdg}) reads
\beq
J = c_{12}c_{23}c_{13}^2s_{12}s_{23}s_{13}{\rm sin}\delta~,
\eeq
and, using the Wolfenstein parameterization, one finds
\beq
        J = A^2\lambda^6\eta\left(1-\lambda^2/2\right)
                        + \mathcal{O}(\lambda^{10})
         \sim 10^{-5}~.\nonumber
\eeq
The empirical value of $J$ is small compared to its
mathematical maximum of $1/(6\sqrt{3})\simeq 0.1$ showing
that \CP  violation is suppressed as a consequence of the strong
hierarchy exhibited by the CKM matrix elements.
Remarkably, to account for \CP violation (see Eq.~(\ref{eq:commu})) 
requires not only a non-zero $J$ but also a non-degenerated 
quark-mass hierarchy. Equal masses for at least two generations 
of up-type or down-type quarks would eliminate the CKM phase.
\vs
Phase convention invariance of the $\VCKM$-transformed quark
wave functions is a requirement for physically meaningful
quantities. Such invariants are the moduli $|V_{ij}|^2$ and
the quadri-products $V_{ij}V_{kl}V_{il}^*V_{kj}^*$ (\cf\
the Jarlskog invariant $J$). Non-trivial higher order
invariants can be reformulated as functions of moduli
and quadri-products~(see, \eg, Ref.~\cite{CPV-TheBook}).
Indeed, Eq.~(\ref{eq:jarlskog}) expresses the fact that, owing to
the orthogonality of any pair of different rows or columns
of $\VCKM$, the imaginary parts of all quadri-products are
equal up to their sign. We will use phase-invariant representations
and formulae throughout this paper.

\section{The Unitarity Triangle}

The allowed region in the $\rho$ and $\eta$ space can be elegantly
displayed by means of the Unitarity Triangle (UT) described
by the {\em rescaled} unitarity relation between the first
and the third column of the CKM matrix (\ie, corresponding to
the \B\ meson system)
\beq
\label{eq:utriangle}
   \frac{V_{ud}V_{ub}^*}{V_{cd}V_{cb}^*}
        + \frac{V_{cd}V_{cb}^*}{V_{cd}V_{cb}^*}
        + \frac{V_{td}V_{tb}^*}{V_{cd}V_{cb}^*} = 0~.
\eeq
Note that twice the area of the {\em non-rescaled}
UT corresponds to the Jarlskog parameter $J$. This identity
provides a geometrical interpretation of the phase convention
invariance of $J$: a rotation of the CKM matrix rotates the
UT accordingly while leaving its area, and hence $J$ is invariant.
It is the remarkable property of the UT in the $B$ system
that its three sides are governed by the same power of $\lambda$
and $A$ (so that the sides of the rescaled UT~(\ref{eq:utriangle})
are of order one), which predicts large \CP-violating asymmetries
\begin{figure}[t]
  \epsfxsize7.7cm
  \centerline{\epsffile{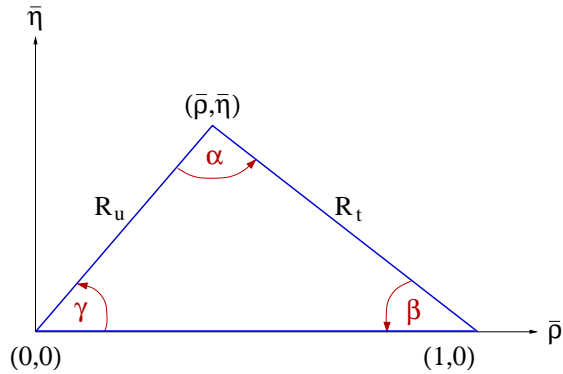}}
  \vspace{-0.0cm}
  \caption[.]{\label{fig:utriangle}\em
        The rescaled Unitarity Triangle in the Wolfenstein
        parameterization.}
\end{figure}
in the $B$ sector. As a comparison, the corresponding
UT for the kaon sector is heavily flattened
\beq
   0 = \frac{V_{ud}V_{us}^*}{V_{cd}V_{cs}^*}
        + \frac{V_{cd}V_{cs}^*}{V_{cd}V_{cs}^*}
        + \frac{V_{td}V_{ts}^*}{V_{cd}V_{cs}^*}
        \sim
        O\left(\frac{\lambda}{\lambda}\right)
        + O\left(1\right)
        + O\left(\frac{A^2\lambda^5}{\lambda}\right)~,
\eeq
exhibiting small \CP  asymmetries.
\begin{figure}[t]
  \epsfxsize\mediumfig
  \centerline{\epsffile{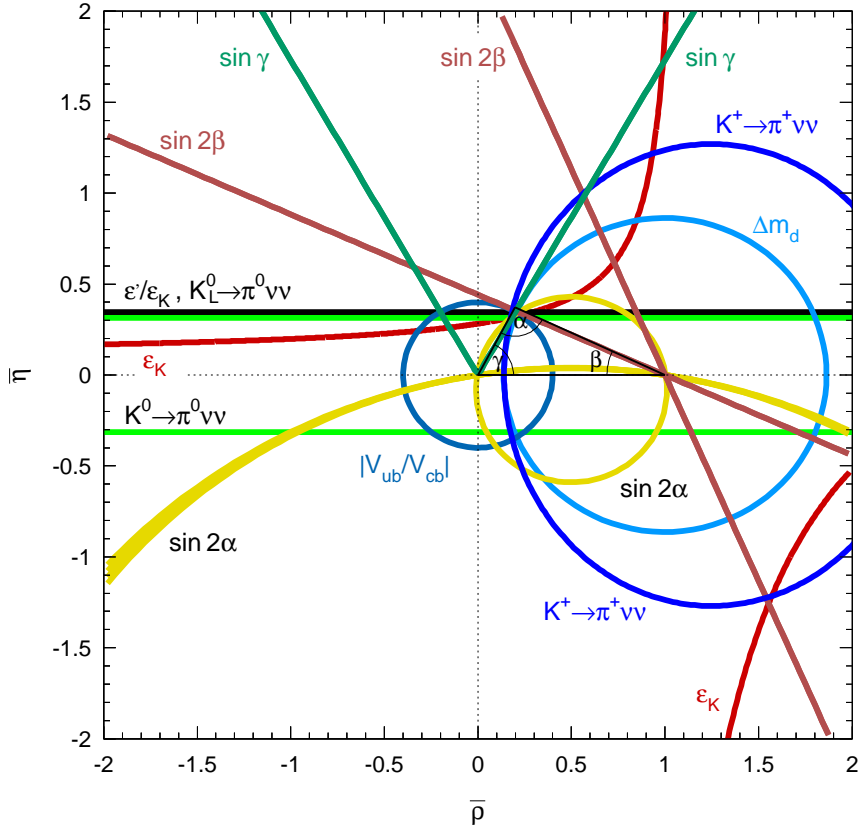}}
  \vspace{-0.5cm}
  \caption[.]{\label{fig:rhoetaall}\em
        Constraints in the unitarity plane for the most relevant
        observables. The theoretical parameters used correspond to some
        ``standard'' set chosen to reproduce compatibility between
        the observables. }
\end{figure}
The UT~(\ref{eq:utriangle}) is sketched in Fig.~\ref{fig:utriangle}
in the complex $\rhoeta$ plane, where the apex is given by the following
phase-convention independent
definition, to all orders in $\lambda$~\cite{buras},
\beqn
\label{eq:rhoetabar}
        \rhobar + i\etabar
        \;\equiv\;-\frac{V_{ud}V_{ub}^*}{V_{cd}V_{cb}^*}~,
\eeqn
of which the inverse reads to all orders\footnote
{
        Expanding Eq.~(\ref{eq:rhoetabar}) in $\lambda$ gives~\cite{buras}
        \beqn
              \rhobar &=&
        \rho
        - \frac{1}{2}\rho\lambda^{2}
        + \left( \frac{1}{2}{A}^{2}\rho
                 - \frac{1}{8}\rho
                 - A^{2}\left(\rho^2 - \eta^{2}\right)
          \right)\lambda^4
        + \mathcal{O}(\lambda^6)~, \\
      \etabar &=&
        \eta
        - \frac{1}{2}\eta\lambda^{2}
        + \left( \frac{1}{2} A^2\eta
                 - \frac{1}{8}\eta
                 -2  A^2\rho\eta
           \right)\lambda^{4}
        + \mathcal{O}(\lambda^6)~.
        \eeqn
}
\beq
\label{eq:rhoetabarinv}
        \rho + i\eta
        \;=\; \frac{\sqrt{1-A^2\lambda^4}(\rhobar+i\etabar)}
                   {\sqrt{1-\lambda^2}\left[1 -
                   A^2\lambda^4(\rhobar+i\etabar)\right]}~.
\eeq
Equation~(\ref{eq:rhoetabarinv}) is the definition used in \ckmfitter.
The sides $R_u$ and $R_t$ of the UT (the third side being
normalized to unity) read to all orders
\beqn
\label{eq:ru}
R_u &=&
        \left|\frac{V_{ud}V_{ub}^*}{V_{cd}V_{cb}^*} \right|
                \;=\; \sqrt{\rhobar^2+\etabar^2}~, \\
\label{eq:rt}
R_t &=&
        \left|\frac{V_{td}V_{tb}^*}{V_{cd}V_{cb}^*}\right|
                \;=\; \sqrt{(1-\rhobar)^2+\etabar^2}~.
\eeqn
The three angles, $\alpha,~\beta,~\gamma$, of the UT are defined by
\beq
\label{eq:utdefinitions}
\alpha = \arg\left[ - \frac{V_{td}V_{tb}^*}{V_{ud}V_{ub}^*} \right]
~,\hspace{0.5cm}
\beta  = \arg\left[ - \frac{V_{cd}V_{cb}^*}{V_{td}V_{tb}^*} \right]
~,\hspace{0.5cm}
\gamma = \arg\left[ - \frac{V_{ud}V_{ub}^*}{V_{cd}V_{cb}^*} \right]~,
\eeq
and the CKM phase in the Standard Parameterization~(\ref{eq:ckmPdg})
reads $\delta = \gamma+A^2\lambda^4\eta+\mathcal{O}(\lambda^6)$.
The relations between the angles and the $\rhobar$, $\etabar$
coordinates, again to all orders in $\lambda$, are given by
\beqn
&&\cos\gamma=\rhobar/R_u\ ,\ \ \sin\gamma=\etabar/R_u~,\\
&&\cos\beta=(1-\rhobar)/R_t\ ,\ \ \sin\beta=\etabar/R_t~,\\
&&\alpha=\pi-\beta-\gamma~.
\eeqn
A graphical compilation of the most relevant present and future
constraints (without errors) is displayed in Fig.~\ref{fig:rhoetaall}.
Some ``standard'' values for the theoretical parameters
are used for this exercise in order to reproduce compatibility
between the constraints.
\vs
Over-constraining the unitary CKM matrix aims at validating
the three-generation SM. The interpretation of these
constraints requires a robust statistical framework which protects
against misleading conclusions. The following part describes
the statistical approach applied for the analysis reported
in this work.

%
%
 \newpage\part{The Statistical Approach \rfit}\setcounter{section}{0}
\markboth{\textsc{Part II -- The Statistical Approach}}
         {\textsc{Part II -- The Statistical Approach}}
\label{sec:statistics}

The statistical analysis performed in this paper is entirely
based on the {\em frequentist} approach \rfit\  described in detail 
in Ref.~\cite{CKMfitter} and recalled below.
\vs
We consider an analysis involving a set of $\Nexp$ measurements 
collectively denoted by $\xexp=\{\xexp(1), \dots, $ $ \xexp(\Nexp)\}$, 
described by a set of corresponding theoretical expressions
$\xthe=\{\xthe(1),$ $\dots,\xthe(\Nexp)\}$. 
The theoretical expressions $\xthe$ are functions of a set 
of $\Nmod$ parameters $\ymod=\{\ymod(1),\dots,\ymod(\Nmod)\}$.
Their precise definition is irrelevant for the present discussion
(\cf\   Section~\ref{sec:standardFit}.\ref{sec:fitInputs} for details)
besides the fact that:
\begin{itemize}

\item{} a subset of $\Nthe$ parameters within the $\ymod$ 
        set are fundamental and free parameters 
        of the theory (\ie, the four CKM unknowns in the 
        SM, the top quark mass, etc.); these are denoted 
        $\ythe$, where $\ythe=\{\ythe(1),\dots,\ythe(\Nthe)\}$.

\item{} the remaining $\NQCD=\Nmod-\Nthe$ parameters are
        due to our present inability to compute precisely 
        strong interaction quantities (\eg, $\fbd$, $B_d$, etc.), and are 
        denoted $\yQCD$, where $\yQCD=\{\yQCD(1),\dots,\yQCD(\NQCD)\}$.

\end{itemize}
There are three different goals of the global CKM analysis:
\begin{enumerate}
\item   within the SM,
        to quantify the agreement between data and the theory, as a whole.

\item   within the SM,
        to achieve the best estimate of the $\ythe$ parameters:
        that is to say to perform a careful metrology of the theoretical 
        parameters.

\item   within an extended theoretical framework, \eg\ Supersymmetry,
        to search for specific signs of New Physics by quantifying the
        agreement between data and the extended theory, and by pinning 
        down additional fundamental and free parameters of the extended
        theory.
\end{enumerate}
These goals imply three distinct statistical treatments all of which
rely on a likelihood function meant to gauge the agreement 
between data and theory. 

\section{The Likelihood Function}
\label{TheLikelihoodFunction}

We adopt a $\chi^2$-like notation and denote
\beq
\label{eq_chi2Function}
   \chi^2(\ymod)\equiv-2\ln(\Lik(\ymod))~,
\eeq
where the likelihood function, $\Lik$ (defined below),
is the product of two contributions: 
\beq
\label{eq_likFunction}
   \Lik(\ymod)=\Likexp(\xexp-\xthe(\ymod))\cdot \Likthe(\yQCD)~.
\eeq
The first term, the experimental likelihood $\Likexp$, 
measures the agreement between $\xexp$ and $\xthe$,
while the second term, the theoretical likelihood $\Likthe$, 
expresses our present knowledge of the $\yQCD$ parameters.
\vs
It has to be recognized from the outset that the $\chi^2$ of 
Eq.~(\ref{eq_chi2Function}) is a quantity that can be misleading. 
In general, using "$\ProbCERN$" the well known routine from the 
CERN library, one {\em cannot} infer a confidence level (CL) from 
the above $\chi^2$ value using
\begin{eqnarray}
\label{eq_probcern}
\CL      &=&     \ProbCERN(\chi^2(\ymod),\Ndof)~, \\
         &=&     \frac{1}{\sqrt{2^{\Ndof}}\Gamma({\Ndof}/2)}
                 \intl_{\chi^2(\ymod)}^\infty 
                 \hmmm e^{-t/2}t^{\Ndof/2-1}\, dt~.
\end{eqnarray}
This is because neither $\Likexp$ nor $\Likthe$
(they are further discussed in the sections below)
are built from purely Gaussian measurements.
\begin{itemize}

\item{} In most cases $\Likexp$ should handle experimental 
        systematics, and, in some instance,
        it has to account for inconsistent measurements.

\item{} In practice, $\Likthe$ relies on hard to quantify educated 
        guesswork, akin to experimental systematic errors,
        but in most cases even less well defined.

\end{itemize}
The first limitation is not specific to the present analysis
and is not the main source of concern. The second limitation is 
more challenging: its impact on the analysis is particularly 
strong with the data presently available. The statistical treatment 
\rfit\   is designed to cope with both of the above limitations.
Notwithstanding its attractive features,
the \rfit\  scheme does not offer a treatment of the problem at hand
free from any assumption: 
an ill-defined problem cannot be dealt with rigorously.
However the \rfit\   scheme extracts the most
out of simple and clear-cut {\it a priori} assumptions.

\subsection{The Experimental Likelihood}
\label{sec:TheExperimentalLikelihood}

The experimental component of the likelihood 
is given by the product
\beq
\label{eq_likExp}
\Likexp(\xexp-\xthe(\ymod))=\prod_{i,j=1}^{\Nexp}\Likexp(i,j)~,
\eeq
where the $\Nexp$ individual likelihood components $\Likexp(i,j)$ 
account for measurements that may be independent 
or not.
\vs\noindent
Ideally, the {\bf likelihood components} 
$\Likexp(i)$ are independent Gaussians
\beq
\label{eq_thegaussian}
\Likexp(i)= 
        {1\over\sqrt{2\pi}\sigexp(i)}\,
        {\rm exp}\left[-{1\over 2}
        \left({\xexp(i)-\xthe(i)\over\sigexp(i)}\right)^2\right]~,
\eeq
each with a standard deviation given by the experimental 
statistical uncertainty $\sigexp(i)$ of the $i^{\rm th}$ 
measurement. However, in practice, one has to deal with 
correlated measurements and with additional experimental 
and theoretical systematic uncertainties. 
\bei
\item   {\em Experimental systematics}
        are assumed to take the form of a possible biasing offset
        the measurement could be corrected, were it known.
        Their precise treatment is discussed in 
        Section~\ref{sec:statistics}.~\ref{sec:likelihoodsAndSysErrors}. 
        In practice,
        these systematics are usually added in quadrature to the 
        statistical errors.     

\item   {\em Theoretical systematics}, when they imply small 
        effects, are treated as the experimental ones. However
        because most theoretical systematics imply large effects
        and affect in a non-linear way the $\xthe$ prediction, 
        most of them are dealt with through the theoretical likelihood 
        component $\Likthe$
        (\cf\   Section~\ref{sec:statistics}.~\ref{sec:theoreticalLikelihood}). 
\eei

\noindent
{\bf Identical observables and consistency:} 
when several measurements refer to the same observable
(\eg, various measurements of $\dmd$) they have to be 
consistent, 
independently of the theoretical framework used for the analysis. 
Similarly, when several measurements refer to different observables
that are linked to the same $\ythe$ parameter, \eg, 
$\Vud$ and $\Vus$, or determinations of $\Vub$ stemming from 
different observables, or measurements of $\stb$ obtained from 
similar $B$ decays, one may {\it decide} to overrule possible 
disagreement by requiring the measurements to be consistent.
By doing so, one is deliberately blinding oneself to possible 
New Physics effects, which may have revealed themselves otherwise.
Clearly, such overruling should be applied with great caution,
and it should be well advertized whenever it occurs.
The method to deal with this {\it imposed} consistency is 
to account for the measurements simultaneously
by merging them into a single component, and applying an
``appropriate'' rescaling method.
\vs
The {\bf normalization} 
of each individual likelihood component is 
chosen such that its maximal value is equal to one.
This is not important for the analysis, but it is 
convenient: it ensures that a measurement does not 
contribute numerically to the overall $\chi^2$ value
if it is in the best possible agreement with theory, 
and that the (so-called) $\chi^2$ takes only positive values.
In the pure Gaussian case, this simply  implies  dropping the 
normalization constant of Eq.~(\ref{eq_thegaussian}):
one then recovers the standard $\chi^2$ definition.

\subsection{The Theoretical Likelihood}
\label{sec:theoreticalLikelihood}

The theoretical component of the likelihood is given by the product
\beq
\label{eq_likThe}
        \Likthe(\yQCD)=\prod_{i=1}^{\NQCD} \Likthe(i)~,
\eeq
where the individual components $\Likthe(i)$ 
account for the imperfect knowledge of the $\yQCD$ 
parameters (\eg, $f_{B_d}$) while more or less accurately 
including known correlations between them (\eg, $f_{B_d}/f_{B_s}$).
Ideally, one should incorporate in $\Likexp$ measurements 
from which constraints on $\yQCD$ parameters can be 
derived.
By doing so, one could remove altogether the theoretical component 
of the likelihood. However usually there is no such measurement: 
the {\it a priori} knowledge on the $\yQCD$ stems rather from 
educated guesswork.
As a result, the $\Likthe(i)$ components are incorporated by 
hand in Eq.~(\ref{eq_likThe}) and they can hardly be treated
as probability distribution functions (PDF). 
In effect, their mere presence in the discussion is a clear 
sign that the problem at hand is ill-defined. It demonstrates 
that here, a critical piece of information is coming neither 
from experimental, nor from statistically limited computations,
but from the minds of physicists. At present,
these components play a leading role in the analysis
and it is mandatory to handle them with the greatest caution.
\vs
In the default scheme, {\bf Range Fit (\rfit)}, we propose that 
the theoretical likelihoods $\Likthe(i)$ do not contribute to the 
$\chi^2$ of the fit while the corresponding $\yQCD$ parameters take 
values within allowed ranges\footnote
{
        Note that the $\yQCD$ parameters can also have errors 
        with (partly) statistical components. Examples for these
        are parameters obtained by Lattice calculations. The treatment 
        of this case is described in 
        Section~\ref{sec:statistics}.\ref{sec:likelihoodsAndSysErrors}. 
} denoted 
$[\yQCD]$. The numerical derivation of these ranges is discussed in 
Sections~\ref{sec:statistics}.\ref{sec:likelihoodsAndSysErrors} and 
\ref{sec:standardFit}.\ref{sec:fitInputs}. Most of them are identified 
to the ranges
\beq
   [\yQCD-\sigma_{\rm syst}\ ,\yQCD+\sigma_{\rm syst}]~,
\eeq
where $\sigma_{\rm syst}$ is the theoretical systematics evaluated 
for $\yQCD$. Hence $\yQCD$ values are treated on an equal footing, 
irrespective of how close they are to the edges of the allowed range.
Instances where even only one of the $\yQCD$ parameters lies outside 
its nominal range are not considered. 
\vs
This is the unique, simple and clear-cut assumption made in the \rfit\  
scheme: $\yQCD$ parameters are bound to remain within {\em predefined} 
allowed ranges. The \rfit\   scheme departs from a perfect frequentist 
analysis only because the allowed ranges $[\yQCD]$ do not extend to the 
whole physical space where the parameters could {\it a priori} take 
their values\footnote
{
   Not all $\yQCD$ parameters need to be given an {\it a priori} 
   allowed range, \eg, values taken by final state strong interaction 
   phases appearing in $B$ decays are not necessarily theoretically 
   constrained.
}.
\vs
This unique and minimal assumption, is nevertheless a strong 
constraint: all the results obtained should be understood as valid
only if all the assumed allowed ranges contain the true values of 
their $\yQCD$ parameters. However, there is no guarantee that this is 
the case, and this arbitrariness should be kept in mind. 

\section{Metrology}
\label{sec:metrology}

For metrology, one is not interested in the quality of the agreement 
between data and the theory as a whole. Rather, taking for granted 
that the theory is correct, one is only interested in the 
quality of the agreement between data and various realizations of the 
theory, specified by distinct sets of $\ymod$ values.
More precisely, as discussed in 
Section~\ref{sec:statistics}.\ref{sec:RelevantandIrrelevantParameters},
the realizations of the theory one considers are under-specified by 
various subsets of so-called relevant parameter values. In the 
following we denote
\beq
        \ChiMinGlob~,
\eeq
the absolute minimum value of the $\chi^2$ function of 
Eq.~(\ref{eq_chi2Function}), obtained when letting all $\Nmod$ 
parameters free to vary. 
\vs
In principle, this absolute minimum value does not correspond to 
a unique $\ymod$ location. This is because the theoretical predictions 
used for the analysis are affected by more or less important 
theoretical systematics. Since these systematics are being handled 
by means of allowed ranges, there is always a multi-dimensional 
degeneracy for any value of $\chi^2$.
\vs
For metrological purposes one should attempt to estimate 
as best as possible the complete $\ymod$ set. In that case, we
use the offset-corrected $\chi^2$
\beq
        \Delta\chi^2(\ymod)=\chi^2(\ymod)-\ChiMinGlob~,
\eeq
where $\chi^2(\ymod)$ is the $\chi^2$ for a given set of model 
parameters $\ymod$. The minimum value of $\Delta\chi^2(\ymod)$ 
is zero, by construction. This ensures that, to be consistent 
with the assumption that the SM is correct, CLs equal to unity 
are obtained when exploring the $\ymod$ space. 
\vs
A necessary condition is that the CL constructed from $\Delta\chi^2(\ymod)$     
provides correct coverage, that is, the CL interval for a parameter           
under consideration covers the true parameter value with a frequency            
of $1-\CL$ if the measurement(s) were repeated many times. This issue           
will be further adressed in several subsections.                                

\subsection{Relevant and Less Relevant Parameters}
\label{sec:RelevantandIrrelevantParameters}

Usually, one does not aim at a metrology of all the $\ymod$ values,
but only in a subset of them. This can be for two distinct reasons:
\bei

\item   the other parameters being deemed less relevant.
        For instance, in the SM, \CP  violation can be summarized 
        by the value taken by the Jarlskog parameter $\J$, or by 
        the value of the \CP-violating phase determined by the 
        parameters $\rho$ and $\eta$ in the Wolfenstein parameterization:
        the other CKM parameters and the $\yQCD$ parameters may 
        thus conceivably be considered of lower interest. 
        
\item   parameters that cannot be significantly constrained
        by the input data of the CKM fit.
        This is the case for most of the non-CKM parameters:
        $\yQCD$ parameters, but also the quark masses, etc.

\eei
In practice, the $\ymod$ parameters often retained as relevant 
for the discussion are $\rhobar$ and $\etabar$. The other parameters 
$\lambda$, $A$, the quark masses (etc.) and all the $\yQCD$ are 
considered as subsidiary parameters, to be taken into account 
in the analysis, but irrelevant for the discussion\footnote
{
        This point of view does not mean that the role of the $\yQCD$
        is irrelevant. In particular, if the agreement between data and 
        theory is not convincing one needs to set CLs in the $\yQCD$
        space.
}. 
In that case, the aim of the metrological stage of the analysis is 
to set CLs in the $\rhoeta$ plane.
\vs
We denote by $\a$ the subset of $\Na$ parameters under discussion 
(\eg, $\a=\arhoeta$) and $\Mu$ the $\Nmu$ remaining $\ymod$ 
parameters\footnote
{
   It is worth stressing that this splitting is arbitrary and 
   that it can be changed at will: for instance one may decide 
   to focus only on $\a=\{\J\}$,
   or to consider $\a=\astastb$ or other experimental observables.
   In practice, constraints on observables that are functions 
   of the $\ymod$ parameters are obtained by means of the technique of
   {\em Lagrange multipliers}.
}. 
{\em The goal is to set CLs in the $\a$ space, irrespective of the 
$\Mu$ values.}
\vs
The smaller the region in the $\a$ space where the CL is sizable 
(above $\CLcut=0.05$, say) the stronger the constraint is. The ultimate 
(and unattainable) goal is to shrink the allowed region to
a point: it would then correspond to the 'true' $\a$. This means that 
one seeks to exclude the largest possible region of the $\a$ space.
To do so, for a fixed value of $\a$, one has to find the $\Mu$ values 
that maximize the agreement between data and theory, and set the CL 
on $\a$ at the value corresponding to this optimized $\Mu$
\beq
\label{TheIncredibleStatement}
    \CL(a)={\rm Max}_\mu\{ \CL(a,\mu)\}~.
\eeq
Proceeding that way, one uses the most conservative estimate for a 
given $\a$ point: this point will be engulfed in the excluded region 
only if $\CL(\a,\Mu)<\CLcut$, $\forall\Mu$. 
As long as the theoretical likelihoods contain the true value of the 
$\yQCD$ parameters, the CL obtained has correct coverage and is to 
be understood as an upper limit of a CL.

\subsection{Metrology of Relevant Parameters}
\label{sec:MetrologyofRelevantParameters}

According to the above discussion, we denote 
\beq
        \chi^2_{\mini ;\,\Mu}(a)~,
\eeq
the minimum value of the $\chi^2$ function of Eq.~(\ref{eq_chi2Function}),
for a fixed value of $\a$, when letting all $\Mu$ parameters free 
to vary. For metrological purposes, we use the offset-corrected 
$\chi^2$
\beq
\label{eq_metrologicalOffset}
  \Delta\chi^2(a)
          =\chi^2_{\mini ;\,\Mu}(a)-\ChiMinGlob~,
\eeq
the minimum value of which is zero, by construction.

\subsubsection{Gaussian Case}
\label{sec:metrology_GaussianCase}

In a Gaussian situation, one directly obtains the CL for $\a$ as 
\beq
\label{eq_conLev}
        \CL(\a)=\Prob(\a)=\ProbCERN(\Delta\chi^2(\a),\Ndof)~,
\eeq
where $\Ndof=\mini(\Neff-\Nmu,\Na)$ and $\Neff$ is the effective
number of constraints (observables). 
\vs
To illustrate the use of $\Ndof$, let us first consider the 
standard CKM fit (see Part~\ref{sec:standardFit}). Several 
observables constrain the $\rhoeta$ plane so that the 
number of degrees of freedom exceeds the dimension of 
the $\a$ space ($\Na$). The offset-corrected $\Delta\chi^2(a)$, 
defined in Eq.~(\ref{eq_metrologicalOffset}), reduces 
the number of degrees of freedom to the dimension $\Na=2$ 
of the $\rhoeta$ plane.
However if one is to consider the constraint of only one
observable, \eg, $\stb$ in the $\rhoeta$ plane, the number of 
degrees of freedom is one, \ie, it is smaller than the dimension
of the $\a$ space. Indeed, given $\stb$ and, \eg, $\rhobar$,
the value of $\etabar$ is fixed. 
\vs
Other cases exist where the situation is less clear-cut: for 
instance, in the presence of penguins, the $\Cpipi$ parameters 
in $\Bz\to\pi^+\pi^-$ decays may be non-zero and hence acquires
some information on the unitarity angle $\alpha$. One would thus 
conclude that the appropriate number of degrees of freedom should 
be $\Ndof=2$. However in comparison with the $\Spipi$ parameter, 
the $\alpha$ constraint from $\Cpipi$ is insignificant, so that
using $\Ndof=1$ is the better approximation.
One concludes that even in a Gaussian case, ill-posed problems 
can occur, which must be individually studied with (toy) Monte
Carlo simulation.

\subsubsection{Non-Gaussian Case}
\label{sec:metrology_nonGaussianCase}

In a non-Gaussian situation,
one has to consider $\Delta\chi^2(a)$ as a test statistic,
and one must rely on a Monte Carlo simulation to obtain 
its expected distribution in order to compute $\CL(\a)$. 
As further discussed in Section~\ref{sec:statistics}.\ref{sec:probingTheSM},
this does not imply taking a Bayesian approach and to make 
use of PDFs for the unknown theoretical parameters $\Mu$.
\vs
For the sake of simplicity, we use Eq.~(\ref{eq_conLev}) in the 
present work with one exception discussed below. This implies 
that the experimental component $\Likexp(\xexp-\xthe(\ymod))$ is 
free from non Gaussian contributions and inconsistent measurements. 
However the $\Delta\chi^2(\a)$ function itself does not have to 
be parabolic. What matters is that the $\Likexp$ components are 
derived from Gaussian measurements, being understood that no 
$\Likthe$ components are present.
Applying Eq.~(\ref{eq_conLev}) using $\Likexp$ may lead to an 
under-coverage of the CL for a branching fraction measurement
with a very small number of signal events. That is, the interval 
belonging to a given CL value constructed in this way covers 
the true branching fraction value with a probability lower than 
$1-\CL$.
\vs 
Under the assumption 
that the measurement is free from background, the probability to 
measure $N_{\rm obs}$ events for a true number of $N_{\rm true}$ 
events is given by the Poissonian probability distribution
\beq
\label{eq:poisson}
        f(N_{\rm obs};N_{\rm true}) \;=\;
                \frac{e^{-N_{\rm true}} N_{\rm true}^{N_{\rm obs}}}
                     {N_{\rm obs}!}~.
\eeq
One prominent example is the measurement of $N_{\rm obs}=2$ rare
$K^+\to \pi^+\nu\bar{\nu}$ events with almost vanishing background 
probability by the E787 collaboration~\cite{E787} (see 
Section~\ref{sec:kaons}.\ref{sec:Kpinn}, ignoring the recent result from 
E949~\cite{E949} in the discussion here). In this case, the experimental 
likelihood $\Likexp(N_{\rm true})$ for a true number of $N_{\rm true}$ 
events given the number of $N_{\rm obs}$ observed events is 
the same Eq.~(\ref{eq:poisson}). The corresponding CL is obtained
by means of the following recipe.
\begin{enumerate}
\item   In the pure Poissonian case, the exact central confidence interval 
        $[a,b]$ at $\CL=2\alpha$ with probabilities 
        $P(n \geq N_{\rm obs};a)=\sum_{n=N_{\rm obs}}^{\infty} f(n;a) =\alpha$ 
        and
        $P(n \leq N_{\rm obs};b)=\sum_{n=0}^{N_{\rm obs}} f(n;b)=\alpha$ is 
        obtained by solving the following equations for $a$ and $b$, 
        respectively:
        \beqn
        \label{eq_poissoncl}
        \alpha      &=& \sum_{n=N_{\rm obs}}^{\infty} 
                        \frac{e^{-a} a^{n}}{n!} 
                        \;=\; 
                        1-\sum_{n=0}^{N_{\rm obs}-1}\frac{e^{-a} a^{n}}{n!}~, 
                        \\[0.2cm]
        \beta       &=& \sum_{n=0}^{N_{\rm obs}} \frac{e^{-b} b^{n}}{n!} ~.
        \eeqn
        Their inverse reads
        \begin{eqnarray}
        \label{eq_calcpoissonlower}
        a       &=&     \frac{1}{2} F_{\chi^{2}}^{-1} 
                                    (\alpha;\Ndof=2 N_{\rm obs})~,
                \\[0.2cm]
        \label{eq_calcpoissonupper}
        b       &=&     \frac{1}{2} F_{\chi^{2}}^{-1} 
                                    (1-\alpha;\Ndof=2 (N_{\rm obs}+1))~,
        \end{eqnarray}
        where $F_{\chi^{2}}$ is the cumulative distribution for 
        a $\chi^{2}$ distribution for $\Ndof$ degrees of freedom. The 
        quantities $F_{\chi^{2}}^{-1}$ can be calculated with
        the CERN library function CHISIN.
        Using Eqs.~(\ref{eq_calcpoissonlower}) and (\ref{eq_calcpoissonupper}) 
        we construct the correct CL as a function of $N_{\rm true}$.

\item   The experimental likelihood, $\Likexp^{\rm inp}$, is
        obtained from the inverse
        $\CL^{-1}=\Prob^{-1}(-2 \ln \Likexp^{\rm inp},1)$.
        In this way, the CKM fit can again use Eq.~(\ref{eq_conLev}) 
        to infer a CL for the Poissonian case with very small statistics.

\end{enumerate}

The situation becomes more complicated if the statistics is very small 
and, in addition, the amount of background is not negligible and possibly 
only known with limited precision. In this case, there are two possible 
ways to proceed. Either the experiment publishes a CL which then can be 
again translated into likelihood function using $\CL^{-1}$ (see above). 
This has been done, for example, by the BNL experiment E949~\cite{E949} (see 
Section~\ref{sec:kaons}.\ref{sec:Kpinn}) the successor of E747.  
Or, if this information is not available, the CL has to be constructed 
by means of a toy Monte Carlo simulation provided that the experimental 
information needed has been published.

\subsubsection{Physical Boundaries}
\label{sec:metrology_physicalBoundaries}

\begin{figure}[t]
  \epsfxsize16.0cm
  \centerline{\epsffile{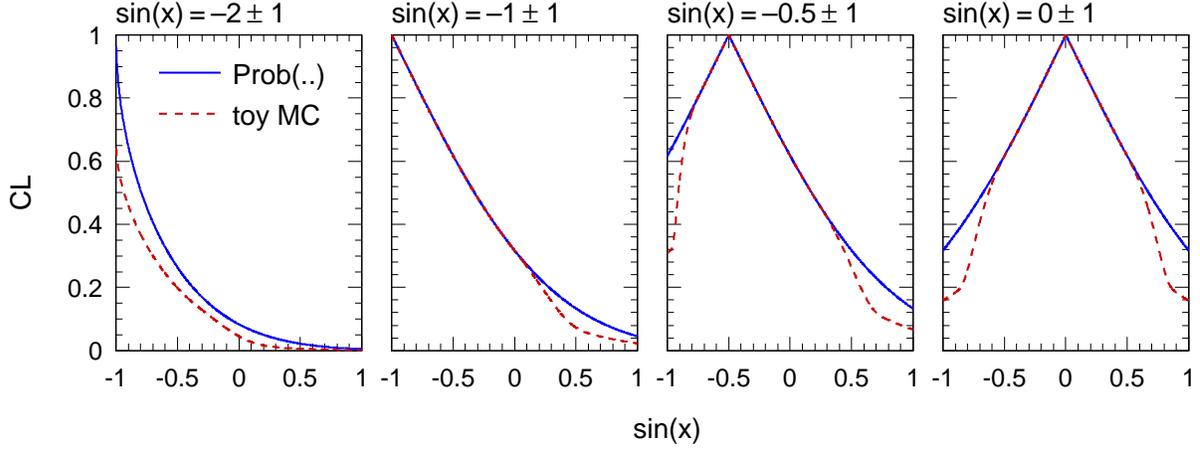}}
  \vspace{-0.5cm}
  \caption[.]{\label{fig:sin2bscan}\em
        Difference between Eq.~(\ref{eq_conLev}) and a Monte 
        Carlo evaluation of the confidence level for various
        measurements of a hypothetical quantity $\sin(x)$.
        The Monte Carlo evaluation takes into account the 
        physical boundaries of the observable.  }
\end{figure}

{\bf Physical boundaries:} in cases where the $\a$ value space is 
bounded, \eg, $\stb\in[-1,1]$, the confidence level $\Prob(a)$ is
modified close to the boundaries, even in a Gaussian case. 
In general, the presence of physical boundaries improves the 
parameter knowledge. The easiest way to derive the appropriate 
CL is to use Monte Carlo techniques. The procedure is as follows:
\begin{enumerate}

\item   choose the coordinate $\a_0$ in the (bounded) $\a$ space 
        at which the $\CL(\a_0)$ shall be determined.

\item   determine for the measurements at hand the offset-corrected
        $\Delta\chi^2(a_0)$ using Eq.~(\ref{eq_metrologicalOffset}).

\item   generate Monte Carlo measurements that fluctuate according
        to the experimental likelihoods $\Likexp$.

\item   determine the global minimum $\ChiMinGlob[{\rm MC}]$ for each set
        of measurements by leaving all $\ymod$ parameters free to 
        vary.

\item   determine the offset-corrected $\Delta\chi^2(\a_0)[{\rm MC}]$ 
        for each set of measurements using 
        Eq.~(\ref{eq_metrologicalOffset}).

\item   from the sample of Monte Carlo simulations, one 
        builds $\F_{\a_0}(\Delta\chi^2(\a_0)[{\rm MC}])$, the distribution 
        of $\Delta\chi^2(\a_0)[{\rm MC}]$, normalized to unity. 

\item   the CL referring to the coordinate $\a_0$ is then given by
        \beq
          \label{eq:toymcdchi2}
          \Prob(\a_0)\;\le\hmm\hmm
                \intl_{\Delta\chi^2\ge\Delta\chi^2(\a_0)}
                \hmmm\hmmm\F_{\a_0}(\Delta\chi^2) \ d\Delta\chi^2~.
        \eeq

\end{enumerate}
An illustration of the difference between a straight application of 
Eq.~(\ref{eq_conLev}) and the accurate Monte Carlo result is given
for various measurements of a hypothetical quantity $\sin(x)$ in 
Fig.~\ref{fig:sin2bscan}. The effects can be significant close to 
the boundaries.
\vs
The inclusion of the physical boundaries in a one-dimensional 
case is semi-analytically realized in \ckmfitter\  as described
in the digression below\footnote
{
\underline{\bf Digression.}
We consider a measurement $\xm\pm1$ of an observable that is 
confined to the interval $[\xmin,\xmax]$, and derive
the confidence level of a test value $\x\in[\xmin,\xmax]$.
The corresponding test statistic is given by
\beq
        \Delta\chi^2=\chitwo-\chi^2_{\rm min}~,
        \hspace{0.5cm}{\rm with}\hspace{0.5cm}
        \chitwo = \left(\x-\xm\right)^2~,
\eeq
and we aim at the solution of the convolution integral
\beq
\label{eq:convInt}
        \CL(\x) 
        = \frac{\intl_{\Delta\chi^2}^{\infty}
                \intl_{-\infty}^{\infty}
                \delta\left(\left(\x-\xmp\right)^2-\delchitwop\right)
                \frac{1}{\sqrt{2\pi}}e^{-\frac{1}{2}(\xmp-\xm)^2}
                d\delchitwop d\xmp}
               {\intl_{0}^{\infty}
                \intl_{-\infty}^{\infty}
                \delta\left(\left(\x-\xmp\right)^2-\delchitwop\right)
                \frac{1}{\sqrt{2\pi}}e^{-\frac{1}{2}(\xmp-\xm)^2}
                d\delchitwop d\xmp}~.
\eeq
The integration of
Eq.~(\ref{eq:convInt}) leads to multiple Heaviside step functions,
so that several cases must be distinguished:
\bei
\item   {\bf\boldmath   Measurement inside the allowed interval
                $\CL(\xmin\le\xm\le\xmax)$:}
\vs
Using $\Prob(\chitwo)\equiv\ProbCERN(\chitwo,1)=\Erfc(\sqrt{\chitwo/2})$ 
the CL obtained ignoring the possibly non-zero value of $\minchitwo$
(the $\chitwo$ returned by the procedure below is offset-corrected into
a $\Delta\chitwo$ for metrology reasons in \ckmfitter), 
and denoting $\P_{\x}[x_1,x_2]$ the probability that $\x$ occurs in 
the range $[x_1,x_2]$ (taken to be negative if $x_1>x_2$)
\beq
     \P_{\x}[x_1,x_2] = 
     \left\{\begin{array}{ll}
        \frac{1}{2}\left(\Erfc\left(|\x-x_1|/\sqrt{2}\right) 
        - \Erfc\left(|\x-x_2|/\sqrt{2}\right)\right)
                & ~{\rm if~} \x< x_1~\AND~\x<x_2 \\[0.2cm]
        \frac{1}{2}\left(\Erfc\left(|\x-x_2|/\sqrt{2}\right) 
        - \Erfc\left(|\x-x_1|/\sqrt{2}\right)\right)
                & ~{\rm if~} \x> x_1~\AND~\x>x_2 \\[0.2cm]
        1 - \frac{1}{2}\left(\Erfc\left(|\x-x_1|/\sqrt{2}\right) 
        + \Erfc\left(|\x-x_2|/\sqrt{2}\right)\right)
                & ~{\rm elsewhere~}  
     \end{array}\right.
\eeq
one obtains for the different domains of $\xt$
\beq
     \CL = 
     \left\{\begin{array}{ll}
        \Prob(\chitwo)
                & ~{\rm if~}\frac{1}{2}(\xm+\xmin)
                            \le\xt
                            \le\frac{1}{2}(\xm+\xmax) \\[0.2cm]
        \frac{1}{2}\Prob(\chitwo)
                & ~{\rm if~}(\xt=\xmin~\OR~\xt=\xmax)~\AND~\xt\ne\xm \\[0.2cm]
        \Prob(\chitwo)
        -\P_{\xt}\left[\frac{1}{2}\left(\xt+\xmin
                                       -\frac{(\xt-\xm)^2}{\xt-\xmin}
                                 \right),
                             2\xt-\xm\right]
                & ~{\rm if~}\xt<\frac{1}{2}(\xm+\xmin) \\[0.2cm]
        \Prob(\chitwo)
        -\P_{\xt}\left[2\xt-\xm,
                 \frac{1}{2}\left(\xt+\xmax-\frac{(\xt-\xm)^2}{\xt-\xmax}
                            \right)\right]
                & ~{\rm if~}\xt>\frac{1}{2}(\xm+\xmax) 
     \end{array}\right.
\eeq

\item   {\bf\boldmath   Measurement below the allowed interval
                $\CL(\xm<\xmin)$:}              
\beq
     \CL = 
     \left\{\begin{array}{ll}
        1
                & ~{\rm if~}\xt=\xmin\\[0.2cm]
        \Prob(\chitwo)-\P_{\xt}\left[2\xt-\xm,
                                     \xt+\sqrt{\Sigma}
                               \right]
                & ~{\rm if~}\left\{\begin{array}{l}
                            \xmin<\xt\le\frac{1}{2}(\xmin+\xmax) 
                            ~\AND   \\
                            \xt+\xmin+\frac{(\xmax-\xt)^2}{\xmin-\xt}\le 2\xm
                            \end{array}\right\}\\[0.5cm]
        \Prob(\chitwo)-\P_{\xt}\left[[2\xt-\xm,
                                     \frac{1}{2}\left(\xt+\xmax
                                                 +\frac{\Sigma}
                                                       {\xmax-\xt}
                                                \right)
                               \right]
                & ~{\rm if~}\left\{\begin{array}{l}
                            \xmin<\xt\le\frac{1}{2}(\xmin+\xmax)
                            ~\AND   \\
                            \xt+\xmin +\frac{(\xmax-\xt)^2}{\xmin-\xt}>2\xm
                            \end{array}\right\}\\[0.5cm]
        \Prob(\chitwo)-\P_{\xt}\left[2\xt-\xm,
                                     \frac{1}{2}\left(\xt+\xmax
                                                 +\frac{\Sigma}
                                                       {\xmax-\xt}
                                                \right)
                               \right]
                & ~{\rm if~}\xt>\frac{1}{2}(\xmin+\xmax)
     \end{array}\right.
\eeq
where $\Sigma\equiv(\xt+\xmin-2\xm)(\xt-\xmin)$.

\item   {\bf\boldmath   Measurement above the allowed interval
                $\CL(\xm > \xmax)$:}
\beq
     \CL = 
     \left\{\begin{array}{ll}
        1
                & ~{\rm if~}\xt=\xmax\\[0.2cm]
        \Prob(\chitwo)-\P_{\xt}\left[\xt-\sqrt{\Xi},
                                     2\xt-\xm
                               \right]
                & ~{\rm if~}\left\{\begin{array}{l}
                            \frac{1}{2}(\xmin+\xmax)\le\xt<\xmax
                            ~\AND   \\
                            \xt+\xmax+\frac{(\xmin-\xt)^2}{\xmax-\xt}\ge2\xm
                            \end{array}\right\}\\[0.5cm]
        \Prob(\chitwo)-\P_{\xt}\left[\frac{1}{2}\left(\xt+\xmin
                                                 +\frac{\Xi}
                                                       {\xmin-\xt}
                                                \right),
                                     2\xt-\xm
                               \right]
                & ~{\rm if~}\left\{\begin{array}{l}
                             \frac{1}{2}(\xmin+\xmax)\le\xt<\xmax
                            ~\AND   \\
                            \xt+\xmax+\frac{(\xmin-\xt)^2} {\xmax-\xt}<2\xm
                            \end{array}\right\}\\[0.5cm]
        \Prob(\chitwo)-\P_{\xt}\left[\frac{1}{2}\left(\xt+\xmin
                                                 +\frac{\Xi}
                                                       {\xmax-\xt}
                                                \right),
                                     2\xt-\xm
                               \right]
                & ~{\rm if~}\xt<\frac{1}{2}(\xmin+\xmax)
     \end{array}\right.
\eeq
where $\Xi\equiv(2\xm-\xt-\xmax)(\xmax-\xt)$.
\eei
}. 
The results are identical to the toy Monte Carlo simulation technique 
introduced above.

%
%
\section{Probing the Standard Model}
\label{sec:probingTheSM}

By construction, the metrological phase is unable to detect 
if the SM fails to describe the data. This is because 
Eq.~(\ref{eq_metrologicalOffset}) wipes out the information 
contained in $\ChiMinGlob$. This value is a measure (a test 
statistics) of the best possible agreement between data and 
theory. The agreement can be quantified by the so-called 
p-value $\Prob(\ChiMinGlob|\rm SM)$: the probability to observe 
a $\chitwo$ as large as or larger than $\ChiMinGlob$ if the 
Standard Model is the correct theory.
Ideally, in a pure Gaussian case, $\ChiMinGlob$ could be turned 
easily into a p-value referring to the SM as a whole in a 
straightforward way
\beq
\label{eq_naiveGaussian}
   \Prob(\ChiMinGlob|{\rm SM})\le
           \ProbCERN(\ChiMinGlob,\Ndof)~,
\eeq
with $\Ndof=\Neff-\Nmod$, if it were a positive value. The whole 
Standard Model being at stake, one should not rely on 
Eq.~(\ref{eq_naiveGaussian}), but use a Monte Carlo simulation 
to obtain the expected distribution of $\ChiMinGlob$. The Monte 
Carlo simulation proceeds as follows:
\begin{enumerate}

\item   determine for the measurements at hand the global minimum
        $\ChiMinGlob$ and the corresponding $\ymod$ values, which
        are assumed to be the true ones\footnote
        {
           As discussed above, in the presence of theoretical
           uncertainties various $\ymod$ realizations may yield identical 
           theoretical predictions. The choice made for a particular 
           $\ymod$ solution (leading to $\ChiMinGlob$) is irrelevant. 
        }
        
\item   generate the $\xexp(i)$ for all measurements $(i)$, following 
        the individual experimental likelihood components $\Likexp(i)$, 
        having reset their central values to the values $\xexp(i)=\xthe(i)$ 
        computed with the above $\ymod$ solution set.

\item   in contrast to the above, the $\Likthe$ component of the 
        likelihood is not modified: their central 
        values are kept to their original settings. This is 
        because these central values are not random numbers,
        but parameters contributing to the definition of $\Lik$.

\item   compute the minimum of the $\chi^2$ by allowing all $\ymod$ 
        to vary freely, as is done in the actual data analysis.

\item   from this sample of Monte Carlo simulations, one 
        builds $\F_{\rm SM}(\chi^2)$, the distribution of $\ChiMinGlob$, 
        normalized to unity. 

\item   the p-value referring to the SM as a whole is then
        \beq
        \label{eq_monteCarlo}
          \Prob(\ChiMinGlob|{\rm SM})\;\le\hmm\hmm\intl_{\chi^2\ge\ChiMinGlob}
                  \hmmm\hmmm\F_{\rm SM}(\chi^2) \ d\chi^2~.
        \eeq

\end{enumerate}

\section{Probing New Physics}
\label{sec:probingNewPhysics}

If the above analysis establishes that the SM cannot accommodate the data, 
that is the p-value $\Prob(\ChiMinGlob|{\rm SM})$ is small, the next step 
is to probe the New Physics (NP) revealed by the observed discrepancy. 
The goal is akin to metrology: 
it is to measure new physical parameters $\yNP$ 
(whose values, for example, are null if the SM holds) 
complementing the set of $\ythe$ parameters of the SM.
The treatment is identical to the one of Section~\ref{sec:statistics}.\ref{sec:metrology}, 
using $\a=\{\yNP\}$. 
The outcome of the analysis is for example a $95\%~\CL$ domain 
of allowed values for $\yNP$ defined, in a first approximation,
from Eq.~(\ref{eq_conLev})
\beq
\label{eq_newPhysicsCL}
\CL(\yNP)=\ProbCERN(\Delta\chi^2(\yNP),\NNP)\ge 0.05~.
\eeq
Even if the SM cannot be said to be in significant 
disagreement with data, it remains worthwhile to perform 
this metrology of new NP for the following reasons:

\begin{itemize}

\item   it might be able to faster detect the first signs 
        of a discrepancy between data and the SM
        if the theoretical extension used in the analysis 
        turns out to be the right one. The two approaches 
        are complementary, the first (\cf, 
        Section~\ref{sec:statistics}.\ref{sec:probingTheSM}) 
        leading to a general 
        statement about the agreement between data and the SM
        independently of any assumption about the NP,
        the second being specific to a particular extension of the 
        SM. In that sense, it is less satisfactory.
        The two approaches can nevertheless disagree:
        the first may conclude that the SM is in 
        acceptable agreement with data, while the second may 
        exclude the SM value $\yNP=0$, and, conversely, 
        the first may invalidate the SM, while the second 
        may lead to a fairly good value of $\CL(\yNP=0)$
        if the extension of the SM under consideration is not 
        on the right track.

\item   the most sensitive observables, and the precision 
        to be aimed at for their determination cannot be derived 
        by any other means than by this type of analysis. When 
        considering new experiments, it is therefore particularly 
        valuable to have a sensitive model of NP, to 
        prioritize the effort and set the precision to be achieved.

\end{itemize}

%
%
\section{Alternative Statistical Treatments}
\label{AlternativeStatisticalTreatments}

Several alternative statistical treatments are available and
the reader is referred to Ref.~\cite{CKMfitter} for a detailed
discussion of the merits and drawbacks of each of the methods.
In the following, we only recall the \erfit\  method as a
conservative extension to \rfit, and briefly comment on 
the use of Bayesian methods.

\subsection{The Extended Conservative Method (\erfit)}
\label{TheExtendedConservativeMethod}

The \rfit\  scheme uses $\Likthe(i)$ functions that take 
only two values: either 1 within the allowed range, or 0 outside,
thereby restricting $\yQCD$ to the range $[\yQCD]$.
Instead, the extended \erfit\  scheme allows intermediate values
between 0 and 1 for $\Likthe(i)$. They are equal to 
1 within $[\yQCD]$ (there, they do not contribute at all to the 
full $\chi^2$, and one recovers the \rfit\  scheme) and drop 
smoothly to 0 outside. These functions are not treated as PDFs
and hence the \erfit\  scheme is not a Bayesian scheme.
\vs
The way the \erfit\  likelihood functions decrease down 
to zero is arbitrary: one needs to define a standard.
The proposed expressions for $\Likthe(i)$ are presented 
in Section~\ref{sec:statistics}.\ref{sec:likelihoodsAndSysErrors}. 
Because \erfit\  acknowledges the fact that the allowed 
ranges should not be taken literally, it offers two advantages 
over \rfit:
\bei

\item   \erfit\   is more conservative than \rfit: by construction,
        a \erfit\  CL is always equal or larger than the corresponding 
        \rfit\  one, and its CL surface in the $\a$ space
        exhibits the same plateau of equal $\CL=1$.

\item   in the case where the SM appears to be ruled out by \rfit,
        the \erfit\  scheme is able to detect the $\yQCD$ parameter(s) 
        beyond the nominal allowed range that would restore an 
        acceptable agreement between data and theory.

\eei
Despite the two above arguments in favor of \erfit, we chose \rfit\
as the standard scheme used in this paper rather than \erfit:
because it uses a simpler and unique prescription to incorporate 
theoretical systematics, it is less prone to be confused with a 
Bayesian treatment. 

\subsection{The Bayesian Treatment}
\label{sec:TheBayesianTreatment}

The Bayesian treatment~\cite{Achille1} considers $\Lik$ 
as a PDF, from which is defined $\F(\a)$, the PDF of $\a$, through 
the convolution
\beq
\label{BayesianII}
   \F(\a)=C \int \Lik(\ymod)\ \delta(\a-\a(\ymod))\ d\ymod~,
\eeq
where the constant $C$ is computed {\it a posteriori} to ensure 
the normalization to unity of $\F(\a)$. In practice, the integral 
can be obtained conveniently by Monte Carlo techniques\footnote
{
   This convenience may sometimes boost the application of Bayesian
   techniques, since no use of sophisticated minimization techniques 
   is necessary.
}. 
For each point in the $\a$ space, one sets a confidence level 
$\CL(\a)$, for example according to:
\beq
\label{eq_CLBayesian}
\CL(\a)\;=\hmm\hmm
        \intl_{\F(\a^\prime)\le\F(\a)}\hmm\hmm \F(\a^\prime) 
        \ d\a^\prime~.
\eeq 
Other definitions for the domain of integration can be 
chosen. 
\vs
New Physics is not meant to be detected by the Bayesian treatment: 
it is aimed at metrology mostly.

\subsection{Comparison with \rfit}

Although the graphical displays appear similar, the Bayesian 
treatment and the \rfit\  scheme are significantly different:
the meaning attached to a given CL value is not the same.
For the Bayesian treatment, the CL is a quantity {\it defined}
for example by Eq.~(\ref{eq_CLBayesian}).
The justification of this definition lies in the understanding 
that a CL value is meant to provide a quantitative measure 
of our qualitative {\it  degree of belief}. Whereas one understands 
qualitatively well what is meant by {\em  degree of belief},
because of its lack of formal definition, one cannot check that 
it is indeed well measured by the CL: the argument is thus circular.
\vs
The key point in the Bayesian treatment is the use of 
Eq.~(\ref{BayesianII}), even though the likelihood contains 
theoretical components. This implies that the $\yQCD$ parameters, 
which stem from theoretical computations, are to be considered as 
random realizations of their true values. The PDFs of these 
``random'' numbers are then drawn from guess-work 
(the $[\yQCD]$ ranges do not fare better with respect to that.).
For self-consistency, if one assumes that a large number of 
theorists perform the same $\yQCD$ computation,
the distribution of their results should then be interpreted
as a determination of the $\yQCD$ PDF. Once injected in 
Eq.~(\ref{BayesianII}), this PDF, the shape of which contains 
no information on nature, 
will be transformed into information pertaining to nature.
This entails to a confusion between what is an 
experimental result and what is a thinking result. Illustrations of 
this are given in the appendix of Ref.~\cite{CKMfitter}\footnote
{
        Methodogical problems that may appear with the use of 
        Bayesian statistics in metrological CKM analyses are outlined
        below.
        \bei

        \item   The convolution of several, arbitrary {\em a priori} 
                theoretical PDFs can lead to the creation of seemingly
                accurate information (the convolved PDF) out of no 
                initial knowledge. A quantification of the uncertainty
                related to this information is impossible.

        \item   All $\ymod$ parameters need {\em a priori} PDFs, also those
                that are to be determined by the fit. For instance, if the
                CKM parameters are the physical unknowns of the global
                fit, the results obtained will depend on the parameterization
                chosen to have a, say, uniform prior. This breaks a 
                fundamental invariant of physics theories. For example, 
                the CKM-related physics depends on whether the CKM 
                matrix is parameterized with the Euler angles \& phase, or  
                with the Wolfenstein parameters.

        \item   It frequently occurs in the phenomenological description
                of $B$ decays that {\em a priori} unknown strong-interaction 
                phases contribute to the $\ymod$ parameters. While this is 
                no problem in \rfit\  (or \erfit), where these parameters
                are free to vary within their $2\pi$ periodicity, it may
                exhibit biases in a Bayesian approach: in particular,   
                in the presence of multiple unknown phases,
                the CL obtained may depend on whether the 
                validity range is chosen to be $[-\pi,\pi]$ or $[0,2\pi]$
                or any other $2\pi$ interval. 
        
        \eei
}.

%
%
\section{Likelihoods and Systematic Errors}
\label{sec:likelihoodsAndSysErrors}

So far, we have reviewed the basic formalism of the \rfit\  scheme. The 
treatment of experimental and theoretical systematics is the subject of 
this section.
\vs
Let $\xo$ be a quantity, which is not a random variable, but which is 
not perfectly known. We will consider two quantities of this type.
\bei
\item   A theoretical parameter which is not well determined
        (\eg, $\xo=f_{B_d}$): the theoretical prediction of an 
        observable depends on $\xo$ (\eg, $\Delta M_{B_d}$).

\item   An experimental bias due to detector/analysis defects:
        the measurement should be corrected for this bias.

\eei
It is the purpose of this section to suggest a prescription 
of how to incorporate the limited knowledge of such quantities 
into the analysis. The standard treatment of this problem relies 
on a $\chi^2$ analysis, which is satisfactory as long as the degree 
of belief we put on the knowledge of the value of $\xo$ is distributed 
like a Gaussian. However this is not necessarily what is meant 
when one deals with systematic errors. Rather, the theorist (resp. 
the experimentalist) {\it may} mean that the prediction (resp. the 
measurement) can take any value obtained by varying $\xo$ 
at will within the range $[\xobar-\zeta\sigxo,\xobar+\zeta\sigxo]$ 
(denoted {\em the allowed range} below, where $\zeta$ is a constant 
scale factor of order unity and $\xobar$ is the expected central 
value of $\xo$), but that it is unlikely that $\xo$ takes its true 
value outside the allowed range. This does {\em not} imply that 
the possible values for  $\xo$ are equally distributed within the 
allowed range: they are not distributed {\em at all}. 
\vs
If a systematic error is given 
such a meaning, then the statistical analysis should treat all 
$\xo$ values within the allowed range on the same footing
(which again does not imply with equal {\em probability}):
this corresponds to the \rfit\  scheme (with $\zeta=1$).
On the other hand, it may be convenient to define specific tails 
instead of sharp cuts, thus allowing the theoretical parameters 
to leave their allowed ranges, if needed: this corresponds to the 
\erfit\  scheme.
\vs
The idea is to move from a pure $\chi^2$ analysis to a log-likelihood 
one, redefining the $\chi^2$ to be
\beq
\label{eq_chisqNew}
\chi^2=
        \left({\xexp-\xthe\over\sigexp}\right)^2
        -2\ln\Hatsyst(\xo)~,
\eeq
where $\Hatsyst(\xo)$, hereafter termed the {\em Hat} function,
is a function equal to unity for $\xo$ within the allowed range. 

\subsection{The Hat Function}
\label{sec:flattenedGaussian}

The Hat function $\Hatsyst(\xo,\kappa,\zeta)$ is a continuous function 
defined as
\beq
\label{eq_flattenedGaussian}
-2\ln\Hatsyst(\xo,\kappa,\zeta)=
        \left\{
        \begin{array}{ll}
        0~, 
                & \forall\xo\in[\xobar\pm\zeta\sigxo] \\
        \left(\displaystyle\frac{\xo-\xobar}{\kappa\sigxo}\right)^2
        -\left(\displaystyle\frac{\zeta}{\kappa}\right)^2~,~~
        &       \forall\xo\notin[\xobar\pm\zeta\sigxo]
        \end{array}
        \right.
\eeq
where the constant $\kappa$ determines the behavior of the function
outside the allowed range. For the \rfit\   scheme $\kappa =0$ is used.
To define a standard $\kappa$ can be chosen to be a function of $\zeta$ 
such that the relative normalization of $\Hatsyst(\xo,\kappa,\zeta)$
(considered here, for the purpose of defining a standard, as a PDF) 
be equal to the one of a Gaussian of width $\sigxo$
\beq
\intl_{-\infty}^{+\infty}\Hatsyst(\xo,\kappa,\zeta)\ d\xo
                \cdot\intl_0^{\zeta/\sqrt{2}}e^{-t^2}\ dt
        = \sqrt{\pi}\,\zeta\sigma_0~.
\eeq
The parameter $\kappa$ is numerically computed as a function of $\zeta$. 
For the limit $\zeta\to 0$ one obtains $\kappa\to 1$,
and the Hat becomes a pure Gaussian. The \erfit\  scheme is defined 
by $\zeta=1$, for which one obtains $\kappa\simeq 0.8$.
\vs
Examples of Hat functions with $\xobar=0$ and $\sigxo=1$ are shown on 
the left plot of Fig.~\ref{fig_loglik}. Being a likelihood and not a PDF, 
$\Hatsyst(\xo)$ needs not be normalized to unity. 

\subsection{Combining Statistical and Systematic Uncertainties}
\label{CombiningStatisticalandSystematicUncertainties}

Having defined $\Hatsyst(\xo)$ one proceeds with the minimization of 
the $\chi^2$ of Eq.~(\ref{eq_chisqNew}) by allowing $\xo$ to vary freely.
\vs
For theoretical systematics, the result depends on the way $\xo$ enters 
$\xthe$, and not much more can be said in generality. 
\vs
For experimental and theoretical systematics
where $\xo$ can be assumed to be an unknown offset\footnote
{
        If systematics take the form of an unknown 
        multiplicative factor, and this is often the case 
        for theoretical uncertainties, a treatment similar 
        to the one discussed here applies.
}: 
the quantity to be confronted to the theoretical prediction $\xthe$ 
is simply $\xexp+\xo$. Omitting the details of straightforward 
calculations, and assuming that $\xobar=0$ (otherwise $\xexp$ 
should be corrected for it), one obtains, after minimization of 
the $\chi^2$ with respect to $\xo$:
\begin{figure}[t]
  \epsfxsize11cm
  \centerline{\epsffile{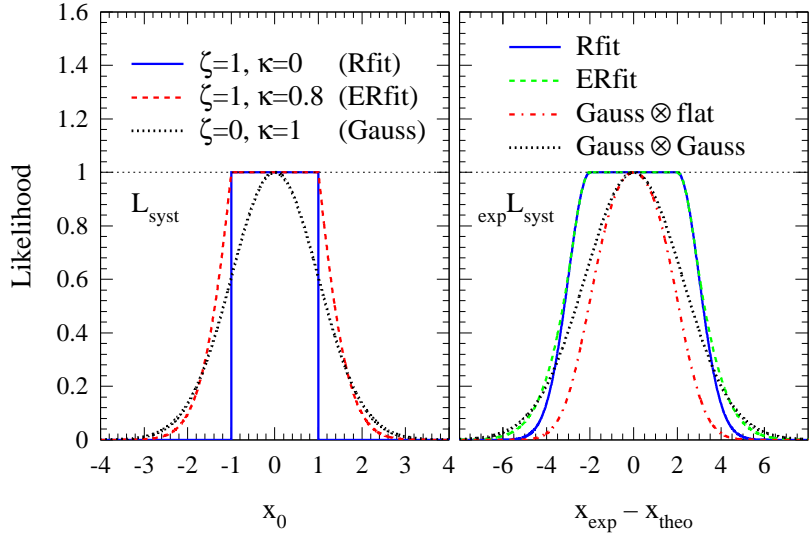}}
  \vspace{-0.2cm}
  \caption[.]{\label{fig_loglik}\em
        \underline{Left:} Hat functions
        ($\xobar=0$ and $\sigxo=1$) used for the {\rm \rfit} 
        scheme, the {\rm \erfit} scheme, and the Gaussian treatment.
        \underline{Right:} combined likelihood
        $\expHatsyst$ (with $\xobar=0$ and $\sigexp=\sigxo=1$)
        for the {\rm \rfit} scheme, the {\rm \erfit} scheme,
        a convolution of a Gaussian with a uniform distribution
        (hence taken as a PDF, following the Bayesian approach)
        and a convolution of two Gaussians.}
\end{figure}

\bei

\item   $\mid\xexp-\xthe\mid\le\zeta\sigxo$~:\hskip .5truecm
        $\chi^2_{\mini ;\,\xo}=0~.$

\item   $\zeta\sigxo\le\mid\xexp-\xthe\mid\le\zeta\sigxo
        (1+({\sigexp\over\kappa\sigxo})^2)$~:\hskip .5truecm
        $\chi^2_{\mini ;\,\xo}
                =\left({\mid\xexp-\xthe\mid
                -\zeta\sigxo\over\sigexp}\right)^2~.$

\item   $\mid\xexp-\xthe\mid\ge\zeta\sigxo
        (1+({\sigexp\over\kappa\sigxo})^2)$~:\hskip .5truecm
        $\chi^2_{\mini ;\,\xo}
                ={(\xexp-\xthe)^2\over\sigexp^2+(\kappa\sigxo)^2}
                -\left({\zeta\over\kappa}\right)^2~.$

\eei
In the limit $\zeta\to 0$ (and hence, $\kappa\to 1$) 
only the third instance is met, and one recovers the usual rule of 
adding in quadrature the statistical and the systematic uncertainties.
Otherwise, the result is non-trivial. An example of the effective 
likelihood
$\expHatsyst(\xexp-\xthe)\equiv{-{1\over 2}\chi^2_{\mini ;\,\xo}}$
(with $\xobar=0$ and $\sigexp=\sigxo=1$) is shown in the right hand 
plot of Fig.~\ref{fig_loglik} for the \rfit\  scheme, the \erfit\  
scheme, a convolution of a Gaussian with a uniform distribution
(hence taken as a PDF, following the Bayesian approach) and a 
convolution of two Gaussians.

%
%
 \newpage\part{The Global CKM Fit}\setcounter{section}{0}
\markboth{\textsc{Part III -- The Global CKM Fit}}
         {\textsc{Part III -- The Global CKM Fit}}
\label{sec:standardFit}

%
%
\section{Introduction}

With the remarkable exceptions of $\sin2\beta$ (see
Section~\ref{sec:input_s2b}) and $\sin2\alpha$ (see
Section~\ref{sec:input_rhorho}), the experimental observables that are
presently used to constrain $\rhoeta$ depend on hadronic matrix
elements, which have to be evaluated at a much smaller energy than the weak interaction
scale. Since the discovery of asymptotic freedom, 
Quantum 
Chromodynamics (QCD) is well established as {\em the} quantum field theory 
of strong interaction. It has been tested to high precision 
in the perturbative regime, where the coupling constant $\as$ is 
small and allows one to build a systematic expansion.
Unfortunately, no general solution of QCD is known, and not even a 
well-controlled approximation is available (at least in an analytical 
form) that would be valid for an arbitrary $\as$.
\vs
While it is far beyond the scope of this introduction to review the
wealth of approaches to non-perturbative QCD, it is useful to recall
a few general techniques to evaluate the matrix elements that are
relevant for quark flavor physics. The theoretical methods can be
classified, somewhat arbitrarily, into four categories: constituent
quark models, QCD sum rules, lattice simulations, and effective
theories of QCD.
\bei

\item   \textit{Constituent quark models}
        comprise, to a first
        approximation, methods that assume a fixed number of particles and treat
        them in the framework of quantum mechanics. Multi-body wave functions,
        which satisfy bound state potential equations, are constructed, and
        external operators that describe flavor transitions are represented in
        terms of constituent quarks and then sandwiched between these wave functions.

\item   \textit{QCD sum rules}
        rely on quark-hadron duality to
        identify a correlation function written in terms of quarks and gluons
        with its representation as a sum over hadronic bound states. The desired
        matrix element is then isolated from the rest of the sum and its
        contribution is controlled in various ways.

\item   \textit{Lattice simulations}
        implement quantum field
        theory on an Euclidean space-time lattice. The path integrals that
        represent correlation functions are then numerically evaluated 
        with Monte Carlo methods.

\item   \textit{Effective theories of QCD} exploit 
        additional symmetries of full QCD in specific kinematic or
        parametric regimes. Matrix elements are then related to others
        that are simpler to compute or to measure, up to corrections that are
        suppressed by the typical symmetry breaking scale.

\eei

\section{Inputs to the standard CKM fit}
\label{sec:fitInputs}

This section provides a compendium of the measurements and SM predictions 
entering the overall constrained CKM fit, denoted {\em standard CKM fit} 
in the following. The corresponding numerical values 
used and the treatment of their uncertainties within \rfit\
are summarized in Table~\ref{tab:ckmInputs}.
In cases where different independent measurements for an input quantity are 
available, we multiply the corresponding (\rfit) likelihoods. Experimental 
and theoretical correlations, if present and known, 
are taken into account if not stated otherwise.

\subsection{$\Vud$} 
\label{sec:input_vud}

        The matrix element $\Vud$ has been extracted by means of 
        three different methods: superallowed nuclear $\beta$-decays, 
        neutron $\beta$-decay and pion $\beta$-decay.
        \vs
        The most precise experimental determination of $\Vud$ 
        comes from lifetime measurements of superallowed nuclear 
        $\beta$-decays with pure Fermi-transitions ($0^+\to 0^+$).      
        The $ft$-value is the product of the integral over the electron energy
        spectrum $f$ and the electron lifetime $\tau$: $ft = f \cdot \tau \cdot \ln{2}$. 
        Its theoretical prediction can be written as 
        \begin{equation}
           ft \cdot (1 + \delta_{R}) 
              \cdot (1 - \delta_{C}) 
              =\frac{K}{2 G_{F}^2 \Vud^2 (1 + \Delta^{V}_{R})}~,
        \end{equation}
        where $G_{F}$ is the Fermi constant (see Table~\ref{tab:ckmInputs}),
        $\delta_{R}$ and $\Delta^{V}_{R}$ are the nucleus-dependent and
        nucleus-independent parts of the radiative corrections, respectively,
        $\delta_{C}$ is the charge-symmetry breaking corrections, and
        $K=2\pi^{3} \ln{2}/m_{e}^{5}$.
        The charge-symmetry breaking corrections, as well as part 
        of the nucleus-dependent radiative corrections, depend on the 
        nuclear structure of the nucleus under consideration. 
        Using the results for nine different superallowed nuclear 
        $\beta$ decays, the average is
        $\Vud=0.9740 \pm 0.0001_{\rm exp}$~\cite{townerhardy}.
        This result is however dominated by theoretical 
        uncertainties, namely $\sigma(\Vud)[\Delta^{V}_{R}] = 0.0004$, 
        $\sigma(\Vud)[\delta_{C}] = 0.0003$ and 
        $\sigma(\Vud)[\delta_{R}] = 0.0001$. 
        Adding these in quadrature results in $\sigma(\Vud) = 0.0005$, 
        whereas adding them linearly (as in the {\rfit} approach) results in
        $\sigma(\Vud) = 0.0008$. 
        \vs        
        A possible enhancement, $\Delta \Vud = +0.0005$, is predicted 
        by a quark-meson coupling model due to a change of charge symmetry 
        violation for quarks inside bound nucleons compared to unbound 
        nucleons~\cite{vudsaito1}. 
        The theoretical error has been enlarged by the PDG~\cite{PDG} 
        by adding linearly the amount of the possible correction to
        the quoted error of $\sigma(\Vud) = 0.0005$, resulting in 
        $\Vud = 0.9740 \pm 0.0010$. 
        Since this correction may be partially contained 
        in the charge-symmetry breaking corrections, and since the effect 
        can be significantly smaller depending on the model used, we do 
        not enlarge the error and use in the fit:
        $\Vud=0.9740 \pm 0.0001_{\rm exp} \pm 0.0008_{\rm theo}$.
        \vs
        Nuclear structure effects do not play a role in neutron 
        $\beta$ decays. However, to extract $\Vud$, one needs to 
        measure the neutron lifetime and the ratio 
        of the axial-vector coupling constant to the vector coupling 
        constant $g_{A}/g_{V}$
        \beq
           \Vud^{2} 
         = \frac{K \ln{2}}
           {G_{F}^2 (1 + \Delta^{V}_{R}) 
                    (1 + 3 (g_{A}/g_{V})^2) 
                  f (1 + \delta_{R}) \tau_{n}}~.
        \eeq
        In contrast to nuclear $\beta$ decays, these measurements are 
        not yet dominated by theoretical uncertainties. The weighted 
        mean for the neutron lifetime measurements is 
        $\tau_{n} = (885.7 \pm 0.7){\rm\, s}$~\cite{townerhardy},
        where the available results are statistically consistent. 
        Recently, the PERKEO-II experiment has measured 
        $g_{A}/g_{V}=-1.2739 \pm 0.0019$~\cite{PERKEO}. 
        Using the world average for the neutron lifetime this translates 
        into 
        $\Vud=0.9717 \pm 0.0013_{g_{A}/g_{V},\tau_{n}} \pm 0.0004_{\rm theo}$. 
        The experimental error on this result is a factor of two 
        smaller than any preceding measurement with high neutron 
        polarization~\cite{PDG}. 
        When considering all data on $g_{A}/g_{V}$ with high neutron 
        polarization, the measurements are not consistent. A rescaling 
        by a factor of 1.6 is therefore applied following the PDG 
        recipe, which results in $g_{A}/g_{V} = -1.2690 \pm 0.0022$, with
        $\Vud=0.9745 \pm 0.0016_{\rm stat} \pm 0.0004_{\rm theo}$~\cite{townerhardy}.
        Since the PERKEO-II result was obtained using a very high 
        neutron polarization and since ${\cal O}(2\%)$ corrections 
        used to extract the final result from data are much 
        smaller than in previous experiments, we only use this result
        in the fit.
        \vs
        The pion $\beta$ decay $\pi^{+} \to \pi^{0} e^{+} \nu_{e}$
        is an attractive candidate to extract $\Vud$ from the 
        branching ratio 
        ${\BR}(\pi^{+} \to \pi^{0} e^{+} \nu_{e})$ 
        and the pion lifetime, since it is mediated by a pure vector 
        transition and does not suffer from nuclear structure effects. 
        However, due to the small branching ratio, 
        ${\BR}(\pi^{+} \to \pi^{0} e^{+} \nu_{e}) 
         = (1.025 \pm 0.034) \times 10^{-8}$~\cite{PDG}, 
        the statistical precision is not competitive with the other 
        methods: $\Vud=0.967 \pm 0.016_{\BR} \pm 0.0005_{\rm theo}$.
        The preliminary result from the PIBETA experiment~\cite{PIBETA}, 
        ${\BR}(\pi^{+} \to \pi^{0} e^{+} \nu_{e}) 
         = (1.044 \pm 0.007_{\rm stat} \pm 0.009_{\rm sys}) \times 10^{-8}$,
        yields $\Vud=0.9765 \pm 0.0055_{\BR} \pm 0.0005_{\rm theo}$,
        which still has a statistical error that is
        a factor of four times larger 
        than the result from neutron decay experiments. It will not be competitive 
        even when the final expected experimental uncertainty of
        $\sigma(\Vud)=0.002$ is reached.
        \vs
        We build a combined likelihood for the $\Vud$ determinations from
        superallowed $\beta$ decays, from neutron $\beta$ decays and from
        the pion $\beta$ decay, taking into account the correlation due to the 
        uncertainty on $\Delta^{V}_{R}$. We obtain the $\CL>5\%$ interval
        $0.9730<\Vud<0.9750$.
 
\subsection{$\Vus$ }
\label{sec:input_vus}

        The analyses of kaon and hyperon semileptonic decays provide 
        the best determination of $\Vus$.
        However, due to theoretical uncertainties from the breakdown 
        of SU(3) flavor symmetry, the hyperon decay data are less 
        reliable~\cite{vusdhk,vusflores}. Although, as pointed out in 
        Ref.~\cite{cabibbo2}, linear SU(3) breaking corrections can be 
        avoided, we do not use results from hyperon decays 
        since the uncertainties on the vector form factor $f_{1}$ in
        these decays have not been fully evaluated yet.
        As a consequence, we only use the value obtained from the vector 
        transitions $K^{+} \to \pi^{0} \ell^{+} \nu_{\ell}$ and 
        $\KL \to \pi^{-} \ell^{+} \nu_{\ell}$. The rates for
        these decays depend on two form factors, $f_{+}(t)$ and 
        $f_{0}(t)$, where $t = (p_{K} - p_{\pi})^{2}$ is the four-momentum
        transfer-squared between the kaon and the pion. 
        Owing to the small electron mass, only $f_{+}(t)$ plays a role 
        in $K_{e3}$ decays whose functional dependence can be 
        extracted from data. The form factor value at 
        zero recoil, $f_{+}(0)$, is calculated within the framework 
        of chiral perturbation theory and is found to be
        $f_{+}^{\Kz\pi^{-}}(0) = 0.961 \pm 0.008$~\cite{vusleutw}.
        The error estimate for this value has been questioned in 
        Ref.~\cite{vuspaschos}. 
        We note that a relativistic constituent quark model, successful 
        in the description of electroweak properties of light mesons,
        gives the consistent result
        $f_{+}^{K^{0}\pi^{-}}(0) = 0.963 \pm 0.004$~\cite{vusjaus}. 
        \vs
        A precise calculation of $f_{+}^{K^{0}\pi^{-}}(0)$ is a difficult
        task. Order $p^6$ contributions in chiral perturbation theory have
        been calculated only recently~\cite{PostSchilcher,BijnensTalavera}. 
        The ${\cal{O}}(p^6)$ calculation contains a ``local'' and a
        ``loop'' 
        contribution leading to a strong cancellation, with the result depending 
        on the renormalization scale. Ref.~\cite{CiriNeufeldPichl} quotes 
        $f_{+}^{K^{0}\pi^{-}}(0)|_{p^6}=-0.001 \pm 0.010$ leading to 
        $f_{+}^{K^{0}\pi^{-}}(0) = 0.981 \pm 0.010$ while emphasizing, 
        however, that further work is needed to clarify whether the 
        uncertainty quoted is realistic.
        A value of $f_{+}^{K^{0}\pi^{-}}(0) = 0.981 \pm 0.010$ would increase
        the deviation from unitarity in the first family.
        It is worthwhile to mention in this context that a recent quenched
        Lattice-QCD calculation obtains
        $f_{+}^{K^{0}\pi^{-}}(0)=0.960\pm0.005_{\rm stat}\pm0.007_{\rm sys}$~\cite{BECIf0}
        in agreement with the Leutwyler-Roos value~\cite{vusleutw}.
        \vs
        Channel-independent and channel-dependent 
        radiative corrections~\cite{vusmarciano,vuswilliams,cirigliano}
        as well as charge-symmetry ($\Kp/\KL$) 
        and charge-independence ($\pi^{-}/\pi^{0}$) breaking 
        corrections~\cite{vusleutw} are applied to 
        compare the branching fraction results from both channels~\cite{PDG}:
        $f_{+}^{K^{0}\pi^{-}}(0) \Vus
         =0.2134\pm 0.0015_{\rm exp}\pm 0.0001_{\rm rad}$ 
        ($K^{+} \to \pi^{0} e^{+} \nu_{e}$) and
        $f_{+}^{K^{0}\pi^{-}}(0) \Vus 
        = 0.2101 \pm 0.0013_{\rm exp} \pm 0.0001_{\rm rad}$
        ($\KL \to \pi^{-} e^{+} \nu_{e}$).
        Their weighted average is
        $f_{+}^{K^{0}\pi^{-}}(0) \Vus = 0.2114 \pm 0.0016$,
        where the error has been rescaled by a factor of 1.6 to account 
        for the inconsistency between neutral and charged kaon
        decay data.
        \vs
        Using $f_{+}^{K^{0}\pi^{-}}(0) \Vus = 0.2114 \pm 0.0016$
        and the Leutwyler-Roos value 
        $f_{+}^{K^{0}\pi^{-}}(0) = 0.961 \pm 0.008$~\cite{vusleutw},
        one obtains
        $\Vus = 0.2200 \pm 0.0017_{\rm exp} \pm 0.0018_{\rm theo}$.
        Recently, the BNL-E865 collaboration measured 
        ${\BR}(K^{+}_{e3})=(5.13\pm 0.02_{\rm stat}\pm 0.09_{\rm sys}\pm 0.04_{\rm norm})\%$~\cite{E865}, 
        which exhibits a $2.2\sigma$ deviation from the
        world average ${\BR}(K^{+}_{e3}) = (4.87 \pm 0.06)\%$~\cite{PDG}.
        The BNL-E865 result translates into 
        $\Vus=0.2285 \pm 0.0023_{\rm exp} \pm 0.0019_{\rm theo}$~\cite{cirigliano} 
        when using $f_{+}^{K^{0}\pi^{-}}(0) = 0.961 \pm 0.008$.
        If this result is confirmed this would imply that the previous
        ${\BR}(\KL{}_{,e3})$ results~\cite{PDG} are incorrect, since 
        it is not likely that such a discrepancy can be explained by 
        isospin breaking effects~\cite{cirigliano}. 
        The KLOE collaboration had presented preliminary 
        results on $K^{0}_{e3}$ and $\KL{}_{\mu 3}$ decays~\cite{CkmKLOE} 
        in agreement with the PDG values for $K^{+}_{e3}$ decays\footnote
        {
                In the meantime, the KLOE $\KS{}_{,e3}$ result has been updated 
                with nearly final systematic error~\cite{KLOEMoriond04}. 
                The result for $\Vus$ is now in 
                reasonable agreement with the BNL-E865 result for $K^{+}_{e3}$ 
                and hence differs from the former determinations using $K^{0}_{e3}$
                decays. The understanding of final state radiation of photons 
                plays a crucial role in these analyses and may become a key
                issue when comparing the results of the various experiments.
                Very recently, the KTeV collaboration has presented a result
                for semileptonic $K_{L}$ branching fractions which gives
                $f_{+}^{K^{0}\pi^{-}}(0) \Vus = 0.2165 \pm 0.0012$~\cite{VusKTeV} 
                in agreement with the BNL-E865 result. However, this result 
                has not yet been included in our average.
        }.
        We use a weighted average of the BNL-E865 result and the former 
        $\Vus$ average, rescale the experimental uncertainty and obtain
        $\Vus = 0.2228 \pm 0.0039_{\rm exp} \pm 0.0018_{\rm theo}$ by
        using the Leutwyler-Roos value 
        $f_{+}^{K^{0}\pi^{-}}(0) = 0.961 \pm 0.008$~\cite{vusleutw}. As 
        mentioned previously there is intense theoretical activity concerning 
        an improved determination of this form factor value.
        \vs
        There are good prospects to clarify the experimental situation 
        in the near future. 
        The KLOE, KTeV and NA48 experiments have the potential to determine
        $K_{l3}$ decays with different experimental techniques. Another 
        promising method to measure $\Vus$ from moments of the strange 
        spectral functions in $\tau$ decays has been proposed in 
        Ref.~\cite{CkmJamin} and might be realized at the $B$ factories
        where more than $10^{8}$ $\tau$ pairs have currently been recorded.

\subsection{$\Vcd$ and $\Vcs$}

        Both the $\Vcd$ and $\Vcs$ matrix elements can be determined 
        from di-muon production in deep inelastic scattering (DIS) of 
        neutrinos and anti-neutrinos on nucleons. In an analysis performed 
        by the CDHS collaboration~\cite{vcdCdhs}, $\Vcd$ and $\Vcs$ are extracted 
        by combining the data from three experiments, CDHS~\cite{vcdCdhs}, 
        CCFR~\cite{vcdCcfr} and CHARM II~\cite{vcdCharmii}, giving
        $\Vcd^2 \times B_{c} = (4.63 \pm 0.34) \times 10^{-3}$, 
        where $B_{c} = 0.0919 \pm 0.0094$~\cite{bolton,ushida,kubota} is 
        the weighted average of semileptonic branching ratios of charmed 
        hadrons produced in neutrino-nucleon DIS.
        This results in $\Vcd = 0.224 \pm 0.014$~\cite{heraB}.
        The average DIS result from CDHS, CCFR and CHARM II is
        $\kappa \Vcs^{2} B_{c} = (4.53 \pm 0.37) \times10^{-2}$, where 
        $\kappa = 0.453 \pm 0.106^{+0.028}_{-0.096}$ is the relative 
        contribution from strange quarks in the sea with respect to 
        $u$ and $d$ quarks, leading to $\Vcs = 1.04 \pm 0.16$~\cite{PDG}.
        \vs
        Similarly to $\Vus$ coming from $K_{e3}$ decays, $\Vcs$ can be extracted 
        from $D_{e3}$ decays. 
        However the theoretical uncertainty in the form factor calculation
        $f_{+}(0) = 0.7 \pm 0.1$~\cite{vcsMontanet} limits its precision 
        to $\Vcs = 1.04 \pm 0.16$~\cite{heraB} 
        (in coincidental agreement with $\Vcs$ from DIS).
        \vs
        Assuming unitarity and using as additional input the constraints 
        on $\Vud$, $\Vus$, $\Vub$, $\Vcd$ and $\Vcb$, $\Vcs$ can also be
        extracted from the following quantities:
        \begin{itemize} 
        \item $R_{c}^{W} = \Gamma(W^{+} \to c\bar{q})/ \Gamma(W^{+} 
        \to {\rm hadrons}) 
        = \sum_{i=d,s,b}|V_{ci}|^{2}/ (\sum_{i=d,s,b;\,j=u,c}|V_{ji}|^{2})$. 
        For the three-generation CKM matrix $R_{c}^{W}$ is expected to be 1/2.
        The measurements~\cite{vcsAleph,vcsDelphi,vcsOpal2} are found to 
        be consistent with this expectation.         
        \item $\Gamma(W \to X_{c} X)= R_{c}^{\rm W} \cdot 
        \BR(W \to {\rm hadrons}) \cdot \Gamma_{\rm tot} \propto 
        \sum_{i=d,s,b}|V_{ci}|^{2}$~\cite{vcsOpal2}. 
        \item $\Gamma(W^{+} \to {\rm hadrons})/\Gamma(W^{+} \to {\rm leptons})
               = \sum_{i=d,s,b;\,j=u,c}|V_{ji}|^{2} 
        \times(1+\as(m_{W})/\pi + \dots)$~\cite{PDG}, 
        for which the experimental result is
        $\sum_{i=d,s,b;\,j=u,c}|V_{ji}|^{2} =
        (2.039\pm0.025)\cdot
        (\BR(W \to \ell \nub_{\ell})\pm 0.001(\as))$~\cite{LEPew}. 
        \end{itemize}
        For the three generation CKM matrix all these quantities have theoretical 
        predictions that are independent of the actual values of 
        the CKM elements involved, so that they cannot be used in a CKM fit. 
        On the other hand, these measurements test the unitarity
        of a three-generation CKM matrix requiring, for 
        example, $\sum_{i=d,s,b;\,j=u,c}|V_{ji}|^{2}=2$.
        \vs
        There are prospects that $\Vcs$, $\Vcd$ and $\Vcs/\Vcd$ will
        be determined at the CLEO-c experiment with unprecedented precision 
        in semileptonic $D$-meson decays to a pseudoscalar meson 
        $D \to P \ell^{+} \nu$~\cite{cleoc}.
        For a $3\invfb$ data sample, the relative errors on 
        $\Gamma(\Dz \to K^{-} \ell^{+} \nu)$ and 
        $\Gamma(\Dz \to \pi^{-} \ell^{+} \nu)$ are expected to 
        be $\sigma(\Gamma) / \Gamma = 1.2\%$ and 
        $\sigma(\Gamma)/ \Gamma = 1.5\%$, respectively. The extraction 
        of $\Vcs$ and $\Vcd$ from these decays
        will require a substantial improvement 
        of the theoretical precision in the form factor calculation, 
        which may be achieved in the forthcoming years by Lattice 
        QCD. A relative uncertainty on the form factor of $3\%$, for 
        example, would then translate into the errors
        $\sigma(\Vcs)/\Vcs  = 1.6\%$ and 
        $\sigma(\Vcd)/\Vcd  = 1.7\%$. 
        The ratio $\Vcs/\Vcd$ will be determined at CLEO-c 
        following two different approaches that are expected to be
        less dependent on theoretical uncertainties. 
        In the first approach, one compares semileptonic decays with
        the same initial state but with different final states as
        $\Gamma(D^{0} \to K^{-} \ell^{+} \nu)$ and 
        $\Gamma(D^{0} \to \pi^{-} \ell^{+} \nu)$. The ratio of branching 
        fractions depends on the product of $|V_{cs}/V_{cd}|^{2}$
        and a form factor ratio that differs from unity only due to
        SU(3) breaking corrections. In the second approach, one compares 
        reactions with different initial states but the same final 
        state, for instance $D_{s}^+ \to \KS \ell^{+} \nu$
        and $D^{+} \to \KS \ell^{+} \nu$.

\subsection{$\Vcb$}
\label{sec:input_vcb}

        In the Wolfenstein parameterization, $\Vcb$ determines
        the parameter $A$ which plays an important
        role for the constraints on $\rhobar$, $\etabar$
        from $\Vub$, $|\epsk|$ and $\dmd$. Its precision also
        has a significant impact on the SM prediction 
        for the rare decays $K \to \pi \nu \nub$. It is most accurately obtained  
        from exclusive $B\to D^{(*)}\ell{\nub}_\ell$ 
        and inclusive semileptonic $b$ decays to charm.
      
        \subsubsection*{Exclusive Decays}

        In the exclusive technique, the differential spectrum $d\Gamma / dw$
        for the decay $B\to D^{(*)}\ell{\nub}_\ell$ 
        is measured, where $w$ is the scalar product of the velocity 
        four-vectors of the $B$ and the $D^{(*)}$ mesons.
        This allows one to extract the product ${\cal F}_{D^*}(1)\Vcb$, 
        where ${\cal F}_{D^*}(w=1)$ is the $B$-to-$D^{(*)}$ form factor 
        at zero-recoil.
        In the heavy quark limit, the form factor is given by the 
        Isgur-Wise function~\cite{IsgurWise}, which is equal to $1$ at $w=1$, but 
        which receives corrections due to the finite $b$ and $c$ quark masses
        that can be calculated in the framework of Heavy Quark Effective 
        Theory (HQET)~\cite{IsgurWise,hqet}. At present, the most precise 
        determination using the exclusive technique comes from the decay 
        $B\to D^{*}\ell{\nub}_\ell$.
        Due to the presence of a soft pion in the $D^{*}$ decay, its 
        reconstruction is less affected by combinatorial background than 
        for a $D$-meson decay. Moreover,
        the phase space function for $B\to D \ell{\nub}_\ell$ drops more rapidly 
        when approaching $w=1$, leading to a larger statistical error. Finally, 
        the calculation of the form factor at zero-recoil 
        is believed to have smaller theoretical uncertainties in the case 
        of a $B$-to-$D^{*}$ transition, since linear $1/m_{Q}$ corrections 
        in the heavy quark mass $m_{Q}$ vanish, a property known as Luke's 
        theorem~\cite{luketheorem}.
        It has been pointed out that the form factor for $B$-to-$D$ 
        transitions may be calculable with good theoretical precision
        despite the presence of $1/m_{Q}$ corrections~\cite{uraltsevfpcp03}.
        \vs
        Previous theoretical determinations of ${\cal F}_{D^*}(1)$ were based 
        either on QCD sum rules (see, \eg, Ref.~\cite{ShifmanSR}) or on 
        HQET, where long-distance contributions had been estimated with
        the use of 
        non-relativistic quark models~\cite{NeubertF1,NeubertFalk}. Both 
        methods obtained values for ${\cal F}_{D^*}(1)$ around $0.9$ with 
        quoted uncertainties of the order of $4\%$, which are however  
        difficult to control.
        Recently, important progress has been achieved through the calculation
        of ${\cal F}_{D^*}(1)$ using Lattice QCD in conjunction with 
        HQET~\cite{KronfeldF1,Harada}.
        Their result, ${\cal F}_{D^*}(1)=0.913^{+0.030}_{-0.035}$~\cite{Hashimoto},
        is used in our fit. It is expected that the uncertainty can 
        be reduced in the forthcoming years. Averaging eight 
        different measurements, the Heavy Flavor Averaging Group (HFAG) obtains 
        ${\cal F}_{D^*}(1)\Vcb = (36.7 \pm 0.8)\times10^{-3}$~\cite{HFAG} 
        and $\rho^{2} = 1.44 \pm 0.14$, where $\rho$ is the slope of the form 
        factor as a function of $w$. The linear correlation coefficient 
        between the two parameters is 0.91. The goodness of the average
        ($\chi^{2}=30$ for 14 degrees-of-freedom, that is $\CL=0.08$)
        indicates an inconsistency among the various measurements, which is
        mainly driven by the somewhat large result obtained by CLEO.
        \vs
        It has been argued that the uncertainties in the Lattice QCD
        calculation of ${\cal F}_{D^*}(1)$~\cite{Hashimoto} can be considered
        as mainly statistical ones~\cite{CernCkmWS}. Following this reasoning
        and using ${\cal F}_{D^*}(1)\Vcb = (36.7 \pm 0.8)\times10^{-3}$~\cite{HFAG},
        we obtain $\Vcb = (40.2^{+2.1}_{-1.8})\times 10^{-3}$, which is used in
        the fit.

        \subsubsection*{Inclusive Decays}
       
        In the inclusive approach, the semileptonic width 
        $\Gamma(B \to X \ell \nu_{\ell})$ is determined 
        experimentally from the semileptonic branching fraction 
        $\BR(B \to X \ell \nu_{\ell}) = (10.90 \pm 0.23)\%
        $~\cite{HFAG} 
        and the $B$-meson lifetime, where the admixture of neutral 
        ($\tau_{\Bz}=(1.534 \pm 0.013)\ps^{-1}$~\cite{HFAG}) and 
        charged ($\tau_{\Bp}=(1.653 \pm 0.014)\ps^{-1}$~\cite{HFAG}) 
        $B$ mesons is understood. Relying on the concept of quark-hadron 
        duality, the theoretical prediction for the semileptonic width is 
        obtained by means of a Operator Product Expansion called Heavy 
        Quark Expansion (HQE~\cite{hqt}), which invokes perturbative 
        corrections and non-perturbative hadronic matrix elements 
        that dominate the theoretical uncertainty. The theoretical 
        expression for the semileptonic rate reads
        \beqn
        \label{eqof-dmd1}
           \Gamma(b \to c) &=& 
                \frac{\GF^2 \Vcb^2 m_{b}^5}{192 \pi^3} \,
                 f\!\left(\frac{m_{c}^2}{m_{b}^2}\right) \nonumber\\
                & & \times\;\left[1 + A\left(\frac{\as}{\pi}\right) 
                    + B\left(\frac{\as^2}{\pi^2} \beta_{0}\right)
                    + C\left(\frac{\Lambda_{\rm QCD}^2}{m_{b}^2}\right)
  + {\cal{O}}\left(\as^2,\frac{\Lambda_{\rm QCD}^3}{m_{b}^3},\frac{\as}{m_{b}^2}\right) 
                  \right]~,
        \eeqn
        where $m_{b}$ is the $b$-quark mass, $f$ corrects for the finite
        charm quark mass $m_{c}$, and the coefficients $A$, $B$ and $C$ are
        functions of hadronic matrix elements and depend on $m_{b}$ and 
        $m_{c}$. Perturbative QCD corrections are known up to order
        $\as^{2} \beta_{0}$. Non-perturbative corrections are suppressed
        by powers of
        $\Lambda_{\rm QCD}/m_{b}$. At order $(\Lambda_{\rm QCD}/m_{b})^2$,
        the hadronic matrix elements can be expressed by the HQET parameters
        $\lambda_{1}$ and $\lambda_{2}$,
        the expectation values of the heavy-quark kinetic energy and the
        chromomagnetic interaction, respectively.
        Additional parameters occur at order $(\Lambda_{\rm QCD}/m_{b})^3$. 
        Alternatively, in the kinetic mass scheme~\cite{KineticMassS}, these 
        matrix elements are given by the parameters $-\mu_{\pi}^2 = \lambda_{1}$ and 
        $\mu_{G}^{2}/3 = \lambda_{2}$, up to higher-order corrections.
        The parameter $\lambda_{2}$ can be obtained from the 
        observed hyperfine splitting in the \B-meson spectrum. 
        The semileptonic width can be written in terms of the \B-meson
        mass $m_{B}$ instead of $m_{b}$ by introducing the non-perturbative 
        parameter $\Lbar$, that is the energy of the 
        light-degrees-of-freedom.
        \vs
        Besides the total semileptonic width, HQE can be used to predict
        sufficiently inclusive differential distributions. 
        Since different regions of the phase space have different 
        sensitivity to the quark masses and the non-perturbative 
        parameters, spectral moments calculated from measured differential 
        distributions can be used to constrain these parameters. However
        moments at too high order cannot be reliably predicted since 
        quark-hadron duality starts to break down.
        In recent years, the parameters $\Lbar$ and 
        $\lambda_{1}$ have been 
        constrained experimentally by measurements of leptonic energy and hadronic 
        mass moments. In addition, the measured photon energy distribution in 
        $B \to X_{s} \gamma$ allows one to extract 
        $\Lbar$~\cite{CLEOXsgamma}.
        So far, the constraints on these parameters from CLEO, DELPHI and \babar\
        provide consistent results, which may be interpreted as a test of the 
        validity of the HQE up to order ${\cal{O}}(1/m_{Q}^{3})$. However no
        global fit taking into account all various measurements has been 
        performed yet\footnote
        {
                Such a global analysis is now available in Ref.~\cite{GLOBALVcb},
                which has been published after completion of this document.
        }. 
        An overview of several $\Vcb$ determinations using input from measured
        moments can be found, \eg, in Ref.~\cite{Schubert} from which we obtain 
        $\Vcb = (42.0 \pm 0.6_{\rm stat} \pm 0.8_{\rm theo})\times10^{-3}$.
        Here, the first error arises from the experimental uncertainties 
        on the branching fraction, the \B-meson lifetime 
        and the fit error from the determination of $\Lbar$ 
        and $\lambda_{1}$, and the second error contains the theoretical 
        uncertainty due to higher-order (${\cal{O}}(1/m_{Q}^{3})$) and $\as$ 
        corrections\footnote
        {
                Very recently, the \babar\  collaboration has been 
                presented precisely measured electron energy and hadronic 
                mass moments~\cite{BABARelemom,BABARhadmom}. The value obtained
                $\Vcb = (41.4 \pm 0.4_{\rm stat} \pm 0.4_{\rm HQE} \pm 0.6_{\rm theo})\times10^{-3}$
                from a fit~\cite{BABARVcbfit} in the kinetic mass 
                scheme~\cite{KineticMassS} to these moments currently
                provides 
                 the most precise single $\Vcb$ determination. 
                This input however has not yet been taken into account
                in our global fit.
        }.
        \vs

        \subsubsection*{Average}
 
        In the fit, we combine the likelihoods of $\Vcb$ from inclusive and 
        exclusive measurements where we assume that they are uncorrelated.

\subsection{$\Vub$}
\label{sec:input_vub}

        The third column element $\Vub$, with additional input from $\Vus$ 
        and $\Vcb$, describes a circle in the
        $\rhoeta$ plane. It can be extracted either from inclusive 
        $B \to X_u\ell^-{\nub}_\ell$ decays, or from exclusive decays
        such as $B\to\pi\ell\nu_\ell$, $B\to\rho\ell\nu_\ell$, 
        $B\to\omega\ell\nu_\ell$ and $B\to\eta\ell\nu_\ell$.

        \subsubsection*{Exclusive Decays}

        In contrast to heavy-to-heavy transitions like $B \to D^{(*)}$,
        there is no heavy quark symmetry argument that allows one to constrain
        the form factor normalization in the heavy-to-light decays 
        $B \to \pi, \rho, ...$~. As a consequence, exclusive determinations
        -- besides being experimentally challenging -- suffer from large 
        theoretical uncertainties in the form factor calculations. From a
        theoretical point of view, one expects that $B\to\pi\ell\nu_\ell$ 
        will ultimately be the most promising mode for an extraction of $\Vub$
        in exclusive decays, since only one form factor function is present in 
        pseudoscalar-to-pseudoscalar transitions (while for instance three 
        different form factor functions have to be calculated for
        $B\to\rho\ell\nu_\ell$ decays). On the other hand, 
        the softer lepton spectrum in $B\to\pi\ell\nu_\ell$ with respect
        to $B\to\rho\ell\nu_\ell$, where the lepton momentum benefits 
        from the polarization of the $\rho$, leads to an enhanced 
        $b\to c$ background contamination in the former decay.
        \vs
        The \babar\  collaboration has published a measurement of the branching 
        fraction 
        $\BR(B\to\rho\ell\nu_\ell)=(3.29\pm0.42_{\rm stat}\pm0.47_{\rm sys}\pm0.60_{\rm theo})\times10^{-4}$~\cite{BABARrholnu}.
        Using several form factor models they extract 
        $\Vub=(3.64\pm0.22_{\rm stat}\pm0.25_{\rm sys}{^{+0.39}_{-0.56}}_{\rm theo})\times10^{-3}$.
        The CLEO collaboration has recently presented a combined analysis of 
        the decays $B\to\pi\ell\nu_\ell$, 
        $B\to\rho\ell\nu_\ell$, $B\to\omega\ell\nu_\ell$
        and $B\to\eta\ell\nu_\ell$~\cite{CLEOpirholnu}.
        Owing to a largely hermetic detector, CLEO also measured
        the rates in three different bins of $q^{2}$, the lepton-neutrino 
        four-momentum-squared. CLEO finds the combined value 
        $\Vub=(3.17\pm0.17_{\rm stat}{^{\,+0.16}_{\,-0.17}}_{\rm sys}{^{\,+0.53}_{\,-0.39}}_{\rm theo})\times10^{-3}$
        for $B\to\pi\ell\nu_\ell$ and $B\to\rho\ell\nu_\ell$.
        Even the single CLEO number for $B\to\rho\ell\nu_\ell$ is
        hard to compare with the \babar\  result~\cite{BABARrholnu} since
        different form factor calculations have been used in both experiments, and
        results from different $q^{2}$ regions have been taken into
        account. A combination of the \babar\  and CLEO numbers is also difficult
        because the applied theoretical uncertainty range by itself
        varies. These problems in mind, we have averaged both results by 
        symmetrizing each of the results with respect to the quoted theoretical
        uncertainties and assuming that they are
        fully correlated between both experiments. With this method, we obtain
        $\Vub = (3.35 \pm 0.20_{\rm exp} \pm 0.50_{\rm theo})\times10^{-3}$.

        \subsubsection*{Inclusive Decays}
       
        Starting from the inclusive semileptonic width
        $\Gamma(B \to X_u\ell^-{\nub}_\ell)$, $\Vub$ 
        can be predicted within the HQE framework with a theoretical 
        uncertainty of approximately $5\%$. However, there is a large
        background from $b\to c$ transitions that is about 50 times larger
        than the $b\to u$ signal. To suppress this background,
        experimental cuts in the three-dimensional phase space have to 
        be applied which introduce additional theoretical uncertainties.
        In a first kind of analyses, $B \to X_u\ell^-{\nub}_\ell$
        decays are separated from $b\to c$ background by accepting 
        leptons with center-of-mass momenta typically larger than 
        $2.2$--$2.3\gevc$, a region which kinematically excludes
        $B \to X_c\ell^-{\nub}_\ell$ decays. However this
        requirement retains only $10$--$15\%$ of the semileptonic
        branching fraction.
        In this endpoint-region, the spectrum is dominated by the 
        so-called {\em shape} function, a non-perturbative object that 
        reflects the Fermi motion of the $b$ quark inside the $B$ meson. 
        Without knowledge of the shape function, the extrapolation of 
        the measured partial branching fraction to the full semileptonic 
        branching fraction is highly model-dependent, a drawback from 
        which the pioneering $\Vub$ determinations by the ARGUS and CLEO 
        collaborations suffered~\cite{vub_cbArgus,vub_cbCleo1,vub_cbCleo2} 
        (see also Refs.~\cite{modACCMM,modISGW,modKS,modWSB}). The problem can
        be circumvented to some extent by measuring the shape function in 
        inclusive $B \to X_{s} \gamma$ decays, so far only published 
        by the CLEO collaboration~\cite{CLEOXsgamma}. 
        Recent lepton endpoint measurements have been presented by CLEO, 
        \babar\  and Belle~\cite{CLEOendpoint,BABARendpoint,Belleendpoint},
        where all analyses are using the $B \to X_{s} \gamma$
        measurement from CLEO~\cite{CLEOXsgamma}.
        There is a discussion in the literature concerning uncertainties
        from subleading shape 
        functions~\cite{SubleadingLeibovich,SubleadingMannel,SubleadingNeubert,SubleadingLuke}).
        From this, one deduces that an additional theoretical uncertainty 
        on $\Vub$ of the order of a few percent may be present.
        One should also note that there could be sizable effects from
        the violation of quark-hadron duality in this small region of
        the phase space, which introduces theoretical uncertainties 
        that are difficult to quantify~\cite{DualityBigiUraltsev}. 
        \vs
        The $b\to c$ background in inclusive decays 
        can also be suppressed by cutting on the 
        hadronic invariant mass $m_{X}$. Accepting only events with $m_X$
        below the $D$ meson mass retains about $70$--$80\%$ of the 
        $B \to X_u\ell^-{\nub}_\ell$ events. Due to detector resolution
        effects, the cut has to be lowered, which again increases the theoretical
        uncertainties from the shape function. The kinematic region sensitive
        to the shape function can be avoided by cutting in addition on $q^{2}$,
        which in turn increases the statistical uncertainty.
        A complication in the high-$q^2$ region is the possible
        significant contribution from annihilation diagrams which cannot be
        computed at present~\cite{Voloshin}.
        A possible way to quantify such annihilation contributions would
        be to determine the branching fraction for charged and
        neutral \B mesons separately. Various analyses reconstructing
        $m_{X}$ have been published in 
        the past~\cite{DELPHImX,BABARmX,CLEOmXq2,BellemX}, some of which 
        apply additional cuts on $q^{2}$~\cite{CLEOmXq2,BellemX}. 
        \vs
        A different approach has been followed by ALEPH~\cite{ALEPHVub} and 
        OPAL~\cite{OPALVub} who extract $\Vub$ from measurements
        of the full semileptonic branching fraction by suppressing the dominant 
        $b\to c$ background by means of a neural network.
        \vs
        For an up-to-date review of $\Vub$ determinations see, \eg,
        Ref.~\cite{HFAG}. Following the HFAG recipe, we have only averaged the 
        results from experiments running on the $\Upsilon(4S)$ and obtain
        $\Vub = (4.45 \pm0.09_{\rm stat,sys} \pm0.56_{\rm theo})\times10^{-3}$.
        Similarly as in Ref.~\cite{Gibbons}, we enlarge the theoretical 
        uncertainty due to possible additional uncertainties from subleading 
        shape function effects, annihilation contributions in the high-$q^{2}$ 
        region and quark-hadron duality violations, resulting in
        $\Vub = (4.45 \pm0.09_{\rm stat,sys} \pm0.68_{\rm theo})\times10^{-3}$.

        \subsubsection*{Average}
        
        In the CKM fit, we use $\Vub$ from inclusive and exclusive
        determinations. Usually, one would combine them by 
        multiplying their corresponding likelihoods. However their 
        agreement is marginal so that the two likelihoods only overlap
        if the theoretical uncertainties are driven 
        to their extreme. Given the fact that several theoretical issues 
        are not settled yet and since $\Vub$ is one of the key ingredients 
        of the CKM fit, we adopt a more conservative treatment: the inclusive 
        and exclusive $\Vub$ central values are averaged and as theoretical
        error is assigned the larger one of both determinations. This gives
        $\Vub = (3.90 \pm0.08_{\rm exp} \pm0.68_{\rm theo})\times10^{-3}$.

\subsection{$|\epsk|$}
\label{sec:input_epsk}

        The neutral kaon system provides constraints on the Unitarity Triangle 
        through $\KzKzb$ mixing, indirect\footnote
        {
                The term ``indirect'' comprises \CP violation in mixing and \CP 
                violation in the interference between decay with and 
                without mixing. In the kaon sector, these types of 
                \CP violation are given by ${\rm Re}\epsk$ and ${\rm Im}\epsk$, 
                respectively, which are of similar size.
        } and 
        direct \CP violation, and the rare decays $\Kp\to\pip\nu\nub$ and (yet 
        unknown) $\KL\to\piz\nu\nub$. Only indirect \CP violation is used 
        in the standard CKM fit since the corresponding 
        matrix element can be obtained by Lattice QCD with accountable 
        systematic uncertainties. The SM prediction for neutral kaon mixing 
        suffers from badly controlled long-distance contributions to the mixing 
        amplitudes (see, however, Ref.~\cite{cata} where $\Delta m_K$ is
        found to be short-distance dominated). Moreover, complicated non-perturbative physics with 
        large hadronic uncertainties prevents us from using the measurement 
        of direct \CP violation.  Rare decays are much cleaner and will
        give precise constraints as soon as they are measured with a
        reasonable accuracy. We refer to Part~\ref{sec:kaons} for a 
        dedicated study of direct \CP violation and rare kaon decays.
        \vs
        The most precise measurement of the \CP-violation parameter $\epsk$
        comes from the ratios of amplitudes, $\eta_{+-}$ and $\eta_{00}$,
        of $\KL$ to $\KS$ decaying to pairs of charged and neutral pions,
        respectively,
        \beq
           \epsk = \frac{2}{3}\eta_{+-} + \frac{1}{3}\eta_{00}~.
        \eeq
        We use the average $\left|\epsk\right|=(2.282 \pm 0.017)\times 10^{-3}$,
        obtained from the PDG values for  
        $\eta_{+-}$ and $\eta_{00}$~\cite{PDG} assuming no 
        phase difference between these amplitude ratios, and taking into account
        the correlation induced by the measurements of $\epe$~\cite{Trippe}.~\footnote{Very recently, the KTeV collaboration has presented a new
        precise result on $|\eta_{+-}|=(2.228 \pm 0.010)\times 10^{-3}$~\cite{KTeVepsk} 
        to be compared to the value of $|\eta_{+-}|=(2.286 \pm 0.017)$~\cite{PDG}
        translating in a $2.9~\sigma$ difference. 
        This new value has not been taken into account yet as an input to
        our fit.
        }
        The phase of $\epsk$ is not considered here as it does not depend
        on the CKM matrix elements.
        Other observables related to $\left|\epsk\right|$, like the charge 
        asymmetries $\delta_L$ in semileptonic $\KL$ decays, 
        $\left|\eta_{+-\gamma}\right|$, or decay-plane asymmetries in 
        $\KL\to \pi^+\pi^-e^+e^-$ decays are not considered in this average, 
        since their precision is not competitive.
        \vs
        Within the SM, \CP  violation is induced 
        by $\Delta S=2$ transitions, mediated by box diagrams.
        They can be related to the CKM matrix elements by means of the 
        local hadronic matrix element
        \beq
        \langle
               \Kzb|(\bar{s}\gamma^{\mu}(1-\gamma^{5})d)^2|\Kz
        \rangle
         = \frac{8}{3} m_{K}^2 f_{K}^2 B_{K}~.
        \eeq
        The neutral kaon decay constant, $f_{K}=(159.8\pm1.5)\mev$~\cite{PDG},
        is extracted from the leptonic decay rate $\Gamma(\Kp\to\mup\nu_{\mu})$,
        assuming negligible
        isospin violation. The most reliable prediction of the 
        ``bag'' parameter $B_{K}$, which parameterizes the deviation with
        respect to the vacuum insertion approximation $B_{K}=1$, 
        is obtained from Lattice QCD. 
        At present, calculations are performed assuming SU(3) symmetry and within
        the quenched approximation, $\ie$, neglecting sea-quark contributions 
        in closed loops, which 
        leads to a substantial reduction in computing time. The world average 
        is $B_{K} = 0.86 \pm 0.06 \pm 0.14$~\cite{Lellouch},
        where the first error combines statistical and accountable systematic
        uncertainties and the second is an estimate of the
        bias
        from the quenched approximation
        and SU(3) breaking. Note that analytical approaches based on the
        large-$N_c$ expansion of QCD find a significantly smaller value for
        $B_{K}$ in the chiral limit~\cite{BK_Nc}. Large chiral
        corrections could play an important r\^ole. At present, $B_K$ is
        the first source of theoretical uncertainty in the SM
        prediction of $\varepsilon_K$, while the coupling
        $|V_{ts}|\sim |V_{cb}|$ is the second one.
        \vs
        Neglecting the real part of the non-diagonal element of the 
        neutral kaon mixing matrix $M_{12}$, one obtains~\cite{bbl}\footnote
        {
                Note the non-trivial CKM dependence in Eq.~(\ref{eq_epsk}),
                which only reduces to a hyperbola at lowest 
                orders in $\lambda$, and for values of $\rhobar$ and 
                $\etabar$ close to the origin.
        }
        \beqn
        \label{eq_epsk}
        |\epsk|
                &=& \frac{G_{F}^2 m_{W}^2 m_{K} f_{K}^2}
                       {12 \sqrt{2} \pi^2 \Delta m_{K}}
                  B_{K}\bigg(
                      \eta_{cc} S(x_{c},x_{c})
                         {\rm Im}\big[(V_{cs} V_{cd}^*)^2\big]
                      + \eta_{tt} S(x_{t},x_{t})
                         {\rm Im}\big[(V_{ts} V_{td}^*)^2\big]\nonumber\\
                & &\hspace{3.5cm}
                      +\; 2\eta_{ct} S(x_{c},x_{t})
                         {\rm Im}\big[V_{cs} V_{cd}^* V_{ts} V_{td}^*\big]
                        \bigg)~,
        \eeqn        
        where 
        $\Delta m_{K}=(3.490 \pm 0.006)\times 10^{-12}\mevcc$~\cite{PDG},
        and where $S(x_i,x_j)$ are the Inami-Lim functions~\cite{InamiLim} 
        \beqn
               S(x)\equiv S(x_i,x_j)_{i=j}
                        &=& x\left(\frac{1}{4} + \frac{9}{4(1 - x)}
                               - \frac{3}{2(1 - x)^2}\right)
                          - \frac{3}{2}\left(\frac{x}{1 - x}\right)^{\!3}
                            \ln(x)~, \nonumber \\
        \label{eq_inamilim}
               S(x_i,x_j)_{i\ne j}  &=&  x_i x_j \Bigg[
                                \left(\frac{1}{4} + \frac{3}{2(1 - x_i)}
                                       - \frac{3}{4(1 - x_i)^2}\right)
                                \frac{1}{x_i - x_j}\ln(x_i)\nonumber \\
                           & &  \hspace{1cm}+\;
                                (x_i\leftrightarrow x_j)
                                - \frac{3}{4}\frac{1}{(1 - x_i)(1 - x_j)}
                                      \Bigg]~,
        \eeqn
        with $x_i=m_i^2/m_W^2$ ($i=c,t$). We use the $\MSbar$ masses\footnote
        {
                We derive the value of $\mtRun(m_t)$ from the newest 
                measurement of the pole mass
                $m_t=(178.0\pm2.7\pm3.3)\gevcc$ by the CDF and D0
                collaborations~\cite{tevatronmtop},  
                where the first error given is statistical
                and the second systematic. We apply the pole-to-$\MSbar$ matching
                at three loops~\cite{mtopmatch1,mtopmatch2,mtopmatch3} with
                five light quark flavors,
                where we neglect the mass of the light quark flavors with 
                respect to the $t$-quark mass. This leads to the perturbative
                series
                \beq
                \frac{\mtRun(m_t)}{m_t} 
                        = 1 - \frac{4}{3}\left(\frac{\as}{\pi}\right)
                          - \:9.12530\left(\frac{\as}{\pi}\right)^{\!2}
                          - \:80.4045
                            \left(\frac{\as}{\pi}\right)^{\!3}~,
                \eeq
                where $\as\equiv\as^{(6)}(m_t)=0.1068\pm0.0018$ (see 
                Table~\ref{tab:ckmInputs}) is the $\MSbar$ strong 
                coupling constant for six active quark flavors at the 
                scale of the pole mass. With this we find
                $\mtRun(m_t)=(167.5\pm4.0\pm0.6)\gevcc$, where the 
                first error is experimental and the second is due
                to the truncation of the perturbative series and 
                the uncertainty on $\as$.
        } $\mtRun(m_t)=(167.5\pm4.0\pm0.6)\gevcc$ 
        and $\mcRun(m_c)= (1.2\pm0.2)\gevcc$, 
        where a conservative error is assigned to the running charm quark mass.
        The parameters $\eta_{ij}$ in Eq.~(\ref{eq_epsk}) are next-to-leading 
        order QCD corrections to the Inami-Lim functions.
        We use the values $\eta_{ct} = 0.47\pm 0.04$
        and $\eta_{tt} = 0.5765\pm 0.0065$~\cite{Nierste,Nierste2}, 
        while for $\eta_{cc}$, the parameter with the largest uncertainty, 
        we use the parameterization~\cite{Nierste}
        \beqn
        \label{eq_etacc}
        \eta_{cc} \simeq \left(1.46\pm \delta_{cc}\right)\left[
                          1 - 1.2 \left( \frac{\mcRun(m_c)}
                                              {1.25\gevcc} -1 \right)        
                                \right]         
                                \left[
                         1 + 52 \left( \as(m_Z) - 0.118  \right)
                                \right] ,
        \eeqn
        with an uncertainty from higher-order corrections parameterized 
        by the $\delta_{cc}$ term
        \beqn
                \delta_{cc} = 0.22\left[
                          1 - 1.8 \left( \frac{\mcRun(m_c)}
                                              {1.25\gevcc}-1
                                  \right)
                                \right]
                                \left[
                         1 + 80 \left( \as(m_Z) - 0.118  \right)
                                \right]. 
        \eeqn
        In this way, Eq.~(\ref{eq_etacc}) agrees with the complete 
        NLO calculation within a few percent.   
        With $\mcRun(m_c)$ given above and $\as(m_Z)=0.118\pm 0.003$, 
        and treating the errors as theoretical systematics, we find for 
        $\eta_{cc}$ the range 1.0 -- 2.6 at $95\%~\CL$.
         
\subsection{$\dmd$}

        The $\BzBzb$ oscillation frequency is determined by the mass difference
        $\dmd$ between the two $\Bz$ mass eigenstates, $B_{H}$ and 
        $B_{L}$. It is defined as a positive number and has been measured 
        by many experiments to an average accuracy of almost 
        $1\%$ (see Table~\ref{tab:ckmInputs}).
        In analogy to $|\epsk|$, $\BzBzb$ oscillation in the SM is driven 
        by effective flavor-changing neutral current (FCNC) processes 
        through $\Delta B=2$ box diagrams. However, in contrast to 
        $|\epsk|$ where the large hierarchy in the Inami-Lim 
        functions is partly compensated by the CKM matrix elements, the 
        $\Delta B=2$ box diagrams are dominated by top quark exchange between 
        the virtual $W^\pm$ boson lines. This simplifies the theoretical
        prediction of $\dmd$ which is given by\footnote
        {
                The CKM factor $|V_{td} V_{tb}^*|^2$ occurring in 
                Eq.~(\ref{eqof-dmd2}) approximately describes a circle 
                around $(1,0)$ in the $\rhoeta$ plane, to which a  
                distortion appears at order $\mathcal{O}(\lambda^{10})$:
                \beqns
                \left|V_{td} V_{tb}^*\right|^2 = 
                        \lambda^6 A^2\left[(1-\rhobar)^2 + \etabar^2\right]
                        + 
                        \lambda^{10} A^4(2\rhobar-1)
                        \left[(1-\rhobar)^2 + \etabar^2\right]
                        +
                        \mathcal{O}(\lambda^{12})~.
                \eeqns
        }
        \beq
        \label{eqof-dmd2}
           \dmd = \frac{\GF^2}{6 \pi^2}\etaB m_{B_d}f_{B_d}^2B_d
                  m_W^2 S(x_t) \left|V_{td} V_{tb}^*\right|^2~,
        \eeq
        where $\etaB = 0.551 \pm 0.007$ (for a review, see Ref.~\cite{bbl})
        is a perturbative QCD correction to the Inami-Lim function $S(x_t)$
        from perturbative QCD. The matrix 
        element $f_{B_d} \sqrt{B_d}$ is taken from Lattice QCD. 
        Much progress has been achieved 
        in this domain recently, where partially unquenched
        calculations are now available. Nevertheless, there is an ongoing discussion
        in the Lattice community whether the extrapolation for the light $d$ quark 
        mass to the chiral-limit is well-understood (see the discussion 
        of $\dms$ below for further details).
        Here, we use the value and errors derived in Ref.~\cite{Becirevic2}
        (\cf\  Table~\ref{tab:ckmInputs}).
        \vs
        For leptonic \B decays and the semileptonic \CP asymmetry 
        $A_{\rm SL}$ discussed in Part~\ref{sec:newPhysics},
        the decay constant $f_{B_d}$ and the bag parameter ${B_d}$ are needed 
        separately. The values and uncertainties used 
        (\cf\  Table~\ref{tab:ckmInputs}) are also taken from  
        Ref.~\cite{Becirevic2}.
       
\subsection{$\dms$} 

        In the SM, the mass difference $\dms$ between the heavy and the light
        $\Bs$ mesons has only a weak relative dependence on the Wolfenstein
        parameters $(\rhobar,\etabar)$. Nevertheless, a measurement 
        of $\dms$ is useful in the CKM fit since it indirectly leads to
        an improvement of the constraint from the $\dmd$ measurement on 
        $\left|V_{td} V_{tb}^*\right|^2$. The SM prediction 
        \beq
        \label{eqof-dms1}
           \dms = \frac{\GF^2}{6 \pi^2}\etaB m_{B_s}\fbs^2 B_s
                  m_W^2 S(x_t) \left|V_{ts} V_{tb}^*\right|^2~,
        \eeq
        can be rewritten as
        \beq
        \label{eqof-dms2}
           \dms = \frac{\GF^2}{6 \pi^2}\etaB m_{B_s} \xi^2 \fbd^2 B_d
                  m_W^2 S(x_t) \left|V_{ts} V_{tb}^*\right|^2~,
        \eeq
        where the parameter $\xi = \fbs \sqrt{B_s}/\fbd \sqrt{B_d}$ quantifies
        SU(3)-breaking corrections to the matrix elements, which can 
        be calculated more accurately in Lattice QCD than the matrix 
        elements themselves. Hence a measurement of $\dms$ improves the knowledge 
        of $\fbd \sqrt{B_d}$. In our previous analysis~\cite{CKMfitter} we 
        used the value $\xi = 1.16 \pm 0.05$. Recently, the uncertainty on 
        $\xi$ from Lattice QCD has been revisited: Lattice calculations
        using Wilson fermions have to work with light quark masses of 
        ${\cal{O}}(100\mevcc)$, so that calculations for $\Bd$ 
        mesons need to be extrapolated to the ``chiral limit'' (this is 
        not necessary for the $\Bs$ due to the heavy strange quark). 
        This process is controversial because of the potential presence 
        of a curvature in the chiral extrapolation curve
        (see \eg\  Refs.~\cite{Booth,SharpeZhang,BernardBlumSoni,Yamada,Kronfeld,Becirevic1}).
        The recent development using ``staggered'' fermions allows one to perform 
        Lattice QCD calculations with significantly smaller light quark masses. 
        So far, these studies do not show a significant enhancement of 
        $\xi$~\cite{Davies} but one should keep in mind that the 
        interpretation of results obtained with the use of staggered fermions 
        is still under discussions (\cf\  \eg\  Ref.~\cite{Becirevic2}).
        Based on a phenomenological analysis, it has been shown in 
        Ref.~\cite{Becirevic1} that in the chiral limit, the double-ratio 
        $(\fbs/\fbd)/(f_{K}/f_{\pi})$ does not differ much from 
        unity resulting in $\xi = 1.21 \pm 0.04 \pm 0.05$, where the second
        error is due to the chiral extrapolation\footnote
        {
                        Note that the latter error is strongly correlated
                        with the one on $\fbd \sqrt{B_d}$, because both
                        have the same source. We neglect this
                        correlation, which may result in an
                        underestimation of the impact of the bound on
                        $\dms$.
        }.
        Interestingly, based on a quark model,
        the Lattice value of $\xi \approx 1.16$ was considered too
        low~\cite{Rosner}, even before the discussion about the chiral 
        extrapolation had started. 
        \vs
        Limits on $\dms$ from the search for $\BszBszb$ oscillation 
        have been obtained by several 
        experiments~\cite{dmsALEPH,dmsDELPHI,dmsOPAL,dmsSLD,dmsCDF}.
        A convenient approach to average various results on $\dms$ is the 
        Amplitude Method~\cite{Roussarie} (see also the exhaustive study 
        in Ref.~\cite{toy}), which consists of introducing an {\it ad hoc} 
        amplitude coefficient, $\Ampli$, placed in front of the cosine 
        modulation term (see Appendix~\ref{sec:amplitudeMethod} for further
        details). The advantage of this indirect probe for oscillation 
        stems from the fact that the dependence on $\Ampli$ is 
        linear and hence the measurement of $\Ampli$ is Gaussian,
        so that merging different experimental measurements is straightforward.
        One can then define the experimental sensitivity for given 
        $\dms$ by $1.645 \times \sigma_{\Ampli}(\dms)$ (found to be 
        $18.7~{\rm ps}^{-1}$~\cite{HFAG}), and a 
        $95\%~\CL$ lower limit for $\dms$, given by the sum of the sensitivity 
        and the central value of the measured amplitude. It is found to be
        $14.4~{\rm ps}^{-1}$~\cite{HFAG}.
        \vs
        The question on how to deduce the confidence level from the 
        available experimental information is crucial to the CKM analysis 
        and has been scrutinized on many 
        occasions~\cite{Roussarie,toy,checchiaLogLik,Achille1,CKMfitter,CernCkmWS}.
        Traditionally, $\BszBszb$ oscillation results have been implemented into 
        fits using $\chi^2_{|1-\Ampli|} = ((1-\Ampli)/\sigma_{\Ampli})^2$ with
        ${\rm CL}(\chi^2_{|1-\Ampli|}) = {\rm Erfc}(|1-\Ampli|/\sigma_{\Ampli}/\sqrt{2}))$~\cite{Roussarie}.
        However this procedure does not properly interpret the information of the 
        amplitude spectrum. 
        For instance, two measured amplitudes $\Ampli_{1}$ and $\Ampli_{2}$, where 
        $\Ampli_{1} > 1$ and $\Ampli_{2} < 1$ but $\Ampli_{1}-1 = 1-\Ampli_{2}$, 
        result in the same confidence level in this approach although it would be 
        natural to assign a larger likelihood for an oscillation to $\Ampli_{1}$ 
        than to $\Ampli_{2}$.
        In Ref.~\cite{CKMfitter}, an alternative procedure, which exploits the 
        information from the sign of $1-\Ampli$ by omitting the modulus in the 
        above definition of $\chi^2_{|1-\Ampli|}$, has been proposed:
        $
        \chi^2_{1-\Ampli} = 2\cdot
        \left[{\rm Erfc}^{-1}
        \left(
                \frac{1}{2}\,{\rm Erfc}
                \left(\frac{1-\Ampli}{\sqrt{2}\sigma_\Ampli}\right)
        \right)
        \right]^{2}.
        $
        As pointed out in Ref.~\cite{CernCkmWS}, this procedure is an approximation 
        and can lead to a bias in presence of a true measurement. The information 
        from the fit to the proper time distributions of mixed and unmixed $\BszBszb$ 
        decays is obtained from the ratio of the likelihood at given frequency $\dms$, 
        ${\cal L}(\Delta m_s)$, to the likelihood at infinity, 
        ${\cal L}(\Delta m_s = \infty)$~\cite{Roussarie,checchiaLogLik,Achille1},
        for which the logarithm reads
        \beq
        \label{eq_likratio}
                2\Delta \ln {\cal L}^{\infty}(\Delta m_s) 
                   =   \frac{(1-\Ampli)^2}{\sigma_{\Ampli}^2} 
                     - \frac{\Ampli^2}{\sigma_{\Ampli}^2}~.
        \eeq
        In Appendix $A$ we propose a frequentist method
        to deduce a confidence level from this information.
        This method is the one used in our CKM fit.

\subsection{$\Vts$}
\label{sec:input_vts}

        Besides $\dms$, the inclusive branching fraction of the radiative 
        decay $B \to X_{s} \gamma$ determines the product $|V_{ts}V_{tb}^*|$.
        Using the measurements from CLEO, ALEPH, \babar  and Belle, 
        Ref.~\cite{AliMisiak} quotes $|V_{ts}V_{tb}^*| = 0.047 \pm 0.008$. 
        Since the precision of this constraint is not competitive
        it is not used in our fit.
        \vs
        The ratio $|V_{td}/V_{ts}|$ can be determined from the ratio of the 
        exclusive rates for the decays $\B\to\rho\gamma$ to 
        $\B\to\Kstar\gamma$, which eliminates the form factor dependencies
        up to SU(3) breaking. Based on the recent $3.5\sigma$ evidence 
        for the decay $\B\to\rho\gamma$ found by Belle~\cite{bellerhogam},
        a phenomenological study has been performed in Ref.~\cite{alirhogam}. 
        The constraint on $|V_{td}/V_{ts}|$ derived from the measurement is
        found to be in agreement with the expectation, but is not (yet) 
        accurate enough to represent a competitive input in the global 
        CKM fit.

\subsection{$\stb$}
\label{sec:input_s2b}

        In $b \to c \bar{c} s$ quark-level decays, the time-dependent 
        \CP-violation parameter $S$ measured from the interference 
        between decays with and without mixing is equal to $\stb$ 
        to a very good approximation. The world average uses measurements from
        the decays $\Bz \to \jpsi \KS$, $\jpsi \KL$, $\psitwos \KS$, 
        $\chicone\KS$, $\etac\KS$ and
        $\jpsi K^{*0}$ $(K^{*0} \to \KS\pi^{0})$ and gives 
        $\stbwa = 0.739 \pm 0.048$~\cite{HFAG}. It is dominated by the 
        measurements from \babar~\cite{BABARstb} and Belle~\cite{Bellestb}.
        In $b \to c \bar{c} d$ quark-level decays, such as 
        $\Bz \to \jpsi \pi^{0}$ or $\Bz \to D^{(*)}D^{(*)}$, unknown contributions 
        from (not CKM-suppressed) penguin-type diagrams carrying a different weak 
        phase than the tree-level diagram compromises the clean extraction of $\stb$.
        As a consequence, they are not taken into account in the $\stb$ average.
        \vs
        Within the SM, decays mediated by the loop transitions 
        $b \to s \bar{q} q$ ($q=u,d,s$),
        such as $\Bz \to \phi \Kz$ or $\Bz \to K^{+} K^{-} \KS$, 
        but also the recently measured $\Bz \to f_0\KS$, as well as $\Bz \to \etapr\KS$ and 
        $\Bz\to\piz\KS$, can be used to extract $\stb$ in a relatively
        clean way (see Refs.~\cite{RFpeng,sonipeng,ligetiquinn,grossrosner,BN} and 
        Refs.~\cite{BABARphiks,Bellephiks,babarallbeta,babarf0ks,babaretaprks,HFAG}
        for the experimental results).
        Due to the large virtual mass scales occurring in the penguin loops,
        additional diagrams with heavy particles in the loops and new
        \CP-violating phases may contribute. Such a measurement of the weak phase
        from mixing-induced \CP violation and the comparison with the SM 
        expectation is therefore a sensitive probe for physics beyond the SM.
        Assuming penguin dominance and neglecting CKM-suppressed amplitudes,
        these decays carry approximately the same weak phase as the decay 
        $\Bz\to\jpsi\KS$. As a consequence, their mixing-induced \CP-violation
        parameters are expected to be $-\eta_f\times\stb$ to a 
        reasonable accuracy in the SM, where $\eta_f$ is the \CP eigenvalue 
        of the final state. Recent measurements from \babar and Belle give 
        conflicting results for the mixing-induced \CP parameter of $\phi\Kz$:
        with the Belle result, 
        $S_{\phi \KS} = -0.96 \pm0.50 ^{+0.09}_{-0.11}$~\cite{Bellephiks},
        indicating a $3.3\sigma$ deviation from the SM,
        while \babar finds (using both $\phi\KS$ and $\phi\KL$)
        $-\eta_{\phi K}S_{\phi K}=0.47\pm0.34^{\,+0.08}_{\,-0.06}$~\cite{BABARphiks},
        which is in agreement with $\stbwa$. Moreover, the world averages
        $-S_{K^{+} K^{-} \KS}=0.54\pm0.18^{\,+0.17}_{\,-0}$ 
        and $S_{\etapr\KS}=0.27\pm0.21$~\cite{HFAG} (\babar\   and Belle
        agree for these) do not show  a significant departure from the 
        charmonium reference, though both tend to support the observation 
        $\stb_{{\rm eff},\,sq\bar q}<\stbwa$. Finally, 
        the recent \babar  measurement
        $-S_{f_0 \KS}=1.62^{+0.51}_{-0.56} \pm{0.10}$~\cite{babarf0ks}
        is in decent agreement with $\stbwa$.
        \vs
        A more detailed numerical discussion of the various $\stb$ results is 
        given in Section~\ref{sec:standardFit}.\ref{sec:sin2bProblem}. 
        At present, we do not include the results from penguin-dominated 
        decays in the $\stb$ average.

\subsection{$\stwoa$}
\label{sec:input_rhorho}

        The measurement of the time-dependent 
        \CP-violating asymmetries in the charmless decay $\Bz \to \rho^{+}\rho^{-}$ 
        allows us to derive a significant constraint on the angle $\stwoa$ using the 
        Gronau--London isospin analysis~\cite{grolon} (extended here to include 
        electroweak penguins), which invokes $\B \to \rho\rho$ decays
        of all charges. While the full isospin analysis requires the 
        measurement of (at least one of) the time-dependent $\CP$ 
        parameters\footnote
        {
                Note that in contrast to $\Bz\to\piz\piz$ both, $C_{\rho\rho,L}^{00}$ 
                and $S_{\rho\rho,L}^{00}$, are experimentally accessible in 
                $\Bz\to\rho^0\rho^0$.
                Their measurement overconstrains the isospin analysis and
                can be used to remove some of the ambiguities on $\alpha$
                (see 
                Section~\ref{sec:charmlessBDecays}.\ref{sec:introductionrhorho}).
        } in
        the color-suppressed decay $\Bz\to\rho^0\rho^0$, the
        available upper limit on its branching fraction can be used to 
        constrain $|\alpha-\alphaeff|$. Albeit analytical bounds have
        been derived for this case~\cite{GrQu,theseJerome,GrLoSiSi}, the numerical
        analysis performed in \ckmfitter\  leads to equivalent results (see
        the detailed discussions of the isospin analysis in 
        Sections~\ref{sec:charmlessBDecays}.\ref{sec:su2pipi} 
        and \ref{sec:charmlessBDecays}.\ref{sec:introductionrhorho}).
        \vs
        For the isospin-related decays we use the branching fractions
        $\BR(\Bz\to \rho^{+}\rho^{-})=(30 \pm 4 \pm 5)\tmsix$~\cite{babarrhorhoprl}
        (see also the first observation and polarization measurement of
        this mode in Ref.~\cite{rhorhoold}),
        $\BR(\Bp\to\rho^+\rho^{0})=(26.4^{\,+6.1}_{\,-6.4})\tmsix$~\cite{HFAG,BABARrhoplusrho0,Bellerhoplusrho0}, 
        and the upper limit at $90\%~\CL$
        $\BR(\Bz\to\rho^0\rho^0)<2.1\tmsix$~\cite{BABARrhoplusrho0} (in the CKM
        fit we use the result 
        $\BR(\Bz\to\rho^0\rho^0)=(0.6^{\,+0.8}_{\,-0.6}\pm0.1)\tmsix$,
        which leads to this limit).
        The $\rho$ mesons in the decays $\Bz \to \rho^{+}\rho^{-}$ and 
        $\Bp\to \rho^+\rho^{0}$ are found to be longitudinally 
        polarized with the longitudinal fractions ($f_{L}\equiv\Gamma_L/\Gamma$):
        $f_{L}^{+-}= 0.99 \pm 0.03 ^{\,+0.04}_{\,-0.03}$~\cite{babarrhorhoprl} and 
        $f_{L}^{+0}= 0.962^{+0.049}_{-0.065}$~\cite{BABARrhoplusrho0,Bellerhoplusrho0}, 
        respectively. As a consequence, the $B \to \rho\rho$ system is 
        actually like the $\B \to \pi\pi$ system. Assuming (conservatively) the 
        relative polarization of the $\rho^{0}$ mesons in $\Bz\to\rho^0\rho^0$
        to be fully longitudinal, and using the \CP asymmetries 
        $S_{\rho\rho,L}^{+-}=-0.19 \pm 0.33 \pm 0.11$ and
        $C_{\rho\rho,L}^{+-}=-0.23 \pm 0.24 \pm 0.14$ measured
        by \babar~\cite{babarrhorhoprl,BABARrhorho} for the longitudinal fraction
        of the $\Bz\to\rho^+\rho^-$ event sample, together with the branching 
        fraction and polarization measurements for the other charges, we 
        obtain constraints on $\stwoa$ as described in 
        Section~\ref{sec:charmlessBDecays}.\ref{sec:introductionrhorho}.
        \vs
        Note that the present analysis neglects non-resonant contributions 
        and possible other $\pi^+\pi^-$ resonances under the $\rho^0$,
        as well as effects from the radial excitations $\rho(1450)$ 
        and $\rho(1700)$ that were found to be significant contributors to the 
        pion form factor in $\epem$ annihilation~\cite{cmd2} and $\tau$ 
        decays~\cite{aleph_vsf}. Also neglected are isospin-violating 
        contributions due to the finite width of the $\rho$~\cite{ligetirhorho},
        as well as electromagnetic and strong sources of isospin 
        violation (see the 
        discussion in Section~\ref{sec:charmlessBDecays}.\ref{sec:introductionrhorho}).

\begin{table}[p]
\begin{center}
\setlength{\tabcolsep}{0.0pc}
{\small
\begin{tabular*}{\textwidth}{@{\extracolsep{\fill}}lcccc}\hline
&&&& \\[-0.3cm]
        &       &       &       \mc{2}{c}{Errors}       \\
  \rs{Parameter}        & \rs{Value $\pm$ Error(s)}
                & \rs{Reference}
                & GS
                & TH \\[0.15cm]
\hline
&&&& \\
  \rs{$\Vud$ (neutrons)}        
                & \rs{$0.9717\pm0.0013\pm0.0004$}
                & \rs{(see text)}
                & \rs{$\star$} & \rs{$\star$} \\
  \rs{$\Vud$ (nuclei)}  
                & \rs{$0.9740\pm0.0001\pm0.0008$}
                & \rs{(see text)}
                & \rs{$\star$} & \rs{$\star$} \\
  \rs{$\Vud$ (pions) }
                & \rs{$0.9765\pm0.0055\pm0.0005$}
                & \rs{\cite{PIBETA}}
                & \rs{$\star$} & \rs{$\star$} \\
  \rs{$\Vus$}   
                & \rs{$0.2228\pm0.0039\pm0.0018$}
                & \rs{see text}
                & \rs{$\star$} & \rs{$\star$} \\
  \rs{$\Vub$ (average)}         
                & \rs{$(3.90 \pm 0.08 \pm 0.68) \times 10^{-3}$}
                & \rs{see text, \cite{HFAG}}
                & \rs{$\star$} & \rs{$\star$} \\
  \rs{$\Vcb$ (incl.)}   
                & \rs{$(42.0\pm 0.6 \pm 0.8)\times10^{-3}$}
                & \rs{see text}
                & \rs{$\star$} & \rs{$\star$} \\
  \rs{$\Vcb$ (excl.)}   
                & \rs{$40.2^{\,+2.1}_{\,-1.8}\times10^{-3}$}
                & \rs{\cite{HFAG,Hashimoto}}
                & \rs{$\star$} & \rs{-} \\
\hline
&&&& \\
  \rs{$|\epsk|$}        
                & \rs{$(2.282\pm0.017)\times10^{-3}$}
                & \rs{\cite{Trippe}}
                & \rs{$\star$} & \rs{-} \\
  \rs{$\dmd$}           
                & \rs{$(0.502 \pm 0.006)~{\rm ps}^{-1}$}
                & \rs{\cite{HFAG}}
                & \rs{$\star$} & \rs{-} \\
  \rs{$\dms$}           
                & \rs{\footnotesize Amplitude spectrum}
                & \rs{\cite{HFAG}}
                & \rs{$\star$} & \rs{-} \\
  \rs{$\stbwa$} & \rs{$0.739 \pm 0.048$}
                & \rs{\cite{HFAG}}
                & \rs{$\star$} & \rs{-} \\
  \rs{$S_{\rho\rho,L}^{+-}$} & \rs{$-0.19 \pm 0.35$}
                & \rs{see text}
                & \rs{$\star$} & \rs{-} \\
  \rs{$C_{\rho\rho,L}^{+-}$} & \rs{$-0.23 \pm 0.28$}
                & \rs{see text}
                & \rs{$\star$} & \rs{-} \\
  \rs{$\BR_{\rho\rho,L}$ all charges}  & \rs{see text}
                & \rs{see text}
                & \rs{$\star$} & \rs{-} \\
\hline
&&&& \\
  \rs{$\mcRun(m_c)$}    & \rs{$(1.2\pm0.2)\gevcc$}
                & \rs{\cite{PDG}}
                & \rs{-} 
                & \rs{$\star$} \\
  \rs{$\mtRun(m_t)$}    
                & \rs{$(167.5\pm4.0\pm0.6)\gevcc$}
                & \rs{\cite{PDG}}
                & \rs{$\star$} 
                & \rs{-} \\
  \rs{$m_{\Kp}$}            
                & \rs{$(493.677\pm0.016)\mevcc$}
                & \rs{\cite{PDG}}
                & \rs{-}
                & \rs{-} \\
  \rs{$\Delta m_K$}     
                & \rs{$(3.490 \pm 0.006)\times 10^{-12}\mevcc$}
                & \rs{\cite{PDG}}
                & \rs{-}
                & \rs{-} \\
  \rs{$m_{B_d}$}        
                & \rs{$(5.2794\pm0.0005)\gevcc$}
                & \rs{\cite{PDG}}
                & \rs{-}
                & \rs{-} \\
  \rs{$m_{B_s}$}        
                & \rs{$(5.3696\pm0.0024)\gevcc$}
                & \rs{\cite{PDG}}
                & \rs{-}
                & \rs{-} \\
  \rs{$m_W$}            
                & \rs{$(80.423\pm0.039)\gevcc$}
                & \rs{\cite{PDG}}
                & \rs{-}
                & \rs{-} \\
  \rs{$G_F$}            
                & \rs{$1.16639\times 10^{-5}\gev{}^{-2}$}
                & \rs{\cite{PDG}}
                & \rs{-}
                & \rs{-} \\
  \rs{$f_K$}            
                & \rs{$(159.8\pm1.5)\mev$}
                & \rs{\cite{PDG}}
                & \rs{-}
                & \rs{-} \\
\hline
&&&& \\
  \rs{$B_K$}            
                & \rs{$0.86\pm0.06\pm0.14$}
                & \rs{\cite{Lellouch}}
                & \rs{$\star$} 
                & \rs{$\star$} \\
  \rs{$\as(m_Z^2)$}
                & \rs{$0.1172\pm0.0020$}
                & \rs{\cite{PDG}}
                & \rs{-} 
                & \rs{$\star$} \\
  \rs{$\eta_{ct}$}      
                & \rs{$0.47\pm0.04$}
                & \rs{\cite{Nierste}}
                & \rs{-}
                & \rs{$\star$} \\
  \rs{$\eta_{tt}$}
                & \rs{$0.5765\pm0.0065$}
                & \rs{\cite{Nierste,Nierste2}}
                & \rs{-}
                & \rs{-} \\
  \rs{$\etaB(\MSbar)$}
                & \rs{$0.551\pm0.007$ }
                & \rs{\cite{bbl}}
                & \rs{-}
                & \rs{$\star$} \\
  \rs{$\fbd\sqrt{B_d}$}
                & \rs{$(228\pm30\pm10)\mev$}
                & \rs{\cite{Becirevic2}}
                & \rs{$\star$} 
                & \rs{$\star$} \\
  \rs{$\fbd$}
                & \rs{$(200\pm28\pm9)\mev$}
                & \rs{\cite{Becirevic2}}
                & \rs{$\star$} 
                & \rs{$\star$} \\
  \rs{$B_d$}
                & \rs{$1.3\pm0.12$}
                & \rs{\cite{Becirevic2}}
                & \rs{$\star$} 
                & \rs{$\star$} \\
  \rs{$\xi$}            
                & \rs{$1.21 \pm 0.04 \pm 0.05$}
                & \rs{\cite{Becirevic2}}
                & \rs{$\star$} 
                & \rs{$\star$} \\
  \rs{$B\to\rho\rho$ amplitude params.}         
                & \rs{all floating}
                & \rs{see text}
                & \rs{-} 
                & \rs{$\star$} \\
\hline
\end{tabular*}
}
\caption{\label{tab:ckmInputs} \em
        Inputs to the standard CKM fit. 
        If not stated otherwise: for two errors given, the 
        first is statistical and accountable systematic and the 
        second stands for systematic theoretical uncertainties.
        The fourth and fifth columns indicate
        the treatment of the input parameters within \rfit: 
        measurements or parameters that have statistical errors 
        (we include here experimental systematics)
        are marked in the ``GS'' column by an asterisk; measurements
        or parameters that have systematic theoretical
        errors, treated as ranges in \rfit, are marked 
        in the ``TH'' column by an asterisk.
        \underline{Upper part:} 
        experimental determinations of the CKM matrix elements.
        \underline{Middle upper part:} 
        \CP-violation and mixing observables.
        \underline{Middle lower part:} 
        parameters used in SM predictions that are 
        obtained from experiment.
        \underline{Lower part:} 
        parameters of the SM predictions obtained from 
        theory.}
\end{center}
\end{table}

\section{Results of the Global Fit}
\label{sec:standardFitResults}

The standard CKM fit includes those observables for
which the Standard Model predictions (and hence the CKM constraints) 
can be considered to be quantitatively under control. We only 
take into account measurements that lead to significant and competitive 
constraints on the CKM parameters. The {\em standard observables} are:
\beqn
\label{eq:standardFitObs}
   \Vus~,\hskip .3truecm   
   \Vud~,\hskip .3truecm   
   \Vub~,\hskip .3truecm   
   \Vcb~,\hskip .3truecm   
   |\epsk|~,\hskip .3truecm   
   \dmd~,\hskip .3truecm   
   \dms~,\hskip .3truecm   
   \stbwa~,\hskip .3truecm   
   \stwoa_{[\rho\rho]}~.
\eeqn
The theoretical uncertainties related to these observables
are discussed in Section~\ref{sec:standardFit}.\ref{sec:fitInputs}.
\vs
Among the observables that are not (yet) considered are the following.
\bei

\item   Measurements of the remaining CKM elements as well 
        as the constraints from the rare kaon decay 
        $K^+\to\pi^+\nu\nub$ and from $\B\to\rho\gamma$ are 
        not (yet) precise enough to improve the knowledge of the \CKM\  
        matrix (\cf\  Section~\ref{sec:standardFit}.\ref{sec:input_vts}). 

\item   The theoretical prediction of direct CPV in kaon decays 
        ($\epe$) is not yet settled (\cf\  
        Section~\ref{sec:standardFit}.\ref{sec:fitInputs}). 

\item   Charmless $B$ decays other than $\B\to\rho\rho$
        also lead to various interesting 
        constraints on the angles of the Unitarity Triangle
        and provide sensitivity to contributions from physics beyond the SM.
        Detailed discussions are given in Part~\ref{sec:charmlessBDecays}.
        In principle, amplitude analyses using SU(2)
        symmetry are theoretically safe (the 
        isospin-breaking contributions can be controlled) so 
        that they will be used in the standard CKM fit once they lead
        to significant results. SU(3)-based analyses
        are in general more constraining, with however the limitation 
        that theoretical uncertainties are more difficult to control.

\item   Various constraints on the angle $\gamma$ can be obtained
        from the comparison of CKM-favored and CKM-suppressed
        $b\to c$ decays (\cf\  Part~\ref{sec:gamma}). The accumulated 
        statistic are, however, not yet sufficient to perform fully
        data-driven analyses and to eliminate theoretical 
        input with uncertain errors.

\eei

We use the observables~(\ref{eq:standardFitObs})
to perform constrained fits to the CKM parameters and related quantities. 
We place ourselves in the framework of the \rfit\ scheme (\cf\  
Part~\ref{sec:statistics}) and hence define the theoretical 
likelihoods of Eq.~(\ref{eq_likThe}) to be one within the allowed 
ranges and zero outside. In other words, we use $\kappa=0$ and 
$\zeta=1$ for the Hat function $\Hatsyst(\xo)$ defined in
Eq.~(\ref{eq_flattenedGaussian}). As a consequence, no hierarchy 
is introduced for any permitted set of theoretical parameters, \ie, 
the $\chi^2$ that is minimized in the fit receives no contribution 
from theoretical systematics. However the theoretical parameters cannot 
trespass their allowed ranges. When relevant, statistical and 
theoretical uncertainties are combined beforehand, following the
procedure outlined in 
Section~\ref{sec:statistics}.\ref{sec:flattenedGaussian}.
Floating theoretical parameters are labelled by an asterisk in 
the ``TH'' column of Table~\ref{tab:ckmInputs}. For parameters 
with insignificant theoretical uncertainties, the errors are propagated 
through the theoretical predictions, and added in quadrature to 
the experimental error of the corresponding measurements\footnote
{
        This procedure neglects the correlations occurring when 
        such parameters are used in more than one theoretical 
        prediction. 
}.

%
%
\subsection{Probing the Standard Model}
\label{sec:compOfTheSM}

\begin{figure}[t]
  \epsfxsize10cm
  \centerline{\epsffile{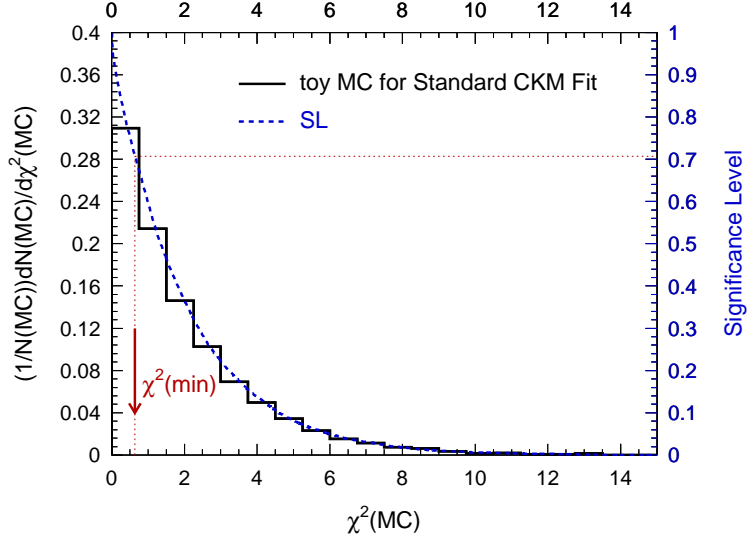}}
  \caption[.]{\label{fig:chi2min}\em
        Simulated $\F(\chi^2)$ distribution and corresponding
        SL curve for the standard CKM fit. The arrow indicates the 
        corresponding minimal $\ChiMinGlob$ found in the analysis.}
\end{figure}
We have demonstrated in Part~\ref{sec:statistics} 
that the metrological phase is intrinsically 
unable to detect a failure of the SM to describe the data. We therefore 
begin the CKM analysis with an interpretation of the test statistics 
$\ChiMinGlob$, which is a probe of the goodness-of-fit test for the SM
hypothesis. 
We perform the toy Monte Carlo simulation described in 
Section~\ref{sec:statistics}.\ref{sec:probingTheSM}.
The standard CKM fit returns after convergence
\beq
   \ChiMinGlob = 0.6~,
\eeq
for the full data set (including $\stbwa$ and $\stwoa_{[\rho\rho]}$,
where the latter has little impact only). We generate the probability 
density distribution 
$\F(\chi^2)$ of $\ChiMinGlob$ by fluctuating the measurements and 
$\yQCD$ parameters according to their non-theoretical errors around 
the theoretical values obtained with the use of the parameter set $\ymodopt$ 
for which $\ChiMinGlob$ is obtained. The resulting toy distribution 
is shown by the histogram in Fig.~\ref{fig:chi2min}. Integrating the 
distribution according to Eq.~(\ref{eq_monteCarlo}) leads to the 
significance level ($\SL$) represented by the smooth curve in 
Fig.~\ref{fig:chi2min}. We find a p-value of
\beq
   \Prob(\ChiMinGlob|{\rm SM}) 
        \le {\SL}(\ChiMinGlob) 
        = 71\%~,
\eeq
for the validity of the SM. One notices that compared to previous 
fits~\cite{CKMfitter}, the ``unitarity problem'' in the first row, 
that is the incompatibility between $\Vud$ and $1-\Vus^2$, becomes 
insignificant with the likelihoods we use for these two quantities 
(see the discussion in Section~\ref{sec:input_vud} and subsequent 
paragraphs). Their average has $\chi^2_{{\rm min};\Vud,\Vus}=0.16$.
\vs
The large p-value of the electroweak sector of the SM when confronted 
with all CKM-related data strongly supports the KM mechanism~\cite{kmmatrix} 
as the dominant source of \CP  violation at the electroweak scale. 
It is the necessary condition that permits us to move to the CKM 
metrology.

\subsection{Metrology of the CKM Phase}
\label{sec:metrologyStandardFit}

\begin{figure}[p]
  \epsfxsize\smallfig
  \centerline{\epsffile{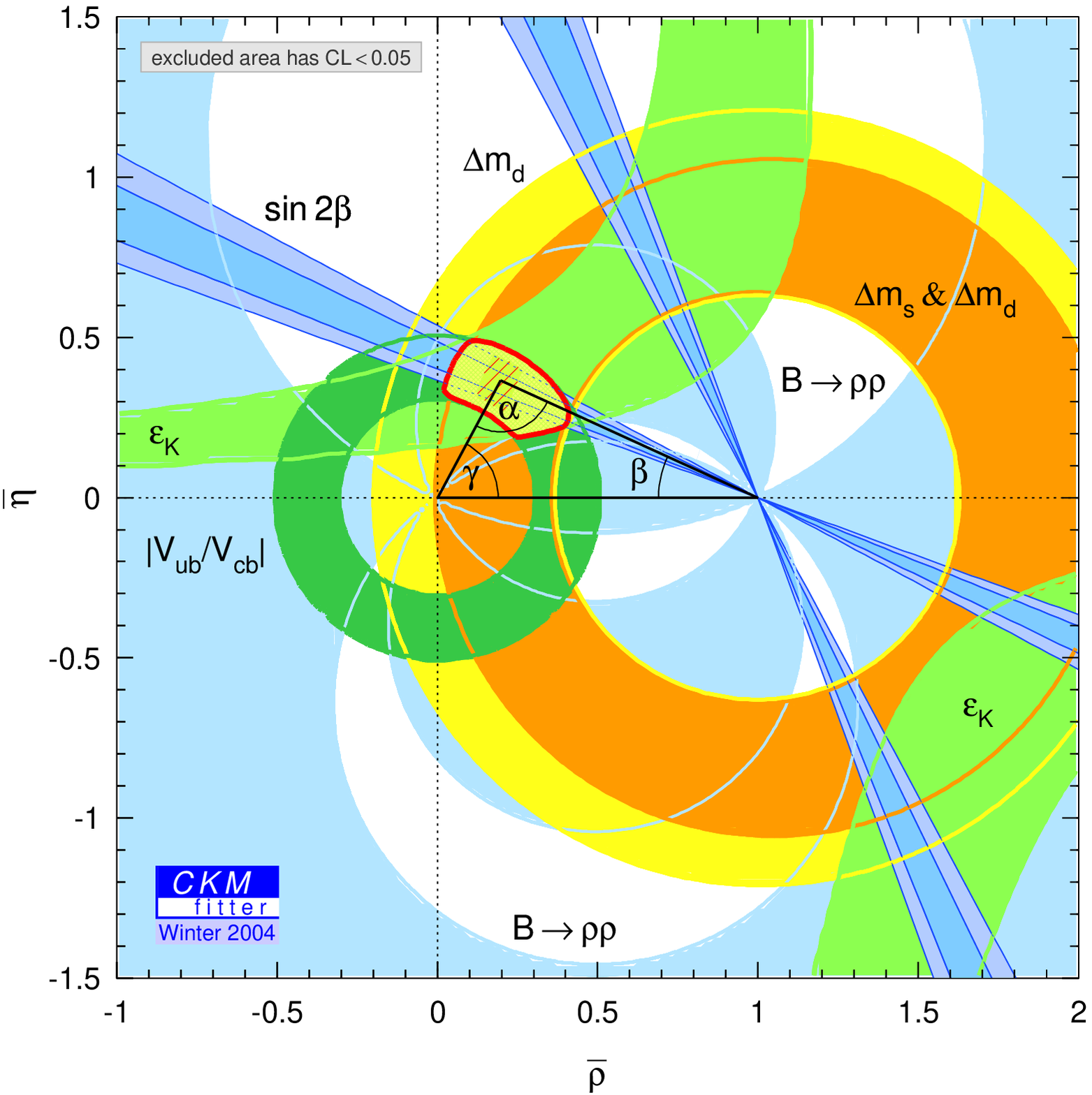}}  
  \vspace{0.4cm}
  \centerline{\epsffile{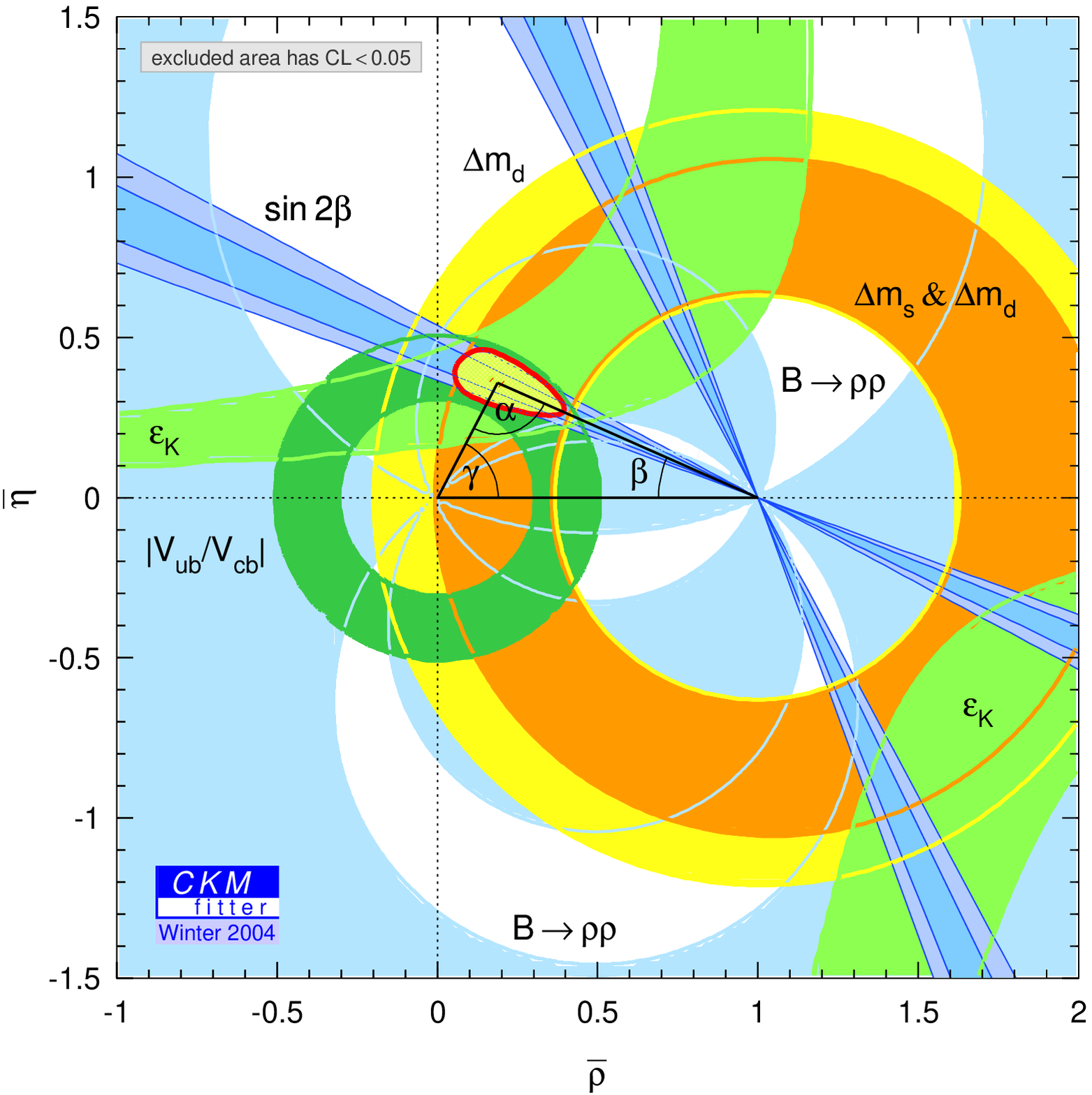}}
  \caption[.]{\label{fig:rhoeta}\em
        Confidence levels in the enlarged $\rhoeta$ plane for the 
        global CKM fit. The shaded areas indicate the regions of
        $\ge5\%$ CLs. For $\stb$ the $\ge32\%$ and $\ge5\%$~CL 
        constraints are shown. The upper (lower) plot excludes
        (includes) the constraints from $\stb$ and $\sta$ in the
        combined fit. The hatched area in the center of the combined 
        fit result indicates the region where theoretical errors 
        dominate. The lower plot corresponds to the standard CKM fit.}
\end{figure}

\begin{figure}[t]
  \epsfxsize\smallfig
  \centerline{\epsffile{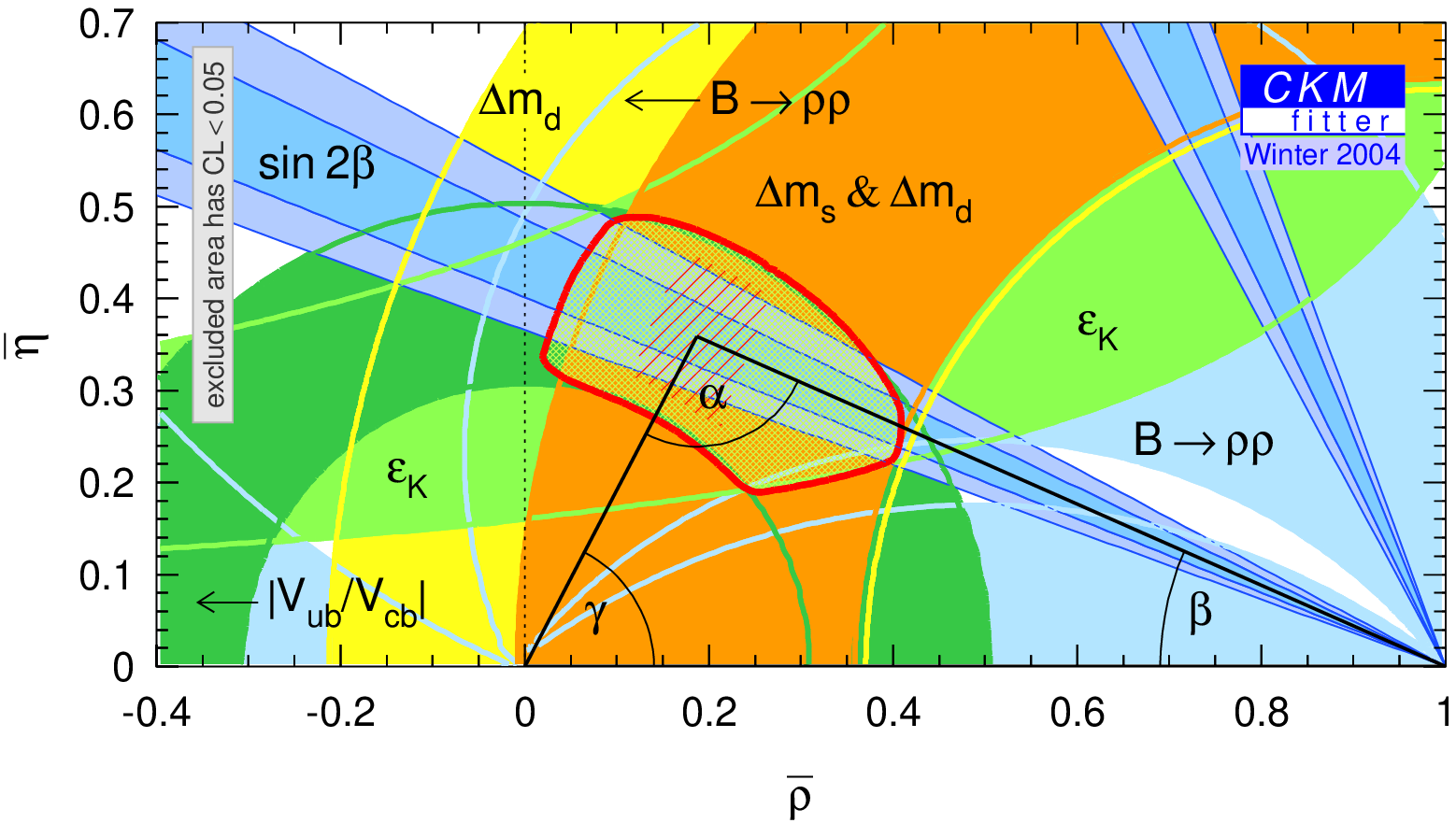}}  
  \vspace{0.4cm}
  \centerline{\epsffile{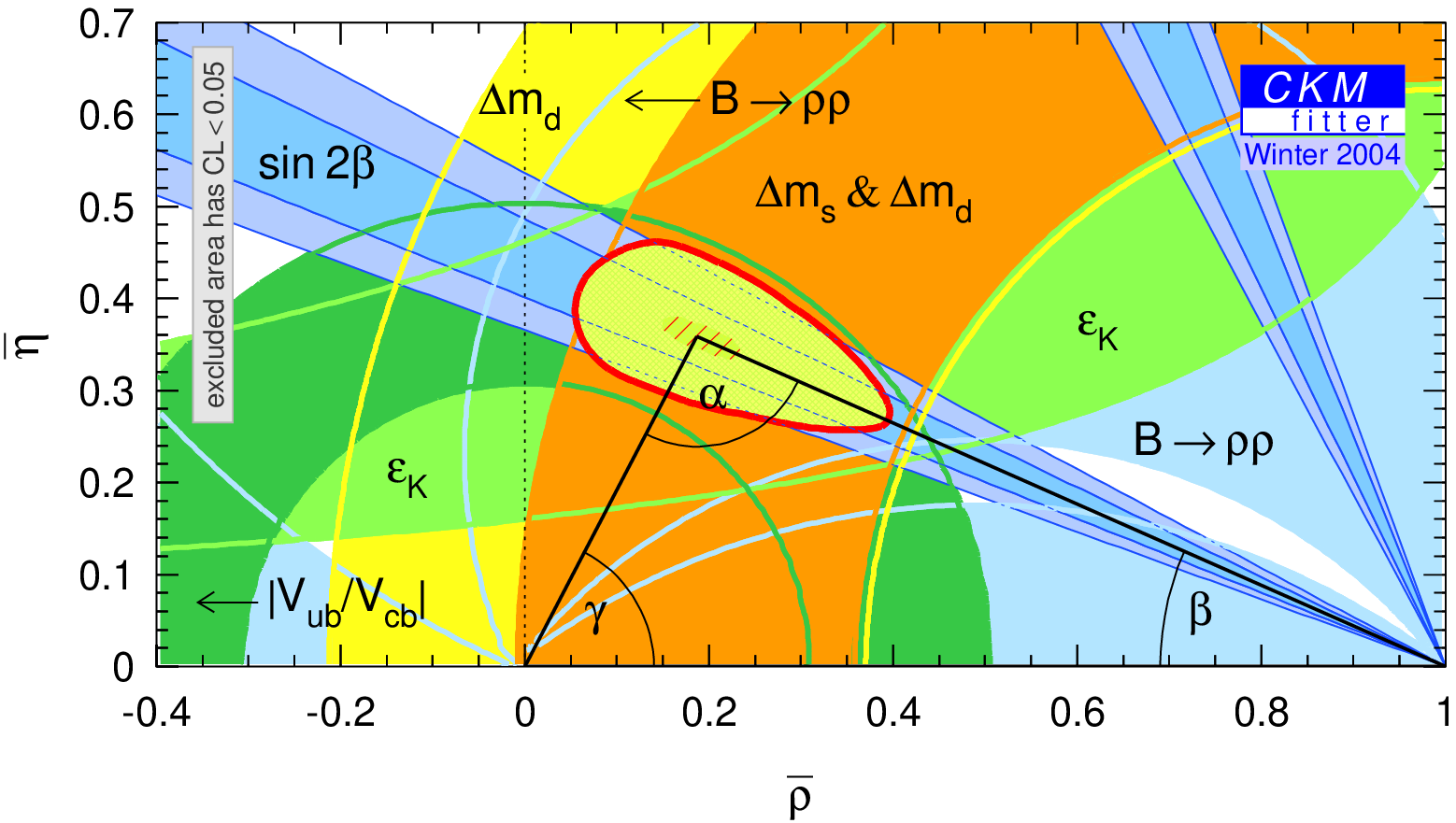}}
  \caption[.]{\label{fig:rhoeta_small}\em
        Confidence levels in the  $\rhoeta$ plane for the 
        global CKM fit. The shaded areas indicate the regions of
        $\ge5\%$ CLs. For $\stb$ the $\ge32\%$ and $\ge5\%$~CL 
        constraints are shown. The upper (lower) plot excludes
        (includes) the constraints from $\stb$ and $\sta$ in the
        combined fit. The hatched area in the center of the combined 
        fit result indicates the region where theoretical errors 
        dominate. The lower plot corresponds to the standard CKM fit.}
\end{figure}

It has become customary to present the constraints on the \CP-violating 
phase in the $\rhoeta$ (unitarity) plane of the Wolfenstein 
parameterization. In the case of such a two-dimensional graphical display,
the $\a$ parameter space 
(see Section~\ref{sec:statistics}.\ref{sec:RelevantandIrrelevantParameters})
is defined by the coordinates $\a=\{x, y\}$ (\eg, 
$\a=\arhoeta$) and the $\Mu$ space by the other CKM parameters $\lambda$ 
and $A$, as well as the $\yQCD$ parameters.
The results of the standard CKM fit are shown in the enlarged $\rhoeta$
plane in Fig.~\ref{fig:rhoeta}, not including (upper plot) and including 
(lower plot) in the fit the world average of $\stb$ and $\stwoa_{[\rho\rho]}$
(see Table~\ref{tab:ckmInputs}). The outer contour of the combined fit 
corresponds to the $5\%$ CL, and the inner contour gives the 
region where $\CL\sim1$ and hence theoretical systematics dominate
(an adjustment of the $\Mu$ parameters can maintain maximal agreement
\ie, the $\ChiMinGlob$ value is reproduced therein).
Also shown are the $\ge 5\%$ CL regions of the individual constraints. 
For $\stb$ both the $\ge 32\%$ and $\ge 5\%$ CL regions are drawn.
A zoom into the region favored by the combined fit is given in 
Fig.~\ref{fig:rhoeta_small}.

\subsection{Numerical Constraints on CKM Parameters and Related Observables}

\begin{figure}[t]
  \epsfxsize16cm
  \centerline{\epsffile{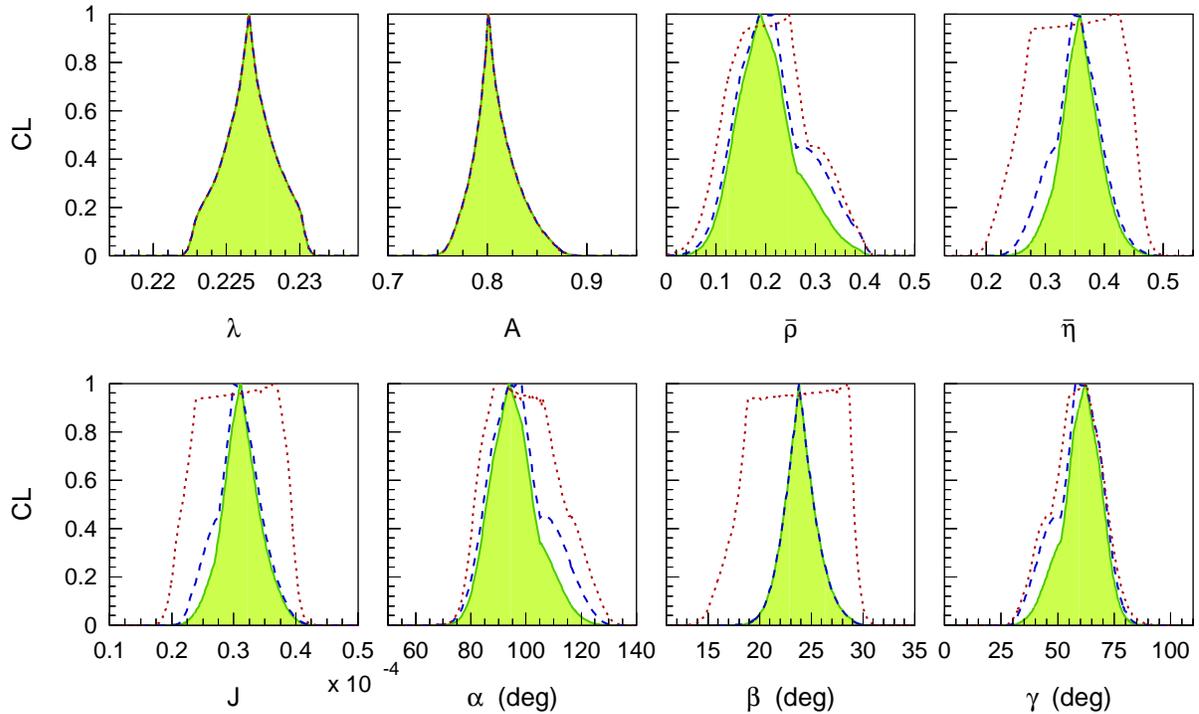}}  
  \caption[.]{\label{fig:ckm1d}\em
        Confidence levels for the Wolfenstein parameters and 
        UT surface and angles obtained from the standard CKM fit (shaded
        areas). The dotted curves give the results without using $\stb$ 
        and $\stwoa_{[\rho\rho]}$ in the fit, while the dashed 
        line excludes only $\stwoa_{[\rho\rho]}$. }
\end{figure}

Using the standard CKM fit inputs~(\ref{eq:standardFitObs}), we derive
one-dimensional numerical constraints for the Wolfenstein parameters,
the CKM matrix elements, branching ratios of rare $K$ and $B$ meson 
decays\footnote
{
        In the SM the branching ratio for the helicity-suppressed decay 
        $\Bp\to\ell^+\nu$ is given by
        \beqns
           \BR(B^+\to\ell^+\nu) 
                = \frac{G_{\rm F}^2m_{B_d}m_\ell^2}{8\pi}
                  \left(1 - \frac{m_\ell^2}{m_{B_d}^2}\right)
                  \fbd^2|V_{ub}|^2\tau_{\Bp}~.
        \eeqns
} as 
well as a selection of theoretical parameters. In the case of such 
one-dimensional displays, the $a$ parameter is defined by the $x$ 
coordinate, and the $\Mu$ space by all the other parameters. 
The Wolfenstein $\lambda$ has a larger error compared to the fit 
presented in Ref.~\cite{CKMfitter} since we enlarged the uncertainty 
on $\Vus$ as discussed in Section~\ref{sec:fitInputs}.
Numerical and graphical results are obtained for CKM fits {\em including} 
$\stbwa$ and $\stwoa_{[\rho\rho]}$. The results are quoted in 
Tables~\ref{tab:fitResults1} and \ref{tab:fitResults2} and some 
representative variables are plotted in Fig.~\ref{fig:ckm1d}
for fits with and without $\stbwa$ and $\stwoa_{[\rho\rho]}$.
The statistical precision of the present result for $\stwoa_{[\rho\rho]}$
is not yet sufficient to give a significant improvement of 
the standard CKM fit (see the outlook into the future given in 
Section~\ref{sec:charmlessBDecays}.\ref{sec:introductionrhorho}).
\vs
The predictions of the rare $W$-annihilation decays $\Bp\to\ell^+\nu$ 
can be compared to the present (yet unpublished) upper limits
$\BR(\Bp\to\tau^+\nu)<4.1\times10^{-4}$~\cite{babarbtaunu}
and $\BR(\Bp\to\mu^+\nu)<6.8\times10^{-6}$~\cite{bellebmunu}.
While the $\mu^+\nu$ limit is still an order of magnitude larger than
the expected value, the experiments approach the sensitivity
required for a discovery of $\Bp\to\tau^+\nu$. It may become one of
the key analyses in the coming years.
 \begin{table}[p]
 \begin{center}
 {\normalsize
\setlength{\tabcolsep}{0.0pc}
\begin{tabular*}{\textwidth}{@{\extracolsep{\fill}}lccc} \hline &&& \\[-0.3cm]
 & & Central value $\pm$ error at given CL & \\
 Quantity & CL = 0.32 & CL = 0.05 & CL = 0.003 \\[0.15cm]
 \hline  &&& \\[-0.3cm]
 $\lambda$                                                             & $   0.2265 ^{\,+   0.0025}_{\,-   0.0023}$ & $^{\,+   0.0040}_{\,-   0.0041}$ & $^{\,+   0.0045}_{\,-   0.0046}$
 \\[0.15cm]
 $A$                                                                    & $    0.801 ^{\,+    0.029}_{\,-    0.020}$ & $^{\,+    0.066}_{\,-    0.041}$ & $^{\,+    0.084}_{\,-    0.054}$
 \\[0.15cm]
 $\rhobar$                                                           & $    0.189 ^{\,+    0.088}_{\,-    0.070}$ & $^{\,+    0.182}_{\,-    0.114}$ & $^{\,+    0.221}_{\,-    0.156}$
 \\[0.15cm]
 $\etabar$                                                           & $    0.358 ^{\,+    0.046}_{\,-    0.042}$ & $^{\,+    0.086}_{\,-    0.085}$ & $^{\,+    0.118}_{\,-    0.118}$
 \\[0.15cm]
 \hline &&&      \\[-0.3cm]
 $J$~~$[10^{-5}]$                                                       & $     3.10 ^{\,+     0.43}_{\,-     0.37}$ & $^{\,+     0.82}_{\,-     0.74}$ & $^{\,+     1.08}_{\,-     0.96}$
 \\[0.15cm]
 \hline &&&      \\[-0.3cm]
 $\stwoa$                                                        & $    -0.14 ^{\,+     0.37}_{\,-     0.41}$ & $^{\,+     0.57}_{\,-     0.71}$ & $^{\,+     0.74}_{\,-     0.82}$
 \\[0.15cm]
 $\stwoa$ {\small (meas. not in fit)}                          & $    -0.29 ^{\,+     0.56}_{\,-     0.46}$ & $^{\,+     0.77}_{\,-     0.65}$ & $^{\,+     0.93}_{\,-     0.70}$
 \\[0.15cm]
 $\stwob$                                                         & $    0.739 ^{\,+    0.048}_{\,-    0.048}$ & $^{\,+    0.096}_{\,-    0.095}$ & $^{\,+    0.124}_{\,-    0.137}$
 \\[0.15cm]
 $\stwob$ {\small (meas. not in fit)}                            & $    0.817 ^{\,+    0.037}_{\,-    0.222}$ & $^{\,+    0.053}_{\,-    0.279}$ & $^{\,+    0.067}_{\,-    0.334}$
 \\[0.15cm]
 $\alpha$~~(deg)                                                       & $       94 ^{\,+       12}_{\,-       10}$ & $^{\,+       24}_{\,-       16}$ & $^{\,+       32}_{\,-       22}$
 \\[0.15cm]
 $\alpha$~~(deg) {\small (meas. not in fit)}                                                      & $       98 ^{\,+       15}_{\,-       16}$ & $^{\,+       26}_{\,-       22}$ & $^{\,+       31}_{\,-       28}$
 \\[0.15cm]
 $\beta$~~(deg)                                                        & $     23.8 ^{\,+      2.1}_{\,-      2.0}$ & $^{\,+      4.5}_{\,-      3.8}$ & $^{\,+      6.0}_{\,-      5.3}$
 \\[0.15cm]
 $\beta$~~(deg) {\small (meas. not in fit)}                           & $     27.4 ^{\,+      1.9}_{\,-      9.2}$ & $^{\,+      2.8}_{\,-     11.1}$ & $^{\,+      3.7}_{\,-     13.0}$
 \\[0.15cm]
 $\gamma\simeq\delta$~~(deg)                                         & $       62 ^{\,+       10}_{\,-       12}$ & $^{\,+       17}_{\,-       24}$ & $^{\,+       23}_{\,-       30}$
 \\[0.15cm]
 \hline &&&      \\[-0.3cm]
 $\sin\theta_{12}$                                                    & $   0.2266 ^{\,+   0.0025}_{\,-   0.0023}$ & $^{\,+   0.0040}_{\,-   0.0041}$ & $^{\,+   0.0045}_{\,-   0.0046}$
 \\[0.15cm]
 $\sin\theta_{13}$~~$[10^{-3}]$                                       & $     3.87 ^{\,+     0.35}_{\,-     0.30}$ & $^{\,+     0.35}_{\,-     0.60}$ & $^{\,+     0.35}_{\,-     0.76}$
 \\[0.15cm]
 $\sin\theta_{23}$~~$[10^{-3}]$                                       & $    41.13 ^{\,+     1.37}_{\,-     0.58}$ & $^{\,+     2.43}_{\,-     1.16}$ & $^{\,+     3.08}_{\,-     1.73}$
 \\[0.15cm]
 \hline &&&      \\[-0.3cm]
 $R_u$                                                                  & $    0.405 ^{\,+    0.035}_{\,-    0.032}$ & $^{\,+    0.077}_{\,-    0.062}$ & $^{\,+    0.093}_{\,-    0.083}$
 \\[0.15cm]
 $R_t$                                                                  & $    0.889 ^{\,+    0.073}_{\,-    0.095}$ & $^{\,+    0.118}_{\,-    0.196}$ & $^{\,+    0.161}_{\,-    0.243}$
 \\[0.15cm]
 \hline &&&      \\[-0.3cm]
 $\Delta m_d$~~$({\rm ps}^{-1})$ {\small (meas. not in fit)}         & $     0.54 ^{\,+     0.26}_{\,-     0.21}$ & $^{\,+     0.62}_{\,-     0.31}$ & $^{\,+     0.94}_{\,-     0.34}$
 \\[0.15cm]
 $\Delta m_s$~~$({\rm ps}^{-1})$                                      & $     17.8 ^{\,+      6.7}_{\,-      1.6}$ & $^{\,+     15.2}_{\,-      2.7}$ & $^{\,+     22.1}_{\,-      3.7}$
 \\[0.15cm]
 $\Delta m_s$~~$({\rm ps}^{-1})$ {\small (meas. not in fit)}         & $     16.5 ^{\,+     10.5}_{\,-      3.4}$ & $^{\,+     17.7}_{\,-      5.7}$ & $^{\,+     23.9}_{\,-      7.2}$
 \\[0.15cm]
 \hline &&&      \\[-0.3cm]
 $\epsk$~~$[10^{-3}]$ {\small (meas. not in fit)}                & $      2.5 ^{\,+      1.6}_{\,-      1.1}$ & $^{\,+      2.4}_{\,-      1.4}$ & $^{\,+      3.1}_{\,-      1.6}$
 \\[0.15cm]
 \hline &&&      \\[-0.3cm]
 $f_{B_d}\sqrt{B_d}$~~(MeV) {\small (theor. value not in fit)}        & $      215 ^{\,+       28}_{\,-       21}$ & $^{\,+       79}_{\,-       31}$ & $^{\,+       79}_{\,-       39}$
 \\[0.15cm]
 $B_K$ {\small (theor. value not in fit)}                                     & $     0.86 ^{\,+     0.26}_{\,-     0.30}$ & $^{\,+     0.57}_{\,-     0.39}$ & $^{\,+     0.90}_{\,-     0.45}$
 \\[0.15cm]
 \hline &&&      \\[-0.3cm]
 $\mtRun(m_t)$~~$({\rm GeV}/c^2)$ {\small (meas. not in fit)}                 & $      165 ^{\,+       48}_{\,-       47}$ & $^{\,+      124}_{\,-       64}$ & $^{\,+      194}_{\,-       77}$
 \\[0.15cm]
 \hline
 \end{tabular*}
 }
 \end{center}
 \vspace{-0.5cm}
 \caption[.]{\label{tab:fitResults1}\em
        Fit results and errors (deviations from central values 
        at confidence levels that correspond to one-, two- and 
        three standard deviations, respectively) using the 
        standard input observables~(\ref{eq:standardFitObs})
        (i.e., including the world average on $\stbwa$).
        For results marked by ``meas. not in fit'', the measurement
        of the corresponding observable has not been included in
        the fit. The input parameters used for this fit are quoted in 
        Table~\ref{tab:ckmInputs}.}
 \end{table}
  
 \begin{table}[p]
 \begin{center}
 {\normalsize
\setlength{\tabcolsep}{0.0pc}
\begin{tabular*}{\textwidth}{@{\extracolsep{\fill}}lccc} \hline &&& \\[-0.3cm]
 & & Central value $\pm$ error at given CL & \\
 Quantity & CL = 0.32 & CL = 0.05 & CL = 0.003 \\[0.15cm]
 \hline  &&& \\[-0.3cm]
 $\Kzpiznn$~~$[10^{-11}]$                                              & $     2.89 ^{\,+     0.84}_{\,-     0.69}$ & $^{\,+     1.71}_{\,-     1.24}$ & $^{\,+     2.41}_{\,-     1.52}$
 \\[0.15cm]
 $\Kppipnn$~~$[10^{-11}]$                                              & $      6.7 ^{\,+      2.8}_{\,-      2.7}$ & $^{\,+      3.7}_{\,-      3.2}$ & $^{\,+      4.6}_{\,-      3.6}$
 \\[0.15cm]
 \hline &&&      \\[-0.3cm]
 $\Bptaun$~~$[10^{-5}]$                                                & $     11.9 ^{\,+      4.5}_{\,-      5.7}$ & $^{\,+     10.4}_{\,-      8.2}$ & $^{\,+     17.9}_{\,-      10.1}$
 \\[0.15cm]
 \hline &&&      \\[-0.3cm]
 $\Bpmun$~~$[10^{-7}]$                                                 & $      4.7 ^{\,+      2.3}_{\,-      1.7}$ & $^{\,+      4.6}_{\,-      2.7}$ & $^{\,+      7.6}_{\,-      3.5}$
 \\[0.15cm]
 \hline &&&      \\[-0.3cm]
 $|V_{ud}|$                                                             & $  0.97400 ^{\,+  0.00054}_{\,-  0.00058}$ & $^{\,+  0.00094}_{\,-  0.00095}$ & $^{\,+  0.00106}_{\,-  0.00106}$
 \\[0.15cm]
 $|V_{us}|$                                                             & $   0.2265 ^{\,+   0.0025}_{\,-   0.0023}$ & $^{\,+   0.0040}_{\,-   0.0041}$ & $^{\,+   0.0045}_{\,-   0.0046}$
 \\[0.15cm]
 $|V_{ub}|$~~$[10^{-3}]$                                                & $     3.87 ^{\,+     0.35}_{\,-     0.30}$ & $^{\,+     0.73}_{\,-     0.60}$ & $^{\,+     0.73}_{\,-     0.76}$
 \\[0.15cm]
 $|V_{ub}|$~~$[10^{-3}]$ {\small (meas. not in fit)}                   & $     3.87 ^{\,+     0.34}_{\,-     0.31}$ & $^{\,+     0.81}_{\,-     0.61}$ & $^{\,+     1.27}_{\,-     0.88}$
 \\[0.15cm]
 $|V_{cd}|$                                                             & $   0.2264 ^{\,+   0.0025}_{\,-   0.0023}$ & $^{\,+   0.0040}_{\,-   0.0041}$ & $^{\,+   0.0045}_{\,-   0.0046}$
 \\[0.15cm]
 $|V_{cs}|$                                                             & $  0.97317 ^{\,+  0.00053}_{\,-  0.00059}$ & $^{\,+  0.00094}_{\,-  0.00097}$ & $^{\,+  0.00106}_{\,-  0.00112}$
 \\[0.15cm]
 $|V_{cb}|$~~$[10^{-3}]$                                                & $    41.13 ^{\,+     1.36}_{\,-     0.58}$ & $^{\,+     2.43}_{\,-     1.16}$ & $^{\,+     3.08}_{\,-     1.73}$
 \\[0.15cm]
 $|V_{cb}|$~~$[10^{-3}]$ {\small (meas. not in fit)}                    & $     41.2 ^{\,+      5.1}_{\,-      5.7}$ & $^{\,+      7.9}_{\,-      5.8}$ & $^{\,+      9.9}_{\,-      5.8}$
 \\[0.15cm]
 $|V_{td}|$~~$[10^{-3}]$                                                & $     8.26 ^{\,+     0.72}_{\,-     0.86}$ & $^{\,+     1.23}_{\,-     1.79}$ & $^{\,+     1.64}_{\,-     2.25}$
 \\[0.15cm]
 $|V_{ts}|$~~$[10^{-3}]$                                                & $    40.47 ^{\,+     1.39}_{\,-     0.62}$ & $^{\,+     2.42}_{\,-     1.21}$ & $^{\,+     3.17}_{\,-     1.78}$
 \\[0.15cm]
 $|V_{tb}|$                                                             & $ 0.999146 ^{\,+ 0.000024}_{\,- 0.000058}$ & $^{\,+ 0.000047}_{\,- 0.000104}$ & $^{\,+ 0.000070}_{\,- 0.000133}$
 \\[0.15cm]
 \hline &&&      \\[-0.3cm]
 $|V_{ud}V_{ub}^*|$~~$[10^{-3}]$                                        & $     3.77 ^{\,+     0.34}_{\,-     0.30}$ & $^{\,+     0.71}_{\,-     0.59}$ & $^{\,+     0.71}_{\,-     0.75}$
 \\[0.15cm]
 $\arg\left[V_{ud}V_{ub}^*\right]$~~(deg)                               & $       62 ^{\,+       10}_{\,-       12}$ & $^{\,+       16}_{\,-       24}$ & $^{\,+       22}_{\,-       31}$
 \\[0.15cm]
 $|V_{cd}V_{cb}^*|$~~$[10^{-3}]$                                        & $     9.31 ^{\,+     0.31}_{\,-     0.15}$ & $^{\,+     0.62}_{\,-     0.34}$ & $^{\,+     0.80}_{\,-     0.49}$
 \\[0.15cm]
 $\arg\left[V_{cd}V_{cb}^*\right]$~~(deg)                               & $-179.9653 ^{\,+   0.0047}_{\,-   0.0042}$ & $^{\,+   0.0091}_{\,-   0.0084}$ & $^{\,+   0.0122}_{\,-   0.0107}$
 \\[0.15cm]
 $|V_{td}V_{tb}^*|$~~$[10^{-3}]$                                        & $     8.24 ^{\,+     0.73}_{\,-     0.85}$ & $^{\,+     1.24}_{\,-     1.78}$ & $^{\,+     1.64}_{\,-     2.24}$
 \\[0.15cm]
 $\arg\left[V_{td}V_{tb}^*\right]$~~(deg)                               & $    -23.8 ^{\,+      2.0}_{\,-      2.1}$ & $^{\,+      3.8}_{\,-      4.5}$ & $^{\,+      5.3}_{\,-      6.0}$
 \\[0.15cm]
 $|V_{td}/V_{ts}|$                                                      & $     0.204 ^{\,+     0.018}_{\,-     0.022}$ & $^{\,+     0.029}_{\,-     0.046}$ & $^{\,+     0.039}_{\,-     0.058}$
 \\[0.15cm]
 \hline &&&      \\[-0.3cm]
 ${\rm Re}\lambda_c$                                                    & $  -0.2204 ^{\,+   0.0022}_{\,-   0.0023}$ & $^{\,+   0.0038}_{\,-   0.0037}$ & $^{\,+   0.0043}_{\,-   0.0041}$
 \\[0.15cm]
 ${\rm Im}\lambda_c$~~$[10^{-4}]$                                       & $    -1.41 ^{\,+     0.18}_{\,-     0.18}$ & $^{\,+     0.34}_{\,-     0.36}$ & $^{\,+     0.44}_{\,-     0.48}$
 \\[0.15cm]
 \hline &&&      \\[-0.3cm]
 ${\rm Re}\lambda_t$~~$[10^{-4}]$                                       & $    -3.04 ^{\,+     0.32}_{\,-     0.31}$ & $^{\,+     0.67}_{\,-     0.60}$ & $^{\,+     0.86}_{\,-     0.80}$
 \\[0.15cm]
 ${\rm Im}\lambda_t$~~$[10^{-4}]$                                       & $     1.41 ^{\,+     0.19}_{\,-     0.17}$ & $^{\,+     0.37}_{\,-     0.34}$ & $^{\,+     0.48}_{\,-     0.44}$
 \\[0.15cm]
 \hline
 \end{tabular*}
 }
 \end{center}
 \vspace{-0.5cm}
 \caption[.]{\label{tab:fitResults2}\em
        Fit results and errors (deviations from central values 
        at confidence levels that correspond to one-, two- and 
        three standard deviations, respectively) using the 
        standard input observables~(\ref{eq:standardFitObs})
        (i.e., including the world average on $\stbwa$).
        The variables in the last four lines are defined by 
        $\lambda_i\equiv V_{id}V_{is}^*$.
        For results marked by ``meas. not in fit'', the measurement
        related to the corresponding observable has not been 
        included in the fit.
        The input parameters used for this fit are quoted in 
        Table~\ref{tab:ckmInputs}.}
 \end{table}
\begin{figure}[p]
  \def\sintwobfig{8.1cm}
  \epsfxsize\sintwobfig
  \centerline{\epsffile{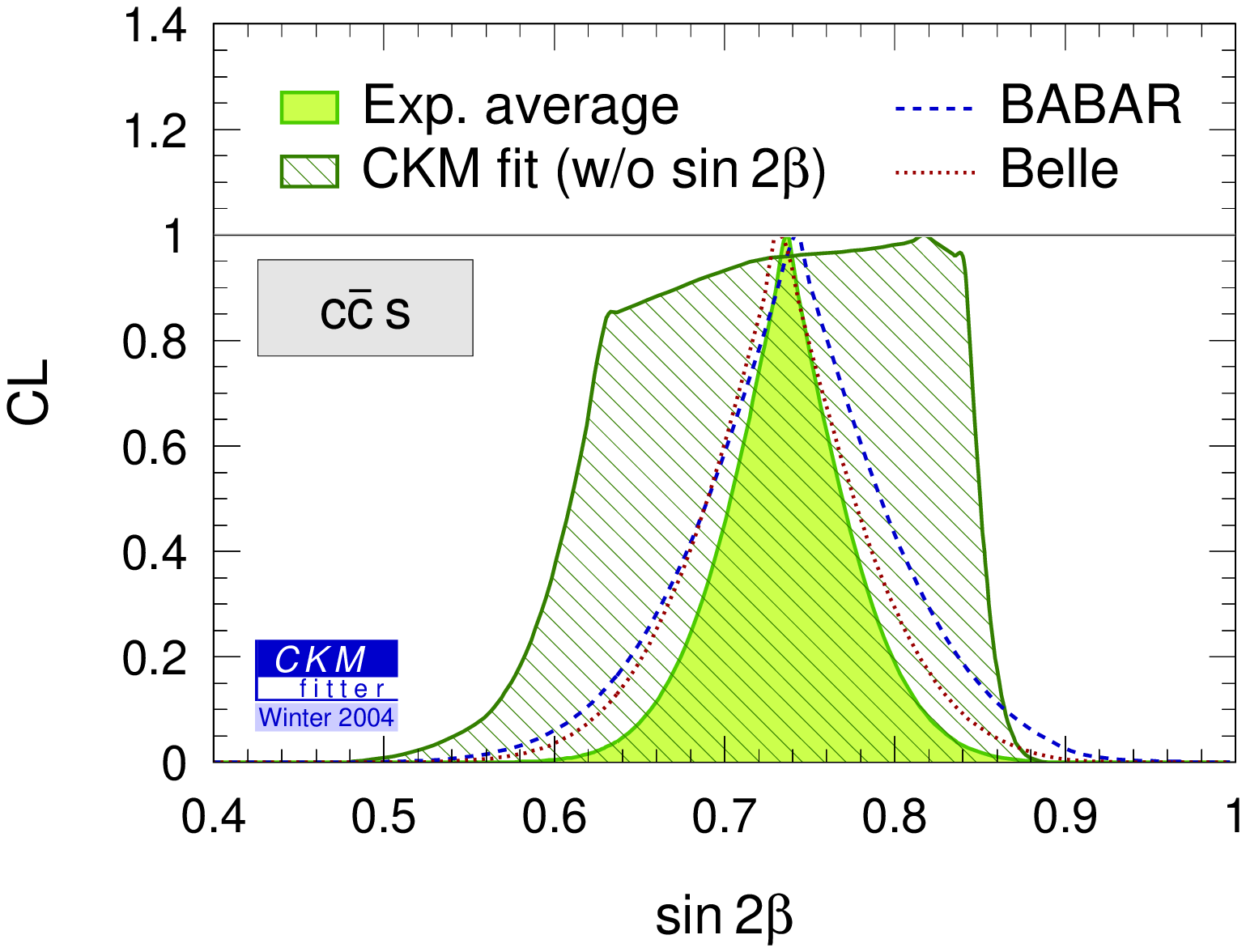}
              \epsfxsize\sintwobfig
              \epsffile{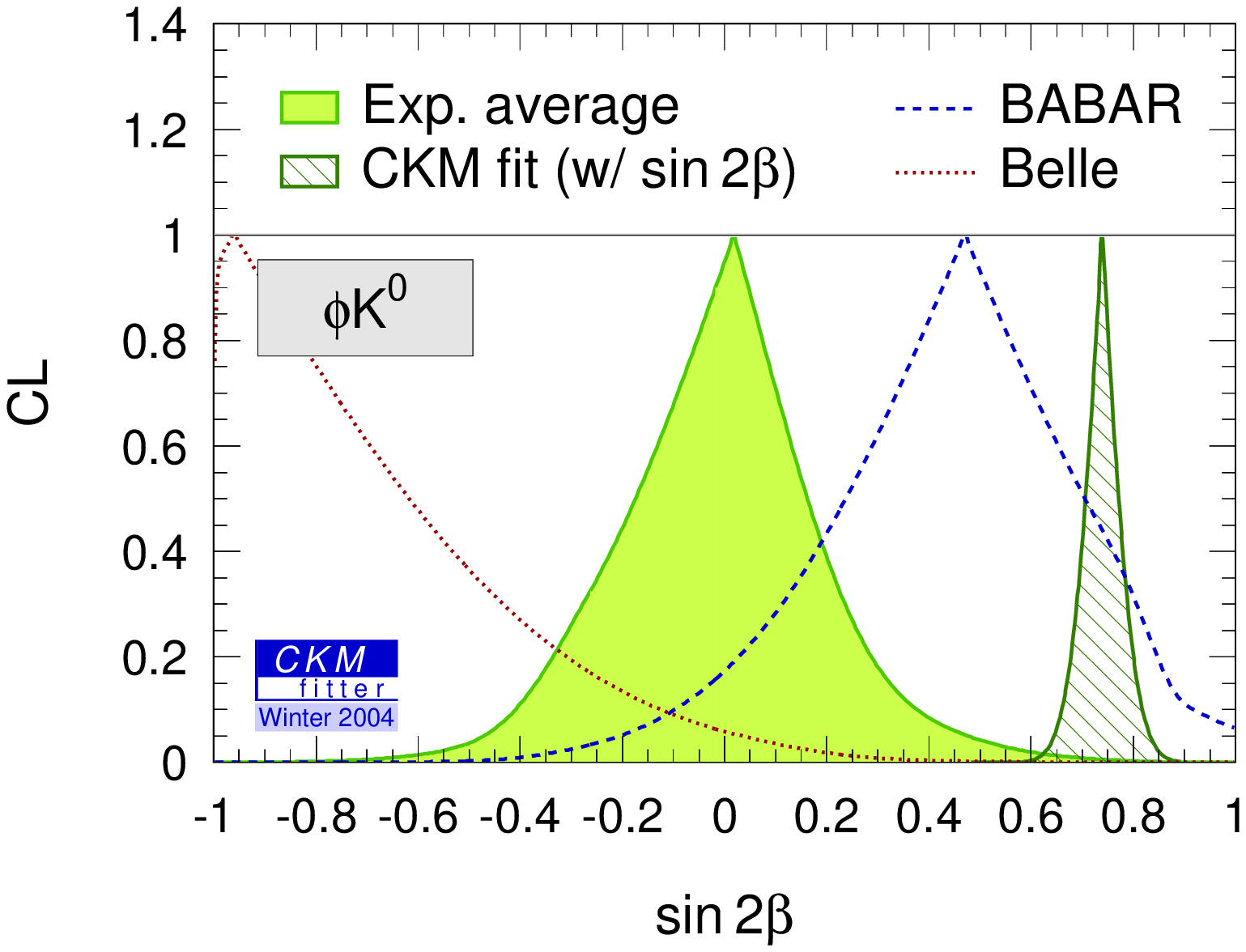}}
  \centerline{\epsffile{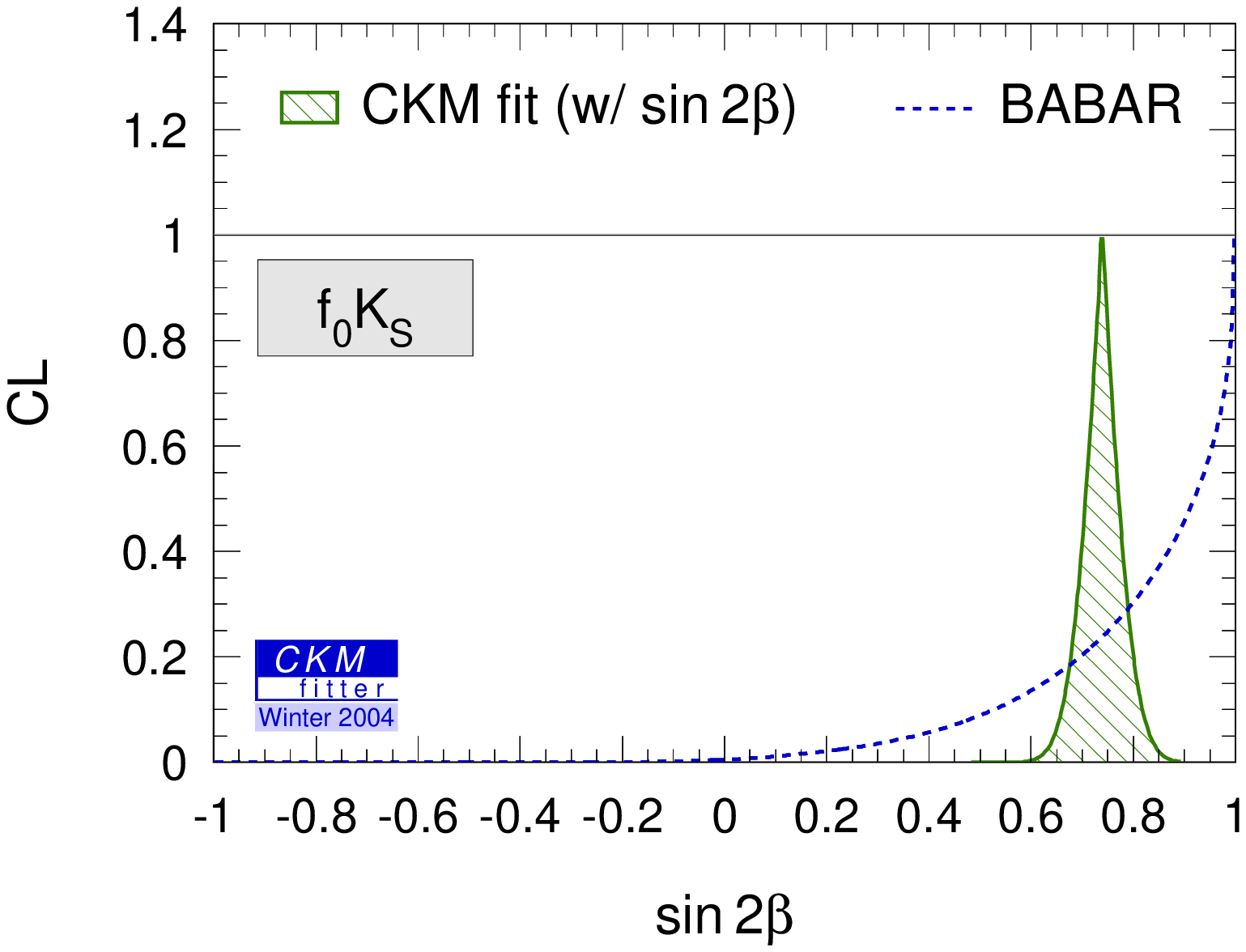}
              \epsfxsize\sintwobfig
              \epsffile{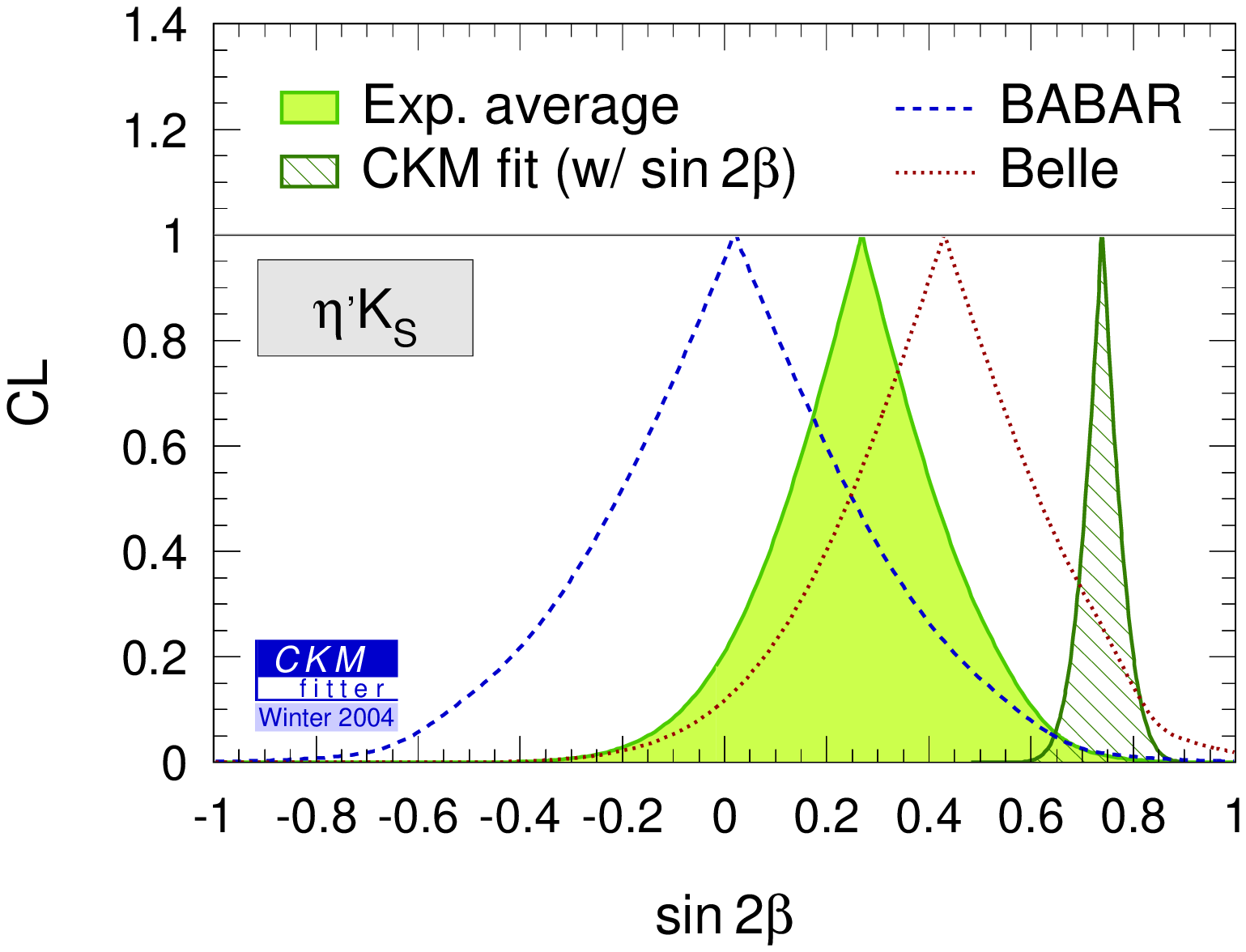}}
  \centerline{\epsffile{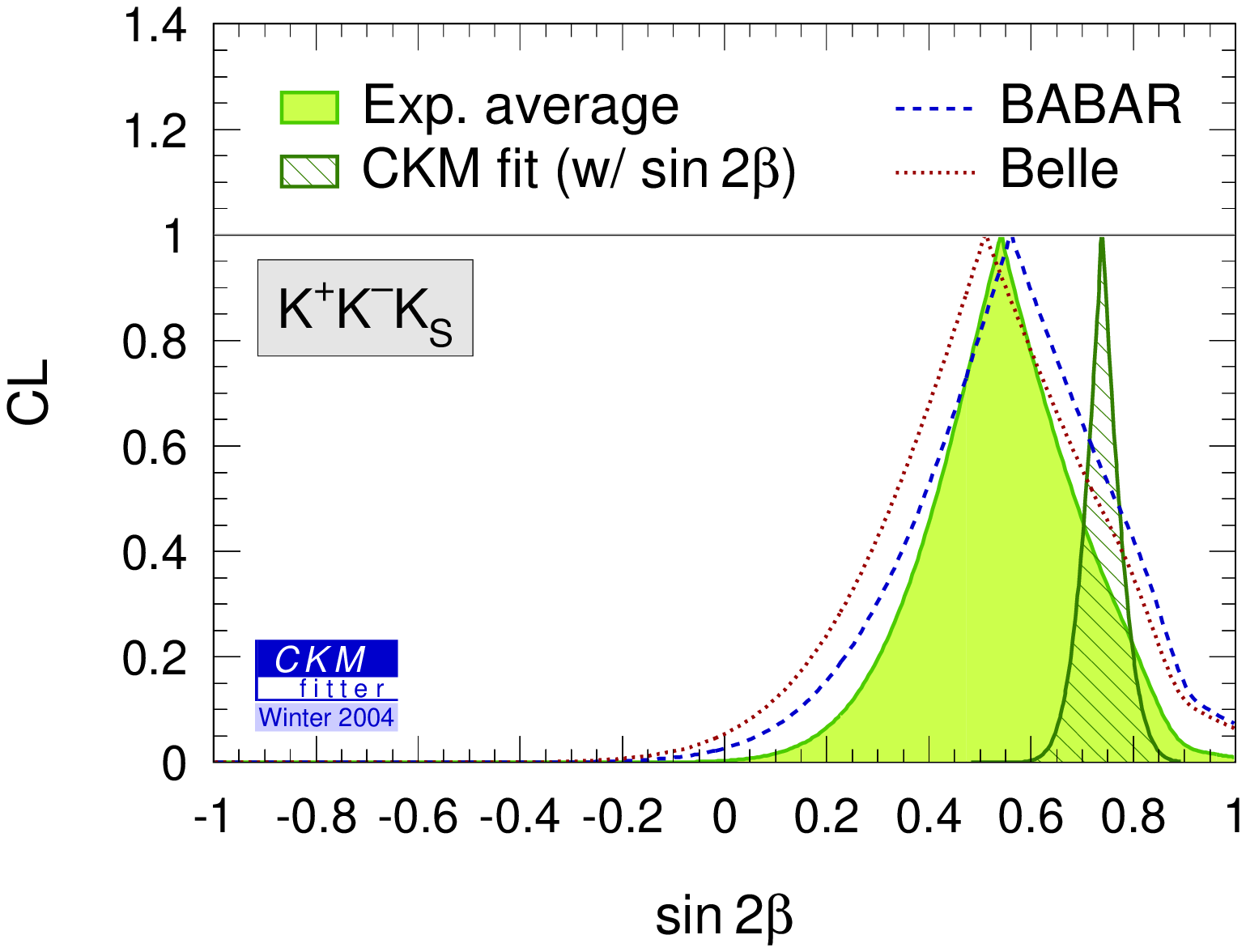}
              \epsfxsize\sintwobfig
              \epsffile{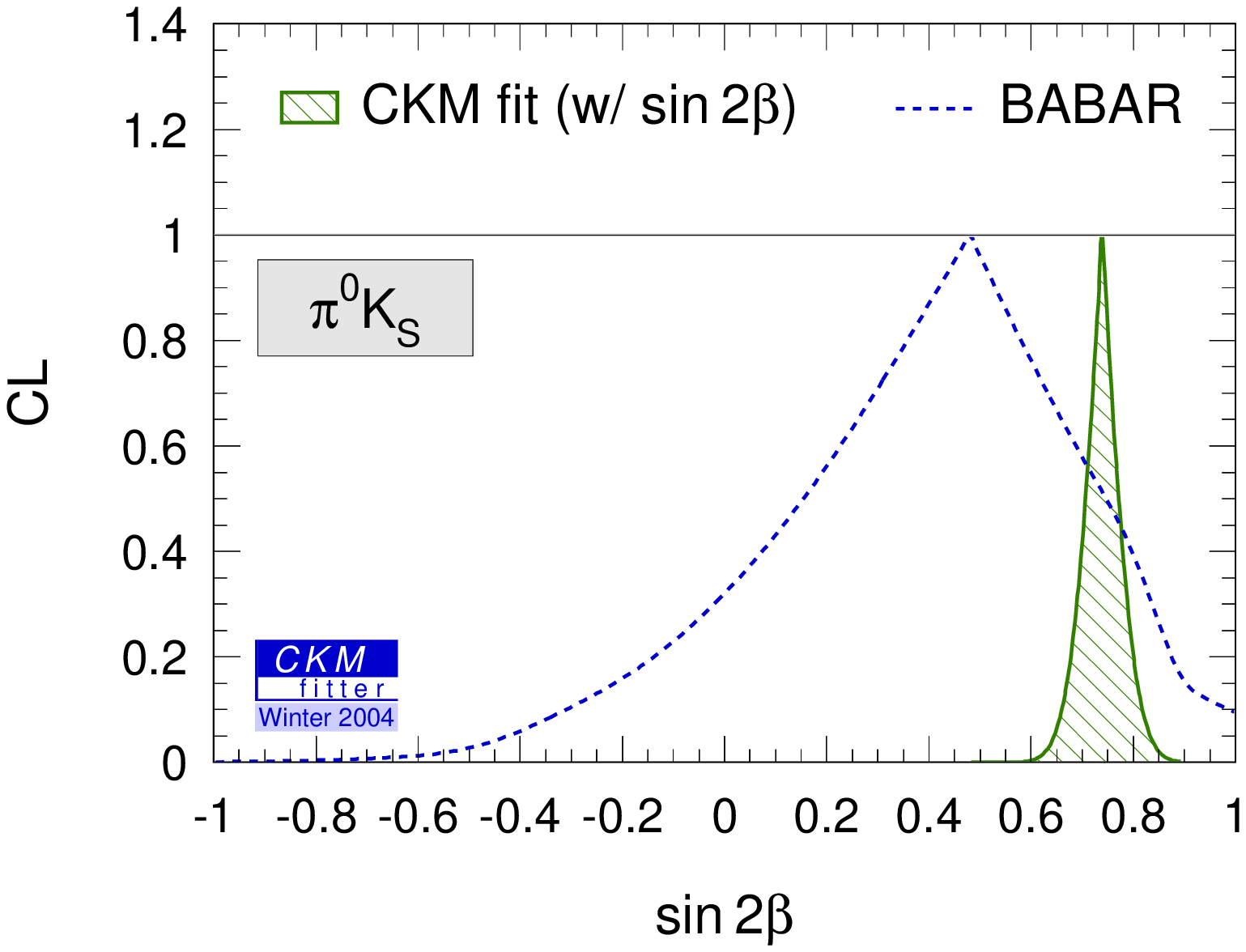}}
  \vspace{-0.4cm}
  \caption[.]{\label{fig:s2b1d}\em
        Confidence levels for the various $\stb_{[{\rm eff}]}$ 
        measurements that are believed to be dominated by 
        a single CKM phase, their averages and the result from 
        the standard CKM fit.}
\end{figure}

\subsection{Is there a $\stb$ Problem in Penguin-Dominated Decays?}
\label{sec:sin2bProblem}

As pointed out in Section~\ref{sec:standardFit}.\ref{sec:fitInputs}, 
penguin-dominated decays like $\phi\Kz$, $\etapr\KS$ and \CP-even-domi\-na\-ted
$K^+K^-\KS$ as well as $\pi^0\KS$  ($b\to s\bar q q$ transitions) show
on average lower experimental $\stb$ values than $b\to c\bar c s$ transitions.
An exception to this is the recent \babar\  measurement using the 
decay $f_0\KS$.
\vs
The interpretation of the non-charmonium decays in terms of $\stb$
has to be done with care, 
since contributions from CKM-suppressed penguins and trees may lead 
to deviations from the leading weak decay phase of up to 
$|\stb_{s\bar qq}-\stb_{c\bar cs}|\sim0.2$ within the 
SM~\cite{BN,ligetiquinn,grossrosner}.
When averaging all penguin as well as charmonia measurements we obtain 
a p-value of $1.7\%$ ($2.4\sigma$). If CKM-suppressed penguins and 
trees can be neglected in $b\to s\bar q q$ transitions this might 
be a hint of an anomaly.
When taking into account the modifications of the $\CL$ due 
to the presence of the non-physical boundaries
(\cf\  Section~\ref{sec:statistics}.\ref{sec:TheExperimentalLikelihood})
the overall p-value decreases to $1.1\%$ ($2.6\sigma$). 
\vs
The individual measurements\footnote
{
        Note that the $C$ coefficients, which vanish if the penguin 
        dominance and the SM assumptions are correct, are left free 
        to vary in the time-dependent fits performed by the experiments. 
        All of them are found to be in reasonable agreement with zero~\cite{HFAG}.
} compared 
to the constraint from the standard CKM fit, not including $\stb$, are 
shown in Fig.~\ref{fig:s2b1d}. The average of the $S_{\phi K}$ measurement
from \babar\  and Belle has a p-value of $4.9\%$, so that more data 
are needed to conclude. If we remove Belle's $S_{\phi\KS}$ from the 
all-mode average, we obtain for the compatibility with the charmonium 
results a p-value of $29\%$, which is 
$1.1\sigma$. Hence without confirmation of Belle's $S_{\phi\KS}$ 
measurement there is no statistical justification to claim evidence 
for New Physics on the basis of the present data. We stress that
a clear sign of New Physics in these penguin decays would be a 
pattern of $\stb_{{\rm eff},\,sq\bar q}$ values that are 
significantly different from $\stbwa$ {\em and} significantly different 
from each other\footnote
{
        Unless some specific symmetry or dynamical mechanism relates 
        the New Physics to SM amplitude ratios in different 
        channels~\cite{nirichep}.
}.
It might be that Belle's $S_{\phi\KS}$ measurement represents a statistical 
fluctuation and that very large New Physics effects are not to be expected, 
which of course does not imply that more precise data will not be
able to give evidence for non-standard contributions if they exist. We 
revisit the $\phi\Kz$ mode in a more general New Physics 
framework in Section~\ref{sec:newPhysics}.\ref{NPdB=1}.

\subsection{Resolving the Two-fold Ambiguity in $2\beta$}
\label{sec:jpsikstar}

In spite of the agreement with the standard CKM fit of one out of 
the four solutions for $\beta$ from the precise $\stb$ measurement 
using charmonium decays, it is still possible that, because of 
contributions from New Physics, the correct value of $\beta$ is one 
of the three other solutions. The measurement of the sign of $\ctb$ 
would reduce the solution space to an indistinguishable two-fold 
ambiguity\footnote
{
        The invariance $\beta\to\pi+\beta$ remains. 
        It cannot be lifted without theoretical 
        input on a strong phase~\cite{quinngrossman}.
}.
The \babar\  collaboration has performed a measurement of 
$\ctb$ in a time-dependent transversity analysis of the 
pseudoscalar to vector-vector decay $\Bz\to \jpsi\Kstarz(\to\KS\piz)$,
where $\ctb$ enters as a factor in the interference between
\CP-even and \CP-odd amplitudes~\cite{babarallbeta}. In principle,
this analysis comes along with an ambiguity on the sign
of $\ctb$ due to an incomplete determination of the strong
phases occurring in the three transversity amplitudes. 
\babar\  resolves this ambiguity by inserting the 
known variation~\cite{lassKpi} of the rapidly moving 
$P$-wave phase relative to the slowly moving $S$-wave 
phase with the invariant mass of the $K\pi$ system 
in the vicinity of the $\Kstarz(892)$ resonance. 
\vs
When fixing the $\stb$ value to the world average, \babar\  finds 
\beq
\label{eq:cos2b}
        \ctb \:=\: 2.72^{\,+0.50}_{\,-0.79}\pm0.27~,
\eeq
where the effect introduced by the variation of $\stb$ within 
its small errors is negligible. \babar\  quotes the probability 
that the true $\ctb$ is positive\footnote
{
        This solution corresponds to reasonably small strong 
        phases between transversity amplitudes, as expected in the 
        factorization approximation~\cite{13r}.
} to 
be $89\%$, where the value is obtained with the use of Monte Carlo 
methods. This is much less than the $3.8\sigma$ exclusion of the mirror 
solutions $\pi/2-\beta$ and $3\pi/2-\beta$, obtained from a 
probabilistic treatment of the result~(\ref{eq:cos2b}), using 
the analytical method described in 
Section~\ref{sec:statistics}.\ref{sec:metrology_physicalBoundaries},
and assuming that the log-likelihood function belonging 
to~(\ref{eq:cos2b}) has parabolic tails. Since the Monte Carlo
evaluation is reliable, we conclude that a Gaussian interpretation 
of the errors given in~(\ref{eq:cos2b}) is flawed.
Due to the lack of a more accurate experimental CL function for $\ctb$ 
(precisely the one obtained from Monte Carlo simulation), we do not 
include the present measurement in the standard CKM fit, although we will
assume $\ctb>0$ in part of our New Physics analysis (see
Part~\ref{sec:newPhysics}). Proposals for alternative determinations of
${\rm sign}(\ctb)$ can be found in Refs.~\cite{13r,7r,15r,17r,18r,19r}.

\section{Conclusions}

The robustness of the Unitarity Triangle fit has been greatly 
improved since the precision measurement of $\stb$ became 
available. It outperforms by far all other contributions in
the combined experimental and theoretical precision. A new 
constraint on $\stwoa$ from the isospin analysis of $B\to\rho\rho$
decays has become available, the theoretical uncertainties of 
which -- though not yet entirely evaluated or known -- seem to
be under control. Its inclusion into the standard CKM fit
already leads to a modest improvement on the knowledge of 
$\alpha$ and $\gamma$. We derive a large 
number of quantitative results on the CKM parameters for 
various parameterizations and related quantities, theoretical 
parameters and physical observables from the standard CKM fit
(see Tables~\ref{tab:fitResults1} and \ref{tab:fitResults2}).
\vs
The goodness-of-fit of the global CKM fit is found to 
be $71\%$. We find that penguin-dominated measurements of (to 
good approximation) $\stb$ are in agreement with the reference
value from $\Bz$ decays into charmonium states. It might turn
out that the large negative $S$ value found by Belle in 
$\Bz\to\phi\KS$ represents a statistical fluctuation. 
A measurement of $\ctb$ in 
$\Bz\to \jpsi\Kstarz(\to\piz\KS)$ decays indicates that the 
$\beta$ solution from the $\stb$ measurement that is favored 
by the standard CKM fit corresponds to the one that occurs in 
$\BzBzb$ mixing.

%
%
  \newpage\part{Constraints from Kaon Physics}\setcounter{section}{0}
\markboth{\textsc{Part IV -- Constraints from Kaon Physics}}
         {\textsc{Part IV -- Constraints from Kaon Physics}}
\label{sec:kaons}

This part presents the CKM constraints from  direct \CP\  violation and
other \CP-related observables in
kaon decays that do not belong to the standard CKM fit.
We refer to Sections~\ref{sec:standardFit}.\ref{sec:input_vus} and 
\ref{sec:standardFit}.\ref{sec:input_epsk} for a discussion of the 
constraints from $K_{e3}$ decays, giving $|V_{us}|$, and from 
indirect \CP violation, respectively. 

Section~\ref{sec:kaons}.\ref{subsec:epe} discusses current experimental and
theoretical status of constraints related to the \CP-violating parameter 
$\epe$. Conventions and input values from Ref.~\cite{Buras03} are used. 
Sections~\ref{sec:kaons}.\ref{sec:Kpinn} and~\ref{sec:kaons}.\ref{sec:K0Lpi0nn} 
discuss the status of measurements of the rare kaon decays  $K^+\to\pi^+\nu\nub$ 
and  $\KL\to\piz\nu\nub$, together with a study of future perspectives. 
For a recent detailed review of these rare kaon decays we refer to 
Ref.~\cite{PajuoMayor}.

\section{\boldmath Direct \CP Violation in the Neutral Kaon System: $\epe$}\label{subsec:epe}

\begin{table}[b]
\begin{center}
\setlength{\tabcolsep}{0.0pc}
\begin{tabular*}{\textwidth}{@{\extracolsep{\fill}}lrl} \hline
&& \\[-0.3cm]
Experiment      &   Value $[10^{-4}]$   & Status
\\[0.15cm]
\hline
&& \\[-0.3cm]
NA31~\cite{na31}        &       $23.0\pm 6.5$   & final
\\
E731~\cite{e731} &       $7.4\pm 5.9$   & final
\\
NA48~\cite{na48} &       $14.7\pm 2.2$  & final
\\
KTeV~\cite{ktev} &       $20.7\pm 2.8$  & $1/2$ data sample 
\\[0.15cm]
\hline
&& \\[-0.3cm]
Average         &       $16.7\pm 1.6$   & $\CL=10\%
$
\\[0.15cm]
\hline
\end{tabular*}
\end{center}
\caption{\em Experimental results for ${\rm Re}(\epe)$. }
\label{tab:eprime} 
\end{table}

A non-zero value for the \CP-violation parameter $\varepsilon^\prime$, 
defined as
\beqn
        \eta_{+-} = \varepsilon + \varepsilon^\prime~,\hspace{1cm}
        \eta_{00} = \varepsilon -2\varepsilon^\prime~,
\eeqn
where $\varepsilon \equiv \epsk$, establishes direct \CP violation
in the neutral kaon system. The corresponding experimental observable 
is ${\rm Re}(\epe)$. The first evidence of direct \CP violation 
in neutral kaons decays was found by the NA31 collaboration~\cite{na31}.
Statistically significant observations were obtained by the 
next-generation of experiments, NA48~\cite{na48} and KTeV~\cite{ktev}.
Table~\ref{tab:eprime} summarizes the available measurements that yield an 
average of ${\rm Re}(\epe)=(16.7\pm 1.6)\times 10^{-4}$, with 
$\chi^2 = 6.3$ for 3 degrees of freedom, that is a 
p-value of $10\%$. 
\vs
The SM prediction of ${\rm Re}(\epe)$ has large uncertainties
because it relies on the precise knowledge of penguin-like hadronic matrix elements. 
Detailed calculations at NLO~\cite{Buras93,Roma93} show that two hadronic
parameters $B_6^{(1/2)}$ (gluonic penguins) and $B_8^{(3/2)}$ 
(electroweak penguins) dominate, where the superscripts 
denote the dominant $\Delta I=1/2$ and $\Delta I=3/2$
contributions, respectively, and refer to the isospin change in the 
$K\to \pi\pi$ transition. It is convenient to express the SM prediction
as a function of the hadronic parameters with the use of the approximate
formula~\cite{Buras03}
\beqn
        {\rm Re}(\epe) = {\rm Im}(V_{td}V_{ts}^*)
                        \left[ 18.7 R_6\left(1-\Omega_{\rm IB}\right)
                                             -6.9 R_8 -1.8 
                        \right] 
                        \frac{\Lambda^{(4)}_{\MSbar}}{340\mev}~,
\eeqn
where $\Omega_{\rm IB} = 0.06\pm 0.08$ corrects for isospin-breaking~\cite{Piketal},
$\Lambda^{(4)}_{\MSbar} = (340 \pm 30)\mev$ is the characteristic QCD 
scale for 4 active quark flavors in the $\MSbar$ scheme, and where $R_6$ 
and $R_8$ are defined by
\begin{figure}[]
  \epsfxsize8.5cm
  \centerline{\epsffile{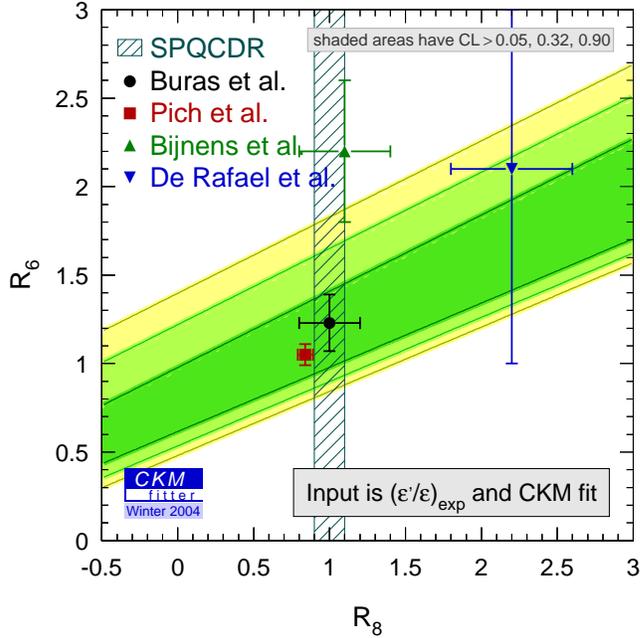}}  
  \vspace{-0.0cm}
  \caption[.]{\label{fig:eprime_b6b8}\em
        Allowed values for the hadronic parameters $R_6$ and $R_8$, using
        the experimental average of ${\rm Re}(\epe)$ and the result on 
        ${\rm Im}(V_{td}V_{ts}^*)$ from the standard CKM fit as input. 
        The symbols with error bars give the theoretical predictions
        taken from Refs.~\cite{Buras03,Piketal,deRafael,Bijnens,SPQCDR}.}
\end{figure}
\beq
        R_6 = B_6^{(1/2)} \left[\frac{121\mev}{\msRun(m_c)}\right]^2~,\hspace{1cm}
        R_8 = B_8^{(3/2)} \left[\frac{121\mev}{\msRun(m_c)}\right]^2~,
\eeq
with the running $s$-quark mass $\msRun(m_c) = (115\pm 20)\mev$.
In the strict large-$N_c$ limit, the hadronic parameters satisfy 
$B_6 = B_8 = 1$.
The quoted values and errors for $\Omega_{\rm IB}$, $\Lambda^{(4)}_{\MSbar}$,
and $\msRun(m_c)$ are taken from Ref.~\cite{Buras03}. In the Wolfenstein 
parameterization one has
${\rm Im}(V_{td}V_{ts}^*)=A^2\lambda^5\etabar + \mathcal{O}(\lambda^7)$.
Even though the experimental value of ${\rm Re}(\epe)$ is known to $10\%
$
accuracy, reliable constraints on $\etabar$ cannot be obtained (not even 
on its sign) due to the present uncertainties assigned to the hadronic 
parameters $R_6$ and $R_8$. Various dynamical effects come into play,
and while it is possible to estimate some of them thanks to appropriate
theoretical methods, it is very difficult to take into account all possible
contributions within a single approach.
\vs
Some consensus has been achieved on the value and error of $R_8$
obtained by lattice QCD. The most precise lattice
calculation~\cite{SPQCDR}  has an accuracy of $\sim 10\%
$. Taking into
account other lattice  results~\cite{CPPACS,RBC}, the conservative
average $R_8 = 1.00 \pm 0.20$  is quoted in Ref.~\cite{Buras03}, and
the value $R_6 = 1.23\pm 0.16$ is   derived from the correlation
between the experimental result for $\epe$  and $R_8$. The computation
of $R_6$ on the lattice is more difficult due to the mixing with
lower dimensional operators. An attempt can be found in Ref.~\cite{RBC};
however this study does not take into account flavor-symmetry breaking
effects that come from quenching artefacts, as stressed in
Ref.~\cite{GP}. As for $R_8$, the
authors of Ref.~\cite{deRafael} argue that again significant contributions 
that vanish in the quenched approximation could spoil the lattice
estimate.
Several analyses
using analytical non-perturbative techniques are also available. An
approach based on dispersion relations to evaluate final state
interaction finds $R_6 = 1.05\pm 0.06$ and  $R_8= 0.84\pm
0.05$~\cite{Piketal}. A chiral perturbation theory calculation  gives
$R_6=2.2\pm 0.4$ and $R_8=1.1\pm 0.3$~\cite{Bijnens}. Another recent
calculation, taking into account ${\cal O}(n_f/N_c)$  corrections to
the large-$N_c$ limit of QCD, finds $R_6=2.1\pm 1.1$ and  $R_8 =2.20\pm
0.40$~\cite{deRafael}. Noticeable disagreement  in both values and
errors among the different approaches is observed.
\vs
It is therefore instructive to study the constraints put upon these 
parameters by the experimental data. Figure~\ref{fig:eprime_b6b8} shows 
the allowed region for $R_6$ versus $R_8$, obtained from the experimental
average for ${\rm Re}(\epe)$ together with the standard CKM fit result for 
${\rm Im}(V_{td}V_{ts}^*)$. The symbols with error bars indicate the 
theoretical calculations. One concludes that the various
theoretical predictions provide estimates for $\epe $ that are in 
agreement with the experimental data, but the present size of the
uncertainties and also the disagreement between the various predictions
for $R_6$ prevents us from using $\epe$ as a constraint in the 
standard CKM fit. In the future, model-independent constraints on $R_6$
could be extracted from the measurement of \CP-violation in kaon decays
to three pions~\cite{GPS_B6}.
\vs
As an exercise we follow the strategy of Ref.~\cite{Buras03}
and use the average lattice QCD value for $R_8$, together with the 
experimental average of ${\rm Re}(\epe)$ and the standard CKM fit, 
to constrain $R_6$. We find the $95\%$ CL range
\beqns
        0.75 \:<\: R_6 \:<\: 1.80~.
\eeqns

\section{Rare Decays of Charged Kaons: $K^+\to\pi^+\nu\nub$} 
\label{sec:Kpinn}

\begin{figure}[t]
  \centerline{\epsfxsize8.1cm\epsffile{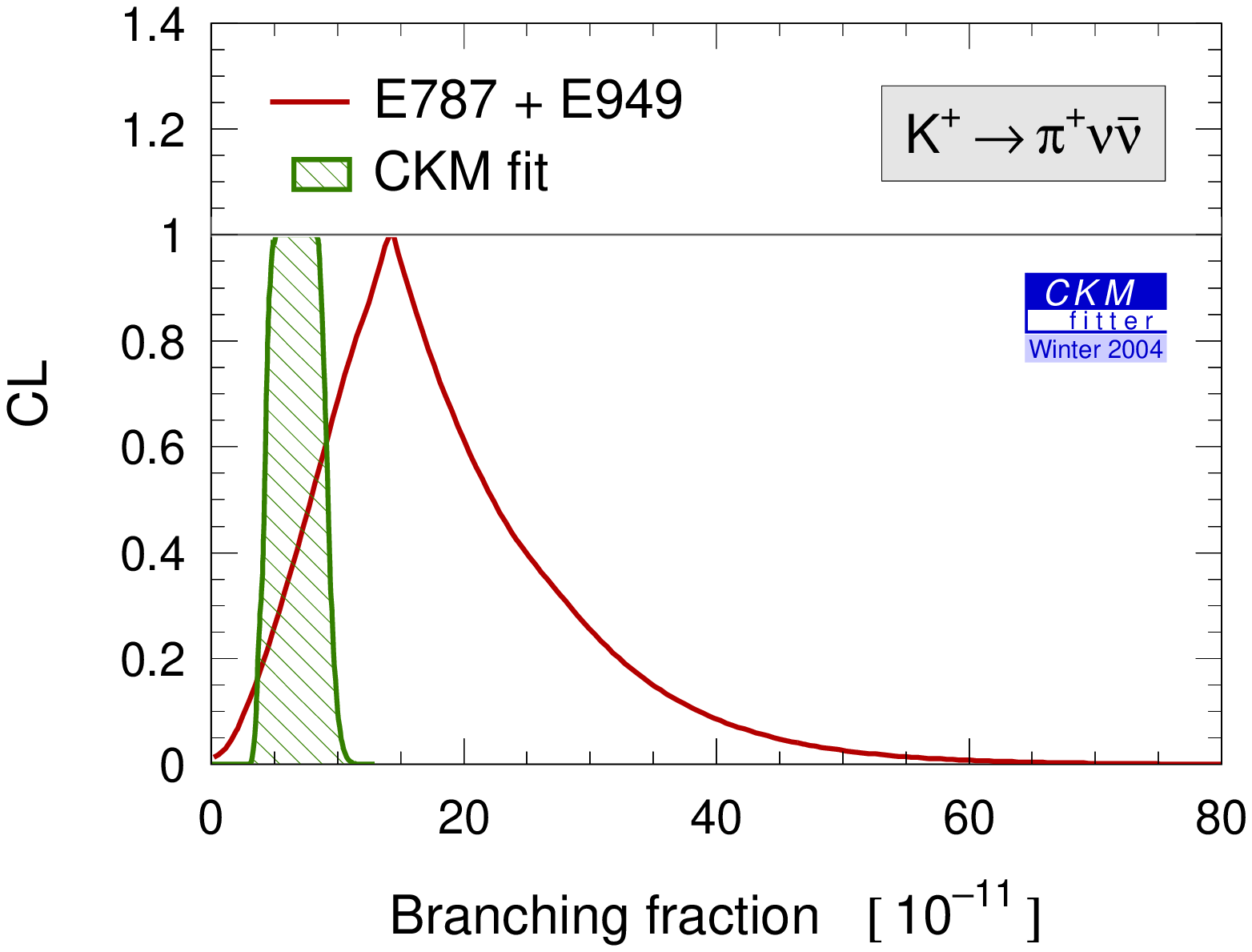}
              \epsfxsize8.1cm\epsffile{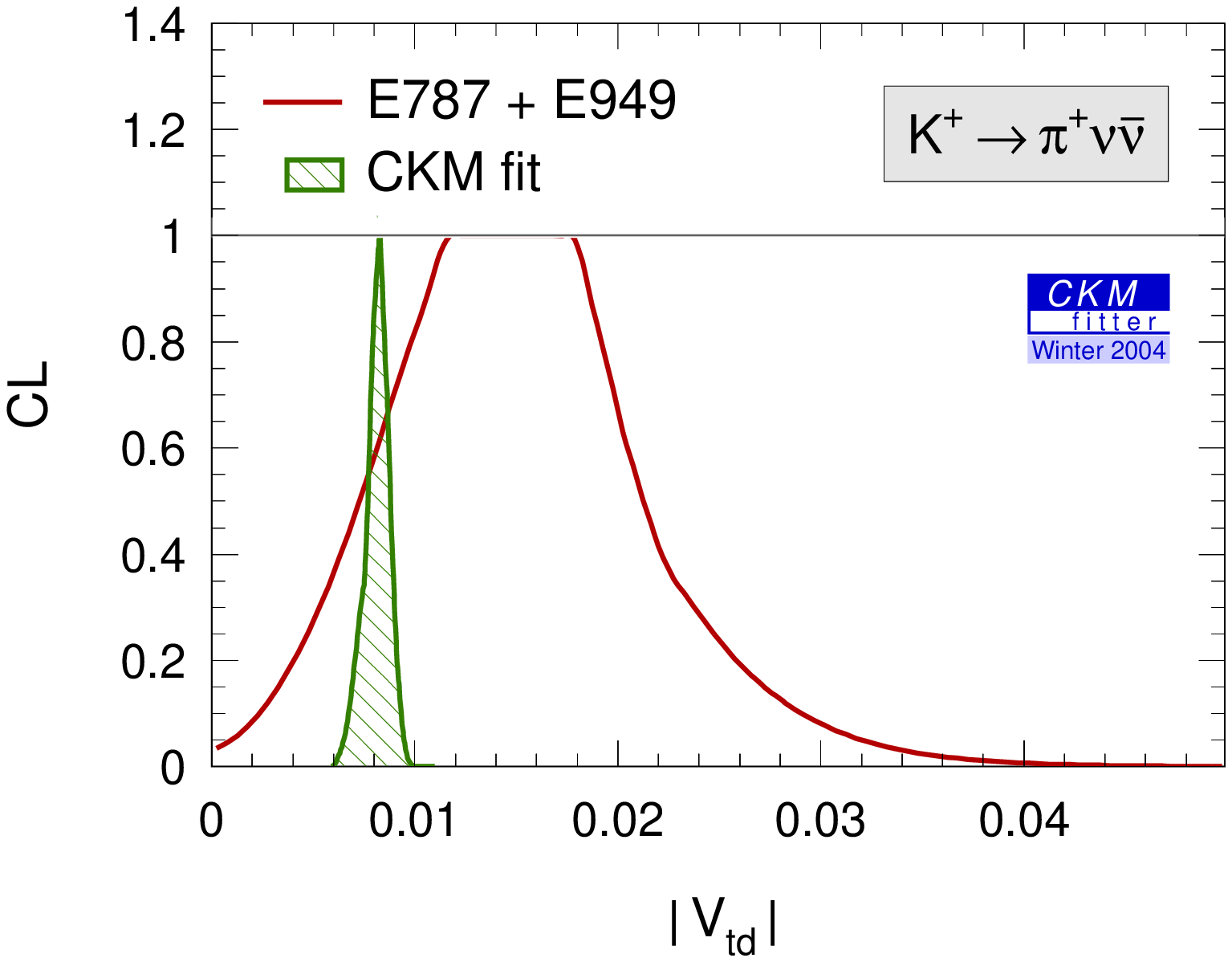}}
  \vspace{-0.5cm}
  \caption{\em\label{fig:kpinnbr}
        Confidence level of $\BR(K^+\to \pi^+\nu\nub)$ (left) and $\Vtd$ 
        (right). The solid lines give to the constraints from the combined
        E787 and E949 measurements, and the hatched areas represent the 
        SM predictions obtained from the standard CKM fit. }
\end{figure}

The BNL-E787 experiment has observed two events of the rare decay
$K^+\to \pi^+\nu\nub$, resulting in the branching fraction
$\BR(K^+\to\pi^+\nu\nub)=(1.57^{+1.75}_{-0.82}) \times 10^{-10}$~\cite{E787},
which due to the small expected background rate ($0.15\pm 0.03$ events) 
effectively excludes the null hypothesis. One additional event has 
been observed near the upper kinematic limit by the successor 
experiment BNL-E949~\cite{E949}. They quote the combined branching 
fraction $\BR(K^+\to\pi^+\nu\nub)=(1.47^{\,+1.30}_{\,-0.89})\times 10^{-10}$.
The left hand plot in Fig.~\ref{fig:kpinnbr} gives the CLs 
for the experimental result (CL obtained from Ref.~\cite{E949}) 
and the SM prediction (see paragraphs below), with input from
the standard CKM fit.
\vs
In the SM, the branching fraction is given by~\cite{BuBu}
\beq
\label{eq_kpppnunu}
    \BR(K^+\to\pi^+\nu\nub)
        \:=\: r_{K^+}
          \frac{3\alpha^2}{2\pi^2}
          \frac{\BR(K^+\to\pi^0 e^+ \nu)}
               {|V_{us}|^2{\rm sin}^4\theta_{\rm W}}
          \sum_{i=e,\mu,\tau}
        \left|\eta_X X_0(x_t)V_{td}V^*_{ts}
              + X_{\rm NL}^{(i)}V_{cd}V^*_{cs}
        \right|^2\ .
\eeq
Here, $r_{K^+}=0.901$ corrects for isospin breaking~\cite{kaonmarciano}, 
$X_0(x_{t})$ (with $x_{t}=\mtRun^2/m_{W}^2$)
is the Inami-Lim function 
\beq
\label{InamiLiKpinn}
        X_0(x) \:=\: \frac{x}{8}
        \left(
                \frac{x+2}{x-1} + \frac{3x-6}{(x-1)^2}\ln{x}
        \right),
\eeq    
corrected by a phenomenological QCD factor $\eta_X=0.994$, which is due to 
the top quark contribution~\cite{InamiLim} to order $\alpha_{s}$,
and the functions $X_{\rm NL}^{(\ell)}$, $\ell = e, \mu , \tau$, contain 
the contributions from $Z^0$ penguin and box diagrams with charm quarks
in the loops, and have been calculated at the next-to-leading log 
approximation~\cite{BuBu}. 
\vs
To illustrate the CKM constraint, we express Eq.~(\ref{eq_kpppnunu}) 
in the Wolfenstein parameters 
\beq
\label{BRkpinn}
        \BR(K^+ \to \pi^+ \nu \nub) \:=\: \kappa_{+} A^{4} X^{2}(x_{t}) 
                                           \frac{1}{\sigma}
        \left[(\sigma \etabar)^{2} + (\rho_{0} - \rhobar)^{2}\right]~,
\eeq
with
\beq
        X(x)     \,=\, \eta_X X_0(x)~, \hspace{1cm}
        \sigma   \,=\, 1 + \lambda^2 + \mathcal{O}(\lambda^4) ~, \hspace{1cm}
        \rho_{0} \,=\, 1 + \frac{P_{0}}{A^{2} X(x_{t})}~.
\label{BRkpinn1}
\eeq
Equation~(\ref{eq_kpppnunu}) provides an almost elliptic constraint in the 
($\rhobar,\etabar$) with the center close to the $(\rhobar=1,\etabar=0)$ 
apex of the Unitarity Triangle. It allows us to extract the CKM matrix 
element $\Vtd$ from the branching fraction measurement. The 
constant $\kappa_{+}$ is defined in Ref.~\cite{BuBu}. It contains a 
$\lambda^8$ dependence so that $\BR(K^+ \to \pi^+ \nu\nub)$ is a function 
of $\left(A\lambda^2\right)^4$, which is constrained by $\Vcb$ and 
experimentally determined from inclusive and exclusive $b\to c\ell \nu$ 
transitions.
\begin{figure}[t]
  \centerline{\epsfxsize8.1cm\epsffile{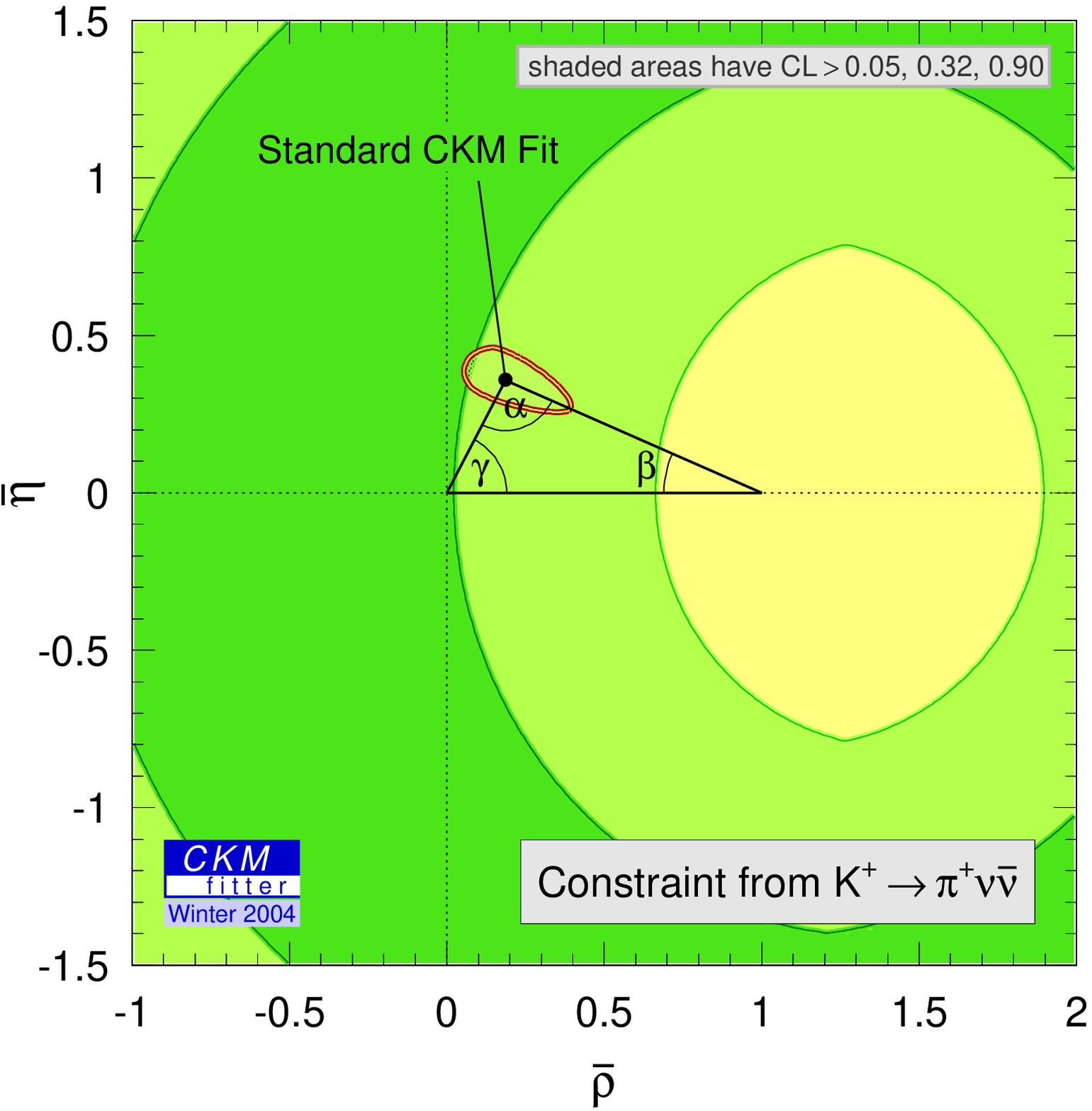}
              \epsfxsize8.1cm\epsffile{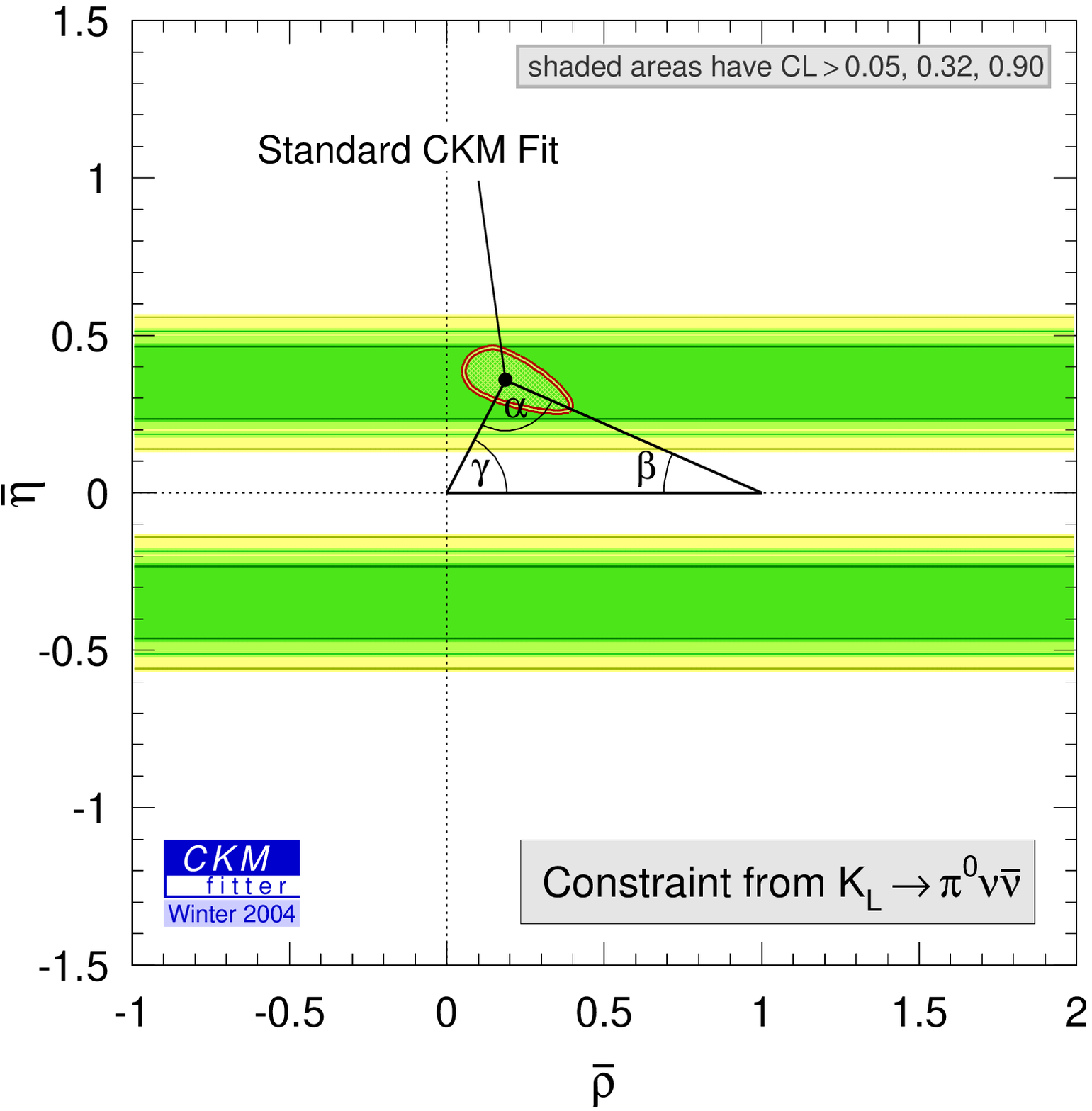}}
  \caption{\em \label{fig:kpinn2d}\label{fig:klpi0nn}
        Constraint in the $\rhoeta$ plane from the combined E787 and E949 
        measurements of $\BR(K^+\to\pi^+\nu\nub)$ (left), and from a 
        hypothetical $\BR(K_L^0\to \pi^0\nu\nub)$ measurement
        with $10\%$ relative error (right). Dark, medium and light 
        shaded areas have $CL>0.90$, $0.32$ and $0.05$, respectively. }
\end{figure}
Finally, the parameter $P_0$ quantifies the charm quark contribution and
is given by
\beq
\label{P_0}
        P_0 \:=\: \frac{1}{\lambda^4}\left[\frac{2}{3} X_{\rm NL}^{e} 
                + \frac{1}{3} X_{\rm NL}^{\tau}\right]~.
\eeq
Theoretical uncertainties on $P_0$ arise from the charm quark mass, the 
renormalization scale dependence and $\Lambda_{\rm QCD}$. 
\vs
The left hand plot in Fig.~\ref{fig:kpinn2d} shows the constraints in the 
$\rhoeta$ plane obtained from the comparison of the experimental result with the SM prediction. Within the large experimental errors, the constraint 
is found to be compatible with the allowed region obtained from the standard 
CKM fit.
\vs
We concentrate in the following on the study of the constraint on $\Vtd$
to evaluate the potential of future $\BR(K^+\to \pi^+\nu\nub)$
measurements. Relative uncertainties on the branching fraction scale 
approximately as $4\sigma ( \Vcb)/\Vcb$ and $2\sigma\left( X_0\right)/X_0$, 
and the relative error on $\Vtd^2$ scales equivalently. Moreover, the charm 
quark contribution~(\ref{P_0}) induces an uncertainty on the center of the
elliptical constraint, which translates into an uncertainty on $\Vtd$.
The right hand plot in Fig.~\ref{fig:kpinnbr} gives the present constraints 
on $\Vtd$ for the combined E787 and E949 measurements, and the standard CKM fit,
respectively. We extrapolate into the future by assuming that the 
branching fraction is equal to the central value obtained in the present 
CKM fit. Other inputs used in this study are $\Vus$ (to fix the Wolfenstein 
parameter $\lambda$), and $\Vub$ together with $\Vcb$ to intersect the 
elliptical constraint in a restricted area of the $\rhoeta$ plane, and to
hence reduce the effect of the uncertainty on the center of the ellipse.
Table~\ref{InputParameters} gives a breakdown of the uncertainties 
contributing to the error of $\Vtd$ for three scenarios:
\bei

\item[(\iA)]    using the present knowledge of the relevant input parameters
                and neglecting the statistical error on the 
                $\BR(K^+ \to \pi^+ \nu\nub)$ measurement;

\item[(\iB)]    assuming a measurement with a statistical precision of
              $10\%$ and an improvement of the relevant uncertainties to a 
                $1\%$ error on $\Vcb$, a $2\gev$ error on the top quark 
                mass, and a $50\mev$ error on the charm quark mass;

\item[(\iC)]    assuming a measurement with a statistical precision of $10\%$, 
                and neglecting all theoretical uncertainties in the prediction 
                of $\BR(K^+ \to \pi^+ \nu\nub)$.

\eei
We conclude from this exercise that, once an accurate branching fraction measurement
becomes available, the quantitative knowledge of the input parameters to the 
SM prediction must be significantly improved so that it does not dominate
the uncertainty on the $\Vtd$ constraint\footnote
{
        One could, of course, argue that instead of $\Vtd$ the parameter of 
        interest is $|V_{td}V_{ts}^*|^2$. However we note 
        that the SM prediction for $\BR(K^+ \to \pi^+ \nu\nub)$ and hence the 
        extraction of $\Vtd$ is not dominated by the uncertainty on $\Vcb$.
        Also, the interest in this mode is twofold: firstly the improvement 
        of the knowledge of the CKM phase, and secondly the search for physics
        beyond the SM. In both cases $\Vtd$ appears to be the appropriate 
        parameter.
}.
For the input values used in the Scenario~(\iA), the charm term is the dominant 
source of uncertainty, mostly due to the charm quark mass and the renormalization 
scale dependence. Since the top quark mass is expected to be measured with 
increasing accuracy at current and future hadron machines, and the error 
used for the charm quark mass is rather conservative, an improved precision
on $\Vcb$ will become mandatory (see Section~\ref{sec:input_vcb}).
\begin{table}[t]
\begin{center}
\setlength{\tabcolsep}{0.0pc}
\begin{tabular*}{\textwidth}{@{\extracolsep{\fill}}lcc}\hline
&& \\[-0.3cm]
                Scenario        &       $\geq 5\%$ CL range on $\Vtd$ $[\times 10^{-3}]$        
                                                                & Half width
        \\[0.15cm]
        \hline
&& \\[-0.3cm]
                (\iA)           &       $6.1$ - $10.5$  & $2.2$ \\
        $\sigma(\mcRun)$ only   &       $6.8$ - $9.9$   & $1.6$ \\
        $\sigma(\mtRun)$ only   &       $7.7$ - $8.9$   & $0.6$ \\
        $\sigma(\Vcb)$ only     &       $7.9$ - $8.7$   & $0.4$ \\[0.15cm] \hline
&& \\[-0.3cm]
                (\iB)           &       $7.0$ - $9.7$   & $1.3$\\[0.15cm] \hline
&& \\[-0.3cm]
                (\iC)           &       $7.2$ - $9.4$   & $1.1$ \\[0.15cm] \hline
&& \\[-0.3cm]
        standard CKM fit        &       $6.5$ - $9.5$   & $1.5$ \\[0.15cm] \hline
\end{tabular*}
\caption{\label{InputParameters}\em 
        Constraints on $\Vtd$ from $\BR(K^+ \to \pi^+ \nu\nub)$ for the three
        scenarios described in the text. The last line gives the result from 
        the present standard CKM fit (Table~\ref{tab:fitResults2}). }
\end{center}
\end{table}

\section{Rare Decays of Neutral Kaons: $\KL \to \pi^{0} \nu \nub$}
\label{sec:K0Lpi0nn}

In the SM, the golden decay $\KL \to \pi^{0} \nu \nub$
proceeds almost entirely through a direct \CP-violating amplitude 
dominated by the top quark contribution. The theoretical prediction 
of the branching fraction is given by~\cite{BuBu}
\beqn
        \BR(\KL \to \pi^{0} \nu \nub) 
        &=& \kappa_{L}  \left(\frac{{\rm Im}[V_{td}V_{ts}^*]}{\lambda^5}
                        \right)^{\!\!2}
                           X^{2}(x_{t}) \nonumber \\[0.2cm]
        &=& \kappa_{L} A^{4}\eta^2 X^{2}(x_{t}) + \mathcal{O}(\lambda^4)~,
\eeqn
where 
$\kappa_{L}=\kappa_{+}(r_{K_L}\tau_{K_L})/(r_{K^+}\tau_{K^+})=(2.12\pm 0.03)\times10^{-10}$~\cite{PajuoMayor},
and where $r_{K_L}=0.944$ accounts for isospin breaking~\cite{kaonmarciano}.
The constant $\kappa_+$ is defined in Ref.~\cite{BuBu}. It contains  
a $\lambda^8$ term so that the branching fraction is again proportional to 
$\Vcb^4$. The constraint in the ($\rhobar,\etabar$) plane obtained 
from a future measurement of $\BR(\KL \to \pi^{0} \nu \nub)$ (here with 
$10\%$ relative uncertainty) corresponds to two horizontal lines
as illustrated on the right hand plot of Fig.~\ref{fig:klpi0nn}.
The relative error on $|\etabar|$ scales with 
$2\sigma\left( \Vcb\right)/\Vcb$ and $\sigma \left(X\right)/X$.
\vs
We study the CL on $|\etabar|$ obtained from a $\BR(\KL \to \pi^{0} \nu
\nub)$ measurement, for the same three scenarios  introduced in the
previous section. We assume that the branching fraction is equal to the
central value from the present CKM fit.  Table~\ref{TabK0L} gives a
breakdown of the uncertainties  contributing to the error of
$|\etabar|$ for the three scenarios defined in the previous section.
The dominant source of uncertainty  on $|\etabar|$ in Scenario (I)
 is introduced by
$\Vcb$. Since the sensitivity to theoretical  uncertainties is reduced,
the constraint on  $|\etabar|$  will remain  statistically limited for
realistic expectations on near-future measurements  of the $\KL \to
\pi^{0} \nu \nub$ branching fraction (with $\sim 30$--$60$  signal
events).
\begin{table}[t]
\begin{center} 
\setlength{\tabcolsep}{0.0pc}
\begin{tabular*}{\textwidth}{@{\extracolsep{\fill}}lcc} \hline
&& \\[-0.3cm]
                Scenario        &       $\geq 5\%$ CL range on $|\etabar|$      
                                                                &       Half width
        \\[0.15cm]
        \hline
&& \\[-0.3cm]
                (\iA)           &       $0.313$ - $0.399$       &  $0.043$
        \\
        $\sigma(\mtRun)$ only   &       $0.333$ - $0.379$       &  $0.023$
        \\
        $\sigma(\Vcb)$ only     &       $0.327$ - $0.385$       &  $0.028$
        \\[0.15cm]
        \hline
&& \\[-0.3cm]
                (\iB)           &       $0.336$ - $0.376$       &  $0.020$
        \\[0.15cm]
        \hline
&& \\[-0.3cm]
                (\iC)           &       $0.317$ - $0.395$       &  $0.039$
        \\[0.15cm]
        \hline
&& \\[-0.3cm]
        standard CKM fit        &       $0.273$ - $0.444$       & $0.086$ \\[0.15cm] \hline
\end{tabular*}
\caption{\label{TabK0L}\em 
        Constraints on $|\etabar|$ from $\BR(\KL\to \pi^0\nu\nub)$,
        for the three scenarios described in the text. The last line gives the 
        result from the present standard CKM fit (Table~\ref{tab:fitResults2}).}
\end{center} 
\end{table}

\section{Conclusions}
        
Despite the success of the experimental effort that lead to a precise 
measurement of ${\rm Re}(\epe)$, the present  situation does 
not allow us to use it as a reliable constraint in the CKM fit without a 
substantial improvement on the theoretical side. 
As for the rare kaon decays, a handful of events are expected to be observed 
by BNL-E949~\cite{E949}, while the CKM project at FNAL~\cite{CKM}, starting 
around 2005, expects to collect about 100 events within a few years of data 
taking. Since these measurements have very small backgrounds, a $\sim 10\%$ 
statistical error on the $\BR(K^+\to \pi^+\nu\nub)$ is expected. The prospects 
for a measurement of the decay $\KL\to\piz\nu\nub$ are more uncertain due to
the enormous experimental challenge. Long-term projects~\cite{JHF} are designed 
to collect high statistics, but even intermediate-statistics branching fraction 
measurements~\cite{KOPIO} may reveal potentially large deviations from the SM,
and are hence of considerable interest.

%
%
 \newpage\part{Constraints on $2\beta+\gamma$ and $\gamma$ from Tree Decays}\setcounter{section}{0}
\markboth{\textsc{Part V -- Constraints on $2\beta+\gamma$ and $\gamma$ from Tree Decays}}
         {\textsc{Part V -- Constraints on $2\beta+\gamma$ and $\gamma$ from Tree Decays}}
\label{sec:gamma}

 \section{\CP-Violating Asymmetries in $\Bz\to D^{(*)\pm}\pi^\mp$ Decays}
\label{sec:dstarpi}

Even though they are not \CP  eigenstates, partially and fully reconstructed 
$\Bz\to D^{(*)\pm}\pi^\mp$ decays are sensitive to the UT angle
$\gamma$ because of the interference between the CKM-favored 
amplitude of the decay $\Bz\to D^{(*)-}\pi^+$ with the doubly 
CKM-suppressed amplitude of $\Bz\to D^{(*)+}\pi^-$.\footnote
{
        This is similar to the situation in $\Bz\to\rho^\pm\pi^\mp$ 
        decays (see Section~\ref{sec:charmlessBDecays}.\ref{sec:introductionrhopi}),
        even if the two amplitudes there are of the same CKM order,
        which considerably increases their potential \CP asymmetries. 
} The relative weak phase between these two amplitudes is $-\gamma$ and, 
when combined with the $\Bz\Bzb$ mixing phase, the total phase 
difference is $-(2\beta+\gamma)$ to all orders in $\lambda$:
\beq
        -(2\beta+\gamma)\:=\: 
                \arg\left[-\frac{V_{td}V_{tb}^*}{V_{td}^*V_{tb}}
                           \frac{V_{cd}^*V_{ub}}{V_{ud}V_{cb}^*}
                    \right]~.
\eeq
The interpretation of the \CP-violation observables in terms of 
the UT angles requires external input on the ratio
\beq
        r^{(*)} \equiv 
        \left|\frac{q}{p}
              \frac{A(\Bzb\to D^{(*)-}\pi^+)}
                   {A(\Bz\to D^{(*)-}\pi^+)}
        \right|~,
\eeq
which can be obtained experimentally from the corresponding 
flavor-tagged branching fractions, or from similar modes that are 
easier to measure. These can be ratios of branching fractions of 
the charged $B^+\to D^{(*)+}\pi^0$ to the neutral CKM-favored 
decay, or ratios involving self-tagging decays with strangeness like 
$B^0\to D_s^{(*)+}\pi^-$. Corrections for SU(3) breaking 
in the latter case generate a significant theoretical uncertainty,
which is generally hard to quantify. Naively, one can estimate
$r^{(*)}\sim|V_{cd}^*V_{ub}/V_{ud}V_{cb}^*|\simeq0.02$. At present,
the most precise semi-experimental determination of $r^{(*)}$
can be obtained from the SU(3)-corrected ratio
\beq
\label{eq:rstar}
        r^{(*)} = \frac{\Vus}{\Vud}
                  \sqrt{\frac{{\cal B}(\Bz\to D_s^{(*)+}\pi^-)}
                             {{\cal B}(\Bz\to D^{(*)-}\pi^+)}
                       }
                  \frac{f_{D^{(*)}}}{f_{D_s^{(*)}}}~.
\eeq
Inserting the corresponding branching fractions and
decay constants leads to~\cite{babardstarpiF}
\beq
        r^* = 0.017^{\,+0.005}_{\,-0.007}~,\hspace{1.0cm}
        r   = 0.014\pm0.004~.
\eeq
In Ref.~\cite{babardstarpiF} a theoretical uncertainty of $30\%
$ of the central value is  attributed in addition to the experimental
errors to each of  the quantities. It accounts for SU(3)-breaking
corrections and  the neglect of $W$-exchange contributions to the
$\Bz\to D^{(*)+}\pi^-$  decay amplitude. However Eq.~(\ref{eq:rstar})
already corrects for the main (factorizable) symmetry breaking; on the
other hand, the exchange diagram is the only possible contribution to
the $D_s^\pm K^\mp$ mode: thus one has roughly
$|\mathrm{exchange}/\mathrm{emission}|^2\sim 
\BR(B^0\to D_s^- K^+)/\BR(B^0\to D^-\pi^+) \sim 1\%$~\cite{DsKref}. 
As a consequence, taking into account the residual non
factorizable SU(3) breaking and the order of magnitude of the exchange
contribution, we estimate the total theoretical uncertainty to be
of the order of $15\%
$ for both $r$ and $r^*$, keeping in mind that a more refined estimate
of this error source will be needed when the statistics increase.
\vs
\babar~\cite{babardstarpiF,babardstarpiP} and Belle~\cite{belledstarpiF} 
use two sets of observables
\beq
        S^{(*)\pm} = 2r^{(*)}\sin(2\beta+\gamma\pm\delta^{(*)})~,
\eeq
where $S^{(*)\pm}$ is the coefficient of the sine term in the 
time evolution of the $\Bz(\Bzb)\to D^{(*)\pm}\pi^\mp$
system, and 
\beqn
        a^{(*)} &\equiv& \frac{1}{2}\left(S^{(*)+} + S^{(*)-}\right) 
                = 2r^{(*)}\sin(2\beta+\gamma)\cos(\delta^{(*)}) ~, \\
        c^{(*)} &\equiv& \frac{1}{2}\left(S^{(*)+} - S^{(*)-}\right) 
                = 2r^{(*)}\cos(2\beta+\gamma)\sin(\delta^{(*)}) ~,
\eeqn
so that $S^{(*)+}=a^{(*)}+c^{(*)}$ and $S^{(*)-}=a^{(*)}-c^{(*)}$. 
These definitions are valid in the limit of small $r^{(*)}$ 
only so that terms of order $r^{(*)\ge2}$ can be neglected and
the cosine coefficient in the time evolution 
is one ($C^{(*)}=(1-r^2)/(1+r^2)\to1$). The relative strong 
phase $\delta^{(*)}$ is unknown and has to be determined simultaneously 
with $2\beta+\gamma$ from the experimental observables.
Due to the disparate strength of the two interfering amplitudes, 
the \CP   asymmetry is expected to be small, so that the possible 
occurrence of \CP  violation on the tag side becomes an important 
obstacle. Tag side CPV is absent for semileptonic $B$ decays (mostly 
lepton tags). The parameter $a^{(*)}$ is independent of tag side CPV. 
\begin{table}[t]
\begin{center}
{\small
\setlength{\tabcolsep}{0.0pc}
\begin{tabular*}{\textwidth}{@{\extracolsep{\fill}}ccccc} \hline
&&&& \\[-0.3cm]
                & \mc{2}{c}{\babar~\cite{babardstarpiP,babardstarpiF}}  
                & Belle~\cite{belledstarpiF}    &  \\[0.15cm]
                & partially reconstructed& fully reconstructed
                & fully reconstructed   & \rs{Average}\\[0.15cm]
\hline 
&&&& \\[-0.3cm]
$a^*$           & $-0.063 \pm 0.024 \pm 0.014 $ 
                                & $-0.068 \pm 0.038 \pm 0.020  $
                & $\ph{-}0.063 \pm 0.041 \pm 0.016 \pm 0.013 $  
                                & $-0.038 \pm 0.021 $ \\[0.15cm]
$c^*$           & $-0.004 \pm 0.037 \pm 0.020 $ 
                                & $\ph{-}0.031 \pm 0.070 \pm 0.033 $
                & $\ph{-}0.030 \pm 0.041 \pm 0.016 \pm 0.030 $  
                                & $\ph{-}0.012 \pm 0.030 $ \\[0.15cm]
\hline
&&&& \\[-0.3cm]
$a$             & -             & $-0.022 \pm 0.038 \pm 0.020  $
                & $-0.058 \pm 0.038 \pm 0.013  $ 
                                & $ -0.041 \pm 0.029 $ \\[0.15cm]
$c$             & -             & $\ph{-}0.025 \pm 0.068 \pm 0.033  $
                & $-0.036 \pm 0.038 \pm 0.013 \pm 0.036   $
                                & $-0.015 \pm 0.044  $ \\[0.15cm]
\hline
\end{tabular*}
}
\end{center}
\vspace{-0.5cm}
  \caption[.]{\label{tab:s2bpg}\em
        Experimental results from time-dependent CP-asymmetry analyses of
        partially and fully reconstructed $\Bz\to D^{*\pm}\pi^\mp$ decays, 
        respectively. The averages are taken from the HFAG~\cite{HFAG}.}
\end{table}
\begin{figure}[p]
  \epsfxsize8.5cm
  \centerline{\epsffile{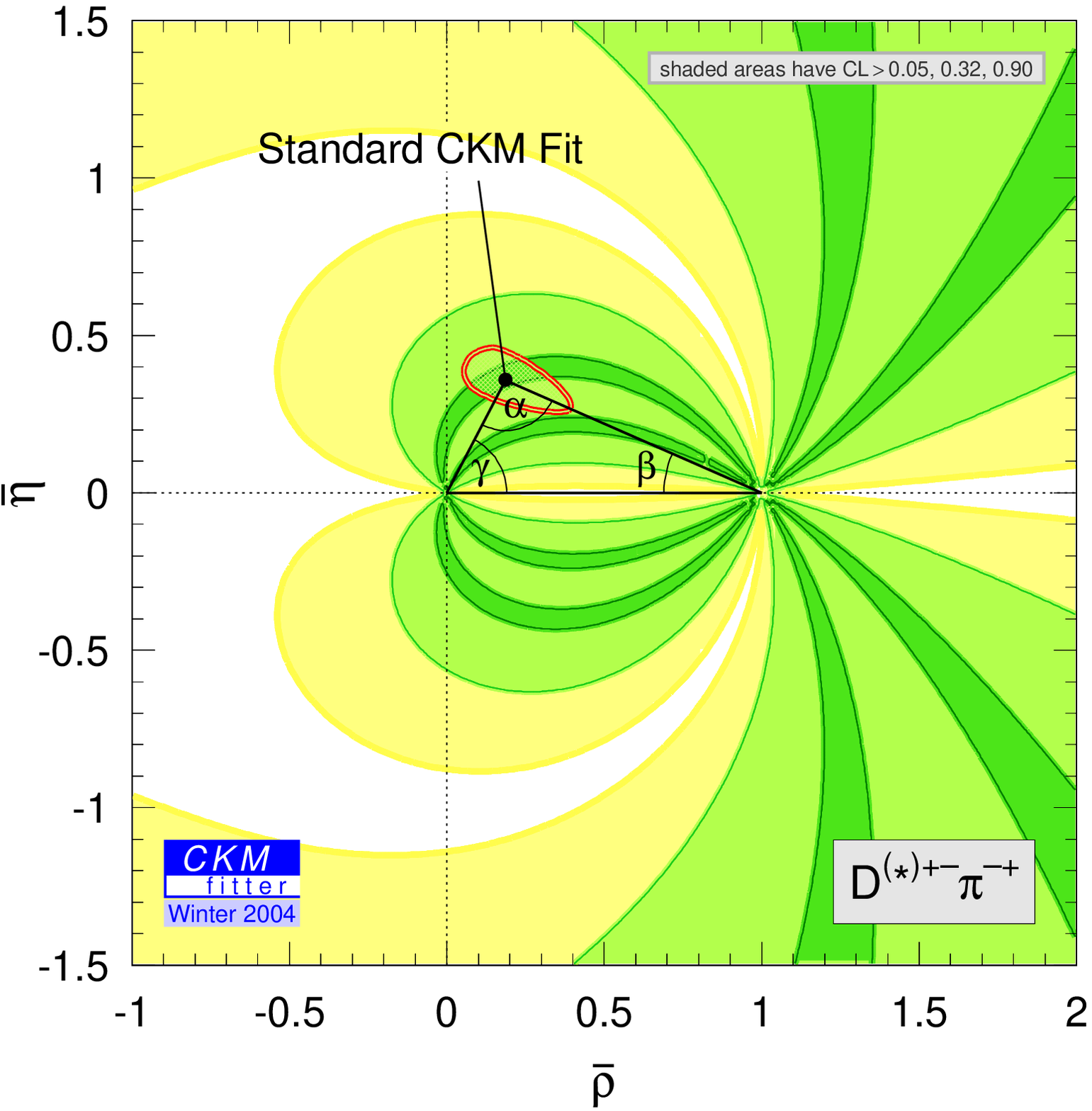}}
  \caption[.]{\label{fig:s2bpg_rhoeta}\em
        Constraint in the enlarged $\rhoeta$ plane from the 
        average measurement of time-dependent \CP-violating asymmetries in 
        $\Bz\to D^{*\pm}\pi^\mp$ decays. The shaded areas indicate
        $CL>5\%$ (light), $CL>32\%$ (medium) and 
        $CL>90\%$ (dark) regions. Also shown is the 
        $CL>5\%$ region of the standard CKM fit.}
  \vspace{1cm}
  \centerline{\epsfxsize8.1cm\epsffile{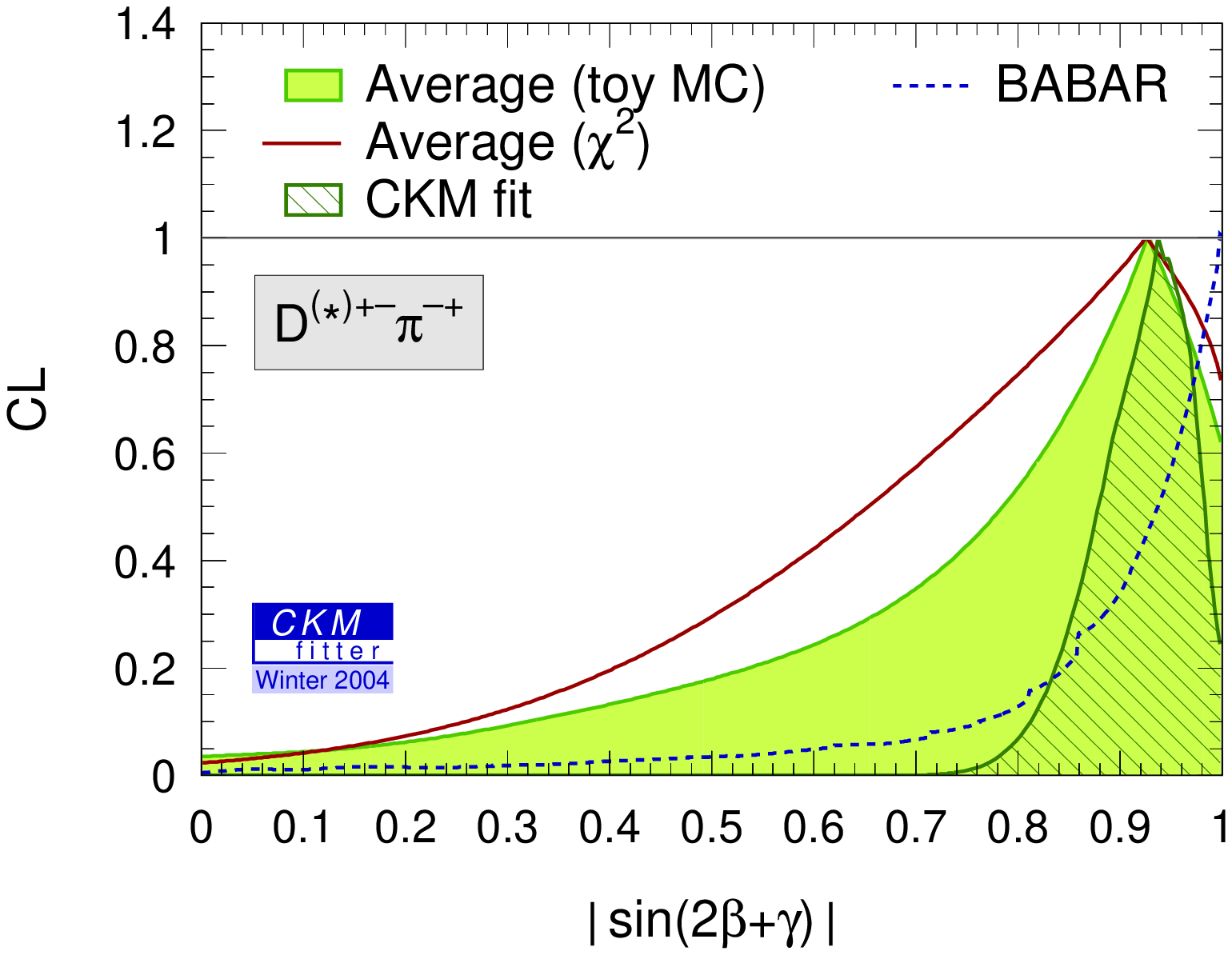}
              \epsfxsize8.1cm\epsffile{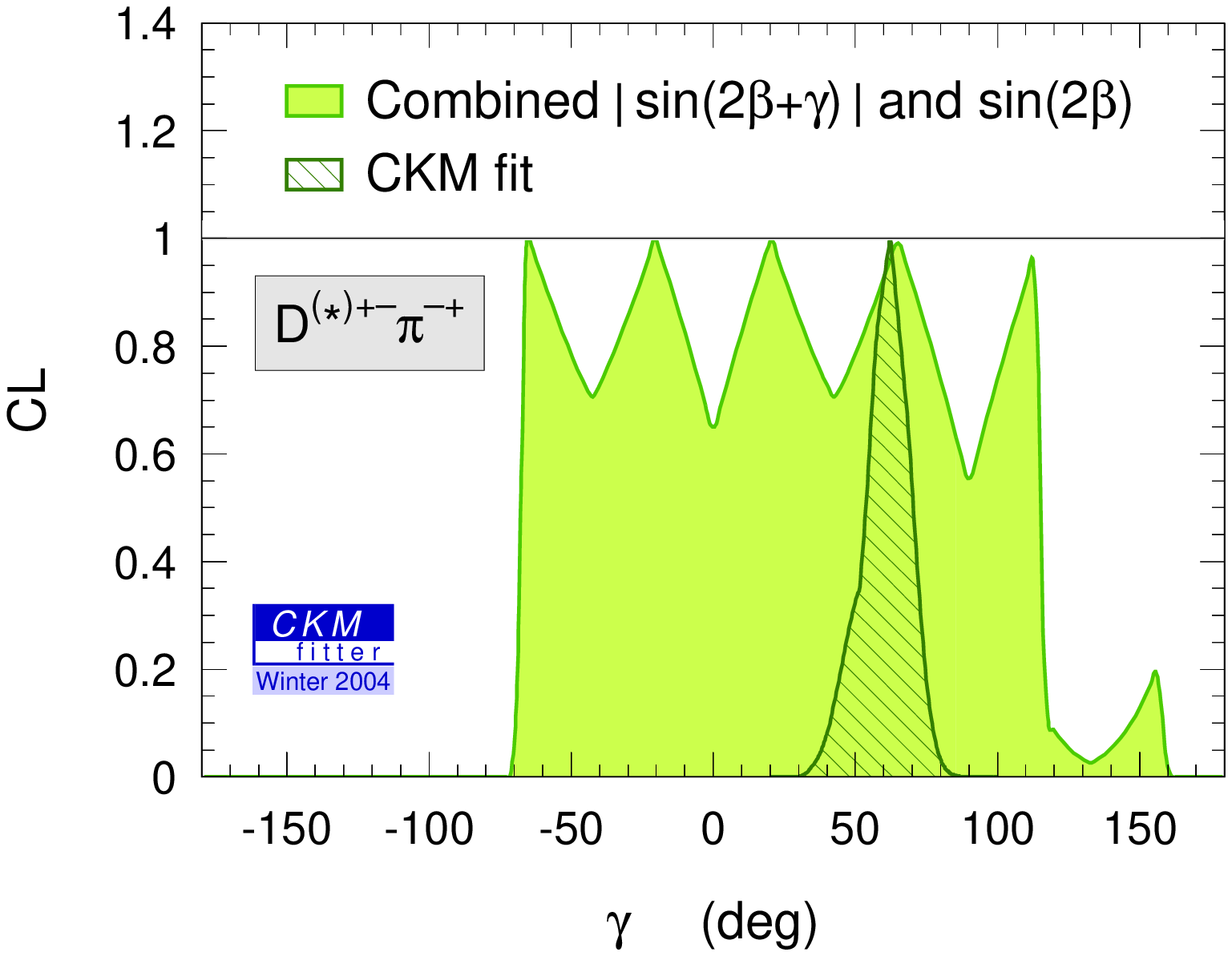}}
  \vspace{-0.5cm}
  \caption[.]{\label{fig:s2bpg1d}\em
        \underline{Left:} 
        confidence level obtained for $|\stbpg|$. The shaded
        area gives the average from \babar\  and Belle obtained
        by means of a toy Monte Carlo simulation (see 
        Section~\ref{sec:statistics}.\ref{sec:metrology_physicalBoundaries}). 
        As a comparison,
        we show by the solid line the approximate result from the 
        $\ProbCERN(\chi^2,1)$ interpretation. Also shown is 
        the result from \babar\ only, which leads to a stronger
        exclusion of small $|\stbpg|$ values due to the somewhat
        propitiously small value of $c^*$ in partially reconstructed 
        $\Bz\to D^{*\pm}\pi^\mp$ decays~\cite{babardstarpiP}.
        Also shown is the prediction from the standard CKM fit.
        \underline{Right:} confidence level obtained for the UT
        angle $\gamma$ when using \babar\  and Belle's results
        on $|\stbpg|$ $($and $|\ctbpg|)$ combined with the world
        average of $\stbwa$. Also shown is the prediction from the 
        standard CKM fit.}
\end{figure}
\begin{figure}[t]  
  \centerline{\epsfxsize8.1cm\epsffile{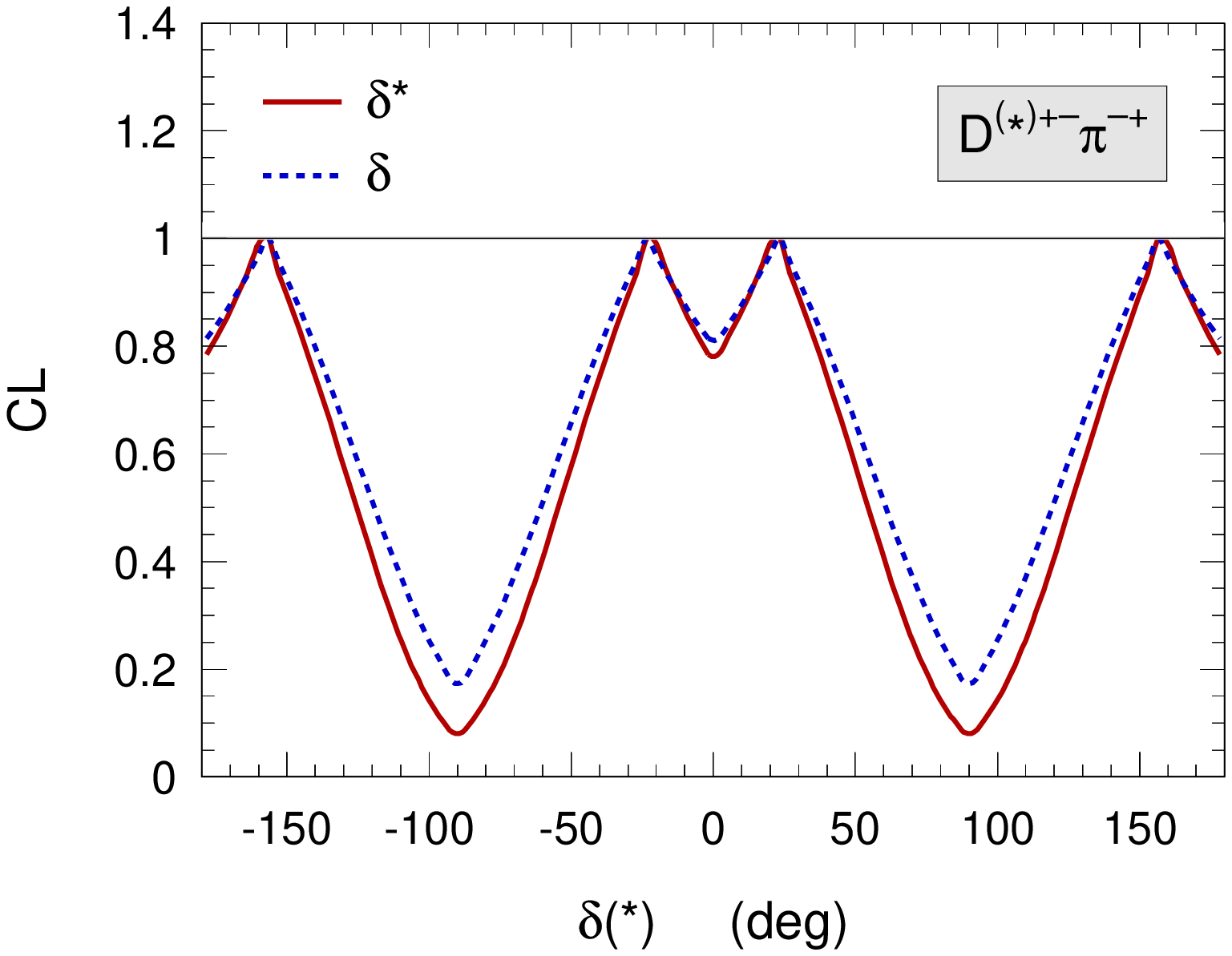}}
  \vspace{-0.5cm}
  \caption[.]{\label{fig:s2bpg1d_delta}\em
        Confidence levels obtained for the strong phases $\delta^{(*)}$
        occurring between the CKM-favored and CKM-suppressed
        branches of $\Bz\to D^{*\pm}\pi^\mp$ decays. The standard
        CKM fit has been used to constrain the weak CKM phases.}
\end{figure}
The experimental results are given in Table~\ref{tab:s2bpg}.
The averages quoted are taken from the HFAG~\cite{HFAG}.
\vs
For each mode the combinations
$\sin(2\beta+\gamma\pm\delta_f)$ are extracted ($\delta_{D\pi}=\delta$ 
and $\delta_{D^*\pi}=\delta^*$). Simple trigonometry shows that, as far
as the \CP angle is concerned, this is
equivalent to the determination of the quantity $|\sin(2\beta+\gamma)|$
up to a two-fold discrete ambiguity. Should the two strong phases
$\delta$ and $\delta^*$ be
different, the discrete ambiguity could in principle be resolved, leaving
a single solution for $|\sin(2\beta+\gamma)|$ (and thus four solutions
for the angle $2\beta+\gamma$ itself in $[0,\pi]$).
\vs
Figure~\ref{fig:s2bpg_rhoeta} shows the confidence level obtained
in the $\rhoeta$ plane. Also shown is the allowed region from 
the standard CKM fit. Good agreement is observed, although the 
statistical significance of the measurement is still weak.
Note that we have used a Gaussian $\ProbCERN(\chi^2,1)$ here 
to evaluate the CL. As seen below, this tends to overestimates 
the CL (and hence to weaken the constraint).
The left hand plot in Fig.~\ref{fig:s2bpg1d} shows the confidence 
level for $|\stbpg|$ obtained from the various measurements. A single
peak is observed, although $\delta$ and $\delta^*$ are very close to
each other (see below), because limited statistics merges the position 
of the two solutions. 
We perform a toy Monte Carlo simulation in order to evaluate 
the goodness of the Gaussian $\ProbCERN(\chi^2,1)$ approximation 
for the confidence level. Significant deviations are observed.
Also shown are the prediction from the standard CKM fit as well
as the result obtained when using the \babar\  results only,
which benefit from a small $|c^*|$ fluctuation in the measurement
with partially reconstructed $B$ decays. The right hand plot
shows the constraint obtained on the UT angle $\gamma$ when
also using the world average of $\stbwa$. For simplicity, we use 
the $\ProbCERN(\chi^2,1)$ approximation for this as well as for 
the upcoming plots. Also shown is the constraint from the 
standard CKM fit. One can use the latter prediction of $|\stbpg|$
together with the $\Bz\to D^{(*)\pm}\pi^\mp$ measurements to
constrain the strong phases $\delta^{(*)}$ as shown in 
Fig.~\ref{fig:s2bpg1d_delta}.
\vs
In summary we conclude that in spite of the considerable experimental 
effort to achieve this first direct constraint on $2\beta+\gamma$,
the present statistical accuracy is insufficient to improve the 
knowledge of the apex in the unitarity plane.
The errors of the present world averages given in Table~\ref{tab:s2bpg}
have to be reduced by a factor of about five (approximately $5\invab$
accumulated luminosity) to be competitive with the standard CKM 
fit on $|\stbpg|$ (assuming the above $15\%
$ uncertainty on $r^{(*)}$).
Such large statistics samples, which are necessary
due to the smallness of the $\CP$-violating asymmetries,
are likely to increase the 
importance of the experimental systematic uncertainties. 
Similar modes like $\Bz\to D^{(*)\pm}\rho^\mp$ must be included
in future to improve the reach of this analysis.

 \section{Dalitz Plot Analysis of $\Bp\to D^{(*)0} \Kp$ Decays}
\label{sec:d0k}

The golden method to measure the angle $\gamma$ at the $B$ factories 
has been proposed by Gronau, London and Wyler 
(GLW)~\cite{gronaulondon,gronauwyler} 
(see also Refs.~\cite{dunietz1,dunietz2}) and extended by Atwood, Dunietz 
and Soni (ADS)~\cite{ADS}. The GLW method consists of reconstructing the $\Dzb$ ($\Dz$) 
occurring in charged $\Bp\to\Dzb\Kp$ ($\Bp\to\Dz\Kp$) decays
as a \CP eigenstate (\eg, $\KS\pi^0$ or $\KS\pip\pim$) so that the 
CKM-favored ($b\to c$) and CKM-suppressed ($b\to u$) transition 
amplitudes interfere. The relative phase between these amplitudes is
$\gamma+\delta$, where $\delta$ is a \CP-conserving strong phase and 
$\gamma$ the weak UT angle. The measurement of the corresponding branching
fractions and \CP-violating asymmetries allows one to simultaneously 
extract $\gamma$ and the strong phase from a triangular isospin
analysis, up to discrete ambiguities, even if the strong phase vanishes,
but with virtually no theoretical uncertainties.
The feasibility of this or related analyses crucially depends on the 
size of the color- and CKM-suppressed $b\to u$ transition (expected 
to be roughly $\rB\sim1/8$, if color-suppression holds). Recently,
the \babar\  collaboration has determined an upper limit for the
amplitude ratio $\rB=|A(\Bp\to\Dz\Kp)/A(\Bp\to\Dzb\Kp)|$ of $0.22$ at 
$90\%
$~CL~\cite{babardk},
which dampens the hope for a performing $\gamma$ analysis using the 
GLW or ADS techniques at the first generation $B$ factories.
\vs
Along the line of Ref.~\cite{d0kidea}, the Belle collaboration
overcomes these difficulties by performing a Dalitz plot analysis of
$\Bp\to \Dz \Kp$ (and  $\Bp\to D^{*0}(\to\Dz\piz)\Kp$) decays followed
by a three-body  $\Dz$ decay to $\KS\pip\pim$~\cite{belled0k}\footnote
{
	 After the completion of this work, an update of the Belle analysis has been
	 submitted~\cite{belled0k2}.
	By means of a frequentist analysis, Belle 
	finds the combined result 
	$\gamma=[77^{\,+17}_{\,-19}({\rm stat})\pm13({\rm
	syst})\pm11({\rm model})]^{\circ}$, 
	which slightly differs from the previous value.
	This modification in the result does however not alter 
	the conclusion drawn from the
	study of New Physics in the present work.
}. The weak
phase  $\gamma$ and the strong phase $\delta$ as well as the magnitude
of the suppressed-to-favored amplitude ratio $\rB$ are extracted from a
fit to the interference pattern between $\Dz$ and $\Dzb$ in the Dalitz 
plot. A large number of intermediate resonances has to be considered to
properly model the full $\KS\pip\pim$ Dalitz plot, where high-statistics 
samples of charm decays can be used to fit the model parameters~\cite{cleod0}. 
Belle determines a probability density function (PDF) for $\phi_3=\gamma$ 
by means of a Bayesian analysis with uniform priors for $\gamma$, $\delta$  
and $\rB$. Single-sided integration of this PDF, and choosing the solution 
that is consistent with the standard CKM fit, results in
\begin{figure}[p]
  \centerline{\epsfxsize9.6cm\epsffile{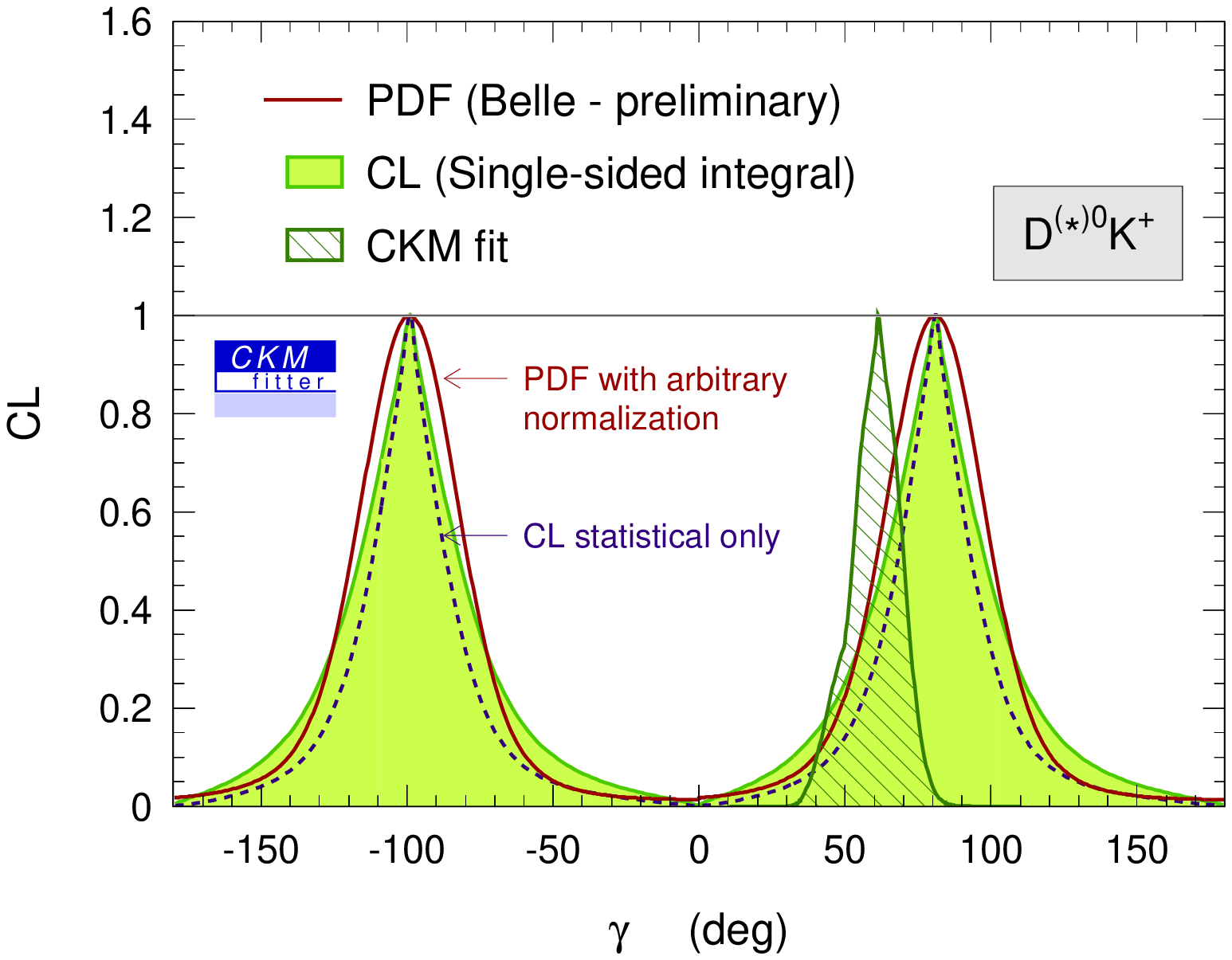} }
  \vspace{-0.4cm}
  \caption[.]{\label{fig:d0k_gamma}\em
	Confidence level of the UT angle $\gamma$ 
	from the Dalitz plot analysis
	of $\Bp\to D^{(*)0}\Kp$ decays~\cite{d0kidea,belled0k}.
	Shown are the experimental PDF (solid line - statistical only),
	found by Belle, and the corresponding
	CL from single-sided integration (shaded area - including 
	systematics, while the 
	dashed line gives the CL obtained when ignoring systematic 
	uncertainties).
	Also given is the constraint from the standard CKM fit.}
  \vspace{1.5cm}
  \centerline{\epsfxsize8.5cm\epsffile{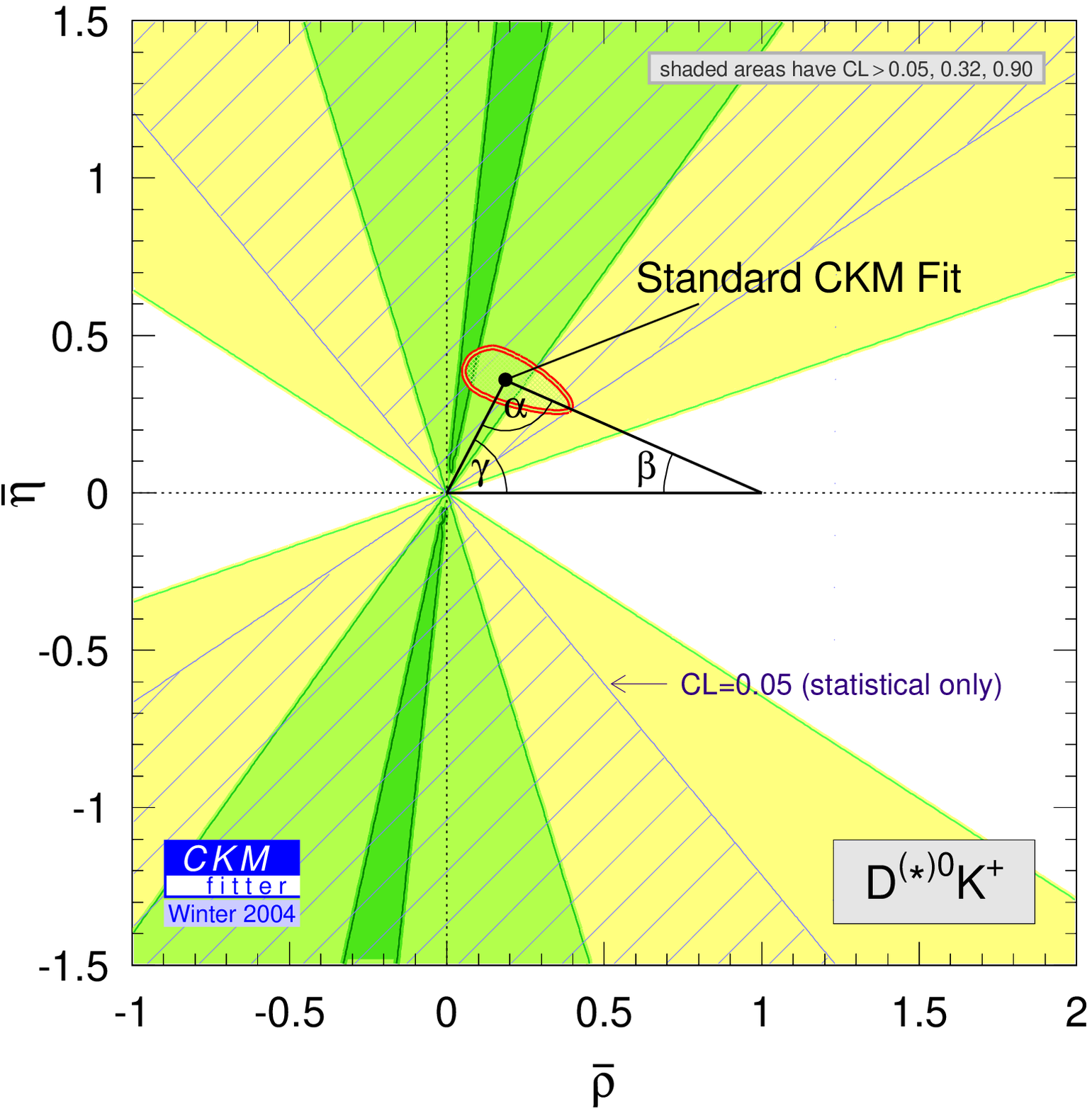}}
  \caption[.]{\label{fig:d0k_rhoeta}\em
	Confidence level in 
	the enlarged $\rhoeta$ plane from the Dalitz plot analysis
	of $\Bp\to D^{(*)0}\Kp$ decays~\cite{d0kidea,belled0k}.
	The shaded areas indicate
	$CL>5\%$ (light), $CL>32\%$ (medium) and 
	$CL>90\%$ (dark) regions. The hatched area indicates the 
	$CL>5\%$ region when ignoring systematic uncertainties.
	Also shown is the $CL>5\%$ constraint from the standard CKM fit.}
\end{figure}
\beq
\label{eq:d0k_gamma}
	\gamma = 81^\circ \pm 19^\circ \pm 13^\circ \pm 11^\circ~,
\eeq
where the first error is statistical, the second systematic and the 
third is due to the amplitude model. The constraint on the second 
solution, which is not  consistent with the standard CKM fit, is obtained
by the transformation $\gamma \to \gamma + \pi$ (the full analysis
actually leads to the determination of $\tan\gamma$). We have 
extracted and integrated the PDF from Ref.~\cite{belled0k} and find 
the confidence levels shown as a function of $\gamma$ and $\rhoeta$ 
in Figs.~\ref{fig:d0k_gamma} and \ref{fig:d0k_rhoeta}. Since systematic 
and model-dependent errors are not included in the PDF, we add them in 
quadrature to the statistical error. With these rather strong assumptions, 
agreement with the standard CKM fit is observed. Belle determines the 
magnitude of the suppressed-to-favored amplitude ratio to be 
$\rB=0.28^{\,+0.09}_{\,-0.11}$, which is slightly
larger than the expectation and than
the $90\% 
$~CL bound set by \babar, though the results are 
well compatible within errors ($23\%
$). Since large values of $\rB$ 
lead to an increased sensitivity to $\gamma$, the error given 
in~(\ref{eq:d0k_gamma}) may increase if the true $\rB$ is significantly
smaller. 
More data are needed to clarify this.
\vs
Since the result is still preliminary, we do not introduce it
in the standard CKM fit. It is however used in our analysis
of New Physics presented in Part~\ref{sec:newPhysics}.

%
%
 \newpage\part{Charmless \B Decays}\setcounter{section}{0}
\markboth{\textsc{Part VI -- Charmless \B Decays}}
         {\textsc{Part VI -- Charmless \B Decays}}
\label{sec:charmlessBDecays}

Unlike $\Bz\to \jpsi \KS$ or other charmonium decays, for which
amplitudes with weak phases that are different  from the dominant tree
phase are doubly CKM-suppressed, multiple weak  phases must be
considered in the analyses of charmless $B$ decays. This complication
makes the extraction of the CKM couplings from the experimental
observables considerably more difficult, and at the same time richer.
\vs
The first section of this part is devoted to the extraction of $\alpha$ 
from the analysis of  $\B\to\pi\pi$
decays, using four different scenarios with increasing theoretical input.
Section~\ref{sec:introductionKPi} describes the constraints obtained 
from the analysis of $B \to K\pi$ modes only. We also study the 
impact from electroweak penguin amplitudes. The extraction of $\alpha$ 
from the pseudoscalar-vector final states, $\rho\pi$ and SU(2) or 
SU(3)-related modes, is presented in Section~\ref{sec:introductionrhopi}. 
Finally, we discuss the isospin analysis of $B \to \rho\rho$ decays in 
Section~\ref{sec:introductionrhorho}, which is similar to the 
$\pi\pi$ system. In most cases we attempt to evaluate the constraints 
obtained with higher luminosity samples.
\vs
Throughout this part, we will assume that \CP violation in mixing
is absent, \ie, $|q/p|=1$, as suggested by the Standard Model
($|\Gamma_{B_H}-\Gamma_{B_L}|\ll \Gamma_{\Bz}$) and confirmed by 
experiment ($A_{\rm SL} = -0.007 \pm 0.013$, see 
Section~\ref{sec:newPhysics}.\ref{ModelIndependentNewPhysics}).

\subsubsection*{Remark on Radiative Corrections}

We point out that the charmless analyses, published by the \babar\  and 
Belle collaborations up to approximately Summer 2003, utilized Monte 
Carlo simulation without treatment of radiative corrections  
in the decays. The simulation is used by the experiments to compute 
selection efficiencies and to predict probability density distributions 
of signal events for the use in maximum-likelihood fits.
%
%
A study based on Ref.~\cite{bib:Gif89} suggests that the branching fractions
of $\Bztopipi$ and $\Bz\to\Kp\pim$ may be underestimated by up to 
$10\%$~\cite{bib:TheseMu}. Since the effects strongly depend on the final 
state and the analysis strategy used, we do not attempt to correct the branching
fraction results here. However one should be aware that this systematic may 
lead to increased branching fractions for modes that decay to light charged 
particles.

 %
%
\section{Analysis of $\B\to\pi\pi$ and SU(3)-Related Decays}
\label{sec:introductionpipi}

%
%
\subsection{Basic Formulae and Definitions}


\subsubsection{Transition Amplitudes}
\label{par:amplitude}

\begin{figure}[t]
  \centerline{
        \epsfxsize5.0cm\epsffile{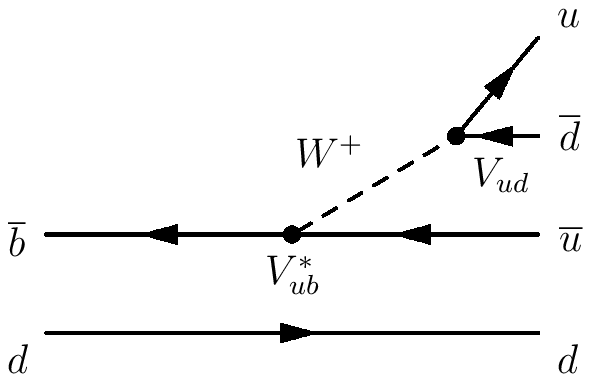}
        \hspace{2cm}
        \epsfxsize5.0cm\epsffile{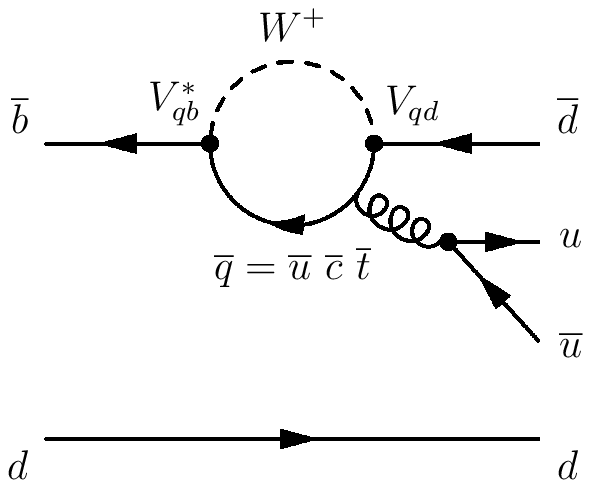}
  }
  \vspace{0.2cm}
  \caption[.]{\label{fig:B0pippim}\em
        Example of tree (left) and penguin (right) diagrams for the decay
        $\Bz\to\pi^+\pi^-$.}
\end{figure}

The general form of the $B^0\to\pi^+\pi^-$ decay amplitude, accounting for the
tree and penguin diagrams that correspond to the three up-type
quark flavors ($u,c,t$) occurring in the $W$ loop (see Fig.~\ref{fig:B0pippim}),
reads
\beq
\label{eq:utgeneral}
   \Apipi \equiv  A(B^0\to\pi^+\pi^-)
                          =   V_{ud}V_{ub}^* M_u
                            + V_{cd}V_{cb}^* M_c
                            + V_{td}V_{tb}^* M_t~,
\eeq
and similarly for the \CP-conjugated amplitude.
One can benefit from the unitarity relation (\ref{eq:utriangle})
to eliminate one of the three amplitudes, resulting in the three 
conventions $\U$, $\C$, $\T$, namely
\beqn
\label{eq:amplitudeConventions}
   \Apipi \:=\:
   \left\{
   \begin{array}{lcccll}
      & & V_{cd}V_{cb}^* (M_c - M_u)& +& V_{td}V_{tb}^* (M_t - M_u)
           & (\U) \\
      V_{ud}V_{ub}^*(M_u - M_c)& +&&& V_{td}V_{tb}^*(M_t - M_c)
           & (\C) \\
      V_{ud}V_{ub}^*(M_u - M_t)& +&   V_{cd}V_{cb}^* (M_c - M_t)  &&
           & (\T)
   \end{array}
   \right.
\eeqn
for which the $u$ ($\U$), $c$ ($\C$) and $t$ ($\T$) amplitude coefficients
have been substituted respectively. 
In the following, we adopt convention $\C$ so that $\Apipi$ reads
\beq
\label{eq:apipi}
        \Apipi = V_{ud}V_{ub}^*\Tpipi + V_{td}V_{tb}^*\Ppipi~,
\eeq
where $\Tpipi$ and $\Ppipi$ are defined by
\beq
\label{eq:defTP:B}
        \Tpipi \equiv M_u - M_c
                \hspace{0.5cm} {\rm and} \hspace{0.5cm}
        \Ppipi \equiv M_t - M_c~.
\eeq
The particular choice of which amplitude to remove
in the definition of a total transition amplitude
is arbitrary\footnote
{
   Note that $M_{u,c,t}$ amplitudes are intrinsically divergent and
   only differences between them lead to finite results
   (see, \eg, Ref.~\cite{BabarPhysBook}).
}
and does not have observable physical implications.
However the convention does modify the contents of
the phenomenological amplitudes $\Tpipi$ and $\Ppipi$. 
We will often refer to  $T^{+-}$ and $P^{+-}$ amplitudes
as ``tree'' and ``penguin'', respectively, although it is implicitly 
understood that both of them receive various contributions of 
distinct topologies, which are mixed under hadronic rescattering.

\subsubsection{\CP-Violating Asymmetries}
\label{par:CPasym}

The time-dependent \CP-violating asymmetry of the $\BzBzb$  system is given by
\beqn
   a_{\CP}(t) &\equiv&
        \frac{\Gamma(\overline B^0(t)\to\pi^+\pi^-)
                   - \Gamma(B^0(t)\to\pi^+\pi^-)}
                  {\Gamma(\overline B^0(t)\to\pi^+\pi^-)
                   + \Gamma(B^0(t)\to\pi^+\pi^-)}
                \nonumber\\[0.2cm]
        &=& \Spipi {\rm sin}(\dmd t) - \Cpipi {\rm cos}(\dmd t)~,
\eeqn
where $\dmd$ is the $\BzBzb$ oscillation frequency and  $t$ is either the
decay time of the $\Bz$ or the $\Bzb$ or, at \B factories running at 
the $\FourS$ mass, the time difference between the \CP  and the tag 
side decays. The coefficients of the sine and cosine terms are given by
\beq
\label{eq:spipicpipicoefficients}
   \Spipi = \frac{2{\rm Im}\lambda_{\pi\pi}}{1 + |\lambda_{\pi\pi}|^2}
        \hspace{0.5cm} {\rm and} \hspace{0.5cm}
   \Cpipi = \frac{1 - |\lambda_{\pi\pi}|^2}{1 + |\lambda_{\pi\pi}|^2}~,
\eeq
where the \CP   parameter $\lambda_{\pi\pi}$ is given by
(we recall that it is assumed $|q/p|=1$)
\beq
\label{eq:lambda}
        \lambda_{\pi\pi}
        = \frac{q}{p} \frac{\Abarpipi}{\Apipi}~,
\eeq
where the phase $\arg[q/p]=2\arg[V_{td}V_{tb}^*]\approx -2\beta$ 
(in our phase convention) arises due to $\BzBzb$ mixing.
We have used in the above equations that $\pi^+\pi^-$ is a \CP
eigenstate with eigenvalue $+1$.
\vs
In the absence of penguin contributions $(P=0)$, Eq.~(\ref{eq:lambda})
reduces to $\lambda_{\pi\pi}=e^{2i\alpha}$ (using the triangle
definition~(\ref{eq:utdefinitions})) and hence
\beq
        \Spipi[\Ppipi=0] = \sta
        \hspace{0.5cm} {\rm and} \hspace{0.5cm}
        \Cpipi[\Ppipi=0] = 0~.
\eeq
In general, the phase of $\lambda_{\pi\pi}$ is modified
by the interference between the penguin and the tree amplitudes.
In addition, the parameter $\Cpipi$ will be non-zero if
\beq
\deltapipi\equiv\arg[P^{+-}T^{+-*}] \ne 0~,
\eeq
hence measuring the occurrence of direct \CP  violation.
Defining an effective angle $\alphaeff$ that incorporates the
phase shift
\beq
\label{eq:lambdaPipi}
        \lambda_{\pi\pi}
        \equiv
        |\lambda_{\pi\pi}|e^{2i\alphaeff}~,
\eeq
and, using $|\lambda_{\pi\pi}|=\sqrt{1-\Cpipi}/\sqrt{1+\Cpipi}$,
one finds
\beq
        \Spipi = D\staeff~,
\eeq
where $D\equiv\sqrt{1-\Cpipi^2}$.
Twice the effective angle $\alphaeff$ corresponds to the relative
phase between the amplitudes $e^{-2i\beta}\Abarpipi$ and $\Apipi$.
It is useful for the following to define the penguin-induced
phase difference
\beq
        \deltaAlpha \equiv\frac{1}{2}(2\alpha-2\alphaeff)~, 
        \hspace{0.5cm}(\deltaAlpha\in [0,\pi])~.
\eeq
We note that the sign of the direct \CP asymmetry is related 
to the $\alpha\to\pi+\alpha$ and $\deltapipi\to\pi+\deltapipi$ ambiguity
through the relation
\beq
\mathrm{sign}(\Cpipi)=\mathrm{sign}(\sin\alpha)\times\mathrm{sign}(\sin\delta)~.
\eeq

\subsubsection{Isospin Related Decays}
\label{par:isospin}

\begin{figure}[t]
  \centerline{
        \epsfxsize5.0cm\epsffile{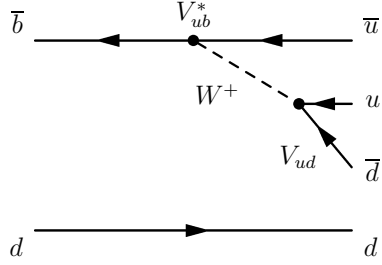}
  }
  \vspace{0.2cm}
  \caption[.]{\label{fig:B0pi0pi0T}\em
        Color-suppressed tree diagram for the decay $B^0\to\pi^0\pi^0$.}
\end{figure}
Owing to isospin invariance of strong interaction, the amplitudes of 
the various $B\to\pi\pi$ decays are related to each other. 
Gronau and London have shown~\cite{grolon} that the
measurements of rates and \CP-violating asymmetries of the charged and two
neutral $\pi\pi$ final states together with the exploitation of their
isospin relations provides sufficient information to extract the angle
$\alpha$ as well as the various $T$ and $P$ amplitudes. Unfortunately,
as far as $\alpha$ is concerned, the general solution is plagued by
an eightfold ambiguity within $[0,\pi]$~\cite{charles}.
\vs
Using convention $\C$, one can write
\beqn
  \sqrt{2}\Apippiz \:\equiv\:  \sqrt{2}A(B^+\to\pi^+\pi^0)
                        &=&   V_{ud}V_{ub}^*\Tpipiz
                            + V_{td}V_{tb}^*\PpippizEW~, \nonumber\\
\label{eq:b0pi0pi0}
  \sqrt{2}\Apizpiz \:\equiv\:  \sqrt{2}A(B^0\to\pi^0\pi^0)
                        &=&   V_{ud}V_{ub}^*\Tcpizpiz
                            + V_{td}V_{tb}^*\Ppizpiz~,
\eeqn
and similarly for the \CP-conjugated
modes. The ${\rm C}$ subscript stands for the color-suppressed amplitude
(see Fig.~\ref{fig:B0pi0pi0T} for the color-suppressed tree
diagram in the decay $B^0\to\pi^0\pi^0$), and the ${\rm EW}$ 
superscript stands for the electroweak penguin amplitude
contributing to $\pi^+\pi^0$. Note that the latter notation only refers
to the $\Delta I=3/2$ electroweak penguin contribution, since the
$\Delta I=1/2$ part is absorbed in the strong penguins. Indeed,
gluonic quark anti-quark production has $\Delta I=0$ so that
QCD penguins can only mediate $\Delta I=1/2$ transitions of the
$b$ quark. As a consequence, the $\Delta I=3/2$ decay $B^+\to\pi^+\pi^0$
has no strong penguin contribution. Applying the isospin
relations~\cite{grolon}
\beqn
  \Apippiz   &=& \frac{1}{\sqrt{2}}\Apipi + \Apizpiz~, \nonumber\\
\label{eq:isospinbar}
  \Abarpippiz   &=& \frac{1}{\sqrt{2}}\Abarpipi + \Abarpizpiz~,
\eeqn
 with $\overline\Apippiz=\Apimpiz$,
one can rearrange the amplitudes~(\ref{eq:b0pi0pi0})
\begin{figure}[t]
  \epsfxsize10cm
  \centerline{\epsffile{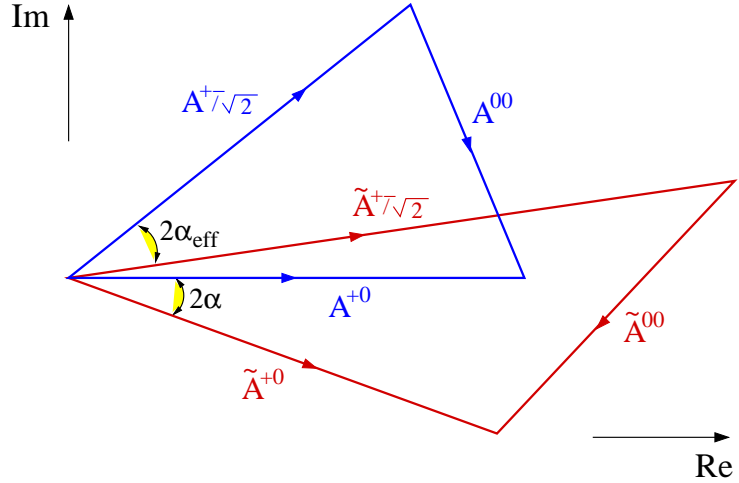}}
  \vspace{-1.0cm}
  \caption[.]{\label{fig:isospin_triangle}\em
        The isospin relations~(\ref{eq:isospinbar}) in the complex plane
        neglecting electroweak penguins.
        Note that the relative $\BzBzb$ mixing phase is included 
        in the $\Bb$ triangle
        ($\tilde {A^{ij}} \equiv e^{-2i\beta} {\overline A^{ij}}$).}
\end{figure}
\beqn
  \sqrt{2}\Apippiz              &=&   V_{ud}V_{ub}^*
                                \left(\Tpipi+\Tcpizpiz
                                \right)
                            + V_{td}V_{tb}^*\PpippizEW~, \nonumber\\
\label{eq:b0pi0pi0iso}
  \sqrt{2}\Apizpiz              &=&   V_{ud}V_{ub}^*\Tcpizpiz
                            - V_{td}V_{tb}^*
                              \left(\Ppipi
                                    - \PpippizEW\right)~.
\eeqn
The isospin relations between the three $\pi\pi$ amplitudes in the complex
plane are drawn in Fig.~\ref{fig:isospin_triangle} for the simplified 
case where electroweak penguins are neglected. They represent two
distinct triangles for the two \CP-conjugated amplitudes.
 Taking into account the phase shift due to $\BzBzb$ mixing,
the relative angle between $\tilde A^{+0}$ and $A^{+0}$ adds
 up to $2\alpha$, as it is shown in the figure.
\vs
Equations~(\ref{eq:isospinbar}) can be considered as exact
to a very good
approximation. Isospin-breaking corrections like, \eg, 
$\piz$--$\eta,\etapr$ mixing~\cite{garnderpipi} 
are expected to be below a few percent. We refer to 
Section~\ref{sec:charmlessBDecays}.\ref{sec:SU2strongbreaking} for
a numerical discussion of isospin-breaking effects in the
$B\to\rho\rho$ case.
Nevertheless, extracting $\alpha$ relying on this unique theoretical 
assumption appears to be difficult at present, given the number of 
ambiguities and the experimental uncertainties. In the following, we 
therefore explore several scenarios which, though still relying on 
Eqs.~(\ref{eq:isospinbar}), involve additional experimental and 
theoretical inputs.

%
%

\subsection{Theoretical Frameworks}
\label{sec:theoFrame}

To extract $\alpha$ from the experimental measurements of the \CP-violating 
asymmetries, we use four different scenarios, with rising theoretical 
assumptions~\cite{ourpipi}:
\begin{itemize}

\item[(\iA)]    using as input $\Spipi$ and $\Cpipi$ as well the
                branching fractions $B\to\pi\pi$ (all charges)
                and strong isospin symmetry SU(2)~\cite{grolon};

\item[(\iB)]    using (\iA) and the branching fraction $B^0\to K^+\pi^-$ 
                together with SU(3)
                flavor symmetry, and
                neglecting OZI-suppressed penguin annihilation
                topologies~\cite{charles};

\item[(\iC)]    using (\iB) and a phenomenological estimate of $|\Ppipi|$
                by means of the decay rate of $B^+\to K^0\pi^+$, and neglecting
                doubly CKM-suppressed penguin and annihilation 
                contributions~\cite{FM2pi,charles,GrRo};

\item[(\iD)]    using $\Spipi$ and $\Cpipi$ and the prediction of the complex 
                penguin-to-tree ratio $\Ppipi/\Tpipi$ in the framework of QCD 
                Factorization~\cite{BBNS0,BBNS}.

\end{itemize}

\subsubsection{Isospin Analysis, Isospin Bounds and Electroweak Penguins}
\label{sec:su2pipi}

It was shown in Ref.~\cite{grolon} that using the \CP-averaged 
branching fractions 
\beqn
   \BRpipi   &\propto& \frac{\tau_{B^0}}{2}\left(|\Apipi|^2   + |\Abarpipi|^2\right)~,
        \nonumber\\[0.2cm]
   \label{eq:cpavBRsPipPi0}
   \BRpippiz  &\propto& \frac{\tau_{B^+}}{2}\left(|\Apippiz|^2 + |\Apimpiz|^2\right)~,
        \\[0.2cm]
   \BRpizpiz &\propto& \frac{\tau_{B^0}}{2}\left(|\Apizpiz|^2 + |\Abarpizpiz|^2\right)~,\nonumber
\eeqn
where $\tau_{B^0}$ and $\tau_{B^+}$ are the neutral and charged $B$ 
lifetimes (\cf~Section~\ref{sec:standardFit}.\ref{sec:input_vcb}), and 
the \CP-violating asymmetries
\beqn
\label{eq:cpm}
\Cpipi&=&
{\mid\Apipi\mid^2-\mid\Abarpipi\mid^2
\over
\mid\Apipi\mid^2+\mid\Abarpipi\mid^2}~,\nonumber\\[0.2cm]
\label{eq:spm}
\Spipi&=&
{2\,{\rm
Im}(\Apipi{^\ast}{\Abarpipi})\over\mid\Apipi\mid^2+\mid\Abarpipi\mid^2}~,\\[0.2cm]
\label{eq:coo}
\Cpizpiz&=&{\mid\Apizpiz\mid^2-\mid\Abarpizpiz\mid^2\over\mid\Apizpiz\mid^2+\mid\Abarpizpiz\mid^2}~,\nonumber
\eeqn
one can  extract the angle $\alpha$, up to discrete ambiguities, provided 
electroweak penguin contributions are negligible ($P^{\rm EW}=0$). 
The geometrical description of the isospin analysis presented in the 
preceding section can be conveniently complemented by the explicit solution 
in terms of $\alpha$~\cite{mufrapipi}
\beq
\tan\alpha=\frac{\sin(2\alphaeff)\bar c+\cos(2\alphaeff)\bar s+s}
        {\cos(2\alphaeff)\bar c-\sin(2\alphaeff)\bar s+c}~,
\eeq
where all quantities on the right hand side can be expressed in term of
the observables as follows:
\beqn
        \sin(2\alphaeff)&=& \frac{\Spipi}{D}~,\nonumber\\[0.2cm]
        \cos(2\alphaeff)&=& \pm\sqrt{1-\sin^2(2\alphaeff)}~,\nonumber\\[0.2cm]
        c               &=&\sqrt{\frac{\tau_{B^+}}{\tau_{B^0}}}\,\frac{\frac{\tau_{B^0}}{\tau_{B^+}}\BRpippiz+\BRpipi(1+\Cpipi)/2-\BRpizpiz(1+\Cpizpiz)}
                           {\sqrt{2\BRpipi\BRpippiz(1+\Cpipi)}}~,\\[0.2cm]
        s               &=&\pm\sqrt{1-c^2}~,\nonumber\\[0.2cm]
        \overline{c}         
        &=&\sqrt{\frac{\tau_{B^+}}{\tau_{B^0}}}\,\frac{\frac{\tau_{B^0}}{\tau_{B^+}}\BRpippiz+\BRpipi(1-\Cpipi)/2-\BRpizpiz(1-\Cpizpiz)}
                           {\sqrt{2\BRpipi\BRpippiz(1-\Cpipi)}}~,\nonumber\\[0.2cm]
        \overline{s}          &=&\pm\sqrt{1-\overline c^2}~.\nonumber
\eeqn
The eightfold ambiguity for $\alpha$ in the range $[0,\pi]$ is made explicit 
by the three arbitrary signs\footnote
{
        \label{foot:pipi_alterparam}
        We may consider an alternative amplitude representation, which makes
the occurrence of the discrete ambiguities more explicit~\cite{mufrapipi}:
 \beqns
         \Apipi         =\mu\a ~, 
                \hspace{0.5cm}
         \Abarpipi      =\mu\abar\ e^{+2i\alphaeff}~, 
                \hspace{0.5cm}
         \Apippiz       =\mu\ e^{i(\Delta-\alpha)}~,  
                \hspace{0.5cm}
         \Abarpippiz    =\mu\ e^{i(\Delta+\alpha)}~, \\
         \Apizpiz       = \mu\ e^{i(\Delta-\alpha)}\left(1-{\a\over\sqrt{2}}\ 
                             e^{+i(\alpha-\Delta)}\right)~,
                \hspace{0.5cm}
         \Abarpizpiz    = \mu\ e^{i(\Delta+\alpha)}\left(1-{\abar\over\sqrt{2}}\ 
                            e^{-i(\alpha+\Delta-2\alphaeff)} \right)~,
 \eeqns
 which satisfy the triangular SU(2) relations~(\ref{eq:isospinbar}),
 and where
 $\mu$, $\a$ and $\abar$ are three unknown real (and positive)
 parameters  which drive the strength of the branching
 fractions, while  $\Delta$ is a phase.
 The phase convention chosen here is such that $\Apm$ is real positive, 
 and the $\overline A$ amplitudes include the $\BzBzb$ mixing phase
        $\arg[q/p]$. 
 With this choice, the phase $\Delta$ is not to be viewed as arising 
 purely from strong interaction since it absorbs the weak phase {\it a priori}
 present in $\Apm$: there is no reason to expect it to be confined to 
 small values.
 In terms of the above parameterization, the observables take the form
 \beqns
        \frac{1}{\tau_{B^0}}
        \BRpipi         &=& \mu^2{1\over 2}(\a^2+\abar^2)~, \\
        \frac{1}{\tau_{B^+}}
        \BRpippiz       &=& \mu^2~, \\
        \frac{1}{\tau_{B^0}}
        \BRpizpiz       &=& \mu^2{1\over 2}
                           \left( 2+{1\over 2}(\a^2+\abar^2)
                                 -\sqrt{2}(\a c+\abar\CCbar) \right)~, \\
        \Cpipi          &=& {\a^2-\abar^2\over\a^2+\abar^2}~, \\
        \Spipi          &=& {2\a\abar\over\a^2+\abar^2}\sin(2\alphaeff)~, \\
        \Cpizpiz        &=& { {1\over 2}(\a^2-\abar^2)-\sqrt{2}(\a\CC-\abar\CCbar)
                                \over
                                2+{1\over 2}(\a^2+\abar^2)-\sqrt{2}(\a\CC+\abar\CCbar)}
                             ~,\\
        \Spizpiz        &=& { 2\sin(2\alpha)+\a\abar\sin(2\alphaeff)
                              -\a\sqrt{2}\sin(\alpha+\Delta)
                              -\abar\sqrt{2}\sin(\alpha-\Delta+2\alphaeff)
                                \over
                              2+{1\over 2}(\a^2+\abar^2)-\sqrt{2}(\a\CC+\abar\CCbar)}~.
 \eeqns
 \newpage
 The eight mirror solutions (for $\alpha$ in $[0,\pi]$) are summarized in 
 the table below. Solutions 5 through 8 are just $\pi/2$ minus 
 solutions 1 through 4. The eight mirror solutions are strictly equivalent
 if no input is added, like $\Spizpiz$ for example ($S_{\rho\rho,L}^{00}$ 
 is experimentally accessible in the decay $\Bz\to\rho^0\rho^0$, see 
 Section~\ref{sec:charmlessBDecays}.\ref{sec:introductionrhorho}).
 \begin{center}
 \setlength{\tabcolsep}{0.0pc}
 \begin{tabular*}{\textwidth}{@{\extracolsep{\fill}}cccc}
 \hline
 &&&\\[-0.3cm]
 Solution   & $\alpha                  $ & $\Delta                  $ & $\alphaeff  $    \\[0.10cm] \hline
 &&&\\[-0.3cm]
 1          & $\alpha                $ & $\Delta                  $ & $\alphaeff $   \\ 
 2          & $\Delta                  $ & $\alpha                $ & $\alphaeff $   \\ 
 3          & $-\alpha+2\alphaeff $ & $-\Delta+2\alphaeff   $ & $\alphaeff $   \\ 
 4          & $-\Delta+2\alphaeff   $ & $-\alpha+2\alphaeff $ & $\alphaeff $    \\[0.1cm]
 5          & $\piotwo-\alpha           $ & $\piotwo-\Delta             $ & $\piotwo-\alphaeff $  \\ 
 6          & $\piotwo-\Delta             $ & $\piotwo-\alpha           $ & $\piotwo-\alphaeff $   \\ 
 7          & $\piotwo+\alpha-2\alphaeff $ & $\piotwo+\Delta-2\alphaeff $ & $\piotwo-\alphaeff $  \\
 8          & $\piotwo+\Delta-2\alphaeff $ & $\piotwo+\alpha-2\alphaeff  $ & $\piotwo-\alphaeff $     \\[0.1cm]
 \hline
 \end{tabular*}
 \end{center}

}. 
The quantity $\Spizpiz$ could also be considered, and would help lifting 
these ambiguities, but its measurement, which could make use of 
$\pi^0$ Dalitz decays for instance,
requires very large statistics, which is not available at present.

\subsubsection*{SU(2) Bounds}

The direct \CP asymmetry $\Cpizpiz$ has not yet been measured
so that the extraction of $\alpha$ itself is not possible and one 
has to derive upper limits on~$\deltaAlpha$ instead. It was first
pointed out by Grossman and Quinn~\cite{GrQu} that a small value for 
the branching fraction to $\pi^0\pi^0$ would mean that the penguin 
contribution cannot be too large. We stress that 
the numerical analysis performed with \ckmfitter\  guarantees the 
optimal use of the available and relevant experimental information, 
once the isospin relations are implemented at the amplitude level.
It is nevertheless instructive to derive analytical bounds on
$\deltaAlpha$. As shown by Grossman--Quinn~\cite{GrQu},
and later rediscussed by one of us~\cite{charles} and
Gronau--London--Sinha--Sinha (GLSS)~\cite{GrLoSiSi}, one obtains the 
inequality
\beq
\label{eq:boundPZ}
   {\rm cos}\,2\deltaAlpha \,\ge\,
        \frac{1}{D}\left(1-2\frac{\tau_{\Bp}}{\tau_{\Bz}}
                            \frac{\BRpizpiz}{\BRpippiz}\right)
        + \frac{\tau_{\Bp}}{\tau_{\Bz}}\frac{1}{D}
                     \frac{\left(\BRpipi-2\frac{\tau_{\Bz}}{\tau_{\Bp}}
                                          \BRpippiz+2\BRpizpiz\right)^2}
                          {4\BRpipi\BRpippiz }~,
\eeq
or, equivalently,
\beq
\label{eq:boundPM}
   {\rm cos}\,2\deltaAlpha \,\ge\,
        \frac{1}{D}\left(1-4\frac{\BRpizpiz}{\BRpipi}\right)
         + \frac{\tau_{\Bp}}{\tau_{\Bz}}\frac{1}{D}
                     \frac{\left(\BRpipi-2\frac{\tau_{\Bz}}{\tau_{\Bp}}
                                          \BRpippiz-2\BRpizpiz\right)^2}
                           {4\BRpipi\BRpippiz }~.
\eeq
The first term on the right hand side of
Eqs.~(\ref{eq:boundPZ}) and~(\ref{eq:boundPM}) corresponds to the limit
considered  in Ref.~\cite{charles}, while the original
Grossman--Quinn bound is obtained when setting $D=1$ in the
first term on the right hand side of Eq.~(\ref{eq:boundPZ}).
\vs
The above bound has interesting consequences on the discrete ambiguity
problem. 
In the limit where $\BRpizpiz$ goes to zero, the GLSS bound~(\ref{eq:boundPZ}) merges
the eight mirror solutions for $\alpha$ (in the range $[0,\pi]$)
in two distinct intervals, each of which containing one quadruplet
of them.
\vs
Following the same line it is possible to derive lower and upper bounds 
on the branching fraction into two neutral pions~\cite{GrLoSiSi}
\beq
\label{eq:GLSSbound}
        \BrooGLSSl\,\le\,\BRpizpiz\,\le\,\BrooGLSSu~,
\eeq
with
\[
        \BrooGLSSlu=\frac{\tau_{\Bz}}{\tau_{\Bp}}
                    \BRpippiz+{1\over 2}\BRpipi \pm 
                    \sqrt{\frac{\tau_{\Bz}}{\tau_{\Bp}}\BRpippiz\BRpipi
                    \left(1+D\right)}~,
\]
where the limits are weakest for $D=1$, that is vanishing direct \CP
violation. Equation~(\ref{eq:boundPZ}) can be rewritten~\cite{mufrapipi}
\begin{equation}
\label{eq:sinmax}
        \sin^2\deltaAlpha \,\le\, 
        \frac{\tau_{\Bp}}{\tau_{\Bz}}\frac{1}{D}
        \frac{\left(\BRpizpiz-\BrooGLSSl\right)
                     \left(\BrooGLSSu-\BRpizpiz\right)}
                    {2\BRpipi\BRpippiz }~,
\end{equation}
which does not provide new information but makes explicit that 
$\alpha=\alphaeff$ if $\BRpizpiz$ {\em reaches either $\BrooGLSSl$ or
$\BrooGLSSu$}. This is the case in the formal limit
$\BRpizpiz\to 0$, which is close to being realized in $B\to\rho\rho$
(see Section~\ref{sec:charmlessBDecays}.\ref{sec:introductionrhorho}).
However for $B\to\pi\pi$ the lower bound $\BrooGLSSl$ in 
Eq.~(\ref{eq:GLSSbound}) is not so small and as soon as $\BRpizpiz$
deviates from it,  $\alphaeff$ can be rather different
from $\alpha$,
as described further below.
\vs
If we assume that $\alpha$ is known, \eg, from the standard CKM fit, 
we obtain the bound on $\BRpizpiz$~\cite{mufrapipi}
\beq
\label{eq:rebound}
        \reBoundl \,\le\, \BRpizpiz  \,\le\, \reBoundu~,
\eeq
with
\beq
\label{eq:Balpha}
        \reBoundlu=\frac{\tau_{\Bz}}{\tau_{\Bp}}
                    \BRpippiz+{1\over 2}\BRpipi \pm 
                    \sqrt{\frac{\tau_{\Bz}}{\tau_{\Bp}}\BRpippiz\BRpipi
                    \left(1+\tilde{D}_\alpha\right)}~,
\eeq
and where
\[
        \tilde{D}_\alpha     = \sqrt{(1-\sin^22\alpha)
                                  (D^2-\Spipi^2)}
                            + \Spipi\,\sin2\alpha~.
\]
Since $\tilde{D}_\alpha\le D$, this bound is tighter than
Eq.~(\ref{eq:GLSSbound}). With known $\alpha$, the \CP asymmetry 
$\Cpizpiz$ is not a free parameter anymore: it can be determined 
using Eq.~(\ref{eq:Balpha})
\beqn
\label{eq:coof}
        \Cpizpiz^{\pm}  &=&
                \frac{1}{\BRpizpiz(1+\tilde{D}_\alpha)} \Bigg[
                -\Cpipi \left(\frac{\tau_{\Bz}}{\tau_{\Bp}}
                 \BRpippiz-{\tilde{D}_\alpha\over 2}\BRpipi-\BRpizpiz\right)
                \nonumber \\[0.2cm]
                        && \hspace{2.4cm}
                \pm\;\sqrt{(\BRpizpiz-\reBoundl)(\reBoundu-\BRpizpiz)
                      (D^2-\tilde{D}_\alpha^2)} \Bigg]~.
\eeqn
There are two solutions of $\Cpizpiz$ for a given $\BRpizpiz$. An 
application of Eq.~(\ref{eq:coof}) is shown in Fig.~\ref{fig:CooBoo}.

\subsubsection*{Electroweak Penguins}

As pointed out by Buras and Fleischer~\cite{BFPEW}
and Neubert and Rosner~\cite{NRPew}, the electroweak penguin 
amplitude $\PpippizEW$ in $B^+\to\pi^+\pi^0$ can be related to the tree 
amplitude in a model-independent way using Fierz transformations
of the relevant current-current operators in the effective Hamiltonian 
$\cal H_{\rm eff}$ for $B\to\pi\pi$ decays
\beqn
\label{eq:ewpeng}
        {\cal H}_{\rm eff} 
        = \frac{G_F}{\sqrt{2}}\left[\sum_{q=u,c}
           V_{qb}V_{qd}^* (c_1 O_1^q + c_2 O_2^q) - 
                \sum_{i=3}^{10} V_{tb}V_{td}^* c_i O_i 
          \right] ~+~ {\rm  h.\;c.}~.
\eeqn
Here $O_1^q$ and $O_2^q$ are tree operators of the Lorentz structure
$(V-A)\times(V-A)$, $O_{3-6}$ are short-distance gluonic penguin
operators, and $O_{7-10}$ are electroweak penguin operators. 
The Lorentz structure of $O_7$ and $O_8$ is $(V-A)\times(V+A)$ while  
$O_9$ and $O_{10}$ are $(V-A)\times(V-A)$. In the limit of isospin 
symmetry, the $\Delta I=3/2$ part of the latter operators is 
Fierz-related to the operators $O_1$ and $O_2$. Since $c_{7,8}$ are 
small compared to $c_{9,10}$, they can be neglected so that one 
obtains
\beqn
\label{EWPratio}
        \frac{\PpippizEW}{\Tpipiz} \simeq 
        -\frac{3}{2}\left(\frac{c_9 + c_{10}}{c_1 + c_2}\right)
        = +(1.35 \pm 0.12)\times 10^{-2} ~.
\eeqn
The theoretical error on the numerical evaluation of this ratio has
been estimated from the residual scale and scheme dependence of the
Wilson coefficients~\cite{bbl}. It also accounts for the neglect of
the  contributions from $O_{7}$ and $O_{8}$~\cite{theseJerome}. One
notices that there is no strong phase difference between $\PpippizEW$
and $\Tpipiz$ so that electroweak penguins do not generate a charge
asymmetry in $B^+\to\pi^+\pi^0$ if this picture holds: this prediction
is in agreement with the present experimental average of the
corresponding asymmetry (see Table~\ref{tab:BRPiPicompilation}). Although in the
SM electroweak penguins in two-pion modes appear to be small,
their inclusion into the full isospin analysis is  straightforward and
will become necessary once high-statistics  data samples are available.

\subsubsection{SU(3) Flavor Symmetry}

\begin{figure}[t]
  \centerline{
        \epsfxsize5.0cm\epsffile{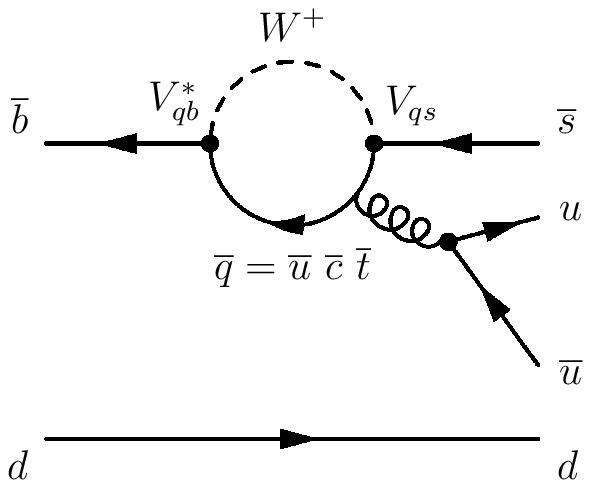}
        \hspace{2cm}
       \epsfxsize5.6cm\epsffile{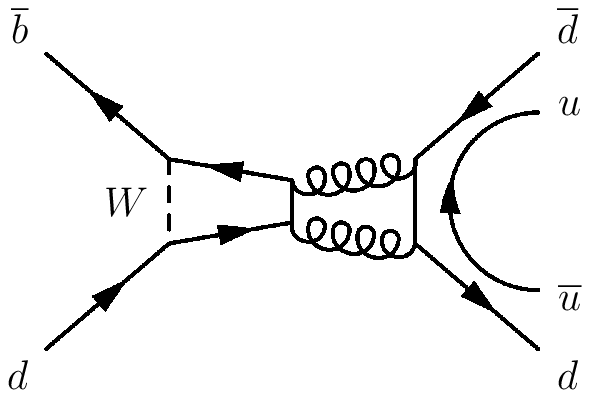}
  }
  \vspace{0.2cm}
  \caption[.]{\label{fig:B0KppimP}\em
        \underline{Left:} penguin diagram for the decay
        $\Bz\to \Kp\pim$. \underline{Right:} OZI-suppressed penguin
        annihilation diagram for the
        decay $\Bz\to\pip\pim$.}
\end{figure}
We extend the use of flavor symmetries to SU(3), considering
the amplitude of the decay $B^0\to K^{+}\pi^{-}$ in convention $\C$
\beq
\label{eq:b0K+pi-}
        \AKpi \:\equiv\:  A(B^0\to K^+\pi^-)
                        \:=\:   V_{us}V_{ub}^{*} \TKpi
                              + V_{ts}V_{tb}^{*} \PKpi~.
\eeq
With the assumption of SU(3) flavor symmetry and neglecting
OZI-suppressed penguin annihilation diagrams (see right hand
diagram in Fig.~\ref{fig:B0KppimP}), which contribute to 
$B^0\to\pi^+\pi^-$ but not to $B^0\to K^{+}\pi^{-}$, the
penguin amplitudes in $B^0\to\pi^+\pi^-$ and $B^0\to K^{+}\pi^{-}$ 
are equal
\beqn
\label{eq:PpipiPKpi}
        \Ppipi=\PKpi~.
\eeqn
As in the isospin symmetry case, one can derive the following
bound~\cite{charles}, which benefits from the CKM enhancement of the
penguin contribution to the $\Bz\to K^+\pi^-$ decay
\beq
\label{eq:Rkpioverpipi}
        \cos2\deltaAlpha  \:\ge\:
                \frac{1}{D}\left(1-2\lambda^2 
                \frac{\BRKpi}{\BRpipi}\right)~,
\eeq
where $\lambda$ is the Wolfenstein parameter.
\vs
Another possibility~\cite{silvwolf,fleischer}, that would eventually give
stronger constraints, would be to identify $\Tpipi$ and $\TKpi$ in
addition to $\Ppipi$ and $\PKpi$. At first sight such an approximation
is similar in spirit to the neglect of OZI-suppressed penguins, because 
it is violated by exchange diagrams only, that are expected to be
power-suppressed\footnote
{
   The term ``power-suppression'' refers to the quantity
   $\Lambda_{\rm QCD}/m_b\ll1$.
}. 
However as shown in 
Ref.~\cite{charles}, an estimate of $\Ppipi$ leads to the determination 
of the shift $\deltaAlpha$, while an estimate of $\Tpipi$ determines
the angle $\alpha$ itself. Hence, as far as $\alpha$ 
is concerned, the error on the estimate of $\Ppipi$ is a second order 
effect, while the error on the estimate of $\Tpipi$ is of leading order. 
We therefore expect the hadronic uncertainties in the relation $\Tpipi=\TKpi$ 
to be potentially more dangerous than in the relation $\Ppipi=\PKpi$. As
a consequence, the ratio $\TKpi/\Tpipi$ is kept unconstrained in our fit.
\vs
SU(3) flavor symmetry is only approximately realized in nature and one
may expect violations of the order of $30\%
$ at the amplitude level.
For example, within factorization the relative size of SU(3) symmetry
breaking is expected to be $(f_{K}-f_{\pi})/f_{K}$, where $f_K$ and $f_\pi$
are the pion and kaon decay constants, respectively.
Notwithstanding, the bound~(\ref{eq:Rkpioverpipi}) can be considered 
conservative with respect to SU(3) breaking, since a correction would 
lead to a stronger bound. For example, assuming factorization
the ratio of branching fractions $\BRKpi/\BRpipi$
would be lowered by $(f_{\pi}/f_{K})^{2}\simeq0.67$. Since the penguin
annihilation contributions, which spoil the relationship between the
$B\to \pi^+\pi^-$
and $B\to K^+\pi^-$ penguin amplitudes, are power and OZI-suppressed,
they
are expected to be small with respect to the dominant SU(3) breaking. 
We will therefore study the constraints derived from
Eq.~(\ref{eq:PpipiPKpi}) as if they were a consequence of strict
SU(3) symmetry, although it is understood that an additional dynamical
assumption is made.

\subsubsection{Estimating $|\Ppipi|$ from $B^{+}\to K^0\pi^+$}
\label{sec:naiveFA}

In addition to the theoretical assumptions of Scenario~(\iB), the 
magnitude $|\Ppipi|$ can be estimated from the branching fraction of the 
penguin-dominated mode $B^{+}\to K^0\pi^+$. Neglecting the doubly 
CKM-suppressed difference between $u$ and $c$ penguins, as well 
as the doubly CKM-suppressed tree annihilation contribution, 
the $\Bp\to \Kz\pip$ transition amplitude reads 
\beq
\label{eq:ampConvKpi} 
        A^{0+}_{K\pi}\equiv A(B^{+}\to K^0\pi^{+}) =
        V_{tb}^{*}V_{ts}\PKzpi~.
\eeq 
Now, if one takes the SU(3) limit and neglects the penguin annihilation
and color-suppressed electroweak penguin contributions, one
has~\cite{FM2pi,charles,GrRo}
\beqn
\label{eq:relPPcharged} 
        |\Ppipi| = 
                   \frac{f_{\pi}}{f_{K}} \frac{1}{R}
                   |\PKzpi|~.
\eeqn 
The first factor on the right hand side corrects for factorizable
SU(3)  breaking, while the second factor, $R=0.95 \pm 0.23$, is a 
theoretical estimate, within the QCD Factorization approach, of the
residual effects that break the relation between $B\to\pi^+\pi^-$ and
$B^+\to K^0\pi^+$ penguin amplitudes\footnote
{
        The numerical value of $R$ and its uncertainty are obtained from 
        Eq.~(\ref{eq:relPPcharged}) by estimating $|\Ppipi|$ and $|\PKzpi|$  
        from the full QCD FA calculation, as described in 
        Section~\ref{sec:charmlessBDecays}.\ref{par:qcdfa}. 
}~\cite{BBNS}.
Our evaluation also includes the uncertainty due to the neglect of
the $V_{us}V_{ub}^*$ contribution to $B^{+}\to K^0\pi^+$. 
As in Scenario~(\iB), the strong phase $\deltapipi$ remains
unconstrained in Scenario~(\iC). The size of the tree amplitude
$|\Tpipi|$ is conveniently deduced from the measurement of $\BRpipi$,
taking advantage of the above estimate of $|\Ppipi|$: the analytical
constraint in the $\rhoeta$ plane cannot be expressed in terms
of the angle $\alpha$ alone, but rather as a degree-four polynomial
equation~\cite{charles}, or as a relation between $\alpha$ and $\gamma$. 
Other methods to estimate the tree amplitude are found in the literature:
\bei

\item   one can use the spectrum of the decay $B^0\to\pi^+\ell^- \bar{\nu}$
        near $q^2 = 0$ (that is the squared effective mass of the recoiling
        $\ell \nu$ system) with theoretical estimates for
        the form factor~\cite{GrRo}, to infer an estimate for the quantity
        $|V_{ub}^*|\times|T_u| $, where $T_u$ is the semileptonic
        amplitude at $q^2=0$.
        The method is not used here as it provides $T_u$,
        which is not simply related to the full $|\Tpipi|$,
        except in the na\"{\i}ve factorization approximation.

\item   according to Eqs.~(\ref{eq:b0pi0pi0iso}) and (\ref{eq:cpavBRsPipPi0}), 
        the branching fraction of the tree-dominated decay $B^+ \to \pi^+\pi^0$ 
        is given by (neglecting electroweak penguins)
        \beq
        2\BRpippiz =
        |V_{ud}V_{ub}^*|^2\left[ |\Tpipi|^2 + |\Tcpizpiz|^2
                + 2
                {\rm Re}\left(\Tpipi {\Tcpizpiz}^*\right) \right]~.
        \eeq
        Using theoretical assumptions on the ratio $|\Tcpizpiz/\Tpipi|$
        one may infer the size of $|\Tpipi|$ from the measured
        branching fraction~\cite{GrRo}.
\eei

\subsubsection{Beyond na\"{\i}ve Factorization}
\label{par:qcdfa}

Considerable theoretical progress to calculate the tree and penguin amplitudes
in $B\to hh^\prime$ with the use of QCD has been achieved in the recent
years. If such calculations reliably predicted the penguin and tree
contributions and their relative strong phase difference, they could be used
to translate a measurement of $\Spipi$ and $\Cpipi$ into a constraint
on the CKM couplings.
\vs
The QCD Factorization Approach (QCD FA)~\cite{BBNS0,BBNS,BN} is based on the
concept of color transparency~\cite{colorTransparency}.  In the heavy
quark limit ($m_b \gg \Lambda_{\rm QCD}$), the decay amplitudes are
calculated by virtue of a new factorization theorem. To leading power
in $\Lambda_{\rm QCD}/m_b$ and in lowest order in perturbation theory,
the result of na\"{\i}ve factorization is reproduced. It is found that
power-dominant non-factorizable corrections are calculable as perturbative corrections
in $\as$ since the interaction of soft gluons with the small
color-dipole of the high-energetic ($W$-emitted) quark-anti-quark pair
is suppressed.
Non-factorizable power-suppressed contributions are neglected within
this framework. However, due to  a chiral enhancement and although they
are formally power-suppressed\footnote
{
        The power-suppression in annihilation diagrams and 
        hard spectator contribution occurs 
        by the ratio  $r^\pi_\chi = 2 m^2_\pi/(m_b(m_u+m_d))$,
        which is numerically of order one.
}, hard-scattering spectator interactions and annihilation diagrams 
cannot be neglected. Since they
give rise to infrared endpoint singularities when computed
perturbatively, they  can only be estimated 
in a model-dependent way. In Ref.~\cite{BBNS} these contributions
are parameterized by two complex quantities, $X_{H}$ and $X_{A}$,
that are logarithmically large but always appear with a relatively small
factor proportional to $\as$.
\vs
The QCD FA has been implemented in \ckmfitter\  and is used in two
different configurations. The first configuration defines a {\it
leading order} (LO)  calculation by neglecting  the non-factorizable
power-suppressed terms,  \ie, the annihilation contribution and the
divergent part of the hard  spectator diagrams ($X_{H}=0$). This
configuration is not fully consistent because the power corrections
that are convergent, once factorized, are kept:  LO QCD FA is very
close to the usual na\"{\i}ve factorization model (see, \eg,
Ref.~\cite{StechNeubert}), and only differs from the latter  by small
convergent radiative corrections. In the second configuration, the full
QCD FA calculations are used and the quantities $X_{H}$ and $X_{A}$ are
parameterized as \cite{BBNS}
\beq
        X_{H,A}= \left(1+ \rho_{H,A}e^{i\phi_{H,A}}\right)
                 \ln\frac{m_B}{\Lambda_h}~,
\eeq
where $\Lambda_h=0.5\gev$,  $\phi_{H,A}$ are free phases 
($-180^\circ<\phi_{H,A}<180^\circ$) and $\rho_{H,A}$ are parameters
 varying within $[0,1]$.
\vs
In addition to $X_{H}$ and $X_{A}$, other theoretical parameters used in
the calculation such as quark masses, decay constants, form factors and
Gegenbauer moments, are varied within the ranges given in Ref.~\cite{BN}.
Therefore, all scenarios defined in Ref.~\cite{BN} are automatically contained
in our results.
\vs
Another approach, denoted pQCD~\cite{KLS}, differs from QCD FA mainly in the
power counting in terms of $\Lambda_{\rm QCD}/m_{b}$. The pQCD approach has
not been implemented in \ckmfitter\  yet and is therefore not considered in 
the following discussion.
\vs
There is an ongoing debate among the experts concerning the reliability of
these calculations. The main concerns are the computation of the chirally
enhanced penguins, the endpoint singularities
in hard spectator interactions and the control of non-factorizable annihilation
contributions (see, \eg, Refs.~\cite{Ciuch,SCETfacto}). As we will see, LO
QCD FA is very predictive but fails to describe the experimental data. 
On the other hand, full QCD FA with parameterized power corrections is quite 
successful within large theoretical uncertainties, but it can no 
longer be viewed as a systematic expansion of QCD.

\subsection{Experimental Input}
\label{par:expinput}

\begin{sidewaystable}[p]
\begin{center}
\setlength{\tabcolsep}{0.0pc}
{\normalsize
\begin{tabular*}{\textwidth}{@{\extracolsep{\fill}}lcccc}\hline
&&&& \\[-0.3cm]
Parameter & \babar\  & Belle & CLEO & Average  \\[0.1cm]
\hline
&&&& \\[-0.3cm]
$\Cpipi$
        & $-0.19\pm0.19\pm0.05$     \cite{pipiBabar}
        & $-0.58\pm0.15\pm0.07$     \cite{pipiBelle}
        & -
        & $-0.46\pm0.13$ \\[0.05cm]
$\Spipi$
        & $-0.40\pm0.22\pm0.03$     \cite{pipiBabar}
        & $-1.00\pm0.21\pm0.07$     \cite{pipiBelle}
        & -
        & $-0.73\pm0.16$ \\[0.15cm]
Correlation coeff.
        & $-0.02       $     \cite{pipiBabar}
        & $-0.29       $     \cite{pipiBelle}
        & -
        & $-0.17       $ \\[0.15cm]
\hline
&&&& \\[-0.3cm]
$\Ckspiz$
        & $0.40^{\,+0.27}_{\,-0.28}\pm0.10$     \cite{BABARK0pi0}
        & -
        & -
        & $0.40^{\,+0.27}_{\,-0.28}\pm0.10$ \\[0.05cm]
$\Skspiz$
        & $0.48^{\,+0.38}_{\,-0.47}\pm0.11$     \cite{BABARK0pi0}
        & -
        & -
        & $0.48^{\,+0.38}_{\,-0.47}\pm0.11$ \\[0.15cm]
\hline
&&&& \\[-0.3cm]
$A_{\CP}(\pi^+\pi^0)$
        & $-0.03^{\,+0.18}_{\,-0.17}\pm0.02$    \cite{BABAR2}
        & $-0.14\pm0.24^{\,+0.05}_{\,-0.04}$    \cite{BELLEACP}
        & -
        & $-0.07\pm0.14$ \\[0.05cm]
$A_{\CP}(K^+\pi^-)$
        & $-0.107\pm0.041\pm0.013$              \cite{BABARBelleLP03}
        & $-0.088\pm0.035\pm0.018$              \cite{BABARBelleLP03}
        & $-0.04 \pm0.16 \pm0.02 $              \cite{CLEO2}
        & $-0.095\pm0.028$ \\[0.05cm]
$A_{\CP}(K^+\pi^0)$
        & $-0.09\pm0.09\pm0.01$                 \cite{BABAR2}
        & $+0.23\pm0.11$                        \cite{BELLEACP}
        & $-0.29\pm0.23\pm0.02$                 \cite{CLEO2}
        & $ 0.00\pm0.07$ \\[0.05cm]
$A_{\CP}(K^0\pi^+)$
        & $-0.05\pm0.08\pm0.01$                 \cite{BABAR3}
        & $+0.07^{\,+0.09\,+0.01}_{\,-0.08\,-0.03}$ \cite{BELLEACP}
        & $+0.18\pm0.24\pm0.02$                 \cite{CLEO2}
        & $+0.02\pm0.06$ \\[0.05cm]
\hline
&&&& \\[-0.3cm]
$\BR(B^0\to\pi^+\pi^-)$
        & $4.7\pm0.6\pm0.2$                     \cite{BABAR1}
        & $4.4\pm0.6\pm0.3$                     \cite{BELLE1}
        & $4.5^{\,+1.4\,+0.5}_{\,-1.2\,-0.4}$   \cite{CLEO1}
        & $4.55\pm0.44$ \\[0.05cm]
$\BR(B^+\to\pi^+\pi^0)$
        & $5.5^{\,+1.0}_{\,-0.9}{\pm0.6}$       \cite{BABAR2}
        & $5.0 \pm1.2\pm0.5           $         \cite{BELLE1}
        & $4.6^{\,+1.8\,+0.6}_{\,-1.6\,-0.7}$   \cite{CLEO1}
        & $5.18^{\,+0.77}_{\,-0.76}$ \\[0.1cm]
&&&& \\[-0.3cm]
$\BR(B^0\to\pi^0\pi^0)$
        & $2.1\pm0.6\pm0.3$                     \cite{BABAR0}
        & $1.7\pm0.6\pm0.2$                     \cite{BELLE1}
        &  $<4.4$                               \cite{CLEO1}
        & $1.90\pm0.47$ \\[0.1cm]
&&&& \\[-0.3cm]
$\BR(B^0\to K^+\pi^-)$
        & $17.9\pm0.9\pm0.7$                    \cite{BABAR1}
        & $18.5\pm1.0\pm0.7$                    \cite{BELLE1}
        & $18.0^{\,+2.3\,+1.2}_{\,-2.1\,-0.9}$  \cite{CLEO1}
        & $18.16\pm0.79$ \\[0.05cm]
$\BR(B^+\to K^+\pi^0)$
        & $12.8^{\,+1.2}_{\,-1.1}\pm1.0$        \cite{BABAR2}
        & $12.0 \pm1.3^{\,+1.3}_{\,-0.9}$       \cite{BELLE1}
        & $12.9^{\,+2.4\,+1.2}_{\,-2.2\,-1.1}$  \cite{CLEO1}
        & $12.6^{\,+1.1}_{\,-1.0} $ \\[0.05cm]
$\BR(B^+\to K^0\pi^+)$
        & $22.3 \pm 1.7\pm1.1$                  \cite{BABAR3}
        & $22.0 \pm 1.9\pm1.1          $        \cite{BELLE1}
        & $18.8^{\,+3.7\,+2.1}_{\,-3.3\,-1.8}$  \cite{CLEO1}
        & $21.8\pm1.4$ \\[0.05cm]
$\BR(B^0\to K^0\pi^0)$
        & $11.4\pm1.7\pm0.8$                    \cite{BABAR3}
        & $11.7\pm2.3^{\,+1.2}_{\,-1.3}$        \cite{BELLE1}
        & $12.8^{\,+4.0\,+1.7}_{\,-3.3\,-1.4}$  \cite{CLEO1}
        & $11.7\pm1.4 $ \\[0.1cm]
&&&& \\[-0.3cm]
$\BR(B^0\to K^+K^-)$
        & $<0.6$                                \cite{BABAR1}
        & $<0.7$                                \cite{BELLE1}
        & $<0.8$                                \cite{CLEO1}
        &  \\[0.1cm]
$\BR(B^+\to K^+\overline{K}^0)$
        & $<2.5$                                \cite{BABAR3}
        & $<3.3$                                \cite{BELLE1}
        & $<3.3$                                \cite{CLEO1}
        &  \\[0.15cm]
$\BR(B^0\to K^0\overline{K}^0)$
        & $<1.8$                                \cite{BABAR3}
        & $<1.5$                                \cite{BELLE1}
        & $<3.3$                                \cite{CLEO1}
        &  \\[0.05cm]
\hline
&&&& \\[-0.3cm]
\end{tabular*}
}
\vspace{-0.3cm}
\caption[.]{\label{tab:BRPiPicompilation} \em
        Compilation of experimental results on the $B\to hh^\prime$ branching
        fractions (in units of $10^{-6}$) and \CP-violating asymmetries.
        Limits are quoted at $90\%$~CL. For the averages we use the 
        results from the HFAG~\cite{HFAG}. $\Ckspiz$ and
        $\Skspiz$ are defined similarly to Eqs.~(\ref{eq:cpm})
        while the direct \CP asymmetries $A_{\CP}$ are defined
        with an opposite sign: $A_{CP}=(|\Abar|^2-|A|^2)/(|\Abar|^2+|A|^2)$.
        Note that missing radiative corrections in the Monte Carlo simulations 
        used by the experiments may lead to underestimated branching fractions for 
        modes with light charged particles in the final state (see remark in 
        the introduction to Part~\ref{sec:charmlessBDecays}).
        The CDF collaboration has presented the preliminary result 
        $A_{\CP}(K^+\pi^-)=0.02\pm0.15\pm0.02$~\cite{BABARBelleLP03,CDFKpiblessed},
        which is however not yet included in the HFAG average. }
\end{center}
\end{sidewaystable}
The experimental values for the time-dependent \CP asymmetries 
measured by \babar~\cite{pipiBabar} and Belle~\cite{pipiBelle} are collected 
in Table~\ref{tab:BRPiPicompilation}. We have reversed the sign of Belle's 
$A_{\pi\pi}=-\Cpipi$ to account for the different convention adopted. Also 
quoted are the statistical correlation coefficients between $\Spipi$ and 
$\Cpipi$ as reported by the experiments. Significant mixing-induced \CP 
violation has been observed by Belle. Averaging \babar\  and Belle,
the no-\CP-violation hypothesis ($\Spipi=0$, $\Cpipi=0$) is ruled out with 
a p-value of $1.2\times10^{-9}$, and deviations of $4.7\sigma$ and $3.7\sigma$ 
from the $\Spipi=0$ and $\Cpipi=0$ hypotheses are observed, respectively\footnote
{
        These exclusion probabilities are estimates only, assuming 
        Gaussian error propagation of the averages.
}. 
We note that $\Cpipi\neq0$ is incompatible with the na\"{\i}ve factorization 
approximation, which predicts no final state interaction phases. 
Since the time-dependent \CP parameters measured by Belle are outside 
of the physical domain, we apply the procedure outlined in 
Section~\ref{sec:statistics}.\ref{sec:metrology_physicalBoundaries} 
to obtain the corresponding CLs within $\Cpipi^2+\Spipi^2\le1$. The results 
are given in Fig.~\ref{fig:scpipi} together with the theoretical predictions
(discussed below).
\vs
\begin{figure}[t]
  \centerline{
        \epsfxsize5.6cm\epsffile{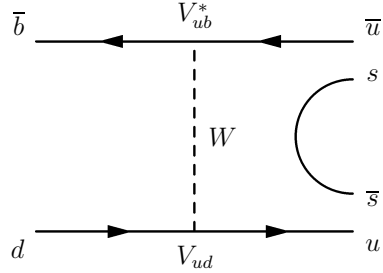}
  }
  \vspace{0.2cm}
  \caption[.]{\label{fig:B0KpKmEch}\em
        $W$ exchange diagram responsible for the decay
        $\Bz\to K^+K^-$.}
\end{figure}
Also given in Table~\ref{tab:BRPiPicompilation} is the time-dependent 
\CP asymmetry in  the $\Bz\to\KS\piz$ decay measured by \babar,
and a compilation of the branching fractions and charge asymmetries (direct 
\CP  violation) of all the $B\to hh^\prime$ modes ($hh^\prime = \pi, K $). 
Most of the rare two-body pseudoscalar-pseudoscalar decay modes have been discovered. 
The unobserved $K\Kbar$ modes are either mediated \via\  power-suppressed $W$
exchange/annihilation diagrams ($\Bz\to \Kp \Km$, $\Bp\to \Kp \Kzb$, see 
Fig.~\ref{fig:B0KpKmEch}) or penguin diagrams ($\Bz\to \Kz \Kzb$, $\Bp\to \Kp \Kzb$)
and hence are expected to be small. The ratio of $\BRKpi/\BRpipi \sim4$
is a strong indication of the presence of penguin diagrams. In effect,
according to Eqs. (\ref{eq:apipi}) and (\ref{eq:b0K+pi-}), if there were 
no penguin, it would be of the order of $\lambda^2$. Charge asymmetries 
are all consistent with zero so far, except for $A_{\CP}(K^+\pi^-)$  which 
differs from zero by $3.4\sigma$. The averages quoted in 
Table~\ref{tab:BRPiPicompilation} are taken from the \HFAG~\cite{HFAG}. 

\subsection{Numerical Analysis of $B\to \pi\pi$ Decays}
\label{sec:resultsPipi}

We ran \ckmfitter\  corresponding to the analysis scenarios (\iA) through 
(\iD), using the inputs from
Table~\ref{tab:BRPiPicompilation}\footnote
{Other recent analyses can be found in
Refs.~\cite{BFRS2,analyses2pi,SCETfacto}.}. If not 
stated otherwise, all plots are produced using the \babar\  and Belle  
averages for $\Spipi$ and $\Cpipi$, as well as the world averages
for all other observables.

\subsubsection{Constraints on $\alpha$ and $\rhoeta$}

\begin{figure}[t]
  \centerline
        {
        \epsfxsize8.1cm\epsffile{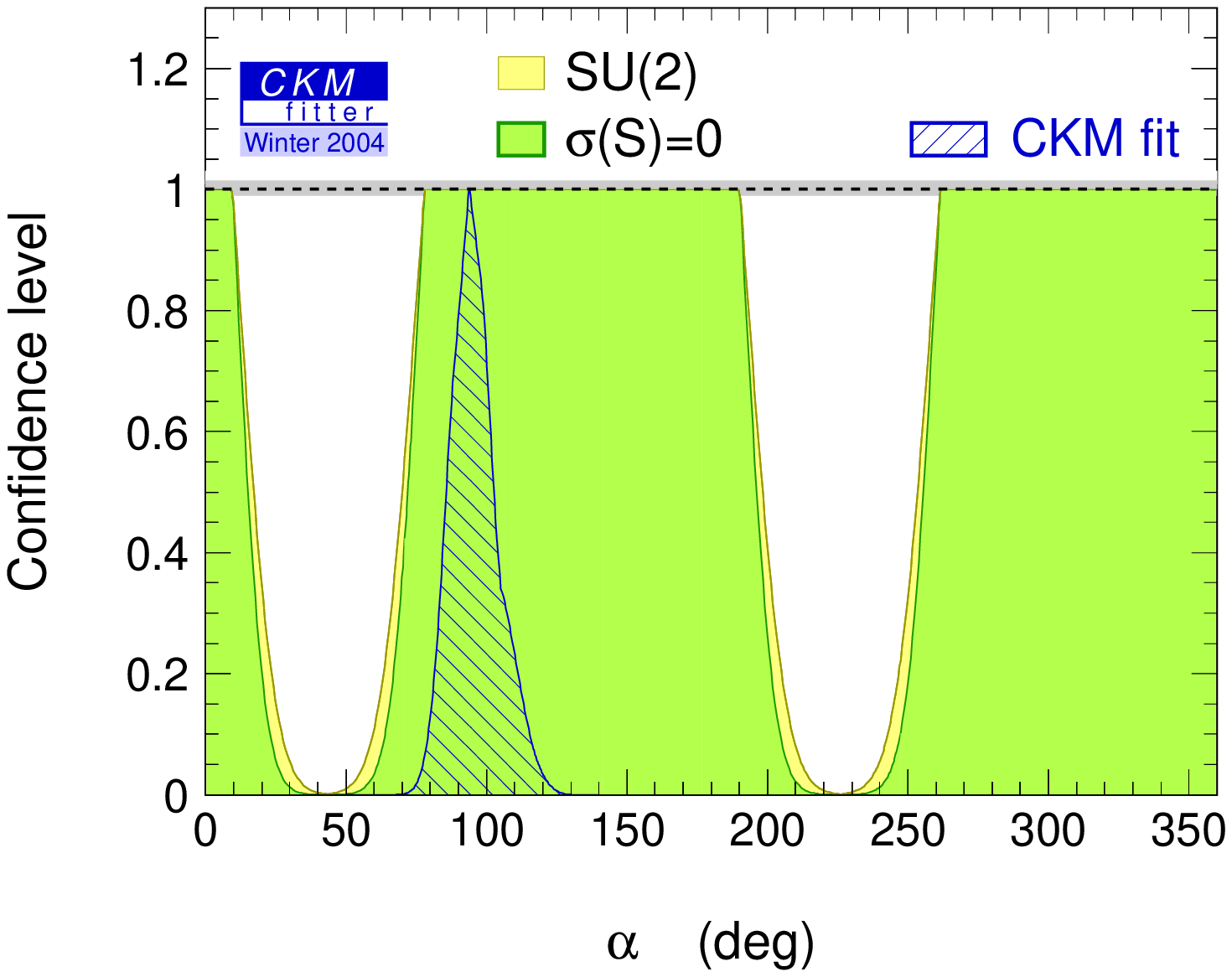}
        \epsfxsize8.1cm\epsffile{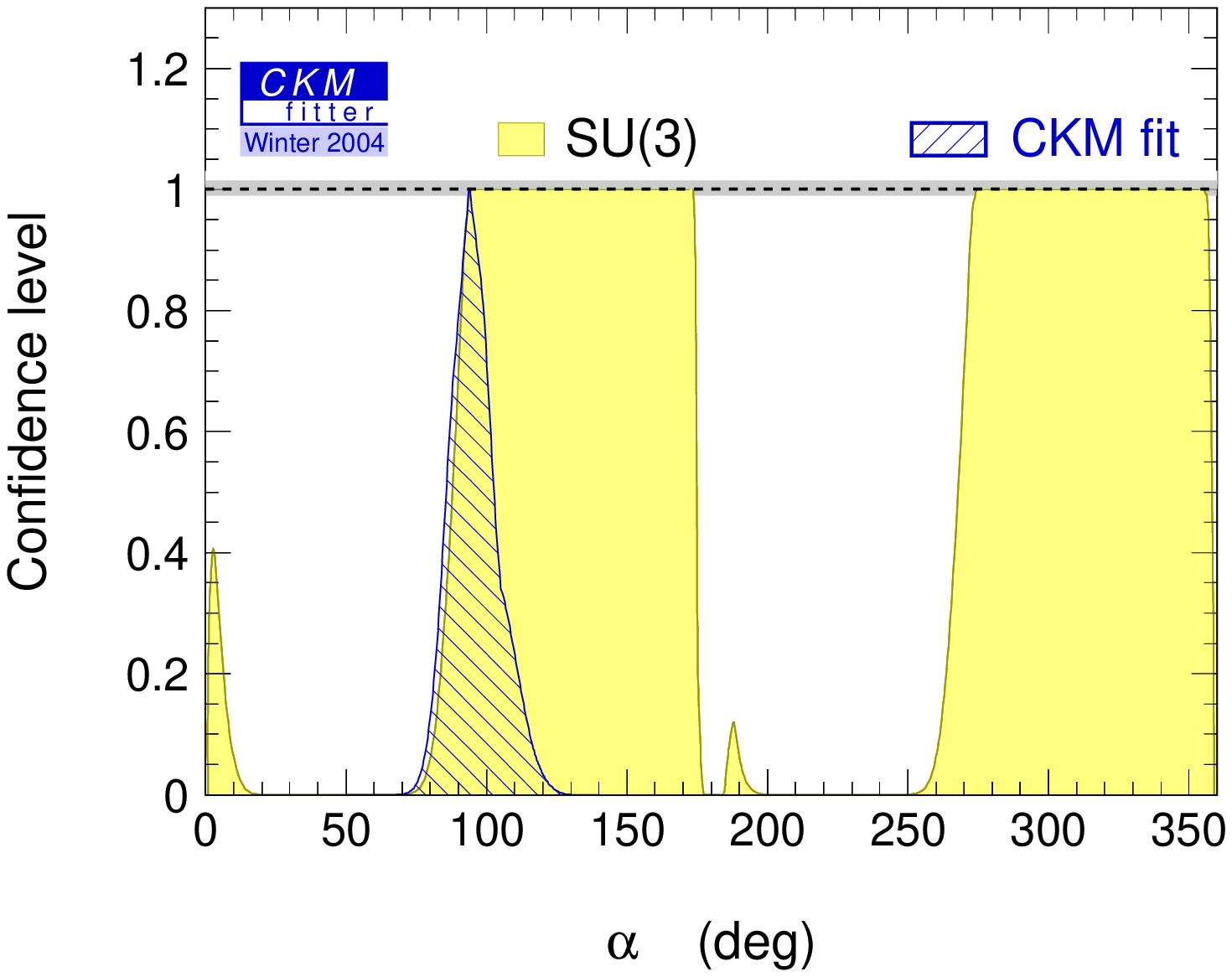}
        }
  \vspace{0.0cm}
  \centerline
        {
        \epsfxsize8.1cm\epsffile{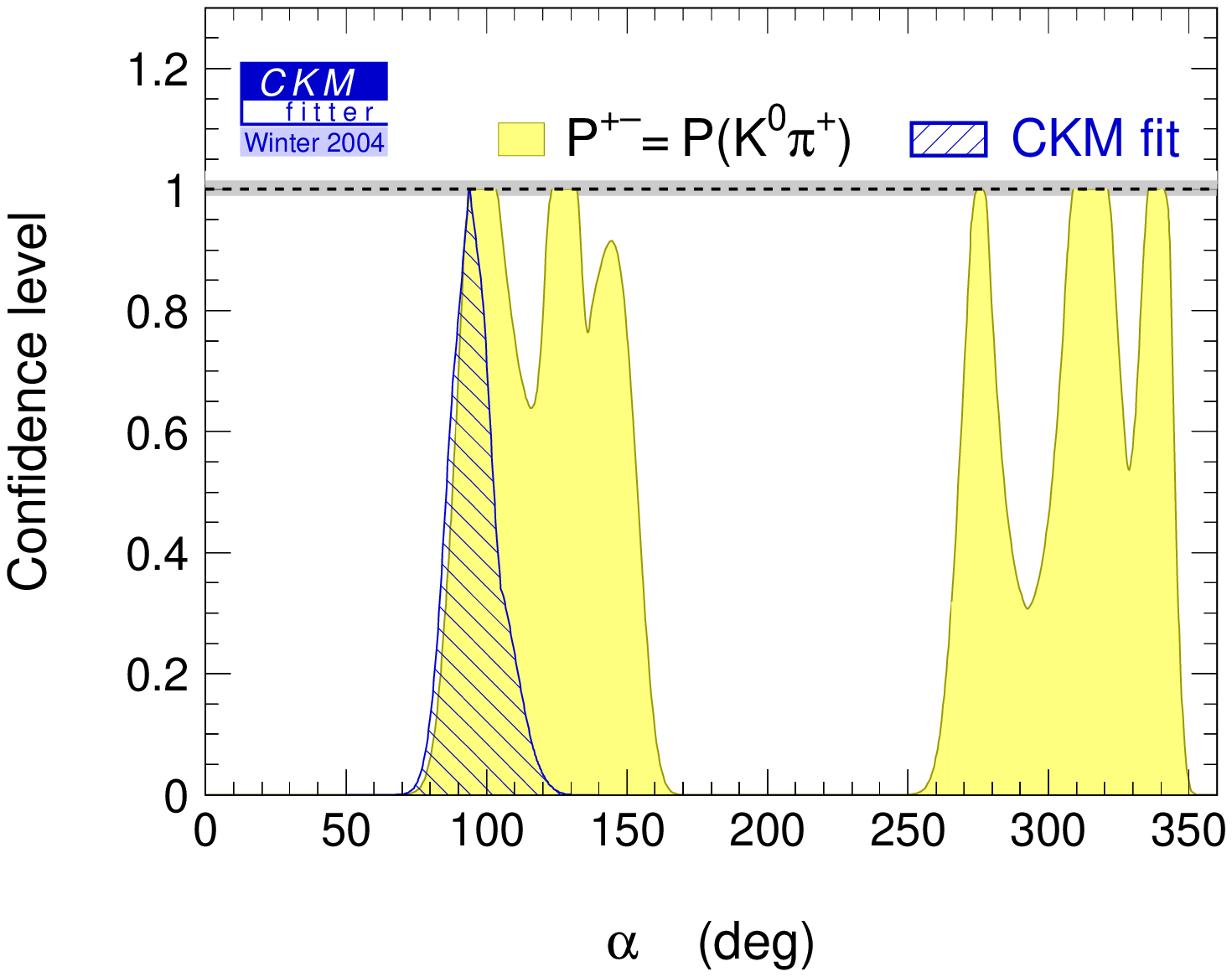}
        \epsfxsize8.1cm\epsffile{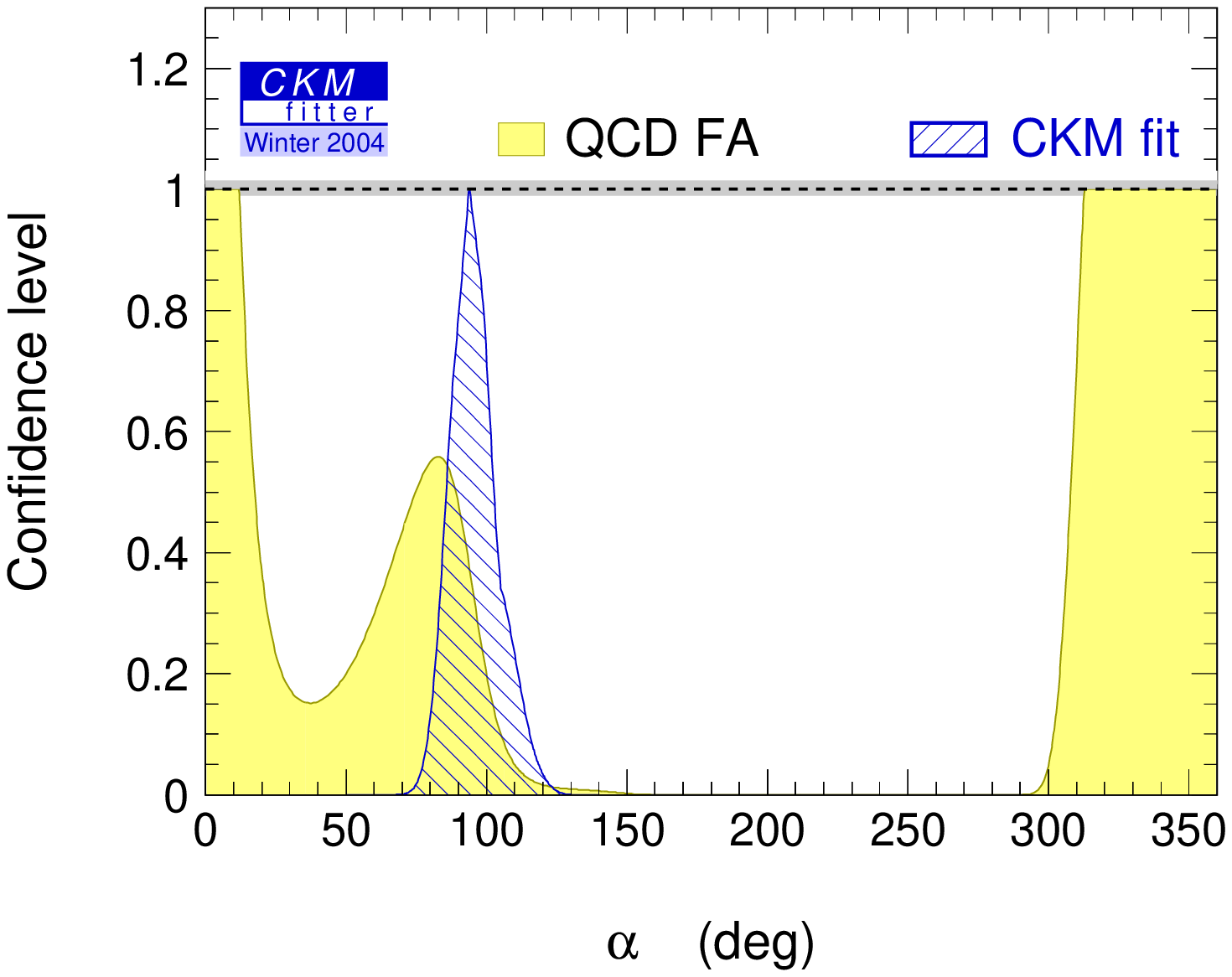}
        }
  \vspace{-0.4cm}
  \caption[.]{\label{fig:theAlphaPlots}\em
        Confidence levels for $\alpha$ for Scenarios (\iA) through (\iD) 
        (\cf\  Section~\ref{sec:charmlessBDecays}.\ref{sec:theoFrame}) of 
        the $B \to \pi\pi$ data. The dark shaded function on the upper 
        left hand plot shows the constraint from SU(2) when the experimental
        uncertainty on $\Spipi$ is set to zero. It hence 
        displays the uncertainty on $|\alpha-\alphaeff|$ due to the penguin
        contribution. Also shown on each plot is the result
        from the standard CKM fit.}
\end{figure}

\begin{figure}[p]
  \centerline
        {
        \epsfxsize8.1cm\epsffile{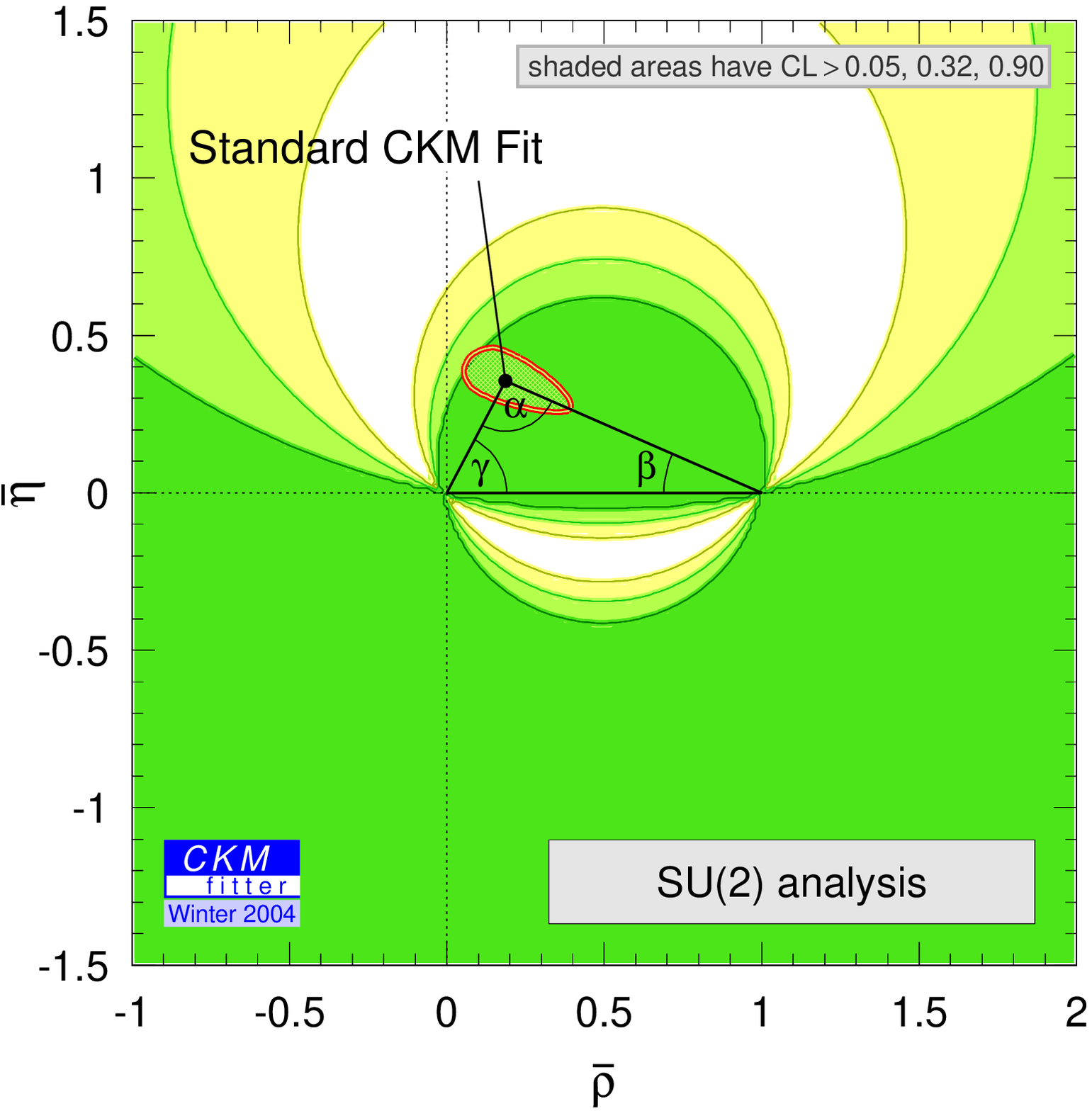}
        \epsfxsize8.1cm\epsffile{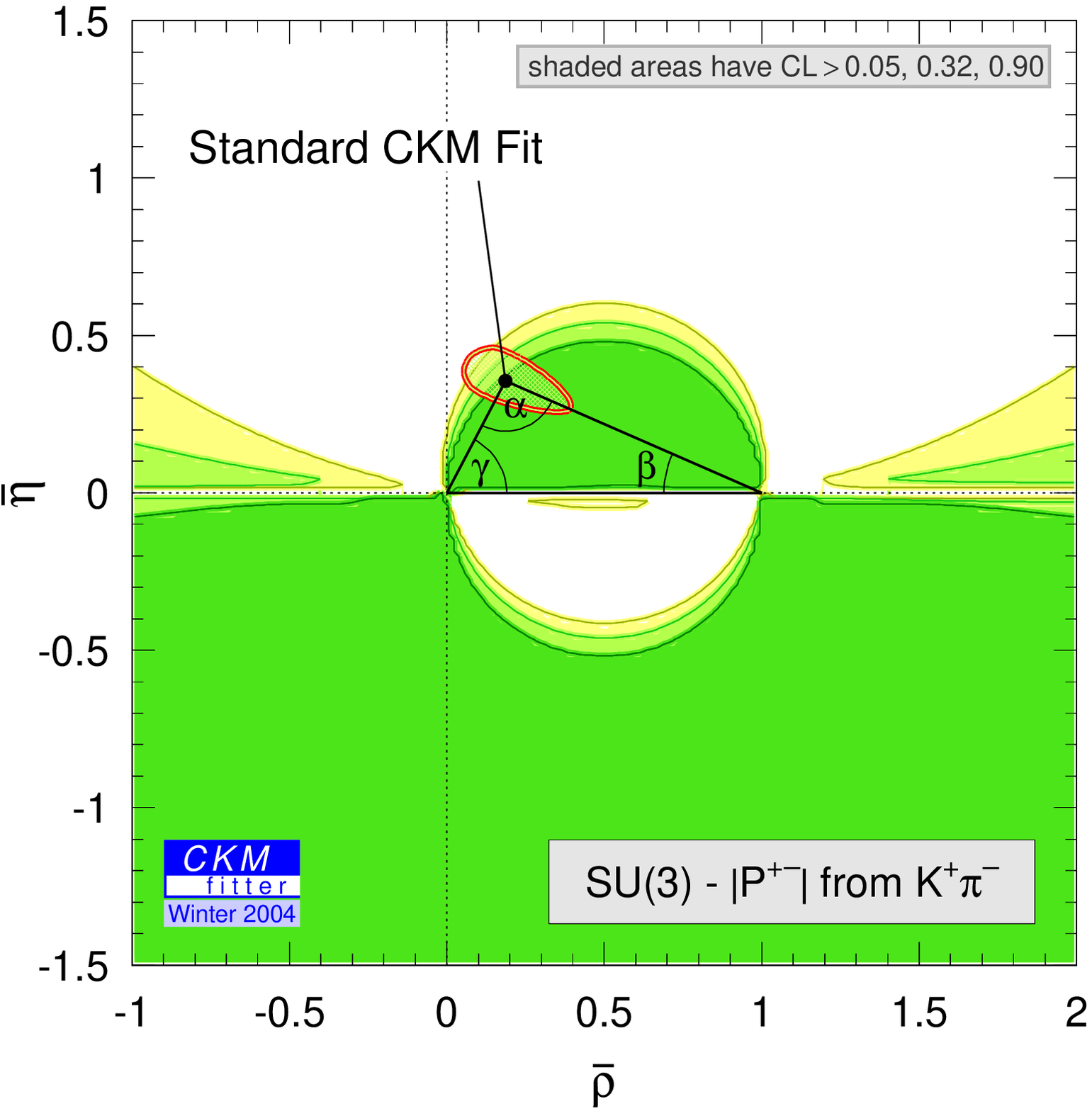}
        }
  \vspace{0.5cm}
  \centerline
        {
        \epsfxsize8.1cm\epsffile{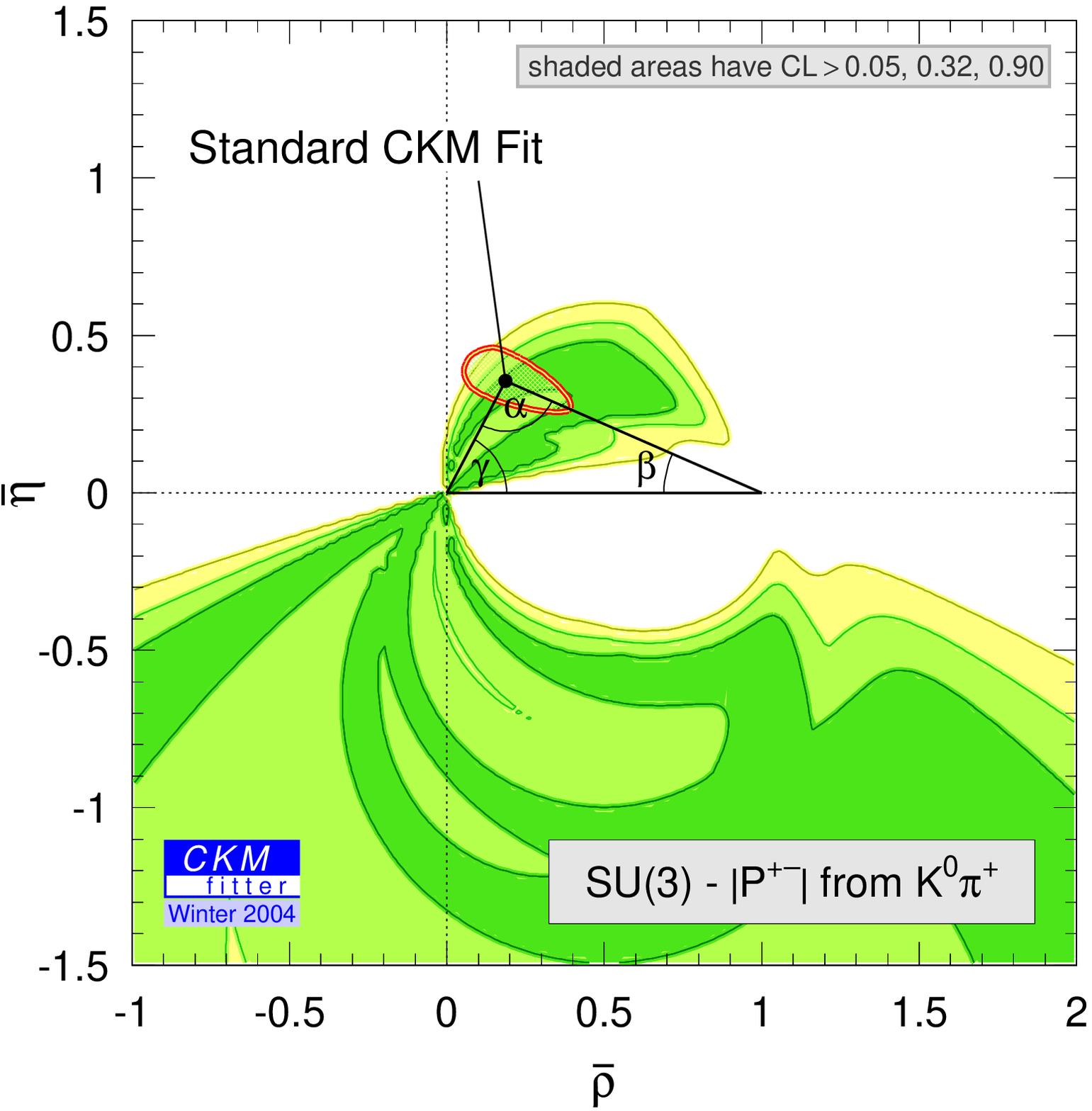}
        \epsfxsize8.1cm\epsffile{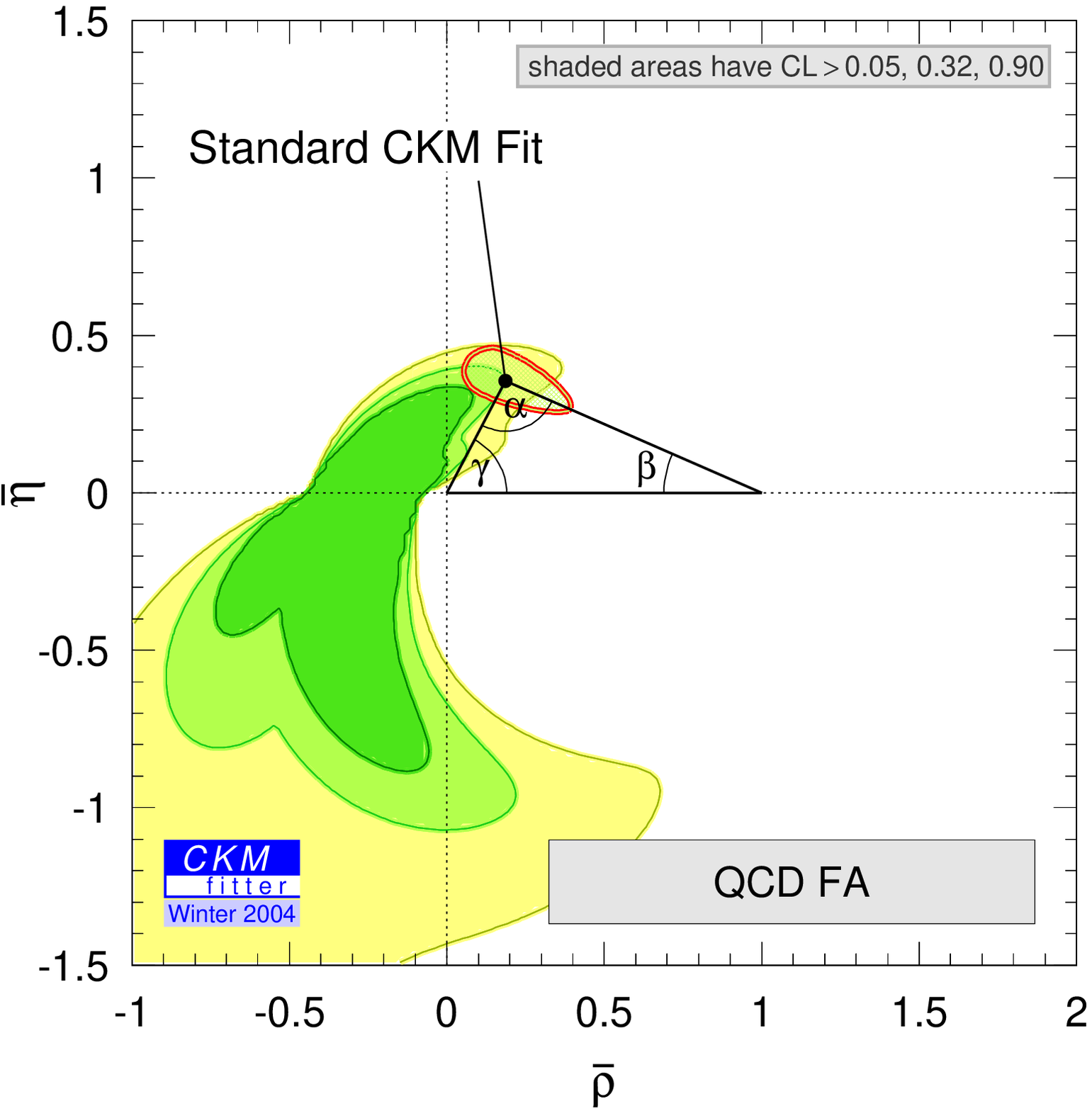}
        }
  \vspace{-0.0cm}
  \caption[.]{\label{fig:theRhoEtaPlots}\em
        Confidence levels in the $\rhoeta$ plane for Scenarios
        (\iA) through (\iD) of the $B \to \pi\pi$ data.
        Dark, medium and light shaded areas have $CL>0.90$,
        $0.32$ and $0.05$, respectively. Also shown
        on each plot is the result from the standard CKM fit. 
        Significant constraints are obtained once the penguin-to-tree 
        ratio is determined with the use of phenomenological or theoretical
        input (Scenarios (\iC) and (\iD)). Consistency with
        the SM is found, in spite of the sensitivity of 
        the data to $b\to d$ transitions that could in principle 
        receive sizable New Physics contributions. }
\end{figure}

The constraints on $\alpha$ and in the $\rhoeta$ plane obtained for 
the various scenarios are plotted in Figs.~\ref{fig:theAlphaPlots} and 
\ref{fig:theRhoEtaPlots} and discussed below\footnote
{
        The presence of non-zero electroweak penguins leads to a small
        modification in the isospin analysis which breaks the relation
        $\CL(\rhobar,\etabar)=\CL(\alpha(\rhobar,\etabar))$. As a consequence,
        the CL versus $\alpha$ is uniform if both $\rhobar$ {\em and} $\etabar$ 
        are free varying variables. As a remedy to this we also use 
        $|V_{ub}/V_{cb}|$ (see Section~\ref{sec:standardFit}.\ref{sec:input_vub})
        in the $\alpha$ scans, which introduces a slight effect on the
        angle itself.
}.
\bei
\item   At present, we achieve essentially no useful constraint from the 
        SU(2) analysis (Scenario~(\iA), upper left hand plots). We find 
        the limit $-54^{\circ}<\deltaAlpha<52^{\circ}$ for $\CL>10\%$, 
        largely dominated by the uncertainty on the contribution from 
        gluonic penguins (see the dark shaded function in the upper 
        left hand plot of Fig.~\ref{fig:theAlphaPlots}). 
        The asymmetry in the limit is due to the 
        contribution from electroweak penguins.

\item   Using in addition SU(3) (Scenario~(\iB)) one begins to rule
        out regions in the $\rhoeta$ plane (upper right hand
        plots). The wide unconstrained arcs correspond to the still
        unfruitful bound $-29^{\circ}<\deltaAlpha< 28^{\circ}$ for $\CL>10\%$.
        These constraints cannot compete with the size of the allowed
        $\rhoeta$ region obtained from the standard CKM fit.

\item   Interesting information is obtained for Scenario~(\iC)
        (lower left hand plots). The overall uncertainty is dominated
        by the errors on $(\Spipi,\Cpipi)$ on the one hand, and on the
        correction factor $R$ (cf. Eq.~(\ref{eq:relPPcharged}))
        on the other hand. For larger statistics
        the latter will limit the accuracy of the
        constraint, unless theoretical progress, together with combined
        fits of many SU(3)-related modes, is able to estimate $R$ more
        precisely.

\item   The constraint is significantly improved when using QCD FA to predict the
        various amplitudes\footnote
        {
                The full QCD FA calculation~(\ref{par:qcdfa}) is used here. 
                See Section~\ref{sec:charmlessBDecays}.\ref{sec:bbnsres} 
                for a discussion of the leading order (LO) approach.
        }.
        The preferred region is found in agreement with the standard CKM
        fit, despite the potential sensitivity of the observables to the
        suppressed $b\to d$ FCNC transitions.
        The main theoretical uncertainty is due to the phenomenological 
        parameters $X_{A}$ and $X_{H}$ \cite{ckmws03}. In particular, the sign of $\etabar$ 
        cannot be constrained because the sign of $\deltapipi$
        is not well predicted by the calculation (see Fig.~\ref{fig:dpot}).

\eei

\subsubsection{Constraints in the ($\Spipi,\Cpipi$) Plane}

\begin{figure}[t]
  \epsfxsize10.5cm
  \centerline{\epsffile{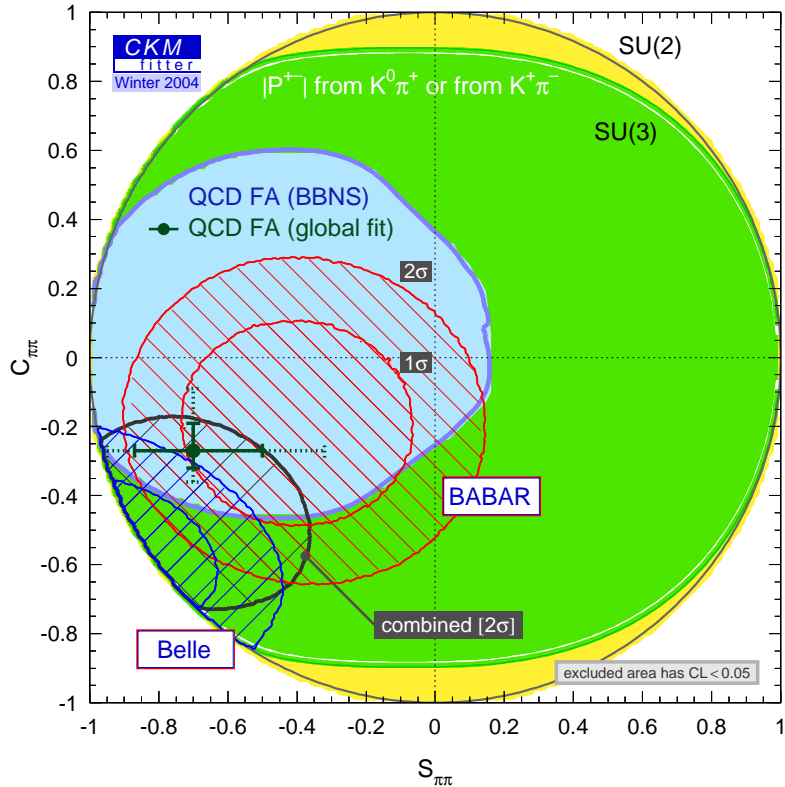}}
  \vspace{-0.0cm}
  \caption[.]{\label{fig:scpipi}\em
        Predictions for $\Spipi$ and $\Cpipi$ for Scenarios (\iA) through 
        (\iD). Drawn are $CL=0.05$ contours. The input values for $\rhobar$
        and $\etabar$ are taken from the standard CKM fit assuming the SM 
        to hold. The dot with error bars gives the QCD FA prediction 
        obtained from the global fit to all $\B\to\pi\pi,\,K\pi$ observables
        (excluding $\Spipi$ or $\Cpipi$ from the fit when 
        determining $\Cpipi$ and $\Spipi$, respectively) presented in 
        Section~\ref{sec:statistics}.\ref{sec:globalQCDfit}.
        For comparison, the CL contours corresponding
        to $1\sigma$ and $2\sigma$ for the experimental results from 
        \babar, Belle and their averages are overlaid. Note that we have 
        applied the statistical method described in 
        Section~\ref{sec:statistics}.\ref{sec:metrology_physicalBoundaries}
        to account for the presence of the physical boundaries
        when computing the CLs. }
\end{figure}
The predictions obtained for $\Spipi$ and $\Cpipi$ for Scenarios (\iA) 
through (\iD) are shown in Fig.~\ref{fig:scpipi}. Input requirements to
these predictions are the values of $\rhobar$ and $\etabar$, as
predicted by the standard CKM fit, the errors of which are properly
propagated in the calculations (see Part~\ref{sec:standardFit}). In
accordance with  the above findings, the present experimental inputs
used in the isospin analysis~(\iA) are not sufficient to constrain
$\Spipi$ and $\Cpipi$. 
Also, little information is obtained from the SU(3) analysis~(\iB) and
Scenario~(\iC), while the QCD FA (\iD) remains the most predictive
framework. 
\vs
To a good approximation, the SU(2) solution in the ($\Spipi,\Cpipi$)
plane  represents a circle of which the center is located at ($\sta,0$)
and of which the radius is given by the penguin-to-tree ratio 
$\rpipi\equiv (R_t/R_u)\times|\Ppipi/\Tpipi|$. The relative strong phase
$\deltapipi$  determines the position on the circle. Consequently, the
large uncertainty  on $\Spipi$ reflects both the relatively weak $\sta$
constraint of the  standard CKM fit and the insufficient knowledge of
$\rpipi$. On the other  hand, the accuracy of the $\Cpipi$ prediction
is determined by $\rpipi$ and,  in case of Scenario~(\iD), by
$\deltapipi$. Values of $\Cpipi$ that are  far from zero, as suggested
by the Belle measurement, are in marginal  agreement with the QCD FA
since it requires both a large relative strong  phase $\deltapipi$ and
a large $\rpipi$. If such a large non-zero value for $\Cpipi$ is
confirmed, it would be a strong hint for significant  rescattering effects,
independently of potential New Physics contributions. 

\subsubsection{Constraints on Amplitude Ratios}
\label{sec:pipidpot}

\begin{figure}[t]
  \centerline{\epsfxsize8.1cm\epsffile{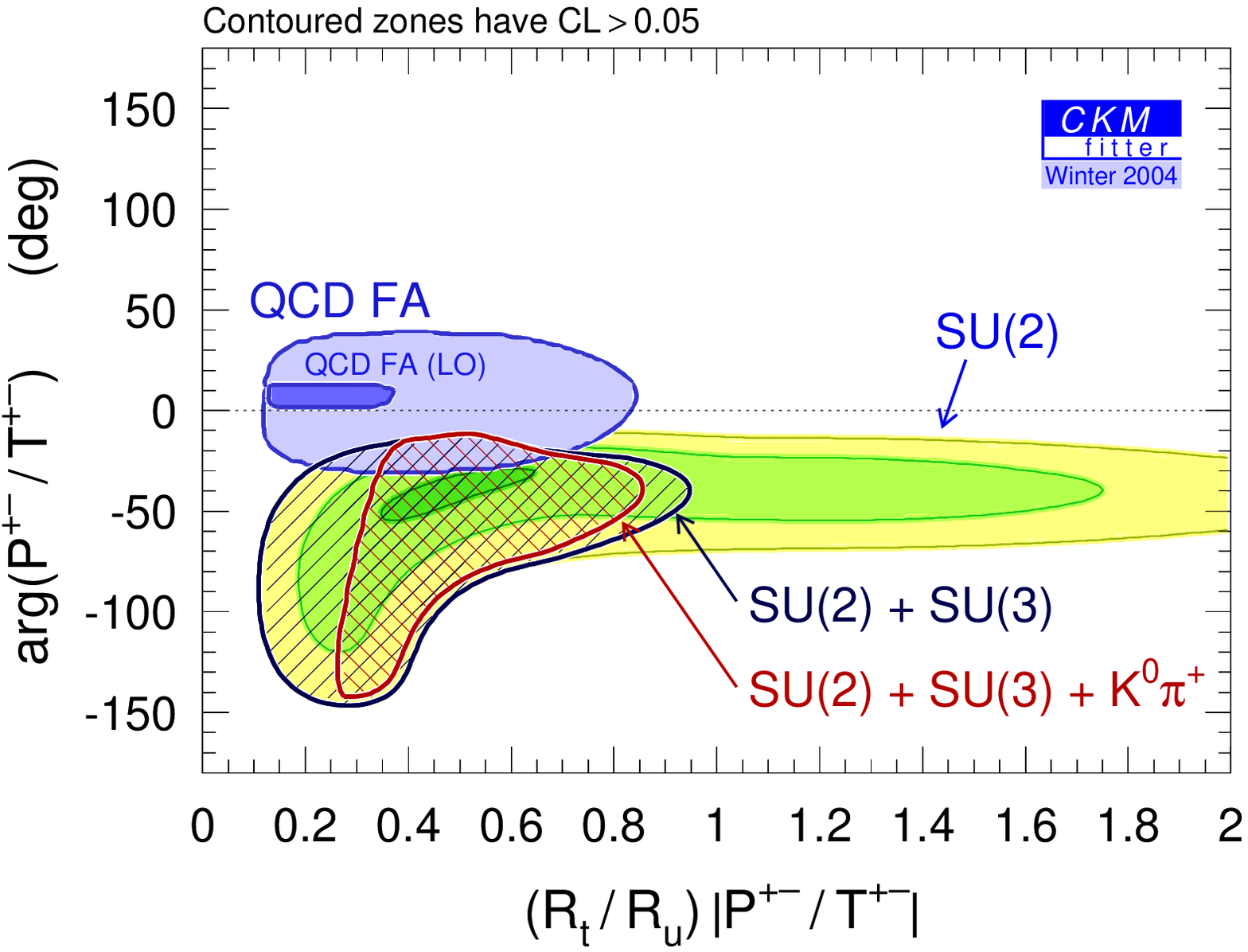}
              \epsfxsize8.1cm\epsffile{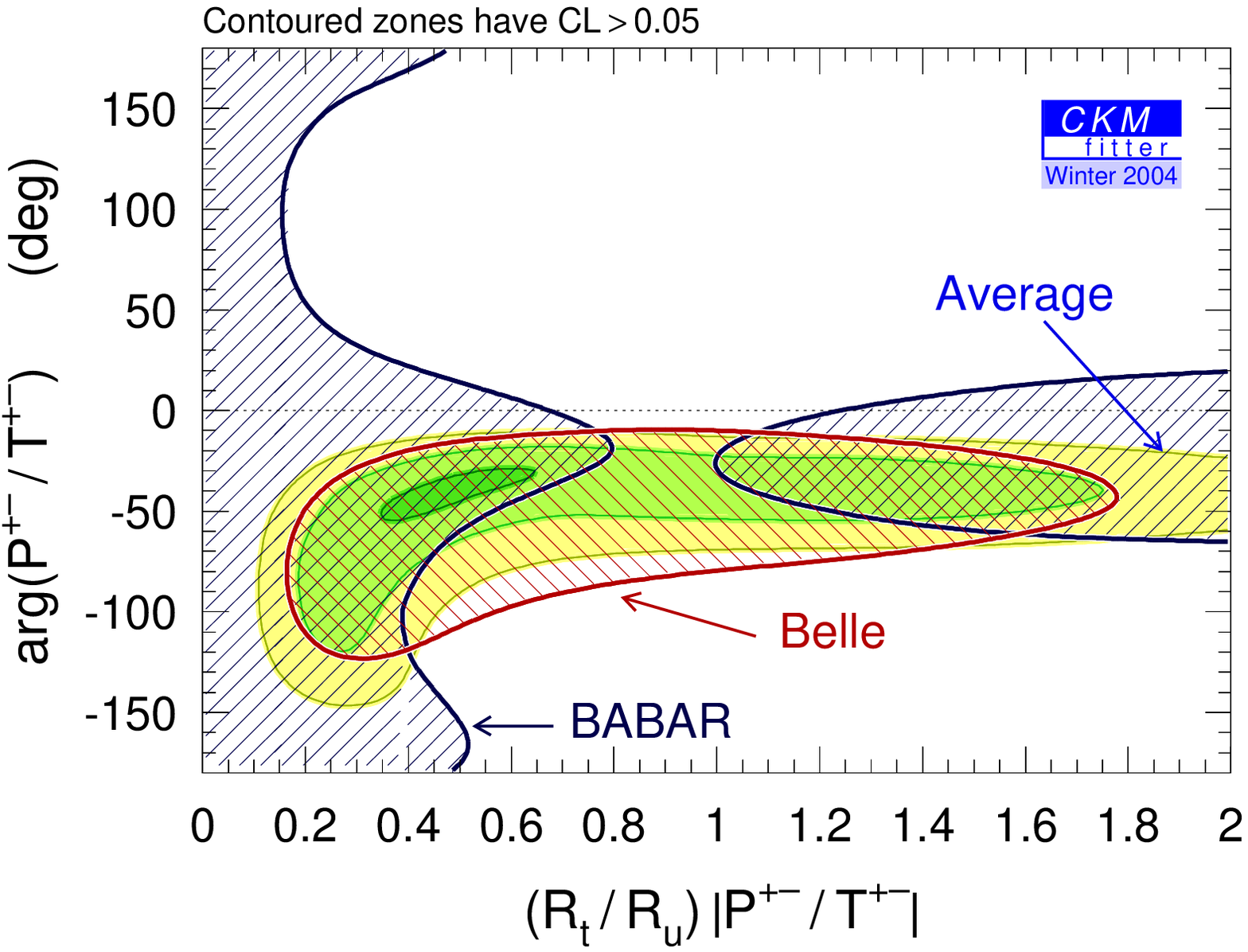}}
  \vspace{-0.0cm}
  \caption[.]{\label{fig:dpot}\em
        Constraints on the penguin-to-tree ratio $\rpipi$ and the 
        relative strong phase $\deltapipi$ in $B \to \pi\pi$ decays, 
        obtained when using as additional input the CKM parameters $\rhobar$ and 
        $\etabar$ from the standard CKM fit. The gradually shaded regions 
        give the CLs for fits of Scenario (\iA) (SU(2)): dark, medium and 
        light shaded areas have $CL>0.90$, $0.32$ and $0.05$, respectively. 
        Also shown are the $5\%$~CL contours obtained for Scenario~(\iB) 
        and ~(\iC).
        The elliptical areas are the prediction from QCD Factorization 
        (Scenario~(\iD)): full calculation (light shaded) and leading order 
        (dark shaded). On the right hand plot, constraints using Scenario
        (\iA) are given individually for \babar, Belle and their averages. }
\end{figure}
\begin{figure}[t]
  \centerline{\epsfxsize8.1cm\epsffile{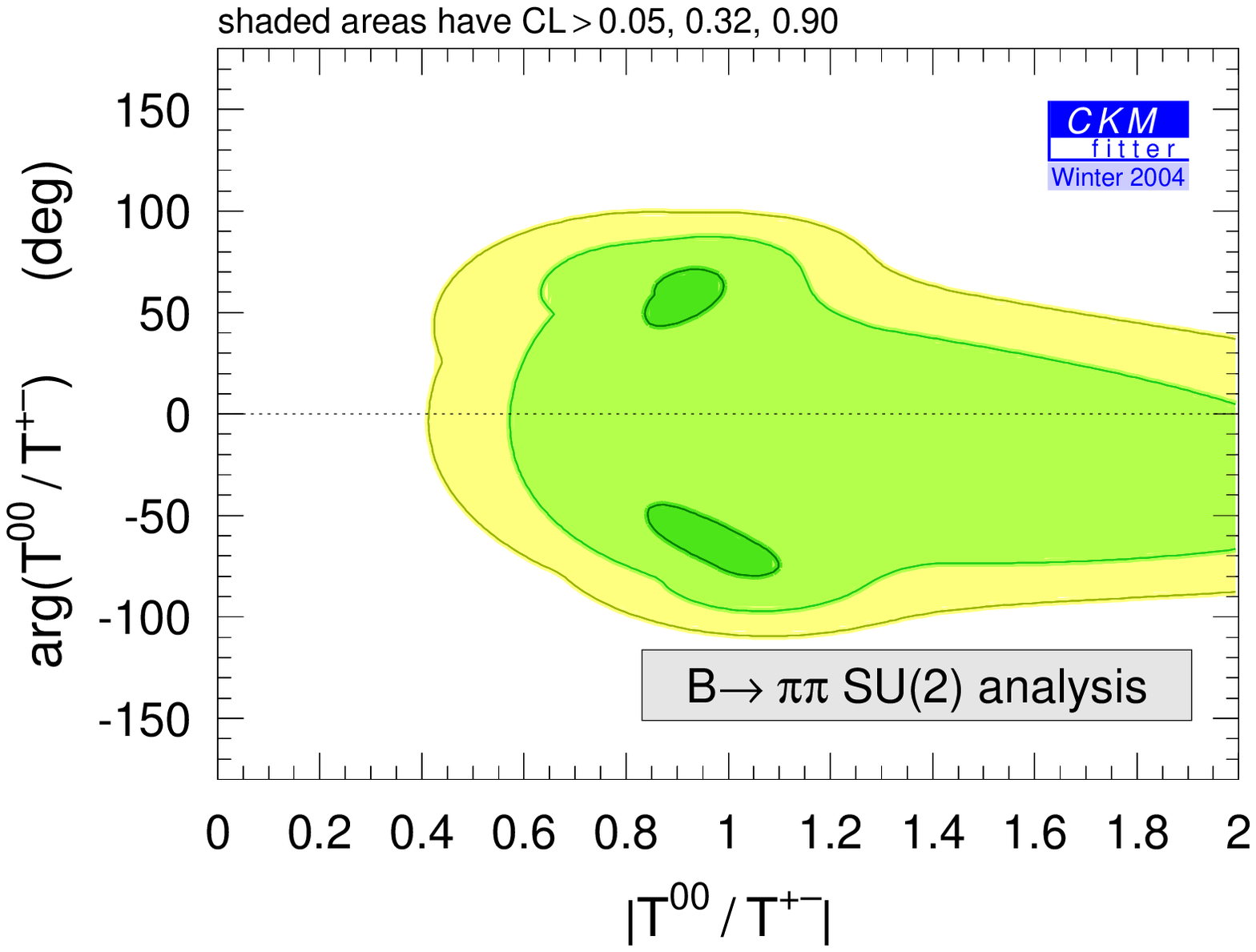}}
  \vspace{-0.0cm}
  \caption[.]{\label{fig:dtotpipi}\em
        Constraints from Scenario~(\iA) (SU(2)) on magnitude and phase
        of the color-suppressed-to-color-allowed ratio $\Tcpizpiz/\Tpipi$ 
        in $B \to \pi\pi$ decays. The CKM parameters $\rhobar$ and 
        $\etabar$ are taken from the standard CKM fit.
        The gradually shaded regions indicate the CLs: dark, medium and 
        light shaded areas have $CL>0.90$, $0.32$ and $0.05$, respectively.}
\end{figure}
One can take another point of view and constrain the unknown
penguin-to-tree ratio $\rpipi$ and its phase $\deltapipi$ using
the standard CKM fit as input. As in the prediction 
of $\Spipi$ and $\Cpipi$ in the previous paragraph, this assumes that 
the experimental measurements are in agreement with the constraints 
obtained on $\rhobar$ and $\etabar$ in the standard CKM fit, \ie, no 
New Physics comes into play. The results are shown in the left hand 
plot of Fig.~\ref{fig:dpot}. The shaded regions give the CLs obtained 
from a fit using Scenario~(\iA) (SU(2)). Significant penguin 
contributions and strong phases are required to accommodate the fit with 
the data. Scenario~(\iB) leads to an exclusion of large values for 
$\rpipi$, while Scenario~(\iC) increases the lower bound. 
We find that the preferred values for $\rpipi$
are in agreement with the allowed regions obtained 
 for Scenario~(\iD) (QCD FA). The right hand plot of 
Fig.~\ref{fig:dpot} shows the Scenario~(\iA) constraints separately 
for \babar, Belle and their average. Large non-zero $\rpipi$ and 
$\deltapipi$ are required by Belle's numbers.
\vs
To test color-suppression, the same procedure is applied to
constrain the color-suppressed-to-color-allowed ratio $\Tcpizpiz/\Tpipi$. 
The resulting CLs are given in Fig.~\ref{fig:dtotpipi}.
For the magnitude we obtain the lower limit $|\Tcpizpiz/\Tpipi|>0.41$ 
for $\CL>5\%$ and a central value of $0.9$, which significantly 
exceeds the na\"{\i}ve $0.2$ expectation from  factorization. Note that a
central value of order one, if confirmed, would challenge the $1/N_c\to
0$ limit of QCD independently of the validity of perturbative factorization.

\subsection{Prospects for the Isospin Analysis}
\label{sec:pipi1000}

\begin{figure}[t]
  \centerline{\epsfxsize8.1cm\epsffile{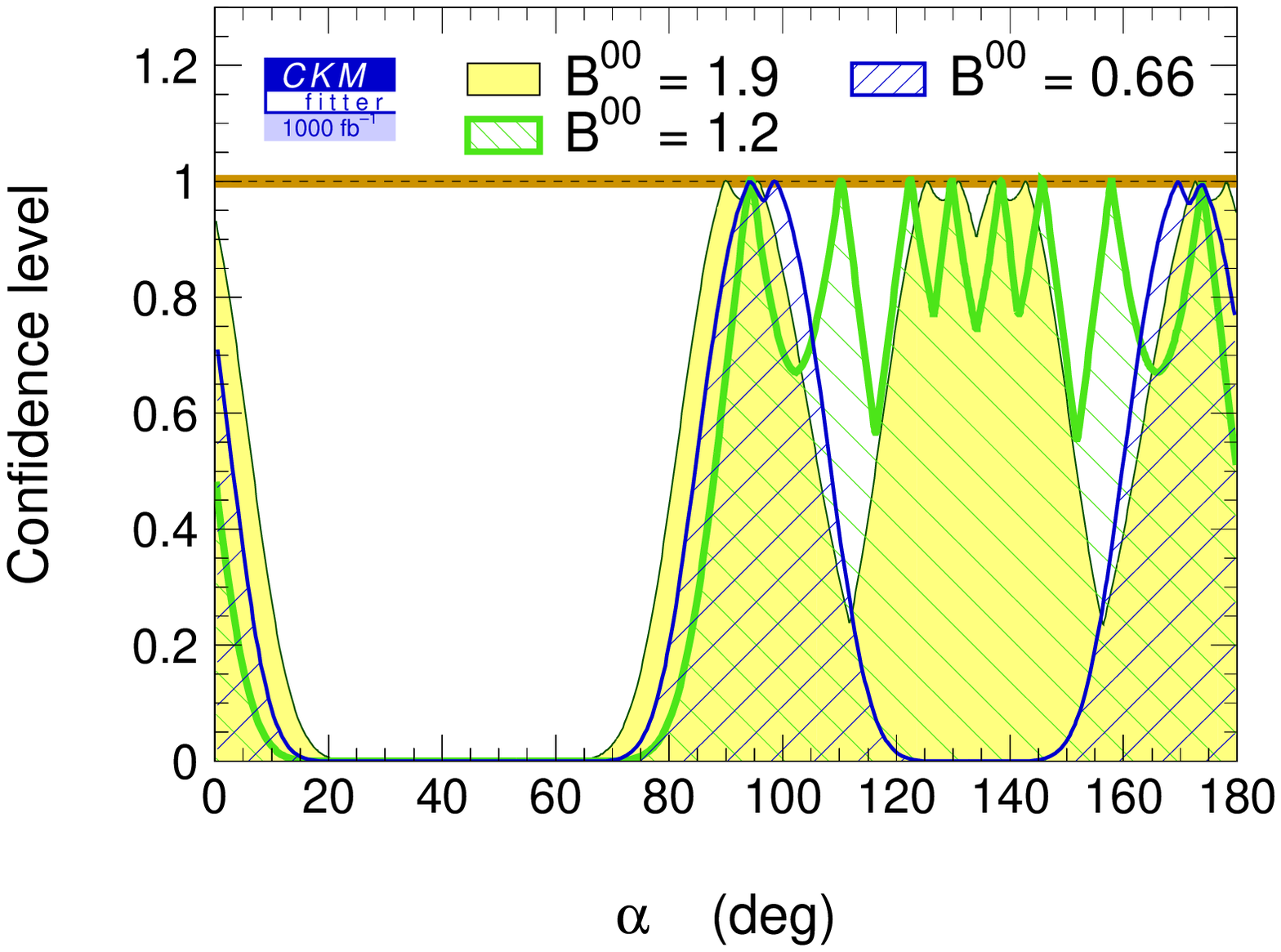}
              \epsfxsize8.1cm\epsffile{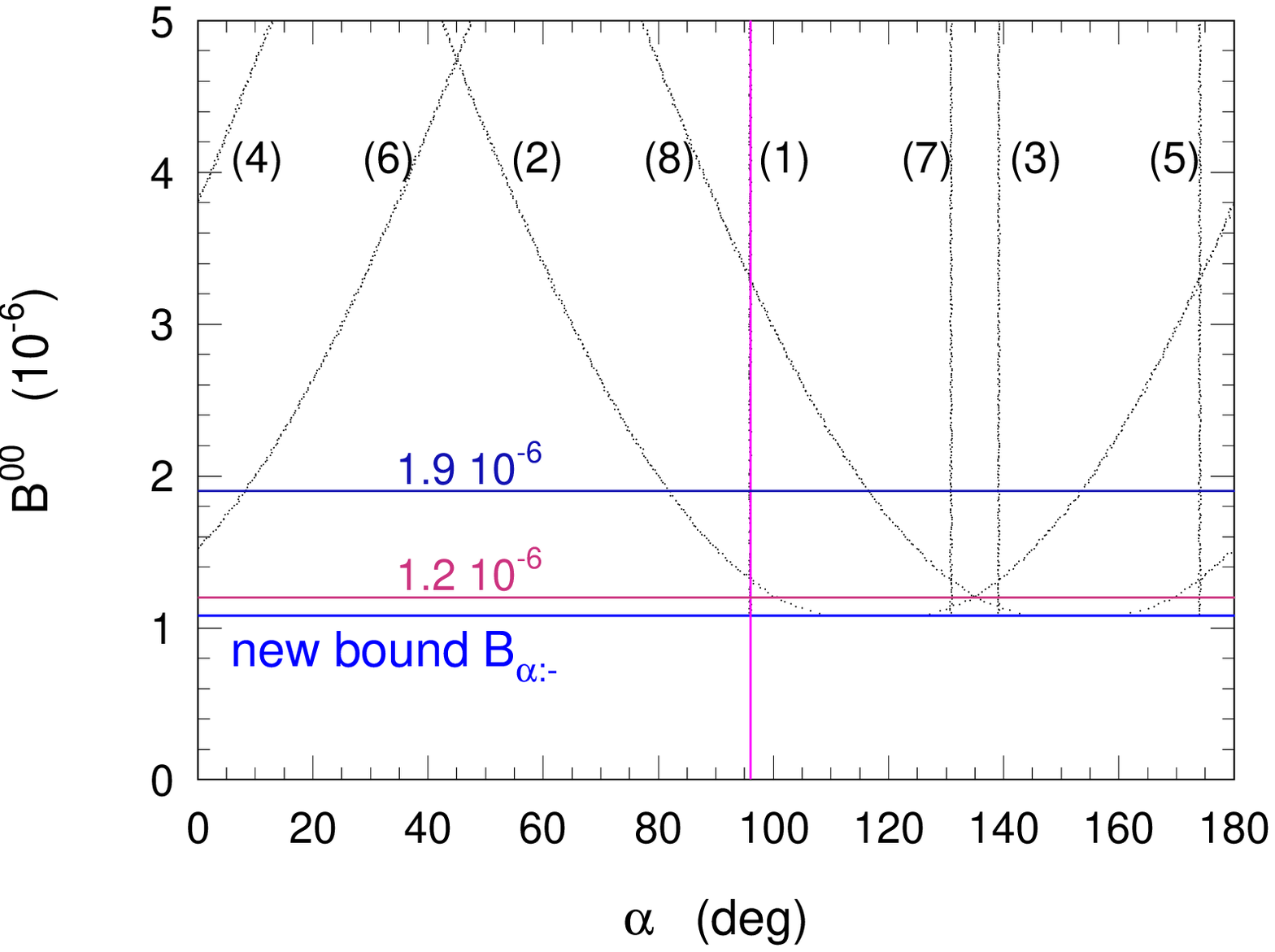}}
  \vspace{-0.5cm}
  \caption[.]{\label{fig:switchyard}\em
        \underline{Left:} confidence level as a function of $\alpha$ for the full 
        isospin analysis (including $\Cpizpiz$)
        at an integrated luminosity of $1\invab$,
        and using the central values and errors given in the text (shaded area).
        Also shown is the CL for $\BRpizpiz=1.2 \times 10^{-6}$ (down-diagonal 
        hatched area) for which the ambiguities move together (see right 
        hand plot), and for $\BRpizpiz=0.66 \times 10^{-6}$ (up-diagonal 
        hatched area), which corresponds to $\BrooGLSSl$, i.e., the ambiguities
        overlap in two quadruplets. The branching ratio values given on the 
        figure are in units of $\tmsix$.
        \underline{Right:} location of the eight mirror solutions as a 
        function of $\BRpizpiz$. The curves refer to the present central values of 
        branching fractions and \CP-violating asymmetries.  The horizontal lines
        indicate the bound~(\ref{eq:rebound}), computed at the input value of 
        $\alpha$, as well as the two branching fractions 
        used for the isospin analyses of the left hand plot. Electroweak penguins
        are neglected.}
\end{figure}
\begin{figure}[!thb]
  \def\thisputaindefiguresize{8.1cm}
  \centerline{\epsfxsize\thisputaindefiguresize\epsffile{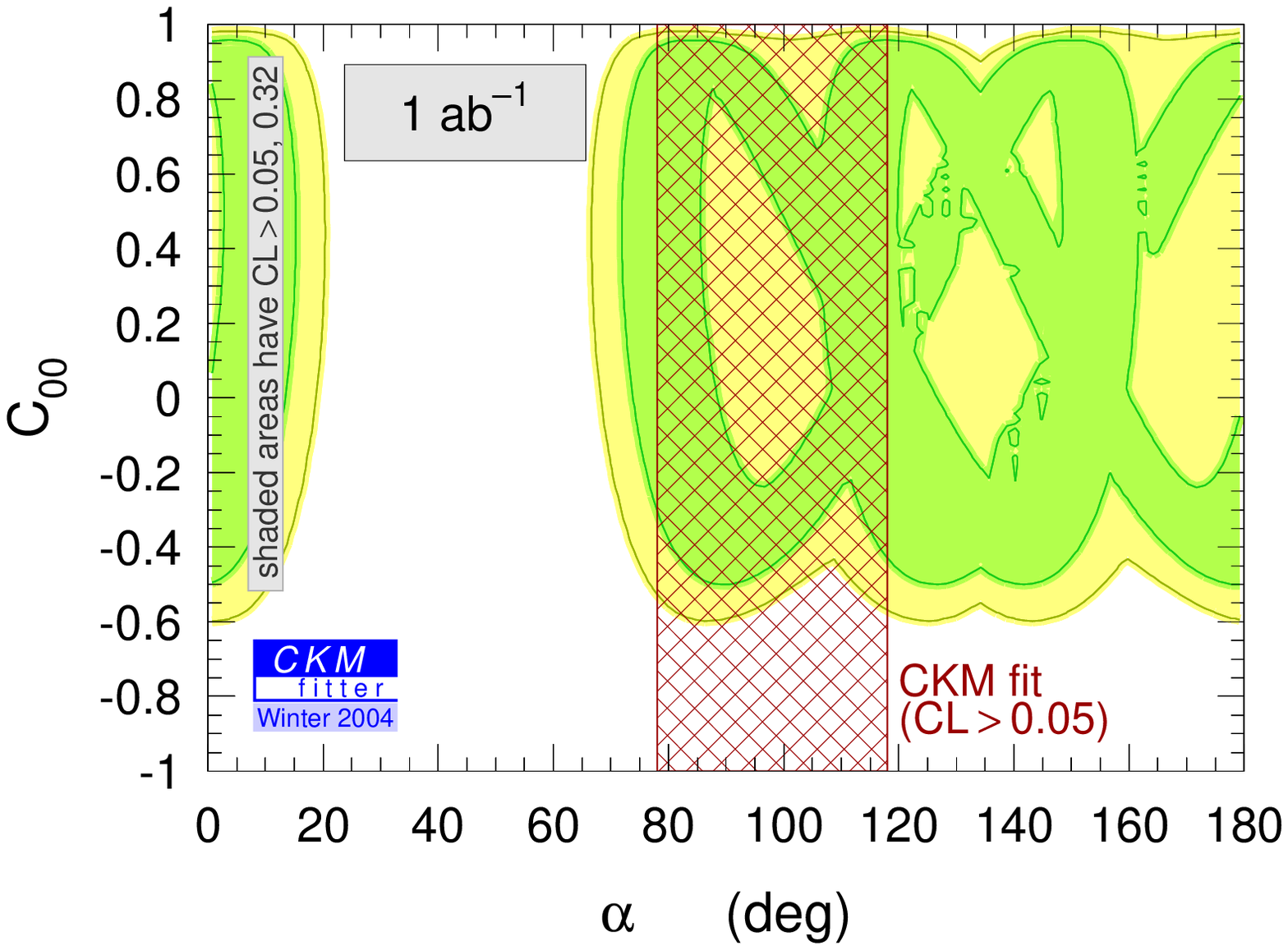}
              \epsfxsize\thisputaindefiguresize\epsffile{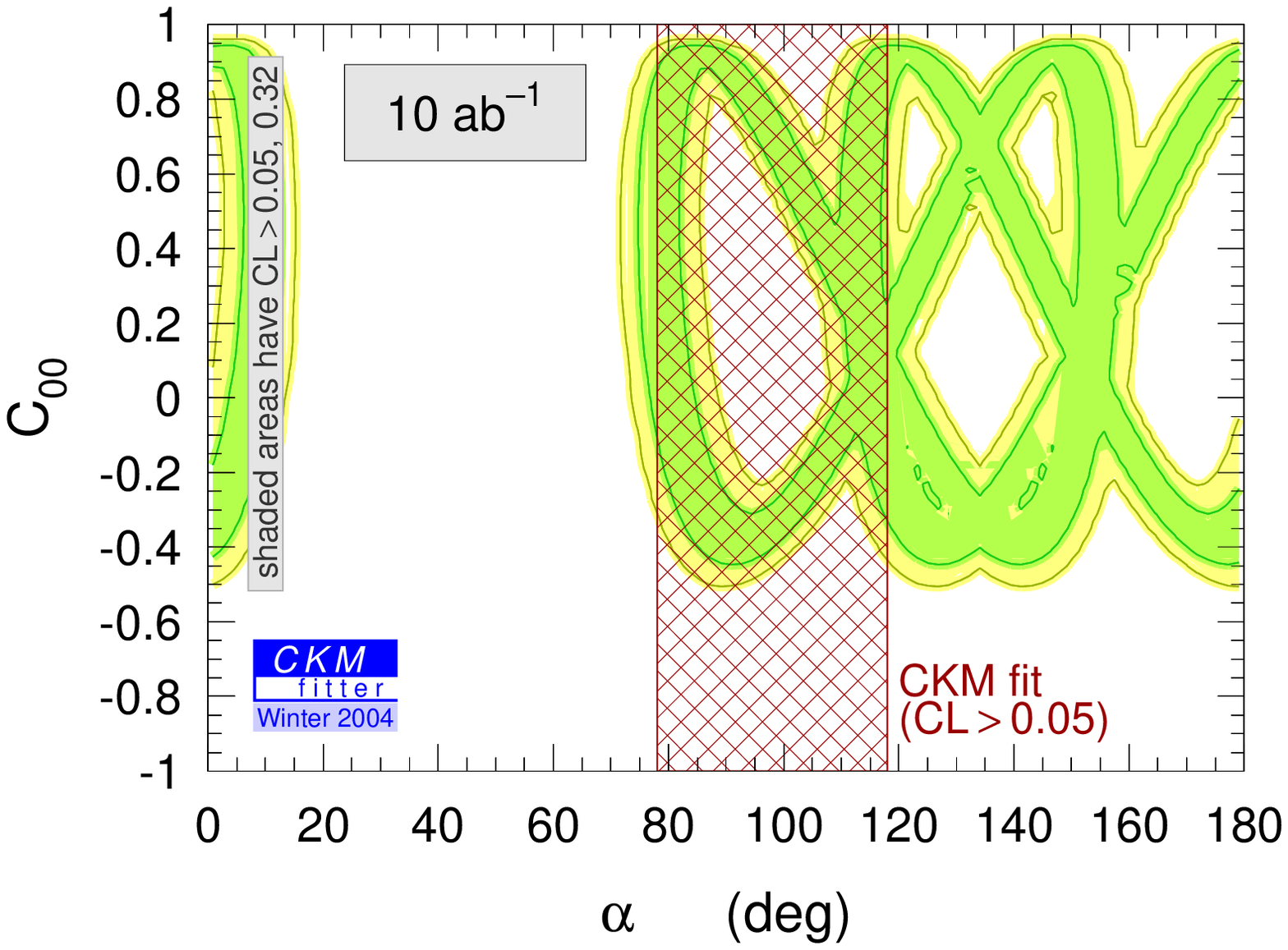}}
  \vspace{-0.0cm}
  \caption[.]{\label{fig:alC00}\em
        Confidence level in the $(\alpha,\Cpizpiz)$ plane at integrated 
        luminosities of $1\invab$ (left) and $10\invab$ (right), respectively. 
        The observable set and errors used are given in the text. }
 \vspace{0.2cm}
  \centerline{\epsfxsize\thisputaindefiguresize\epsffile{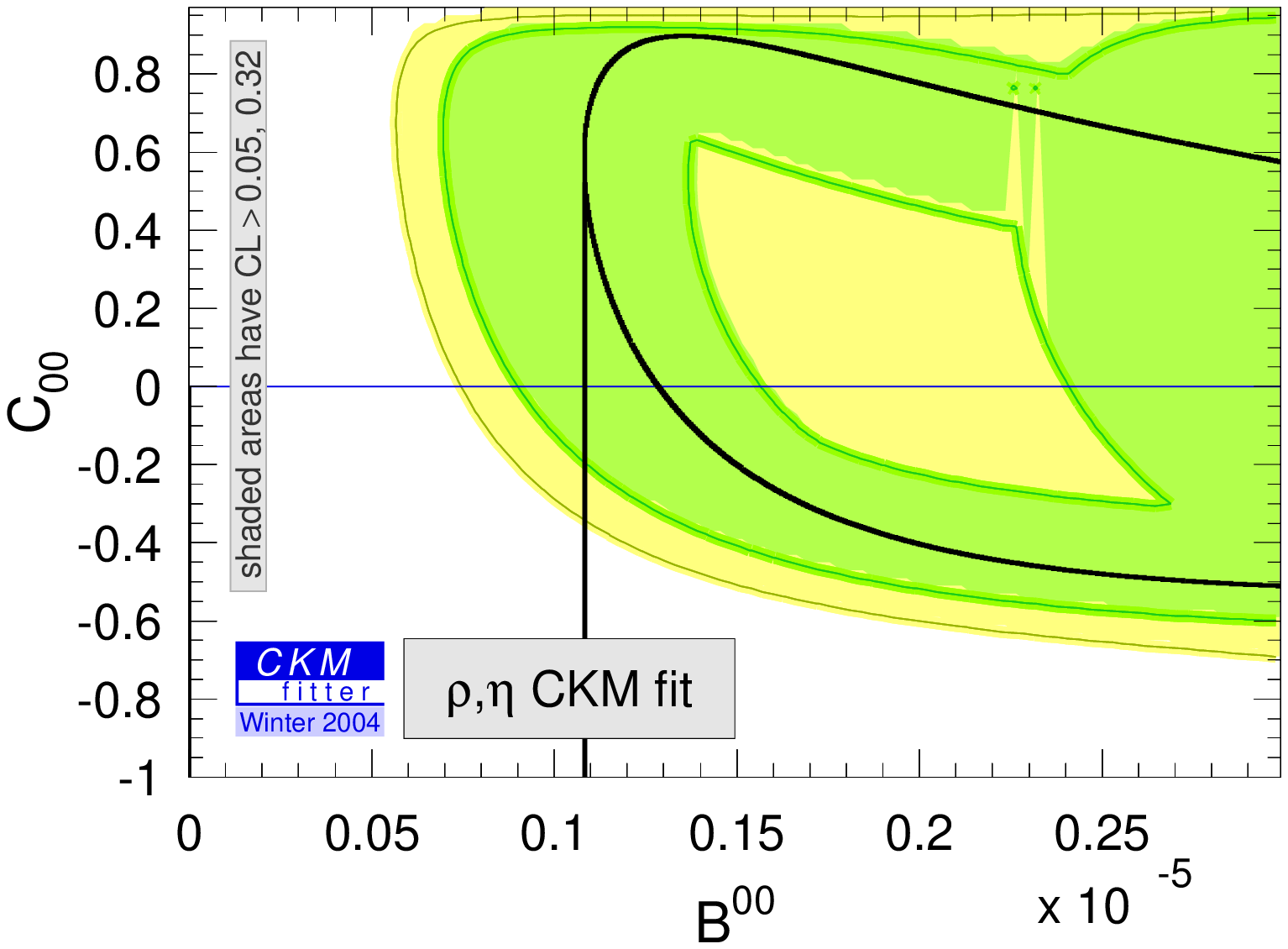}
              \epsfxsize\thisputaindefiguresize\epsffile{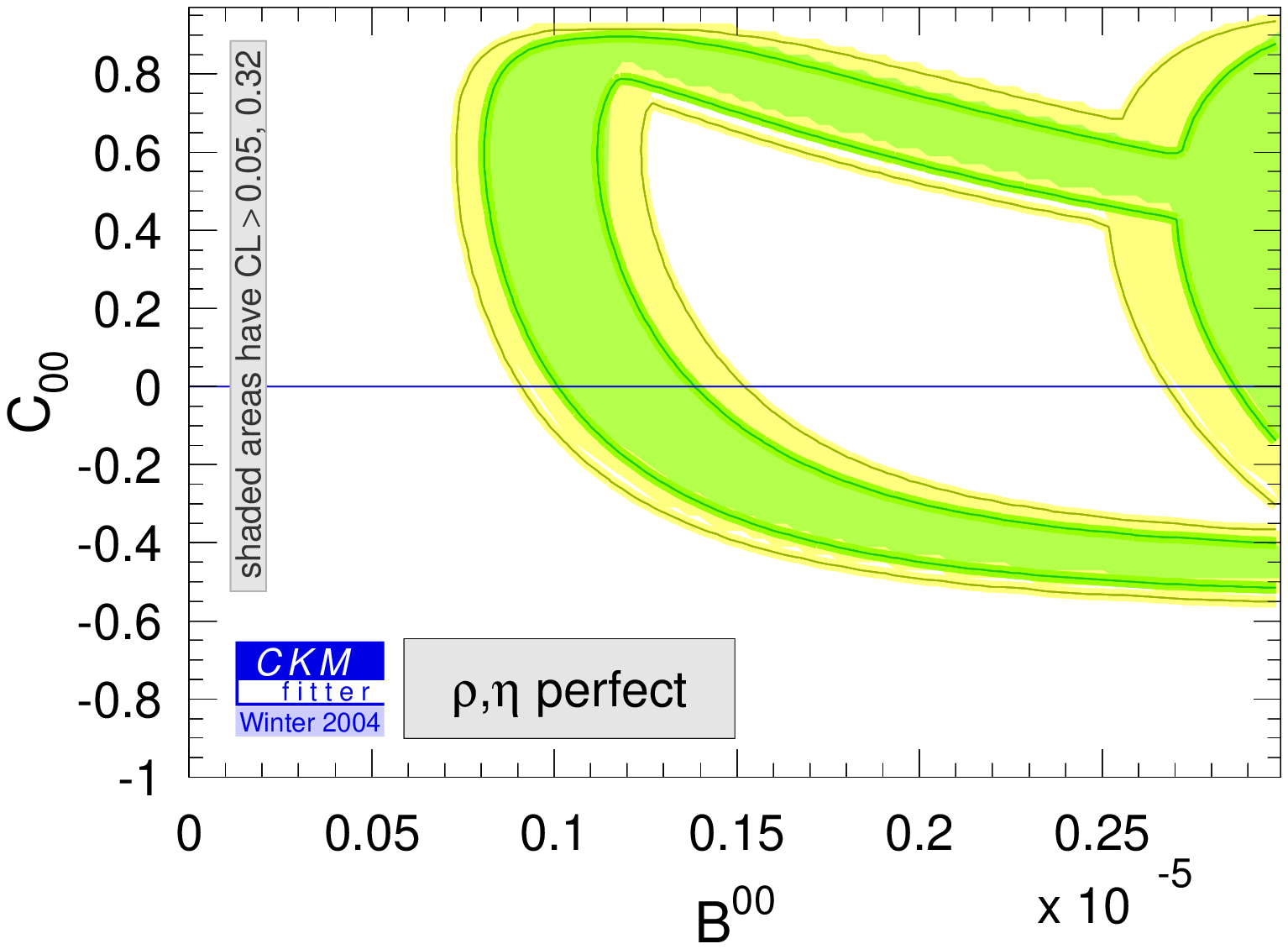}}
  \vspace{-0.0cm}
  \caption[.]{\label{fig:CooBoo}\em
        Confidence level in the $(\BRpizpiz,\Cpizpiz)$ plane at an 
        integrated luminosity of $1\invab$. The left hand plot uses the 
        standard CKM fit as input including the present uncertainties
        on $\alpha$, while the right hand plot assumes perfect knowledge 
        of $\alpha$. Superimposed on the left hand plot is the analytical 
        function $\Cpizpiz(\BRpizpiz)$~(\ref{eq:coof}), where electroweak 
        penguins are neglected which explains the difference 
        with the CL function.
        The vertical line represents the bound~(\ref{eq:rebound}).}
\end{figure}
The preceding sections have shown that, at present, relevant
information on $\alpha$ requires input from flavor symmetry
other than SU(2) and/or theoretical assumptions, the accuracy of
which is hard to determine. However the ultimate goal of the
experimental effort should be a model-independent determination
of $\alpha$. This prejudice given, we shall attempt an outlook
into the future to assess the performance of a full isospin
analysis, where $\Cpizpiz$ is determined in a time-integrated
measurement by the experiments.
\vs
Figure~\ref{fig:switchyard} (left) shows the CL of the angle $\alpha$ for 
the following set of observables (branching fractions are given in units 
of $10^{-6}$):
\beqns
\begin{array}{rclrcl}
    \BRpipi     &=&      4.55 \pm 0.17 \pm 0.09 ~,   & \hspace{1.0cm}
    \Spipi      &=&     -0.73 \pm 0.07 \pm 0.02 ~,\\[0.1cm]
    \BRpimpiz   &=&      5.18 \pm 0.28 \pm 0.16 ~,   & \hspace{1.0cm}
    \Cpipi      &=&     -0.46 \pm 0.06 \pm 0.03 ~,\\[0.1cm]
    \BRpizpiz   &=&      1.90 \pm 0.20 \pm 0.09 ~,   & \hspace{1.0cm}
    \Cpizpiz    &=&     -0.37 \pm 0.24 \pm 0.03 ~, 
\end{array}
\eeqns
where we have kept the central values of the present experimental results.
For the parameter $\Cpizpiz$ we choose one out of the two solutions preferred 
by the data when inserting $\alpha$ from the standard CKM fit. The statistical 
errors are extrapolated to an integrated luminosity of $1\invab$. 
For the systematic uncertainties we assume an optimistic development:
the branching ratios are dominated by uncertainties due to the 
reconstruction of neutrals ($2.5\%$ per $\piz$), while the \CP parameters 
are dominated by the unknown \CP violation on the tag side. One observes
the characteristic eightfold ambiguity within $[0,\pi]$, where the position
of the peaks depends in particular on $\BRpizpiz$ (see comments below). Although
the allowed region for $\alpha$ largely exceeds the one obtained by the 
standard CKM fit, significant $\alpha$ domains are excluded and the peaking 
structure provides metrological information when combined with other 
$\alpha$ measurements. We also note that due to the significant penguin
pollution in the $\pi\pi$ system, contributions from New Physics may be 
present in the data.
\vs
As outlined in Section~\ref{sec:charmlessBDecays}.\ref{sec:su2pipi}, the 
central value of $\BRpizpiz$ drives the position of the discrete 
ambiguities for $\alpha$. The location of the eight mirror solutions as 
a function of $\BRpizpiz$ are shown on the right hand plot of 
Fig.~\ref{fig:switchyard}. The curves refer to the present central values of 
branching fractions and \CP-violating asymmetries. The horizontal lines
indicate the bound~(\ref{eq:rebound}) as well as the two values used 
for the isospin analyses represented in the left hand plot (apart from 
the nominal setup given above, a second set is used with $\BRpizpiz=1.2\tmsix$ 
and the corresponding value $\Cpizpiz=0.13$, and with all other parameters 
kept unchanged). The quality of the metrological constraint on $\alpha$ 
depends on how much the different solutions overlap. The worst case occurs
when several mirror solutions gather around the true value of $\alpha$ 
within a distance of about $\sigma(\alpha)$. As a consequence we note that 
large values of $\BRpizpiz$ can lead to a better metrology. 
\vs
In a third extrapolation we study the best-case scenario, where  
$\BRpizpiz$ is chosen to be equal to one of the GLSS bounds~(\ref{eq:GLSSbound}). 
While the upper bound is excluded by experiment, the lower bound, 
$\BrooGLSSl=0.66\tmsix$, may still be reached. We choose 
$\BRpizpiz=(0.66\pm0.12\pm0.03)\tmsix$, as well as 
$\Spipi = -0.25\pm0.07\pm0.02$ and $\Cpizpiz=0.75\pm0.39\pm0.03$,
to achieve consistency between the observable set, SU(2), and
the standard CKM fit. The modified $\Spipi$ value (with respect to
the previous extrapolations) ensures that $\BrooGLSSl=\reBoundl$,
which is required for overlapping ambiguities\footnote
{
        One notices in the up-diagonal hatched function of the left hand 
        plot in Fig.~\ref{fig:switchyard} that the ambiguities do not 
        exactly overlap. This is because electroweak penguins are
        neglected in the bounds~(\ref{eq:GLSSbound}), (\ref{eq:rebound}),
        while they are taken into account in the numerical 
        analysis used to produce the plots.     
}. The chosen set
of observables is only marginally consistent with the present 
measurements. The resulting CL for $\alpha$ is given by the up-diagonal 
hatched function in the left hand plot of Fig.~\ref{fig:switchyard}.
The $1\sigma$ precision on $\alpha$ for this scenario is found to 
be $14^\circ$. This study provides an illustration of how precise 
the measurement of $\alpha$ could turn out to be in the coming years. 
However, one should keep in mind that if $\BRpizpiz$ is not equal to, 
but only close to $\BrooGLSSl$, the metrology is spoilt~\cite{mufrapipi}.
\vs
Figure~\ref{fig:alC00} shows the CLs in the $(\alpha,\Cpizpiz)$ plane at 
integrated luminosities of $1\invab$ (left) and $10\invab$ (right), where
we have used the same parameter configuration as in the above discussion,
with the exception of $\Cpizpiz$ which is not used. It is assumed in the 
extrapolation that the systematic uncertainties do not decrease any further 
beyond $1\invab$. As for $\BRpizpiz$, 
one observes that (for given $\BRpizpiz$ and $\Cpipi$) the ambiguity pattern 
for $\alpha$ depends strongly on $\Cpizpiz$. An extraction of $\alpha$ with
an accuracy of a few degrees should be within the reach of a next generation 
$B$ factory. 
\vs
The parameter plane ($\BRpizpiz,\Cpizpiz$) is convenient to immediately display
the consistency between the measurements in the $\pi\pi$ system and the 
standard CKM fit, because it avoids the problem of multiple solutions.
The left hand plot of Fig.~\ref{fig:CooBoo} represents the expectation for 
an integrated luminosity of $1\invab$, using the standard CKM fit as input. 
Very large luminosities will be needed in order to significantly uncover a 
potential disagreement with the SM. 
The right hand plot of Fig.~\ref{fig:CooBoo} is obtained assuming in addition
that $\rhobar$ and $\etabar$ be exactly known and fixed to their present 
central values (\cf\   Table~\ref{tab:fitResults1}). 

\subsection{Predicting the $\Bs\to\Kp\Km$ Branching Fraction 
            and \CP-Violating Asymmetries}
\label{sec:bskk}

It has been pointed out by Pirjol~\cite{pirjolKK} and
Fleischer~\cite{fleischer} that one can
use SU(3) symmetry\footnote
{
    More precisely, only the U-spin subgroup of SU(3)
    ($s\leftrightarrow d$ exchange) is needed. However the accuracy of
    this approximate symmetry is not expected to be significantly better
    than full SU(3). For example, the decay constants $f_{\pi^+}$ and
    $f_{K^+}$ have the same value in either U-spin or SU(3) limit,
    although experimentally $f_{K^+}/f_{\pi^+}\simeq 1.22$~\cite{PDG}.
} to 
relate the amplitudes in $\Bs \to \Kp\Km$ and $\Bz \to \pip\pim$ decays. 
The $\Bs \to K^+K^-$ amplitude is given by
\beq
\label{eq:bsk+k-}
   A(\Bs\to K^+K^-) =   V_{us}V_{ub}^*\Tkk
                      + V_{ts}V_{tb}^*\Pkk~,
\eeq
and using SU(3) symmetry, one can identify
\beqn
\label{eq:uspin}
                    \Tkk    &=&   \Tpipi~, \nonumber\\
                    \Pkk    &=&   \Ppipi~,
\eeqn
which leads to the relation~\cite{pirjolKK,fleischer}
\begin{figure}[t]
  \centerline{\epsfxsize10cm\epsffile{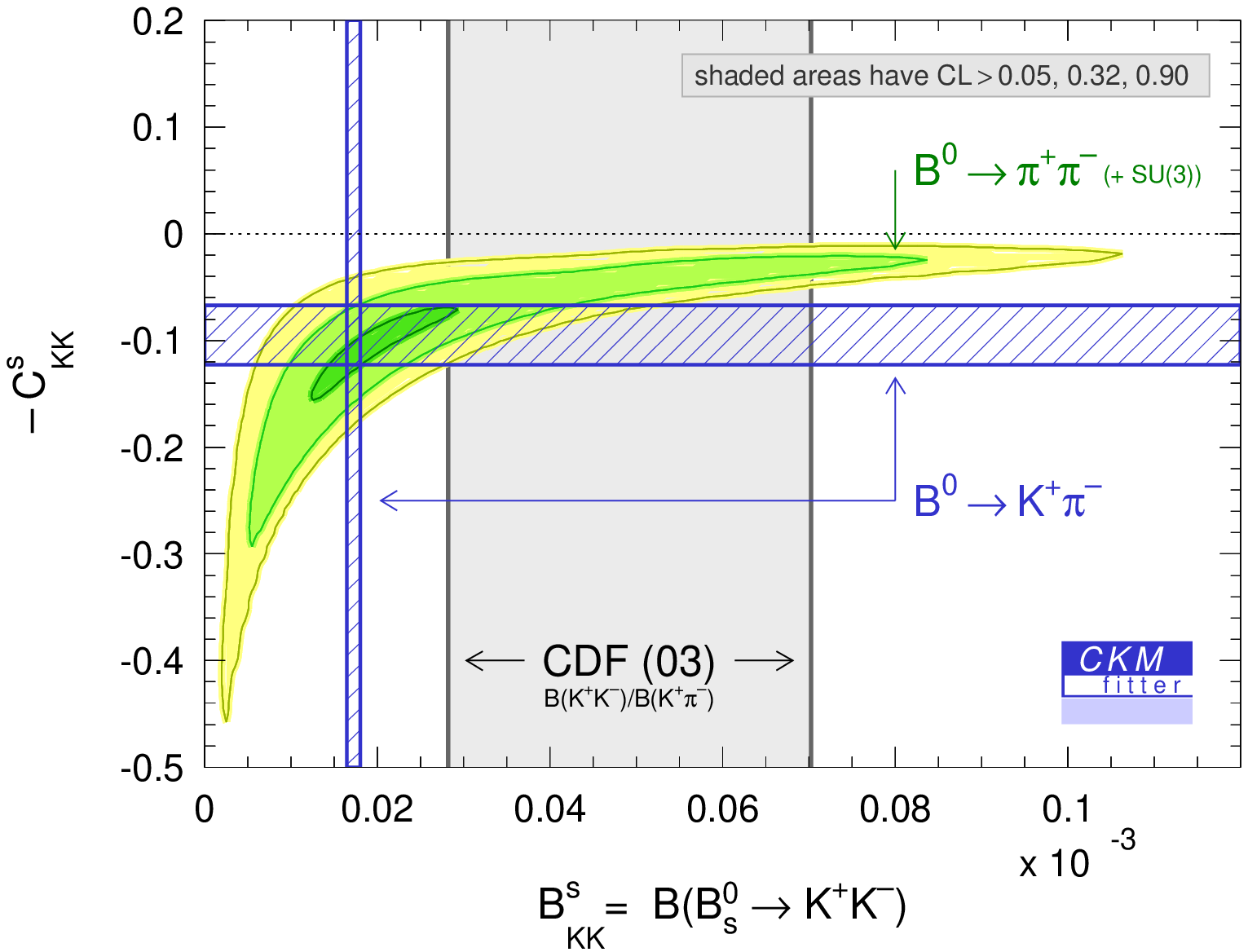}}
  \vspace{-0.0cm}
  \caption[.]{\label{fig:bskk}\em
        Confidence level in the $(\BRkk,-\Ckk)$ plane obtained 
        from $\B\to\pi\pi$ data and the standard CKM fit, assuming SU(3) 
        symmetry. Also shown is the preliminary result on the 
        ratio $\BR(\Bs\to\Kp\Km)/\BR(\Bz\to\Kp\pim)$ from 
        CDF~\cite{BABARBelleLP03,CDFKpiblessed} and the 
        branching fraction (corrected by the lifetime ratio
        $\tau_{\Bs}/\tau_{\Bz}$) and direct \CP asymmetry 
        in $\B\to\Kp\pim$.
        The bands indicate $1\sigma$ error ranges. }
\end{figure}
\beq
\label{eq:hyp}
        \frac{\Ckk \, \BRkk}{\tau_{\Bs}} + 
           \frac{\Cpipi\,\BRpipi}{\tau_{\Bz}} \:=\: 0~,
\eeq
where $\Ckk=C(\Bs\to K^+K^-)$ and $\BRkk$ denote the direct \CP-violation asymmetry and
branching fraction of $\Bs \to \Kp\Km$, respectively. The ratio of the
$\Bs$  to the $\Bz$ lifetimes is $0.951\pm0.038$~\cite{HFAG}.
Constraining $\Tpipi$ and $\Ppipi$ with the $B \to \pi\pi$ branching
fractions and  the $\Spipi$ and $\Cpipi$ measurements, and including
the standard CKM fit to predict the CKM elements, one obtains the
hyperbolic shape in the  $(\BRkk,-\,\Ckk)$ plane shown in
Fig.~\ref{fig:bskk}. We find the $\CL>5\%$  ranges (see also
Ref.~\cite{BFRS2})
\beqn
   \BRkk        & = & (5- 91)\tmsix~, \nonumber\\
   \Ckk         & = &  0.02-0.32~.\nonumber
\eeqn
These predictions can be compared with the branching fraction and \CP
asymmetry found for $\Bz\to K^+\pi^-$ (see Table~\ref{tab:BRPiPicompilation}): 
assuming SU(3) and neglecting all (tree and penguin) exchange topologies they
are expected to be equal. Agreement is observed as illustrated 
in Fig.~\ref{fig:bskk}. Also shown in the figure is the preliminary result
from the CDF collaboration~\cite{BABARBelleLP03,CDFKpiblessed} on the ratio
$\BR(\Bs\to\Kp\Km)/\BR(\Bz\to\Kp\pim)=2.71\pm0.73\pm0.35\pm0.08$,
where the first error is statistical, the second due to the ratio of
$\Bz$ and $\Bs$ production in $b$ jets, and the third is systematic.
The error bands shown are at $1\sigma$. Following the same procedure, 
we also predict the mixing-induced $\CP$ asymmetry in $\Bs\to\Kp\Km$ 
decays to be, for $\CL>5\%$,
\beq
     \Skk \;=\; 0.12  -     0.27~.\nonumber
\eeq

\section{Tests of QCD Factorization in $B\to\pi\pi,\,K\pi$ decays} 
\label{sec:bbnsres}
In this Section, we present several fits of the $K\pi$ and $\pi\pi$ data
to the calculation of hadronic matrix elements within the QCD
Factorization approach~\cite{BBNS0,BBNS,BN}.

\subsection{QCD Factorization at Leading Order}

\begin{figure}[p]
  \centerline
        {
        \epsfxsize8.1cm\epsffile{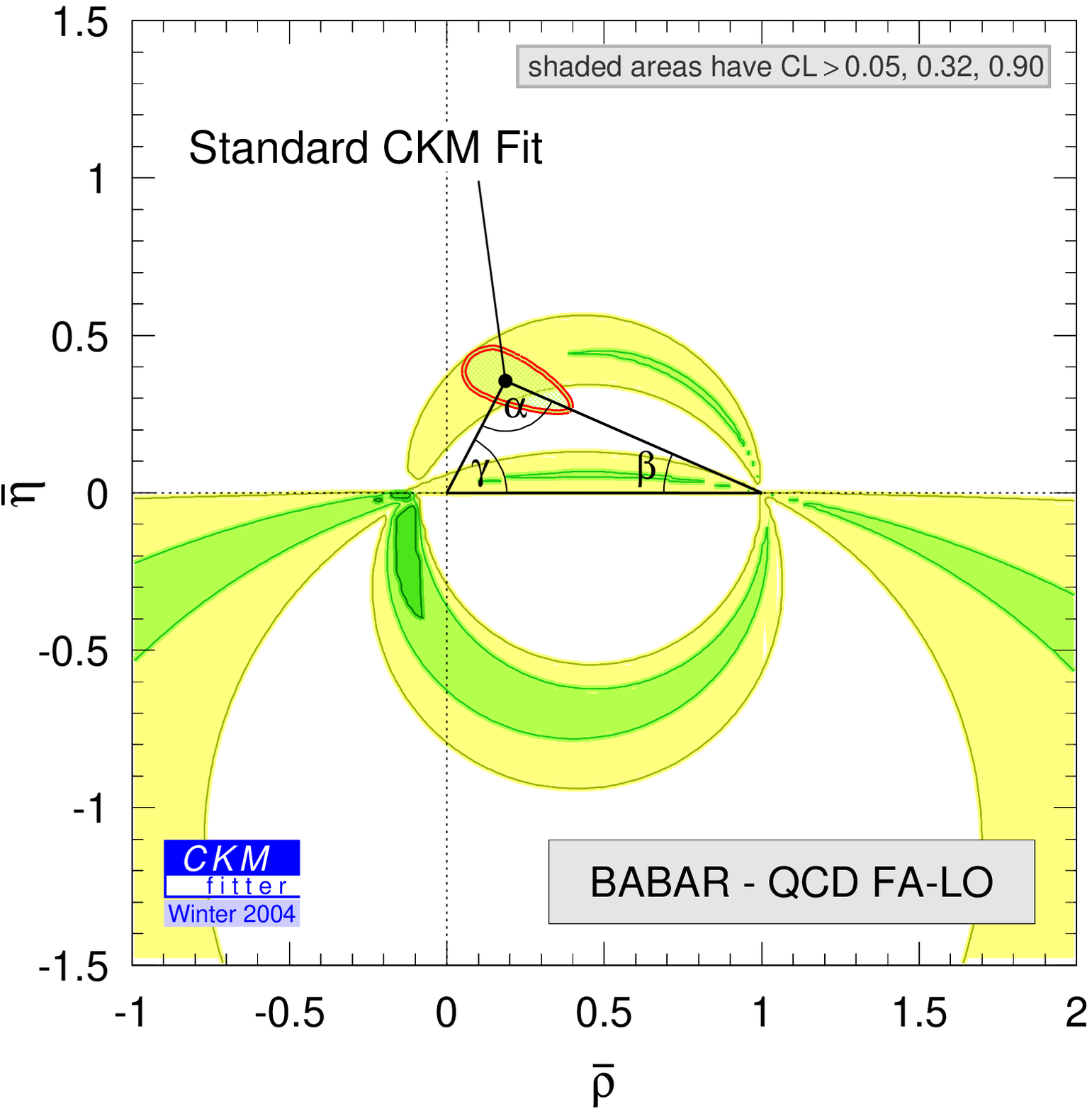}
        \epsfxsize8.1cm\epsffile{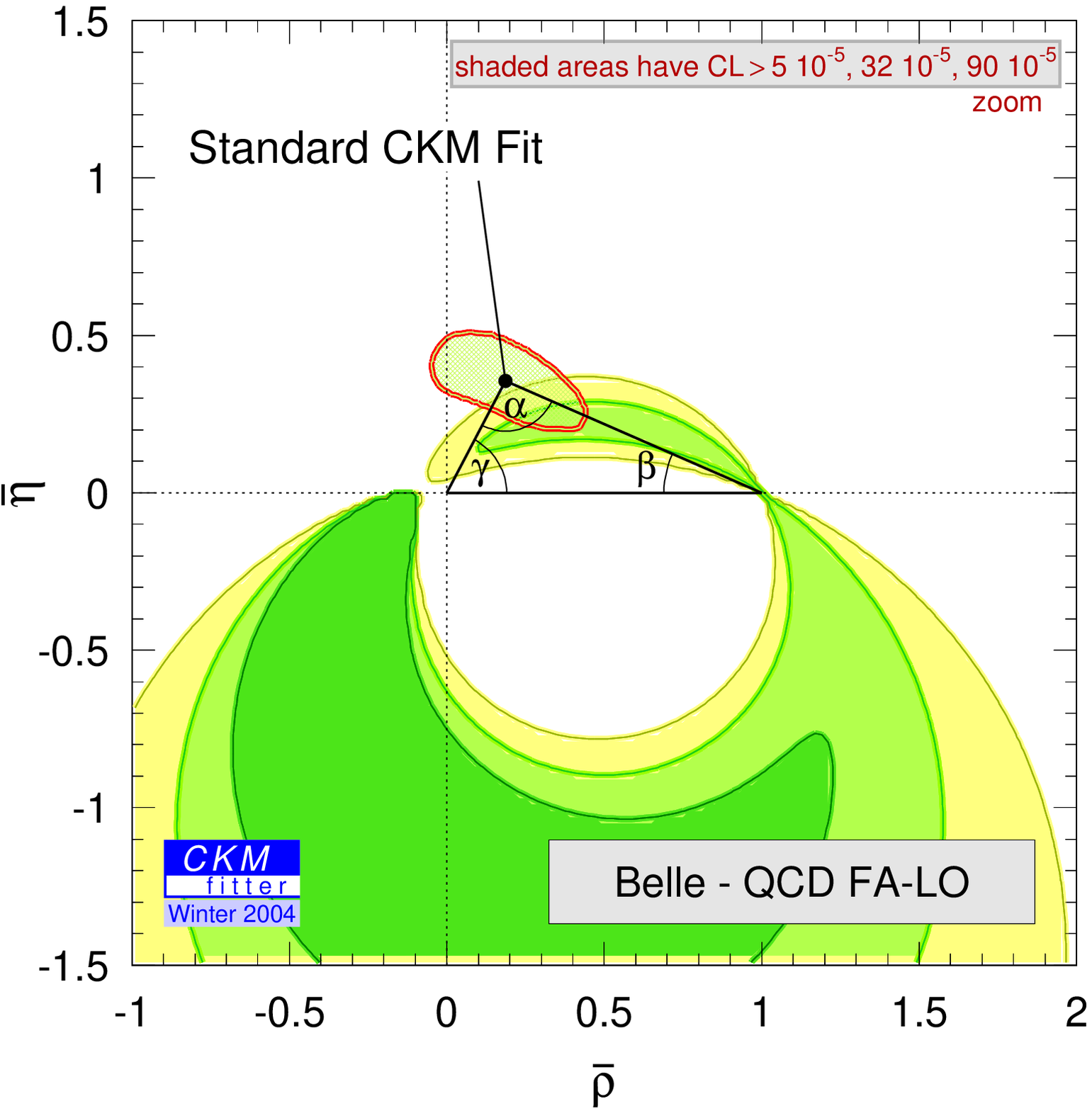}
        }
  \vspace{0.5cm}
  \centerline
        {
        \epsfxsize8.1cm\epsffile{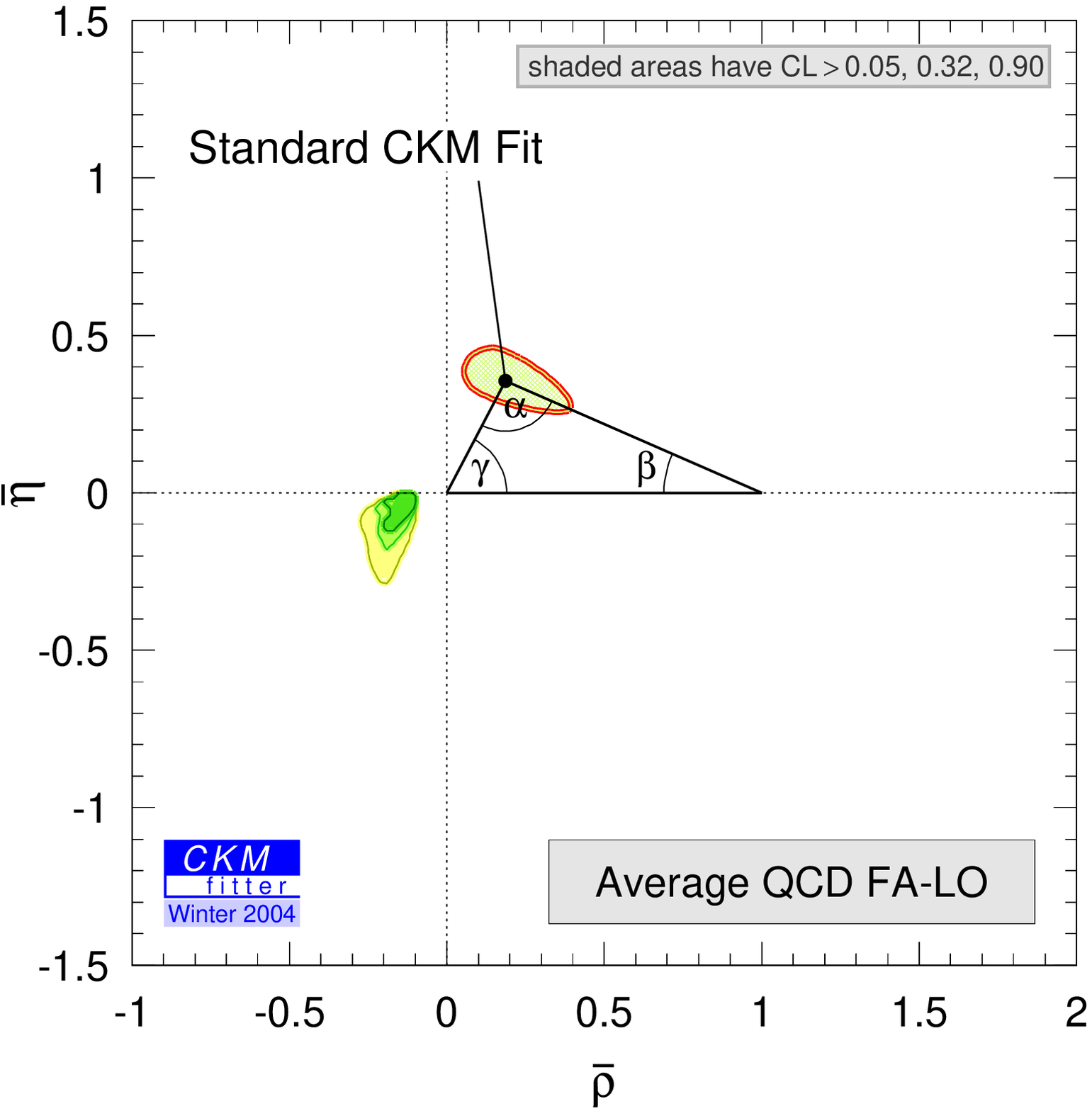}
        \epsfxsize8.1cm\epsffile{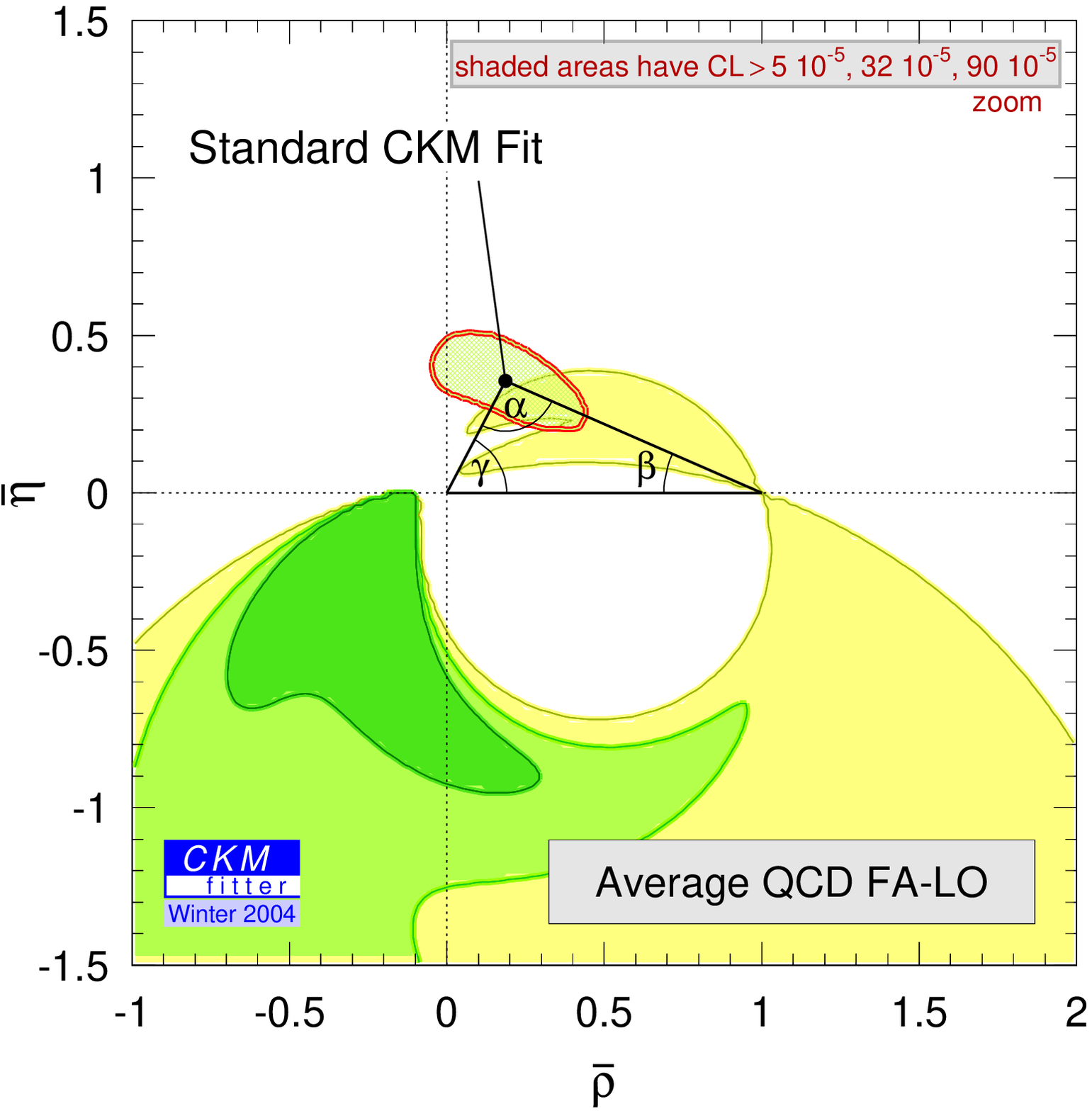}
        }
  \vspace{-0.0cm}
  \caption[.]{\label{fig:BBNSLO}\em
        Confidence levels in the $\rhoeta$ plane for the QCD FA at leading
        order using $\Spipi$ and $\Cpipi$ from \babar\  (upper left), 
        Belle (upper right) and their averages (bottom).
        \underline{Left:} dark, medium and light shaded areas have $CL>0.90$,
        $0.32$ and $0.05$, respectively. \underline{Right:} the maximum 
        CL is set to $10^{-3}$: dark, medium and light shaded areas  
        have $CL>90\times10^{-5}$, $32\times10^{-5}$ and $5\times10^{-5}$. 
        Also shown on each plot is the result from the standard CKM fit.}
\end{figure}
All results using QCD FA presented in the previous Section were obtained
with the full calculation~\cite{BBNS} as defined in 
Section~\ref{sec:charmlessBDecays}.\ref{par:qcdfa}. 
Given the poor knowledge of the parameters $X_{A}$ and $X_{H}$, 
one may examine whether a leading order calculation (see 
Section~\ref{sec:charmlessBDecays}.\ref{par:qcdfa} 
for the exact definition) is sufficient to describe the data.
Figure~\ref{fig:dpot} shows the QCD FA predictions for $\rpipi$
and $\deltapipi$ using both approaches. The uncertainty in the
full QCD FA calculation is dominated by the unknown parameters
$X_{A}$ and $X_{H}$. The leading order calculation predicts
a small positive phase $\deltapipi$ and a moderate ratio $\rpipi$.
\begin{figure}[p]
  \centerline
        {
        \epsfxsize8.1cm\epsffile{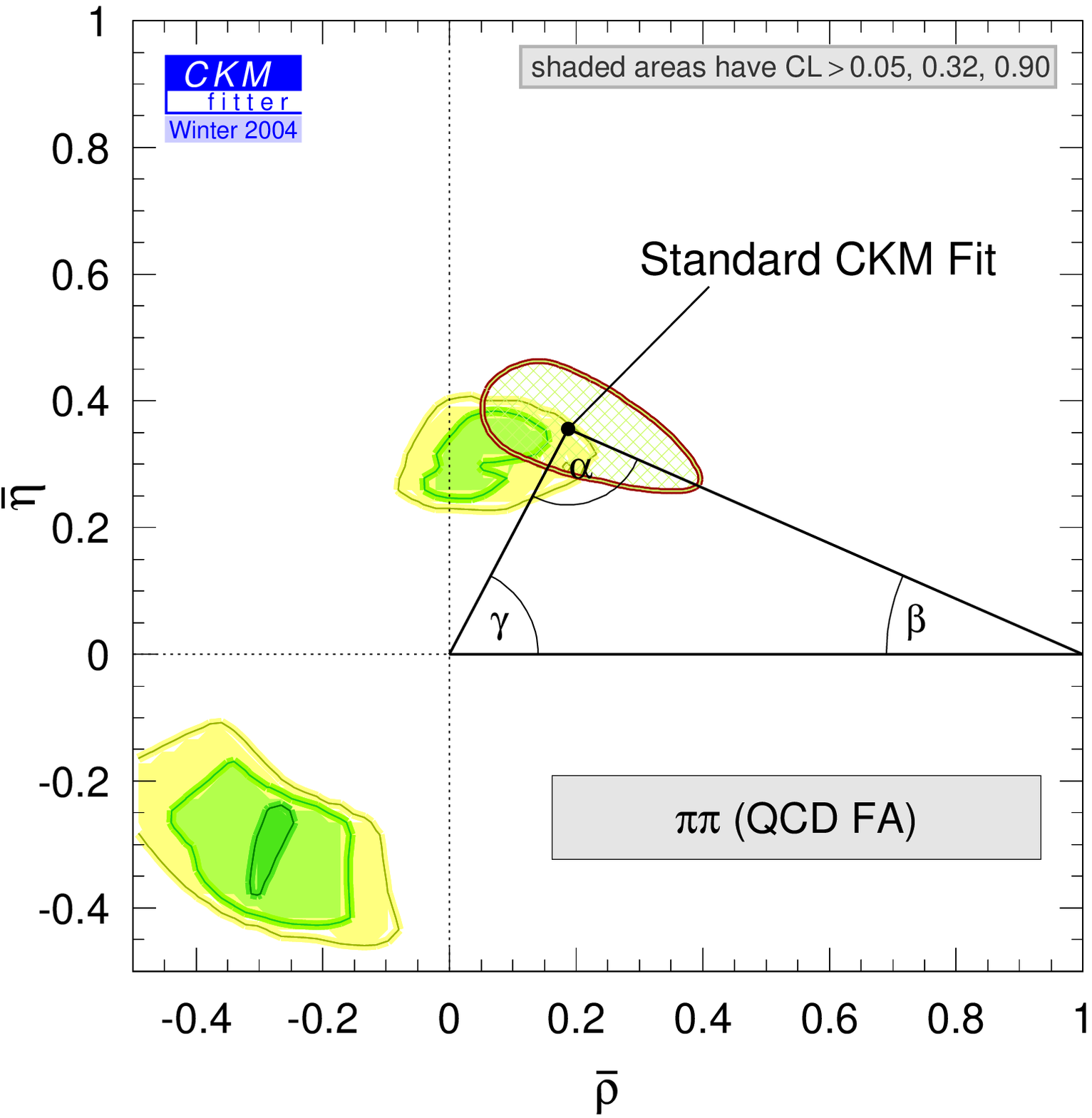}
        \epsfxsize8.1cm\epsffile{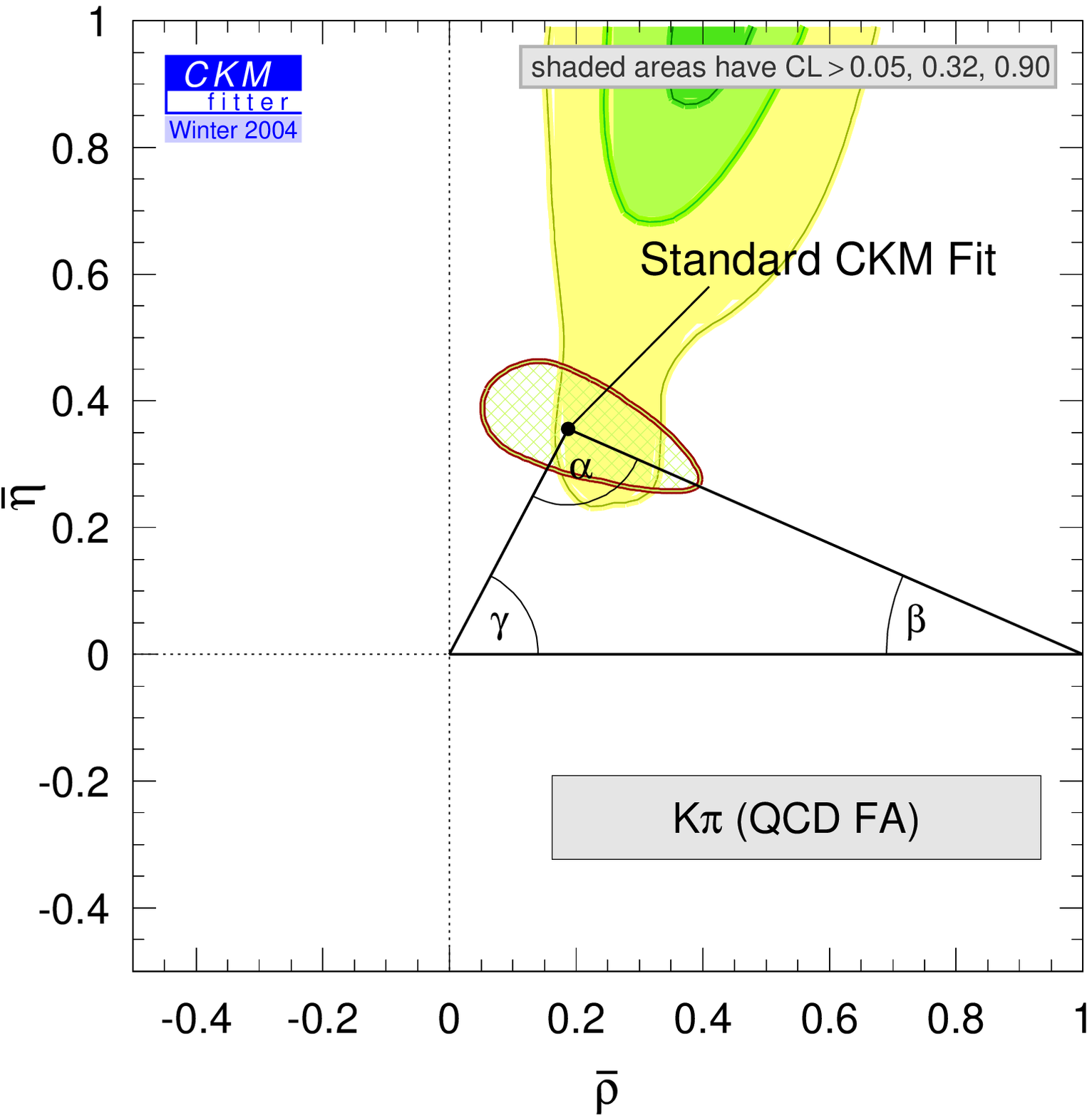}
        }
  \vspace{0.5cm}
  \centerline
        {
        \epsfxsize8.1cm\epsffile{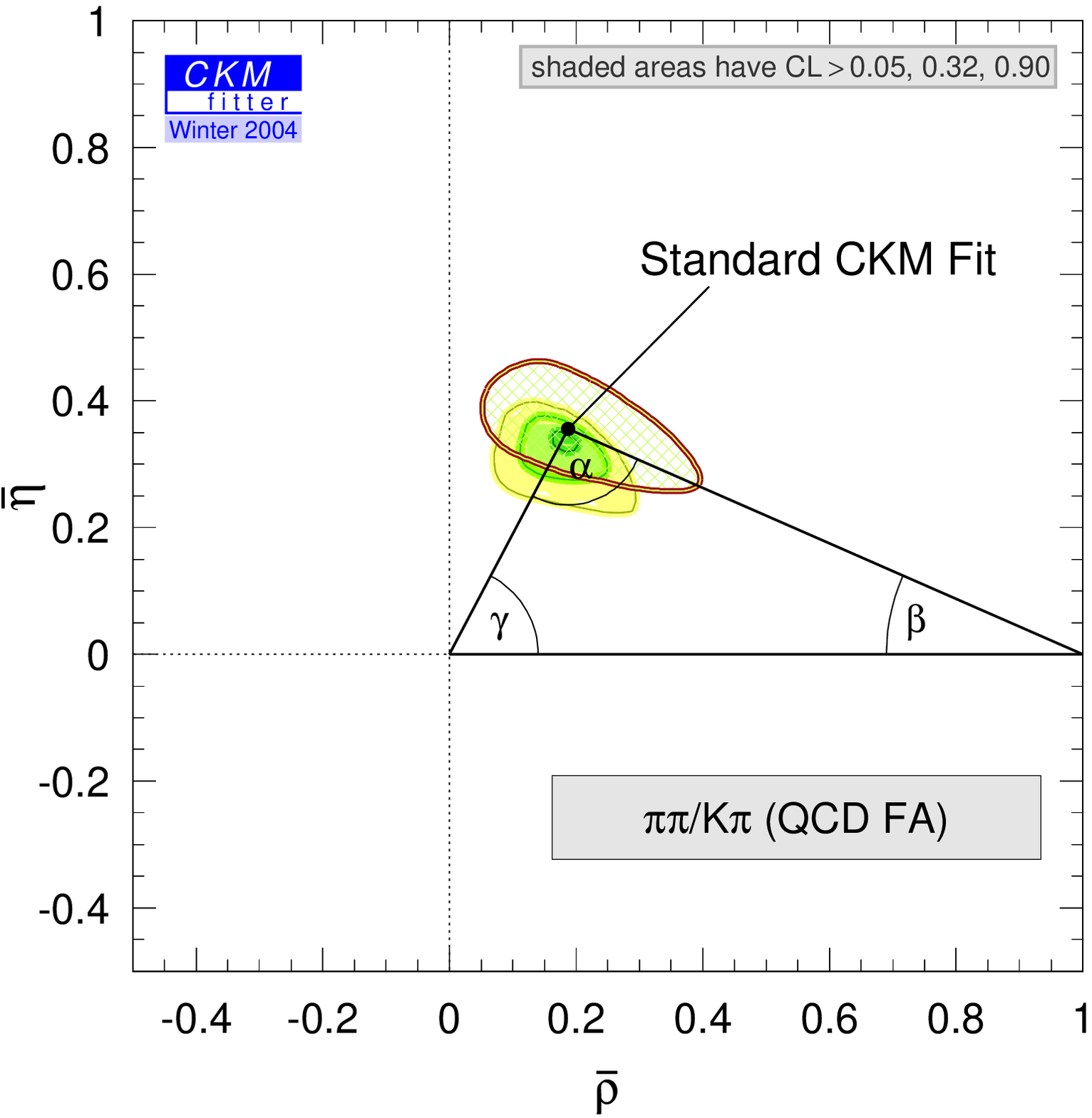}
        }
  \vspace{-0.0cm}
  \caption[.]{\label{fig:bbnsPP}\em
        Confidence level in the $\rhoeta$ plane for Scenario~(\iD): 
        using branching fractions and \CP-violating asymmetries of 
        $B \to \pi\pi$ decays (upper left hand plot), $B \to K\pi$ decays
        (upper right hand plot) and all together (lower plot).
        Dark, medium and light shaded areas have $CL>0.90$,
        $0.32$ and $0.05$, respectively. Also shown is the constraint 
        from the standard CKM fit.  }
\end{figure}
\vs
\begin{figure}[t]
  \centerline{\epsfxsize10cm\epsffile{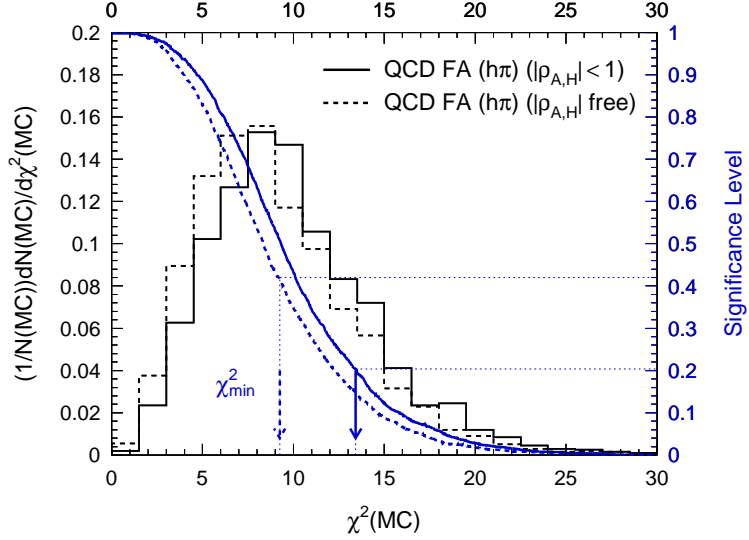}}
  \vspace{-0.0cm}
  \caption[.]{\label{fig:bbnsToy}\em
        Distributions of $\chi^2_{\rm min}$ from toy Monte Carlo experiments 
        and corresponding significance level 
        curves for full QCD FA fits to combined $B\to \pi\pi$ 
        and $B\to K\pi$ results (solid line). \babar\  and Belle averages are 
        used for $\Spipi$ and $\Cpipi$. The dashed lines give the results for 
        unbound parameters $\rho_A$ and $\rho_H$ (see text for details). }
\end{figure}
The constraints in the ($\rhobar,\etabar$) plane using the LO prediction 
are shown in Fig.~\ref{fig:BBNSLO} for $\Spipi$ and $\Cpipi$ from
\babar\  and Belle separately as well as their averages. 
In all three cases, the preferred region is located in the 
negative $\etabar$ half-plane since $\deltapipi$ is predicted positive 
and $\Cpipi$ is found to be negative by both experiments. 
Whereas the \babar\  results are compatible with the standard CKM fit, 
the agreement with Belle is at the $10^{-4}$ level. The average of 
\babar\  and Belle exhibits a compatibility with the standard CKM fit 
at the $5\times10^{-5}$ level.

\subsection{The Full $\pi\pi$ and $K\pi$ Fit}
\label{sec:bbnsfit}

We perform a global fit of the QCD FA to all branching fractions ($\pi\pi$ and
$K\pi$) and \CP asymmetries given in Table~\ref{tab:BRPiPicompilation}. 
Since the leading order calculation cannot describe the data, we use
full QCD FA here.
\vs
The upper plots of Fig.~\ref{fig:bbnsPP} show the CLs in the $\rhoeta$
plane when using only the $B\to\pi\pi$ branching fractions and \CP
asymmetries (left), and when using only the $B\to K\pi$ data (right). The 
constraint from $\Cpipi$ and $\Spipi$ shown in Fig.~\ref{fig:theRhoEtaPlots} 
is now reduced to two distinct zones in the first and the third quadrant of 
the $\rhoeta$ plane. The $K\pi$ measurements prefer large positive values of 
$\etabar$. Given the present experimental accuracy, the compatibility of the 
$K\pi$ data (using QCD FA) with the standard CKM fit is at the $20\%$ level.
\vs
The combined fit to all $B\to \pi\pi,\,K\pi$ observables in the 
$\rhoeta$ plane is shown in the lower plot of Fig.~\ref{fig:bbnsPP}. 
The preferred area is found in excellent agreement (p-value
for the $\chi^2_{\rm min}$ of $21\%$) with the standard CKM fit and has 
competitive precision. We find 
\beqn
\rhobar   &=&     0.182  ^{\,+0.045}_{\,-0.047} \left[ ^{\,+0.089}_{\,-0.092}\right]~,\\
\etabar   &=&     0.332  ^{\,+0.032}_{\,-0.036} \left[ ^{\,+0.056}_{\,-0.081}\right]~,
\eeqn
where the errors outside (inside) brackets are at $1\sigma$ ($2\sigma$). For the UT angle 
$\gamma$, we find
\beq
\gamma =  \left( 62 ^{\,+6}_{\,-9} \left[ ^{\,+12}_{\,-18}\right] \right)^\circ~.
\eeq
Since to leading order in the Cabibbo angle $\lambda$ 
the $K\pi$ system is independent of the CKM phase, the constraint on $\rhoeta$ 
from the combined $\pi\pi,\,K\pi$ fit is dominated by the $\pi^+\pi^-$ 
observables. The $\chi^2_{\rm min}$ amounts to $13.4$ and is dominated by 
contribution from the $K\pi$ data. The corresponding $\chi^2_{\rm min}$ 
distribution is given in Fig.~\ref{fig:bbnsToy} (solid line histogram). The 
dashed histogram is obtained with unbound parameters $\rho_{H,A}$ (we recall
that in full QCD FA they are constrained within $[0,1]$). The $\chi^2$
probability (p-value) 
rises to $42\%$ at $\rho_A=1.4$ and $\rho_H=9.2$, where such 
large values cannot be considered as corrections anymore. In other
words, two additional free parameters parametrizing power corrections
suffice to reconcile the QCD factorization approach with the data.
The measured 
branching fractions for $\Bz \to \Kp\pim$ and $\Bz \to \Kz\piz$ are in 
marginal agreement with QCD FA (\cf\  Table \ref{tab:qcdfaPredi}). Removing 
each of these branching fractions from the nominal ($\rho_{H,A}<1$) combined 
$\pi\pi,\,K\pi$ fit decreases the $\chi^2_{\rm min}$ from $13.4$ to $7.3$ 
and from $13.4$ to $8.5$, respectively.

\begin{table}[t]
\begin{center}
\setlength{\tabcolsep}{0.0pc}
\begin{tabular*}{\textwidth}{@{\extracolsep{\fill}}lcccc}\hline
&&&&\\[-0.3cm]
 & \multicolumn{3}{c}{Fit Result} & Full Prediction \\
 Quantity & Central value $\pm \CL = 0.32$ & $\pm \CL = 0.05$ 
        & $\Delta \chi^2_{\rm min}$ & $\CL>5\%$ range \\[0.15cm]

\hline

&&&&\\[-0.3cm]
$\Cpipi$   
        & $-0.27^{\,+0.08}_{\,-0.05}$ & $^{\,+0.18}_{\,-0.09}$ & $1.0$ 
           & $-0.45$--$0.59$  \\[0.15cm]
$\Spipi$   
        & $-0.70^{\,+0.20}_{\,-0.17}$ & $^{\,+0.38}_{\,-0.25}$ & $0.0$ 
           &  $-1.00$--$0.11$ \\[0.15cm]

\hline

&&&&\\[-0.3cm]
$\Ckspiz$   
        & $0.039^{\,+0.028}_{\,-0.036}$ & $^{\,+0.045}_{\,-0.085}$ & $1.4$ 
           & $-0.36$--$ 0.56$   \\[0.15cm]
$\Skspiz$   
        & $0.827^{\,+0.037}_{\,-0.028}$ & $^{\,+0.071}_{\,-0.056}$ & $0.7$ 
           &  $0.63$--$1.00$ \\[0.15cm]

\hline
&&&&\\[-0.3cm]
$A_{\CP}(\pi^+\pi^0)$~~$[10^{-4}]$ 
        & $-5.2^{\,+1.6}_{\,-0.4}$    & $^{\,+3.6}_{\,-0.9}$ & $0.2$ 
           & $-6$--$15$ \\[0.15cm]
$A_{\CP}(K^+\pi^-)$   
        &$-0.100^{\,+0.035}_{\,-0.007}$    & $^{\,+0.063}_{\,-0.013}$ & $0.0$ 
           & $-0.23$--$0.26$ \\[0.15cm]
$A_{\CP}(K^+\pi^0)$                  
        &$-0.035^{\,+0.042}_{\,-0.038}$    & $^{\,+0.072}_{\,-0.131}$ & $0.2$ 
           &  $-0.40$--$0.55$ \\[0.15cm]
$A_{\CP}(K^0\pi^+)$                  
        & $0.0018^{\,+0.0036}_{\,-0.0041}$ & $^{\,+0.0059}_{\,-0.0044}$ & $0.1$ 
           & $-0.005$--$0.048$ \\[0.15cm]

\hline
&&&&\\[-0.3cm]
$\BR(\Bz \to \pi^+\pi^-)$   
        & $3.80^{\,+0.79}_{\,-0.40}$    & $^{\,+1.92}_{\,-0.90}$ & $0.7$ 
           & $1.7$--$24.8$ \\[0.15cm]
$\BR(B^+ \to \pi^+\pi^0)$   
        & $7.4^{\,+0.9}_{\,-1.2}$    & $^{\,+2.0}_{\,-2.2}$ & $2.5$ 
           & $2.6$--$17.0$ \\[0.15cm]
$\BR(\Bz \to \pi^0\pi^0)$   
        & $1.05^{\,+0.31}_{\,-0.29}$    & $^{\,+0.68}_{\,-0.50}$ & $2.3$ 
           & $0.2$--$4.4$ \\[0.15cm]

$\BR(\Bz \to K^+\pi^-)$   
        & $22.4^{\,+1.2}_{\,-1.4}$    & $^{\,+2.3}_{\,-3.0}$ & $6.1$ 
           & $2.1$--$74.3$ \\[0.15cm]
$\BR(B^+ \to K^+\pi^0)$   
        & $10.87^{\,+0.74}_{\,-0.62}$    & $^{\,+1.69}_{\,-1.22}$ & $2.0$
           &  $0.6$--$45.5$ \\[0.15cm]
$\BR(B^+ \to K^0\pi^+)$   
        & $21.1^{\,+1.0}_{\,-1.2}$    & $^{\,+1.9}_{\,-3.2}$ & $0.1$ 
           & $1.5$--$86.0$ \\[0.15cm]
$\BR(\Bz \to K^0\pi^0)$   
        & $8.36^{\,+0.56}_{\,-0.44}$    & $^{\,+1.36}_{\,-0.86}$ & $4.9$ 
           & $0.7$--$37.0$ \\[0.15cm]
\hline
\end{tabular*}
\caption[.]{\em  \label{tab:qcdfaPredi}
        QCD FA fit results and predictions of $\pi\pi$ and $K\pi$ 
        branching fractions and \CP-violating asymmetries. Left hand 
        part: for each quantity, a full QCD FA fit is performed
        to all $\pi\pi$ and $K\pi$ data but the one that is predicted.
        Central values and $CL=0.32$ and $CL=0.05$ uncertainties 
        are quoted in the two first columns; the third column gives 
        the contribution of the quantity (when included in the fit) to
        the overall $\chi^2$. Right hand part: raw predictions
        from QCD FA without constraints from data. In both configurations, 
        the CKM parameters obtained from the standard CKM fit are
        also included. Branching fractions are given in units of $10^{-6}$.} 
\end{center}
\end{table}
\begin{table}[t]
\begin{center}
\setlength{\tabcolsep}{0.0pc}
{\footnotesize
\begin{tabular*}{\textwidth}{@{\extracolsep{\fill}}lccccccccccccccc}\hline
&&&&&&&&&&&&&&& \\[-0.2cm]
                        & $\Cpipi$ & ${\cal A}^{+0}_{\pi\pi}$ & ${\cal A}^{+-}_{K\pi}$ & ${\cal A}^{0+}_{K\pi}$ & ${\cal A}^{+0}_{K\pi}$ & $\Ckspiz$ & $\Skspiz$ & $\BR^{+-}_{\pi\pi}$ & $\BR^{+0}_{\pi\pi}$ & $\BR^{00}_{\pi\pi}$ & $\BR^{+-}_{K\pi}$ & $\BR^{0+}_{K\pi}$ & $\BR^{+0}_{K\pi}$ & $\BR^{00}_{K\pi}$ \\[0.1cm]
$\Spipi$                &$  -0.44 $&$  -0.45 $&$  -0.39 $&$  +0.10 $&$  -0.07 $&$  -0.15 $&$  +0.39 $&$  +0.04 $&$  -0.06 $&$  +0.08 $&$  +0.07 $&$  -0.28 $&$  +0.34 $&$  -0.43 $\\
$\Cpipi$                &$  +1.00 $&$  +0.47 $&$  +0.79 $&$  -0.49 $&$  +0.55 $&$  -0.35 $&$  -0.21 $&$  +0.48 $&$  +0.01 $&$  -0.29 $&$  -0.22 $&$  +0.22 $&$  -0.09 $&$  +0.05 $\\
${\cal A}^{+0}_{\pi\pi}$&  -       &$  +1.00 $&$  +0.23 $&$  +0.02 $&$  -0.24 $&$  +0.43 $&$  +0.13 $&$  +0.20 $&$  +0.36 $&$  +0.11 $&$  -0.10 $&$  +0.08 $&$  -0.28 $&$  +0.26 $\\
${\cal A}^{+-}_{K\pi}$  &  -       &  -       &$  +1.00 $&$  -0.66 $&$  +0.72 $&$  -0.47 $&$  -0.21 $&$  -0.02 $&$  +0.01 $&$  +0.02 $&$  -0.00 $&$  +0.19 $&$  +0.05 $&$  +0.05 $\\
${\cal A}^{0+}_{K\pi}$  &  -       &  -       &  -       &$  +1.00 $&$  -0.74 $&$  +0.71 $&$  +0.13 $&$  -0.04 $&$  -0.07 $&$  -0.03 $&$  +0.16 $&$  -0.10 $&$  -0.08 $&$  +0.13 $\\
${\cal A}^{+0}_{K\pi}$  &  -       &  -       &  -       &  -       &$  +1.00 $&$  -0.93 $&$  -0.14 $&$  +0.03 $&$  +0.05 $&$  +0.04 $&$  -0.12 $&$  +0.13 $&$  +0.31 $&$  -0.29 $\\
$\Ckspiz$               &  -       &  -       &  -       &  -       &  -       &$  +1.00 $&$  -0.11 $&$  +0.07 $&$  +0.02 $&$  +0.00 $&$  -0.15 $&$  +0.10 $&$  +0.35 $&$  -0.37 $\\
$\Skspiz$               &  -       &  -       &  -       &  -       &  -       &  -       &$  +1.00 $&$  +0.13 $&$  +0.75 $&$  +0.66 $&$  +0.09 $&$  -0.24 $&$  -0.09 $&$  +0.06 $\\
$\BR^{+-}_{\pi\pi}$     &  -       &  -       &  -       &  -       &  -       &  -       &  -       &$  +1.00 $&$  -0.01 $&$  -0.51 $&$  -0.00 $&$  +0.26 $&$  +0.05 $&$  +0.09 $\\
$\BR^{+0}_{\pi\pi}$     &  -       &  -       &  -       &  -       &  -       &  -       &  -       &  -       &$  +1.00 $&$  +0.83 $&$  +0.04 $&$  +0.03 $&$  -0.21 $&$  +0.29 $\\
$\BR^{00}_{\pi\pi}$     &  -       &  -       &  -       &  -       &  -       &  -       &  -       &  -       &  -       &$  +1.00 $&$  +0.14 $&$  -0.12 $&$  -0.10 $&$  +0.17 $\\
$\BR^{+-}_{K\pi}$       &  -       &  -       &  -       &  -       &  -       &  -       &  -       &  -       &  -       &  -       &$  +1.00 $&$  +0.74 $&$  +0.57 $&$  +0.57 $\\
$\BR^{0+}_{K\pi}$       &  -       &  -       &  -       &  -       &  -       &  -       &  -       &  -       &  -       &  -       &  -       &$  +1.00 $&$  +0.46 $&$  +0.63 $\\
$\BR^{+0}_{K\pi}$       &  -       &  -       &  -       &  -       &  -       &  -       &  -       &  -       &  -       &  -       &  -       &  -       &$  +1.00 $&$  -0.27 $\\[0.1cm]
\hline
\end{tabular*}
}
\end{center}
\vspace{-0.3cm}
\caption[.]{\label{tab:qcdfaPrediCorr} \em
        Linear correlation coefficients for the QCD FA fit results
        given in Table~\ref{tab:qcdfaPredi}. The ${\cal A}^{ij}_{pq}$ 
        stand for the direct \CP-asymmetry parameters $A_{\CP}(p^iq^j)$,
        and the $\BR^{ij}_{pq}$ denote the corresponding branching fractions.
        Note that these correlations are not of experimental origin, but due 
        to the uncertainties in the theoretical parameters. }
\end{table}

\subsection{Data driven Predictions for the $\pi\pi$ and $K\pi$ System}
\label{sec:globalQCDfit}

\begin{figure}[p]
  \centerline{\epsfxsize8.1cm\epsffile{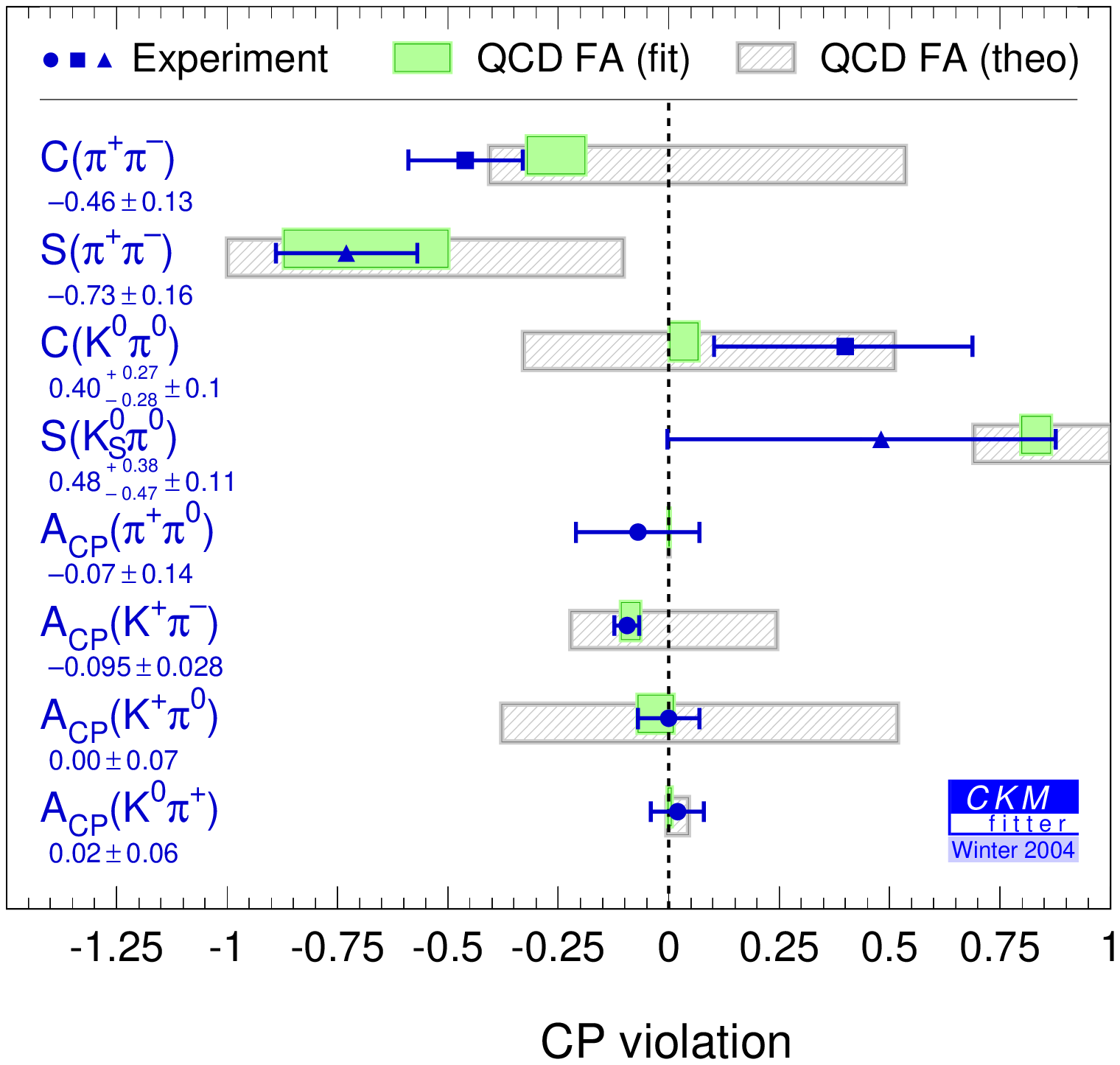}
              \epsfxsize8.1cm\epsffile{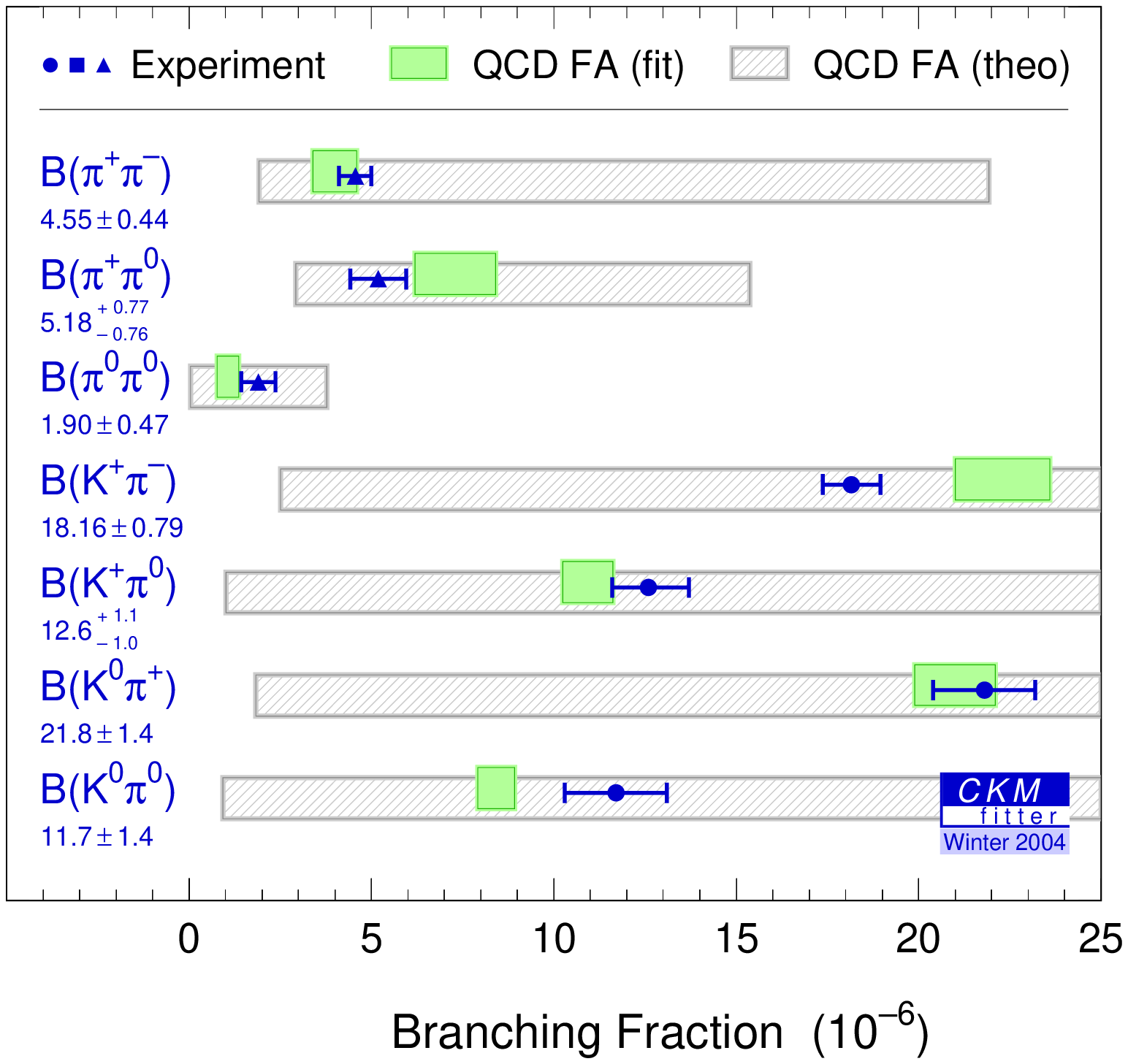}}
  \vspace{-0.2cm}
  \caption[.]{\label{fig:pipikpi_qcdfa_cpbr}\em
        Comparison of the results from the global QCD FA fit to 
        $\B\to\pi\pi,\,K\pi$ data (shaded boxes) and the unconstrained
        QCD FA predictions with experiment for the \CP-violating asymmetries
        (left) and the branching fractions (right) in $B\to h h^\prime$
        ($h,h^\prime=\pi,K$) decays. The predictions are obtained
        ignoring the measurement associated with the observable 
        in the fit. The experimental results are the world averages quoted
        in Table~\ref{tab:BRPiPicompilation} and the theory values are those
        from Table~\ref{tab:qcdfaPredi}. All theory predictions use 
        the standard CKM fit result as input. The error 
        ranges shown correspond to $1\sigma$.}
  \vspace{1.0cm}
  \centerline{\epsfxsize8.1cm\epsffile{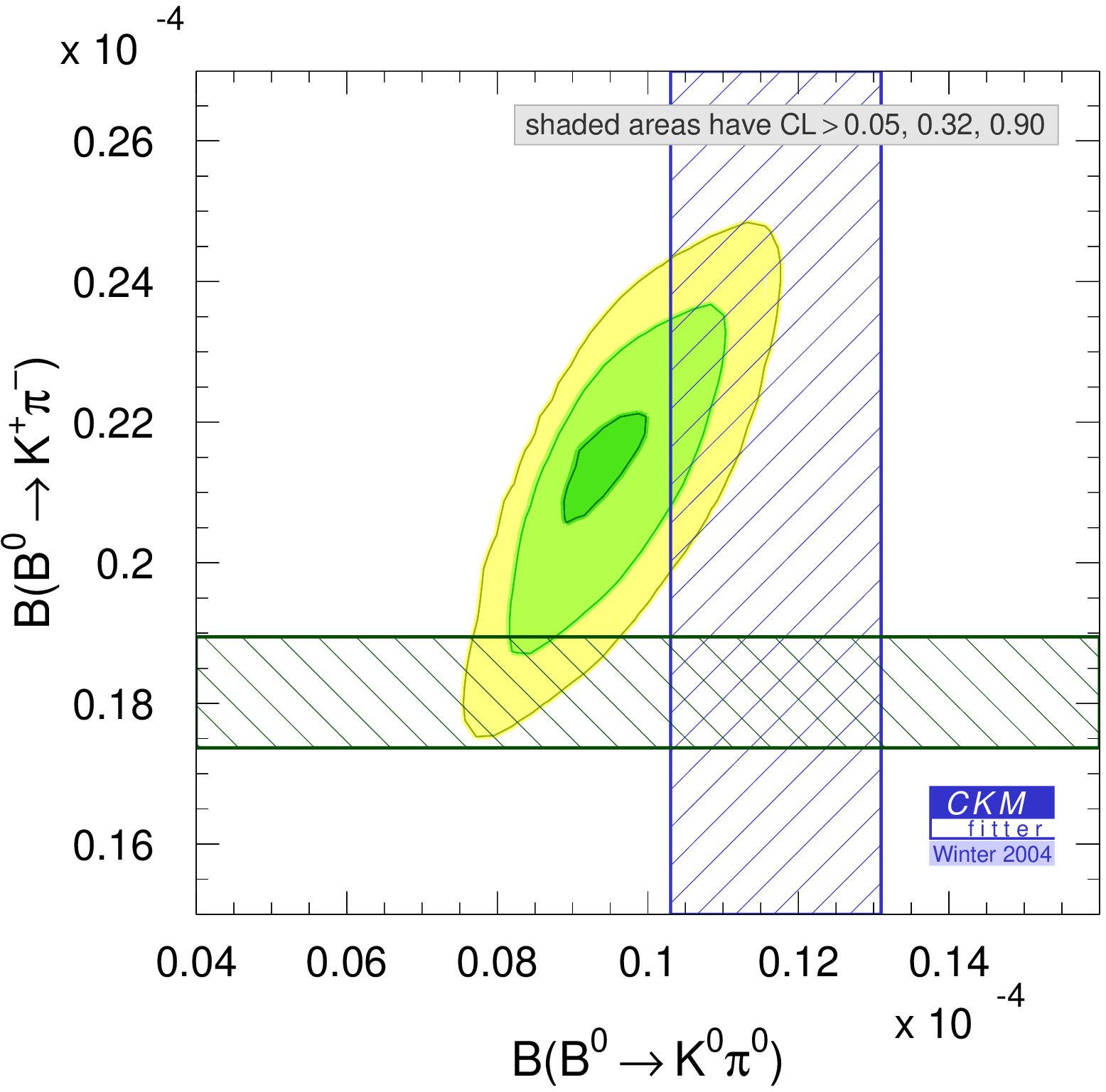}}
  \vspace{-0.0cm}
  \caption[.]{\label{fig:kpi_br_2d}\em
        Confidence level in the $(\BR(\Kz\piz),\BR(\Kp\pim))$ plane obtained 
        from the global QCD FA fit to $\B\to\pi\pi,\,K\pi$ data ignoring the 
        measurements associated with the observables shown.
        Dark, medium and light shaded areas have $CL>0.90$,
        $0.32$ and $0.05$, respectively. The hatched areas indicate 
        the $1\sigma$ error bands of the experimental results.}
\end{figure}

In the spirit of likelihood projections, we study the predictions of
the combined $\pi\pi,\,K\pi$ QCD FA fit on each observable, ignoring 
the measurement associated with the observable in the fit. The results
are hence unbiased by the actual measurement. In addition to the
charmless  data we include the standard CKM fit here. The predictions
obtained are  summarized in Table~\ref{tab:qcdfaPredi} and plotted in
comparison with  the experimental values in
Fig.~\ref{fig:pipikpi_qcdfa_cpbr}. Also given  are the results obtained
from QCD FA and the standard CKM fit alone, for  which in most cases
the uncertainty largely exceeds the experimental precision\footnote
{
        The predictions of Ref.~\cite{BN} from the same inputs
        appear to be much more
        precise, due to the treatment of the uncertainties on the
        theoretical parameters,
        which the authors vary independently and finally add in
        quadrature. On the contrary, our approach \rfit\ amounts to scan
        democratically the whole parameter space. While the (commonly
        found) approach of Ref.~\cite{BN} likely underestimates the
        overall uncertainty, \rfit\ may overestimate it, by ignoring
        possible fine-tuning configurations where many theoretical
        parameters take extreme values in the allowed ranges.
}. Instead when the fit is constrained by the experimental data, the combined
constraints determine rather precisely the parameters that are only
loosely bounded by the theory.  The
predictions of the full QCD FA fit are accurate and found to be  in
good agreement with the measurements, with the exception of the  above
mentioned $\BR(\Bz \to \Kp\pim)$ and $\BR(\Bz \to \Kz\piz)$, for which
however the discrepancy does not exceed $2.5\sigma$ at present. Due to
the common uncertainties on the theoretical
parameters, the results  for the branching fractions exhibit
significant correlations, which have to  be taken into account when
interpreting the results. For completeness, we give the linear
correlation coefficients evaluated with toy Monte
Carlo simulation\footnote
{
        The procedure is as follows: for each toy experiment, the 
        experimental observables are fluctuated within their Gaussian
        experimental errors; the full QCD FA fit is performed, and the
        variance between the fit results for the observables is computed.
        The coefficients quoted in Table~\ref{tab:qcdfaPrediCorr}
        correspond to 500 toy experiments.
} in 
Table~\ref{tab:qcdfaPrediCorr}. As an example, the CLs
of $\BR(\Bz \to \Kp\pim)$ versus $\BR(\Bz \to \Kz\piz)$ are plotted in 
Fig.~\ref{fig:kpi_br_2d} (the correlation coefficient is $+0.57$) and 
compared with the measurements. A potential increase in the  
experimental value for $\BR(\Bz \to \Kp\pim)$ (due to radiative corrections
that were previously neglected, see the remark in the introduction to 
Part~\ref{sec:charmlessBDecays}) could help to reconcile theory and 
experiment. One observes a significant positive correlation between the 
direct \CP-violation parameters in $\pip\pim$ and $\Kp\pim$ decays, 
which is expected from SU(3) symmetry. We note in addition that
\bei

\item   the prediction for the not yet measured direct asymmetry
        in the $\Bz \to \pi^0\pi^0$ decay is 

       \beq
         \Cpizpiz = 0.06 ^{\,+0.10}_{\,-0.12} \left[ ^{\,+0.37}_{-0.24} \right]~,
       \eeq
        where the errors outside (inside) brackets are at $1\sigma$ ($2\sigma$);

\item   a large negative value for $\Spipi$ is predicted, in contrast to
        the small value $\sin2\alpha$ (see Table~\ref{tab:fitResults1}) that would be obtained
        from the standard CKM fit in the no-penguin case;

\item   the deviation between $\stbeff\equiv\Skspiz$ measured in
        $\Bz\to\KS\piz$
        and the charmonium reference is $\stbeff-\stb=0.09\pm0.04$; 

\item   the experimental evidence for negative direct \CP violation 
        in $\Kp\pim$ is consistent with the (precise) expectation;

\item   the qualitative picture (hierarchy) of the branching fractions is 
        understood in the SM.

\eei
Despite this success and given that both experimental and theoretical 
uncertainties are still large, particular care is mandatory when analyzing 
possible anomalies in $\B\to K\pi$ decays. 
In Section~\ref{sec:charmlessBDecays}.\ref{sec:introductionKPi} we revisit 
the $K\pi$ system with the use of a more phenomenological approach.


\section{Phenomenological Analysis of $\B\to K\pi$ Decays}
\label{sec:introductionKPi}
%
%

The decays $B\to K\pi$ have received considerable attention in the 
recent literature~\cite{GRKpi03,BFRS1,BFRS2,KpiBarger} since a fit
to the data leads to an apparent violation of the approximate sum rule
derived in Refs.~\cite{LipkinSR}. Although the errors remain large, it has been argued
by the  authors of Ref.~\cite{BFRS1,BFRS2,KpiBarger} that a better phenomenological
description  could be achieved by including non-standard contributions
in electroweak penguins, that is, $\Delta I=1$ $b\to s$ transitions.
\vs
In this section we discuss the implications of the experimental results 
on strong isospin symmetry in $B \to  K \pi$ decays, by performing fits
to the data under various dynamical hypotheses. We interpret the numerical results
by comparing them to those of the $\pi\pi$ modes, and give our understanding 
of the present situation.

\subsection{Experimental Input}
\label{InputsKPi}

We use the branching fractions and charge asymmetries for the $B \to  K \pi$ 
modes given in Table~\ref{tab:BRPiPicompilation} as inputs to our fits. 
The \CP-averaged branching fractions are defined by
\beqn
        \BR^{ij} &\propto& 
        \frac{\tau_{B^{i+j}}}{2}\left(|\Aij|^2 + |\Abij|^2\right)~,
\eeqn
and the four \CP-violating asymmetries by
\beqn
        {\cal A}^{ij} &=& 
        \frac{|\Abij|^2 - |\Aij|^2}{|\Abij|^2 + |\Aij|^2}~,
\eeqn
where $(i,j)=(+,-)~,(0,+)~, (+,0)~, (0,0)$ and $i+j$ is the charge of the 
$B$ meson. Note that ${\cal A}^{0+}$ is zero by definition in some of
the approximations considered below. The \CP asymmetries
$\Ckspiz=-{\cal A}^{00}$ and $\Skspiz$ are 
defined by
\beqn 
\label{KPiC00S00}
        \Ckspiz &=& 
                \frac{1-|\lambda_{ \KS \piz}|^2}        
                     {1+|\lambda_{ \KS \piz}|^2}~,\\[0.3cm]
        \Skspiz &=& 
                \frac{2 {\rm Im} \lambda_{ \KS \piz}}
                     {1+|\lambda_{\KS \piz}|^2}~,
\eeqn 
where 
$\lambda_{\KS \piz}= -\exp\left\{i\arg[(V_{td}V_{tb}^*)^2]\right\}\Azzb/\Azz$ 
in our phase convention.

\subsection{Transition Amplitudes}
\label{amplitudeKPi}

Using the unitarity relation~(\ref{eq:unitarity}) and adopting
convention $\C$ (\cf\  Section~\ref{sec:charmlessBDecays}.\ref{par:amplitude}), 
each 
$B \to  K^i \pi^j$ decay amplitude can be parameterized by 
$V_{us}V_{ub}^*$, $V_{ts}V_{tb}^*$ and two complex quantities denoted 
$T^{ij}$ and $P^{ij}$. For example, for $\Bz\to \Kp\pim$ one has
\beq
        A^{+-} \equiv  A(B^0 \to  \Kp \pim) 
        =   V_{us}V_{ub}^* T^{+-} + V_{ts}V_{tb}^* P^{+-}~, 
\eeq
and similarly for the other modes.
The amplitudes $T^{ij}$ and $P^{ij}$ implicitly include strong phases
while the weak phases are explicitly contained in the CKM factors.
An important difference with respect to the $\pi\pi$ modes is that the CKM
ratio $|V_{ts}V_{tb}^*/(V_{us}V_{ub}^*)|\sim 50$ enhances considerably the
contribution of loops with respect to tree topologies: this implies a
potentially better sensitivity to unknown virtual particles, and thus to
New Physics, but at the same time this involves more complicated
hadronic dynamics.
\vs
Without loss of generality, 
the complete $B \to K\pi$ system can be parameterized by eight 
amplitudes and the CKM couplings $V_{us}V_{ub}^*$ and 
$V_{ts}V_{tb}^*$. In the following, we will assume 
isospin symmetry, so that there is a quadrilateral relation between
these amplitudes.

\subsection{Isospin Relations and Dynamical Scenarios}
\label{par:IsoKPi}

Using strong isospin invariance, the $B\to K \pi$ amplitudes satisfy 
the relations~\cite{NQmethod}
\beqn
\label{eq:isospinKPi}
        \Azp  + \sqrt{2} \Apz   &=& \sqrt{2} \Azz  + \Apm ~, \\
        \Azpb + \sqrt{2} \Apzb  &=& \sqrt{2} \Azzb + \Apmb~.
\eeqn
Note that for the \CP-conjugate amplitudes one reads for instance
$\Azpb \equiv A(\Bm \to  \Kzb \pim)$ and accordingly for the other 
charges.

\subsubsection{Neglecting Electroweak Penguin but not Annihilation
Diagrams}

\begin{figure}[t]
  \centerline
        {
        \epsfxsize8.1cm\epsffile{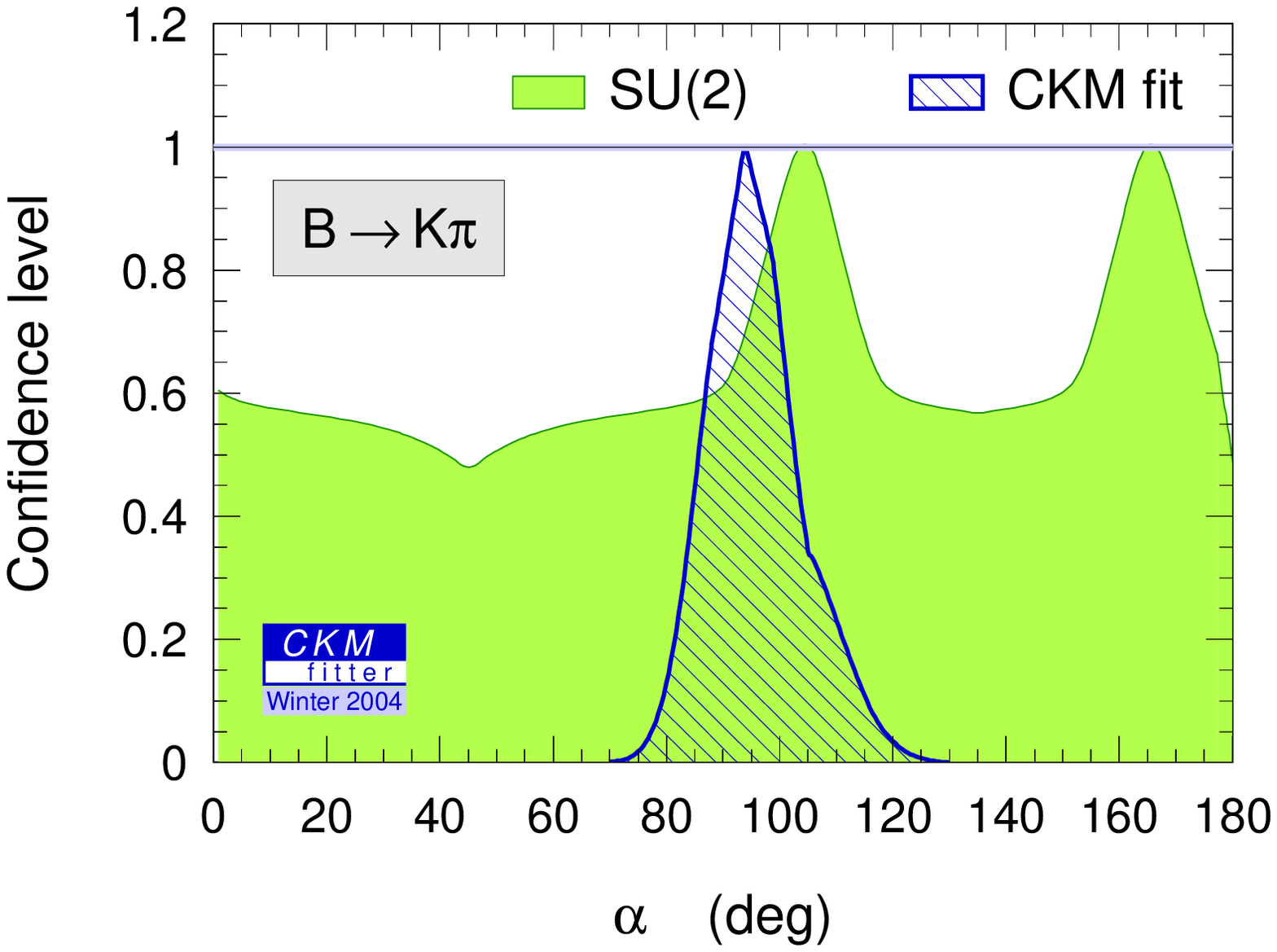}
        \epsfxsize8.1cm\epsffile{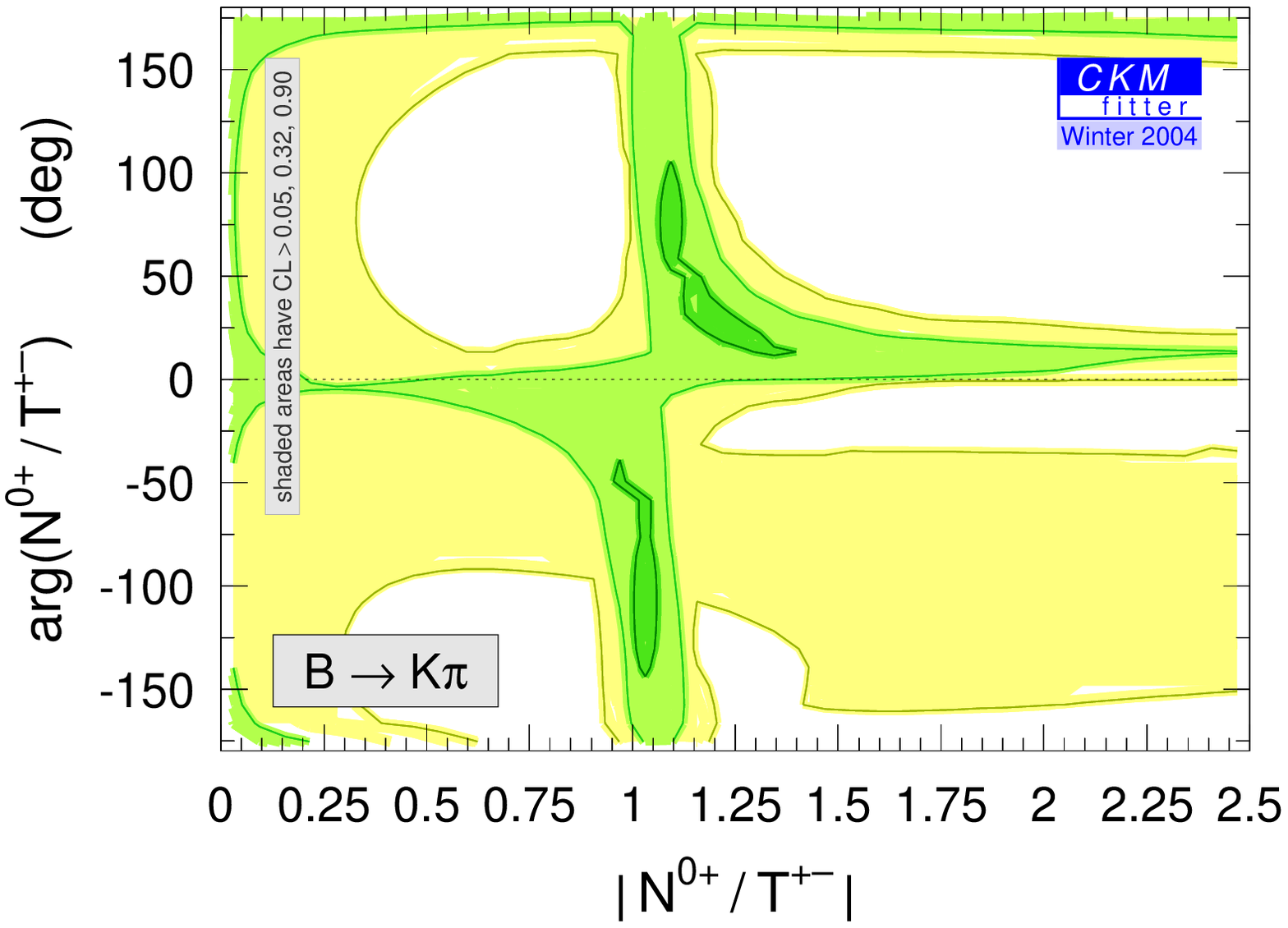}
        }
  \vspace{-0.5cm}
  \caption[.]{\label{KpiAlpha}  \em
        \underline{Left:} 
        confidence level as a function of the CKM angle $\alpha$,
        in the Nir--Quinn approximation (no electroweak penguins). The
        hatched region represents the constraint from the standard CKM
        fit. \underline{Right:} 
        constraint on the annihilation-to-tree amplitude
        ratio $|N^{0+}/T^{+-}|$, as defined in Eq.~(\ref{eq:bKpiIso}).
        Electroweak penguins are neglected.}
\end{figure}

In the absence of electroweak penguin diagrams the isospin analysis 
leads to two additional constraints since the two quadrilaterals 
share the $I=3/2$ amplitude as common diagonal, with a length 
determined from the branching fractions, while the second 
diagonals bisect each other~\cite{NQmethod}:
\beqn
\label{eq:isospinKPiDiag}
        \Apm + \sqrt{2} \Azz            
        &=& \tilde{A}^{+-}+ \sqrt{2} \tilde{A}^{00}~, \\
\label{eq:isospinKPiLenght}
        \sqrt{2} \Azz+ \sqrt{2} \Apz  
        &=& \sqrt{2} \tilde{A}^{00}+ \sqrt{2} \tilde{A}^{+0}~,
\eeqn
where $\tilde{A}= \exp\left[2i\arg(V_{us}V_{ub}^*)\right] \,\Abar$.
The argument goes as follows: since 
gluonic penguins are $\Delta I=0$ transitions, in the absence
of electroweak penguins  the amplitude
$A_{\Delta I=1, I_f=3/2}$ of the transition from $I_i=1/2$ to $I_f=3/2$
is proportional to $V_{us}V_{ub}^*$ and hence
\beq
        A_{1,3/2} \:=\: \tilde{A}_{1,3/2}~.
\eeq
Note that Eqs.~(\ref{eq:isospinKPiDiag}) and
(\ref{eq:isospinKPiLenght})  separately hold for the $T^{ij}$ and
$P^{ij}$. Under these assumptions it is possible to describe the full
$B \to  K \pi$ system with four  out of the seven complex quantities
$T^{ij}$ and $P^{ij}$. In the following  parameterization, we use
$P^{+-}$ and the three tree amplitudes  $T^{+-}$, $N^{0+}=T^{0+}$ and
 $T_\mathrm{C}^{00}=T^{00}$. The
notation $N^{0+}$ refers to the fact that the tree contribution to the
$\Kz\pip$ mode has an annihilation topology (it also receives
contributions from long-distance $u$ and $c$ penguins). Since
$B^0\to \Kz\piz$ is  color-suppressed, its tree amplitude is denoted $T_\mathrm{C}^{00}$.
\beqn
   \Apm     &=&    V_{us}V_{ub}^*\,  T^{+-}
                            + V_{ts}V_{tb}^*\, P^{+-}~,\nonumber\\
  \Azp              &=&  V_{us}V_{ub}^*\,  N^{0+} 
                        -V_{ts}V_{tb}^*\, P^{+-}~,\nonumber\\
\label{eq:bKpiIso}
\sqrt{2}  \Apz      &=&   V_{us}V_{ub}^* \,( T^{+-} + T_\mathrm{C}^{00} -N^{0+} )
                        +V_{ts}V_{tb}^*\, P^{+-}~,\\
\label{eq:b0pi0K0iso}
  \sqrt{2} \Azz     &=&    V_{us}V_{ub}^*\,  T_\mathrm{C}^{00}
                            - V_{ts}V_{tb}^*\, P^{+-}~.\nonumber
\eeqn
In the absence of electroweak penguins, it is possible to  invert the
expressions for the amplitudes and to extract the eight unknown
quantities: $|V_{us}V_{ub}^* ~T^{+-}|$, $|V_{us}V_{ub}^* ~N^{0+}|$,
$|V_{us}V_{ub}^* ~T_\mathrm{C}^{00}|$, 
$|V_{ts}V_{tb}^* ~P^{+-}|$, three relative strong phases and the 
weak phase $\alpha$ from the experimental observables\footnote
{
        The dependence with respect to $\alpha$ comes
        from the interference between $V_{us}V_{ub}^*$ in the $\Delta I=1$
        amplitude with $V_{td}V_{tb}^*$ in the $\Bz\Bzb$ mixing.
} (the so-called Nir--Quinn
method~\cite{NQmethod}). However, as was stressed in Ref.~\cite{lavoura}, 
the discrete ambiguity problem is even more delicate
than in the $\pi\pi$ case, because the relative angles between the
amplitudes are not well constrained by the quadrilateral construction.
\vs
The constraint obtained on the angle $\alpha$ is shown in
Fig.~\ref{KpiAlpha}. Although it peaks near the value from the
standard CKM fit, the constraint is weak. Very large
statistics would be required for a meaningful determination of $\alpha$
by this method. More interesting, perhaps, is the constraint on the
annihilation-to-emission ratio, represented by the quantity
$|N^{0+}/T^{+-}|$ given on the right hand plot of Fig.~\ref{KpiAlpha}: 
although this ratio is expected to be suppressed from the point of view 
of QCD factorization (see the next Section), large values (of order one)
cannot be excluded. Note that large contributions from annihilation 
topologies, if extrapolated to $\Bp\to \Kp\Kzb$ in the SU(3) limit, would
eventually enter in conflict with the experimental bounds~\cite{BN}.
\begin{figure}[t]
  \centerline
        {
        \epsfxsize8.1cm\epsffile{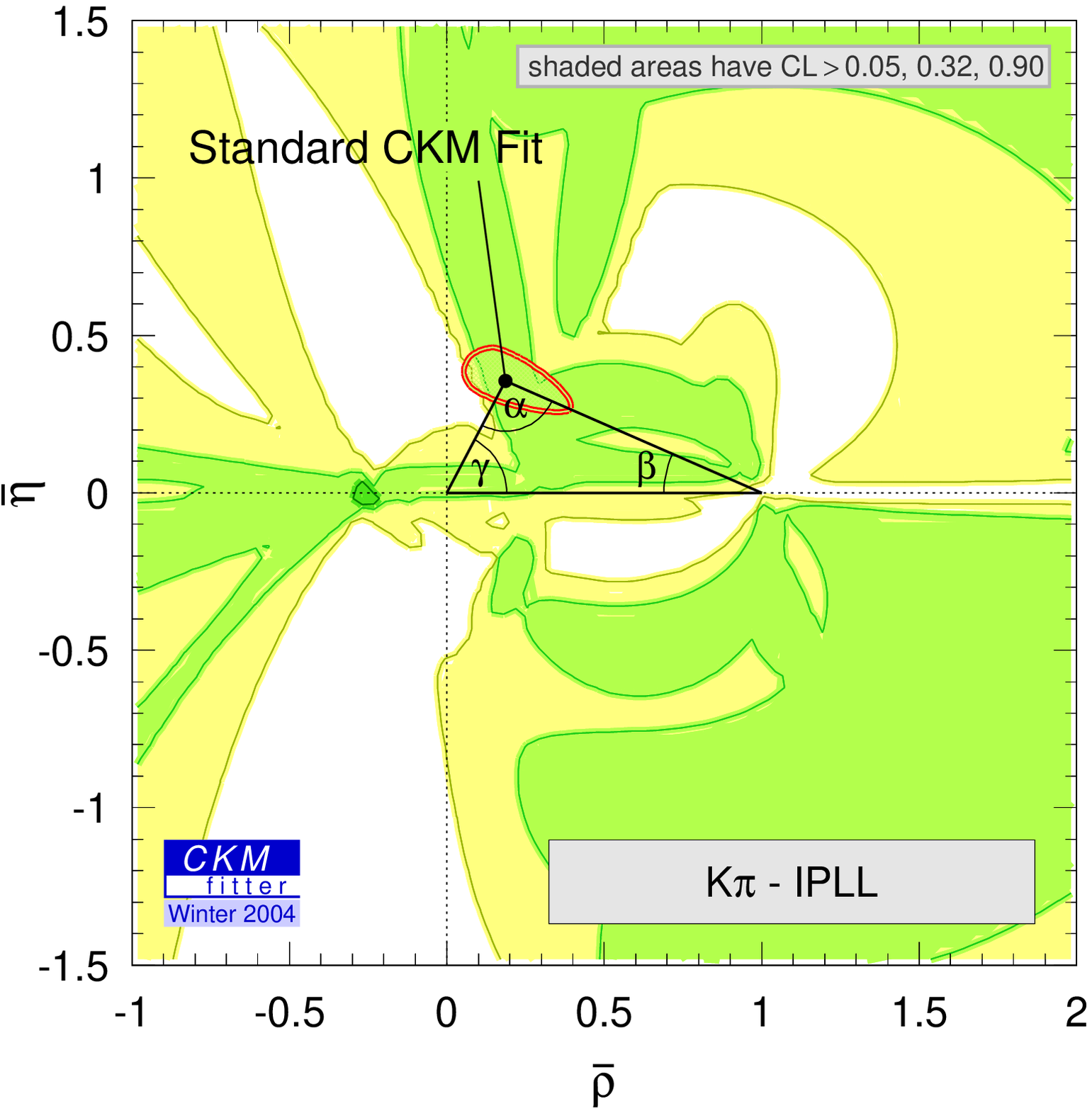}
        \epsfxsize8.1cm\epsffile{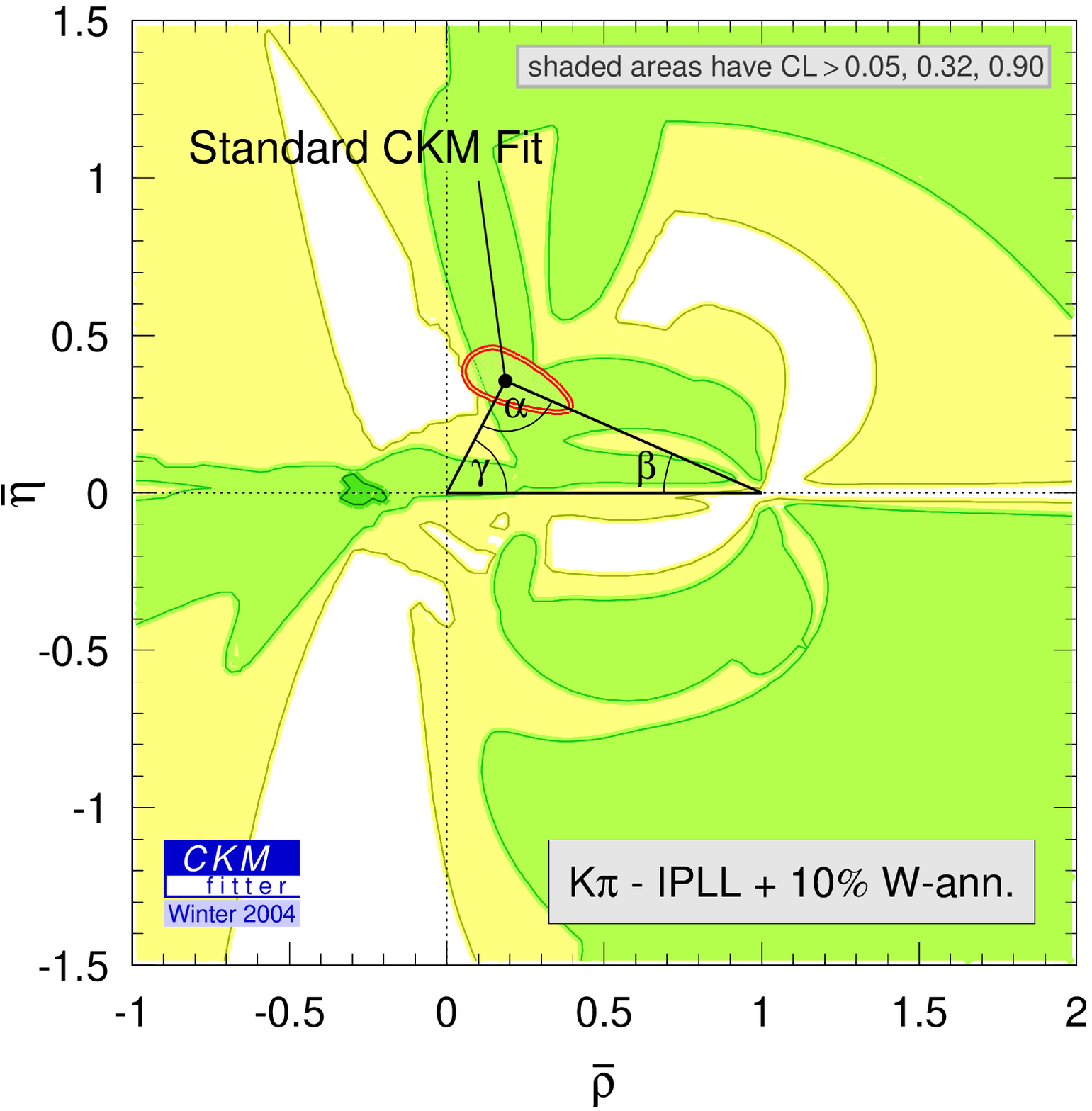}
        }
  \vspace{-0.0cm}
  \caption[.]{\label{thekpiRhoEtaPlots} \em
        Confidence level in the $\rhoeta$ plane for the isospin analysis 
        of $B\to K \pi$ decays within the framework of
        Ref.~\cite{london} (see text). In the left hand plot $N^{0+}$ is
        set to zero, while in the right hand plot it is allowed to vary
        freely within 10\% of the $T^{+-}$ dominant amplitude.  
        Dark, medium and light shaded areas have $CL>0.90$, 
        $0.32$ and $0.05$, respectively. Also shown on each plot is 
        the prediction from the standard CKM fit.}
\end{figure}
\begin{figure}[t]
  \vspace{0.5cm}
  \centerline
        {
        \epsfxsize8.1cm\epsffile{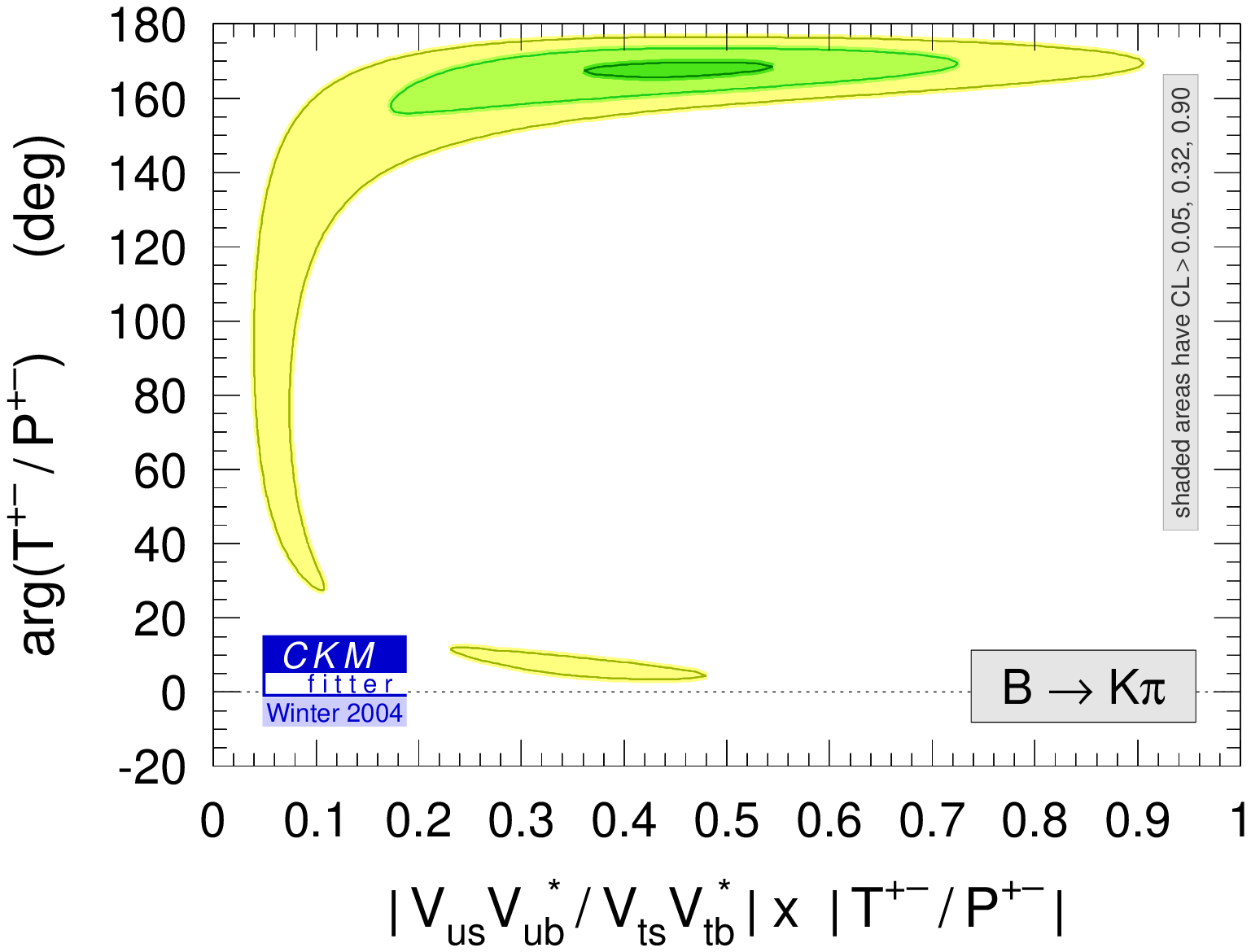}
        \epsfxsize8.1cm\epsffile{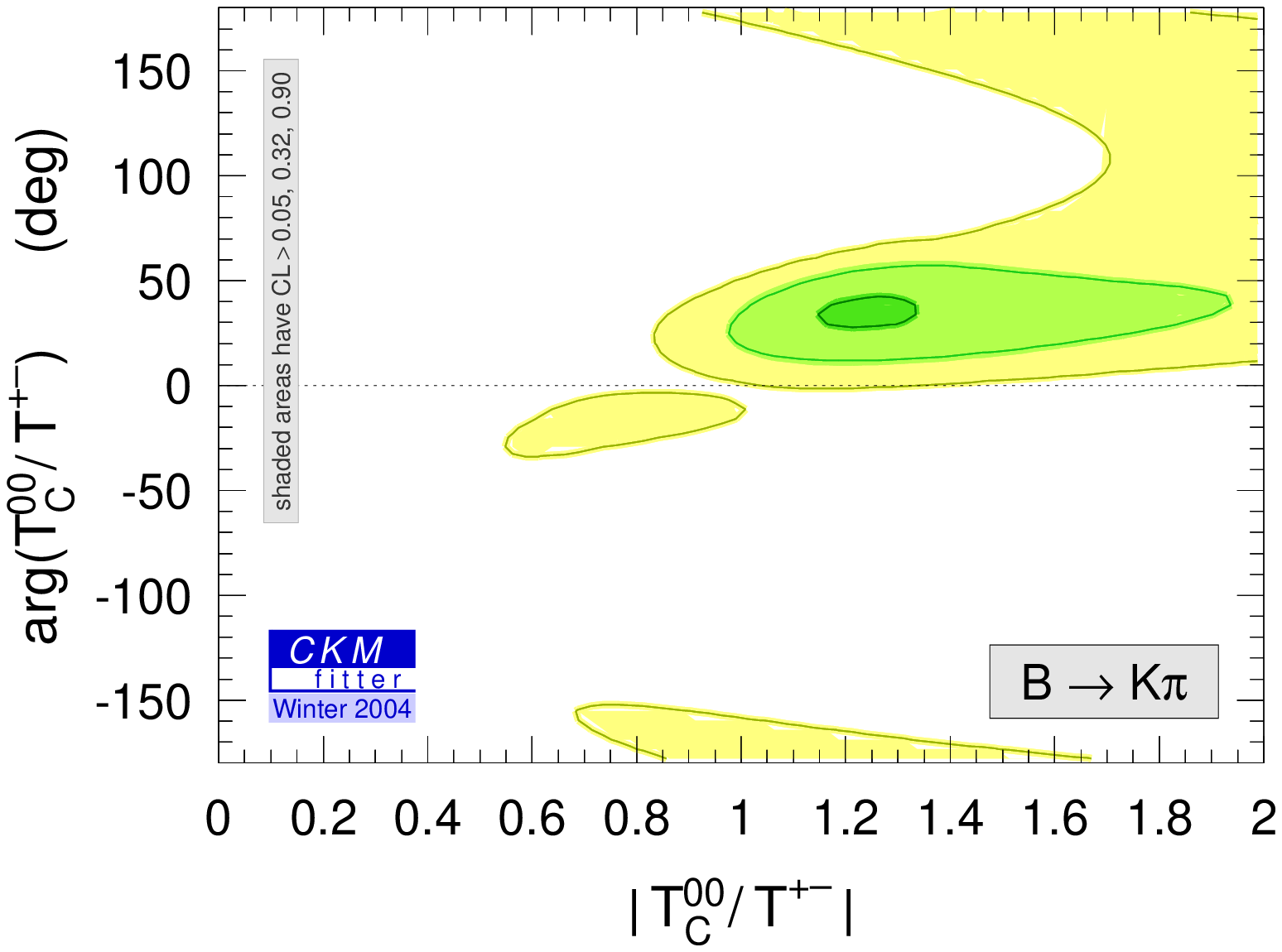}
        }
  \vspace{-0.5cm}
  \caption[.]{\label{thekpidpotPlots} \em
        Constraints on the complex tree-to-penguin amplitude ratio in 
        $\Bz\to \Kp\pim$ decays (left), and the complex 
        color-suppressed-to-color-allowed amplitude ratio (right),
        obtained when using as additional input the CKM parameters $\rhobar$ and 
        $\etabar$ from the standard CKM fit. The gradually shaded regions 
        give the CLs within the framework of Ref.~\cite{london}: dark, medium and 
        light shaded areas have $CL>0.90$, $0.32$ and $0.05$, respectively. }
\end{figure}

\subsubsection{Standard Model Electroweak Penguins and Vanishing 
        Annihilation Topologies}
\label{sec:IPLL}

Since the pioneering work of Nir and Quinn it has been realized that
gluonic penguins are likely to dominate in $B\to K\pi$, and that
CKM-enhanced electroweak penguins would even compete with
CKM-suppressed $T$-type amplitudes~\cite{messGronauEtAl}. Thus one
can add two $\Delta I=1$ amplitudes $P^\mathrm{EW}$ and 
$P^\mathrm{EW}_\mathrm{C}$ that come with the CKM factor\footnote
{
        The $\Delta I=0$ contribution from electroweak penguins can 
        be absorbed in the $P^{+-}$ amplitude.
} $V_{ts}V_{tb}^*~$:
\beqn
   \Apm     &=&    V_{us}V_{ub}^*\,   T^{+-}
                   + V_{ts}V_{tb}^*\,  P^{+-}~,\nonumber\\
  \Azp              &=&  V_{us}V_{ub}^*\,   N^{0+}
                      +  V_{ts}V_{tb}^*\, (-P^{+-}+ P^\mathrm{EW}_\mathrm{C})~,  
                        \nonumber\\
\label{eq:Kpigeneral}
\sqrt{2}  \Apz      &=&   V_{us}V_{ub}^*\,  ( T^{+-} + T_\mathrm{C}^{00}
-N^{0+})
                        + V_{ts}V_{tb}^*\, (P^{+-}+ P^\mathrm{EW} -P^\mathrm{EW}_\mathrm{C})~,\\
  \sqrt{2} \Azz     &=&    V_{us}V_{ub}^*\,   T_\mathrm{C}^{00}
                         + V_{ts}V_{tb}^*\, (-P^{+-} + P^\mathrm{EW})~,\nonumber
\eeqn
where $P^\mathrm{EW}_\mathrm{C}$ is expected to be color-suppressed with respect to
$P^\mathrm{EW}$.
\vs
The general parameterization~(\ref{eq:Kpigeneral}) involves 11 hadronic
parameters and the CKM couplings, which cannot be extracted from the
nine independent $K\pi$ observables. In the SU(3) symmetry limit, the
$P^\mathrm{EW}$ amplitude can be expressed model-independently in terms of the
sum $T^{+-}+T_\mathrm{C}^{00}$, just the same way as for 
$\pi\pi$~\cite{NRPew}: this removes two hadronic parameters, which is 
however not yet
enough to close the system. Hence \textit{without an additional
dynamical assumption one cannot extract a correlation in the
$(\rhobar,\etabar)$ plane from the $K\pi$ observables alone.}
\vs
Faced with this problem the authors of Ref.~\cite{london}  proposed 
to neglect all exchange and annihilation topologies and showed that in
principle the apex of the Unitarity Triangle can be determined, up to
discrete ambiguities (see also Ref.~\cite{GRKaPiReview} for a short
review of other approaches analyzing the $K\pi$ system). The additional 
hypotheses are
\bei
\item   negligible annihilation and long-distance penguin
        contributions to $\Bp\to\Kzb\pip$:
        \beq
        \label{ann}
                N^{0+} = 0~.
        \eeq
\item   SU(3) limit for the color-allowed electroweak penguin amplitude:
        \beq
        \label{PEW}
                P^\mathrm{EW} =  
                        R^+ \left(T^{+-} + T_\mathrm{C}^{00}\right)~.
        \eeq
\item   SU(3) limit and negligible exchange contributions
        for the color-suppressed electroweak penguin amplitude:
        \beq
        \label{PEWC}
                P^\mathrm{EW}_\mathrm{C} =
                        \frac{R^+}{2} \left(T^{+-} + T_\mathrm{C}^{00}\right) -
                        \frac{R^-}{2} \left(T^{+-} - T_\mathrm{C}^{00}\right)~.
        \eeq
\eei
In the above equations $R^+$ and $R^-$ are constants given by\footnote
{
        The numerics for $R^+$ and $R^-$ has been worked out as
        in the $\pi\pi$ case~(\ref{EWPratio}), while the 
        correlation between them is neglected. Equation~(\ref{PEWC}) 
        has been first derived in Ref.~\cite{GPY}.
}
\beqn
\label{R+R-}
        R^+     &=&     -\frac{3}{2}\frac{c_9+c_{10}}{c_1+c_2}
                 \;=\;  +(1.35 \pm 0.12)~10^{-2}~,\nonumber\\
        R^-     &=&     -\frac{3}{2}\frac{c_9-c_{10}}{c_1-c_2}
                 \;=\;  +(1.35 \pm 0.13)~10^{-2}~.
\eeqn
The phenomenological fit is thus expressed in terms of $T^{+-}$,
$P^{+-}$, $T^{00}_\mathrm{C}$ and $\rhoeta$, that is five hadronic
parameters and two CKM parameters. Figure~\ref{thekpiRhoEtaPlots} (left) 
shows the constraints on the unitarity plane obtained within this approach. 
The intricate shape of the CLs is mainly due to the convolution of the 
Nir--Quinn constraints on the CKM angle $\alpha$ and the explicit CKM 
dependence of the electroweak penguins in Eqs.~(\ref{PEW})-(\ref{PEWC}).
\vs
Using the same framework and the standard CKM fit as an additional 
input, we have performed a constrained fit of the tree-to-penguin and 
color-suppressed-to-leading-tree amplitude ratios (Fig.~\ref{thekpidpotPlots}
left and right, respectively). Because of the $3.3\sigma$ evidence of 
negative direct \CP asymmetry in $\Bz\to \Kp\pim$, the relative 
tree-to-penguin phase (left hand plot) is positive. It is surprising 
that the measurements seem to indicate that the expected double CKM suppression
of the tree-to-penguin ratio is well compensated by a large ratio of
the hadronic matrix elements (Fig.~\ref{thekpidpotPlots}, left), which 
tends to contradict the trend observed in the $\pi\pi$ system. We 
further discuss this point in 
Section~\ref{sec:charmlessBDecays}.\ref{sec:unbazarcettesection}. Another
striking feature of the fit results, which has been already observed in
$\piz\piz$ versus $\pip\pim$, is the  value of the color-suppressed
amplitude as compared to the leading-tree one (Fig.~\ref{thekpidpotPlots}, 
right): order one is preferred, and a zero value for this amplitude is 
excluded.

\subsubsection{Including a Correction to the No-Rescattering Assumption}

To correct the assumption of a negligible $V_{us}V_{ub^*}$ term in 
$\Bp\to \Kz\pip$, we attempt to include an estimate for it
into the amplitude parameterization. Model-dependent contributions 
to this term have been evaluated in the QCD FA formalism~\cite{BBNS,BN} 
and are found to be around $10 \%
$ in magnitude with respect to 
the leading $T^{+-}$ amplitude. Since this estimate
is fairly uncertain\footnote
{
        Annihilation topologies are expected to be suppressed by 
        $\Lambda_\mathrm{QCD}/m_b$. The authors of 
        Ref.~\cite{BBNS} estimate them from a
        hard-scattering point of view, which results in a
        stronger,  model-dependent suppression proportional to
        $\as\Lambda_\mathrm{QCD}/m_b$.
},
we assign a $100\%
$ theoretical error to $|N^{0+}|$ and let its phase $\delta^{0+}$
vary in the fit:
\beqn 
\label{eq:IPLLcorr}
        N^{0+} &=& (0.1 \pm 0.1) ~ |T^{+-}| ~ e^{i \delta^{0+}}~.
\eeqn
The transition amplitudes then read as in Eq.~(\ref{eq:Kpigeneral}).
Note that in this model, the expressions for the electroweak penguin
remain unchanged with respect to Eqs.~(\ref{PEW})--(\ref{PEWC}). While this
is true for $P^\mathrm{EW}$ (up to a small correction coming from $Q_{7,8}$
operators), it is incorrect for $P^\mathrm{EW}_\mathrm{C}$,
which receives a contribution from exchange topologies that cannot be
expressed in terms of the $K\pi$ amplitudes alone~\cite{GPY}. We assume
that the effect of this approximation is negligible.
\vs
The constraints obtained in the unitarity plane are shown on the right
hand plot of Fig.~\ref{thekpiRhoEtaPlots}. The relaxed framework does 
not lead to major differences with respect to the $N^{0+}=0$ hypothesis
(left hand plot of Fig.~\ref{thekpiRhoEtaPlots}). We stress that no 
significant constraint is obtained on the phase $\delta^{0+}$, and that 
the fit converges systematically towards the maximal allowed value for 
$\left|N^{0+}\right|$ in Eq.~(\ref{eq:IPLLcorr}). With much improved experimental 
accuracy, the $N^{0+}=0$ assumption will become crucial to obtain
meaningful constraints on $\rhobar$ and $\etabar$.

\subsubsection{Constraining the Electroweak Penguins}
\begin{figure}[t]
  \centerline
        {
        \epsfxsize8.1cm\epsffile{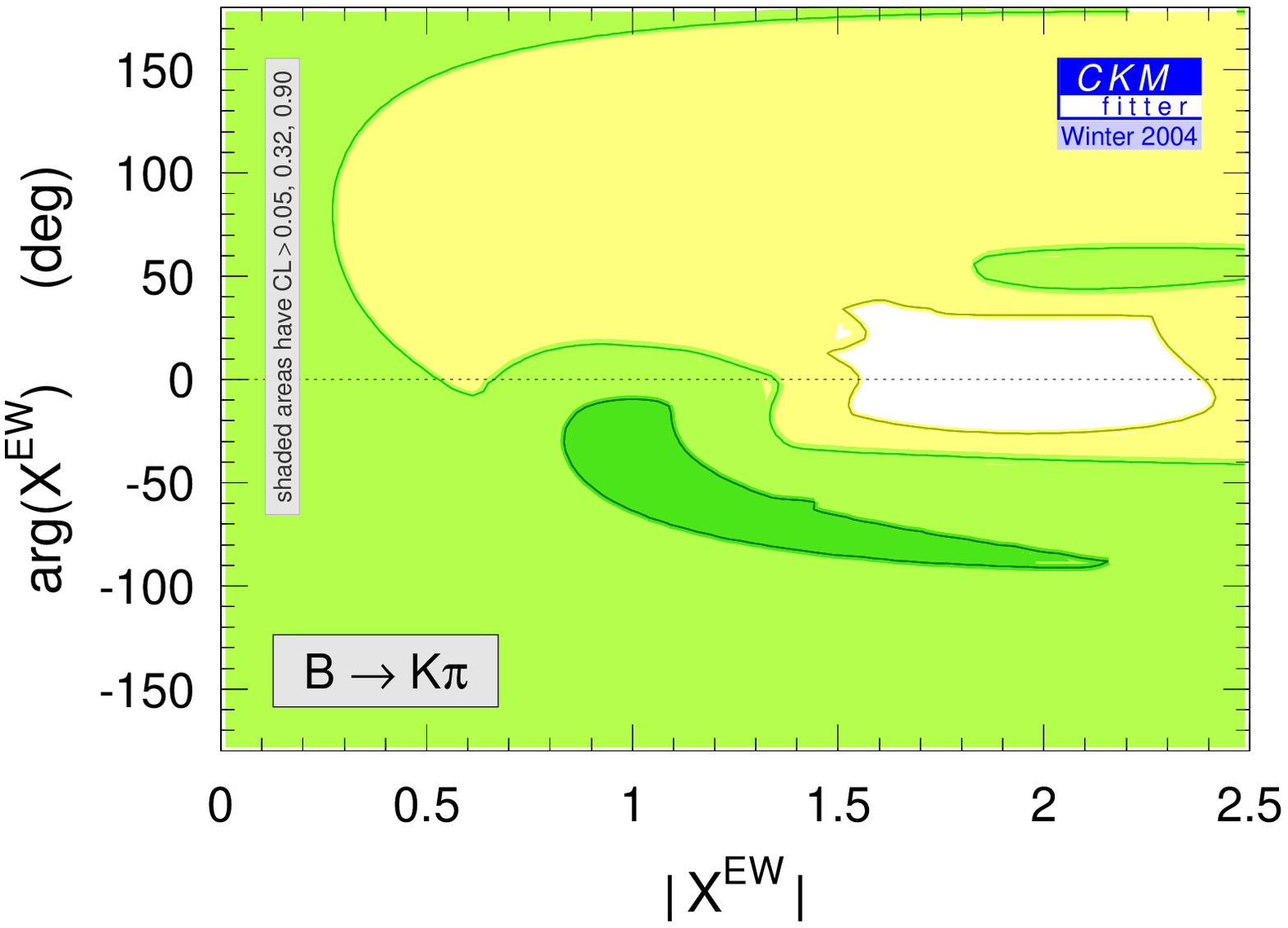}
        \epsfxsize8.1cm\epsffile{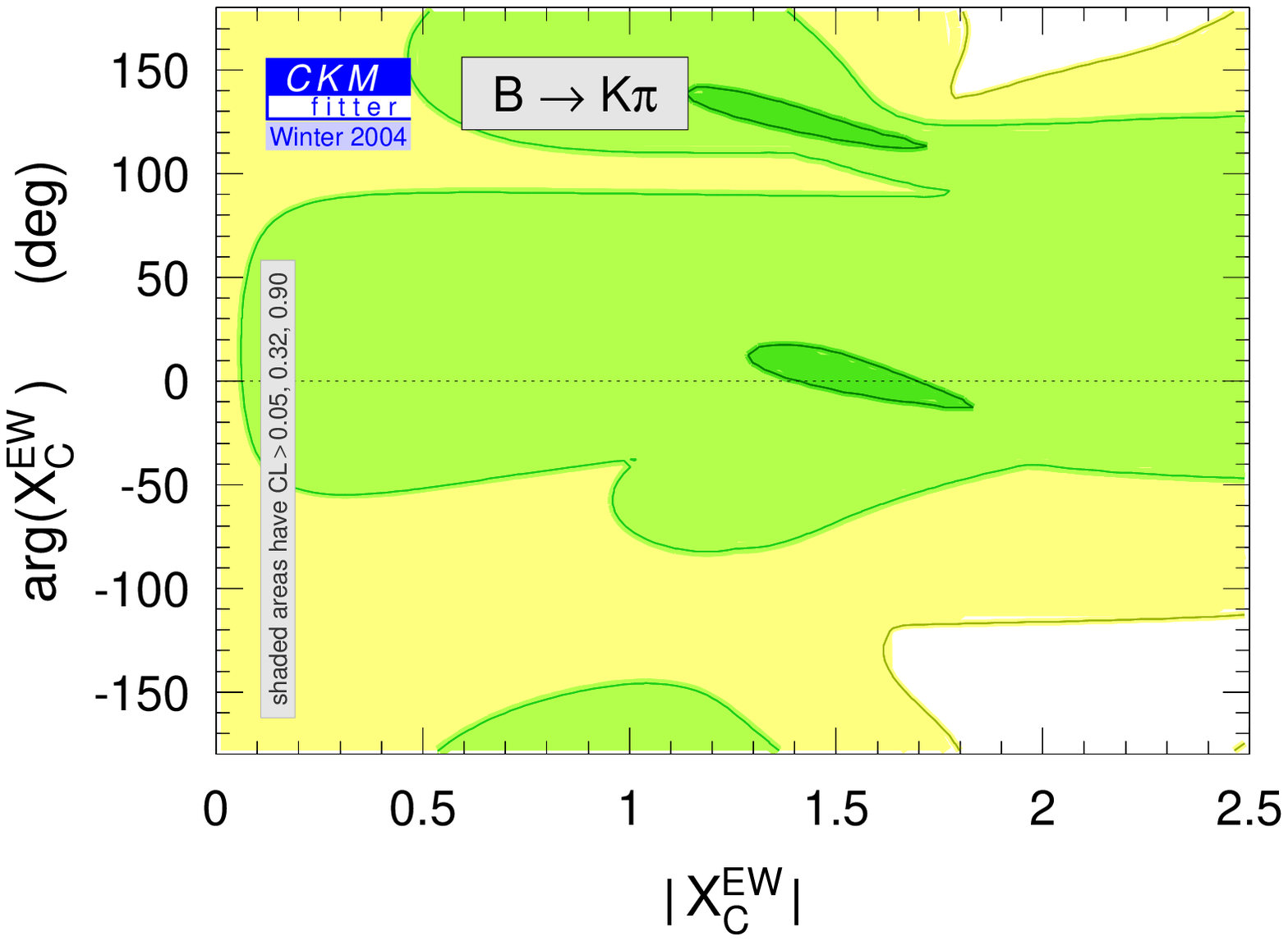}
        }
  \vspace{-0.5cm}
  \caption[.]{\label{thekpiXewPlots} \em
        Constraints on the complex quantities $X^\mathrm{EW}$ (left, where
        $X^\mathrm{EW}_\mathrm{C}$ is fixed to 1) and  
        $X^\mathrm{EW}_\mathrm{C}$ (right, where
        $X^\mathrm{EW}$ is fixed to 1) defined in Eq.~(\ref{eq:kpi_Xew})
        and  (\ref{eq:kpi_XewC}), respectively. Dark, medium and light 
        shaded areas have $CL>0.90$, $0.32$ and $0.05$, respectively.
        The standard CKM fit is used as input to obtain these plots. }
\end{figure}
To investigate whether the data together with the standard CKM fit can 
reveal indirect evidence of electroweak penguin contributions, we correct
Eqs.~(\ref{PEW})--(\ref{PEWC}) by introducing two new complex parameters 
$X^\mathrm{EW}$ and $X^\mathrm{EW}_\mathrm{C}$
\beqn
\label{eq:kpi_Xew}
        P^\mathrm{EW}   &=& 
                X^\mathrm{EW}\left[ R \left(T^{+-} + T_\mathrm{C}^{00}\right)\right]~,\\
\label{eq:kpi_XewC}
        P^\mathrm{EW}_\mathrm{C}        &=& 
                X^\mathrm{EW}_\mathrm{C}\left[ R T_\mathrm{C}^{00}\right]~,
\eeqn
where we have imposed $R^+=R^-\equiv R$. Within the SM and
under the assumptions already explicitly stated, we expect 
$X^\mathrm{EW}\approx X^\mathrm{EW}_\mathrm{C}\approx 1$. We introduce 
two new fit scenarios: a first where $X^\mathrm{EW}\equiv1$ and 
$X^\mathrm{EW}_\mathrm{C}$ is free to vary in the fit (magnitude {\em and} 
phase), and a second where conversely $X^\mathrm{EW}_\mathrm{C}\equiv1$ 
and $X^\mathrm{EW}$ is let free. 
\vs
The CLs (using the standard CKM fit as input) found for
$X^\mathrm{EW}$  (left) and $X^\mathrm{EW}_\mathrm{C}$ (right) are
given in  Fig.~\ref{thekpiXewPlots}. The constraints on both electroweak
penguin  contributions are marginal and essentially any value of the
relevant  parameters can accommodate the data. This said, one notices
that the  standard values $X^\mathrm{EW}=X^\mathrm{EW}_\mathrm{C}=1$
correspond to large CL regions. We also point out that current
data do not  particularly favor a zero value for the strength of the
color-suppressed electroweak  penguin $X^\mathrm{EW}_\mathrm{C}$.
Hence, from our point of view, it is not justified in
Eq.~(\ref{eq:Kpigeneral}) to neglect  $P^\mathrm{EW}_\mathrm{C}$ while
keeping $P^\mathrm{EW}$, as is done by the authors of Ref.~\cite{BFRS1,BFRS2}:
both electroweak penguin amplitudes may be comparable in
magnitude.
 
\subsection{$K\pi$ Observables from $\pi\pi$ Hadronic Parameters}

In Section~\ref{sec:charmlessBDecays}.\ref{sec:IPLL} we have found that the present data point towards
noticeably different values for the hadronic parameters $T^{+-}$, $P^{+-}$
and $T^{00}_\mathrm{C}$ in the $K\pi$ system, compared to the
$\pi\pi$ one, whereas they are expected to be equal in the
SU(3) limit, if annihilation and exchange topologies are negligible.
Another manifestation of the same trend has been explored in
Ref.~\cite{BFRS1,BFRS2}, where the authors compute the $K\pi$ observables with 
the use of the hadronic parameters found in fits to $\pi\pi$ decays.
The pattern obtained that way differs from the one 
observed in the $K\pi$ data. We repeat this exercice with \ckmfitter, 
using the following ratios of \CP-averaged branching fractions\footnote
{
        The ratio $R_{nc}$ is denoted $R$ in Ref.~\cite{BFRS1,BFRS2}.
}
\beqn
\label{defR_BFRS}
        R_{nc} &=& \frac{\tau_{\Bp}}{\tau_{\Bz}}\,
                   \frac{\BR(\Bz\to \Kp\pim)+\BR(\Bzb\to \Km\pip)}
                        {\BR(\Bp\to \Kz\pip)+\BR(\Bm\to\Kzb\pim)}
            \;=\; 0.91^{\,+0.08}_{\,-0.07}
                ~~[1.2\sigma]~,\nonumber\\
        R_{n } &=& \frac{1}{2}\frac{\BR(\Bz\to \Kp\pim)+\BR(\Bzb\to \Km\pip)}
                            {\BR(\Bz\to\Kz\piz)+\BR(\Bzb\to\Kzb\piz)}
            \;=\; 0.78^{\,+0.11}_{\,-0.09}
                ~~[1.8\sigma]~,\\
        R_{c } &=& 2 \frac{\BR(\Bp\to \Kp\piz)+\BR(\Bm\to \Km\piz)}
                       {\BR(\Bp\to\Kz\pip)+\BR(\Bm\to\Kzb\pim)}
            \;=\; 1.16^{\,+0.13}_{\,-0.11}
                ~~[1.4\sigma]~,\nonumber
\eeqn
where the numbers in brackets indicate the departure (in  standard
deviations) from one, the value predicted by gluonic  penguin
dominance. Note in this context that the ratio of two Gaussian
quantities (like branching fractions)
does {\em not} have a Gaussian probability density
(see, \eg, the discussion in Appendix C of Ref.~\cite{CKMfitter}).
\vs
We assume in the following the same (strong) hypotheses as in 
Ref.~\cite{silvwolf,BFRS1,BFRS2}, namely exact SU(3) symmetry and neglect of 
all annihilation and exchange topologies. This allows us to identify
\begin{figure}[t]
  \centerline
        {
        \epsfxsize8.1cm\epsffile{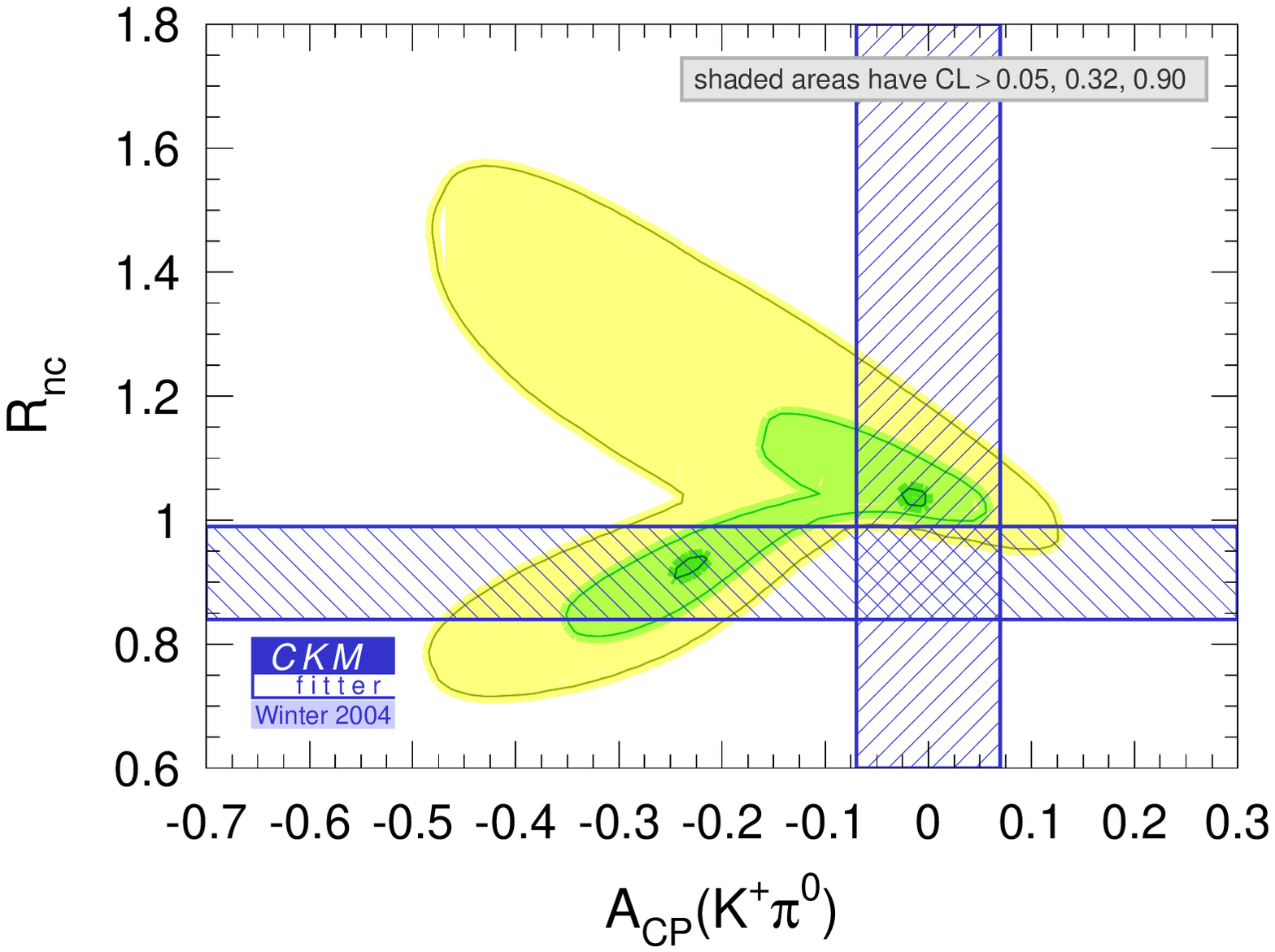}
        \epsfxsize8.1cm\epsffile{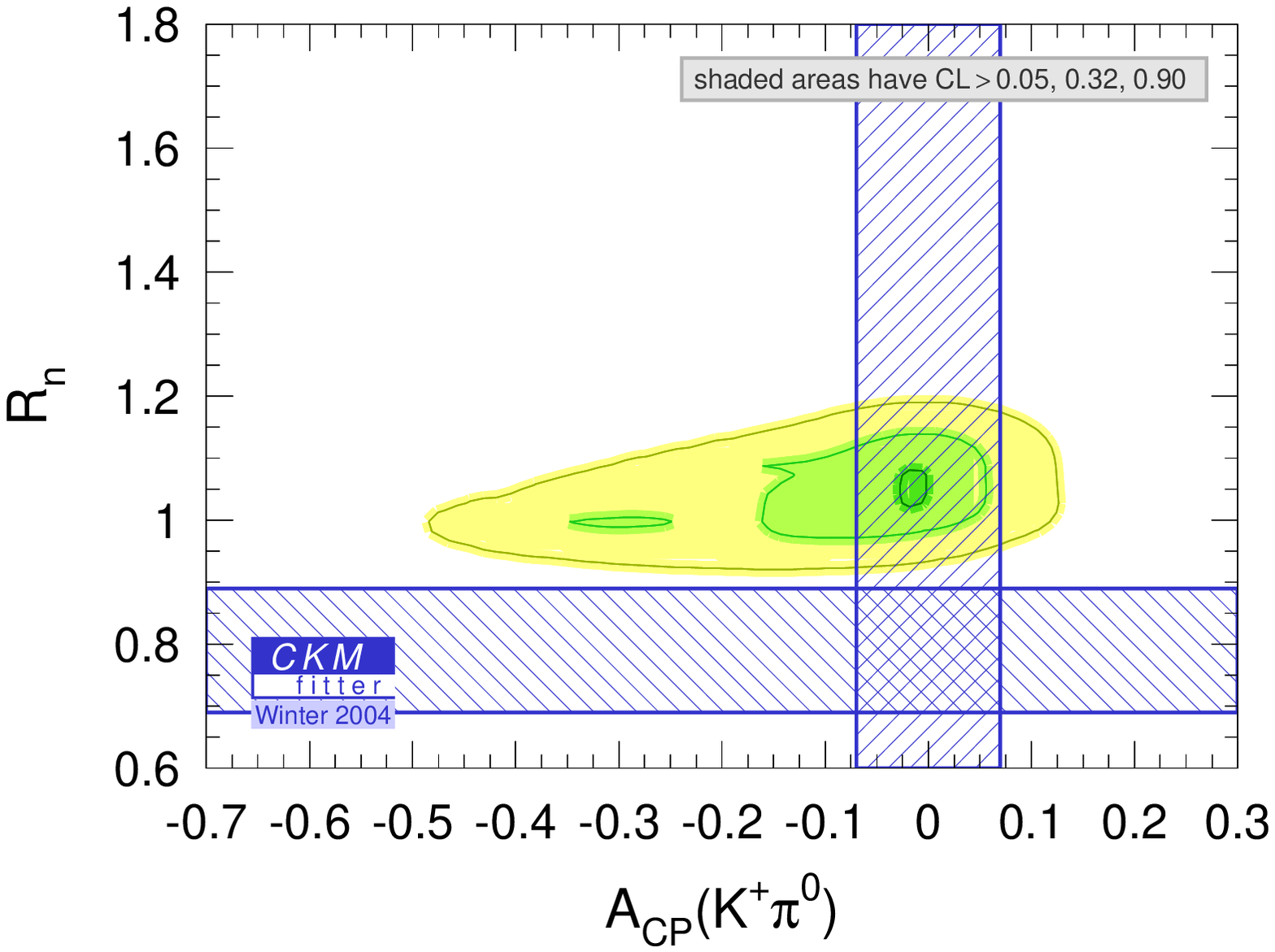}
        }
  \vspace{0.0cm}
  \centerline
        {
        \epsfxsize8.1cm\epsffile{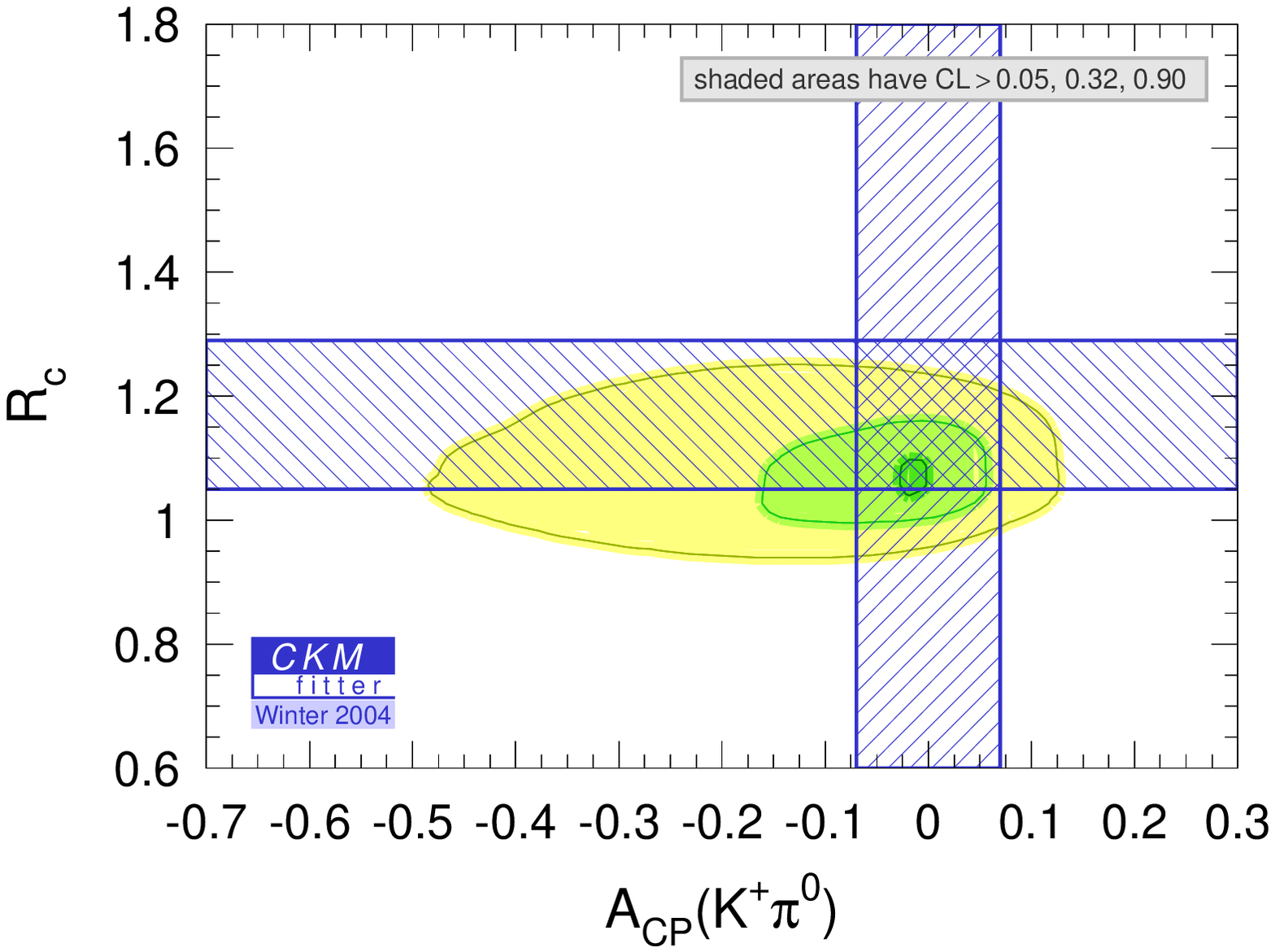}
        \epsfxsize8.1cm\epsffile{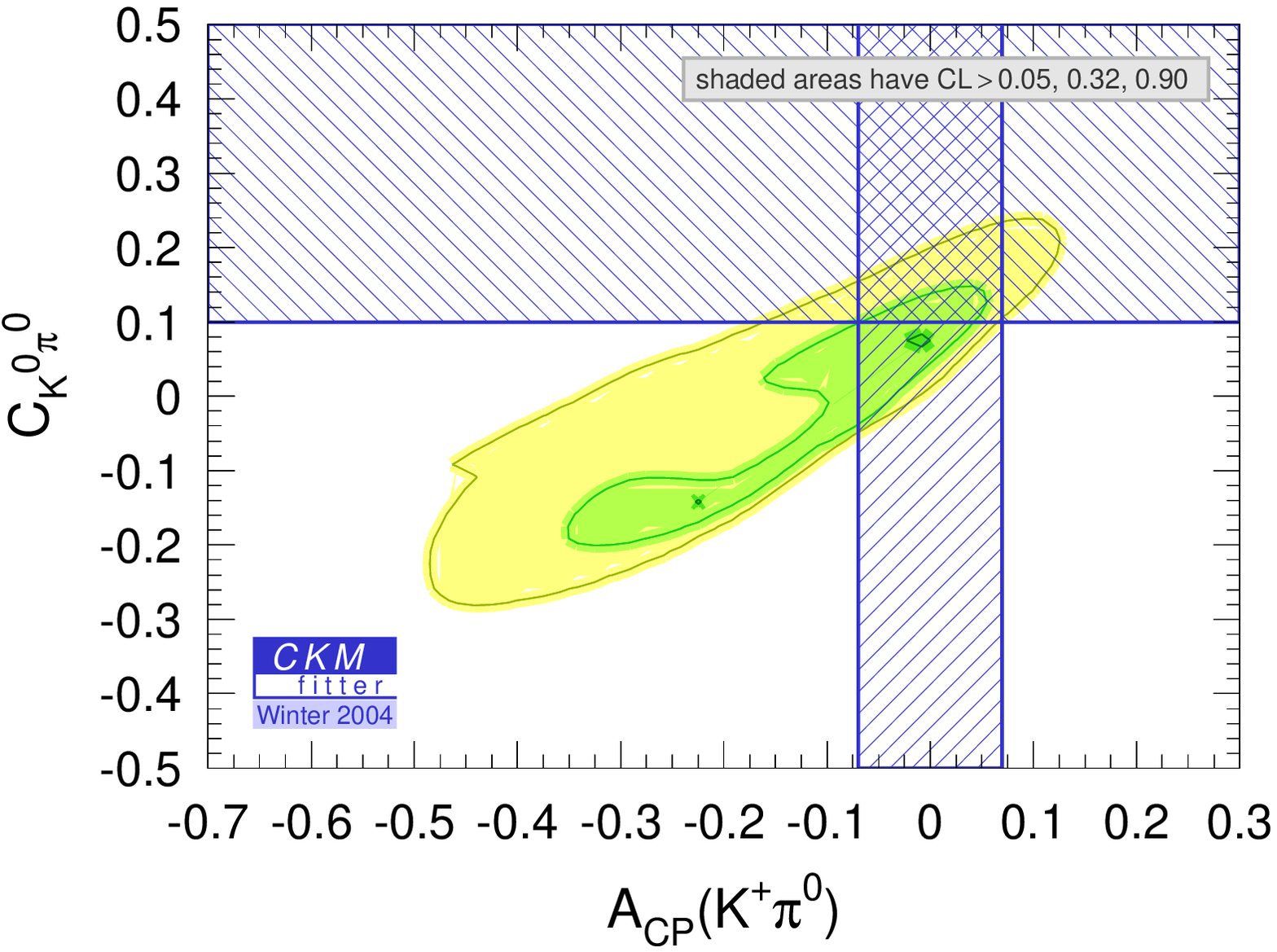}
        }
  \vspace{-0.0cm}
  \caption[.]{\label{fig:KpiUsePipi}\em
        Ratios of branching fractions~(\ref{defR_BFRS}) versus the
        direct \CP asymmetry parameter in $\Bp\to\Kp\piz$ decays as
        predicted from $\B\to\pi\pi$ and the standard CKM fit,
        assuming SU(3) flavor symmetry and neglecting all 
        annihilation and exchange diagrams. Dark, medium and light 
        shaded areas have $CL>0.90$, $0.32$ and $0.05$, respectively.
        The hatched bands indicate the $1\sigma$ regions of the 
        corresponding measurements. (See also Fig.~\ref{fig:bskk}).}
\end{figure}
\beq
\label{crazy}
        T_{K\pi}^{+-}=T_{\pi\pi}^{+-}~,\ \ \ \ 
        P_{K\pi}^{+-}=P_{\pi\pi}^{+-}~,\ \ \ \ 
        T_{K\pi}^{00}=T_{\pi\pi}^{00}~,
\eeq
while electroweak penguins remain estimated according to
Eqs.~(\ref{PEW})--(\ref{PEWC}). The $\pi\pi$ tree and penguin
amplitudes are extracted from the corresponding data using the standard
CKM fit, along  the line described in
Section~\ref{sec:charmlessBDecays}.\ref{sec:pipidpot}.  The $K\pi$
observables evaluated that way are shown on Fig.~\ref{fig:KpiUsePipi}, 
where the $R_{nc}$, $R_{n}$ and $R_{c}$ ratios, as well as the \CP
asymmetry  $C_{\Kz\piz}$, are represented as functions of the \CP
asymmetry  $A_{\CP}(\Kp\piz)$. The overall normalization of the
branching fractions as given by, \eg, $\BR(\Kp\pim)$, and the \CP
asymmetry $A_{\CP}(\Kp\pim)$  can be read off
Fig.~\ref{fig:bskk}\footnote
{
        As already pointed out, in the absence of annihilation and exchange 
        topologies, the amplitudes for $\Bz\to\Kp\pim$ and $\Bs\to\Kp\Km$ 
        are equal: hence one may read $\Bz\to\Kp\pim$
        instead of $\Bs\to\Kp\Km$ on Fig.~\ref{fig:bskk}.
}.
The experimental values are indicated by the hatched $1\sigma$ error
bands.
\vs
The two-fold discrete ambiguity (Fig.~\ref{fig:KpiUsePipi})
corresponds to the two possible solutions for the phase of 
$T^{00}_\mathrm{C}/T^{+-}$ in the $\pi\pi$ system (see
Fig.~\ref{fig:dtotpipi}). The negative one for the latter is 
preferred by the measurement of $A_{\CP}(\Kp\piz)$, which is 
consistent with zero. While there is agreement between the predicted 
values of $R_c$ and $C_{\Kz\piz}$ and the corresponding measurements,
a larger $R_n$ is found, which emphasizes the 
somewhat large branching fraction of $\Bz\to\Kz\piz$ (and 
confirms the findings of Refs.~\cite{LipkinSR,BN,GRKpi03,BFRS1,BFRS2}).
As a consequence, we 
find that the predicted $R_{nc}$ exceeds the measurement, which is 
due to the contribution of $P^\mathrm{EW}_\mathrm{C}$. Putting the
latter amplitude to zero like Ref.~\cite{BFRS1,BFRS2} would decrease the
central value of
$R_{nc}$ to~$\sim 0.94$ (while not changing the two other ratios),
 in better agreement with the experimental
value.
\vs
Despite the intriguing aspect of the plots in Fig.~\ref{fig:KpiUsePipi},
we stress that the present experimental accuracy does not allow 
us to draw definite conclusions. To assess the overall compatibility, 
we have performed a global $\pi\pi$ and $K\pi$ fit with the 
assumptions~(\ref{crazy}) and with the input of the standard CKM fit,
resulting in a decent p-value of $25\%
$: this is significantly better than the $1.6\%
$ compatibility ($2.4\sigma$) that is found from the evaluation of an
approximate sum rule~\cite{GRKpi03}. This is however consistent with the
findings of Ref.~\cite{CGRS_PP}.
\textit{To date we cannot exclude the hypotheses~(\ref{crazy}) within
the SM.}
\vs
Nevertheless, if more precise data confirm the present pattern of the
$K\pi$ modes with respect to the $\pi\pi$ ones, a challenge would be
given to the theory. Various effects could come into play.
\bei

\item   Significant annihilation and exchange topologies and/or SU(3)
        breaking (in other words, non-trivial rescattering effects): 
        while large contributions of this type are unlikely, 
        one has learned from $D$ meson decays, from $\Bz\to D_s^- \Kp$, 
        and from unexpectedly large color-suppressed transitions 
        from beauty to charm, that the calculation of the
        heavy meson non-leptonic decays  is very difficult. As for
        charmless final states, fits using QCD FA
        (Section~\ref{sec:charmlessBDecays}.\ref{sec:bbnsres}) 
        require non-vanishing power corrections, even if they 
        finally turn out to be moderate. Hopefully, with the decrease
        of the experimental bounds on the suppressed $B\to K\Kb$ modes, 
        multichannel studies will provide more information.

\item   Experimental effects: absolute measurements of rates
        represent difficult analyses. For example, radiative
  corrections  to the decays with charged particles in the final states
  have not  been taken into account so far by the experiments. Their
  inclusion is expected to lead to increased branching fractions of
  modes with  (light) charged particles in the final state (\cf\  the
  introduction  to Part~\ref{sec:charmlessBDecays}). As a consequence,
  the ratios $R_{nc}$ and $R_n$ should increase, which could improve the
  agreement with the indirect constraints from the $\pi\pi$ system
  (see Fig.~\ref{fig:KpiUsePipi}).

\item   New Physics in loop-dominated amplitudes: according to the quantum
        numbers of the new field, one may have anomalies in $b\to d$, $\Delta
        I=1/2$ (gluonic) and/or $\Delta I=3/2$ (electroweak) penguin amplitudes,
        and/or in $b\to s$, $\Delta I=0$ (gluonic) and/or $\Delta I=1$
        (electroweak) penguin amplitudes. This would require the
        introduction of new parameters, which would need
        sufficiently accurate data to be fitted.
\eei
The authors of Ref.~\cite{BFRS1,BFRS2} have studied the latter possibility
within a specific class of New Physics models, where the hierarchy
between $b\to d$ and $b\to s$ transitions is the same as in the SM
($\sim \lambda$), and where New Physics only enters as an enhanced 
$\Delta S=\Delta I=1$ electroweak penguin amplitude $P^\mathrm{EW}$
with a single (unknown) weak phase. This scenario is minimal in the
sense that only one magnitude and one phase has to be adjusted in order
to describe the data. As a by-product of its simplicity,
its ``naturalness'' can be questioned:
there is  no obvious reason to exclude the possibility that the failure
of  color-suppression in the $\pi\pi$ modes is a consequence of
significant  non-standard corrections; moreover, the complete NP effect
may not be proportional to a single \CP-violating phase\footnote
{
        The most general parameterization would imply to introduce two 
        new \CP-violating phases, because the arbitrary sum
        $\sum_i M^\mathrm{NP}_ie^{\pm i\phi_i}$ can be rewritten
        as $\tilde
        M_1^\mathrm{NP}e^{i\pm\tilde\phi_1}+ \tilde
        M_2^\mathrm{NP}e^{i\pm\tilde\phi_2}$, where the $M_i$ are
        \CP-conserving complex numbers and the $\phi_i$ are \CP-odd.
}, and finally, the assumption that $P^\mathrm{EW}$ is real with respect 
to the sum $T^{+-}+T^{00}_C$, as it is in the SM\footnote
{
        For this to be true, one would have to show that NP does not
        enhance the coefficients        $c_{7,8}$, see 
        Eq.~(\ref{PEW}).
}, may be significantly violated.

 \section{Analysis of $\B\to\rho\pi$ Decays}
\label{sec:introductionrhopi}

In this section we discuss the phenomenological implications of the
experimental  results from the \babar\  collaboration~\cite{rhopipaper}
on  time-dependent \CP-violating asymmetries in
$\Bz\to\rho^\pm\pi^\mp$  decays. We use amplitude relations based on 
flavor symmetries as explicit theoretical input to dynamical 
constraints\footnote
{
        The present analysis does not include a discussion of 
        results on pseudoscalar-vector modes
        from QCD Factorization~\cite{BN,PV_dueal,PV_orsay,PV_brits} or 
        SU(3) symmetry including all SU(3) multiplets~\cite{PV_pheno}, 
        since this goes beyond the scope of this paper. A dedicated 
        work on this will be forthcoming.
}. 
The analysis is restricted to the quasi-two-body representations of
$\Bz\to (\pi^\pm\piz)\pi^\mp$ 
decays, corresponding to distinct bands in the three-pion Dalitz plot. 
Wrong charge assignments due to the finite width of the $\rho$ 
resonance and interference effects between different two-body 
states are neglected. It has been pointed out in Ref.~\cite{lalrhopi} 
that this neglect induces biases on the observed \CP and 
dilution parameters that can amount to up to $8\%
$ depending on the 
size of the $\Bz\to\rho^0\piz$ amplitude\footnote
{ 
        Whereas the main part of the interference region between 
        the $\rho^+$ and $\rho^-$ has been removed from the analysis, 
        the ones between charged and neutral $\rho$'s are
        kept~\cite{rhopipaper}.
}.
In the future full Dalitz plot \CP analyses~\cite{SnyderQuinn,BabarPhysBook},
these model-dependent uncertainties will be significantly reduced.
\vs
As in all charmless analyses related to the UT angle $\alpha$, the 
phenomenological effort focuses on the determination of the penguin 
contribution to the transition amplitudes. We approach the problem 
following a similar hierarchical structure as in 
Section~\ref{sec:charmlessBDecays}.\ref{sec:introductionpipi}.
\bei
\item[(\iA)]    
        Using as input the present (yet incomplete)
        measurements or bounds on
        branching fractions of the modes involved
        with the SU(2) analysis. Contributions from
        electroweak penguins are neglected\footnote
        {
          \label{foot:PewRhopi}
           The treatment advertised in Eq.~(\ref{eq:ewpeng})
           cannot be directly translated to the $\rho\pi$
           system, if considered as a two-body decay. Only
           in the case of a full Dalitz plot analysis, that
           allows one to extract the
           $\Delta I=3/2$ tree amplitude, is it possible to take
           into account EW penguin contributions in a model-independent
           way~\cite{theseJerome}.
           Since we learned from the $\pi\pi$
           system that these effects are small, we
           can choose to neglect them for the  numerical
           discussion of present $\rho\pi$ \CP  results.
        }.
        We also extrapolate the isospin analysis to future integrated
        luminosities of $1\invab$ and $10\invab$, using educated 
        guesses for measurements and experimental errors.

\item[(\iB)]    
        Using (\iA) and the rates of $\Bz\to K^{*+}\pi^-$
        and $\Bz\to \rho^-K^+$ decays together with $SU(3)$
        flavor symmetry and neglecting OZI-suppressed penguin 
        annihilation topologies. 

\item[(\iC)]    
        Using (\iB) and phenomenological estimates
        of $|\Ppm|$ and $|\Pmp|$ from the rates of
        $B^+\to K^{*0}\pi^+$ (measured) and
        $B^+\to \rho^+K^{0}$ (upper limit) decays, respectively, together
        with $SU(3)$ flavor symmetry and neglecting annihilation and
        long-distance penguin topologies.

\eei

\subsection{Basic Formulae and Definitions}

We follow the conventions adopted in 
Section~\ref{sec:charmlessBDecays}.\ref{sec:introductionpipi}
and use the unitarity of the weak Hamiltonian to eliminate 
the charm quark loop out of the penguin diagrams ($\C$ convention)
in the transition amplitudes.
The complex Standard Model amplitudes of the relevant processes
represent the sum of complex tree ($T$) and penguin ($P$)
amplitudes with different weak and strong phases.
The corresponding diagrams for the decay $B^0\to\rho^\pm\pi^\mp$
are the same as for $B^0\to\pi^+\pi^-$ (see examples in Fig.~\ref{fig:B0pippim}). 
The transition amplitudes read
\beq
\label{eq:b0rho+pi-}
\begin{array}{rcccl}
\Apm &\equiv&  A(\Bz\to\rho^+\pi^-)
                        &=&   V_{ud}V_{ub}^*\Tpm
                            + V_{td}V_{tb}^*\Ppm~, \\
\Amp &\equiv&  A(\Bz\to\rho^-\pi^+)
                        &=&   V_{ud}V_{ub}^*\Tmp
                            + V_{td}V_{tb}^*\Pmp~, \\
\Apmb &\equiv&  A(\Bzb\to\rho^+\pi^-)
                        &=&   V_{ud}^*V_{ub}\Tmp
                            + V_{td}^*V_{tb}\Pmp~, \\
\Ampb &\equiv&  A(\Bzb\to\rho^-\pi^+)
                        &=&   V_{ud}^*V_{ub}\Tpm
                            + V_{td}^*V_{tb}\Ppm~,
\end{array}
\eeq
where the $\rho$ meson is emitted by the $W$ boson
in the case of $\Apm$ and $\Ampb$, while it
contains the spectator quark in the case of
$\Amp$ and $\Apmb$. 
\vs
The time-dependent \CP asymmetries is given by
\beqn
\label{eq:rhopiCPasym}
   a_{\CP}^\pm(\dmt) &\equiv&
        \frac{\Gamma(\Bzb{}(\dmt)\to\rho^\pm\pi^\mp)
                   - \Gamma(\Bz{}(\dmt)\to\rho^\pm\pi^\mp)}
                  {\Gamma(\Bzb{}(\dmt)\to\rho^\pm\pi^\mp)
                   + \Gamma(\Bz{}(\dmt)\to\rho^\pm\pi^\mp)}
                \nonumber\\[0.2cm]
        &=& (\Srhopi\pm\dSrhopi) \,\sin(\dmd \dmt)
        - (\Crhopi \pm \dCrhopi) \,\cos(\dmd \dmt)~,
\eeqn
where the quantities $\Srhopi$ and $\Crhopi$ parameterize mixing-induced 
\CP violation and flavor-dependent direct \CP violation, respectively.
The parameters $\dSrhopi$ and $\dCrhopi$ are \CP-conserving:
$\dSrhopi$ is related to the strong phase difference between
the amplitudes contributing to $\Bz \to\rho^\pm\pi^\mp$ decays, while 
$\dCrhopi$ describes the asymmetry (dilution) between the rates 
$\Gamma({\Bz} \to{\rho^+\pi^-}) + \Gamma({\Bzb} \to {\rho^-\pi^+})$ and
${\Gamma(\Bz} \to {\rho^-\pi^+}) + \Gamma({\Bzb} \to {\rho^+\pi^-})$.
Owing to the fact that $\Bz\to\rho^\pm\pi^\mp$ is not a \CP eigenstate, 
one must also consider the time- and flavor-integrated charge asymmetry
\beqn
\label{eq:Acprhopi}
    \Acp
    &\equiv& \frac{|\Apm|^2 + |\Apmb|^2 - |\Amp|^2 - |\Ampb|^2}
             {|\Apm|^2 + |\Apmb|^2 + |\Amp|^2 + |\Ampb|^2}~,
\eeqn
as another source of possible direct \CP violation.
\vs
We reorganize the experimentally convenient, namely
uncorrelated, direct \CP-violation parameters $\Crhopi$ and $\Acp$
into the physically more intuitive quantities $\Acppm$, $\Acpmp$,
defined by
\beqn
\label{eq:Adirpm}
    \Acppm &\equiv& \frac{|\kappm|^2-1}{|\kappm|^2+1}
    \;=\; -\frac{\Acp+\Crhopi+\Acp\dCrhopi}{1+\dCrhopi+\Acp\Crhopi}
    ~,\\[0.2cm]\nonumber
\label{eq:Adirmp}
    \Acpmp &\equiv& \frac{|\kapmp|^2-1}{|\kapmp|^2+1}
    \;=\; \frac{\Acp-\Crhopi-\Acp\dCrhopi}{1-\dCrhopi-\Acp\Crhopi}
    ~,
\eeqn
where
\beq
\label{eq:kapparhopi}
    \kappm \equiv \frac{q}{p}\frac{\Ampb}{\Apm}~,\hspace{1.cm}
    \kapmp \equiv \frac{q}{p}\frac{\Apmb}{\Amp}~,
\eeq
so that $\Acppm$ ($\Acpmp$) involves only diagrams where the $\rho$
meson is emitted by the $W$ boson (contains the spectator quark).
\vs
In analogy to the $\pi\pi$ system, we introduce the effective weak 
angles that reduce to $2\alpha$ in the absence of penguins
\beq
\label{eq:alphaeffpm}
    2\alphaeffpm \equiv \arg{\kappm}  ~,
    \hspace{1.5cm}
\label{eq:alphaeffmp}
    2\alphaeffmp \equiv \arg{\kapmp}  ~,
\eeq
so that $S_{\rho\pi}$ and $\Delta S_{\rho\pi}$, which are given by
\beqn
        S+\Delta S_{\rho\pi}    
        &=&
                \frac{2\mathrm{Im}\lambda^{+-}}{1+|\lambda^{+-}|^2}~,\\\nonumber
        S-\Delta S_{\rho\pi} 
        &=&
                \frac{2\mathrm{Im}\lambda^{-+}}{1+|\lambda^{-+}|^2}~,
\eeqn
where\footnote
{
        The $\lampmmp$ involve only one $\rho\pi$ charge combination, 
        but both amplitude types $T(P)^{+-}$ and $T(P)^{-+}$. They are 
        insensitive to direct CPV but their imaginary part is directly 
        related to the weak phase $\alpha$ (though complicated by strong 
        phase shifts and penguins). The quantities $\kappmmp$ involve both $\rho\pi$ 
        charges, but only one amplitude type, corresponding to whether 
        ($[-+]$) or not ($[+-]$) the $\rho$ has been produced involving 
        the spectator quark. Their moduli are linked to direct CPV while 
        their phases measure effective weak angles $\alphaeffmppm$ defined 
        further below.

        Compared to \CP  eigenstates, the $\lampm$ and
        $\lammp$ do not have the desired properties under \CP
        transformation, \ie, $\lampm\ne 1$ or $\lammp\ne 1$ does
        not automatically entail \CP  violation. These inequalities 
        are a necessary but not a sufficient condition. Appropriate
        $\lambda$'s can be easily constructed.
        They should reflect the \CP   and the flavor specific
        character of $\Bz\to\rho^\pm\pi^\mp$ decays. A
        possible definition is
        \[
            \tilde\lamCP \equiv \lampm\cdot \lammp~,\hspace{1cm}
            \tilde\lamFL \equiv \lampm / \lammp~,
        \]
        where $\tilde \lamCP\ne1$  in case of direct or mixing induced
        \CP  violation, and for example $\tilde\lamFL=0$ for
        the case that $\Bz\to\rho^+\pi^-$ is a flavor
        eigenstate. A more practical definition is given by:
        \beqn
        \label{eq:dt}
            |\lamCP|^2
            &\equiv&
            \frac{|\lampm|^2 + |\lammp|^2 + 2 |\lampm|^2 |\lammp|^2}
                 {2 + |\lampm|^2 + |\lammp|^2}~,\nonumber\\[0.3cm]
            |\lamFL|^2
            &\equiv&
            \frac{1 + 2|\lampm|^2 + |\lampm|^2 |\lammp|^2}
                 {1 + 2|\lammp|^2 + |\lammp|^2 |\lampm|^2}~,\nonumber\\[0.3cm]
            {\rm Im}\lamCP
            &\equiv&
            \frac{{\rm Im}\lampm(1 + |\lammp|^2) + {\rm Im}\lammp(1 + |\lampm|^2)}
                 {2 + |\lampm|^2 + |\lammp|^2}~,\nonumber\\[0.3cm]
            {\rm Im}\lamFL
            &\equiv&
            \frac{{\rm Im}\lampm(1 + |\lammp|^2) - {\rm Im}\lammp(1 + |\lampm|^2)}
                 {1 + 2|\lammp|^2 + |\lammp|^2 |\lampm|^2}~,\nonumber
        \eeqn
        which has the desired properties, since if:
        \bei
        \item   $\Bz\to\rho^+\pi^-$ is flavor eigenstate, \eg,
                $\Amp=\Apmb=0$ so that $\lampm=0$ and $\lammp=\infty$,
                one has $\lamFL=0$ with maximal dilution. The mode is
                self-tagging as is, \eg, $B^0\to\rho^-K^+$, and
                no mixing-induced CPV can occur.

        \item   $\Bz\to\rho^\pm\pi^\mp$ behaves like a \CP  eigenstate, \ie,
                $|\Apm|=|\Amp|$ and $|\Apmb|=|\Ampb|$ so that
                $|\lampm|=|\lammp|=|\lambda|$, one has $|\lamFL|=1$ with minimal 
                dilution, and $\lamCP=\lambda$. One could hence disregard the
                charge of the $\rho$ in the analysis and just look at the time-dependent
                asymmetry between $\Bz\to(\rho\pi)^0$ and $\Bzb\to(\rho\pi)^0$.
        \eei
        Note that in the presence of penguins, the value of $\lamFL$ does
        depend on the weak phase $\alpha$.
}
\beq
\label{eq:lambdarhopi}
    \lampm \equiv \frac{q}{p}\frac{\Apmb}{\Apm}~,\hspace{1.cm}
    \lammp \equiv \frac{q}{p}\frac{\Ampb}{\Amp}~,\hspace{1.cm}
\eeq
can be rewritten as
\beqn
\label{eq:saeffanddeltapm}
   S + \dSrhopi &=& \sqrt{1-(C+\dCrhopi)^2}\,\sin(2\alphaeffpm + \delpmmp)~,\\
\label{eq:saeffanddeltamp}
   S - \dSrhopi &=& \sqrt{1-(C-\dCrhopi)^2}\,\sin(2\alphaeffmp -
   \delpmmp)~,\nonumber
\eeqn
with\footnote
{
        The alternative parameterization introduced in Ref.~\cite{mufrapipi} 
        (see Footnote~\ref{foot:pipi_alterparam} in
        Section~\ref{sec:charmlessBDecays}.\ref{sec:su2pipi}) can be 
        extended to the $\B\to\rho\pi$ system. Restricted to
        the $\Bz\to\rho^\pm\pi^\mp$ amplitudes, one has
        \beqn
            \Atpm       &=& \mu\apm  e^{-i\delpmmp/2}~, \nonumber\\
            \Atmp       &=& \mu\amp  e^{+i\delpmmp/2}~, \nonumber\\
            (q/p)\Atpmb &=& \mu\apmb e^{+i(2\alphaeffpm + \delpmmp/2)}~, \nonumber\\
            (q/p)\Atmpb &=& \mu\ampb e^{+i(2\alphaeffmp - \delpmmp/2)}~,\nonumber
        \eeqn
        where $\mu$ (overall scale) and the $\apm$, \dots, are real
        numbers and where we have rotated the amplitudes by the
        (arbitrary) global phase
        $A^{ij}_\prime = A^{ij} e^{i(\arg[\Apm{}^*]-\delpmmp/2)}$.
}
$\delpmmp=\arg[\lampm{\kapmp}^*]=\arg[{\lammp}^*\kappm]=\arg[\Amp{\Apm}^*]$.
\vs
In the absence of penguin contributions $(\Ppm=\Pmp=0)$, one has
$|\kappm|=|\kapmp|=|\lampm\lammp|=1$, that is 
$\alphaeffpm=\alphaeffmp=\alpha$, so that
the observables reduce to simple functions of $\alpha$,
$\delpmmp$ and $\hat r$:
\beqn
    \Srhopi &=& \frac{2\hat r}{1+\hat r^2}\,\sin2\alpha\,\cos\delpmmp~,
    \nonumber\\
    \dSrhopi&=& \frac{2\hat r}{1+\hat r^2}\,\cos2\alpha\,\sin\delpmmp~, \nonumber\\
    \Crhopi &=& 0~, \\
    \dCrhopi&=& \frac{1 - \hat r^2}{1 + \hat r^2}~,\nonumber\\
    \Acp    &=& 0~, \nonumber\\[0.2cm]
    \Acppm  &=& \Acpmp = 0~,\nonumber
\eeqn
with $\hat r=|\Tmp/\Tpm|$. Hence $\alpha$ can be determined 
up to an eightfold ambiguity within $[0,\pi]$\footnote
{
    In the absence of penguin contributions, the eight solutions
    for $\alpha$ and $\delpmmp$
    satisfying Eqs.~(\ref{eq:saeffanddeltapm}) and (\ref{eq:saeffanddeltamp})
    read~\cite{Sophie}
    \beqns
    \begin{array}{ccccccccc}
    \hline
    \alpha&\to  &\pi/4-\delpmmp/2\;,
            &\pi/2+\alpha\;,
            &3\pi/4-\delpmmp/2\;,
            &\pi/4+\delpmmp/2\;,
            &\pi/2-\alpha\;,
            &3\pi/4+\delpmmp/2\;,
            &\pi-\alpha\; \\
    \delpmmp&\to    &\pi/2-2\alpha\;,
            &\pi+\delpmmp\;,
            &3\pi/2-2\alpha\;,
            &-\pi/2+2\alpha\;,
            &-\delpmmp\;,
            &-3\pi/2+2\alpha\;,
            &\pi-\delpmmp\; \\
    \hline
    \end{array}
    \eeqns
}.
If furthermore $\Bz\to\rho^\pm\pi^\mp$ represents an effective \CP
eigenstate $(\Tpm=\Tmp)$, the parameters simplify to $\Srhopi=\sta$ and
$\dSrhopi=\Crhopi=\dCrhopi=\Acp=0$, hence reproducing the zero-penguin
case in $\Bz\to\pi^+\pi^-$ decays. If the relative strong phase vanishes,
but there exists a non-zero dilution $\rTpm\ne0$ (\ie, $\dCrhopi\ne0$), 
one has $\Srhopi=\sqrt{1-\dCrhopi^2}\sta$.
\vs
Branching fractions are in general given by the sum of the contributing
squared amplitudes, where final states ($\rho\pi$ charges) are summed and
initial states ($B$ flavors) are averaged. The $\Bz\to\rho^\pm\pi^\mp$
branching fraction reads
\beq
\label{eq:BRrhopi}
   \BRall \;\propto\; \frac{\tau_{\Bz}}{2}
        \left(|\Apm|^2 + |\Ampb|^2 + |\Amp|^2 + |\Apmb|^2
        \right)~.
\eeq

\subsubsection*{Observables from Other Modes}

The various analyses discussed here involve SU(2) and SU(3) 
flavor partners of the signal mode $\Bz\to\rho^\pm\pi^\mp$. We use
branching fractions ($\BR$) as well as charge asymmetries
($\cal A$) for the charged $B$ and self-tagging channels. They are
defined by
\beqn
\label{eq:BRs}
   \BR_{hh^\prime}&\equiv& \BR(B\to h h^{\prime})  \;\propto\;
    \frac{\tau_B}{2}\left(|A_{hh^\prime}|^2 + |\overline A_{hh^\prime}|^2\right)~,
    \\[0.2cm]
\label{eq:A}
   {\cal A}_{hh^\prime}  &\equiv&
    \frac{|\overline A_{hh^\prime}|^2 - | A_{hh^\prime}|^2}
             {|\overline A_{hh^\prime}|^2 + | A_{hh^\prime}|^2}~,
\eeqn
where $\tau_B$ denotes the lifetime of the decaying $B$ meson (neutral
or charged). More details are given in the following.

\subsection{Experimental Input}
\label{sec:theData}

\begin{table}[!t]
\begin{center}
\setlength{\tabcolsep}{0.0pc}
{\normalsize
\begin{tabular*}{\textwidth}{@{\extracolsep{\fill}}lcccc}\hline
&&&& \\[-0.3cm]
Obs. & \mc{1}{c}{\babar} & \mc{1}{c}{Belle}
        & \mc{1}{c}{CLEO} & \mc{1}{c}{Average}    \\[0.1cm]
\hline
&&&& \\[-0.3cm]
\rule[-2.7mm]{0mm}{4mm}
$\BRall$
    & $22.6\pm1.8\pm2.2$    \cite{rhopipaper}
    & $29.1^{\,+5.0}_{\,-4.9}\,\pm 4.0$ \cite{BELLErhopi}
    & $27.6^{\,+8.4}_{\,-7.4}\pm 4.2$   \cite{CLEOrhopi}
    & $24.0 \pm 2.5$ \\[0.05cm]
\rule[-2.7mm]{0mm}{4mm}
$\Acp$
    & $-0.114  \pm0.062\pm0.027$       \cite{rhopipaper}
    & -
    & -
    & $-0.114  \pm0.067$ \\[0.05cm]
\rule[-2.7mm]{0mm}{4mm}
$\Srhopi$
    & $-0.13\pm0.18\pm0.04$          \cite{rhopipaper}
    & -
    & -
    & $-0.13\pm0.18$ \\[0.05cm]
\rule[-2.7mm]{0mm}{4mm}
$\dSrhopi$
    & $0.33\pm0.18\pm0.03$          \cite{rhopipaper}
    & -
    & -
    & $0.33\pm0.18$ \\[0.05cm]
\rule[-2.7mm]{0mm}{4mm}
$\Crhopi$
    & $0.35  \pm0.13\pm0.05$        \cite{rhopipaper}
    & -
    & -
    & $0.35  \pm0.14$ \\[0.05cm]
\rule[-2.7mm]{0mm}{4mm}
$\dCrhopi$
    & $0.20\pm0.13\pm0.05$     \cite{rhopipaper}
    & -
    & -
    & $0.20\pm0.14$ \\[0.05cm]
\rule[-2.7mm]{0mm}{4mm}
$\Acppm$
    & $-0.18\pm0.13\pm0.05$    
    & -
    & -
    & $-0.18\pm0.14$ \\[0.05cm]
\rule[-2.7mm]{0mm}{4mm}
$\Acpmp$
    & $-0.52^{\,+0.17}_{\,-0.19}\pm0.07$    
    & -
    & -
    & $-0.52^{\,+0.18}_{\,-0.20}$ \\[0.05cm]
\hline
&&&& \\[-0.3cm]
\rule[-2.7mm]{0mm}{4mm}
$\BRpz$
    & $11.0 \pm 1.9 \pm 1.9$            \cite{rhopiBRpaper,rhopiCpaper}
    & $13.2 \pm 2.3^{\,+1.4}_{\,-1.9}$  \cite{BELLErho+pi0}
    & $<43$          \cite{CLEOrhopi}
    & $12.0 \pm 2.0$ \\[0.15cm]
\rule[-2.7mm]{0mm}{4mm}
$\Acppz$
    & $0.23 \pm 0.16 \pm 0.06$      \cite{rhopiBRpaper,rhopiCpaper}
    & $0.06 \pm 0.19 \pm^{\,+0.04}_{\,-0.06}$      \cite{BELLErho+pi0}
    & -
    & $0.16 \pm 0.13$ \\[0.05cm]
\hline
\rule[-2.7mm]{0mm}{8mm}
$\BRzp$
    & $9.3\pm 1.0\pm0.8$            \cite{rhopiBRpaper,rhopiCpaper}
    & $8.0^{\,+2.3}_{\,-2.0}\pm0.7$     \cite{BELLErhopi}
    & $10.4^{\,+3.3}_{\,-3.4}\pm2.1$    \cite{CLEOrhopi}
    & $9.1\pm 1.1$ \\[0.05cm]
\rule[-2.7mm]{0mm}{4mm}
$\Acpzp$
    & $-0.17 \pm 0.11 \pm 0.02$     \cite{rhopiBRpaper,rhopiCpaper}
    & -
    & -
    & $-0.17 \pm 0.11$ \\[0.05cm]
\hline
\rule[-2.7mm]{0mm}{8mm}
$\BRzz$
    & $0.9\pm0.7\pm0.5(<2.9)$       \cite{rhopiBRpaper,rhopiCpaper}
    & $5.1\pm1.6\pm0.9$     \cite{BELLErhopi00}
    & $<5.5$                \cite{CLEOrhopi}
    & $1.7\pm0.8$ \\[0.05cm]
\rule[-2.7mm]{0mm}{4mm}
$\Czz$
    & -
    & -
    & -
    & - \\[0.05cm]
\rule[-2.7mm]{0mm}{4mm}
$\Szz$
    & -
    & -
    & -
    & - \\[0.05cm]
\hline
\rule[-2.7mm]{0mm}{8mm}
$\BRrhoK$
    & $7.3^{\,+1.3}_{\,-1.2}\pm1.3$     \cite{rhopipaper}
    & $15.1^{\,+3.4}_{\,-3.3}\,^{\,+2.4}_{\,-2.6}$
                        \cite{BELLEKKpi}
    & $16.0^{\,+7.6}_{\,-6.4}\pm2.8$    \cite{CLEOrhopi}
    & $9.0\pm1.6$ \\[0.05cm]
\rule[-2.7mm]{0mm}{4mm}
$\AcprhoK$
    & $0.18  \pm0.12\pm0.08$        \cite{rhopipaper}
    & $0.22^{\,+0.22}_{\,-0.23}\pm0.02$ \cite{BELLEKKpi}
    & -
    & $0.19  \pm0.12$ \\[0.05cm]
\hline
\rule[-2.7mm]{0mm}{8mm}
$\BRKstrpi$
    & -
    & $14.8^{\,+4.6}_{\,-4.4}\,^{\,+2.8}_{\,-1.3}$
                        \cite{BELLEKKpi}
    & $16^{\,+6}_{\,-5}\pm2$        \cite{CLEOKstrpi}
    & $15.3^{\,+4.1}_{\,-3.5}$ \\[0.05cm]
\rule[-2.7mm]{0mm}{4mm}
$\AcpKstrpi$
    & -
    & -
    & $0.26^{\,+0.33}_{\,-0.34}\,^{\,+0.10}_{\,-0.08}$
                        \cite{CLEOKstrpi}
    & $0.26 \pm 0.34$ \\[0.05cm]
\hline
\rule[-2.7mm]{0mm}{8mm}
$\BRrhoKC$
    & -
    & -
    & $<48$                 \cite{CLEOKstpiC}
    & $<48$ \\[0.05cm]
\hline
\rule[-2.7mm]{0mm}{8mm}
$\BRKstrpiC$
    & $15.5^{\,+1.8}_{\,-1.5}\pm4.0$    \cite{BabarKstpiC}
    & $8.5^{\,+0.9}_{\,-1.1}\pm 0.9$    \cite{BelleKstpiC}
    & $7.6^{\,+3.5}_{\,-3.0}\pm 1.6$    \cite{CLEOrhopi}
    & $9.0^{\,+1.3}_{\,-1.2}$ \\[0.05cm]
\hline
\end{tabular*}
}
\caption[.]{\label{tab:BRRhoPicompilation} \em
    Compilation of results (from data up to Winter 2004) on
    $B\to \rho \pi$ branching fractions (in units of $10^{-6}$)
    and \CP  asymmetries as well as dilution parameters.
    Limits are quoted at $90\%$ confidence level (CL).
    Also given are the branching fractions and direct \CP
    asymmetries for the modes $B^0\to K^{*+}\pi^-$,
    $B^0\to\rho^-K^+$, $B^+\to K^{*0}\pi^+$  and $B^+\to\rho^+K^0$
    related to $\rho\pi$ via SU(3) flavor symmetry.}
\end{center}
\end{table}
\begin{table}[t]
\centering
\setlength{\tabcolsep}{0.0pc}
\begin{tabular*}{\textwidth}{@{\extracolsep{\fill}}lrrrrrr}
\hline
&&&&&& \\[-0.3cm]
                &$\Acp$ & $\Crhopi$     & $\dCrhopi$    & $\Srhopi$     & $\dSrhopi$    & $N_{\rho\pi}$ \\[0.15cm]
$\Acp$          &100    &$-7.6$         &$-6.7$         &$-3.1$         &$\ph{-}1.5$    &$\ph{-}3.9$    \\
$\Crhopi$       &       & 100           &$\ph{-}13.9$   &$-7.7$         &$-10.0$        &$-7.4$ \\
$\dCrhopi$      &       &-              &100            &$-9.5$         &$-7.7$         &$-6.8$ \\
$\Srhopi$       &       &-              &-              &100            &$\ph{-}22.9$   &$\ph{-}1.2$    \\
$\dSrhopi$      &       &-              &-              &-              &100            &$-3.0$ \\[0.15cm]
\hline
\end{tabular*}
\caption{\em Correlation coefficients (in $\%$) between the
        parameters in the time-dependent
        fit to $B^0\to\rho^\pm\pi^\mp$ decays as
        measured by \babar~\cite{rhopipaper}. Note that 
        the correlations between events yields and 
        the $\CP$ parameters are taken from the branching
        fraction analysis~\cite{rhopipaper}.}
\label{tab:correlationsBABAR}
\end{table}
The present (Winter 2004) results (including world averages taken
from the HFAG~\cite{HFAG}) for the branching fractions and \CP-violating
asymmetries of all $B\to\rho\pi$ decays are given in 
Table~\ref{tab:BRRhoPicompilation}. Also given are the results 
for the modes $B^0\to K^{*+}\pi^-$ and $\Bz\to\rho^-K^+$, which are 
the SU(3) partners of the decays $\Bz\to\rho^+\pi^-$ and 
$\Bz\to\rho^-\pi^+$, respectively. There is some disagreement 
on possible evidence for the decay $\Bz\to\rho^0\pi^0$, which 
has not been seen by \babar\  whereas Belle finds a large 
 central value for the branching fraction that 
may indicate a large color-suppressed tree amplitude or significant
penguin contributions.

\subsubsection{Direct \CP Violation}

The direct \CP  asymmetries $\Acppm$ and $\Acpmp$ (see 
Table~\ref{tab:BRRhoPicompilation}) have been computed from
Eqs.~(\ref{eq:Adirpm}) and (\ref{eq:Adirmp}), using the linear 
correlation coefficients given in Table~\ref{tab:correlationsBABAR}. 
We find a linear correlation coefficient between $\Acppm$ and $\Acpmp$ 
of $0.51$. Confidence levels in the $(\Acppm,\Acpmp)$ plane 
are shown in the left hand plot of Fig.~\ref{fig:Apm_BR_2d}. 
The \babar\  experiment finds some indication for direct \CP  
violation (approximately 2.5 standard deviations including systematics)
mainly in the modes involving the $\Amp$ and $\Apmb$ decay amplitudes.
This result is rather unexpected since, if confirmed, it would require
sizable penguin contributions to these amplitudes, 
with a hierarchy opposite to
the na\"{\i}ve factorization expectation~\cite{aleksanRR}
\beq
    \Pmp \ll \Ppm \lsim P^{+-}_{\pi\pi}~.
\eeq
QCD factorization predicts potentially large corrections to the above
hierarchy~\cite{BN}; on the other hand, because the strong phases are
suppressed, direct \CP violation
 in $\Bz\to\rho^\pm\pi^\mp$ is  expected to remain below $10\%
 $.

\subsubsection{Charge-flavor Specific Branching Fractions}

The yields and \CP-violation results are expressed in the basis 
defined in Eqs.~(\ref{eq:rhopiCPasym}), (\ref{eq:Acprhopi}) and
(\ref{eq:BRrhopi}). Since it is complete, we can transform it to 
any other complete basis, \eg, the branching fractions of the four 
individual tag-charge contributions.
The individual (\ie, not $B$-flavor-averaged) branching fractions
\beqn
\begin{array}{rclcrcl}
    \BRipm  &=& \BR(B^0\to\rho^+\pi^-)~,&\hspace{0.7cm}&
    \BRimp  &=& \BR(B^0\to\rho^-\pi^+)~,\\[0.2cm]
    \BRipmb &=& \BR(\overline B^0\to\rho^+\pi^-)~,
                          &\hspace{0.7cm}&
    \BRimpb &=& \BR(\overline B^0\to\rho^-\pi^+)~,
\end{array}
\eeqn
are obtained \via\
\beq
\label{eq:indRates}
    \BR_{\rho^Q\pi^{-Q}}(f,Q)
    = \frac{1}{2}\left(1 + Q \Acp\right)
              \left(1 + f \cdot \left(\Crhopi+ Q \dCrhopi\right)\right)
    \BRall~,
\eeq
with the $B^0$ flavors $f(B^0)=1$, $f(\Bzb)=-1$, and
the $\rho$ charges $Q(\rho^\pm)=\pm1$. Adding statistical and
systematic errors in quadrature, we find (in units of $10^{-6}$):
\beqn
\label{eq:rhopiflavorBRs}
\begin{array}{rclcrcl}
    \BRipm  &=& 16.5 ^{\,+ 3.1} _{\, -2.8}~,&\hspace{0.7cm}&
    \BRimp  &=& 15.4 ^{\,+ 3.2} _{\, -2.9} ~,\\[0.2cm]
    \BRipmb &=& 4.8  ^{\,+ 2.6} _{\, -2.3}~,&\hspace{0.7cm}&
    \BRimpb &=& 11.4 ^{\,+ 2.8} _{\, -2.6}~,
\end{array}
\eeqn
and the correlation coefficients
\beq
    \begin{array}{lccc}
\rule[-2.7mm]{0mm}{8mm}
            &\BRimp &\BRipmb &\BRimpb \\
\rule[-2.7mm]{0mm}{8mm}
    \BRipm      & -0.17  &-0.47  & -0.14  \\
\rule[-2.7mm]{0mm}{4mm}
    \BRimp      & 1 & -0.08  & -0.40 \\
\rule[-2.7mm]{0mm}{4mm}
    \BRipmb     & - &1  & -0.06  \\
    \end{array}
\eeq
One notices a significant lack of $\Bzb\to\rho^+\pi^-$ decays in 
the results~(\ref{eq:rhopiflavorBRs}).
We can infer from these numbers the $B$-flavor-averaged 
branching fractions (in units of $10^{-6}$)
\begin{figure}[t]
  \centerline{
        \epsfxsize7.9cm\epsffile{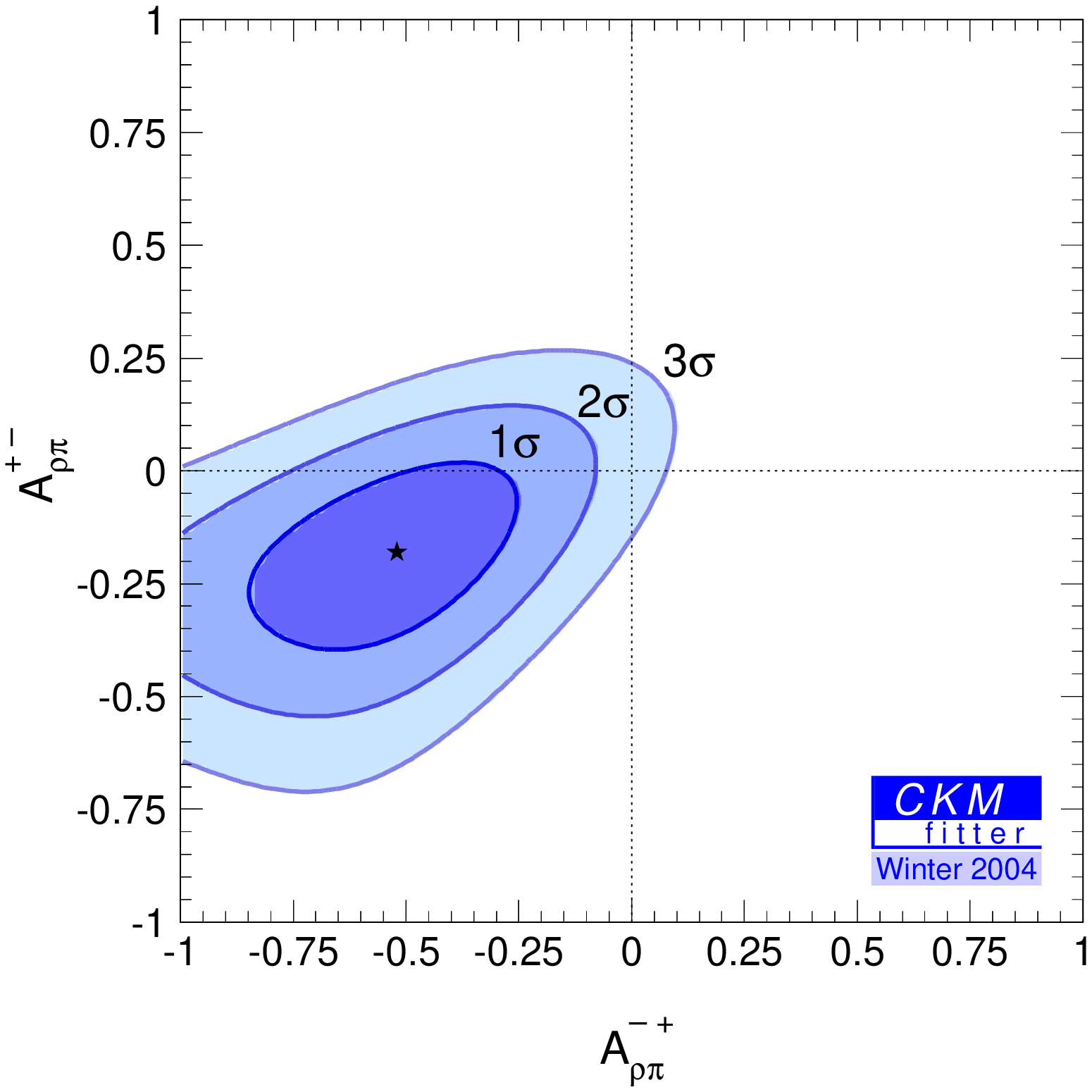}
        \epsfxsize7.9cm\epsffile{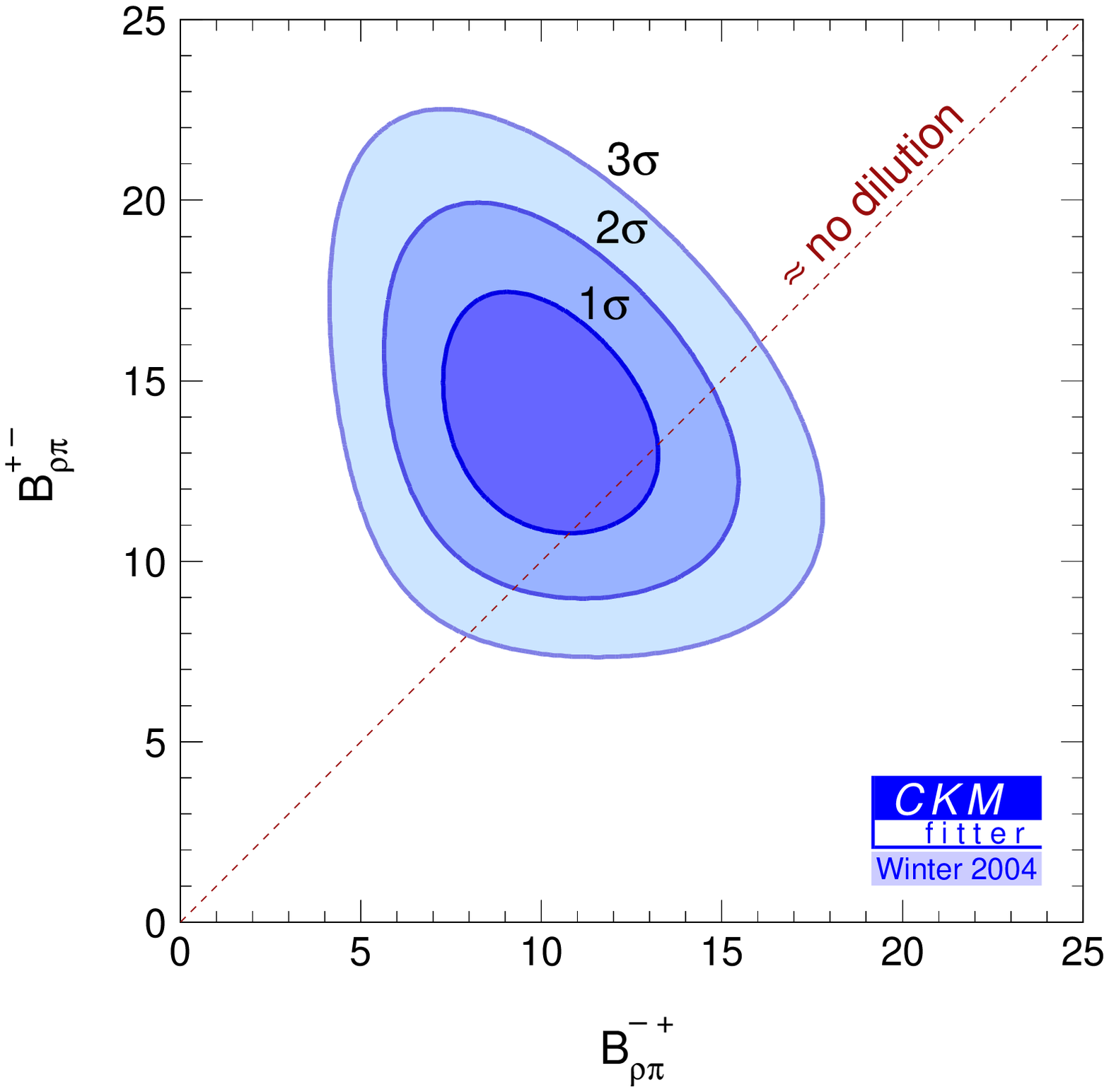}
  }
  \caption[.]{\label{fig:Apm_BR_2d}\em
        Confidence level in the $\Acppm$ versus $\Acpmp$ (left plot)
        and $\BRpm$ versus $\BRmp$ (right plot) planes. Shown are the 
        $1\sigma$ (dark shaded), $2\sigma$ (medium shaded) and the 
        $3\sigma$ (light shaded) regions. The dashed line in the 
        right hand plot approximately indicates vanishing dilution 
        ($\dCrhopi=0$, neglecting $\Acp \Crhopi\ne0$).}
\end{figure}
\beqn
    \BRpm &\equiv& \frac{1}{2}\left(\BRipm + \BRimpb\right)
    \;=\; \frac{1}{2}\left(1 + \dCrhopi + \Acp\Crhopi\right)\BRall
    \;=\; 13.9^{\,+ 2.2} _{\, -2.1}~,  \nonumber\\[0.2cm]
\label{eq:summedBR}
    \BRmp &\equiv& \frac{1}{2}\left(\BRimp + \BRipmb\right)
    \;=\; \frac{1}{2}\left(1 - \dCrhopi - \Acp\Crhopi\right)\BRall
    \;= \; 10.1^{\,+ 2.1} _{\, -1.9}~,
\eeqn
with a linear correlation coefficient of $-0.28$ between $\BRpm$
and $\BRmp$. The CLs of the rates~(\ref{eq:summedBR})
are depicted in the right hand plot of Fig.~\ref{fig:Apm_BR_2d}. 
The branching fractions $\BRpm$ and $\BRmp$ correspond to transitions where
the $\rho$ meson is emitted by the $W$ boson or originates from the
spectator interaction, respectively. Simple form factor arguments
predict that $\BRpm$ should be larger than $\BRmp$,  
which is reproduced by experiment.
\vs
Also given in Table~\ref{tab:BRRhoPicompilation} are the $\rho\pi$ flavor
partners. Since $\rho^0\pi^0$ is a \CP   eigenstate (in the two-body decay 
approximation), its sine and cosine coefficients, $\Szz$, $\Czz$,
can be measured in a time-dependent analysis, provided that the
experimental sensitivity is sufficient. The other $\rho\pi$
modes are charged so that they provide two observables,
one of which describes \CP violation. The decays $B^0\to K^{+(*)} h^-$ 
are self-tagging so that they also provide two observables.

\subsection{Penguins}
\label{sec:penguins}

\begin{figure}[t]
  \centerline{
        \epsfxsize8.1cm\epsffile{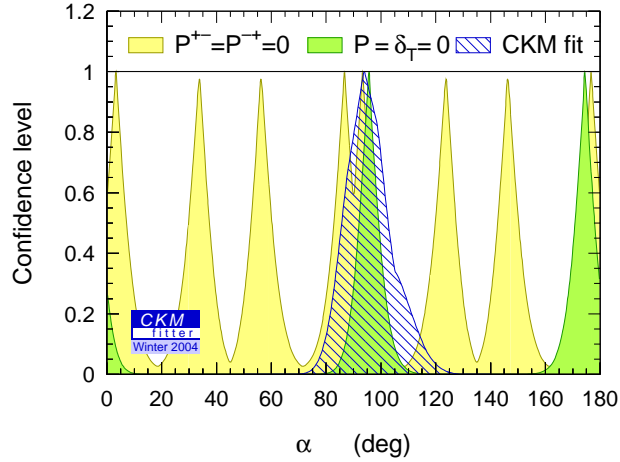}
  }
  \vspace{-0.4cm}
  \caption[.]{\label{fig:alphaNoPeng}\em
        Constraint on the UT angle $\alpha$ using the \CP
        and branching fraction results for $B^0\to\rho^\pm\pi^\mp$
        decays and assuming vanishing penguin contributions. Indicated
        by the dark-shaded areas are the solutions when further
        fixing the relative strong phase $\delpmmp$ to zero.
        Also shown is the constraint from the standard CKM fit.}
\end{figure}
The experimental results on $K^*\pi$ and $K\rho$ modes summarized in Table~\ref{tab:BRRhoPicompilation}
indicate that large penguin contributions may be present in the
SU(3)-related
$\Bz\to\rho^\pm\pi^\mp$ decay amplitudes. On the other hand, within QCD
Factorization, the penguin-to-tree ratio for both charge combinations
is  predicted to be significantly 
smaller (a factor of three) than for $\Bz\to\pip\pim$ decays~\cite{BN}.

\subsubsection{Zero-Penguin Case}

As an exercise, we assume here that the penguin amplitudes $\Ppm$ 
and $\Pmp$ are zero so that $\alphaeffpm=\alphaeffmp=\alpha$ (\cf\   
Eqs.~(\ref{eq:saeffanddeltapm}) and \ref{eq:saeffanddeltamp}). The 
compatibility of a theory without penguins with the $\Bz\to\rho^\pm\pi^\mp$ 
data is marginal. We find $\chi^2=8.6$ for two degrees of freedom which 
corresponds to a CL of $0.014$ ($2.5\sigma$ - which
is equal to the significance of direct \CP violation).
\vs
Figure~\ref{fig:alphaNoPeng} shows the constraint on $\alpha$ obtained 
in this simplified setup. The eightfold ambiguity within $[0,\pi]$
arises due to the unknown strong phase $\delpmmp$. Although vanishing
penguin amplitudes are not a likely scenario, it allows us to assess 
the statistical power of the present data: if all strong phases were 
known, $\alpha$ could be determined with an accuracy of $5.4^\circ$ 
per solution. Further setting $\delpmmp=0$ leads to a twofold ambiguity,
one of which is in agreement with the standard CKM fit.

\subsubsection{Constraining the Penguins}

\begin{figure}[t]
  \epsfxsize11cm
        \epsfxsize8.1cm\epsffile{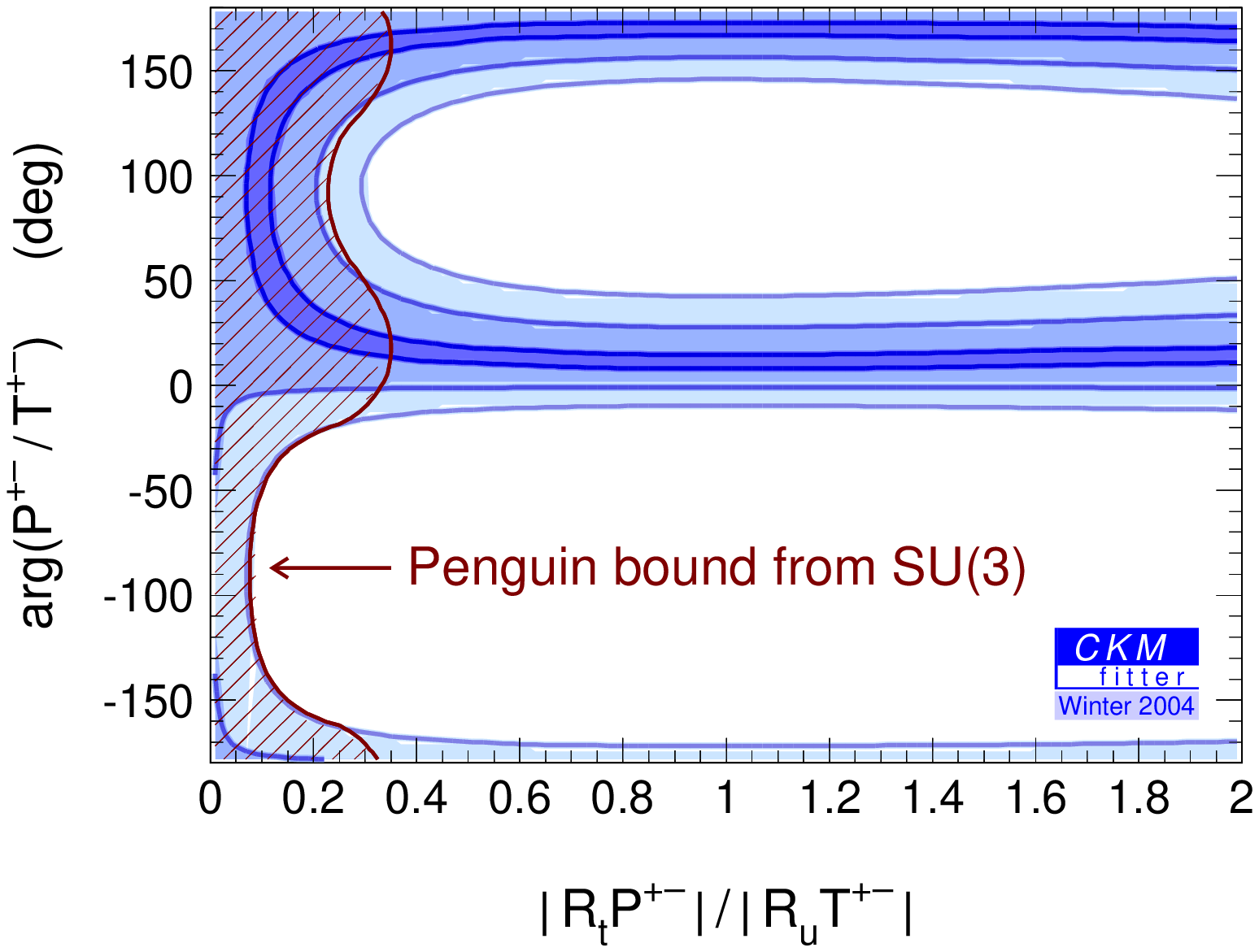}
        \epsfxsize8.1cm\epsffile{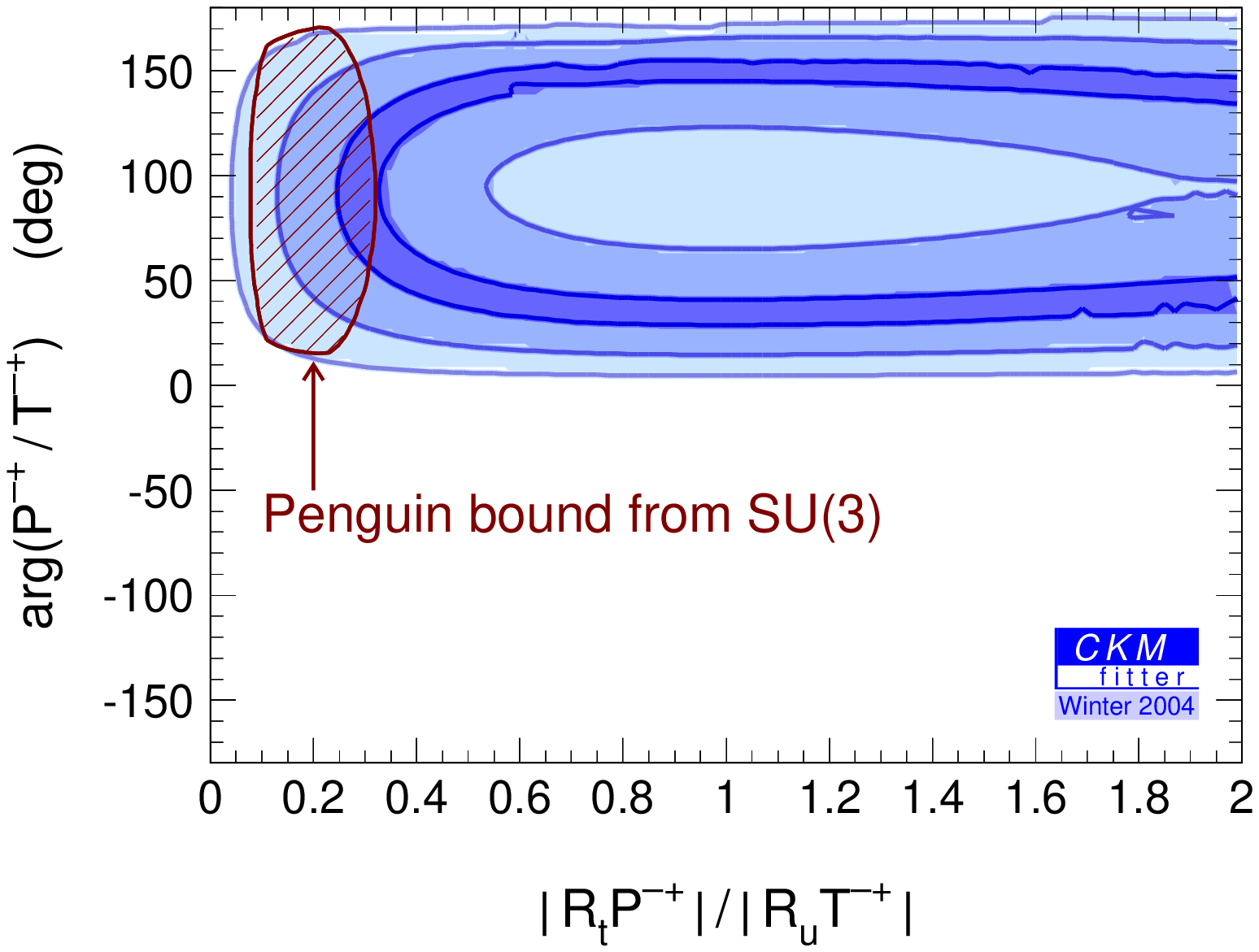}
  \vspace{-0.8cm}
  \caption[.]{\label{fig:ptpm_2d}\em
        Confidence levels in the complex planes
        $|R_t/R_u|(P^{+-}/T^{+-})$ 
        (left) and $|R_t/R_u|(P^{-+}/T^{-+})$ (right),
        using $\alpha$ from the standard CKM fit as input. 
        Dark, medium and light shaded areas have $CL>0.90$,
        $0.32$ and $0.05$, respectively. Shown by the hatched areas 
        are the corresponding $CL>0.05$ regions obtained when using 
        in addition SU(3) flavor symmetry (Scenario~(\iB),
        see Section~\ref{sec:su3}). }
\end{figure}
In analogy to the $\Bz\to\pip\pim$ case, we can constrain the penguin 
contributions by inserting the value of $\alpha$ from the standard CKM 
fit. Whereas
 the information from the other charges (\via\   isospin) is used 
there, we restrict the analysis to $\Bz\to\rho^\pm\pi^\mp$ here.
Figure~\ref{fig:ptpm_2d} gives the 
CLs obtained in the $P^{+-}/T^{+-}$ complex plane (left hand plot) 
as well as the  $P^{-+}/T^{-+}$ complex plane (right hand plot). Due to 
the negative values of the direct \CP-violating 
asymmetries~(\ref{eq:Adirpm})
(\cf\  Table~\ref{tab:BRRhoPicompilation}), positive strong phases
are preferred for both ratios. Since the decays governed by the $\Amp$ 
and $\Apmb$ amplitudes (for which the $\rho$ meson contains the spectator quark)
exhibit larger direct CPV, more sizable penguin-to-tree ratios 
are required here. The characteristic hyperbolic shapes of the
constraints is due to the fact that direct \CP violation is the product
of penguin-to-tree ratio and strong phase difference. Also shown in
Fig.~\ref{fig:ptpm_2d} are the CL $>0.05$ regions obtained when using 
SU(3) flavor symmetry and neglecting OZI-suppressed annihilation
terms, which due to the CKM-favored penguin amplitudes 
provides improved bounds (see the discussion in Section~\ref{sec:su3}).

\subsection{SU(2) Symmetry}
\label{sec:su2}

Similarly to the Gronau-London analysis in $\B\to\pi\pi$ and longitudinally
polarized $\B\to\rho\rho$ decays~\cite{grolon}, the full 
isospin analysis of $\B\to\rho\pi$ decays allows one to constrain the 
angle $\alpha$ up to discrete ambiguities~\cite{oldsnyderetal}. 
However, instead of a triangular isospin relation, a pentagon has 
to be determined in the complex plane, which reduces the
sensitivity to $\alpha$. 

\subsubsection{Isospin Analysis}

The SU(2) flavor decomposition of the neutral and charged $B\to\rho\pi$
amplitudes is given, \eg, in Refs.~\cite{oldsnyderetal,BabarPhysBook}. 
Here we recall only the relevant relations, which complete 
Eq.~(\ref{eq:b0rho+pi-})
\beqn
\label{eq:theRho00}
        2\Azz \:\equiv\:  2A(\Bz\to\rho^0\pi^0)
        &=&
            V_{ud}V_{ub}^*
            T^{00}_\mathrm{C}
                         -\; V_{td}V_{tb}^*
            \left(\Ppm + \Pmp\right)~,\nonumber\\[0.2cm]
\label{eq:theRho0+}
    \sqrt{2}\Azp \:\equiv\:  \sqrt{2}A(B^+\to\rho^0\pi^+)
        &=&
            V_{ud}V_{ub}^*\Tzp
            - V_{td}V_{tb}^*
            \left(\Ppm - \Pmp\right)~, \\[0.2cm]
\label{eq:theRho+0}
    \sqrt{2}\Apz \:\equiv\:  \sqrt{2}A(B^+\to\rho^+\pi^0)
        &=&
            V_{ud}V_{ub}^*
            \left(T^{+-}+T^{-+}+T^{00}_\mathrm{C}-T^{0+}\right)\nonumber\\
        &&    +\; V_{td}V_{tb}^*
            \left(\Ppm - \Pmp\right)~, \nonumber
\eeqn
and equivalently for the \CP-conjugated amplitudes. The amplitudes 
satisfy the pentagonal relations
\beqn
\label{eq:pentagon}
    \sqrt{2}\left(\Apz+\Azp\right)      &=& 2\Azz+\Apm +\Amp~, \nonumber\\
    \sqrt{2}\left(\Apzb+\Azpb\right)    &=& 2\Azzb+\Apmb +\Ampb~.
\eeqn
As in the case of $B\to\pi\pi$, the color-suppressed mode
$\Bz\to\rho^0\pi^0$ constrains the 
(sum of the) penguin contributions to $\Bz\to\rho^\pm\pi^\mp$.
The above isospin relations take advantage of the fact that
QCD penguins can only mediate $\Delta I=1/2$ transitions
in the SU(2) limit. Electroweak penguins, that we neglect here 
(see Footnote~\ref{foot:PewRhopi} in 
Section~\ref{sec:charmlessBDecays}.\ref{sec:introductionrhopi}),
can have $\Delta I=3/2$ and would thus lead to additional terms 
proportional to $V_{td}V_{tb}^*$.
\vs
Information counting results in 12 unknowns (6 complex amplitudes
and the weak phase $\alpha=\pi-\beta-\gamma$ minus one arbitrary
global phase), and 13 observables for the complete SU(2) analysis.
%
%
The isospin analysis constrains the weak phase $\alpha$ up
to discrete ambiguities, which however are not necessarily
degenerate thanks to the fact that the system is over-determined 
(all observables are experimentally accessible).
This is similar to the $\B\to\rho\rho$ system (\cf\  
Section~\ref{sec:charmlessBDecays}.\ref{sec:introductionrhorho}).

\subsubsection{SU(2) Bounds}

Using the SU(2) relations and the \CP-averaged branching fractions,
one can derive simple bounds on the deviation from $\alpha$ induced 
by the penguin contributions~\cite{theseJerome,quinnsilva}
\beq
\label{eq:SU2bound}
    |\alpha - \alphaeffave|
    \;\le\;\frac{1}{2}\arccos\left[\frac{1}{\sqrt{1-\Acpave^2}}
                \left(1-4\frac{\BRzz}{\BRpmave}\right)
              \right]~,
\eeq
where we use the \CP-averaged quantities
\beqn
     2\alphaeffave &\equiv&
        \arg\left[\frac{q}{p}\,\frac{\Apmb+\Ampb}{\Apm+\Amp}\right]~,
        \\[0.3cm]
\label{eq:Acpave}
    \Acpave &\equiv&\frac{|\Apmb+\Ampb|^2 - |\Amp+\Apm|^2}
                             {|\Apmb+\Ampb|^2 + |\Amp+\Apm|^2}~,
        \\[0.3cm]
\label{eq:BRave}
    \BRpmave  &\propto&
        \frac{\tau_{\Bz}}{4}\left(|\Apm+\Amp|^2 + |\Ampb+\Apmb|^2\right)~.
\eeqn
The latter two cannot be experimentally determined in a quasi-two-body
analysis since they involve the relative phases $\arg[\kapmp{\lampm}^*]$
and $\arg[\kappm{\lampm}^*]$, which depend on the interference between 
the two charge states $\rho^+\pi^-$ and $\rho^-\pi^+$ in the Dalitz plot. 
Indirect isospin constraints on $\BRpmave$ and $\Acpave$ using the current 
results (Table~\ref{tab:BRRhoPicompilation}) are insignificant~\cite{lalrhopi}.
As a consequence, no useful constraint on  $|\alpha - \alphaeffave|$ is 
obtained from the bound~(\ref{eq:SU2bound}).
\vs
The presently available experimental information is insufficient
to obtain a meaningful constraint on $\alpha$. We thus attempt
to give an outlook to future integrated luminosities accumulated
at the $B$ factories.

\subsubsection{Prospects for the Full Isospin Analysis}
\label{sec:su2Complete}

\begin{figure}[t]
  \centerline{
        \epsfxsize8.1cm\epsffile{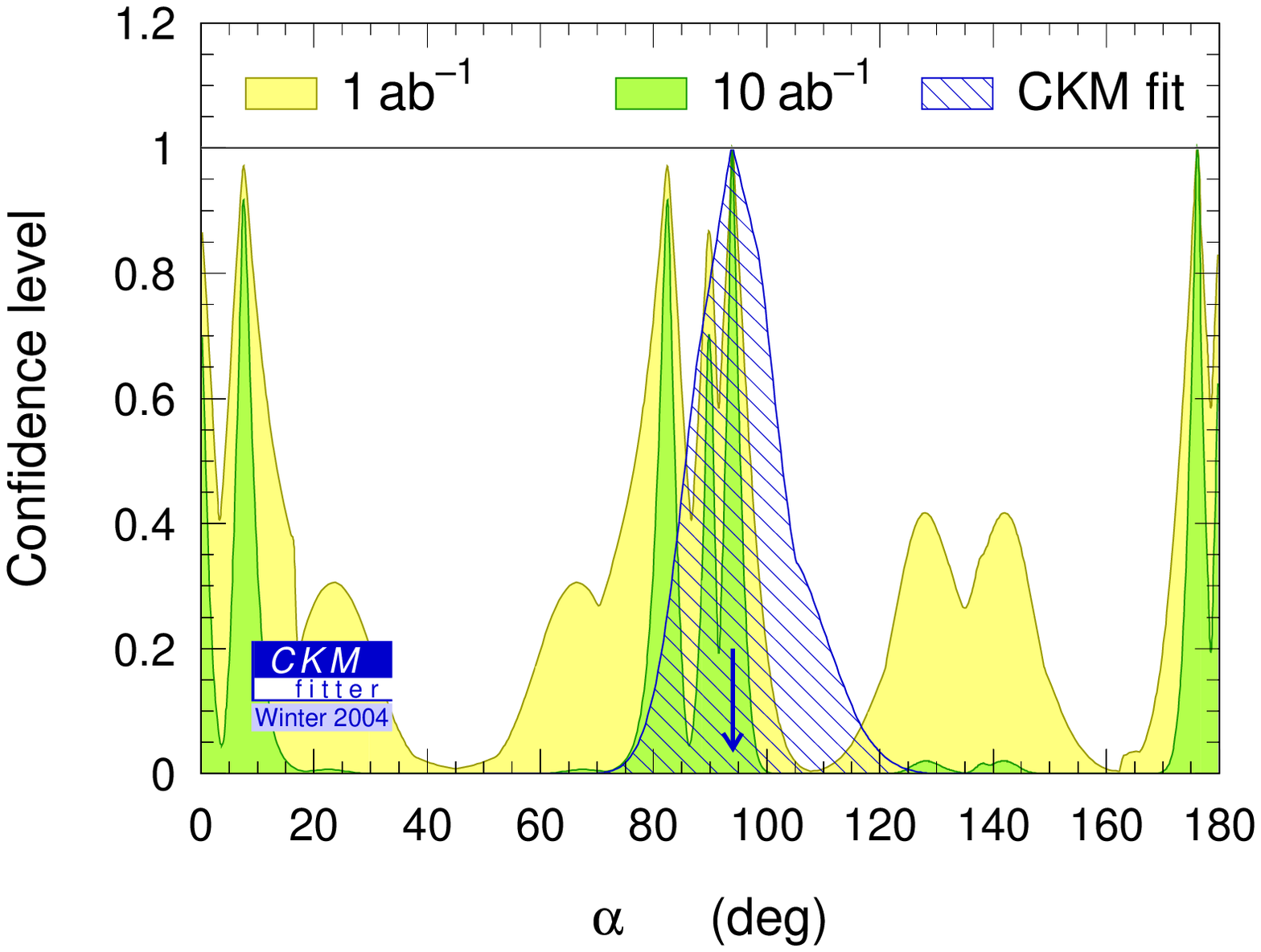}
        \epsfxsize8.1cm\epsffile{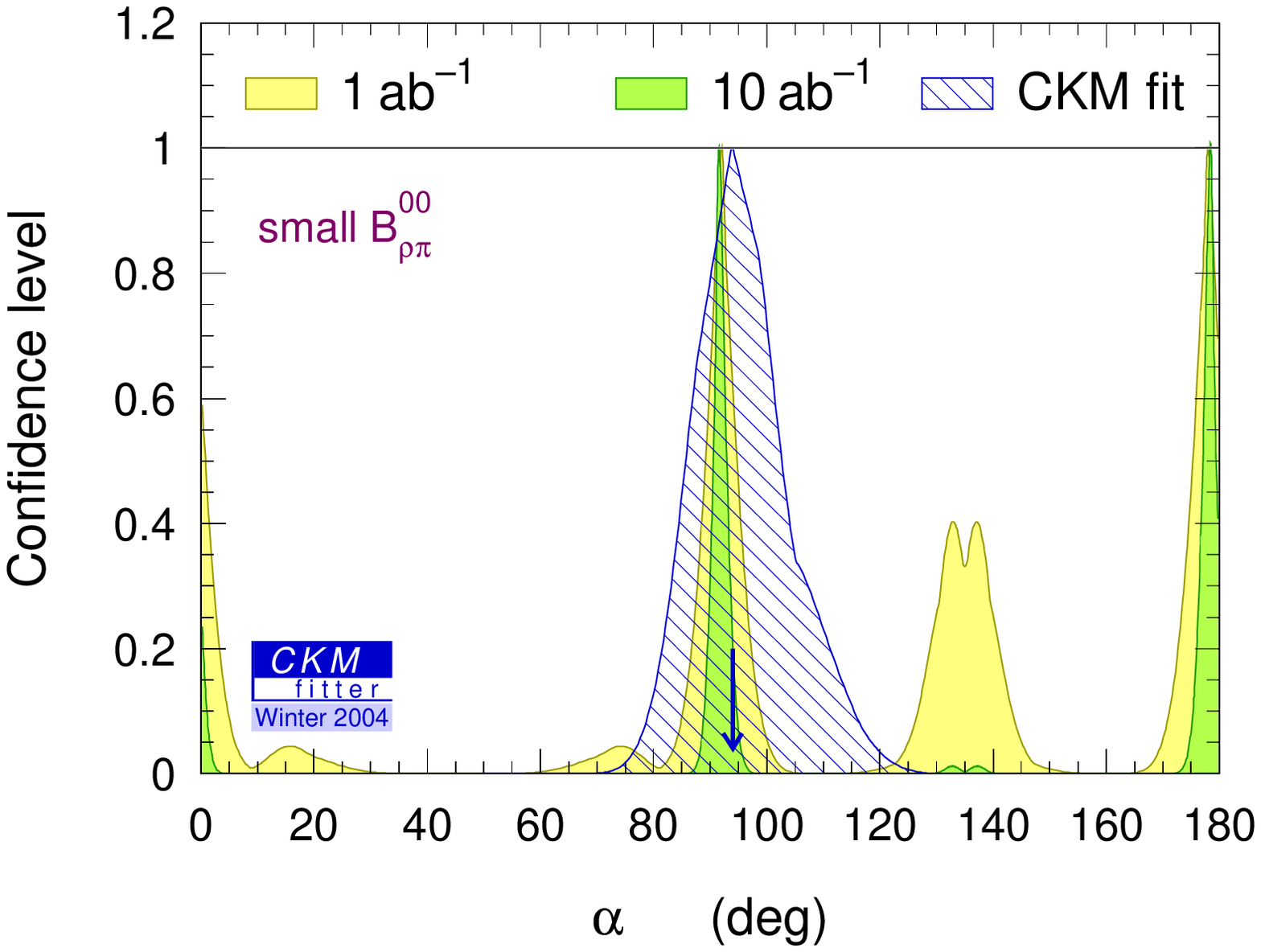}
             }
  \vspace{-0.4cm}
  \caption[.]{\label{fig:su2}\em
        \underline{Left:} 
        constraint on $\alpha$ using the full $B\to\rho\pi$ isospin analysis
        and assuming projections into future integrated luminosities of
        $1\invab$ and $10\invab$.
        The assumptions made on the generating amplitudes are given in
        the text. The arrow indicates the true value of $\alpha$ used
        for the generation of the toy observables. The hatched area shows 
        the constraint obtained from the present standard CKM fit. 
        \underline{Right:} same as left hand plot but with a ten times 
        smaller $\BRzz$.}
\end{figure}
To perform an educated study of the full isospin analysis, we assume 
$\alpha=94^\circ$ and a set of generating decay amplitudes
chosen (arbitrarily) to approximately reproduce the experimental results given 
in Table~\ref{tab:BRRhoPicompilation}. We obtain from these amplitudes the
inputs
{\small
\beqns
\begin{array}{rclrclrcl}
    \BRall  &=&                22.8 \pm 0.82~[0.26]~,   &
    \Acp    &=&               -0.12 \pm 0.025~[0.13]~,  &
    \Czz    &=&     \phantom{-}0.66 \pm 0.162~[0.056]~,\\[0.1cm]
    \BRzz   &=&     \phantom{2}1.2  \pm 0.19~[0.06]~,   &
    \Crhopi &=&     \phantom{-}0.41 \pm 0.052~[0.029]~, &
    \Szz    &=&               -0.69 \pm 0.222~[0.072]~,\\[0.1cm]
    \BRpz   &=&     \phantom{2}8.6  \pm 0.76~[0.24]~,   &
    \dCrhopi&=&     \phantom{-}0.20 \pm 0.052~[0.029]~, &
    \Acppz  &=&               -0.29 \pm 0.051~[0.028]~, \\[0.1cm]
    \BRzp   &=&                16.5 \pm 0.40~[0.13]~,   &
    \Srhopi &=&               -0.12 \pm 0.065~[0.026]~, &
    \Acpzp  &=&               -0.16 \pm 0.032~[0.013]~, \\[0.1cm]
    &&&
    \dSrhopi&=&     \phantom{-}0.30 \pm 0.065~[0.016]~,
\end{array}
\eeqns}$\!\!$
where the errors are extrapolated to integrated luminosities of 
$1\invab$ $[10\invab]$. The statistical errors are assumed to scale with the 
inverse of the square-root of the integrated luminosity. The systematic 
uncertainties are dominated by the limited knowledge of the
backgrounds arising from other \B decays. Because this knowledge, however,
improves when more data become available, the related uncertainty is assumed
to decrease like the statistical errors.
We neglect possible irreducible systematics from tracking and neutral reconstruction 
efficiencies or other effects. 
For the \CP parameters, we assume the systematics decrease with 
the square root of the luminosity up to $1\invab$ due to an improved 
knowledge of the \CP content of the primary \B-background modes, but then 
do not decrease any further since unknown effects, like \CP
violation on the tag side, become dominant.
\vs
We derive the constraints on $\alpha$ shown in the left hand plot
of Fig.~\ref{fig:su2}. A wide range of solutions exists besides 
the true value $\alpha$, indicated by the arrow. 
The right hand plot of Fig.~\ref{fig:su2} shows the CLs obtained
for $\alpha$ when the branching fraction of $\Bz\to\rho^0\pi^0$ is 
below the experimental sensitivity ($\BRzz=0.1\times10^{-6}$). The 
constraint improves compared to the previous case. 
\vs
The conclusions drawn from this exercise are similar to what has been 
observed in $\Bz\to\pip\pim$ decays: unless the branching fraction 
$\BRzz$ is very small (even smaller than expected from the
color-suppression mechanism), very large statistics is needed to 
significantly constrain $\alpha$ from
$\rho\pi$ data alone using the quasi-two-body isospin analysis. 
It is effectively beyond the reach of the first generation \B factories.

\subsection{SU(3) Flavor Symmetry}
\label{sec:su3}

Similarly to the studies in $\Bz\to h h^\prime$ decays, one can use
SU(3) flavor symmetry and dynamical hypotheses
to obtain additional information on the 
penguin amplitudes contributing to $\Bz\to\rho^\pm\pi^\mp$.

\subsubsection{Estimating $|\Ppm|$ and $|\Pmp|$
                from  $B^0\to \rho^- K^+$ and 
                $B^0\to K^{*+}\pi^-$}

\begin{figure}[t]
  \epsfxsize8.5cm
  \centerline{\epsffile{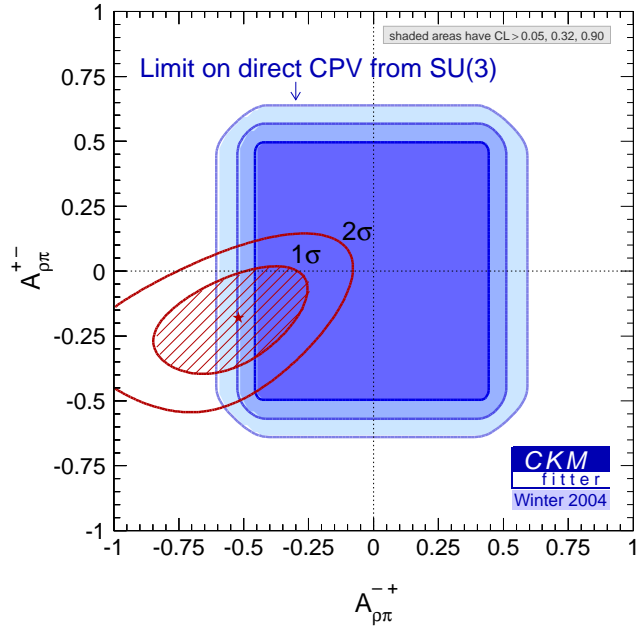}}
   \vspace{-0.0cm}
   \caption[.]{\label{fig:rhopi_su3_apm_2d}\em
        The direct \CP-violating asymmetries $\Acppm$ versus $\Acpmp$
        bound using SU(3) flavor symmetry and the measured flavor-specific
        branching ratios~(\ref{eq:summedBR}) as the 
        only input. Dark, medium and light shaded areas have $CL>0.90$, 
        $0.32$ and $0.05$, respectively. Also shown are the experimental 
        values found by \babar. }
\end{figure}

As proposed in Ref.~\cite{theseJerome}, the penguin amplitudes in
$B^0\to\rho^\pm\pi^\mp$ can be more effectively constrained with the use of the 
corresponding charge states in $b\to u\bar u s$ transitions, namely the
decays $B^0\to K^{*+}\pi^-$ and $B^0\to\rho^-K^+$ (Scenario~(\iB))
for which the amplitudes read
\beq
\label{eq:b0Kpm}
\begin{array}{rcccl}
        A_{K^*\pi}^{+-} &\equiv&  A(B^0\to K^{*+}\pi^-)
                        &=&   V_{us}V_{ub}^{*}T_{K^*\pi}^{+-}
                            + V_{ts}V_{tb}^{*} P_{K^*\pi}^{+-}~, \\
\label{eq:b0Kmp}
        A_{\rho K}^{-+} &\equiv&  A(B^0\to \rho^- K^+)
                        &=&   V_{us}V_{ub}^{*}T_{\rho K}^{-+}
                            + V_{ts}V_{tb}^{*}P_{\rho K}^{-+}~.
\end{array}
\eeq
Under the assumption of SU(3) flavor symmetry, and neglecting
OZI-suppressed penguin annihilation diagrams (see right hand
diagram in Fig.~\ref{fig:B0KppimP}), which contribute
to $B\to\rho\pi$ but not to $B\to \rho K,K^*\pi$, the 
penguin amplitudes in Eq.~(\ref{eq:b0Kpm}) and those entering 
$\Apm$ ($\Amp$) are equal (Scenario~(\iB)):
\begin{figure}[t]
  \centerline{\epsfxsize8.1cm\epsffile{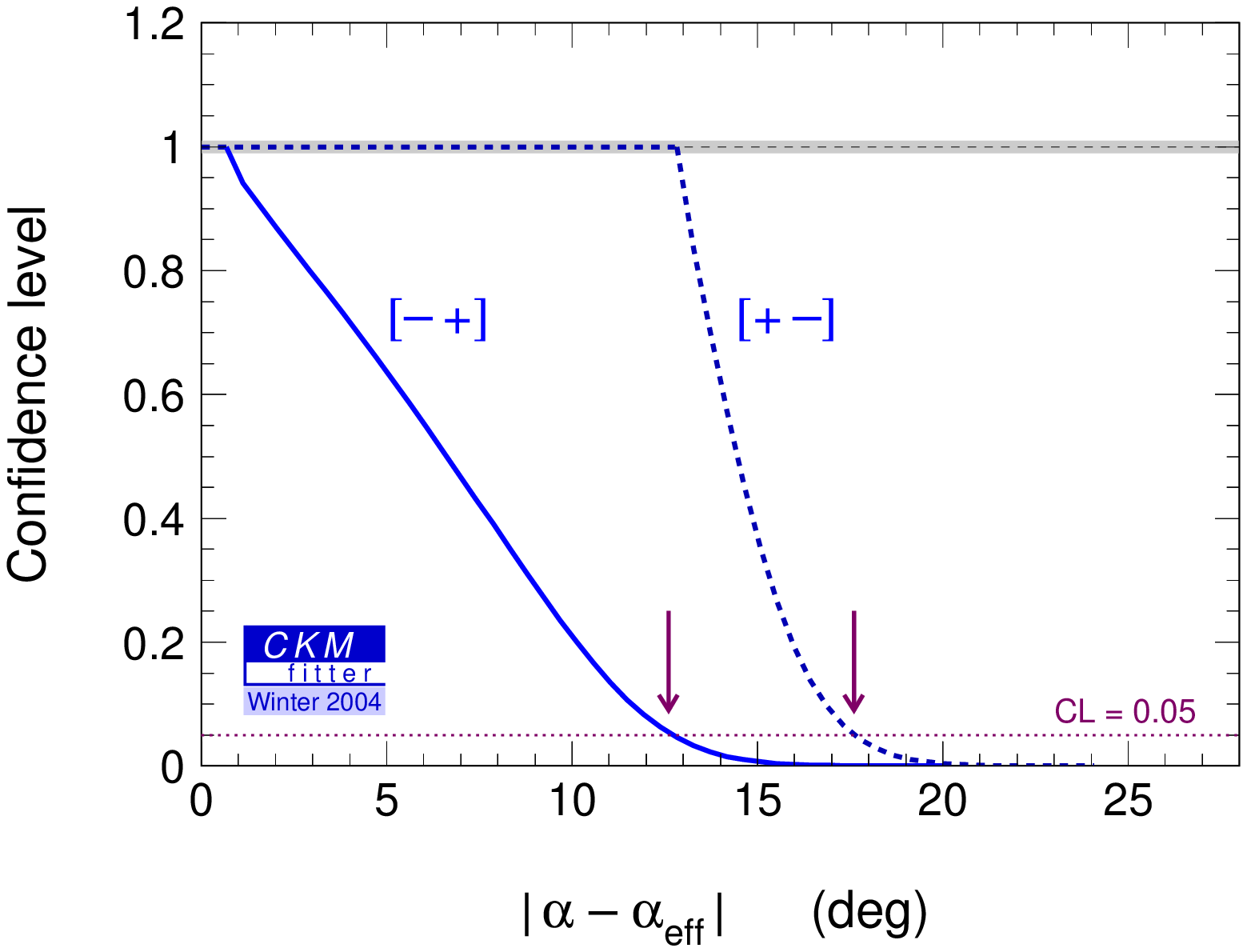}
              \epsfxsize8.1cm\epsffile{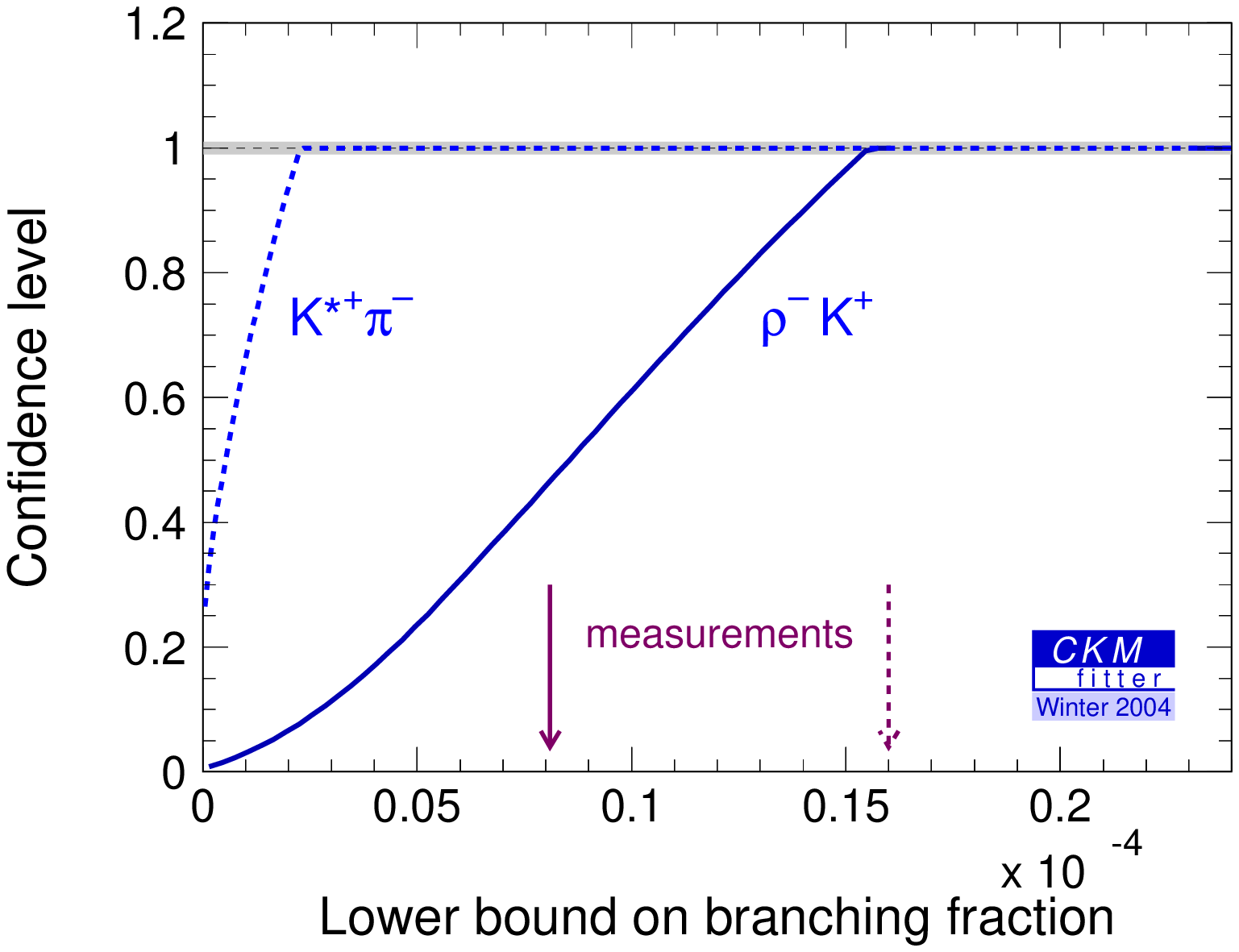}}
  \vspace{-0.4cm}
  \caption[.]{\label{fig:boundspm}\em
        \underline{Left:} 
        Confidence levels for the SU(3) bounds~(\ref{eq:SU3boundpm}) 
        (dashed line) and (\ref{eq:SU3boundmp}) (solid line). 
        The arrows indicate the $CL=0.05$
        crossing values given in Eq.~(\ref{eq:su3bounds}).
        \underline{Right:}
        Lower bounds on the branching fractions $\BRKstrpi$ (dashed
        line) and $\BRrhoK$ (solid line) obtained with the use of SU(3)
        symmetry and inserting $\alpha$ from the standard CKM fit.
        The arrows indicate the corresponding experimental
        values (same line types). }
\end{figure}
\beq
\label{eq:PpiPK}
        \Ppm\:=\:P_{K^*\pi}^{+-}~,\hspace{1cm}
        \Pmp\:=\:P_{\rho K}^{-+}~.
\eeq
This leads to the bounds~\cite{theseJerome}
\beqn
\label{eq:SU3boundpm}
    |\alpha - \alphaeffpm|
    &\le&\frac{1}{2}\arccos\left[\frac{1}{\sqrt{1-{\Acppm}^2}}
                \left(1-2\lambda^2\frac{\BRKstrpi}{\BRpm}\right)
              \right]~, \\
\label{eq:SU3boundmp}
    |\alpha - \alphaeffmp|
    &\le&\frac{1}{2}\arccos\left[\frac{1}{\sqrt{1-{\Acpmp}^2}}
                \left(1-2\lambda^2\frac{\BRrhoK}{\BRmp}\right)
              \right]~,
\eeqn
where $\lambda$ is the Wolfenstein parameter. Compared to the 
bound~(\ref{eq:SU2bound}), the above SU(3) bounds benefit
from the relative CKM enhancement (suppression) of the 
penguin (tree) amplitudes in the strange modes with respect 
to the $ b\to u$ transitions.  The left-hand plot of
Fig.~\ref{fig:boundspm} shows the CLs for $|\alpha - \alphaeffpm|$ 
and $|\alpha - \alphaeffmp|$ obtained with the use of the results 
for the branching fractions given in Table~\ref{tab:BRRhoPicompilation}.
At $95\%$~CL, we find\footnote
{
        Note that the numerical analysis performed with \ckmfitter\  
        does not explicitly involve 
        Eqs.~(\ref{eq:SU3boundpm}) and (\ref{eq:SU3boundmp}), 
        since the full amplitude parameterizations are implemented.
        Fitting all experimental results to these amplitudes, and
        using the SU(3) constraints~(\ref{eq:PpiPK}), automatically
        reproduces the analytical bounds. 
}
\beq
\label{eq:su3bounds}
     |\alpha - \alphaeffpm| < 17.6^\circ~,\hspace{1cm}
     |\alpha - \alphaeffmp| < 12.6^\circ~,
\eeq
which are more restrictive than the corresponding bounds obtained in
the $\pi\pi$ system (see Section~\ref{sec:charmlessBDecays}.\ref{sec:resultsPipi})
\vs
Owing to the fact that non-zero direct CPV requires sizable
penguin contributions,
we can reverse the procedure and infer lower limits on $\BRKstrpi$
and $\BRrhoK$, using SU(3) and inserting $\alpha$ from the standard
CKM fit. The CLs obtained for both branching fractions 
are shown in the right hand plot of Fig.~\ref{fig:boundspm}. The 
arrows indicate the corresponding measurements. As expected, the 
relatively large \CP  asymmetry $\Acpmp$ requires an increased 
$\BRrhoK$, while no useful lower bound is obtained for $\BRKstrpi$. 
Although compatible within the rather large experimental errors 
($\CL=0.46$), one can conclude from this observation that 
if the SU(3) picture holds within the SM, improved statistics is expected 
to give a lower value of $|\Acpmp|$ than the current one, since 
$\BRrhoK$ is already known to good precision.
Following the same line, we have attempted to obtain a lower bound
on the branching fraction $\BRzz$, which is however insignificant
at present~\cite{lalrhopi}.
In turn, we have attempted in Fig.~\ref{fig:rhopi_su3_apm_2d} to use 
the \CP-conserving branching fractions $\BRpm$ and $\BRmp$ as the only 
experimental input (or equivalently $\BRall$, $\dCrhopi$ and the 
\CP-conserving product $\Crhopi\cdot\Acp$) as well as their SU(3) 
partners $\BRKstrpi$ and $\BRrhoK$, to infer bounds\footnote
{
        The analytical SU(3) bounds on the direct \CP  asymmetries read
        \beqns
        \Acppm < \sqrt{1 - \left(1-2\lambda^2\frac{\BRKstrpi}{\BRpm}\right)^{\!2}}~,
        \hspace{1cm}
        \Acpmp < \sqrt{1 - \left(1-2\lambda^2\frac{\BRrhoK}{\BRmp}\right)^{\!2}}~.
        \eeqns
} on 
$\Acppm$ and $\Acpmp$. Since branching fractions are not sensitive to 
the sign of direct CPV, the figure is symmetric around the zero axes. 
We determine the allowed domains ($\CL>0.05$)
\begin{figure}[t]
  \centerline{
        \epsfxsize8.2cm\epsffile{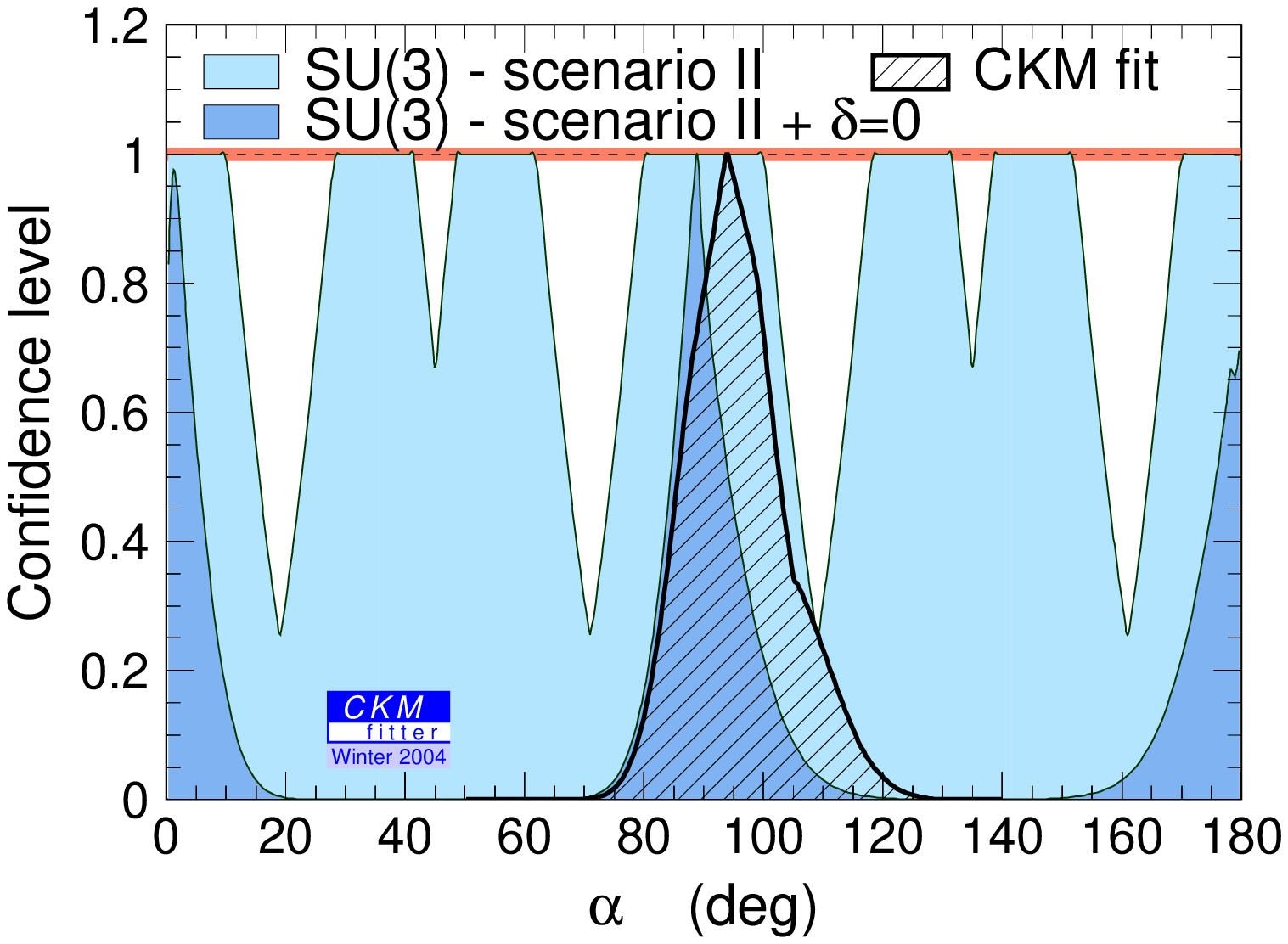}
        \epsfxsize8.2cm\epsffile{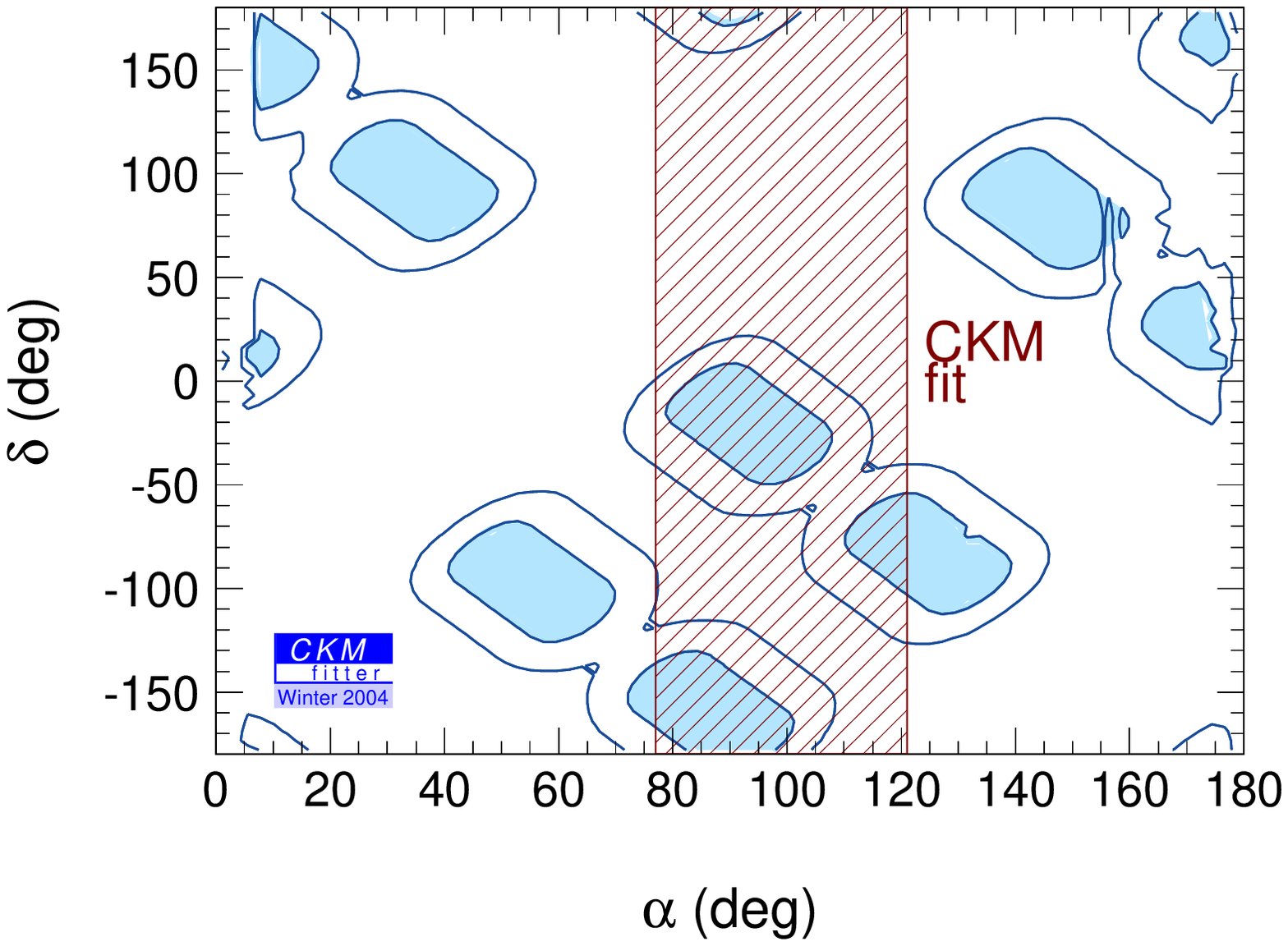}
  }
  \vspace{-0.8cm}
  \caption[.]{\label{fig:su3}\em
        \underline{Left:} confidence levels obtained for $\alpha$ 
        with the use of SU(3) flavor symmetry within Scenario~(\iB).
        The light shaded curve gives the nominal constraint,
        while the dark shaded one corresponds to
        the solutions one would obtain if the
        relative strong phase $\delpmmp$ were known and equal to zero.
        The hatched area shows the
        constraint obtained from the CKM fit using the standard
        constraints (\cf\  Part~\ref{sec:standardFit}).
        \underline{Right:} correlation between $\alpha$ and $\delpmmp$.
        The shaded areas indicate $CL=0.32$ domains and the solid lines
        show the $CL=0.05$ regions. The periodicity is
        $\Delta \alpha=45^\circ$ and $\Delta\delpmmp=90^\circ$.
        The hatched area depicts the
        ${CL}\le0.05$ allowed region for $\alpha$ obtained from
        the CKM fit using the standard constraints
        (\cf\  Part~\ref{sec:standardFit}).}
\end{figure}
\beq
        |\Acppm| < 0.64~,\hspace{1cm}
        |\Acpmp| < 0.59   ~.
\eeq

Using the results given in Table~\ref{tab:BRRhoPicompilation} and
the relations~(\ref{eq:PpiPK}), we can set CLs for
$\alpha$. We obtain six ambiguous solutions shown by the light shaded 
region in the left hand plot of Fig.~\ref{fig:su3}. The widths of the 
plateaus represent the uncertainties $|\alpha -\alpha_{\rm eff}^{+-(-+)}|$
determined by the bounds~(\ref{eq:SU3boundpm}) and~(\ref{eq:SU3boundmp}).
The bounds on $|\alpha -\alpha_{\rm eff}^{+-(-+)}|$ are not good
enough to resolve all the eight ambiguities,  which are
partially merged. The $\chi^{2}_{\rm min}=0.3$ for the best fit is 
satisfactory. Fixing arbitrarily the penguin amplitudes to zero 
results in the significantly worse $\chi^{2}_{\rm min}=8.8$. Also
shown in the figure is the solution obtained when fixing the 
relative strong phase $\delpmmp$ to zero. Good agreement with the 
standard CKM fit is observed.
\vs
It is interesting to correlate $\alpha$ with the relative strong
phase $\delpmmp$ between $\Amp$ and $\Apm$ (\cf\
Eq.~(\ref{eq:saeffanddeltapm})).
The corresponding CLs are shown in the right hand plot of
Fig.~\ref{fig:su3}. We observe a structure of distinctive
islands, and, when using the SM constraint (\cf\  Table~\ref{tab:fitResults1})
on $\alpha$, we conclude that values of $\delpmmp=0,\pi$ are
preferred, one of which ($\delpmmp=0$) is in conformity with 
expectations from factorization.
If $\delpmmp$ were given by theory or determined experimentally through
a Dalitz plot analysis~\cite{SnyderQuinn}, SU(3) symmetry would result in
a useful constraint on $\alpha$. 

\subsubsection{Estimating $|\Ppm|$ and $|\Pmp|$
                from $B^+\to \rho^+ K^0$ and 
                $B^+\to K^{*0}\pi^+$}

The magnitude of the penguin amplitudes $|\Ppm|$ and $|\Pmp|$ can be 
estimated from the branching fraction of the penguin-dominated decays
$B^+\to \rho^+ K^0$ and $B^+\to K^{*0}\pi^+$ (Scenario~(\iC)). Neglecting 
the doubly CKM-suppressed $u$ penguins and annihilation diagrams, the 
transition amplitudes for these modes are given by
\beqn
\label{eq:ampChargedPV}
        \Apz_K\equiv A(B^{+}\to \rho^+ K^{0})   
                &=& V_{tb}^{*}V_{ts}\PrhoKz~,\nonumber\\
        \Azp_K\equiv A(B^{+}\to K^{*0} \pi^{+}) 
                &=& V_{tb}^{*}V_{ts}\PKstzpi~.
\eeqn
Correspondingly to the relation~(\ref{eq:relPPcharged}) and within the
same hypotheses, one has
\beqn
        |\Ppm| &=&  \frac{1}{\sqrt{r_{\tau}}}
                \frac{f_{\rho}}{f_{K^*}} \frac{1}{\RqcdP}
                |\PKstzpi|,\nonumber\\
        |\Pmp| &=&  \frac{1}{\sqrt{r_{\tau}}}
                \frac{f_{\pi}}{f_{K}} \frac{1}{\RqcdM}
                |\PrhoKz|, 
\eeqn
where the correction factors $\RqcdP$ and $\RqcdM$ are both fixed to 
$R = 0.95 \pm 0.23$ (for simplicity we choose the same value as in 
Eq.~(\ref{eq:relPPcharged})), but they vary independently in the fit within 
their theoretical errors. For the $\rho(770)$ and the $K^*(892)$ decay 
constants we use $f_\rho=(209 \pm 1)\mev$ and $f_{K^*}=(218 \pm 4)\mev$~\cite{BN},
respectively.
\vs
The branching fractions of these modes are given 
in Table~\ref{tab:BRRhoPicompilation}.
Figure~\ref{fig:II_IIItheo} shows the constraints obtained on $\alpha$
for Scenario~(\iC). 
Since only an upper limit exists for $B^+\to \rho^+ K^0$, for 
illustration purpose we assign the value (and error)
of $\BR(B^0\to \rho^- K^+)$
to that branching fraction. Note that the number of peaks doubles from 
eight to sixteen when using $B^+\to \rho^+ \Kz$ since it determines the 
magnitude of the penguin amplitude rather than an upper limit only (as 
does Scenario~(\iB)). We draw the following conclusions from this exercise:
\begin{figure}[t]
  \centerline
        {
        \hspace{-0.3cm}\epsfxsize8.4cm\epsffile{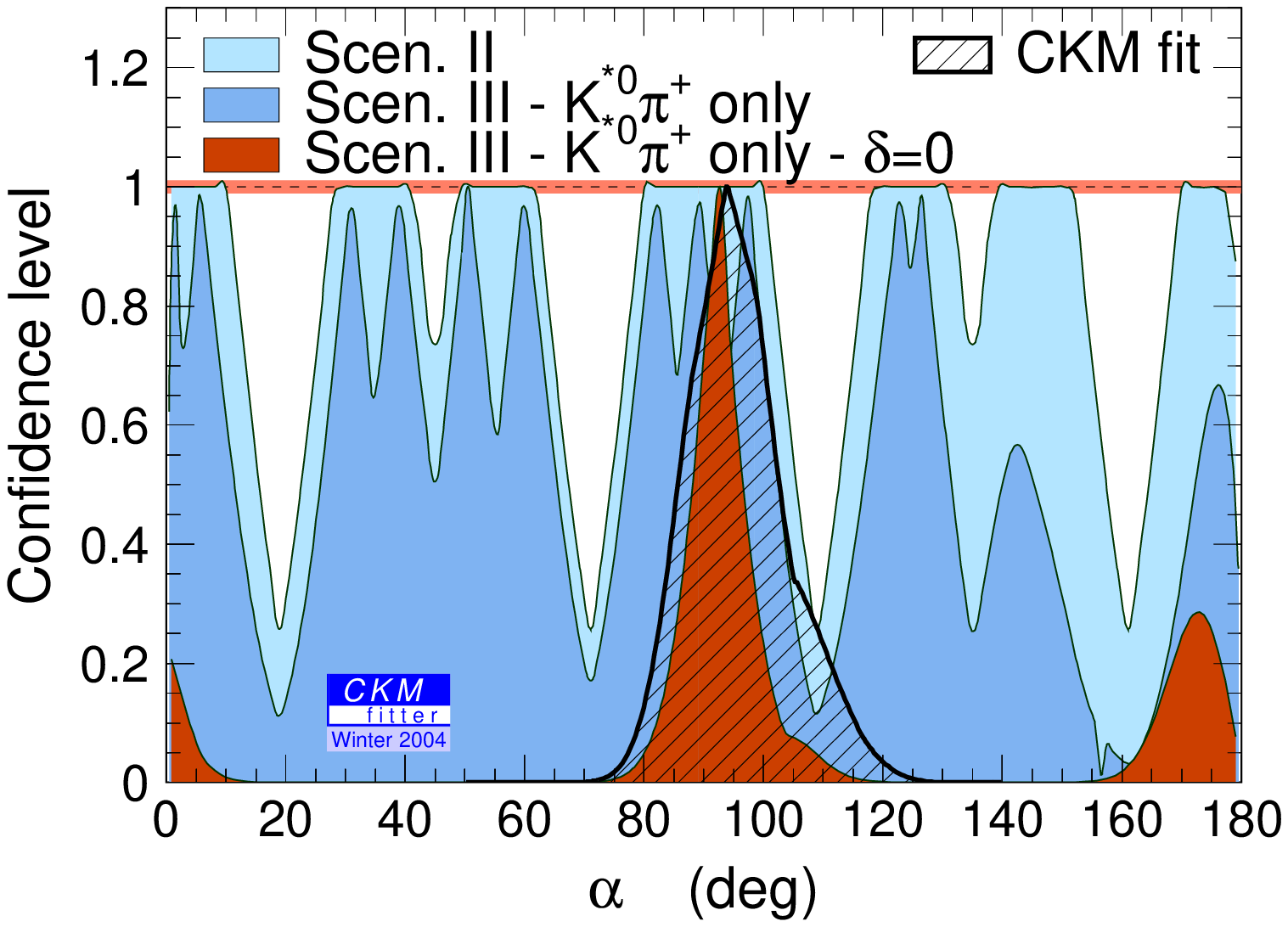}
        \hspace{-0.4cm}\epsfxsize8.4cm\epsffile{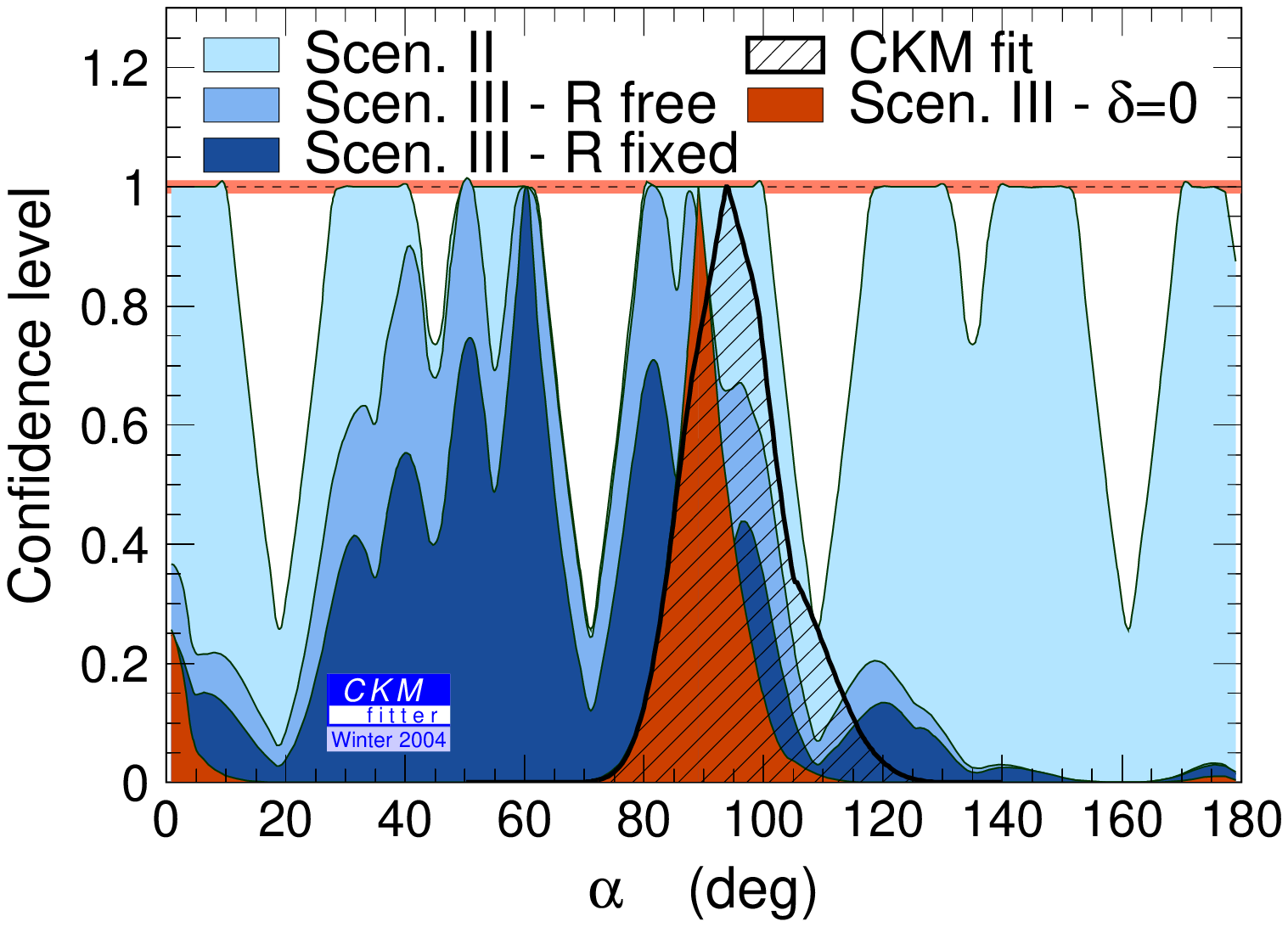}
        }
  \vspace{-0.8cm}
  \caption[.]{\label{fig:II_IIItheo}\em
        Confidence levels obtained for $\alpha$ with the use of SU(3) 
        flavor symmetry.
        \underline{Left}: Scenario~(\iB) (light shaded, same as in 
        left hand plot of Fig.~\ref{fig:su3}), Scenario~(\iC) but only
        using $B^+\to K^{*0}\pi^+$ to constrain $|\Ppm|$ (medium shaded),
        and Scenario~(\iC) as previous case and assuming in addition
        that the relative strong phase $\delpmmp$ is zero (dark shaded).
        \underline{Right}: Scenario~(\iB)  (light shaded, same as left plot); 
        the medium and medium-dark shaded curves correspond to 
        Scenario~(\iC) with SU(3) constraints on both penguins amplitudes
        from $B^+\to K^{*0}\pi^+$ and $B^+\to \rho^+ K^0$, where we assume 
        for that latter mode (which has not yet been seen) that the 
        branching fraction is equal to the one of $B^0\to \rho^- K^+$.
        Medium shaded is for $R$ that varies freely within its
        uncertainties, while medium-dark shaded is for fixed $R$.
        The dark shaded curve gives the result of the 
        previous case when assuming $\delpmmp=0$.       
        In both plots, the hatched area indicates the constraint 
        obtained from the standard CKM fit. }
\end{figure}
\begin{itemize}

\item   the constraint on $\alpha$ will be improved when the measurement
        of the branching ratio to $\rho^+ K^0$ is available.

\item   the uncertainties on the correction factors $\RqcdP$ and $\RqcdM$
        do not much degrade the constraints on $\alpha$, \ie, theoretical       
        uncertainties appear to be subdominant for Scenario~(\iC).

\item   in all SU(3) scenarios, a precise determination of $\alpha$ would
        be obtained if the relative strong phase $\delpmmp$ were known
        (for example from a Dalitz plot analysis~\cite{SnyderQuinn}
        or from theory~\cite{BN}), in which case the r\^ole of the
        errors on the ratios $R^\pm$ may become important.
        However, without this additional input, the constraints 
        obtained by the different approaches are insignificant.

\end{itemize}

%

 \section{Analysis of $\B\to\rho\rho$ Decays}
\label{sec:introductionrhorho}

The isospin analysis of $B \to \rho\rho$ decays leads to the extraction
of $\alpha$ in a way similar to $B \to \pi\pi$
decays~\cite{oldsnyderetal}.   The specific interest of these channels
lies in the potentially small penguin  contribution, which is
theoretically expected~\cite{aleksanRR} and  indirectly confirmed by
the smallness of the experimental upper bound on the $\Bz \to\rho^0\rho^0$ 
branching fraction~\cite{BABARrhoplusrho0}  with respect
to the branching fractions of $\Bz \to \rho^+\rho^-$ and  
$\Bp \to\rho^+\rho^0$~\cite{BABARrhoplusrho0,Bellerhoplusrho0,babarrhorhoprl,HFAG}.
In addition, and more importantly for a precision measurement, both
direct and mixing-induced \CP-violating asymmetries of the color-suppressed
decay $\Bz \to\rho^0\rho^0$ are experimentally accessible.
\vs
The analysis is complicated by the presence of three helicity states for
the two vector mesons. One corresponds to longitudinal polarization and
is \CP-even. Two helicity states are transversely polarized and  are
admixtures of \CP-even and \CP-odd amplitudes. 
In the transversity frame~\cite{transv}, one amplitude accounts for             
parallel transverse polarization and the other for perpendicular                
transverse polarization of the vector mesons, with respect to the
transversity axis: the first is \CP-even             
and the second \CP-odd. Hence the perpendicular polarization (if not            
identified or zero) dilutes the \CP measurement.
By virtue of the helicity suppression, the fraction of transversely polarized 
$\rho$ mesons is expected to be of the order of 
$(\Lambda_\mathrm{QCD}/m_B)^2\sim2\%
$ in the factorization approximation~\cite{BN,kaganVV} supplemented by
the symmetry relations between heavy-to-light form factors in the
asymptotic limit~\cite{LEET}. This has 
been confirmed by experiment in 
$\Bp\to\rho^+\rho^0$~\cite{BABARrhoplusrho0,Bellerhoplusrho0}
and
$\Bz\to\rho^+\rho^-$~\cite{babarrhorhoprl} decays that are both found to 
be dominantly longitudinally polarized. As a consequence, one is allowed to
restrict the SU(2) analysis described in 
Section~\ref{sec:charmlessBDecays}.\ref{sec:su2pipi},
without significant loss of sensitivity, to the longitudinally polarized 
states of the $\B\to\rho\rho$ system. This also applies to the treatment of 
electroweak penguins.
\vs
The isospin analysis relies on the separation of the tree-level
amplitudes ($I=0$ and $I=2$) from the penguin-type amplitudes ($I=0$),
since Bose  statistics ensures that no odd isospin amplitude is present
in two  identical meson final states. It has been pointed out in 
Ref.~\cite{ligetirhorho} that due to the finite width of the $\rho$
meson,  $I=1$ contributions can occur in $\B\to \rho\rho$ decays.
Although no  prediction is made, one may expect these to be of the
order of  $(\Gamma_\rho/m_\rho)^2\sim4\% 
$. In the following, we first neglect $I=1$  contributions, and later 
present a preliminary study including these effects
in Section~\ref{sec:charmlessBDecays}.\ref{sec:SU2strongbreaking}. Also
neglected is isospin violation due to the strong interaction as well
as  effects from the interference with the radial excitations of the 
$\rho$, with other $\pi\pi$ resonances and with a non-resonant
component; in the future these effects may be studied by the
experiments since they depend on the specific analysis requirements. 

\subsection{Theoretical Framework and Experimental Input}

Due to the lack of experimental information, an SU(3) analysis is not
feasible at present (the branching fractions of the SU(3) partners
$\Bz\to\Kstarp\rho^-$ and the penguin-dominated $\Bp\to\Kstarz\rho^+$
have not been published yet). We hence restrict the numerical analysis
to isospin symmetry corresponding to Scenario~(\iA) (\cf\ 
Section~\ref{sec:charmlessBDecays}.\ref{sec:theoFrame}), which is 
however significantly more fruitful than for the $\B\to\pi\pi$ system.
\vs
The experimental results used here are collected in 
Section~\ref{sec:standardFit}.\ref{sec:input_rhorho}. The main ingredient
is the measurement of $\staeff$ from the time-dependent \CP and polarization 
analysis of $\Bz\to\rho^+\rho^-$ decays performed by 
\babar~\cite{BABARrhorho,babarrhorhoprl}. 

\subsection{Numerical Results}

\begin{figure}[t]
  \centerline{\epsfxsize8.1cm\epsffile{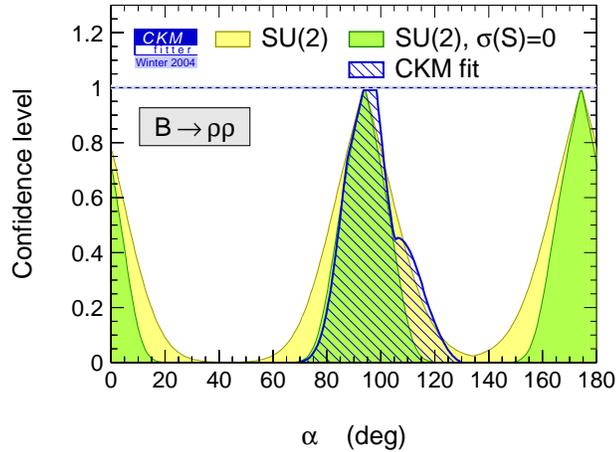}}
  \vspace{-0.5cm}
  \caption[.]{\label{fig:AlphaRhorho}\em
        Confidence levels for $\alpha$ from the SU(2) analysis of the 
        $B \to \rho\rho$ system (light shaded). The dark-shaded function 
        is obtained by setting the error on $S_{\rho\rho,L}^{+-}$
        to zero. It hence represents the
        uncertainty due to the penguin contribution ($\deltaAlpha$). Also  
        shown is the prediction from the standard CKM fit (hatched area),
        which includes the world average for $\stb$ but ignores 
        the $B \to \rho\rho$ data.
}
\end{figure}

\begin{figure}[p]
  \centerline
        {
              \epsfxsize8.1cm\epsffile{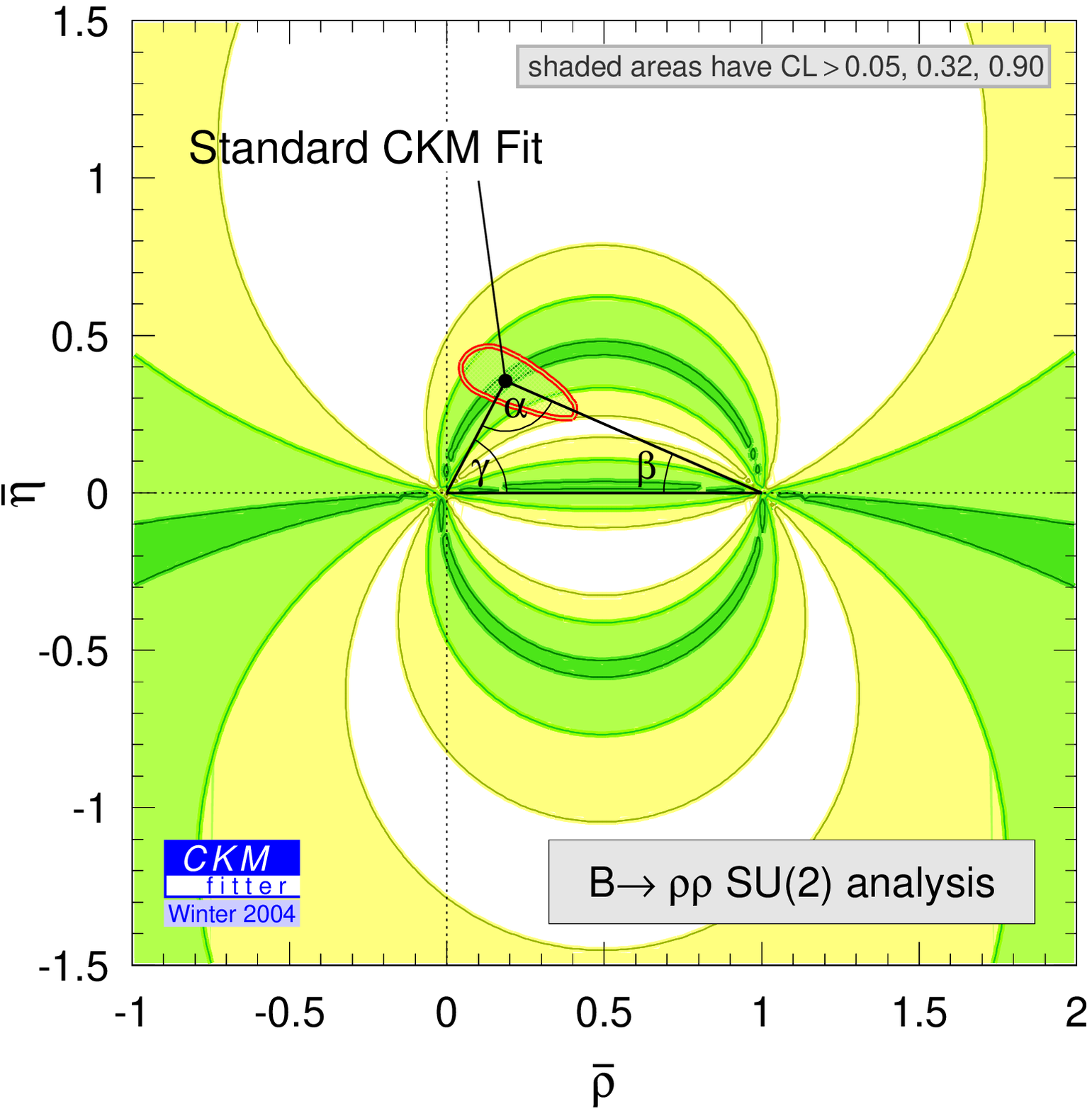}
              \epsfxsize8.1cm\epsffile{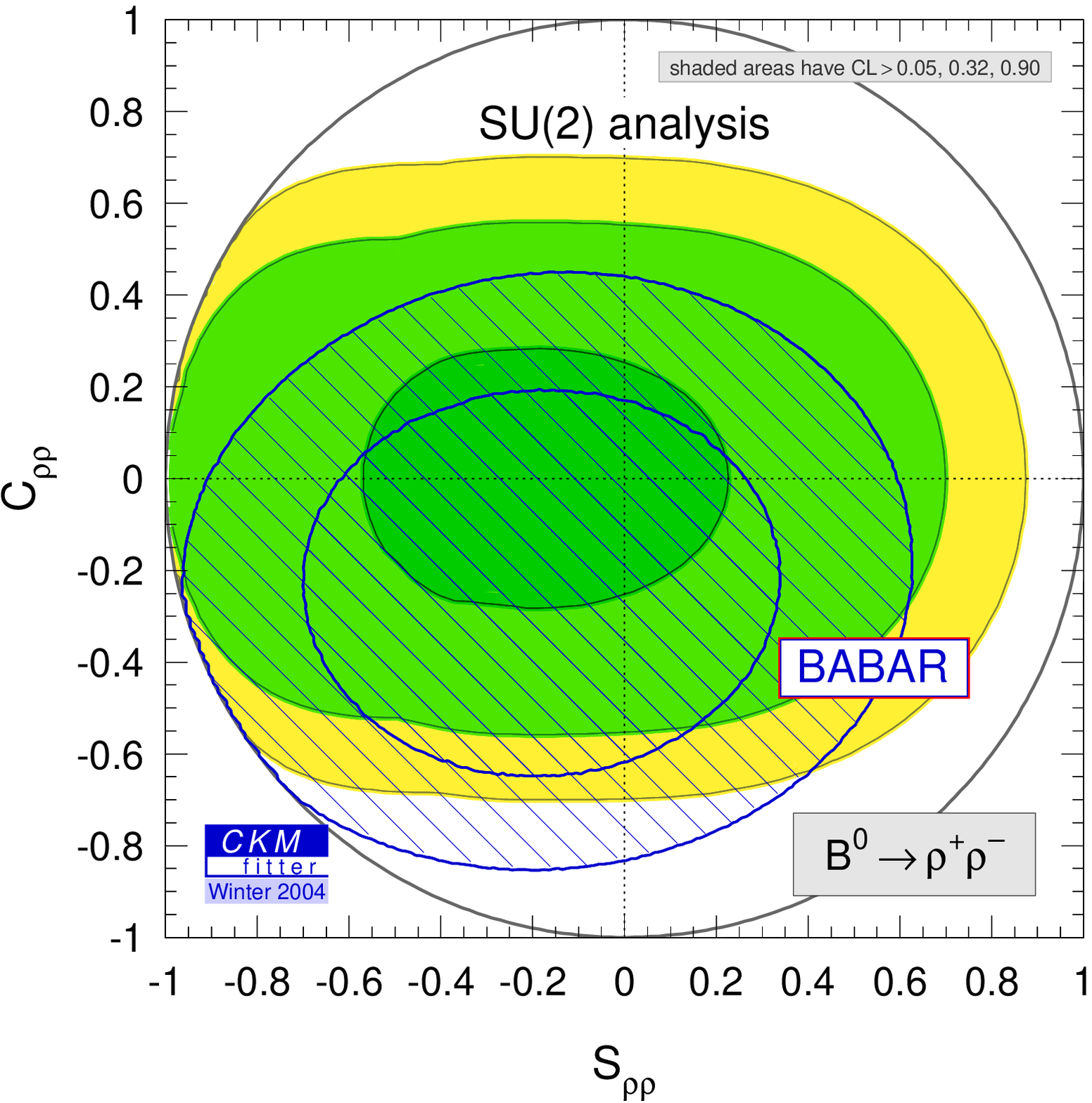}
        }
  \vspace{-0.0cm}
  \caption[.]{\label{fig:scrhorho}\em
        Confidence levels from the SU(2) analysis of the  $B \to \rho\rho$
        system. Dark, medium and light shaded areas 
        have $CL>0.90$, $0.32$, $0.05$, respectively.
        \underline{Left:} constraints in the $\rhoeta$ plane. 
        Overlaid is the prediction from the standard CKM fit,
        which includes the world average for $\stb$ but not
        the $B \to \rho\rho$ data. 
        \underline{Right:} constraints in the 
        ($S_{\rho\rho,L}^{+-},C_{\rho\rho,L}^{+-}$) plane, with the
        input values for $\rhobar$ and $\etabar$ taken from the
        standard CKM fit. The circle 
        indicates the physical limit 
        $S_{\rho\rho,L}^{+-}{}^2+C_{\rho\rho,L}^{+-}{}^2=1$, and the 
        hatched area represents the CL of the \babar\  measurement for 
        which the presence of the physical boundary has been properly 
        taken into account (see 
        Section~\ref{sec:statistics}.\ref{sec:metrology_physicalBoundaries}).
        The contours correspond to $1\sigma$ and $2\sigma$,
        respectively. }
  \vspace{1cm}
  \centerline
        {
        \epsfxsize8.0cm\epsffile{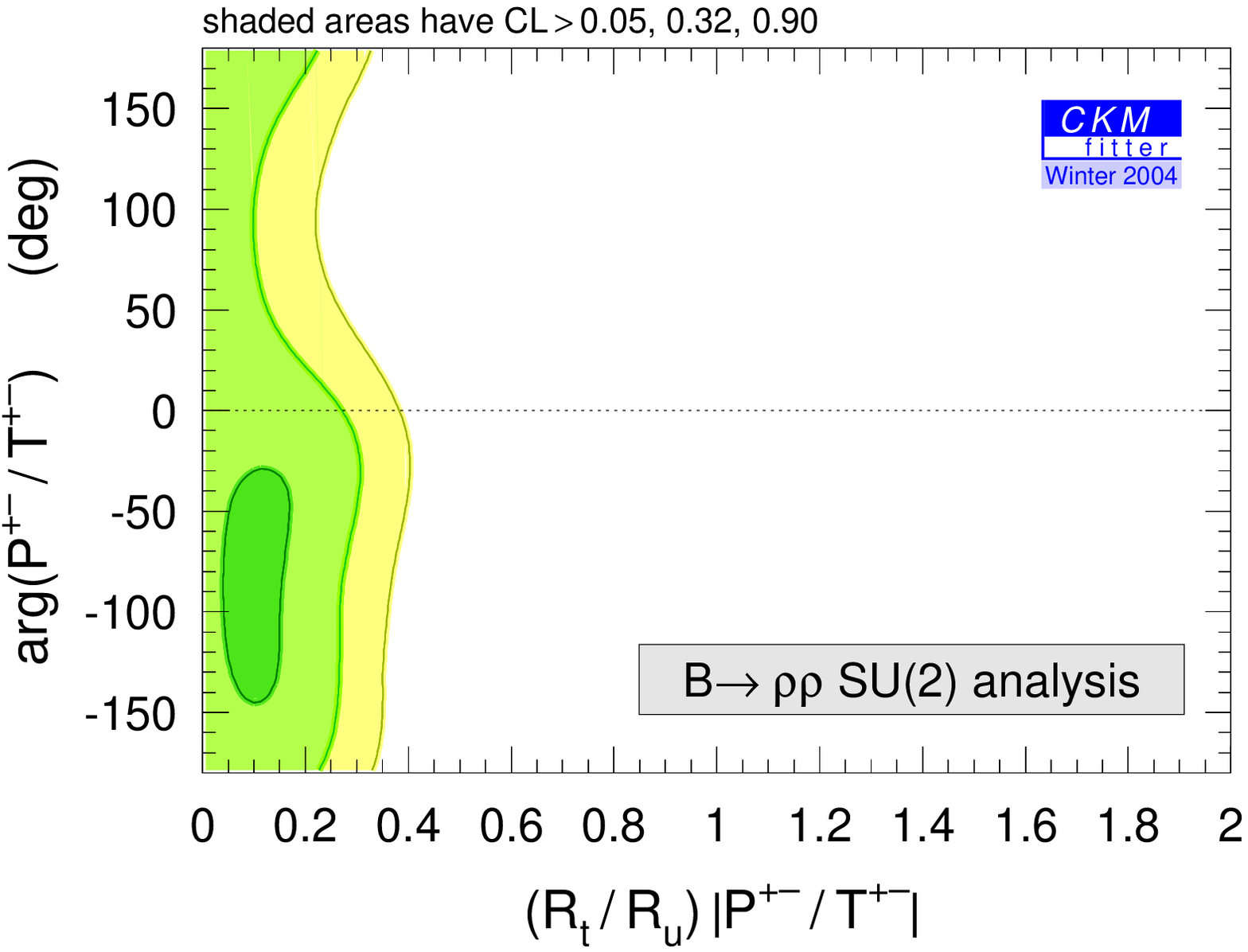}
        \epsfxsize8.0cm\epsffile{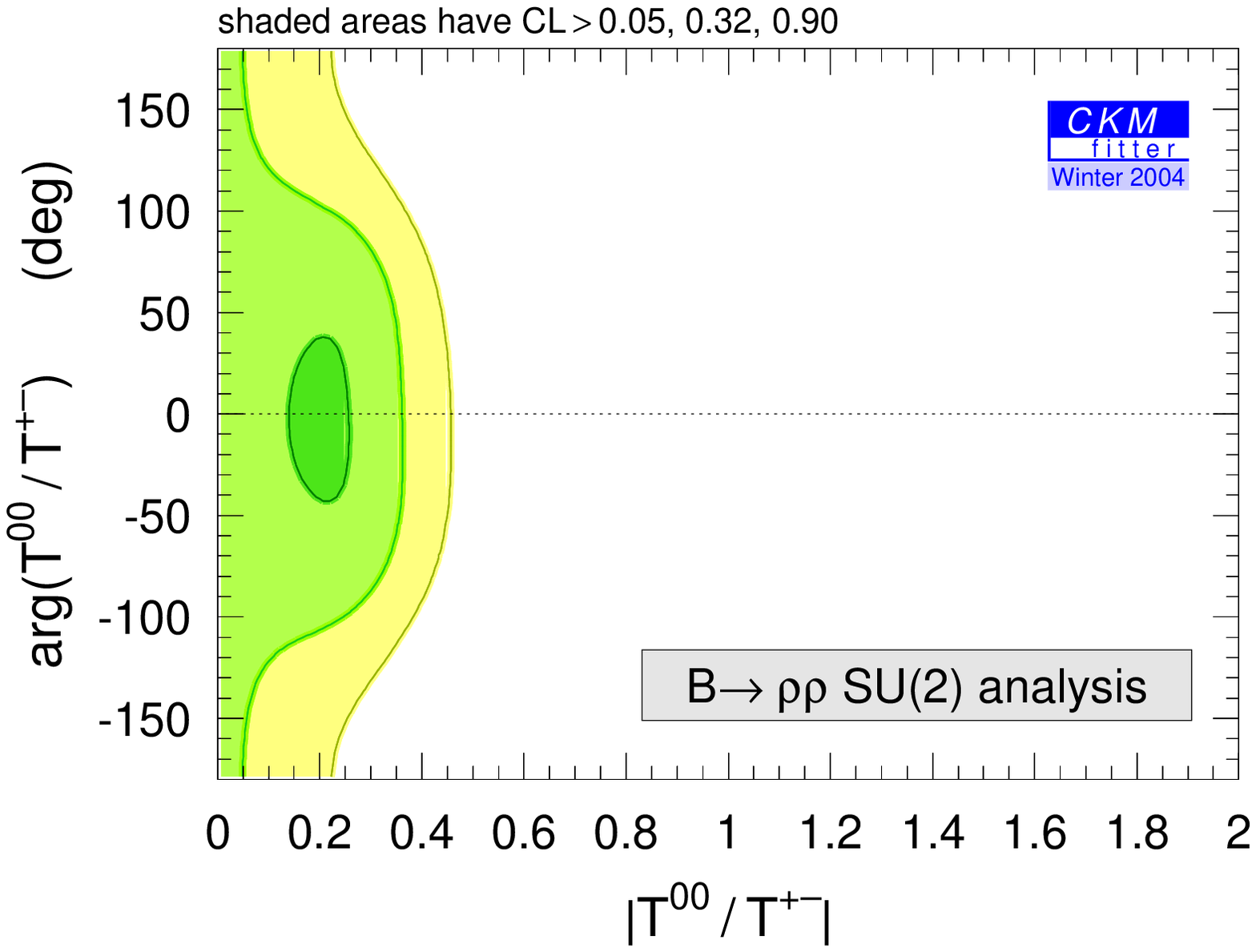}
        }
  \vspace{-0.0cm}
  \caption[.]{\label{fig:ResRhorho}\em
        Constraints on the penguin-to-tree ratio 
        $(R_t/R_u)|\Ppipi/\Tpipi|$ and
        the relative strong phase $\arg[P^{+-}T^{+-*}]$ (left), and on the 
        color-suppressed to color-allowed ratio $|\Tcpizpiz/\Tpipi|$ and the 
        relative strong phase $\arg[\Tcpizpiz T^{+-*}]$ (right), obtained from the 
        SU(2) analysis and using the result from the standard CKM fit as input.
        Dark, light and very light shaded areas have $CL>0.90$, $0.32$, $0.05$, 
        respectively. }
\end{figure}
\begin{figure}[t]
  \centerline{\epsfxsize8.0cm\epsffile{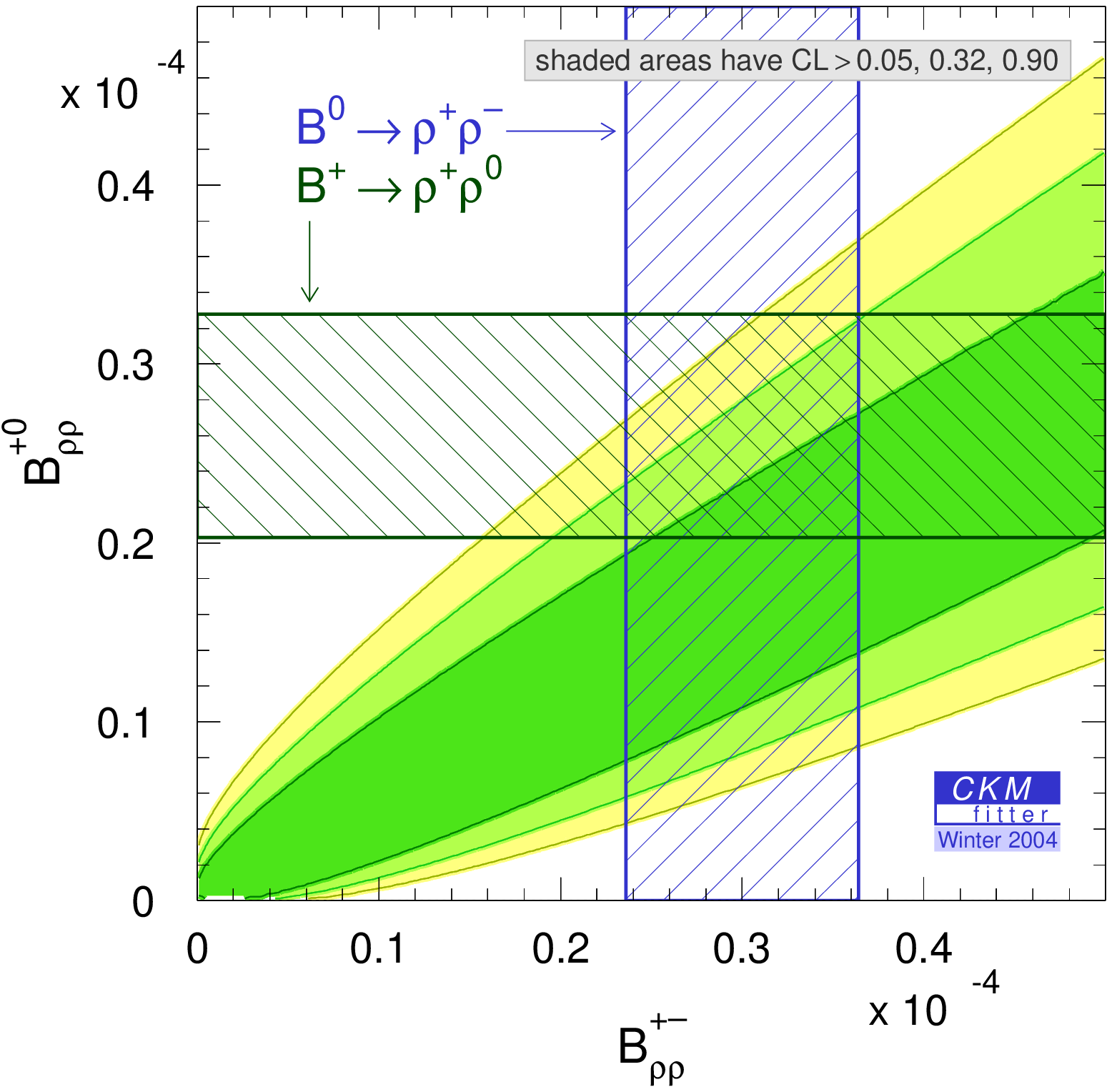}
              \epsfxsize8.0cm\epsffile{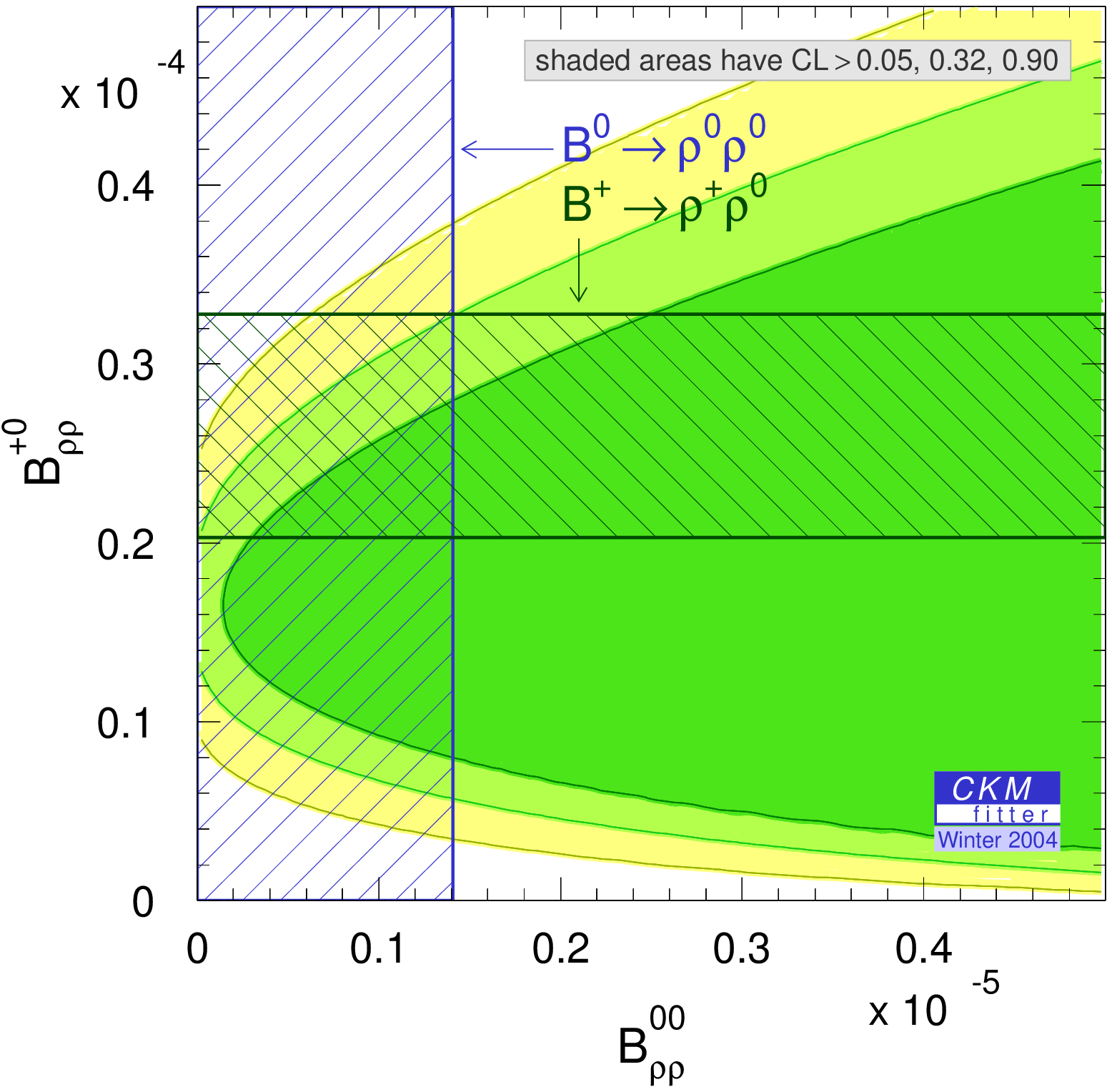}}
  \vspace{-0.0cm}
  \caption[.]{\label{fig:rhorho_br_2d}\em
        Confidence levels for $\BR(\Bp\to\rho^+\rho^0)$ versus 
        $\BR(\Bz\to\rho^+\rho^-)$ (left) and $\BR(\Bp\to\rho^+\rho^0)$  
        versus $\BR(\Bz\to\rho^0\rho^0)$ (right), obtained from the isospin 
        analysis ignoring the branching fractions that are determined,
        and with the use of $\alpha$ from the standard CKM fit. The 
        hatched areas indicate the $1\sigma$ bands of the corresponding
        measurements.}
\end{figure}

We 
present the CLs obtained from the isospin analysis of $B \to \rho\rho$ decays 
as a function of $\alpha$ (Fig.~\ref{fig:AlphaRhorho}) and in the $\rhoeta$ 
plane (left hand plot of Fig.~\ref{fig:ResRhorho}). On both figures, the 
standard CKM fit (excluding the $B \to \rho\rho$ data) is superimposed
exhibiting agreement with the $B \to \rho\rho$ constraints and,
remarkably, a comparable 
precision. The p-value of the measurements within the isospin framework 
(and the result from the standard CKM fit) is found to be $70\%
$. The hatched curve in Fig.~\ref{fig:AlphaRhorho} is obtained by setting 
the error on $S_{\rho\rho,\,L}^{+-}$ to zero, thus reflecting the present 
uncertainty due to the penguin pollution (given the measurement 
of $C_{\rho\rho,\,L}^{+-}$ and of the branching ratios). From this, 
the $90\%$~CL bound on $\deltaAlpha$ is found to be
\beqn
\label{eq:dalpha_rhorho}
        -20^{\circ} \:<\: \deltaAlpha \:<\: 18^{\circ}~.
\eeqn
Converting this into a result for $\alpha$ and taking the solution that 
is in agreement with the standard CKM fit, we find
\beqns
\label{eq:alpha_rhorho}
        \alpha \:=\: 
                 \left(94 \:\pm\: 12\:\left[^{\,+28}_{\,-25}\right] 
                     \:\pm\: 13\:\left[19\right]
                 \right)^\circ ~,
\eeqns
where the first errors are experimental and the second due to the
penguin uncertainty. For CL~$>10\% 
$ the overall  uncertainty on
$\alpha$ amounts to $19^\circ$. Errors in brackets above are given at
 CL~$>5\%
 $ ($2\sigma$). The penguin error contains a systematic uncertainty of
$0.2^\circ$ (interpreted as a theory error by  \rfit) from the 
treatment of the electroweak penguins (see Eq.~(\ref{EWPratio})  in
Section~\ref{sec:charmlessBDecays}.\ref{sec:su2pipi}). The full shift 
on $\alpha$ introduced by $\PpippizEW$ amounts to $-2.1^\circ$. Taking
this into account our result agrees with the \babar\ 
result~\cite{BABARrhorho}. The slightly larger experimental error here
is explained by the difference in the central value of $\alpha$.
\vs
Since the \CP-violating asymmetries in $B^0 \to \rho^0\rho^0$ have not
been  measured, the isospin analysis is incomplete and one expects
plateaus in  the CL as a function of $\alpha$. The width of these
plateaus is driven by the GLSS bound~(\ref{eq:boundPZ}). However, peaks are  
observed in Fig.~\ref{fig:AlphaRhorho}, which is a consequence of the 
relative values of the three branching fractions: the tight upper limit on 
$\BR(\Bz\to\rho^0\rho^0)$ implies that the sum of the color-suppressed
and the penguin amplitudes is small 
so that $\BR(\Bp\to \rho^+\rho^{0}) \sim 0.5\times
\BR(\Bz\to\rho^+\rho^-)$, which is not confirmed by the central values
of the measurements ($\BR(\Bp\to
\rho^+\rho^{0})/\BR(\Bz\to\rho^+\rho^-)=0.88^{\,+0.21}_{\,-0.15}$).
This ``incompatibility'' (which is well covered by the present experimental 
errors) lifts the degeneracy in the (infinite) solution space of $\alpha$.
See Fig.~\ref{fig:rhorho_br_2d} for representations of the predictions obtained
for $\BR(\Bp\to \rho^+\rho^{0})$ versus $\BR(\Bz\to\rho^+\rho^-)$ and
$\BR(\Bp\to\rho^+\rho^0)$ versus $\BR(\Bz\to\rho^0\rho^0)$ from 
the isospin analysis combined with the standard CKM fit. The hatched
areas give the $1\sigma$ bands of the present experimental averages.
\vs
Using the standard CKM fit as input (excluding $B \to \rho\rho$
therein)  and the $\B\to\rho\rho$ branching fraction measurements, one
can obtain predictions for $S_{\rho\rho,L}^{+-}$ and
$C_{\rho\rho,L}^{+-}$ by means of  the isospin analysis. The
corresponding CLs are shown in the right hand plot of
Fig.~\ref{fig:scrhorho} together with the measurement from \babar. Due
to the favorable bound on $\deltaAlpha$~(\ref{eq:dalpha_rhorho}), the 
SU(2) prediction turns out to be meaningful, in sharp contrast to the 
corresponding $\pi\pi$ case (\cf\  Fig.~\ref{fig:scpipi}).

\subsubsection*{Constraints on Amplitude Ratios}
\label{sec:rhorhodpot}

By inserting $\alpha$ from the standard CKM fit we derive constraints on the 
complex amplitude ratios $(R_t/R_u)\Ppipi/\Tpipi$ and $\Tcpizpiz/\Tpipi$. 
Their CLs in polar coordinates are given in Fig.~\ref{fig:ResRhorho}.
The smallness of the penguin-to-tree ratio becomes manifest on the left hand 
plot (although rather large values are still possible). 
The magnitude of the color-suppressed-to-color-allowed ratio (right
hand plot)  is found to be of the order of $0.2$, in agreement
with the na\"{\i}ve  expectation. Comparing these plots with
Figs.~\ref{fig:dpot} and \ref{fig:dtotpipi} in the $\pi\pi$ system
seems to suggest a non-trivial dynamical mechanism, which 
drives the behavior of the penguin and color-suppressed
amplitudes differently with respect to the leading tree in $VV$ 
and $PP$ modes. See 
Section~\ref{sec:charmlessBDecays}.\ref{sec:unbazarcettesection} 
for further discussion of the amplitude ratios.

\subsection{Prospects for the Isospin Analysis}
\label{sec:rhorho_prospects}

\begin{figure}[t]
  \centerline{\epsfxsize8.1cm\epsffile{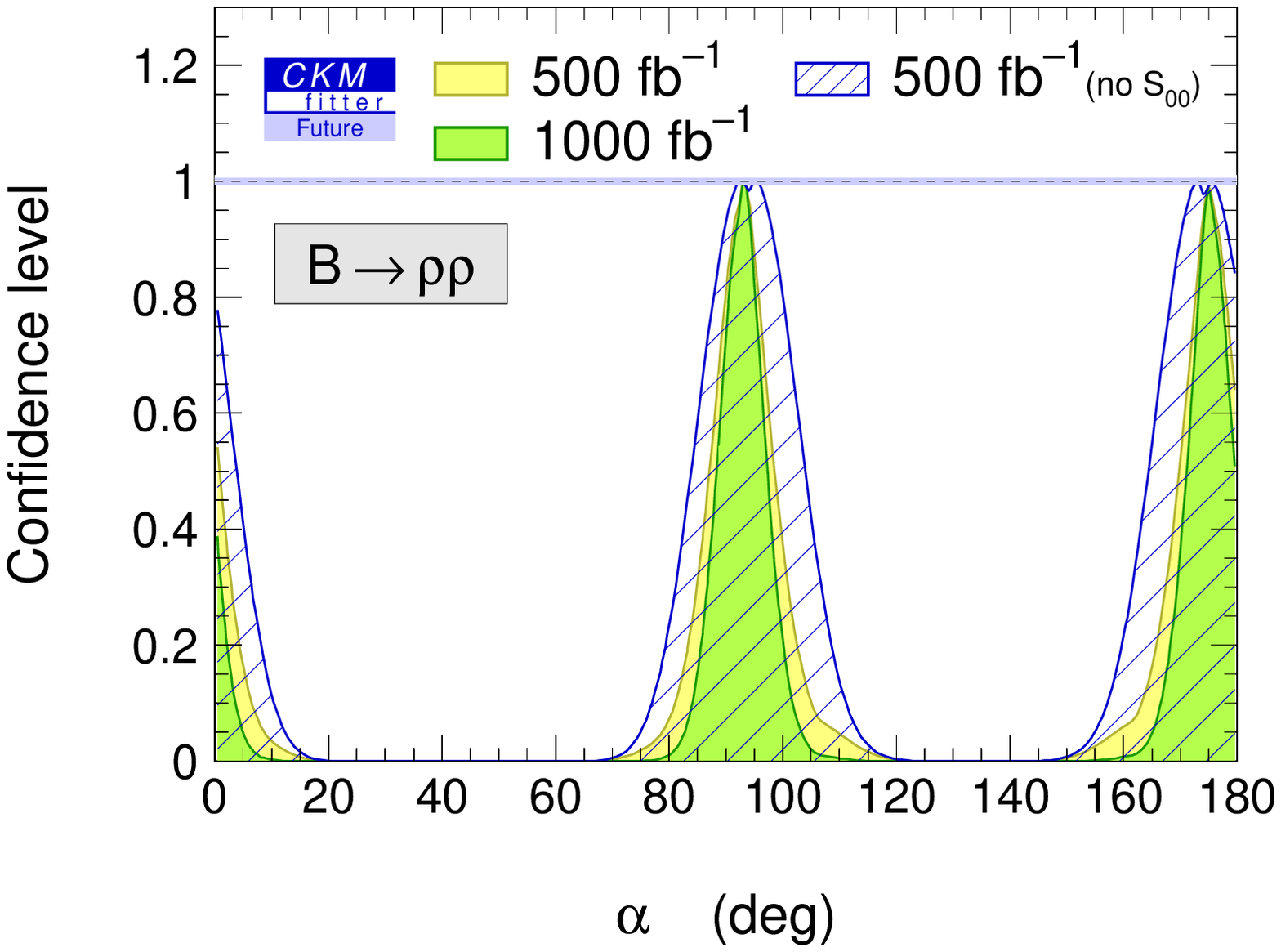}
              \epsfxsize8.1cm\epsffile{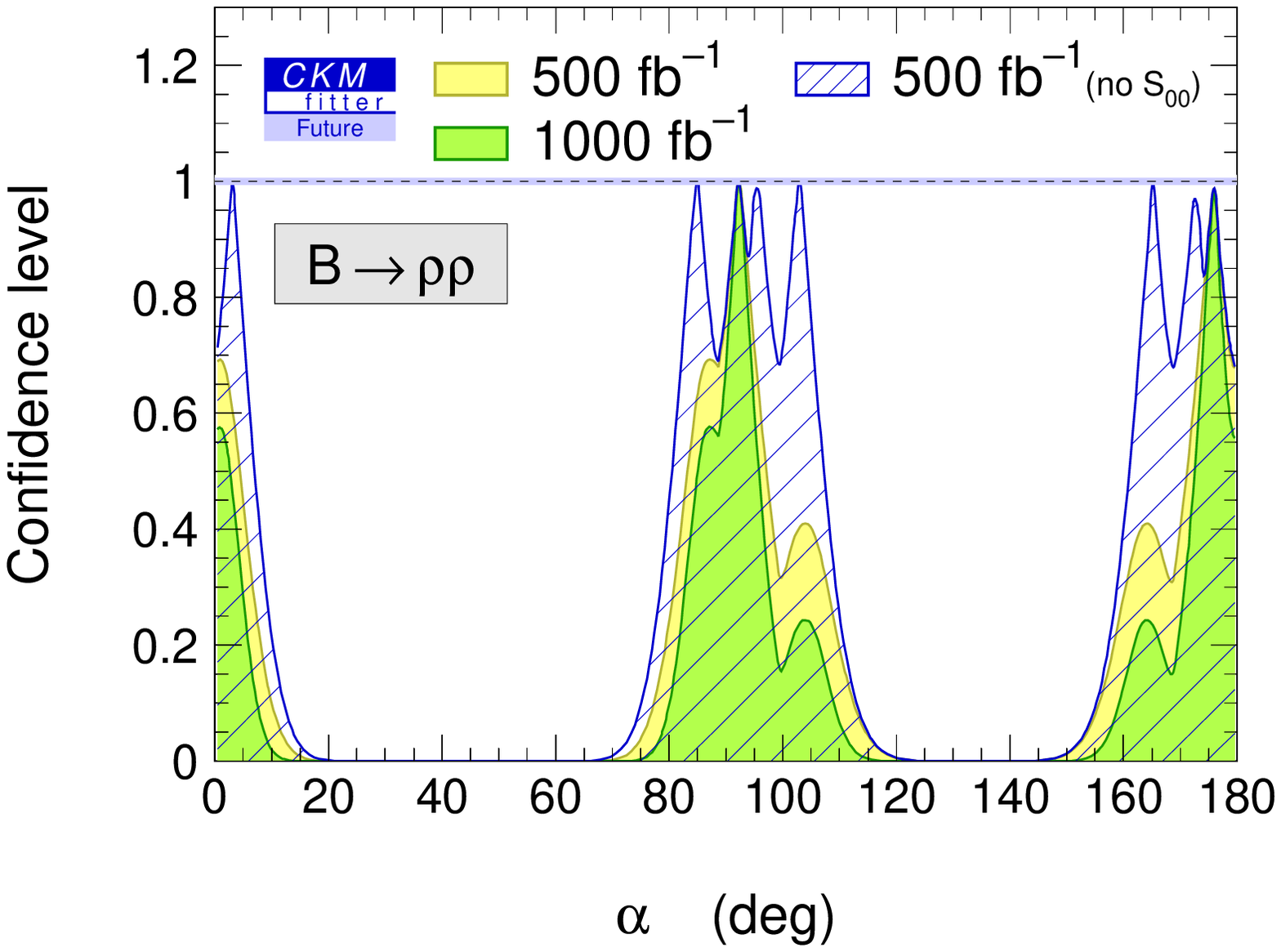}}
  \vspace{-0.5cm}
  \caption[.]{\label{fig:AlphaRhorhoFuture}\em
        Confidence level for $\alpha$ from the SU(2) analysis of the 
        $B \to \rho\rho$ system, extrapolated at integrated luminosities of 
        $500\invfb$ (light shaded) and $1\invab$ (dark shaded). The 
        two scenarios defined in the text are shown: the left (right) 
        hand plot corresponds to a relatively large (small) 
        value for $\BR_{\rho\rho}^{00}$, each one compatible with a 
        subset of the current measurements. To
        illustrate the impact of the $\Spizpiz$ measurement, we also show
        the result at $500\invfb$ ignoring the latter input (hatched).}
\end{figure}

As for $\pi\pi$ and $\rho\pi$, we attempt extrapolations of the
isospin analysis to future integrated luminosities of $500\invfb$ and
$1\invab$. Note that the results strongly depend on the underlying
physics, the knowledge of which is insufficient up to now.
We understand the scenarios studied here are likely to be optimistic. 
\vs
In the first scenario considered, with the exception of 
$\BR(\Bz\to\rho^0\rho^0)$, all measurements are
extrapolated with the use of the present central values. The branching
fraction of $\Bz\to\rho^0\rho^0$ is increased to $1.3\tmsix$ to ensure
the compatibility  of all observables with the isospin
relations~(\ref{eq:isospinbar}). It is  assumed to be dominated by
longitudinal polarization with a longitudinal  fraction of
$f_L^{00}=0.976$ (simple arithmetic average of $f_L^{+-}$ and 
$f_L^{+0}$). We further assume that both time-dependent \CP asymmetries
in  $\Bz\to\rho^0\rho^0$ are measured, and set them to
$S_{\rho\rho,L}^{00}=0.05$ and  $C_{\rho\rho,L}^{00}=0.70$, the
preferred solutions of the standard CKM fit. The statistical and
systematic uncertainties are extrapolated  according to the luminosity
increase. For $\BR(\Bz\to\rho^0\rho^0)$,  $\BR(B^+ \to\rho^+\rho^0)$
and $f_L^{+0}$, we further reduce the extrapolated errors by a factor
of $1.3$ reflecting the improvement (at fixed statistics) in the most
recent $\Bz \to\rho^+\rho^-$ analysis with respect to the previous
one~\cite{rhorhoold}. The errors for
$S_{\rho\rho,L}^{00}$,  $C_{\rho\rho,L}^{00}$ and $f_L^{00}$ are 
estimated from $S_{\rho\rho,L}^{+-}$, $C_{\rho\rho,L}^{+-}$ and
$f_L^{+-}$,  respectively: they are scaled to the expected number of
$\Bz\to\rho^0\rho^0$  events (taking into account the different
selection efficiencies),  except for the systematic uncertainties on
$S_{\rho\rho,L}^{00}$ and   $C_{\rho\rho,L}^{00}$, which are roughly
estimated from the  present values on $S_{\rho\rho,\,L}^{+-}$ and
$C_{\rho\rho,L}^{+-}$. We obtain the extrapolations
\beqns
\begin{array}{rclrcl}
    \BR_{\rho\rho}^{+-}         &=&                     30.0\pm1.6  ~[1.1]\pm2.0  ~[1.4]~,   &
    f_L^{+-}                    &=&                     0.990\pm0.012  ~[0.008]\pm0.014  ~[0.010]~,\\[0.1cm]
    \BR_{\rho\rho}^{+0}         &=&                     26.4\pm1.6  ~[1.1]\pm1.6  ~[1.2]~,   &
    f_L^{+0}                    &=&                     0.962\pm0.014  ~[0.010]\pm0.011  ~[0.008]~,\\[0.1cm]
    \BR_{\rho\rho}^{00}         &=&                     1.30\pm0.14  ~[0.10]\pm0.09  ~[0.06]~,   &
    f_L^{00}                    &=&                     0.976\pm0.030  ~[0.021]\pm0.035  ~[0.025]~, \\[0.3cm]
    S_{\rho\rho,L}^{+-}         &=&                     -0.19\pm0.16  ~[0.11]\pm0.05  ~[0.04]~, &
    C_{\rho\rho,L}^{+-}         &=&     -0.23\pm0.11  ~[0.08]\pm0.07  ~[0.05]~,\\[0.1cm]
    S_{\rho\rho,L}^{00}         &=&                     \phantom{-}0.05\pm0.39  ~[0.28]\pm0.08  ~[0.06]~, &
    C_{\rho\rho,L}^{00}         &=&                     \phantom{-}0.70\pm0.28  ~[0.20]\pm0.10  ~[0.07]~,   
\end{array}
\eeqns
where first errors given are statistical and second systematic. Errors outside
[inside] the brackets are extrapolated to $500\invfb$ [$1\invab$] integrated
luminosity.
\vs
The results of the isospin analyses corresponding to these inputs are
shown in  Fig.~\ref{fig:AlphaRhorhoFuture} (left hand plot). To illustrate 
the impact of the $S_{\rho\rho,L}^{00}$ measurement, we also give the result 
at $500\invfb$ ignoring $S_{\rho\rho,L}^{00}$ in the fit (hatched function). 
If neither $S_{\rho\rho,L}^{00}$ nor $C_{\rho\rho,L}^{00}$  were measured, 
a very similar constraint would be obtained as in the latter fit: only the 
tiny  double-bumps in
the hatched curves reflect the impact of $C_{\rho\rho,L}^{00}$. As for
$\pi\pi$, very large statistics would be needed to resolve the 
discrete ambiguities when relying on $C_{\rho\rho,L}^{00}$ only. The
main additional  information on $\alpha$ is hence provided by
$S_{\rho\rho,L}^{00}$, which is due to its linear dependence on $\sta$
(see the analytical solutions of the  isospin analysis derived in
Footnote~\ref{foot:pipi_alterparam} in
Section~\ref{sec:charmlessBDecays}.\ref{sec:su2pipi}). With this scenario, 
the (symmetrized) $1\sigma$ (resp. $2\sigma$) uncertainty  on $\alpha$ is 
expected to be of the order of $8^\circ$ at $500\invfb$ and $6^\circ$ 
at $1\invab$ (resp. $16^\circ$ and $10^\circ$), which is small enough 
so that the residual SU(2)-breaking effects discussed below have to
be taken into account.
\vs
In a second scenario,  the central value of $\BR_{\rho\rho}^{00}$
is fixed to the present one ($0.6\tmsix$) and the branching fraction of 
the $B^+\to\rho^+\rho^0$ mode is decreased to $17\tmsix$, which is the 
value preferred by the present isospin analysis when using the standard 
CKM fit as input (see Fig.~\ref{fig:rhorho_br_2d}). The new set of 
extrapolations for the $\rho^+\rho^0$ and $\rho^0\rho^0$ modes are 
\beqns
\begin{array}{rclrcl}
    \BR_{\rho\rho}^{+0} &=&  17.0\pm1.3  ~[0.9]\pm1.3  ~[1.0]~,   &
    f_L^{+0}            &=&  0.962\pm0.017  ~[0.012]\pm0.014  ~[0.010]~,\\[0.1cm]
    \BR_{\rho\rho}^{00} &=&  0.60\pm0.09  ~[0.07]\pm0.06  ~[0.04]~,   &
    f_L^{00}            &=&  0.976\pm0.044  ~[0.031]\pm0.051  ~[0.037]~, \\[0.3cm]
    S_{\rho\rho,L}^{00} &=&  0.05\pm0.57  ~[0.41]\pm0.12  ~[0.09]~, &
    C_{\rho\rho,L}^{00} &=&  0.70\pm0.41  ~[0.29]\pm0.15  ~[0.10]~,   
\end{array}
\eeqns
where again errors outside [inside] the brackets are extrapolated to 
$500\invfb$ [$1\invab$] integrated luminosity.
The constraints on $\alpha$ resulting from these extrapolations are shown in
Fig.~\ref{fig:AlphaRhorhoFuture} (right hand plot). The (symmetrized) $1\sigma$ 
(resp. $2\sigma$) errors on $\alpha$ obtained with this setup amount to 
$13^\circ$ at $500\invfb$ and $7^\circ$ at $1\invab$ (resp. $19^\circ$ and 
$16^\circ$). This is significantly worse than for the previous scenario, 
in which one benefits from an almost optimal bound on $\deltaAlpha$ since 
$\BR_{\rho\rho}^{00}\approx\BrooGLSSl=1.25\tmsix$,
while for the second scenario the lower bound $\BrooGLSSl=0.20\tmsix$
is rather different\footnote
{
        The values for $\BrooGLSSl$ corresponding to the different scenarios
        are determined from Eq.~(\ref{eq:GLSSbound}), 
        where we have used as inputs the branching fractions of the 
        longitudinally polarized $\rho$'s. Note that the upper bounds
        $\BrooGLSSu$ correspond in both scenarios to very large 
        $\BR_{\rho\rho}^{00}$ that are excluded by experiment.  
} from 
the central value of $\BR_{\rho\rho}^{00}$. Furthermore, due to the 
small $\BR_{\rho\rho}^{00}$ the precision on $S_{\rho\rho,L}^{00}$ is 
insufficient to effectively suppress the ambiguities.
\vs
An important source of systematic uncertainty on the time-dependent
\CP-violating  asymmetries is the \CP violation exhibited by the
\B-related  background modes~\cite{babarrhorhoprl}, that may be hard to
measure in the  near future. To study its impact, we keep the size of
this systematic error unchanged in the extrapolation, while all other
systematic errors are appropriately scaled  with the increasing
luminosity. Applying this procedure to all time-dependent  \CP
asymmetries, no significant deterioration of the accuracy on $\alpha$
is observed.

\subsection{Breaking of the Triangular Relation in $B\to\rho\rho$}
\label{sec:SU2strongbreaking}

Reference~\cite{garnderpipi} presents a study of isospin-breaking
effects in $B\to\pi\pi$ that come from the strong interaction, through
the $\piz$--$\eta$--$\eta'$ mixing. These effects break the triangular
relation~(\ref{eq:isospinbar}) and entail a systematic error on the
angle $\alpha$. The size of this error depends on the actual values of
the non-leptonic matrix elements, and on the relative amount of the
$I=0$ component in the $\pi^0$ bound state, which could be of order
$1$--$2\% 
$. In $B\to\rho\rho$ decays, besides a similar effect due to the
$\rho$--$\omega$ mixing, it is argued in Ref.~\cite{ligetirhorho} that
a $I=1$ $\rho\rho$ state could be generated by the finite width of the
$\rho$. Although the latter effect does not break isospin symmetry in
the sense that it does not vanish in the $m_u=m_d$ limit, it can be
parameterized the same way as above, through the introduction of an
additional amplitude in the isospin triangle~(\ref{eq:isospinbar}).
\vs
Hence we model a possible breaking of the closure of the isospin relation
by a contribution $\Delta\Apz$ to the $\Apz$ amplitude given by
\beq
\label{eq:rhorho_isobreak}
        \sqrt{2}\,\Delta\Apz \:=\: 
                        V_{ud}V_{ub}^*\, \Delta_T\Tpm 
                              + V_{td}V_{tb}^*\, \Delta_P\Ppm~.
\eeq
\begin{figure}[t]
  \centerline{\epsfxsize8.0cm\epsffile{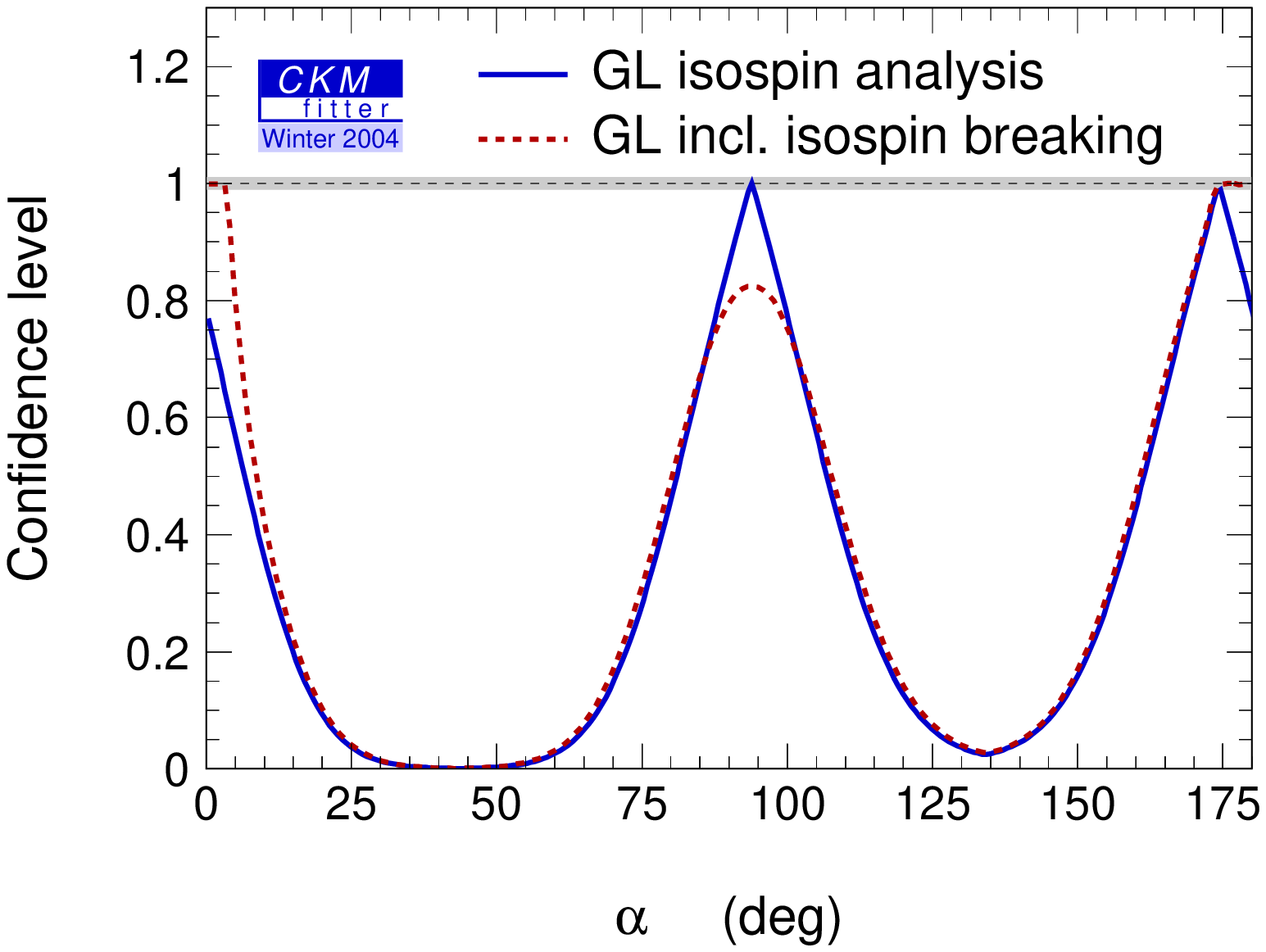}
              \epsfxsize8.0cm\epsffile{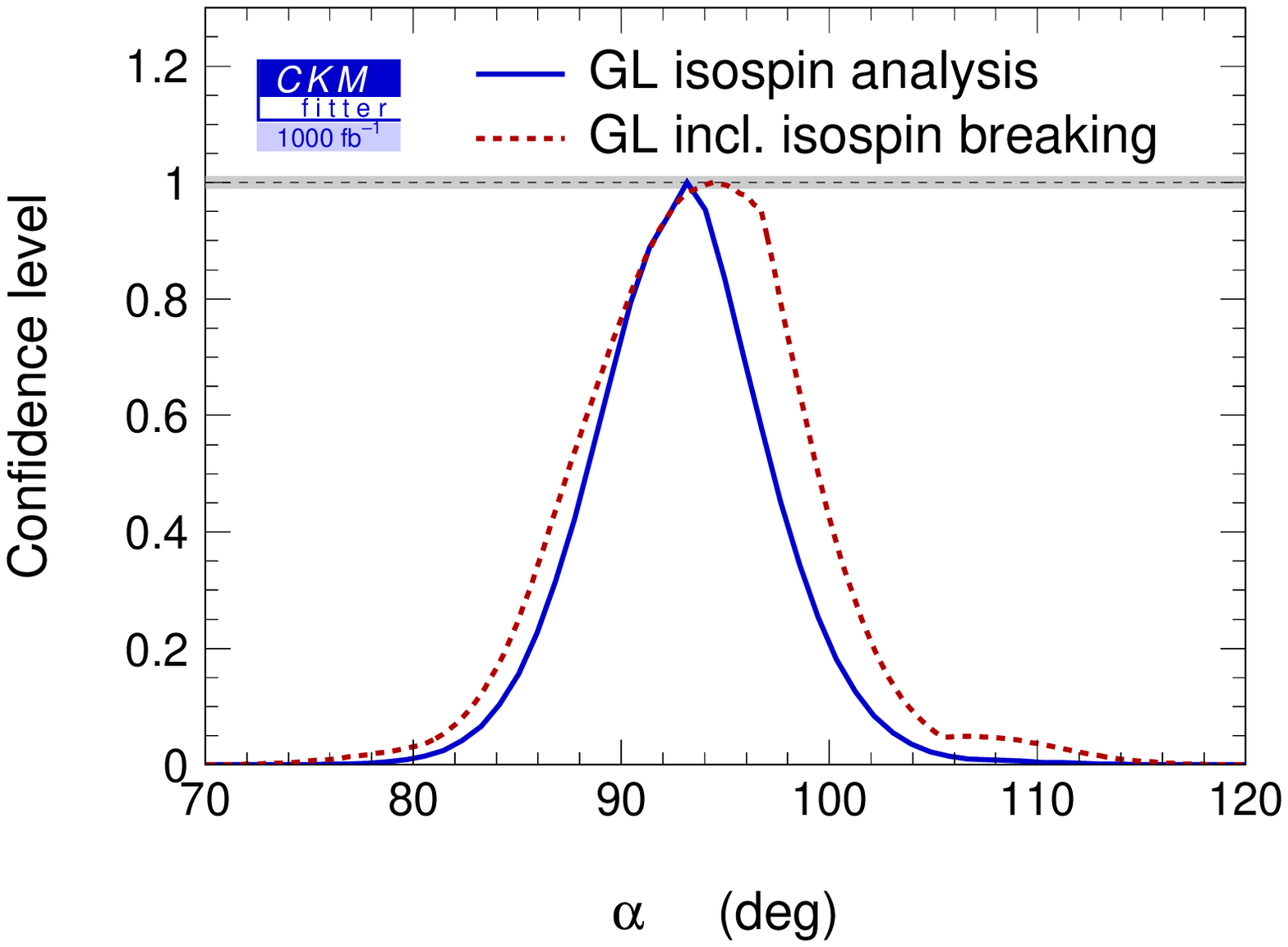}}
  \vspace{-0.5cm}
  \caption[.]{\label{fig:rhorho_alpha_isobreak}\em
        Confidence levels for $\alpha$ from the $\B\to\rho\rho$ isospin 
        analysis where isospin-breaking corrections (with the 
        exception of electroweak penguins) are neglected (solid 
        lines), and when including a relative $4\%
        $ correction with arbitrary phases
        on the amplitude level (dashed lines), according 
        to Eq.~(\ref{eq:rhorho_isobreak}).
        The left hand plot gives the results for the present experimental
        situation (bound on $\BR(\Bz\to\rho^0\rho^0)$), while the 
        right hand plot uses the extrapolation to $1\invab$, defined
        in Section~\ref{sec:rhorho_prospects}. It
        includes the full isospin analysis with available measurements 
        of $S_{\rho\rho,L}^{00}$ and $C_{\rho\rho,L}^{00}$.
        }
\end{figure}
Note that for arbitrary values of the relative 
coefficients $\Delta_T$ and $\Delta_P$, the above equation together with
Eqs.~(\ref{eq:apipi}) and~(\ref{eq:b0pi0pi0}) is the most general
parameterization of $B\to\pi\pi$ and $B\to\rho\rho$ decays within and
beyond the Standard Model (see Footnote~\ref{mostgalparam} in
Section~\ref{sec:newPhysics}.\ref{NPdB=1}). In this Section however, 
we assume that $\Delta_T$ and $\Delta_P$ are unknown within a magnitude 
of up to $(m_\rho/\Gamma_\rho)^2\sim4\%$~\cite{ligetirhorho} and with
arbitrary strong phases\footnote
{
        The effect of hadronic isospin-breaking effects
        is expected to be smaller than the one due to the finite width
        of the $\rho$.
}. The effect on $\alpha$ is given in 
Fig.~\ref{fig:rhorho_alpha_isobreak}. The left hand plot shows
the present experimental situation (bound on $\BR(\Bz\to\rho^0\rho^0)$),
where the solid line indicates the isospin analysis where 
isospin-breaking contributions with the exception of electroweak
penguins are neglected (same as Fig.~\ref{fig:AlphaRhorho}),
and the dashed line corresponds to the replacement 
$\Apz\to\Apz+\Delta\Apz$. The right hand plot shows the corresponding
constraints for the first scenario, defined in
the previous section, at $1\invab$.
\vs
The observed systematic uncertainty on $\alpha$ depends on whether or 
not the full isospin analysis is applied. It is small for the GLSS bound
(equivalent to the isospin analysis with upper limit on 
$\BR(\Bz\to\rho^0\rho^0)$). However, significant effects can occur
once the full isospin analysis is performed. We estimate the size 
of the uncertainty for the setup we have tested to be of the order 
of $3^\circ$. Notwithstanding, one should keep in mind that the 
resulting effect depends, on the one hand, on the relative size of the
additional amplitude~(\ref{eq:rhorho_isobreak}), and, on the other hand,
on the particular solution of the isospin analysis for the contributing 
amplitudes. It must therefore be (re-)estimated for the set of 
measurements that is at hand. Furthermore,
a careful experimental analysis may partially disentangle these
effects~\cite{ligetirhorho}.

 \section{Comparison of Amplitude Ratios}
\label{sec:unbazarcettesection}

Numerical values for the ratios of reduced tree and penguin amplitudes
for the $\pi\pi$, $\rho\rho$, $K\pi$ and $\rho\pi$ systems,
assuming the standard CKM fit and specific hadronic hypotheses (see
table caption) are given in Table~\ref{tab:pot}. 
Before discussing in more detail the
results, two reservations are in order: the overall (mainly experimental) 
uncertainties on these ratios are still large, and within the $2\sigma$ 
errors no specific conclusions can be drawn from the comparison of the 
four systems. The amplitude ratios are obtained assuming strict isospin 
symmetry for $\pi\pi$ and $\rho\rho$, and stronger hadronic hypotheses for
$K\pi$ and $\rho\pi$. The results from the latter two systems should 
therefore be interpreted with care. The results of the fit of the QCD  
Factorization on the $\pi\pi$ and $K\pi$ data (see Section~\ref{sec:bbnsfit})
are also reported.
\vs
In the $\pi\pi$ system, the measurement of the time-dependent
\CP asymmetry requires a potentially large penguin-to-tree ratio 
to be in agreement with the standard CKM fit. This was expected 
after the first measurement of the surprisingly large branching 
fraction of $\Bz\to\Kp\pim$. However a puzzling feature is that 
the $K\pi$ data alone prefer a small value for the same penguin-to-tree
ratio, and that the branching fraction for the $K^+\pi^-$ mode is 
somewhat smaller than the
theoretical expectation (see Table~\ref{tab:qcdfaPrediCorr}).
Various sources for this discrepancy have been discussed 
in Section~\ref{sec:charmlessBDecays}.\ref{sec:introductionKPi}.
\vs
Another feature of the $\pi\pi$ , $K\pi$ and, to a lesser
extent, $\rho\pi$ modes, is the apparent
significant violation of the color-suppression concept. While from 
the point of view of the $1/N_c\to 0$ limit this suppression is 
formally of order $1/N_c\sim 0.3$, na\"{\i}ve semi-perturbative 
counting predicts a further cancellation, leading to
the well known $a_2\lsim 0.2$ ``universal'' factor~\cite{StechNeubert}.
This property of the so-called ``class~II'' decays, according to the
classification of Ref.~\cite{StechNeubert}, remains partially true in
the QCD factorization formalism~\cite{BBNS0} although the latter 
admits the possibility of large corrections~\cite{BN}. There is
evidence from Table~\ref{tab:pot} that the present data suggest
that the color-suppression mechanism is ineffective. Different 
manifestations of violation of the color-suppression concept,
 and dependence of the $a_2$ factor with respect 
to the involved particles, have been observed in simpler $B$ decays\footnote
{
        Not to mention the large $1/N_c$ effects in kaon and $D$-meson 
        decays.
} without penguin contributions, \eg, $B\to \jpsi K^{(*)}$ and $\Delta C=1$ 
transitions~\cite{zito}.
\vs
The results for the amplitude ratios fitted simultaneously on the
$\pi\pi$ and $K\pi$ measurements within the framework of QCD
factorization are also given in Table~\ref{tab:pot}. We find that this
more predictive approach (compared to the fits based only on flavor
symmetry) re-establishes the good agreement between the
penguin-to-tree ratios in the $\pi\pi$ and $K\pi$ systems: this can be
interpreted as the consequence of the smallness of the annihilation and
exchange contributions estimated in this approach. However
larger $|T^{00}_C/T^{+-}|$ ratios are found, although with large
errors.
\vs
\begin{table}[t]
\begin{center}
\setlength{\tabcolsep}{0.0pc}
\begin{tabular*}{\textwidth}{@{\extracolsep{\fill}}lccccc}\hline
&&&&&\\[-0.3cm]
&\multicolumn{4}{c}{Central value $\pm$ error at given CL}&\\[0.15cm]
&\multicolumn{2}{c}{$\PoTpipi$}&\multicolumn{2}{c}{$\ToTpipi$}&\\[0.15cm]
Mode            
        & CL = 0.32 & CL = 0.05 & CL = 0.32 & CL = 0.05 & Method\\[0.15cm]
\hline
&&&&\\[-0.3cm]
$B \to \pi\pi$   
        & $0.23^{+0.41}_{-0.10}$        & $^{+0.81}_{-0.16}$    & $0.98^{+0.58}_{-0.30}$        & $^{+1.54}_{-0.49}$ & SU(2)\\[0.15cm]
$B \to \rho\rho$ 
        & $0.05^{+0.07}_{-0.05}$        & $^{+0.12}_{-0.05}$    & $0.21^{+0.11}_{-0.15}$        & $\pm 0.21$ & SU(2)\\[0.15cm]
\hline
&&&&&\\[-0.3cm]
$B \to K\pi$    
        & $0.04^{+0.03}_{-0.01}$        & $^{+0.14}_{-0.04}$    & $1.22^{+0.32}_{-0.16}$        & $^{+1.20}_{-0.32}$ & SU(2)+ no annihil./exch.  \\[0.15cm]
$B \to \rho\pi${\small$[+-]$}
        & $0.03^{+0.09}_{-0.03}$        & $^{+0.11}_{-0.03}$    &
        $0.48^{+0.14}_{-0.16}$        & $^{+0.25}_{-0.48}$ & SU(3)+ no
        OZI-peng.\\[0.15cm]
$B \to \rho\pi${\small$[-+]$}
        & $0.10^{+0.02}_{-0.03}$        & $^{+0.03}_{-0.06}$    &
        $0.57^{+0.17}_{-0.18}$        & $^{+0.33}_{-0.57}$ & SU(3)+ no
        OZI peng.\\[0.15cm]
\hline
\hline
&&&&&\\[-0.3cm]
$B \to \pi\pi$ 
        & $0.18^{+0.01}_{-0.03}$        & $^{+0.03}_{-0.05}$    & $1.17\pm 0.20$        & 
$\pm0.41$ & QCD FA combined fit\\[0.15cm]
$B \to K\pi$  
        & $0.17^{+0.01}_{-0.03}$        & $^{+0.03}_{-0.05}$    & $1.52^{+0.42}_{-0.47}$        &  $^{+0.69}_{-0.71}$ & QCD FA combined fit\\[0.15cm]
\hline
&&&&&\\[-0.3cm]
\end{tabular*}
\caption[.]{\em  \label{tab:pot}
        Magnitudes of penguin-to-tree ($\PoTpipi$) and
        color-suppressed-to-color-allowed  ($\ToTpipi$) amplitude
        ratios obtained for the four charmless  decay modes studied in
        this part. For the purpose of this comparison, the CKM elements 
        are not included in the ratios, but their input is taken from 
        the standard CKM fit. We denote by $B \to \rho\pi${\small$[+-]$}
        $(${\small$[-+]$}$)$  the branch where the $\rho$ $(\pi)$
        is emitted by the $W$. For the $\pi\pi$ and $\rho\rho$
        modes, only isospin symmetry is assumed. Since SU(2) is
        insufficient at present, the annihilation and exchange contributions 
        are neglected for the $K\pi$ ratios, and the SU(3) partners 
        $\Kstarp\pim$ and $\rho^-\Kp$ are used to constrain the corresponding
        penguin amplitudes for the $\rho\pi$ modes (see text). The last
        two lines give the results of the combined QCD  FA fit to the
        $\pi\pi$ and $K\pi$ data (Section~\ref{sec:charmlessBDecays}.\ref{sec:bbnsfit}). The
        strong phases obtained in this framework are: in the
        $\pi\pi$ system, ${\rm arg}(P^{+-}/T^{+-})= (-29^{\,+5}_{\,-2})^\circ$ and 
        ${\rm arg}(T^{00}_\mathrm{C}/T^{+-})=(146^{\,+6}_{\,-2})^\circ$; and in the 
        $K\pi$ system, 
        ${\rm arg}(P^{+-}/T^{+-})=(-28^{\,+5}_{\,-1})^\circ$ and 
        ${\rm arg}(T^{00}_\mathrm{C}/T^{+-})=(-25^{\,+8}_{\,-5})^\circ$.}
\end{center}
\end{table}$\!\!$
Note that our definition of $T^{00}_\mathrm{C}$ implicitly 
contains long-distance penguin and exchange contributions. Although 
the latter are $1/N_c$ suppressed as well, and there is no
model-independent distinction between the different topologies that 
are mixed by rescattering phenomena, it may occur that a number of
relatively small corrections constructively interfere in
$T^{00}_\mathrm{C}$ and destructively in $T^{+-}$ to eventually give
a globally large effect, which could explain the observed 
$T^{00}_\mathrm{C}/T^{+-}$ ratio~\cite{BN,BFRS2}.
\vs
Finally we stress that although the large penguin and  color-suppressed
amplitudes in the $\pi\pi$ channels likely come from the same type of
non-trivial hadronic dynamics, $B\to\pi^0\pi^0$ cannot be a pure
penguin mode. Indeed, were it the case, in the SU(3) limit and
neglecting electroweak penguin contributions $\BR(B^0\to\pi^0\pi^0)$ would be equal
to $\BR(B^0\to\Kz\Kzb)/2$, which is disfavored by the current
data (see Table~\ref{tab:BRPiPicompilation}). This is somewhat unfortunate
because $\BR(B^0\to\pi^0\pi^0)\sim\BR(B^0\to\Kz\Kzb)/2$ would imply a stronger
constraint on $\alpha$ from the $\B\to\pi\pi$ isospin analysis, since 
it would closer approach the GLSS bound $\BrooGLSSl$~(\ref{eq:GLSSbound}).
\vs
In na\"{\i}ve factorization, there is a clear hierarchy
between penguins in $PP$, $PV$ and $VV$
modes~\cite{aleksanRR}. This is due to the Dirac structure of
$(V-A)(V+A)$ penguin operators, which do not contribute when the meson
that does not receive the spectator quark (the ``upper'' meson) is a
vector, as in
$B^0\to\rho^+\pi^-$ and $B^0\to\rho^+\rho^-$. Similarly, these
operators contribute
constructively (resp. destructively) with $(V-A)(V-A)$ penguin operators
when the upper meson is a pseudoscalar and the lower meson is a
pseudoscalar (resp. vector), as in $B^0\to\rho^-\pi^+$ (resp.
$B^0\to\pi^+\pi^-$). This expectation seems to be in agreement with our
fits to the present
data (see Table~\ref{tab:pot}). However, as a consequence of the fact
that $(V-A)(V+A)$ operators are formally power-suppressed in the full QCD
factorization approach, the above simple hierarchy may receive large
corrections~\cite{BN}. Hopefully, a more detailed dynamical analysis will
be possible when the measurements of the strange $PV$ and $VV$ channels
become more complete and precise.
\vs
The present pattern of amplitude ratios in the different decays, if
confirmed when the experimental errors decrease, might challenge 
theoretical approaches that are based on the factorization of
non-leptonic matrix elements. Na\"{\i}vely suppressed contributions,
such as charming penguins~\cite{charmingpenguins} or other type of
rescattering effects (see, e.g., Ref.~\cite{rescattering} for an
example of a final state interaction that does not vanish in the
$1/m_b\to\infty$ limit),  could finally contribute at leading order:
the approach of Ref.~\cite{SCETfacto}, based on the SCET effective
theory, might be able to handle these difficult problems in a more
systematic way. At present, it seems that it would be difficult to keep
all these effects small while maintaining agreement with the central
values of the experimental observables, unless one is willing to fine-tune
all the observed ``anomalies'' with New Physics contributions. An
interesting question, among others, concerns the behavior of
rescattering effects with respect to the isospin or SU(3) quantum
numbers of the relevant amplitudes. More data and theoretical work are
needed to answer this.

 \section{Conclusion}
\label{sec:charmlessConclusions}

Due to the significant \CP asymmetries on one hand, and the presence
of loop-induced transitions on the other hand, charmless \B decays
can be used for precision measurements of \CP violation within the SM, 
and they are sensitive probes of physics beyond the SM.
\vs
We have studied $\B\to\pi\pi,K\pi$ decays using various phenomenological
approaches with different dynamical assumptions. These include SU(2)
and SU(3) flavor symmetries and QCD Factorization. 
An extra section has been devoted to the phenomenological analysis of 
$\B\to K\pi$ decays due to the peculiarity of the observed branching 
fraction pattern. Constraints on $\rhoeta$ from these decays are 
weak since the sensitivity to the CKM phase through the tree amplitude 
is CKM-suppressed with respect to $\B\to\pi\pi$. However the $K\pi$ 
modes represent a rich field to test flavor symmetry, QCD Factorization 
and to search for manifestations of New Physics.
For the analysis of $\B\to\rho\pi$ decays we have applied SU(2) and SU(3) 
symmetry, and SU(2) symmetry is used for the $\B\to\rho\rho$ system, mostly
because the branching fractions of the relevant SU(3) partners are not yet 
well known. Due to the powerful bound on the penguin pollution in 
$\Bz\to\rho^+\rho^-$ using the upper limit on $\Bz\to\rho^0\rho^0$, a 
significant constraint
on $\alpha$ can be derived from the measurement of $\staeff$ in a 
time-dependent analysis of $\Bz\to\rho^+\rho^-$ performed by the \babar\  
collaboration. The $\pi\pi$ and $\rho\pi$ 
systems do not (yet) provide useful constraints from the corresponding 
isospin analyses, because of the poor sensitivity to the penguin contribution 
($\pi\pi$) and the lack of a full Dalitz analysis ($\rho\pi$). 
\vs
More specifically, we find for the $\pi\pi$ system that:
\bei
\item   hints of a large penguin contribution and a large violation
	of the color-suppression mechanism are found with 
        $|\Ppipi/\Tpipi| = 0.23^{+0.41}_{-0.10}$ and
	$|T^{00}/T^{+-}|=0.98^{+0.58}_{-0.30}$
	so that the SU(2) 
        upper bound fails to provide a significant constraint on 
        $\deltaAlpha=\alpha-\alphaeff$, for which we find ($\CL>10\%
	$):
        $-54^{\circ}<\deltaAlpha<52^{\circ}$.

\item   a somewhat better bound is obtained from the SU(3) analysis
	neglecting OZI- and power-suppressed penguin topologies,
        $-29^{\circ}<\deltaAlpha< 28^{\circ}$, with a weak constraint
        on $\alpha$.

\item   at an extrapolation to $1\invab$, exclusion areas for
        $\alpha$ can be obtained with the $\B\to\pi\pi$ isospin analysis. 
        However a precise measurement of $\alpha$ from the $\pi\pi$ 
        system alone will likely require larger amounts of data ($\sim10\invab$) 
        that could be reached at a next generation \B factory.

\item   useful information in the $\rhoeta$ plane is obtained 
        with partial input from QCD Factorization: either to gauge 
        the uncertainty on SU(3) breaking, or to obtain an 
        estimate of the tree and penguin matrix elements
	(magnitudes and phases). 
        The constraints obtained in both cases are 
        in agreement with the standard CKM fit.

\item   the full calculation of QCD Factorization (taking into account 
        model-dependent power-suppressed terms) is required to accommodate 
        $\rhoeta$ extracted from the \CP measurements
        with the standard CKM fit. A leading order calculation (close to
        na\"{\i}ve factorization) leads only to the marginal compatibility with
        a p-value of $5\times10^{-5}$. By further constraining the 
        full model with all $\pi\pi$ and $K\pi$ branching fractions and 
        \CP-violating asymmetries measured so far, one finds the allowed 
        region in the $\rhoeta$ plane in striking agreement with the standard 
        CKM fit (with comparable precision) and an overall p-value of $21\%
	$.
        
\item   we compute projections on the $\pi\pi$, $K\pi$ observables from the 
        global QCD FA fit, where all observables but the one that is
        projected upon are included in the fit (as well as the standard
        CKM fit). The results are unbiased data driven predictions and exhibit 
        high precision. The agreement with the measurements is 
        satisfying, with the notable exceptions of the branching fractions for 
        $\Bz\to\Kz\piz$ and $\Bz\to\Kp\pim$. The 
        corresponding predictions from QCD FA alone (without experimental 
        input to constrain the theory parameters) suffer from much larger
        uncertainties, which is by part due to the conservative \rfit\  
	treatment of the theoretical systematics.

\item   using SU(3) symmetry we predict the branching fraction and 
        \CP-violating asymmetries in $\Bs\to\Kp\Km$ decays from the 
        $\B\to\pi\pi$ measurements and the standard CKM fit. We find 
        the $95\%$~CL ranges $0.02<\Ckk<0.32$ and $0.12<\Skk<0.27$.
        Only a weak constraint can be derived for $\BRkk$, for which 
	however the correlation with $\Ckk$ is strong.

\eei
For the $K\pi$ system we note:
\bei

\item   the ``historical'' proposal by Quinn and Snyder to use $K\pi$
        modes and isospin symmetry for the extraction of $\alpha$ in the
        absence of electroweak penguins leads only to a weak constraint with the
        present data. The subtle quadrilateral construction would need very
        precise measurements to become meaningful, while at the same time there
        are convincing arguments that electroweak penguin contributions cannot
        be neglected.

\item   the recent proposal to constrain the apex of the UT
        assuming isospin symmetry, neglecting all annihilation and
        long-distance penguin diagrams, and evaluating electroweak penguins in
        terms of tree amplitudes in the SU(3) limit, is expected to provide 
	meaningful results when the data become more precise.
        However the dynamical assumptions behind this method should be
        investigated further, since they are crucial for the result.
        In particular we observe that the fit systematically prefers a
        non-zero $V_{us}V_{ub}^*$ contribution to $B^+\to K^0\pi^+$, although
        theory predicts a small effect.

\item   within the same set of dynamical assumptions, we determine the
        allowed range of amplitude ratios. Color-suppression appears to be
        significantly violated as in the $\pi\pi$ case, while the 
	penguin-to-tree ratio is, quite
        paradoxically, smaller than in the $\pi\pi$ system. This is in
        na\"{\i}ve
        contradiction with the idea that the large branching ratios to
        $K\pi$ with respect than the ones to $\pi\pi$  are  evidence for large
        penguins in $B\to PP$ transitions. Complicated hadronic mechanisms
        and/or New Physics effects in either $b\to d$ or $b\to s$ 
	transitions might be at
        the origin of this intriguing pattern. However present experimental
        uncertainties still exhibit a decent agreement with SU(3)
	between $\pi\pi$ and
        $K\pi$ in the no-rescattering limit.
 
\item   the experimental uncertainties hinder us from obtaining significant
        constraints on electroweak penguin contributions, even in the most
        restrictive theoretical scenario where annihilation and exchange topologies
        are entirely ignored. The SM expectation for
        both color-allowed and color-suppressed (the latter one should not be
        neglected) electroweak penguins can describe the data;
        furthermore the feasibility of a more general study including 
        arbitrary NP contributions in either gluonic or electroweak penguins 
        is not clear, again due to the lack of experimental precision.

\eei
The following conclusions can be drawn from the analysis
of the two-body $\rho\pi$ system:
\bei
\item   scenarios using SU(2) as only input do not constrain $\alpha$ at the first 
        generation \B factories if $\BR(\Bz\to\rho^0\pi^0)$ is not significantly
        smaller than expected from color-suppression. The reason for this  
        failure is merely a problem of experimental 
        precision to resolve $\alpha$ in the pentagon. 
        Setting arbitrarily all strong phases to zero and removing the 
        penguins leads to a value for $\alpha$ that is in agreement with 
        the standard CKM fit with a statistical uncertainty of $5.4^\circ$.

\item   within SU(3) symmetry and neglecting OZI- and
	power-suppressed penguin contributions, we observe some disagreement between the 
        bound on direct \CP-violating asymmetries obtained from 
        $\BR(\Bz\to\rho^- \Kp)$ for the $\Bz\to\rho^-\pip$ branch,
        and the central value of the measurement. While this is not 
        conflicting within the present experimental errors, it requires
        a reduction of the observed $\Acpmp$ with more data, if 
	the SM and SU(3) picture holds.

\item   within the same SU(3)-based hypotheses, we obtain the bounds
        $|\alpha - \alphaeffpm| < 17.6^\circ$ and
        $|\alpha - \alphaeffmp| < 12.6^\circ$ at $95\%$~CL.

\item   SU(3) flavor symmetry does not help to significantly 
        constrain $\alpha$ when all theory parameters are free to
        vary since the eightfold ambiguity due to the unknown relative
        strong phase $\delpmmp$ remains. Using the standard CKM fit as 
        input leads to the preferred values $\delpmmp\approx0,\pm\pi$,
        which is compatible with the no-rescattering expectation.
	
\item   all the approaches that we have studied suffer from the lack of
	knowledge of the interference phase between the two charged
	$\rho$'s, which generate discrete ambiguities.
	In particular, a powerful constraint on $\alpha$, competitive
	with the standard CKM fit, would be
	obtained 
        from a Dalitz plot analysis together with the SU(3) constraints
	from penguin-dominated partners.

\begin{figure}[t]
  \centerline{\epsfxsize10.cm\epsffile{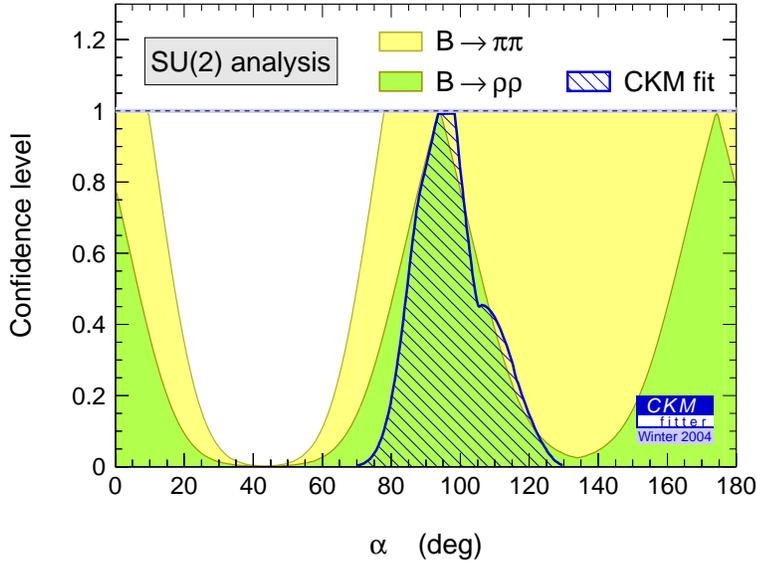}}
  \vspace{-0.5cm}
  \caption[.]{\label{fig:AlphaPiPiRhorho}\em
        Confidence level from the SU(2) analysis of $\B\to\pi\pi$
        (light shaded) and $B \to \rho\rho$ (dark shaded) decays as a 
        function of $\alpha$. Also shown is the prediction from the 
        standard CKM fit (hatched area), which includes the world 
        average of $\stb$ but excludes $B\to\rho\rho$. }
\end{figure}
\eei
We conclude from the SU(2) analysis of the $\rho\rho$ system that
\bei

\item   even without a significant measurement of $\BR(\B\to\rho^0\rho^0)$
        a useful constraint on $\alpha$ is obtained.
        The smallness of the theoretical uncertainties allows us to
        include the measurements from the $\B\to\rho\rho$ system 
	into the standard CKM fit.

\item   the success of the $\rho\rho$ system is threefold: $(i)$ due
        to the small mass of the $\rho$ with respect to the \B, the $\rho$
        mesons have dominant longitudinal polarization (\CP-even) with 
        respect to their decay axis; $(ii)$ small penguin contributions 
        and color-suppression improve the $\deltaAlpha$ bounds in absence 
	of the full isospin 
        analysis, and $(iii)$ the capability to measure $S_{\rho\rho,L}^{00}$
        (not possible for $S_{\pi\pi}^{00}$) significantly enhances the 
        sensitivity of the full isospin analysis to $\alpha$, once 
        $\B\to\rho^0\rho^0$ has been observed.

\item   the present uncertainty due to the penguin pollution is
        $-20^{\circ}<\deltaAlpha<18^{\circ}$ for $\CL>10\%
	$ and the
        total uncertainty on $\alpha$ is $33^\circ$ at two standard
        deviations. Including electroweak penguins induces a shift of 
        $-2.1^\circ\pm0.2^\circ$ on $\alpha$.

\item   two (possibly optimistic) attempts to extrapolate the results from
	the present central values and errors into 
        the future lead to expected $2\sigma$ errors on $\alpha$ of 
        approximately $16^\circ$--$19^\circ$ at $500\invfb$ and 
	$10^\circ$--$16^\circ$ at 
	$1\invab$ integrated luminosities, for the solution that is 
	compatible with the standard CKM fit. 

\item   we study the finite width of the 
	$\rho$ and isospin-breaking effects using a simple ansatz. With 
	a breaking of the triangular relation at the 4\% level,
	the corresponding uncertainty on $\alpha$ 
	is found to be of the order of $3^\circ$ for the full isospin analysis,  
	which however depends on the actual values of the $\B\to\rho\rho$
	observables.
	Systematic effects from $\pi\pi$ resonances other than 
        the $\rho(770)$ and/or non-resonant background have been 
        neglected in this study but may become important when the 
	precision on $\alpha$ increases.

\eei

%
%
 \newpage\part{New Physics in \B Transitions}\setcounter{section}{0}
\markboth{\textsc{Part VII -- New Physics in \B Transitions}}
         {\textsc{Part VII -- New Physics in \B Transitions}}
\label{sec:newPhysics}
\section{General Remarks}

Despite weak inconsistencies in $\stb$ from penguin-dominated 
modes (see Section~\ref{sec:standardFit}.\ref{sec:input_s2b}) 
and in $\B\to K\pi$ decays 
(Section~\ref{sec:charmlessBDecays}.\ref{sec:introductionKPi}), 
the SM is able to accommodate the data from the $B$-meson and 
kaon systems within the present experimental and theoretical 
uncertainties.
Hence there is no need (yet) to introduce contributions from 
physics beyond the SM. However, this does not necessarily mean 
that New Physics (NP) contributions are totally absent. 
It is thus interesting to investigate how far today's experiments 
can constrain NP parameters.  
\vs
A large variety of specific NP models exists in the literature,
but for the purpose of a global CKM fit, one should adopt a 
parameterization that is as model-independent as possible. The 
results obtained under general assumptions may then be used to 
draw conclusions upon more specific classes of models.
\vs
The NP analysis we are performing below proceeds in two steps:
\bei 

\item   in the first step, we list the observables that are expected 
        to be dominated by the SM contributions, according to a specific
        assumption we make on the nature 
        of the potential NP. These observables are used to construct a 
        model-independent Unitarity Triangle~\cite{uut},
	 followed by a constrained fit on NP contributions in $\BzBzb$ mixing.

\item   in the second step, the result of this fit is used as an input
        to probe NP in \B decays with sizable contributions from $b \to d$ 
        or $b\to s $ gluonic penguins.

\eei
Because the present experimental errors are still large, and since 
several key modes are not yet well known, we do not attempt 
to perform an exhaustive numerical analysis as we did for the SM 
fit. In some cases we use a rather ``aggressive'' interpretation of the
experimental results, which is justified in view of the expected
improvement of the measurements in the near future. The studies presented hereafter are to be viewed as preliminary 
proposals, which nevertheless allow us to draw instructive conclusions.

 \section{New Physics in $\Delta B=2$ Transitions}
\label{ModelIndependentNewPhysics}

With the use of dimensional arguments~\cite{FleischerIsidoriMatias},
one finds that in a large class of models NP contributes mainly to the 
$\BzBzb$ mixing amplitude ($\Delta B =2$). We will hence allow for arbitrary 
NP corrections to the mixing, while however keeping the possibility that
also the decays ($\Delta B =1$) are non-standard.
\vs
New Physics effects in $\BzBzb$ mixing can be described model-independently 
by two additional parameters, $r_{d}^2$ and $2\theta_{d}$, with the 
definition~\cite{soares,GrossmanNirWorah}
\beq
 \label{NP-BBbarmixing}
 r_{d}^2\,e^{i 2\theta_{d}} \:=\:
 \frac{\langle \Bz | {\mathcal{H}}_{\rm eff}^{\rm full}| \Bzb\rangle}
      {\langle \Bz | {\mathcal{H}}_{\rm eff}^{\rm SM}| \Bzb\rangle}~,\hspace{0.5cm}
\eeq
where ${\mathcal{H}}_{\rm eff}^{\rm full}$ comprises NP and SM
contributions and where ${\mathcal{H}}_{\rm eff}^{\rm SM}$ contains
only the SM contribution.  The SM values for these parameters are
$r_{d}^2 = 1$ and $2\theta_{d} = 0$. 
\vs
We elaborate an analysis allowing to 
constrain both CKM ($\rhobar$ and $\etabar$) and NP ($r_{d}^2$ and 
$2\theta_{d}$) parameters related to flavor-changing processes.
It uses observables
from the \B-meson system only since they are more sensitive to $\rhobar$ 
and $\etabar$ than those obtained from the kaon system. It is inspired 
by similar previous analyses~\cite{GrossmanNirWorah,KN,FleischerIsidoriMatias}.

\subsection{Basic Assumption on New Physics and Physical Inputs}

As in most model-independent NP parameterizations, we assume that
NP contributions to tree-mediated decays are negligible. More specifically,
we require that \textit{decay transitions with four flavor changes (\ie,
$b\to q_1\bar q_2q_3$, $q_1\neq q_2\neq q_3$) are dominated by the SM (SM4FC)}. 
Hence the CKM parameters related to these decays are extracted within
the SM, with the presence of the additional parameters coming 
from NP in $\BzBzb$ mixing.  Here we assume that the
        unitarity of the CKM
        matrix still holds in the presence of NP, in order to ensure that
        the SM contribution to the $\Bz\Bzb$ mixing
        keeps its usual expression as a function of
        $(\rhobar,\etabar)$ and other theoretical parameters.   
\vs
The observables allowing us to constrain the SM and NP parameters within 
this well-defined assumption are listed below:
\bei

\item   $\Vub$ and $\Vcb$ from $b\to u$ and $b\to c$ semileptonic decays,
        which 
        are the same as in the SM 
        (Section~\ref{sec:standardFit}.\ref{sec:fitInputs}).

\item   the constraint on $\tan\gamma$ from the Dalitz plot analysis 
        of $\Bp\to \Dz \Kp$ decays (the CL on $\gamma$ obtained in this 
        analysis has to be interpreted with care, as discussed in 
        Section~\ref{sec:gamma}.\ref{sec:d0k}).

\item   the \CP-asymmetry measurement in $b\to c\bar ud$ and $b\to u\bar cd$ 
        non-leptonic decays (Section~\ref{sec:gamma}.\ref{sec:dstarpi}),
        which determines $|\sin(2\beta+2\theta_d+\gamma)|$. We use
        the CL determined by the toy simulation described in 
        Section~\ref{sec:gamma}.\ref{sec:dstarpi}, because it gives a
        stronger constraint than the Gaussian approximation.

\item   the $\Delta I=3/2$ amplitude of $b\to u\bar ud$ transitions is 
        standard within the \textit{SM4FC} assumption\footnote
        {
                The Gronau-London isospin analysis allows to isolate 
                the $\Delta I=3/2$ amplitude 
                (see Section~\ref{sec:charmlessBDecays}.\ref{par:isospin}), 
                which is proportional to $V_{ud}V_{ub}^*$ in the \textit{SM4FC} 
                hypothesis (up to a small and computable $V_{td}V_{tb}^*$ 
                electroweak penguin contribution that we take into
                account). This holds independently of the magnitude and 
                (\CP-conserving plus \CP-violating) phase of the $\Delta I=1/2$ amplitude, 
                which {\em a priori} receives contributions from both the SM 
                (tree and penguin diagrams) and the NP.
        }: 
        hence, the isospin analysis of $B\to\pi\pi$ and/or longitudinally polarized
        $B\to\rho\rho$ decays and the Dalitz plot analysis of $\Bz\to(\rho\pi)^0$ 
        can be used to extract the quantity $\sin(2\beta+2\theta_d+2\gamma)$ 
        (\cf\  Section~\ref{sec:charmlessBDecays}.\ref{par:isospin}). Since the
        constraint from the $\pi\pi$ system is rather weak at present and the 
        $\rho\pi$ Dalitz plot analysis is not yet available, we will only use 
        $\rho\rho$ in the following (\cf\
        Section~\ref{sec:charmlessBDecays}.\ref{sec:introductionrhorho}).

\item   the mixing-induced \CP asymmetry in $b\to c \bar{c} s$ transitions 
        (\eg, $\Bz\to\jpsi\Kz$) determines $(2\beta+2\theta_d)$, provided 
        that NP contributions to the decay amplitudes of these transitions 
        are negligible (for general arguments see, e.g.,
        Ref.~\cite{FMpsiK}).
        Although this hypothesis does not belong to the \textit{SM4FC} rule, generic 
        non-standard corrections to $b\to c \bar{c} s$ amplitudes are likely 
        to be small, because these modes are dominated by $V_{cb}^*V_{cs}$ 
        SM tree amplitudes: QCD penguins that would receive contributions 
        from new particles in the loop are dynamically suppressed by the 
        weak coupling of, \eg, the $\jpsi$ to gluons, while $Z$-penguin 
        effects are expected to be at most at the level of 
        the current experimental uncertainty~\cite{AtwoodHiller}.

\item   the constraint on the sign of $\cos(2\beta+2\theta_d)$ 
        from the time- and angular-dependent analysis of $\Bz\to\jpsi\Kstarz$
        decays (\cf\  Section~\ref{sec:standardFit}.\ref{sec:jpsikstar}). 
        We use the Gaussian interpretation of 
        the experimental result, which  essentially imposes 
        $\cos(2\beta+2\theta_d) > 0$, since it gives a stronger
        constraint than the (correct) Monte-Carlo simulation.

\item   the $\BzBzb$ oscillation frequency, as given by 
        $\dmd =r_{d}^{2}\times\dmd^{\rm SM}$.
\begin{table}[t]
\begin{center}
\setlength{\tabcolsep}{0.0pc}
\begin{tabular*}{\textwidth}{@{\extracolsep{\fill}}lcc}\hline
&& \\[-0.3cm]
Parameter & Value $\pm$ Error & Reference \\[0.15cm]
\hline
&& \\[-0.3cm]
  $\as(m_{b})$
                & $0.22$
                & \cite{LaplaceLigetiNirPerez}\\
  $K_1$
                & $-0.295$ 
                & \cite{LaplaceLigetiNirPerez} \\
  $K_2$
                & $1.162$ 
                & \cite{LaplaceLigetiNirPerez} \\
  $m_{b}^{\rm pole}$
                & $4.8\pm0.1_{\rm theo}$ 
                & \cite{LaplaceLigetiNirPerez} \\
  $z$
                & $0.085\pm0.010_{\rm theo}$ 
                & \cite{LaplaceLigetiNirPerez} \\[0.15cm]
\hline
&& \\[-0.3cm]
  $A_{\rm SL}$
                & $(-0.7 \pm 1.3) \times 10^{-2}$
                & see text\\[0.15cm]
\hline
\end{tabular*}
\caption{\label{tab:NPinputs} \em
        Inputs used to predict the semileptonic \CP-violating asymmetry 
        $A_{\rm SL}$, which are not already defined in Table~\ref{tab:ckmInputs}.
        The errors given are treated as systematic theoretical 
        uncertainties within \rfit. If no error is given 
        the uncertainty of the corresponding quantity is neglected. The
        last line quotes the experimental average.}
\end{center}
\end{table}

\item   the \CP-violating charge asymmetry in semileptonic \B decays
        $A_{\rm SL}$ defined by
        \beq
              A_{\rm SL} \:\equiv\:
              \frac{\Gamma(\Bzb(t)\to\ell^{+}X)-\Gamma(\Bz(t)\to\ell^{-}X)}
                   {\Gamma(\Bzb(t)\to\ell^{+}X)+\Gamma(\Bz(t)\to\ell^{-}X)}~.
        \eeq
        In the presence of NP in mixing its theoretical prediction 
        reads~\cite{SandaXing,RandallSu,CahnNir,BarenboimEyalNir,LaplaceLigetiNirPerez}:
        \beq
        \label{asl}
        A_{\rm SL} \:=\: - \Re\left(\frac{\Gamma_{12}}{M_{12}}\right)^{\!\rm SM}
                           \frac{\sin{2\theta_{d}}}{r_{d}^2}
                         + \Im\left(\frac{\Gamma_{12}}{M_{12}}\right)^{\!\rm SM} 
                           \frac{\cos{2\theta_{d}}}{r_{d}^2}~,
        \eeq
        where $\Gamma_{12}$ and $M_{12}$ are respectively the absorptive 
        and dispersive parts in the $\BzBzb$ mixing amplitude. 
        \vs
        The theoretical prediction of $\left(\Gamma_{12}/M_{12}\right)^{\rm SM}$ 
        at leading order~\footnote
        {
                Next-to-leading order calculations of $A_{\rm SL}$ have been
                performed in Refs.~\cite{Beneke:2003az,Ciuchini:2003ww}. We do
                not use these results in our analysis since experimental errors
                dominate at present. With increasing precision the NLO results
                must be included.
        }
        reads~\cite{LaplaceLigetiNirPerez}:
        \beqn
        \label{asltheo}
        \lefteqn{
          \left( \frac{\Gamma_{12}}{M_{12}} \right)^{\!\rm SM} 
        =
        - \frac{4\pi m_{b}^{2}}
               {3 m_{W}^{2}\etaBb S_{0}\left(m_{t}^{2}/m_{W}^{2}\right)}
               \left[\left(K_{1} + \frac{K_{2}}{2}\right)
                     + \left(\frac{K_{1}}{2} - K_{2}\right)
                          \frac{m_{B}^{2}-m_{b}^{2}}{m_{b}^{2}
                     B_{}}\right.} \nonumber\\
        &&\hspace{5.7cm}
          \left. +\; \left(K_{2}-K_{1}\right)
                     \left(\frac{5 B^{'}_{S}}{8 B_{d}}
                           + 3z \frac{1 - \rhobar - i \etabar}
                                     {(1-\rhobar)^2+\etabar^2}
                     \right)
               \right]\:,~~~~~~~~~~~
        \eeqn
        where $K_1$ and $K_2$ are Wilson coefficients, 
        $z \equiv m_{c}^{2}/m_{b}^2$ and
        $\etaBb=\etaB\times[\as(m_b)]^{-6/23}(1+(\as/4\pi)(5165/3174))$.
        The corresponding input values are given in Table~\ref{tab:NPinputs}.
        \vs
        We find the experimental value $A_{\rm SL} = -0.007 \pm 0.013$ as an 
        average of several measurements:
        the direct determination of $A_{\rm SL}$~\cite{OPALASL,CLEOASL,ALEPHASL,BABARASL} 
        is dominated by the \babar\  result 
        $A_{\rm SL} = 0.005 \pm 0.012\pm 0.014$~\cite{BABARASL}.
        The \babar\  Collaboration also measured the quantity
        $|q/p|=1.029\pm0.013\pm0.011$ with a fit to fully reconstructed 
        \B decays~\cite{BABARDGAMMA}. This translates into 
        $A_{\rm SL}=(1-|q/p|^4)/(1+|q/p|^4)=-0.057\pm0.033$.

\eei
A summary of the observables used in the NP fit is given in 
Table~\ref{tab:NPinputs0}. The number of independent constraints is
sufficient to constrain both $(\rhobar,\etabar)$ on the one hand, and
$(r_d^2,2\theta_d)$ on the other hand, up to discrete ambiguities.
\vs
We
do not include the \CP-violation parameter $\varepsilon_K$  because it does
not improve the constraint on $\rhoeta$, unless  possible NP contributions
to $\KzKzb$ mixing are negligible, which is not {\it a priori} known.  We do not
consider input from $s\to d$ and $b\to s$  transitions either, which are therefore
left free in the fit within 
and beyond the SM. 

\begin{table}[t]
\begin{center}
\setlength{\tabcolsep}{0.0pc}
\begin{tabular*}{\textwidth}{@{\extracolsep{\fill}}lll}\hline
&&\\[-0.3cm]
Constraint & SM \& NP dependence & Numerical value \\[0.15cm]
\hline
&&\\[-0.3cm]
$|V_{cb}|$ and $|V_{ub}|$       & $|V_{cb}|$ and $|V_{ub}|$     
        & Section~\ref{sec:standardFit}.\ref{sec:fitInputs} \\
$\Bp\to D^{(*)0}\Kp$ Dalitz plot analysis               & $\tan\gamma$
        & Section~\ref{sec:gamma}.\ref{sec:d0k}. \\
$\Bz\to D^{(*)\pm}\pi^\mp$ \CP asymmetries 
                                & $|\sin(2\beta+2\theta_d+\gamma)|$ 
        & Section~\ref{sec:gamma}.\ref{sec:dstarpi} (``toy'')\\
$B\to\rho\rho$ isospin analysis & $\sin(2\beta+2\theta_d+2\gamma)$ 
        & Section~\ref{sec:charmlessBDecays}.\ref{sec:introductionrhorho} \\
$\Bz\to\jpsi K^0$ \CP asymmetry & $\sin(2\beta+2\theta_d)$ 
        & Section~\ref{sec:standardFit}.\ref{sec:fitInputs} \\
$\Bz\to\jpsi K^{*0}$ time-dependent angular analysis & $\cos(2\beta+2\theta_d)$
        & Section~\ref{sec:standardFit}.\ref{sec:fitInputs} (``Gauss.'') \\
$\dmd$                          & $\dmd^{\rm SM}r_d^2$ 
        & Section~\ref{sec:standardFit}.\ref{sec:fitInputs} \\
$A_{\rm SL}$                    & Eqs.~(\ref{asl})--(\ref{asltheo}) 
        & Table~\ref{tab:NPinputs} \\[0.15cm]
\hline
\end{tabular*}
\caption{\label{tab:NPinputs0} \em
        Inputs to the fit with free New Physics contributions to $\BzBzb$ 
        mixing, and their dependence with respect to the SM and NP flavor
        changing parameters. Discussions on the
        discrete ambiguities occurring in the measurements of $\tan\gamma$,
        $|\sin(2\beta+2\theta_d+\gamma)|$ and
        $\sin(2\beta+2\theta_d+2\gamma)$ are given in the corresponding 
        sections.}
\end{center}
\end{table}

\subsection{Results}

\begin{figure}[p]
  \centerline
        {
        \epsfxsize8.1cm\epsffile{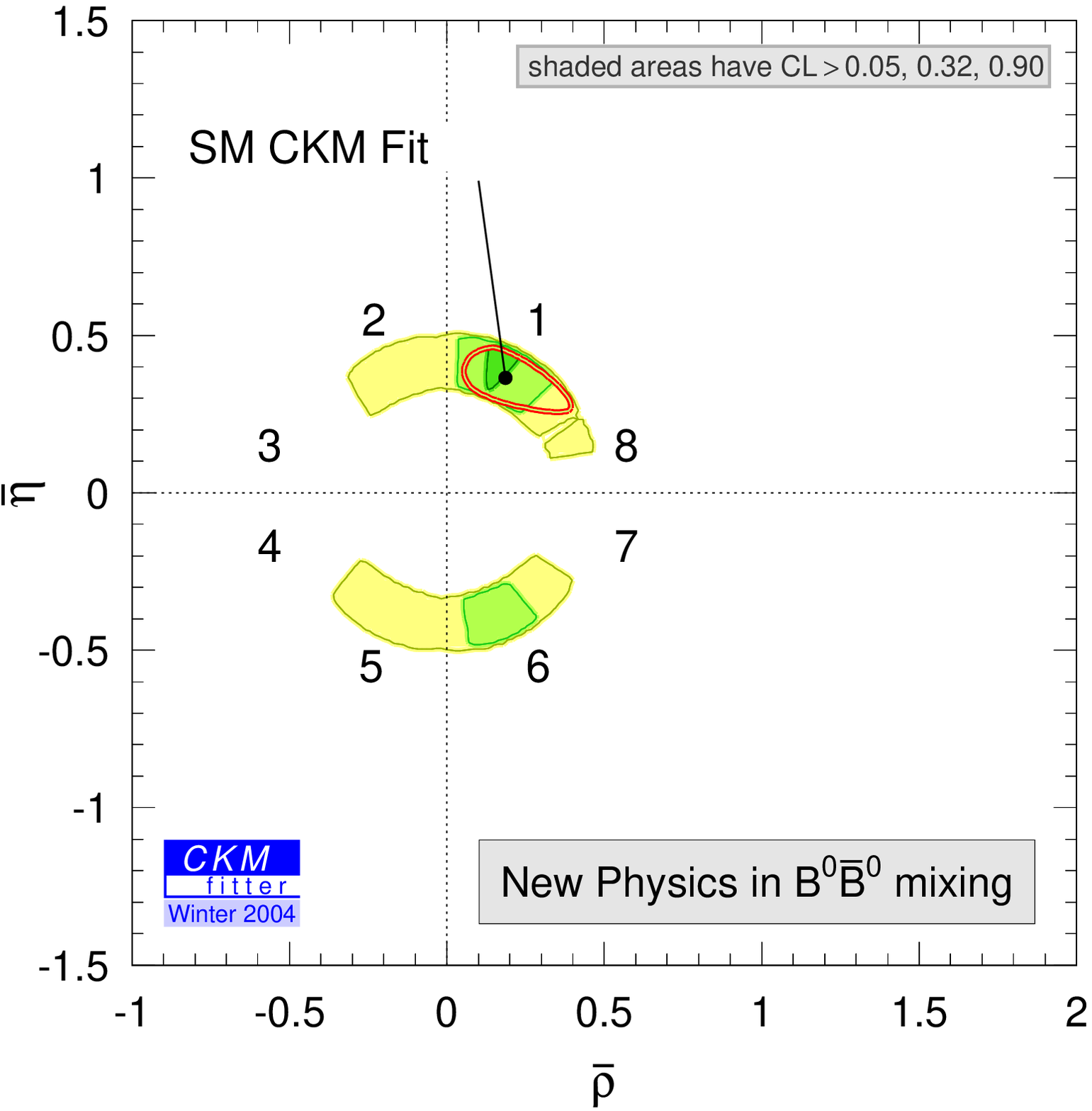}
        \epsfxsize8.1cm\epsffile{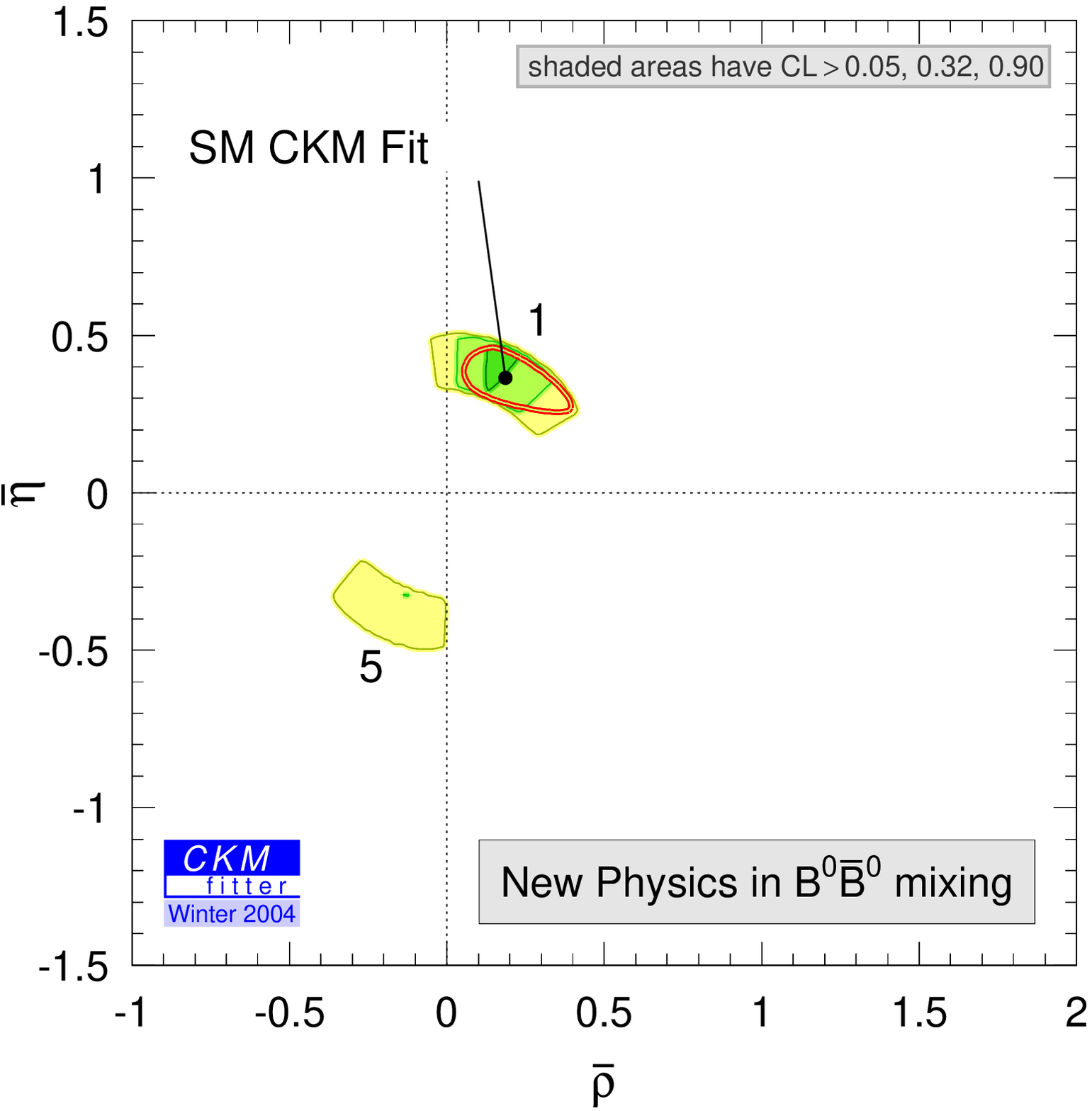}
        }
  \vspace{1.0cm}
  \centerline
        {
    \epsfxsize8.1cm\epsffile{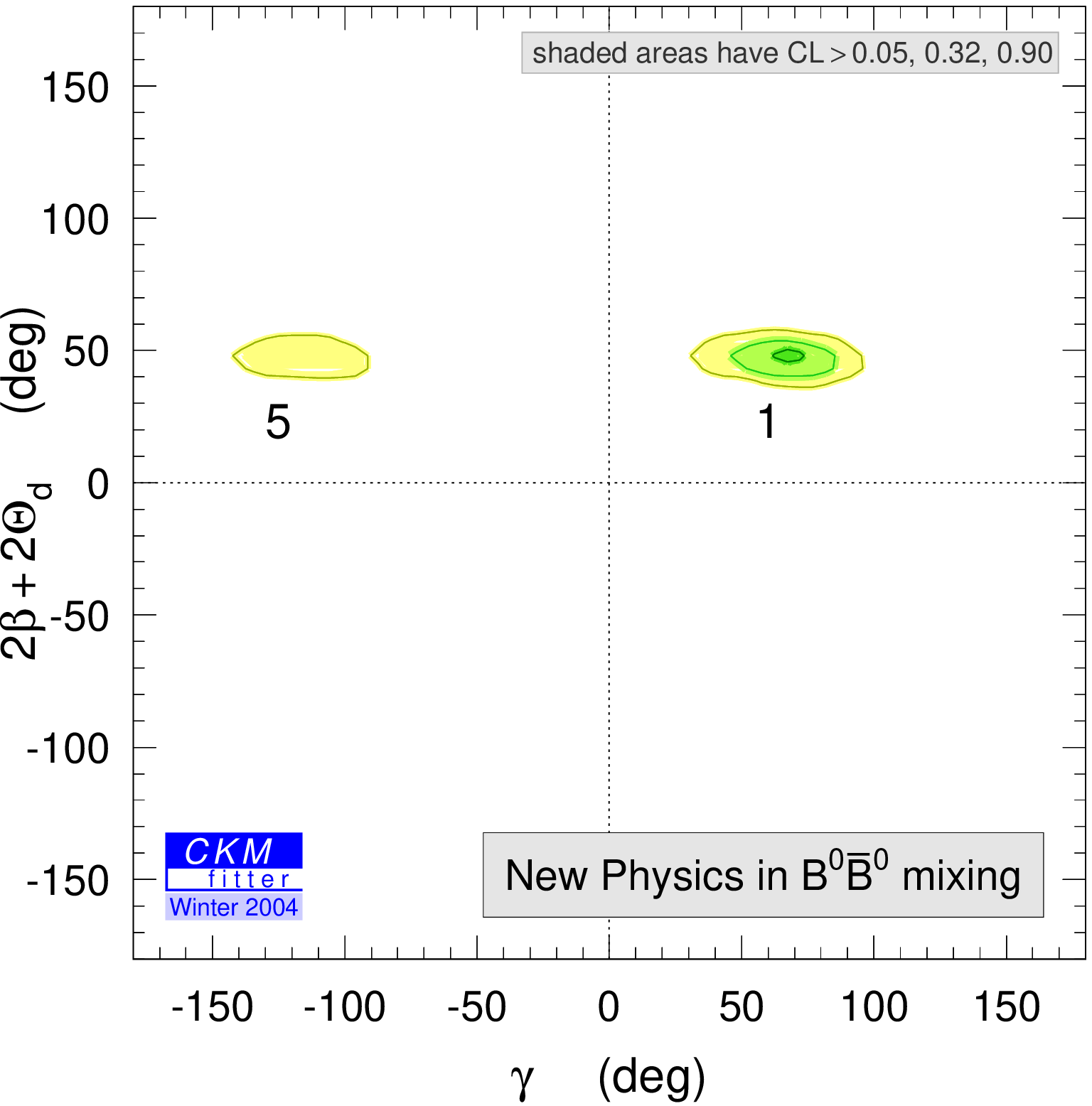}
        }
  \vspace{0.5cm}
  \caption[.]{\label{fig:NPmixing_rhoeta}\em
        Constraints in the $\rhoeta$ (top) and
        $(\gamma,2\beta+2\theta_d)$ (bottom) planes
         from the fit in the framework
        of New Physics in $\BzBzb$ mixing. The top right and bottom
        plots
        give the corresponding constraint when using only the two solutions
        $\beta+\theta_d$ and $\pi+\beta+\theta_d$
        that are favored by the $\cos(2\beta+2\theta_d)>0$ evidence in 
        $\Bz\to\jpsi\Kstar$
        decays (Section~\ref{sec:standardFit}.\ref{sec:jpsikstar}). }
\end{figure}

We perform a global fit using the inputs from Table~\ref{tab:NPinputs0}
with  $r_{d}^2$ and $2\theta_{d}$ left free to vary. The constraint
obtained in the  $\rhoeta$ plane when excluding the sign measurement of
$\cos(2\beta+2\theta_d)$  is shown in the left hand plot of
Fig.~\ref{fig:NPmixing_rhoeta}.  There are eight solutions for the CKM
angle~$\gamma$, numbered from 1 to 8,  that lie on a circle determined
by $|V_{ub}/V_{cb}|$. Solutions 3, 4 and 7 have CLs below $5~\%
$ and
are therefore not shown on this plot. The other individual solutions 
are not well separated yet due to the  experimental uncertainties in
the determination of $\gamma$ from $D^{(*)0}\Kp$, $D^{(*)\pm}\pi^\mp$
and $\rho\rho$.  If we impose in addition the constraint\footnote
{
        As described in Section~\ref{sec:standardFit}.\ref{sec:jpsikstar},
        this is an optimistic assumption, since the $\cos(2\beta)>0$
        result found in Ref.~\cite{babarallbeta} is not yet 
        statistically significant.
}
$\cos(2\beta+2\theta_d)>0$, derived from the analysis of $\jpsi\Kstar$ 
decays~\cite{babarallbeta}, we obtain the right hand plot of 
Fig.~\ref{fig:NPmixing_rhoeta} where four out of eight solutions are
further suppressed.

The CL for $\etabar = 0$ in these fits depends crucially on the inputs from
Table~\ref{tab:NPinputs0}, in particular on the current inputs from the
$\Bp\to D^{(*)0}\Kp$ Dalitz plot analysis ($\tan\gamma$),  the $\Bz\to
D^{(*)\pm}\pi^\mp$ \CP asymmetries ($|\sin(2\beta+2\theta_d+\gamma)|$), the
$B\to\rho\rho$ isospin analysis\footnote
{
        Even if $C_{\rho\rho,L}^{+-}$ significantly departed from zero, 
        the solution $\eta=0$ would still be allowed by the full
        isospin analysis
        in $B^0\to\rho\rho$. This is because
        the isospin analysis only constrains the weak phase of the 
        $\Delta I=3/2$ amplitude, and not the one of the 
        $\Delta I=1/2$ amplitude, which could come from a different source of
        \CP\  violation. In the SM, these two phases are nevertheless 
        correlated since they are related to common CKM couplings.
}
($\sin(2\beta+2\theta_d+2\gamma)$) and  the
$\Bz\to\jpsi K^{*0}$ time-dependent transversity analysis
($\cos(2\beta+2\theta_d)$).  With these inputs the possibility that \CP
violation is absent in the SM  is quite unlikely. With additional
conjectures on the NP's nature, one  could obviously improve the
constraints: for example, with the assumption  that NP is negligible in
the decay, any non-zero direct \CP-asymmetry measurement implies
$\etabar\ne 0$ (\eg, the measurement of $\epe>0$). Generally speaking,
the solution $\etabar=0$ is not a natural value in  the SM: the region
of very small $\etabar$ corresponds to a fine-tuning  scenario, in
which the SM generates vanishingly small \CP violation,  whereas large
NP couplings are needed to accommodate the data.  See however
Ref.~\cite{Krueger} for an example of multi-Higgs-doublet model that
naturally predicts a real-valued CKM matrix. We note that
at that time less
experimental  input was available and consequently a real-valued CKM
matrix could not  be excluded.
\begin{figure}[t]
  \centerline{
        \epsfxsize8.1cm\epsffile{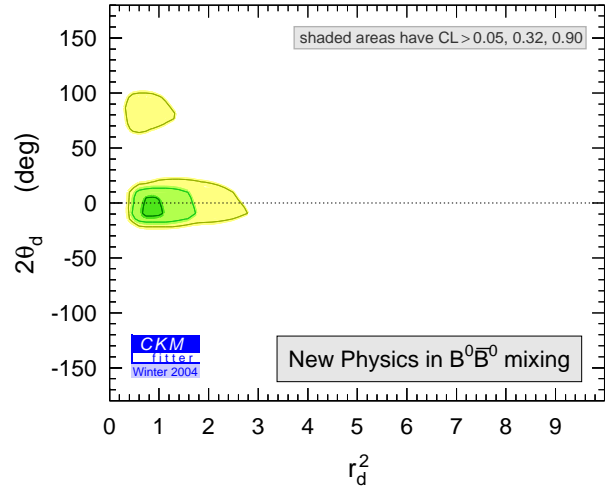}
  }
  \vspace{0.0cm}
  \caption[.]{\label{fig:NPmixing_thetnprd2}\em
        Confidence level in the $(2\theta_{d},r_{d}^{2})$ plane
        obtained as a result of the global CKM fit that includes New 
        Physics in $\BzBzb$ mixing.}
\end{figure}
\vs
Remarkably, Solution 1 in Fig.~\ref{fig:NPmixing_rhoeta} is not
only consistent with the standard (SM) CKM fit but also has the largest 
CL: the SM solution is clearly preferred, while however
 the mirror solution 5 cannot be
excluded at the $1\sigma$ level. At first sight, this is not surprising 
since most of the observables used in the fit are expected to be dominated 
by SM contributions. Nevertheless this leads to an important consequence, 
namely that NP corrections to $\BzBzb$ and $\KzKzb$ mixing are likely to 
be small. In the case of $\BzBzb$, this is illustrated in 
Fig.~\ref{fig:NPmixing_thetnprd2} showing the constraints in the 
$(r_{d}^2,2\theta_{d})$ plane: the SM solution $r_d^2=1$ and $2\theta_d=0$ 
is favored. Values for $r_{d}^{2}$ as large as 2--3 cannot be excluded
yet, which means that in principle order one NP contributions to the
mixing are allowed.
Still, the model-independent
constraint on $r_{d}^{2}$ is much better than in the previous similar
analyses; the uncertainty will decrease with better precision 
on $\Delta{m_d}$ and $A_{\rm SL}$ and, in particular, on $\fbdbd$. 
This highlights the need for improved determinations of the parameter 
$\fbdbd$ both from theory (\eg, from improved Lattice 
QCD calculations) and from experiment (\eg, from constraining 
$\Vub f_{B_{d}}$ by a rate measurement of $\B \to \ell \nu_{\ell}$
decays). The constraint on $r_{d}^{2}$ would also be improved with 
a better knowledge of the angle $\gamma$.
\vs
With the constraint on $\rhoeta$ shown in the right plot in 
Fig.~\ref{fig:NPmixing_rhoeta}, we determine the contribution to
$|\epsk|$ coming from the Standard Model. For Solution 1 we find  $1.3
\times 10^{-3} < |\epsk|_\mathrm{SM} < 5.0 \times 10^{-3}$
for CL $>5\%
$. This  can be compared to the experimental value, $|\epsk| = 2.282
\times 10^{-3}$,  and the constraint $1.1 \times 10^{-3} <
|\epsk|_\mathrm{CKM} < 4.9 \times 10^{-3}$ at  CL $>5\%
$ (Table~\ref{tab:fitResults1}, 
Section~\ref{sec:standardFit}.\ref{sec:metrologyStandardFit}) obtained
from the  standard CKM fit. That is, neither in the framework of NP in
$\BzBzb$ mixing nor in the SM, one can exclude NP contributions to
$\KzKzb$ mixing of order  $100\%
$ due to the uncertainties on the bag parameter $B_{K}$ and also,
to some extent, on the charm quark mass $m_{c}$  (\cf\  
Table~\ref{tab:ckmInputs},
Section~\ref{sec:standardFit}.\ref{sec:metrologyStandardFit}). Solution 5
leads to negative values for $|\epsk|_\mathrm{SM}$ and could only
accommodate the measurement in the presence of NP effects in
$\KzKzb$ mixing that, in addition to being large, would have 
a sign opposite to the SM contribution.
\vs
It is interesting to note that no solution is obtained for $2\theta_{d}=\pi$. 
In Minimal Flavor Violation (MFV) NP models
(see, \eg,
Refs.~\cite{MFV})
a single real parameter, $F_{tt}$, is needed in addition to those of the SM 
to describe model-independently all the observables. In the SM, the
value of this parameter is
$F_{tt}|_\mathrm{SM}=S(m_{t}^2/m_{W}^2)=2.41$, where $S(m_{t}^2/m_{W}^2)$ 
is the Inami-Lim function in the $\Bz\Bzb$ mixing amplitude.
Our fits exclude a negative sign for $F_{tt}$ corresponding to 
$2\theta_{d}=\pi$ leaving $F_{tt}>0$ as the preferred solution. The 
parameter $F_{tt}$ is related to our parameterization by 
$r_{d}^{2}=|F_{tt}|/S(m_{t}^2/m_{W}^2)$.
With the inputs from Table~\ref{tab:NPinputs0} we obtain the range 
$1.03 < F_{tt} < 4.41$ for CL $>10\%
$. In MFV models the NP contribution 
to $\Bz\Bzb$ mixing is directly related to the NP contribution to $\Kz\Kzb$ 
mixing. When taking into account the $\epsK$ constraint we obtain
$1.03 < F_{tt} < 4.18$ for CL $>10\%
$. In addition, the modification of 
$\Dms$ in MFV models is the same as in $\dmd$. When also taking into 
account $\Dms$ in the fit, the bounds on $F_{tt}$ are further tightened: 
we find $1.18 < F_{tt} < 4.01$ for CL $>10\%
$.
\vs
An interesting question is how the non-standard Solution 5 can be
excluded, independently of the argument based on $|\epsK|$ given above.
A reduction of the uncertainties in the $\Bp\to D^{(*)0}\Kp$ Dalitz plot 
analysis, in the $\Bz\to D^{(*)\pm}\pi^\mp$ \CP asymmetries 
 and in the $B\to\rho\rho$ isospin 
analysis will not help in this 
respect~\cite{quinngrossman}, since all these constraints are invariant under 
the transformation $\gamma \rightarrow \gamma+\pi$.
However, discriminating the Solutions 1 and 5 would be possible if the 
measurement of $A_{\rm SL}$ could be significantly improved. 
\vs
In contrast to the analysis performed in
Ref.~\cite{FleischerIsidoriMatias},  we study NP contributions to
$\BzBzb$ mixing model-independently, \ie,   without the neglect of NP
in the decay and without any dynamical assumption.  Better constraints
can be expected, for instance, when precise measurements  of $\gamma$
from tree-level decays become available.

 \section{New Physics in $\Delta B=1$ Decays}
\label{NPdB=1}

Flavor-changing neutral currents that occur only at the loop level in
the SM receive large corrections in many generic New Physics 
scenarios~\cite{revueNP}. In this section we present constraints 
on $b\to d$ and $b\to s$ transitions in a model-independent framework.

\subsection{$B\to PP$ Modes: $\Bz\to\pip\pim$ versus $\Bp\to\Kz\pip$}

The constraints on the angles $\gamma$ and $2\beta+2\theta_d$, obtained 
in the $\Delta B=2$ analysis of the previous section, is used 
to constrain possible NP contributions in $\Delta B=1$ transitions. For 
this purpose, we fit the magnitude of the penguin amplitude $|P^{+-}|$ 
occurring in $\Bz \to \pip\pim$. More precisely, we define the ratio of 
$b\to d$ to $b\to s$ transitions by
\beq
\label{defRatio}
        r_{\pi\pi}^P\:\equiv\: 
        \sqrt{\frac{\tau_{B^+}}{\tau_{B^0}}
        \frac{\mathrm{PS}\left|V_{cs}V_{cb}^*P^{+-}_{\pi\pi}\right|^2}{\BR(K^0\pi^+)}}~,
\eeq
where $\mathrm{PS}$ stands for the usual two-body phase space factor and
where the penguin amplitude is defined, in contrast to
Section~\ref{sec:charmlessBDecays}.\ref{par:amplitude}, in the $\T$
convention
\beq\label{Apipi}
        A(\Bz\to\pip\pim) \:=\: V_{ud}V_{ub}^*T^{+-}_{\pi\pi}
        +V_{cd}V_{cb}^*P^{+-}_{\pi\pi}~.
\eeq
Although at first sight Eq.~(\ref{Apipi}) relies on the SM and CKM 
unitarity, it remains valid in the presence of arbitrary NP contributions, 
since any new amplitude with a new \CP-violating phase can be 
decomposed into two independent CKM couplings\footnote
{\label{mostgalparam}
        This can be seen as follows. Let us denote by 
        $M_{\rm NP}e^{i\phi_{\rm NP}}$ an arbitrary NP amplitude with a
        \CP-violating phase $\phi_{\rm NP}$. One finds the identity
        $M_{\rm NP}e^{i\phi_{\rm NP}}
        =V_{ud}V_{ub}^*T_{\rm NP}+V_{cd}V_{cb}^*P_{\rm NP}$ with
        $T_{\rm NP}=M_{\rm NP}\Im\left[e^{i\phi_{\rm NP}}V_{cd}^*V_{cb}\right]/
        \Im\left[V_{ud}V_{ub}^*V_{cd}^*V_{cb}\right]$ and
        $P_{\rm NP}=M_{\rm NP}\Im\left[V_{ud}V_{ub}^*e^{-i\phi_{\rm NP}}\right]/
        \Im\left[V_{ud}V_{ub}^*V_{cd}^*V_{cb}\right]$. 
        The whole effect of NP in the decay amplitude is to modify the ``$T$''-type and ``$P$''-type amplitudes with respect to their SM values.
}.
In other words, Eq.~(\ref{Apipi}) is the most general parameterization
of the decay amplitude both within and beyond the SM.
\vs
In the SM the ratio $r_{\pi\pi}^P$ is expected to be of order one: it would be
equal to one if SU(3) symmetry were exact and in the limit of vanishing
annihilation/exchange and electroweak penguin topologies (\cf\  
Section~\ref{sec:charmlessBDecays}.\ref{sec:naiveFA}). Thus, any large
($\gsim 30\%
$) deviation would be a hint of non-standard particles occurring
in the gluonic or electroweak penguin loops. Using Eqs.~(\ref{Apipi}) 
for the decay and (\ref{NP-BBbarmixing}) for the mixing, one can express 
$r_{\pi\pi}^P$ in terms of the experimental observables and the angles 
 $\gamma$ and $2\beta+2\theta_d$
\beq\label{magic}
        r_{\pi\pi}^P\:=\:\left[\frac{\tau_{B^+}}{\tau_{B^0}}
                    \frac{\BR(\pi^+\pi^-)}{2\lambda^2\BR(K^0\pi^+)}
                \,
        \frac{1-\sqrt{1-\Cpipi^2}\cos(2\beta+2\theta_d+2\gamma+2\alpha_{\rm eff})}
             {\sin^2\gamma}\right]^{\frac{1}{2}}\,.
\eeq
\begin{figure}[t]
  \centerline{
    \epsfxsize8.1cm\epsffile{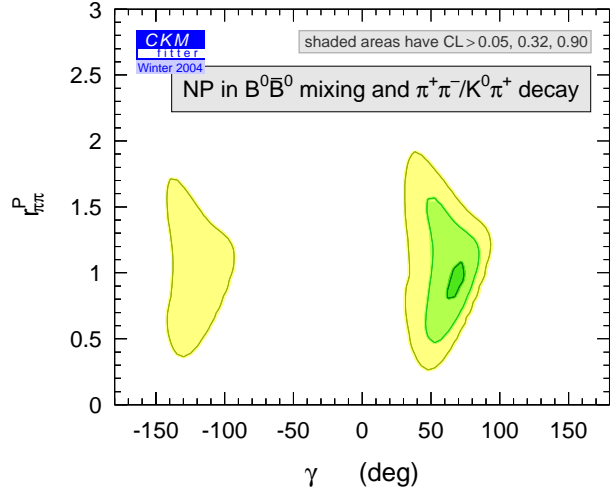}
  }
  \vspace{0.2cm}
  \caption[.]{\label{fig:PpipioversqrtK0pi_r}\em
        Confidence level in the $(\gamma,r_{\pi\pi}^P)$ plane obtained
        by the fit including NP in $\BzBzb$ mixing and in $b\to d,s$
        transitions.}
\end{figure}
Figure~\ref{fig:PpipioversqrtK0pi_r} shows the 
confidence level as a function of $\gamma$ and $r_{\pi\pi}^P$, using the following  input 
quantities:
\bei

\item   the constraints on $\gamma$ and $2\beta+2\theta_d$ obtained from the
        New Physics fit in $\BzBzb$ mixing as shown in the bottom plot of
        Fig.~\ref{fig:NPmixing_rhoeta}.

\item   the \CP-violating asymmetries in $B \to \pi^+\pi^-$: $\Cpipi$ and
        $\Spipi=\sqrt{1-\Cpipi^2}\staeff$ 
        (Table~\ref{tab:BRPiPicompilation}, 
        Section~\ref{sec:charmlessBDecays}.\ref{par:expinput}).

\item   the branching fractions of the three $B \to \pi\pi$ modes 
        (Table~\ref{tab:BRPiPicompilation}, 
        Section~\ref{sec:charmlessBDecays}.\ref{par:expinput}),
        assuming isospin symmetry.

\item   the branching fraction of $B^+ \to K^0\pi^+$
        (Table~\ref{tab:BRPiPicompilation}, 
        Section~\ref{sec:charmlessBDecays}.\ref{par:expinput}).
\eei
The resulting constraints on $r_{\pi\pi}^P$  prefer an order one
value. Since we expect deviations up to $\pm 30~\%
$
from one due to the violations of the relation between $\pi^+\pi^-$ and
$K^0\pi^+$ penguin amplitudes, 
non-standard corrections could be as large, in principle,
 as the SM contribution.
More precise measurements of the observables in the $\pi\pi$ system would
significantly reduce the allowed domain for $r_{\pi\pi}^P$, while the
$\gamma$ input from the NP fit in $\BzBzb$ mixing is found to be less crucial.

\subsection{$B\to VP$ Modes: $B\to\phi K^{0}$ versus $\Bp\to K^{*0}\pip$}

The $b\to d$ to $b\to s$ penguin ratio can be studied in
vector-pseudoscalar channels using, for example, the $\rho\pi$ 
modes compared to the $K^*\pi$ and $K\rho$ partners. 
However a Dalitz plot analysis of the $\pi^+\pi^-\pi^0$ three-body 
decay is necessary to extract the penguin amplitudes. Hence we 
focus on the ratio of two $b\to s$ transitions, represented by 
the decays $\Bz\to\phi K^{0}$ and $\Bp\to K^{*0}\pip$ (see also
Ref.~\cite{GRphiK}). The first is
particularly interesting in view of the marginal agreement between
\babar and Belle in the measurement of the \CP-asymmetry\footnote
{
        We use here the notation $S_{\phi K}$ and
        $C_{\phi K}$ for both decays $\Bz\to\phi \KS$ and $\Bz\to\phi \KL$,
        where the relative sign in $\phi \KS$ with respect to $\phi \KL$
        is taken into account when the results of both channels for 
        $S_{\phi K}$ are averaged. NP effects that could spoil the relation
        between the two decays are expected to be highly
        suppressed~\cite{GKLKSvsKL}.
} $S_{\phi K}$; general studies of this decay can be found in
Refs.~\cite{phiKgal}.
\vs
We define the $\Bz\to\phi K^{0}$ amplitude in the $\T$ convention by
\beq
\label{AphiKs}
        A(B^0\to\phi K^{0}) \:=\: V_{us}V_{ub}^*P_{\phi K}^u
         + V_{cs}V_{cb}^*P_{\phi K}^c~,
\eeq
and the corresponding penguin ratios by
\beq
        r_{\phi K}^{c} \:\equiv\:
        \sqrt{\frac{\tau_{B^+}}{\tau_{B^0}}
        \frac{\mathrm{PS}\left|V_{cs}V_{cb}^*P_{\phi K}^c\right|^2}{\BR(K^{*0}\pi^+)}}~,\ \ \ \ \ \ \
        r_{\phi K}^{u/c} \:\equiv\: \left|\frac{P_{\phi K}^u}{P_{\phi K}^c}\right|~.
\eeq
In the SM, $r_{\phi K}^{c}$ is expected to be close to one, if SU(3) is
a good symmetry and if electroweak penguins and annihilation topologies
are negligible. Long-distance $u$- and $c$- penguins are expected
to be suppressed by $1/m_b$ according to  QCD FA~\cite{BBNS0}. Thus a
value of $r_{\phi K}^{c}$ (resp. $r_{\phi K}^{u/c}$)
that differs significantly from one would
point towards non-standard contributions in electroweak  penguins
(resp. in either gluonic or
electroweak  penguins\footnote
{
        As pointed out in Ref.~\cite{ligetiquinn}, it 
        is not possible from the present data, and from theoretical 
        arguments derived with the use of strict SU(3), to exclude a 
        value of the ratio $r_{\phi K}^{u/c}$ as large as $\sim 10$ in the SM. 
        Such an extreme value would point to very large non-perturbative 
        rescattering effects. However we stress that the natural
        expectation in the SM is $r_{\phi K}^{u/c}\sim 1$. More data and a 
        better understanding of rescattering effects in $B$ decays 
        will help to clarify the situation.
}
). Note also that $P_{\phi K}^u$ appears in
Eq.~(\ref{AphiKs}) together with a $\lambda^2$-suppressed factor.
Thus the natural order of magnitude of the ratio $r_{\phi K}^{u/c}$ in
the presence of a $b\to s$ NP amplitude that competes with the SM
contribution is $1/\lambda^2$.
\vs
The explicit expressions for $r_{\phi K}^c$ and $r_{\phi K}^{u/c}$ 
in terms of the observables are
\beqn
r_{\phi K}^c&=&\left[\frac{\tau_{B^+}}{\tau_{B^0}}\frac{\BR(\phi
K^0)}{2\BR(K^{*0}\pi^+)}
\,
\frac{1-\sqrt{1-C_{\phi K}^2}\cos(2\beta+2\theta_d+2\gamma-2\beta_{\rm eff})}
{\sin^2\gamma}\right]^{\frac{1}{2}},\label{magicVP1}\\
r_{\phi K}^{u/c}&=&\frac{1}{\lambda}\left|\frac{V_{cb}}{V_{ub}}\right|\left[\frac
{1-\sqrt{1-C_{\phi K}^2}\cos(2\beta+2\theta_d-2\beta_{\rm eff})}
{1-\sqrt{1-C_{\phi K}^2}\cos(2\beta+2\theta_d+2\gamma-2\beta_{\rm eff})}\right]
^{\frac{1}{2}}.\label{magicVP2}
\eeqn
We set CLs on the quantities $r_{\phi K}^c$ and $r_{\phi K}^{u/c}$, where 
we distinguish between the \babar\  and Belle results for $C_{\phi K}$ 
and $S_{\phi K}$ since they lead to different implications. The fit inputs 
used are:
\bei
\begin{figure}[t]
  \centerline
        {
        \epsfxsize8.0cm\epsffile{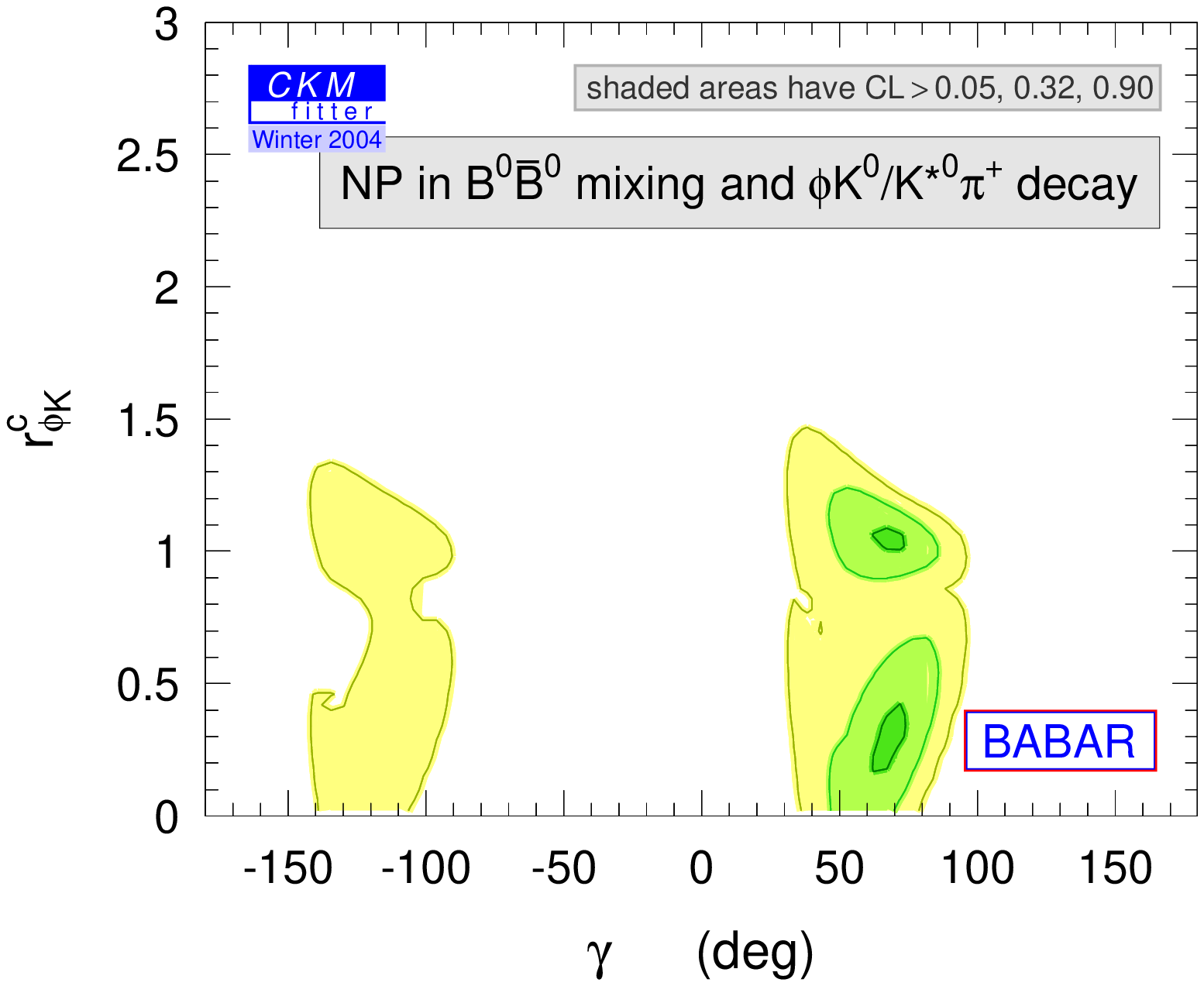}
        \epsfxsize8.0cm\epsffile{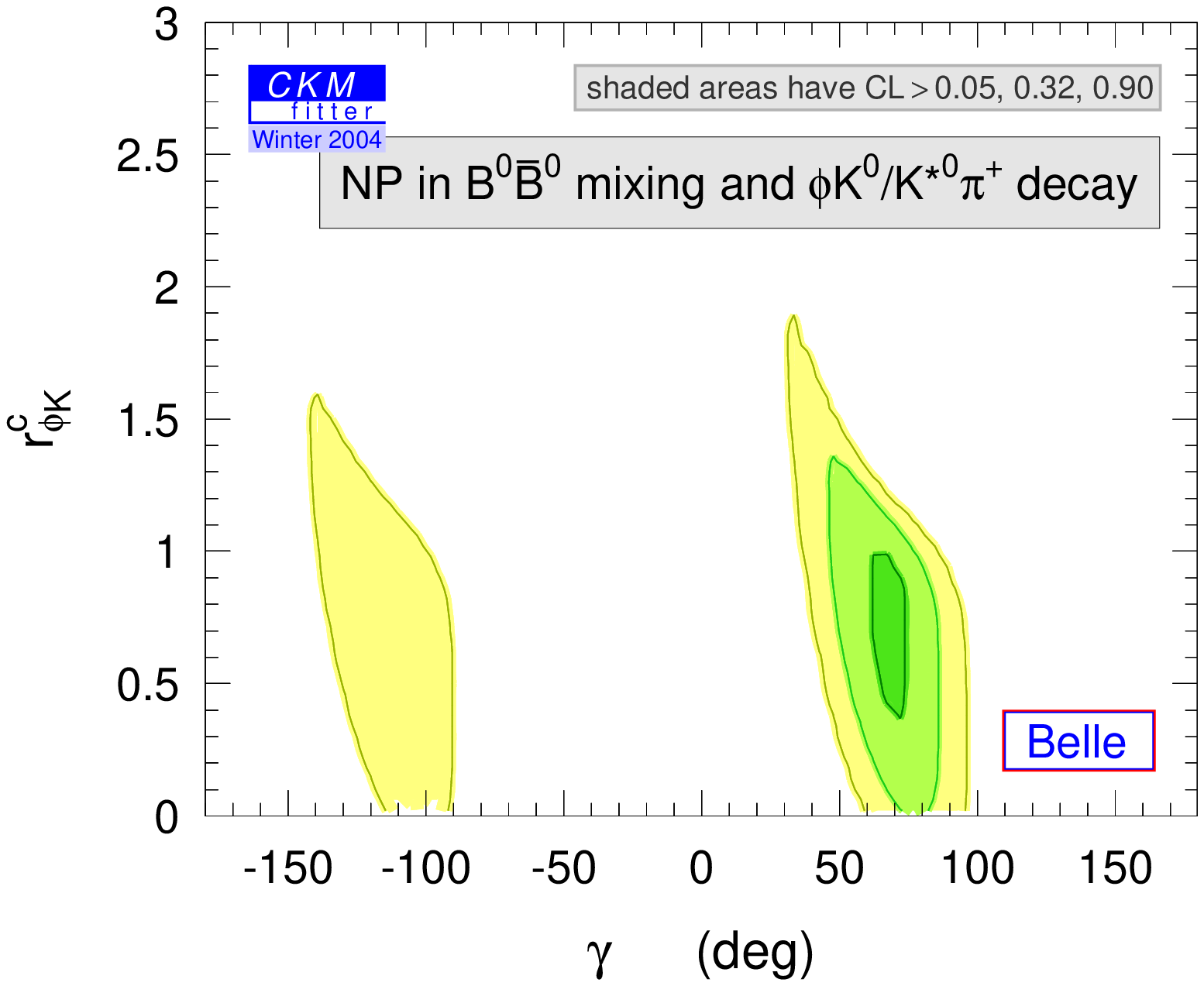}
        }
  \vspace{0.5cm}
  \centerline
        {
        \epsfxsize8.0cm\epsffile{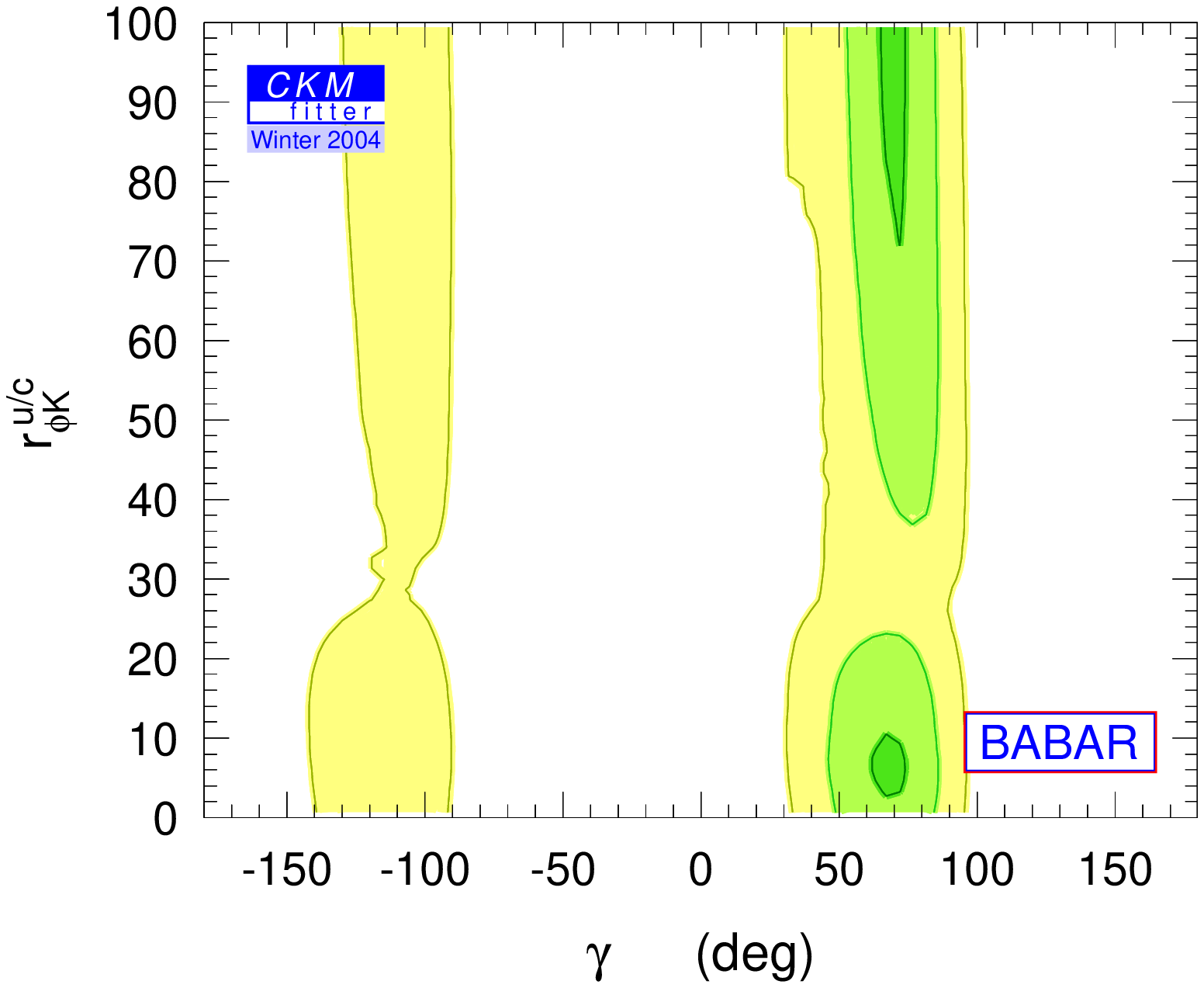}
        \epsfxsize8.0cm\epsffile{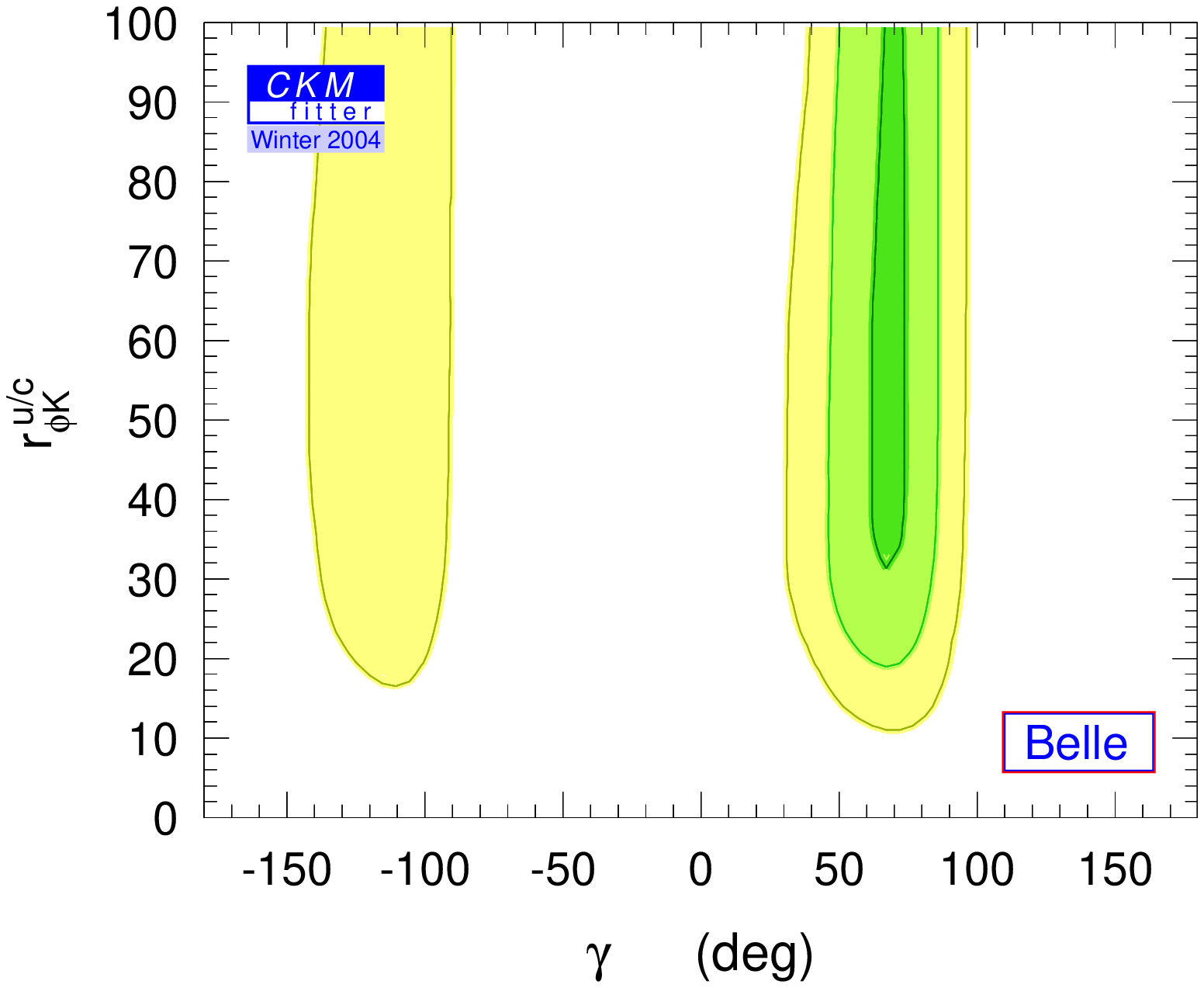}
        }
  \vspace{0.0cm}
  \caption[.]{\label{fig:rc_ruc}\em
        Confidence levels in the $(\gamma,r_{\phi K}^c)$ (upper) and
        $(\gamma,r_{\phi K}^{u/c})$ (lower)
        planes for \babar\  (left) and Belle (right), obtained
        by the fit including NP in $\BzBzb$ mixing and in $b\to s$
        transitions. On the \babar\ plots, small (resp. large)
        values for $r_{\phi K}^c$ (resp. $r_{\phi K}^{u/c}$) correspond
        to the solution $\cos{2\beta_{\mathrm{eff}}}<0$, and conversely.
        On the Belle plots, the two solutions are merged.}
\end{figure}

\item   the constraints on $\gamma$ and $2\beta+2\theta_d$ obtained from the
        NP fit in $\BzBzb$ mixing as shown in the left hand plot of
        Fig.~\ref{fig:NPmixing_rhoeta}.

\item   the branching fraction 
        $\BRPhiKz = (8.3^{+1.2}_{-1.0}) \tmsix$~\cite{HFAG},
        and the \CP asymmetries in 
        $\Bz \to \phi \Kz$, $C_{\phi K}$ and $S_{\phi K}=\sqrt{1-C_{\phi K}^2}\stbeff$. 
        \babar\  measures:
        $S_{\phi K}= 0.47 \pm 0.34^{+0.08}_{-0.06}$,
        $C_{\phi K}= 0.01 \pm 0.33 \pm 0.10$~\cite{BABARphiks}, and 
        Belle finds:
        $S_{\phi K}=-0.96 \pm 0.50^{+0.09}_{-0.11}$,
        $C_{\phi K}= 0.15 \pm 0.29 \pm 0.08$~\cite{Bellephiks,HFAG},
        where the first errors given are statistical and the second systematic.

\item   the branching fraction $\BRKstrpiC$ (Table \ref{tab:BRRhoPicompilation}, 
        Section~\ref{sec:charmlessBDecays}.\ref{sec:theData}).

\eei

The CL on $(\gamma,r_{\phi K}^c)$ for $S_{\phi K}$ and $C_{\phi K}$
measured  by \babar\  (upper left hand plot in Fig.~\ref{fig:rc_ruc}) shows
two solutions for  $r_{\phi K}^c$ due to the the twofold
ambiguity\footnote{
In the $\pi^+\pi^-$ case (see the preceding section), the second
solution for $\cos2\alphaeff$ is suppressed by the additional
constraints coming from the other $\pi\pi$ branching fractions.
} on
$2\beta_{\mathrm{eff}}$. One of the  solutions
($\cos{2\beta_{\mathrm{eff}}}>0$) is in agreement with $r_{\phi
K}^c$  being one as expected if non-standard electroweak penguins are
absent. In the case of the $S_{\phi K}$ and $C_{\phi K}$ results from 
Belle (upper right hand plot in Fig.~\ref{fig:rc_ruc}) the two mirror
solutions cannot be distinguished due  to the $S_{\phi K}$ value being
close to minus one. Also in this case, the constraint on  $r_{\phi K}^c$ is
in agreement with one, which recalls the remark in
Ref.~\cite{hillerphiK} that the measured branching ratio to $\phi K$ is 
compatible with what is theoretically expected.
\vs
In the case of $r_{\phi K}^{u/c}$, one of the solutions
($\cos{2\beta_{\mathrm{eff}}}>0$) for the \babar\  measurement (lower
left hand plot in Fig.~\ref{fig:rc_ruc}) is consistent  with order one values
whereas the other solution 
($\cos{2\beta_{\mathrm{eff}}}<0$) prefers large values. On the contrary, for
the Belle result (lower right hand plot in Fig.~\ref{fig:rc_ruc}), rather 
large values for $r_{\phi K}^{u/c}>10$ are
preferred indicating non-standard gluonic or electroweak penguins.
Since $r_{\phi K}^c$ is found to be compatible with one, these large
$r_{\phi K}^{u/c}$ values suggest that the anomaly, if there, may stem  
from gluonic penguins rather than from electroweak penguins, in contrast 
to some proposals in the literature~\cite{AtwoodHiller,PEW_phiK}. It is 
worthwhile to note that the constraints $r_{\phi K}^{u/c}$ and 
$r_{\phi K}^{c}$ can be significantly improved in the future by reducing 
the experimental uncertainties  on $S_{\phi K}$ and $C_{\phi K}$, and that 
again the $\gamma$ input is less crucial here.

 \section{Conclusion}
We have studied the constraints from present data on the amplitude
parameters  in the presence of arbitrary New Physics contributions to
$\Kz\Kzb$ and $\Bz\Bzb$ mixing,  and to $b\to d$ and $b\to s$ penguin
transitions. The construction of a  model-independent Unitarity
Triangle is not (yet) precise enough to exclude sizable  non-standard
corrections to the mixing, in contrast to a prejudice commonly found 
in the literature. 
\vs
The above statement should be softened in view of the great success of
the standard CKM fit. Although this success could be accidental, the
more general description including NP contributions is not particularly
satisfying since it does not improve the fit while adding new, unknown
parameters. Notably, the preferred region is consistent with the SM
values for the NP parameters in $\Kz\Kzb$ and $\Bz\Bzb$ mixing. It might still be
that NP contributions to $\Kz\Kzb$ and $\Bz\Bzb$ mixing may still be 
present, but can only be
uncovered  if the uncertainties on the inputs are significantly
reduced. This  situation will improve in the future as soon as accurate
determinations  of the angles $\alpha$ and $\gamma$ from tree-dominated
decays become  available, the sign of $\cos(2\beta+2\theta_d)$ is fully settled
and improved determinations of the parameter $\fbdbd$ are obtained.
\vs
We have proposed a fully model-independent parameterization of
$\Delta B=1$ decays. Present errors are large, thus
excluding any definitive statement. A more precise measurement of the
observables of the $B\to\pi\pi$ decays, in particular the
time-dependent \CP-asymmetry in  $\pi^+\pi^-$, would
greatly improve the constraint on the $b\to d$ to $b\to s$ amplitude
ratio, while the input on the angle $\gamma$ from the model-independent
UT fit is less crucial. Still it can be said that potential NP
contributions cannot exceed 100\%
of the SM strength of the $b\to d$ transition. More complete data on the
$b\to u\bar u d$ decays to $VP$ and $VV$ final states would provide valuable
independent information. In addition, our fits to the $B^0\to\phi K^0$ \CP 
asymmetries measured by \babar\  and Belle show that the Belle measurement 
slightly prefers NP contributions with gluonic, rather than electroweak, 
penguin quantum numbers. However, the discrepancy with the \babar\  
result prevents us from drawing a firm conclusion.

 %
%
 \newpage\part{Conclusion}\setcounter{section}{0}
\markboth{\textsc{Part VIII -- Conclusion}}
         {\textsc{Part VIII -- Conclusion}}
\label{sec:conclusions}
During five years of successful running of the \B factories, the
experiments \babar and Belle have produced a wealth of results, which
greatly extend the knowledge  on \B physics acquired at the precursor
experiments ARGUS, CLEO, the LEP collaborations  as well as CDF.
Also using the measurement of indirect \CP violation  in the neutral
kaon system, these experiments succeeded to predict \CP violation in
the \B system, namely $\stb$, with a good precision and far prior to its
direct measurement. Today, however, the measurement of  $\stb$
surpasses in precision the indirect determination. It represents  the
primary constraint on the Unitarity Triangle, and the only one that is
theoretically fully under control. Yet, the determination of $\rhobar$,
$\etabar$ still requires input from measurements for which the
theoretical predictions suffer from notable hadronic uncertainties.
While these are reasonably well controlled, as far as the 
corresponding matrix elements can be computed with
Lattice  QCD, they exhibit large errors. It is the goal of the \B
factories to reduce this dependence on the strong interaction theory
by means of  direct and precise measurements
of the three Unitarity Triangle angles and the two sides. 
\vs
Since $\alpha$ and $\gamma$ are linked to CKM-suppressed $b\to u$
transitions,  large statistics samples are required for their
measurement. Encouraging  results have been presented recently by
\babar on the measurement of $\staeff$  from a time-dependent analysis
of $\Bz\to\rho^+\rho^-$. Using its SU(2) partners  measured by \babar
and Belle, one can determine $\alpha$ with a precision of  $19^\circ$
at $90\% 
$~CL, limited by the unknown penguin contribution. Due to the
weak sensitivity to the penguin contamination of the isospin
relations in $\B\to\pi\pi$, the results on  $\staeff$ from the
measurement of time-dependent \CP asymmetry are less  constraining.
The analysis of $\B\to\rho\pi$ decays mainly lacks  the information
from the Dalitz plot on the strong phase between the  $\rho^+\pi^-$ and
$\rho^-\pi^+$ states of the $\Bz$ decay. The near future  will clarify
the achievable precision on $\alpha$ with these modes,  which strongly
depends on the  underlying decay dynamics. Large \B-related
backgrounds, which are unknown  to some extend at present, complicate
the experimental analysis of  $\B\to\rho\rho$ and (to a lesser degree)
$\B\to\rho\pi$ decays, and hence  produce sizable systematic
uncertainties. The fruitful competition among  the leading experiments
provides important redundant measurements for these modes. The extraction of
$\gamma$ from the interference of $b\to c$ with $b\to u$  transitions
is even more challenging due to the disparateness of the amplitude 
sizes, which suppresses either
the observable \CP-violating asymmetries or the total rate.  A
significant measurement of $\gamma$ in a single mode requires larger
data  samples than those presently available. One way out of this is to
combine  measurements from a large number of different modes.
\vs
Two-body charmless \B decays into pions and kaons are particularly
convenient for phenomenological analyses since all modes, apart from
those dominated by suppressed annihilation, exchange or 
$b\to d$ penguin amplitudes, 
have been measured. Also, the simplicity of the experimental signature reduces 
the systematic uncertainties. Penguin contributions, even 
in decays without net strangeness in the final state, make them to 
potentially sensitive probes of physics beyond the Standard Model. 
\vs
Similarly, the measurement of mixing-induced or direct \CP violation 
in modes that are dominated by $b\to s$ penguin amplitudes enjoy rising interest 
and are among the most anticipated results of the \B factories. We note 
that a claim for New Physics not only requires that at least
one of these modes departs from the Standard Model reference value, 
but also that they disagree among themselves if one wants to avoid 
a fine-tuning scenario. Due to the (inspiring) dissonance from the 
many models introducing New Physics phenomena and predicting specific 
effects on the observables, it is difficult to investigate New Physics 
in a systematic way. We have therefore chosen to build a general 
parameterization of generic New Physics amplitudes that interact in $\BzBzb$ 
mixing and/or penguin \B decays. The global CKM fit 
allows us to derive constraints on these generic New Physics amplitudes. 
Specific New Physics models have then to be in accordance with the
allowed generic variable space. 
\vs
We summarize in the following the main developments and results 
described in this work.
\bei

\item   All results are obtained with the use of the software package 
        \ckmfitter\  that employs statistical analysis tools based on 
        the frequentist approach \rfit. 
        We have extended the analysis 
        to take into account one- and two-dimensional physical 
        boundary conditions as they occur in \CP-asymmetry measurements.

\item    Among the main results of this paper are the numerical 
        (Tables~\ref{tab:fitResults1} and \ref{tab:fitResults2}) and 
        graphical (Figs.~\ref{fig:chi2min}, \ref{fig:rhoeta}, 
        \ref{fig:rhoeta_small} and \ref{fig:ckm1d}) representations 
        of the global CKM fit. The values of the Wolfenstein parameters 
        $\lambda$, $A$, $\rhobar$, $\etabar$ are found to be in agreement with the 
        results from our previous analysis (2001), and 
        their $1\sigma$ errors have changed by relative $+14\%
	$, $-58\%
	$, 
        $-45\%
	$ and $-64\%
	$, respectively, mainly due to the experimental
        improvements on $\stb$ and $\Vcb$. We find for the apex of
        the Unitarity Triangle, the coordinate 
        $\rhobar=0.189^{\,+0.088}_{\,-0.070}$ and 
        $\etabar=0.358^{\,+0.046}_{\,-0.042}$. For the goodness of the 
        global CKM fit within the Standard Model, we find a p-value of $71\%
	$.

\item   We have analyzed observables from rare kaon decays related to 
        $\rhobar$, $\etabar$ and derived constraints on the hadronic parameters 
        $B_6$, $B_8$, related to $\epe$. We discuss the present and future 
        constraints in the unitarity plane from the rare decays 
        $\Kp\to\pip\nu\nub$ and $\KL\to\piz\nu\nub$.

\item   The constraints on $2\beta+\gamma$ and $\gamma$ respectively from
        the \CP analyses of $\Bz\to D^{(*)\pm}\pi^\mp$ and $\Bp\to\Dz\Kp$ 
        decays are displayed. The present experimental errors 
        are still too large to be competitive with the other measurements 
        used in the global CKM fit.

\item   Results on charmless \B decays to $h h^\prime$ ($h,h^\prime=\pi,K$) 
        are studied in four different scenarios based on SU(2) and SU(3) 
        flavor symmetries as well as QCD Factorization. Useful constraints 
        on $\alpha$ are only obtained with significant theoretical input. A 
        global fit of QCD Factorization to all available $\pi\pi, K\pi$ 
        observables leads to an acceptable overall description 
        (p-value of $21\%
	$), however with large non-factorizable corrections, 
        and is remarkably predictive (\cf\  Fig.~\ref{fig:pipikpi_qcdfa_cpbr}). 
        Two predictions show deviations from the measurements: the
        branching fractions for $\Bz\to\Kp\pim$ and $\Bz\to\Kz\piz$, which 
        come out somewhat large and small, respectively. The discrepancy 
        does however not exceed 2.5 standard deviations in the worse case.
        The constraint on $\rhobar$, $\etabar$ obtained from this fit is in 
        agreement with the global CKM fit and competitive in precision. 
        We do not observe significant hints for deviations from the Standard 
        Model in these decays. Using SU(3) symmetry, we predict the 
        branching fraction and \CP-violating asymmetries in $\Bs\to\Kp\Km$ 
        decays. 

\item   A specific section has been dedicated to the study of $\B\to K\pi$ decays, 
        where we analyze the impact of electroweak penguins in discussing the
        recent literature. Using a phenomenological parameterization and various
        dynamical hypotheses, we find that the current data do not
        significantly constrain electroweak parameters, neither hadronic
        amplitude ratios. In particular, the determination of the parameters
        related to electroweak penguins is not possible at present. We do not
        observe an unambiguous sign of New Physics, whereas central values
        of the parameters show  evidence
        for potentially large non-perturbative rescattering effects. It is not
        clear to us whether a unified theoretical approach would be able to
        explain the whole set of observables in the $\pi\pi$ and $K\pi$
	systems, if future measurements confirm
        the present pattern.

\item   Results on charmless \B decays to $\rho\pi$ and their SU(3) partners
        are studied with the use of SU(2) and SU(3) flavor symmetry. The 
        discrete ambiguities due to the unknown relative strong phase between 
        the $\Bz\to\rho^+\pim$ and $\Bz\to\rho^-\pip$ amplitudes obstruct
        useful constraints on $\alpha$ in these modes. A Dalitz plot analysis 
        is required to measure this phase. Within SU(3) symmetry and
	neglecting certain suppressed topologies, bounds on 
        direct \CP-violating asymmetries are derived. It is found that the 
        present amount of direct \CP violation measured by the parameter 
        $\Acpmp$ tends to violate this bound, suggesting 
        that the size of the effect is a statistical fluctuation that is 
        expected to reduce with the availability of more data.

\item   The isospin analysis of 
        the $\B\to\rho\rho$ system provides a useful constraint on $\alpha$,
        which is found to be in agreement with the expectation from the 
        global CKM fit. Including isospin-breaking corrections from 
        electroweak penguins, and choosing the solution that is preferred
	by the CKM fit, we derive from the \babar\  measurement of 
        time-dependent \CP asymmetries in $\Bz\to\rho^+\rho^-$ decays
        $\alpha=(94\pm12\:[^{\,+28}_{\,-25}]\pm13\:[19])^\circ$. Here the 
        first errors given are experimental, the second due to the penguin
        uncertainty, and the errors in brackets are at $2\sigma$.
        If color suppression holds and if penguins
        are small, we expect that the present branching fraction measured 
        for $\Bp\to\rho^+\rho^0$ should reduce with more data in order to 
        close the isospin triangles. The potential to measure mixing-induced
        \CP violation in $\Bz\to\rho^0\rho^0$ promises a brighter future,
        to reduce the uncertainty $\alpha-\alpha_{\rm eff}$ due to the unknown
        penguin pollution in $\rho^+\rho^-$, than it can be expected
        for $\pi\pi$. We have studied a simple extension of the isospin
        analysis to account for possible isospin-breaking effects. We find
        that the systematic uncertainties on $\alpha$ for the full isospin
        analysis can be of the order of $3^\circ$, depending on the 
        amplitude structure of the decays.

\item   We refer to 
        Section~\ref{sec:charmlessBDecays}.\ref{sec:charmlessConclusions} 
        for a more detailed summary of the results on all charmless \B decays 
        studied in this paper.

\item   We have studied the present data constraints on the amplitude parameters 
        in the presence of arbitrary New Physics contributions to $\Bz\Bzb$ 
        mixing and to $b\to d$ and $b\to s$ penguin transitions. The construction 
        of a model-independent Unitarity Triangle is not precise enough to exclude 
        sizable non-standard corrections to the mixing, which appears to be 
	somehow in
        contrast to a prejudice  found in the literature. The situation 
        will improve in the future as soon as accurate determinations of the 
        angles $\alpha$ and $\gamma$ from tree dominated decays become
	available, and the theoretical errors on the lattice matrix
	elements relevant for the mixing are reduced. For $\Delta B=1$ transitions we have shown that a
	general parameterization may give significant model-independent
	constraints on
	potential New Physics contributions, when the \CP-violation 
	measurements become more precise.

\eei
The outstanding role of \B physics in the quest for a better understanding of
\CP violation in the Standard Model and beyond, as well as for the precise 
metrology of the off-diagonal CKM matrix elements, is assured by the continuous
rise of the peak luminosity at the \B factories. New and competitive results
from the Tevatron experiments are expected soon, in particular the highly 
important measurement of $\Bs\Bsb$ oscillation. We have attempted to extrapolate 
the results leading to the Unitarity Triangle angle $\alpha$ up to luminosities
of $1\invab$ (and $10\invab$ in some cases). It seems reasonable to expect
that a determination of $\alpha$, dominated by $\B\to\rho\rho$,
to an error of about $6^\circ$ or better (not including isospin-breaking effects)
can be achieved towards the end of the 
first generation \B-factory program, with an expected integrated luminosity of
combined roughly $2\invab$. Plausible extrapolations for the angle $\gamma$ 
are more difficult since all measurements in the beauty-to-charm sector
crucially depend on the ratios of the corresponding 
CKM-suppressed-to-CKM-favored amplitudes, which are only approximately known 
at present. 
\vs
In summary, we can hope for a  precise metrology of the Unitarity 
Triangle angles within a few years, but---in view of the present results---we 
do not expect it to be sufficiently accurate to reveal inconsistencies with 
the CKM picture. Therefore, besides $\Delta m_s$,
major attention is directed to the forthcoming 
measurements of \CP-violation parameters in penguin-dominated modes.
The near future will show whether the current pattern turns into a significant 
deviation from the expectation, or if the discrepancies fall behind the 
theoretical uncertainties that are expected in these modes.

 \section*{\large Acknowledgements}

We are indebted to 
Martin Beneke,
Achim Denig,
Urs Langenegger,
Matthias Neubert,
Yoshi Sakai
and
Klaus Schubert
for informative discussions. We thank Helen  Quinn and Zoltan Ligeti
for an animated exchange on the SU(3) relations.
The study on rare kaon decays has benefited from valuable correspondence
with Shaomin Chen and David Jaffe. Santi Peris pointed out two
interesting references.
Tom Trippe kindly produced updated input values for the kaon mixing parameters.
Fruitful conversations with David London and Tobias Hurth
on the isospin study of $B\to K\pi$ 
decays are gratefully acknowledged. 
We thank Phil Clark, Malcolm John, David Kirkby and Robert McPherson for 
the constructive and knowledgeable help in our quest for the right title.
JC acknowledges partial support from
EC-Contract HPRN-CT-2002-00311 (EURIDICE).


 \clearpage
 \appendix
 {\small
 \newpage\part{Appendix}\setcounter{section}{0}
\markboth{\textsc{Appendix}}
         {\textsc{Appendix}}

 %
%
\newcommand\Ps{{{P_s}}}
\def\P{{{P}}}
\newcommand\Pdot{\dot{\P}}
\newcommand\Pdotdot{{\ddot{\P}}}
\newcommand\xs{x_s}
\newcommand\xsmes{{x_{s}^{\rm mes}}}
\newcommand\xshyp{{x_{s}^{\rm hyp}}}
\newcommand\frs{f_{ s}}
\newcommand\Gt{{{G}_t}}
\newcommand\tmes{t_{\rm mes}}
\newcommand\dtmes{{ d}t_{\rm mes}}
\newcommand\Liklo{{\cal L}_{\rm lo}}
\newcommand\Liknlo{{\cal L}_{\rm nlo}}
\newcommand\Liknnlo{{\cal L}_{\rm nnlo}}
\newcommand\aLik{{\overline{\cal L}}}
\newcommand\aLiklo{{\overline{\cal L}_{\rm lo}}}
\newcommand\aLiknlo{{\overline{\cal L}_{\rm nlo}}}
\newcommand\aLiknnlo{{\overline{\cal L}_{\rm nnlo}}}
\newcommand\K{{\cal K}}
\newcommand\EK{{\cal E}}
\newcommand\ao{ a_0}
\newcommand\at{ a_3}
\newcommand\nlo{{\rm nlo}}
\newcommand\nnlo{{\rm nnlo}}
\newcommand\Phix{\Phi_{0}}
\newcommand\Philo{\Phi_{\rm lo}}
\newcommand\Phinlo{\Phi_{\nlo}}
\newcommand\phix{\varphi_{\aLik}}
\newcommand\philo{\varphi_{\aLik:lo}}
\newcommand\phinlo{\varphi_{\aLik:\nlo}}
\newcommand\phinnlo{\varphi_{\aLik:\nnlo}}
\newcommand\philik{\varphi_{\Lik}}
\newcommand\dPhi{{\delta\Phi}}
\newcommand\Proba{{\cal P}}
\newcommand\cst{{\rm const}}
\newcommand\sAmp{{\sigma[\Ampli]}}
\newcommand\sAmptwo{{\sigma^2[\Ampli]}}
\newcommand\Ns{n_{ s}}
\newcommand\f{{ f}}
\newcommand\z{{ z}}
\newcommand\azero{\alpha_0}
\newcommand\aone{\alpha_1}
\newcommand\atwo{\alpha_2}
\newcommand\athree{\alpha_3}
\newcommand\afour{\overline{\alpha_4}}
\newcommand\aaone{\overline{\alpha_1}}
\newcommand\aatwo{\overline{\alpha_2}}
\newcommand\aathree{\overline{\alpha_3}}
\newcommand\aafour{\overline{\alpha_4}}

\section{Statistical Significance of $\Bs\Bsb$ oscillation}

The purpose of this appendix is to evaluate the statistical significance 
of the world average results on $\Bs\Bsb$ oscillation. At present,
the world average likelihood as a function of $\dms$ exhibited 
a roughly parabolic behavior at $\dms\simeq 17\ps^{-1}$.
Following an analytical approach, we address two questions:
\bei

\item   what is the \pdf\  of a likelihood measurement of $\dms$
        and what is the confidence level (CL) as a function of 
        $\dms$ to be associated with an observation 
        obtained with the current level of sensitivity;

\item   what is the expected likelihood behavior and 
        how reliable it is to use the likelihood to infer CLs.

\eei

\subsection{Definitions and Proper Decay Time Modeling}

Using the simplified framework of Ref.~\cite{Roussarie}, we denote
for a homogeneous event sample:

\bei

\item   $\Ps_\pm$ the (true) time distribution (in unit of the $B_s$ 
        lifetime $\tau_B$) of mixed ($\Ps_-$) and unmixed ($\Ps_+$) events,
        given by
        \beq
                \Ps_\pm = \frac{1}{2}e^{-t} (1\pm\cos(\xs t))~,
        \eeq
        with $\xs=\dms\tau_b$, 

\item   $\w$ the mistag rate, and $D=1-2\w$ the corresponding 
        dilution factor;

\item   $\frs$ the fraction of signal events in the sample;

\item   The background is assumed to:

        \bei

        \item   follow the same exponential distribution as the signal, 

        \item   be purely of the unmixed type,

        \item   be affected by the same mistag rate;

        \eei

\item   $\Gt$ the detector resolution function for the time measurement
        $t\to\tmes$;

        It is assumed to be a Gaussian of zero mean and time dependent width
        \beqn
                \sigma          &=& \sqrt{a + b t^2} \\
                \Gt(\tmes-t)    &=& \frac{1}{\sqrt{2\pi}\sigma}
                                    \exp{
                                        \left( -\frac{1}{2}
                                                \left(\frac{\tmes-t}{\sigma}
                                                \right)^{\!\!2}
                                        \right)
                                        }~,
        \eeqn
        with $a$ accounting for the decay length measurement and $b$ 
        accounting for the momentum measurement  
        \beq
                a  = \left(\frac{m}{p}\frac{\sigma_L}{c\tau_B}
                        \right)^{\!\!2}\!\!~, \hspace{1cm}
                b  = \left(\frac{\sigma_p}{p}
                        \right)^{\!\!2}\!\!~.
        \eeq

\eei
With these notations the proper time distribution of events,
classified as mixed or unmixed, read
\beqn
        \P_-(\tmes)     &=&
        \left(\frs\frac{1}{2}\left(1-D\cos(\xs t)\right)
                           +(1-\frs)\w\right)e^{-t}\otimes\Gt~, \\
        \P_+(\tmes)     &=&
        \left(\frs\frac{1}{2}\left((1+D\cos(\xs t)\right)
                           +(1-\frs)(1-\w)\right)e^{-t}\otimes\Gt~.
\eeqn
Taken together, these distributions are normalized to unity
\beq
\label{Normalization}
        \int\limits_{-\infty}^{+\infty} (\P_-+\P_+) \dtmes = 1~.
\eeq

\subsection{Measurement}

The $\xs$ measurement is assumed to be performed with the use of 
the log-likelihood 
\beq
\label{loglikelihooddefinition}
        \Lik(\xs)=\sum_-\ln(\P_-)+\sum_+\ln(\P_+)~,
\eeq
where the first (second) sum runs over mixed (unmixed) events.
The measured value of $\xs$ ($\xsmes$) is defined to be the one
maximizing $\Lik(\xs)$
\beq
        \left.\frac{\partial\Lik(\xs)}{\partial \xs}
        \right|_{\xs=\xsmes} = 0~.
\eeq
The outcome of the experiment $\xsmes$ is a random number, which,
for large enough statistics, follows a Gaussian \pdf\ 
\beq
\label{KeyFormula}
        \Proba(\xsmes\mid\xs)\equiv\Philo^{\xs}(\xsmes)
                = {1\over\sqrt{2\pi}\Sigma(\xs)}
                  \exp{ \left(-\frac{1}{2}
                          \left({\xsmes-\xs\over\Sigma(\xs)}
                          \right)^{\!\!2}
                        \right)
                      }~,
\eeq
where the standard deviation $\Sigma(\xs)$ is given by 
the second derivative of $\Lik$, through the integral $A$
\beqn
\label{Aintegral}
        (\sqrt{N}\Sigma(\xs))^{-2}      &=&
                \int\limits_{-\infty}^{+\infty} 
                \left({(\Pdot_-)^2\over\P_-}+{(\Pdot_+)^2\over\P_+}
                \right)\dtmes\equiv A(\xs)~, \\
        \Pdot_\pm                       &=&
                {\partial\P_\pm\over\partial\xs} \\
                                        &=&
                \mp\frs\frac{1}{2}D\ t\sin(\xs t)e^{-t}\otimes\Gt~.
\eeqn
Here $N$ is the total number of mixed and unmixed events, and the 
integrals are performed with the use of the {\em true} value of $\xs$,
not the measured one\footnote
{
        If the event sample is not homogeneous but is an admixture 
        of $\Ns$ homogeneous subsamples, each with a detector 
        resolution function $\Gt^i$, a signal fraction $\frs^i$, 
        a mistag rate $\w^i$, and representing a fraction $\f_i$ 
        of the overall sample of $N$ events, the corresponding 
        time distribution are denoted $\P_\pm^i$ (the factor $\f_i$ 
        being not included). The $A$ integral (as well as other 
        integrals introduced below) is then to be replaced by the 
        weighted sums
        \beq
                A(\xs) = \sum_{i=1}^{\Ns}\f_iA^i(\xs)~.
        \eeq
}.
\vs
It follows from Eq.~(\ref{KeyFormula}) that one may set a confidence 
level $\CL(\xshyp)$ on a given $\xs$ hypothetical value $\xshyp$ using 
the $\chi^2$ law
\beqn
\label{CLGauss}
        \CL(\xshyp)     &=&
                \int\limits_<\Philo^{\xs}(\xsmes^\prime)d \xsmes^\prime
                =\ProbCERN(\chi^2,1)~, \\
\label{chi2true}
        \chi^\xs(\xsmes)&=&
                {\xsmes-\xshyp\over\Sigma(\xshyp)}~,
\eeqn
where the integral is performed over the $\xsmes^\prime$ domain where
$\Philo^{\xs}(\xsmes^\prime)<\Philo^{\xs}(\xsmes)$,
that is to say where $\chi^\xs(\xsmes^\prime)>\chi^\xs(\xsmes)$.

\subsubsection{Parabolic Behavior}
\label{ParabolicBehaviorSection}

If the log-likelihood is parabolic nearby its maximum
\beq
        \Lik(\xshyp)\simeq
        \Lik(\xsmes)+\frac{1}{2}
        \left.\frac{\partial^2\Lik}{\partial\xs^2}
        \right|_{\xs=\xsmes}\left(\xshyp-\xsmes\right)^2~,
\eeq 
then, in the vicinity of $\xsmes$, $\Sigma(\xshyp)\simeq cst=\Sigma(\xsmes)$,
and one can evaluate $\Sigma$ as the second derivative of the experimental 
log-likelihood, taken at the measured value $\xsmes$. In effect
\beqn
        -\left.\frac{\partial^2\Lik}{\partial\xs^2}\right|_{\xs=\xsmes}
        &=&
                -\left(\sum_-\left({\Pdotdot_-\P_- - (\Pdot_-)^2\over \P_-^2}
                             \right)^{\!\!2}
                      +\sum_+\left({\Pdotdot_+\P_+ - (\Pdot_+)^2\over \P_+^2}
                             \right)^{\!\!2}
                 \right) \\
        &&\stackrel{(N\to\infty)}{\longrightarrow}
                NA(\xs)=\Sigma^{-2}~,
\eeqn
where $\Pdotdot_\pm$ denotes the second derivative with respect to $\xs$
\beqn
        \Pdotdot_\pm 
        &=&
                {\partial^2\P_\pm\over\partial\xs^2}\\
        &=&
                \mp\frs\frac{1}{2}D\ t^2\cos(\xs t)e^{-t}\otimes\Gt~,
\eeqn
which however does not appear in the final expression thanks to 
Eq.~(\ref{Normalization}), and assuming that \hbox{$\xsmes=\xs$} 
(which is true for $N\to\infty$).
\vs
Equivalently, one can evaluate $\Sigma$ by locating the value of 
$\xshyp$ which yields a drop of $-1/2$ of the log-likelihood,
for the experiment at hand, or one can compute directly the $\chi^2$ 
using the approximation
\beq 
\label{chi2Lik}
        \chi^2(\xshyp) 
        = \left({\xsmes-\xshyp\over\Sigma(\xshyp)}\right)^{\!\!2}
          \simeq \:2(\Lik(\xsmes)-\Lik(\xshyp))\equiv\tilde\chi^2(\xshyp)~.
\eeq
\noindent
{\footnotesize
{\small\bf Bayesian point of view} 
\vs
Because of the simplicity of the above relations, one may introduce 
the concept of the {\em \pdf\  of the true value of $\xs$} by remarking 
that, if $\xsmes$ is viewed as a non-random number (the actual outcome 
of a finalized single experiment) while the true value of $\xs$ is taken 
to be a random number, the object
\beq
\label{TheBayesianpdf}
        \Proba(\xs\mid\xsmes)\equiv\Philo^{\xsmes}(\xs)~,
\eeq
allows to define
\beqn
\label{CLGaussBayesian}
        \CL(\xshyp)     &=&     \ProbCERN(\chi^2,1)~, \\
\label{chi2hyp}
        \chi            &=&     \frac{\xshyp-\xsmes}{\Sigma(\xsmes)}~,
\eeqn
which is numerically identical to the one of Eq.~(\ref{CLGauss})
---
if $\Sigma(\xsmes)=\Sigma(\xshyp)$
---
but with a completely different reading:
one states that the CL of $\xshyp$ is given by
\beq
        \CL(\xshyp)
        = \int\limits_<\Proba(\xsmes\mid\xshyp^\prime)d \xshyp^\prime~,
\eeq
where the integral is performed over the $\xshyp^\prime$ domain where
$\Proba(\xsmes\mid\xshyp^\prime)<\Proba(\xsmes\mid\xshyp)$\,.
}

\subsubsection{Non-Parabolic Behavior}

Obviously, for large enough $\xshyp$, the approximation 
$\Sigma(\xshyp)\simeq \Sigma(\xsmes)$ breaks down since the sensitivity 
of the experiment vanishes due to the finite vertex resolution, \ie,
 $\Sigma(\xshyp\to\infty)\to\infty$.
It follows that the likelihood is not parabolic for large enough $\xshyp$.  
The vanishing sensitivity makes $\chi^2$, as defined by Eq.~(\ref{chi2true}),
a poor test statistics to probe for large $\xs$ values. Furthermore,
as discussed in Section~\ref{NLO} to infer from the $\chi^2$ value the 
correct $\CL(\xshyp)$ is not a straightforward task: Eq.~(\ref{CLGauss}) 
does not apply (\ie, it is not a real $\chi^2$) because 
Eq.~(\ref{KeyFormula}) is a poor approximation.
\vs
The redefinition of the $\chi^2$ using the right hand side of Eq.~(\ref{chi2Lik})
provides a more appropriate test statistics to deal with large values 
of $\xshyp$. Whereas Eq.~(\ref{CLGauss}) does not apply, $\tilde\chi^2$ 
is capable of ruling out $\xshyp$ values lying beyond the sensitivity 
reach\footnote
{
        The rejection of $\xshyp$ values beyond the sensitivity reach 
        is not a paradox: it uses the fact that large values are unlikely 
        to yield an indication of a clear signal, especially at low values 
        of $\xs$.
}
(if $\Lik(\xsmes)$ value is large enough) provided one computes the CL 
using
\beq
        \CL(\xshyp) \:= \!\!
                \int\limits_{\tilde\chi^2(\xshyp)}^\infty
                \!\!\!\Psi^\xshyp
                (\tilde\chi^{2\prime})\ D\tilde\chi^{2\prime}~,
\eeq
where $\Psi^\xshyp$ is the \pdf\  of the $\tilde\chi^2$ test 
statistics, for $\xs=\xshyp$, to be obtained with the use of toy Monte Carlo.
A tempting shortcut is to bypass the toy Monte Carlo simulation and 
to assume that the approximation 
\beq
\label{Blunt}
        \CL(\xshyp)\approx\ProbCERN(\tilde\chi^2(\xshyp),1)~,
\eeq
remains valid although the approximation of Eq.~(\ref{chi2Lik}) is known 
to break-down.
\vs

\noindent
{\footnotesize
{\small\bf Bayesian point of view} 
\vs
The $\xshyp$ '\pdf' introduced in Eq.~(\ref{TheBayesianpdf}) can be 
redefined as
\beq
        \Proba(\xshyp\mid\xsmes)\equiv\philik^{\xsmes}(\xshyp)
        =
        \cst\times\exp\left(\Lik(\xshyp)-\Lik(\xsmes)\right)~,
\eeq
where the constant should be such that $\philik$ is normalized 
to unity when integrated over $\xshyp$. In the present case, 
such a constant does not exist because 
\beq
        \lim\limits_{\xshyp\to\infty}\!\!\!\!\Lik(\xshyp)
        \:=\:\hbox{finite constant}~,
\eeq
and hence $\philik$ itself tends asymptotically towards a constant.
We will consider below the average value of $\philik$ as computed 
using the average likelihood which one would obtain. Ignoring 
statistical fluctuations the average function is denoted 
\beq
        \phix^{\xs}(\xshyp)
        =
        N\int(\P_-^{\xs}\ln(\P_-^{\xshyp})+\P_+^{\xs}
        \ln(\P_+^{\xshyp}))\dtmes~,
\eeq
and its leading order, next-to-leading order and next-to-next-to 
leading order approximations are denoted $\philo$, $\phinlo$ and 
$\phinnlo$. The numerical value of the ratio 
\beq
        R_{\rm tail} \equiv \frac{\phix(\infty)}{\phix(\xs)}~,
\eeq
which vanishes exponentially with $N$, is a measure of how 
non-Gaussian the likelihood is.
}

\subsection{Experimental Constraint}

The question arises as to how to incorporate experimental constraints 
derived from the $\B_s$ mixing analysis into a global CKM fit. 
A possibility is to add to the (twice)log-likelihood of the global 
fit the term 
\beq 
\label{chi2Bayesian}
        \chi^2(\xshyp) = 2(\Lik(\infty)-\Lik(\xshyp))~,
\eeq
or equivalently to multiply the likelihood ${L}$ by the ratio
\beq 
\label{LikBayesian}
        \Delta L
        =
        \frac{\Proba(\xshyp\mid\xsmes)}{\Proba(\infty\mid\xsmes)}~,
\eeq
where the constant denominator is introduced here for convenience only.
Since the $\Lik$ function is defined up to an irrelevant additive 
constant, using Eq.~(\ref{chi2Bayesian}) or Eq.~(\ref{chi2Lik}) 
amounts to making the same approximation, which is guaranteed to be 
correct, for large enough statistics, and in the vicinity of $\xsmes$.
\vs
The question remains to determine under which conditions on $N$ 
and $\xshyp$ the approximation is 
\begin{enumerate}

\item   {\bf obviously valid:} that is to say to determine the domain of 
        validity of the leading order ($N\to\infty$)
        key-formula Eq.~(\ref{KeyFormula}). To answer this question,
        one should compute its next-to-leading order (NLO) correction 
        terms: Section~\ref{NLO} is devoted to that.

\item   {\bf non-obviously valid:} that is to say to determine whether or 
        not, even though the key-formula does not apply, 
        Eq.~(\ref{chi2Bayesian}) provides nevertheless a means to 
        compute the CL with an acceptable accuracy: Section~\ref{LIKsection} 
        discusses that.

\end{enumerate}

\subsection{Next-to-Leading Order Key-Formula}
\label{NLO}

The next-to-leading order key-formula can be written as\footnote
{
        In the course of the computation, the $\at$ correction term 
        appears in the exponential, as indicated. However
        the formula is correct up to the next-to-leading order only,
        and the $\at$ term can be brought down to the level of the 
        $\ao$ term without affecting this. Although it would guaranty 
        the proper normalization of $\Phinlo$ to unity, this 
        simplification is not done below. 
}
\beqn
\label{KeyNLOFormula}
        \Phinlo^{\xs}(\xsmes)   &=&
                \Philo^{\xs}(\xsmes)\ e^{-\at^{\xs}\chi^3}
                (1+\ao^{\xs}\chi)~, \\
\label{derivativeofSigma}
        \ao^{\xs}               &=&
                {2B-C\over 2A}{1\over\sqrt{N A}}
                =-\dot\Sigma~, \\
        \at^{\xs}               &=&
                {3B-C\over 6A}{1\over\sqrt{N A}}~,
\eeqn
where $A(\xs)$ is the integral defined in Eq.~(\ref{Aintegral}), 
and $B(\xs)$ and $C(\xs)$ are the two new integrals
\beqn
        B(\xs)  &=&
                \int\limits_{-\infty}^{+\infty} 
                \left({\Pdot_-\Pdotdot_-\over\P_-}
                     +{\Pdot_+\Pdotdot_+\over\P_+}
                \right)\dtmes~, \\
        C(\xs)  &=&
                \int\limits_{-\infty}^{+\infty} \left({(\Pdot_-)^3\over\P_-^2}
                + {(\Pdot_+)^3\over\P_+^2}\right)\dtmes~.
\eeqn
The integral $C$ tends to be small because, $(i)$ the two 
contributions have opposite signs, and $(ii)$ the denominator 
is of order two: it follows that $\at\simeq\ao/2$. The right hand side 
of Eq.~(\ref{derivativeofSigma}) links the next-to-leading order 
correction terms $\ao$ and $\at$ to the dependence on $\xs$ of $\Sigma$.
When $\Sigma$ depends significantly on $\xs$ the key-formula breaks down:
not only is the standard treatment of Section~\ref{ParabolicBehaviorSection} 
invalid (and the Bayesian treatment mathematically unjustified), but the 
well-known formula Eq.~(\ref{CLGauss}) itself becomes incorrect, even if 
one uses the correct $\Sigma(\xs)$.
\vs
The expression Eq.~(\ref{KeyNLOFormula}) is identical to Eq.~(\ref{KeyFormula}) 
for small $\chi$ values. Although it extends the range of validity to 
larger $\chi$ values, it cannot be trusted too far away from the origin,
where higher order corrections start to play a role. In particular, 
$\Phinlo$ becomes negative (hence meaningless) for $\chi>-\ao^{-1}$ 
($\ao$ is negative since it is equal to minus the derivative of $\Sigma$
with respect to $\xs$).
\vs
Since $\Phi$ is sizeable only insofar $\chi\sim {\cal O}(1)$ the 
next-to-leading order terms, when relevant, are of the form
$N^{-{1\over2}}\times [{\rm ratio\ of\ integrals}]$. Hence they are 
negligible for large enough $N$ and for a small enough ratio of integrals.
\vs
The most likely value for $\xsmes$ is no longer $\xs$, and a non-zero 
value of $B$ leads to a ${\cal O}(1/N)$ bias in the measurement. 
The expected value of $\xsmes$ reads
\beq
        \langle\xsmes\rangle
        =
        \xs -\left({B(\xs)\over A}{\Sigma\over 2}\right)\Sigma~.
\eeq
The bias is negligible (in unit of $\Sigma$) if the event sample is 
large enough, \ie\  if $N\gg B^2/(4A^3)$. To next-to-leading order,
the double-sided CL reads
\beq
\label{CLnlo}
        \CL_\nlo(\xshyp)
        =
        \int\limits_< \Phinlo^{\xshyp}(\xsmes^\prime) d\xsmes^\prime~,
\eeq
where the integral is performed over the $\xsmes^\prime$ domain where
$\Phinlo^{\xshyp}(\xsmes^\prime)<\Phinlo^{\xshyp}(\xsmes)$.

\subsection{Using the Likelihood Function}
\label{LIKsection}

\subsubsection{Average Likelihood Shape}

To next-to-leading order, and in the vicinity of the true $\xs$ value, the
average log-likelihood function takes the form (\cf\  Section~\ref{NLOLik})
\beq
\label{NLOAverageLik}
        \aLiknlo(\xshyp)        \simeq
                \aLik(\xs)-\aaone\chi +\aatwo\chi^2-\aathree\chi^3~,
\eeq
with
\beq
        \aaone   = 0~,                  \hspace{1cm} 
        \aatwo   = -\frac{1}{2}~,       \hspace{1cm} 
        \aathree = -{3B-2C\over 6A}{1\over\sqrt{NA}}\simeq-\at
                \simeq\frac{1}{2}\dot\Sigma~, 
\eeq
and
\beq
        \chi \equiv{\xs-\xshyp\over\Sigma(\xs)}~.
\eeq
Although the above expression reaches its maximum at $\chi=0$,
this does not contradict the fact that $\xsmes$ is a biased 
estimator of $\xs$: because of the statistical fluctuations, the 
first term of the expansion is non-zero for a given experiment
(\cf\   Section~\ref{NLOLik}).

\subsubsection{Amplitude Formalism}
\label{sec:amplitudeMethod}

It was shown in Ref.~\cite{Roussarie} that the log-likelihood function
$\Lik(\xs)$ can be retrieved from the functions $\Ampli(\xs)$ and 
$\sAmp(\xs)$ defined as the measurement and the uncertainty on the 
measurement of an {\it ad hoc} amplitude coefficient $\Ampli$ placed 
in front of the cosine modulation term
\beqn
\label{AizedFormulae}
        \P_-(\tmes)[\Ampli]     &=&
                \left(\frs\frac{1}{2}\left(1-D\Ampli\cos(\xs t)\right)
                        +(1-\frs)\w
                \right)e^{-t}\otimes\Gt~, \\
        \P_+(\tmes)[\Ampli]     &=&
                \left(\frs\frac{1}{2}\left(1+D\Ampli\cos(\xs t)\right)
                        +(1-\frs)(1-\w)
                \right)e^{-t}\otimes\Gt~.
\eeqn
Restated in the framework of the present work, the advantage of this 
indirect probe of the oscillation phenomenon stems from the fact that 
the dependence on $\Ampli$ is linear and hence the correction terms of 
the NLO key-formula vanish: the measurement of $\Ampli$ is purely Gaussian,
and it follows that merging different experimental measurements is 
straightforward.
\vs
The result established in Ref.~\cite{Roussarie} takes the form
\beq
\label{RoussarieFormula}
        \Lik^\xs(\xshyp) 
        =
        {\Ampli(\xshyp)-\frac{1}{2}\over\sAmptwo}+\Lik^\xs(\infty)~.
\eeq
It can be shown to be an excellent, though approximate, relationship 
by introducing the objects
\beqn
        \EK_-           &=&
                \left(\frs\frac{1}{2}+(1-\frs)\w\right)e^{-t}\otimes\Gt~, \\
        \EK_+           &=&
                \left(\frs\frac{1}{2}+(1-\frs)(1-\w)\right)e^{-t}\otimes\Gt~,\\
        \K_-(\xs)       &=&
                -{1\over\EK_-}
                \left(  \frs\frac{1}{2}
                        D\cos(\xs t)e^{-t}\otimes\Gt
                \right)~,\\
        \K_+(\xs)       &=&
                \phantom{-}{1\over\EK_+}
                \left(  \frs\frac{1}{2}D\cos(\xs t)e^{-t}\otimes\Gt
                \right)~,
\eeqn
Eq.~(\ref{AizedFormulae}) takes the form
\beq
        \P_\pm(\xshyp) = \EK_\pm (1+\Ampli\K_\pm (\xshyp))~.
\eeq
The two $\K_\pm (\xshyp)$ objects bear the properties:
\bei

\item   $\mid\K_\pm\mid\le 1$, in principle.

\item   $\mid\K_\pm\mid\ll 1$, in practice. 
        This is because the $D$ coefficient is usually smaller than one, 
        hence higher powers of $\K$ are suppressed, but also because when 
        considering large enough $\xs$ values the convolution with the
        finite detector response $\Gt$ washes out the cosine modulation 
        \beq
                \lim\limits_{\xshyp\to\infty}\!\!\!\!\K_\pm(\xshyp)
                =0~.
        \eeq

\eei
Hence the log-likelihood of Eq.~(\ref{loglikelihooddefinition})
(with or without $\Ampli$) can be expanded to the second order in $\K_\pm$
(omitting here the $\pm$ index distinguishing mixed and unmixed events)
\beqn
        \Lik(\xshyp:\Ampli)     &=& 
                \sum\ln(\P(\xshyp)) \\
                                &=& 
                \sum\ln(\EK)+\sum\ln(1+\Ampli\K(\xshyp))\\
                                &\simeq&
                \Lik^\xs(\infty)+ \sum(\Ampli\K(\xshyp)
                -\frac{1}{2}\Ampli^2\K^2(\xshyp) )~.
\eeqn
The log-likelihood we are interested in is
\beq
\label{LikKinExpansion}
        \Lik(\xshyp)=\Lik(\xshyp:\Ampli\equiv 1) \simeq
                \Lik^\xs(\infty)+\sum(\K(\xshyp)-\frac{1}{2}\K^2(\xshyp) )~.
\eeq
The derivative of the log-likelihood used to compute $\Ampli(\xshyp)$ is
\beq
        {\partial\Lik(\xshyp:\Ampli)\over\partial\Ampli}
        =
        \sum(\K(\xshyp)-\Ampli\K^2(\xshyp))~,
\eeq
from which one obtains
\beq
\label{AmpExpression}
        \Ampli(\xshyp) 
        = 
        {\sum\K(\xshyp)\over\sum\K^2(\xshyp)}
        \pm {1\over\sqrt{\sum\K^2(\xshyp)}}~,
\eeq
where the expression for the uncertainty neglects higher order
$\K^{n\ge 3}$ terms. Using Eq.~(\ref{AmpExpression}) in 
Eq.~(\ref{LikKinExpansion}) one recovers Eq.~(\ref{RoussarieFormula}).
Equation~(\ref{AmpExpression}) yields a slightly biased estimator of
$\Ampli$ because the higher order terms do not exactly cancel out,
even on the average\footnote
{
        They would cancel if the background were affecting in the 
        same way $\P_-$ and $\P_+$.
}. 
This bias is negligible for all values of $\xshyp$ but for 
$\xshyp\simeq\xs$ where, although it remains small, it becomes 
noticeable.

\subsection{Discussion on Numerical Examples}

For illustration, we use here numbers that correspond to the present 
level of sensitivity of the world average
\begin{table}[t]
{\footnotesize
\begin{center} 
\begin{tabular*}{\textwidth}{@{\extracolsep{\fill}}ccccc}      \hline
&&&& \\[-0.25cm]
$\dms$ $({\rm ps}^{-1})$ & 10               &  17             &   20             &25       \\[0.1cm] \hline
&&&& \\[-0.25cm]
$\Sigma$ $({\rm ps}^{-1})        $      &  0.33            &  1.38           &  2.67            & 7.36    \\[0.1cm] \hline
&&&& \\[-0.25cm]
$A                      $       &  $0.77\ 10^{-2}$ & $0.44\ 10^{-3}$ &  $0.11\ 10^{-3}$ &$0.15\ 10^{-4}$   \\ 
$B/A                   $        & -0.18            & -0.19           &  -0.21           &-0.17    \\ 
$C/A                   $        &  0.03            & -0.002          &  -0.007          & 0.005   \\ 
$D/A                   $        &  1.40            &  0.53           &  0.65            &  1.31   \\ 
$E/A                   $        &  0.08            &  0.005          &  0.001           &  0.00   \\ 
$F/A                   $        &  0.05            &  0.007          &  0.002           &-0.001   \\ 
$G/A                   $        & -1.27            & -0.51           &  -0.30           &-0.38    \\[0.1cm] \hline
&&&& \\[-0.25cm]
$\ao                     $      & -0.06            & -0.26           &  -0.55           &-1.26    \\ 
$\at                     $      & -0.03            & -0.13           &  -0.28           & -0.62   \\[0.1cm] \hline
&&&& \\[-0.25cm]
$\aathree                 $     &  0.03            &  0.13           &   0.27           & 0.63    \\ 
$\aafour                  $     &  0.004           &  0.04           &  -0.22           &-5.48    \\[0.1cm] \hline
&&&& \\[-0.25cm]
$\dms[\rm max]  $               &  15.1            &  22.2           &  24.9            & 30.8    \\ 
$R_{\rm tail}$           & $3.\ 10^{-6}$      &  0.12            &  0.37    & 0.75   \\[0.1cm] \hline
\end{tabular*}
\caption{\em\small\label{tab:Values}
         Numerical values entering into the NLO \pdf\  $\Phinlo$ 
        and NNLO average likelihood, for four values of $\dms$.
        The definition of the four integrals $D$, $E$, $F$ and $G$ 
        are given in Section~\ref{NLOLik}. The value above which 
        $\Phinlo$ becomes negative is  
        $\dms[\rm max]=\dms-\ao^{-1}\Sigma\simeq \dms-A/B\simeq\dms+5(\ps^{-1})$. 
        The ratio $R_{\rm tail}=\phix(\infty)/\phix(\dms)$ provides 
        a measure of how far from its Gaussian limit the likelihood is.}
\end{center}
}
\end{table}
\begin{figure}[p]  
  \centerline{\epsfxsize8cm\epsffile{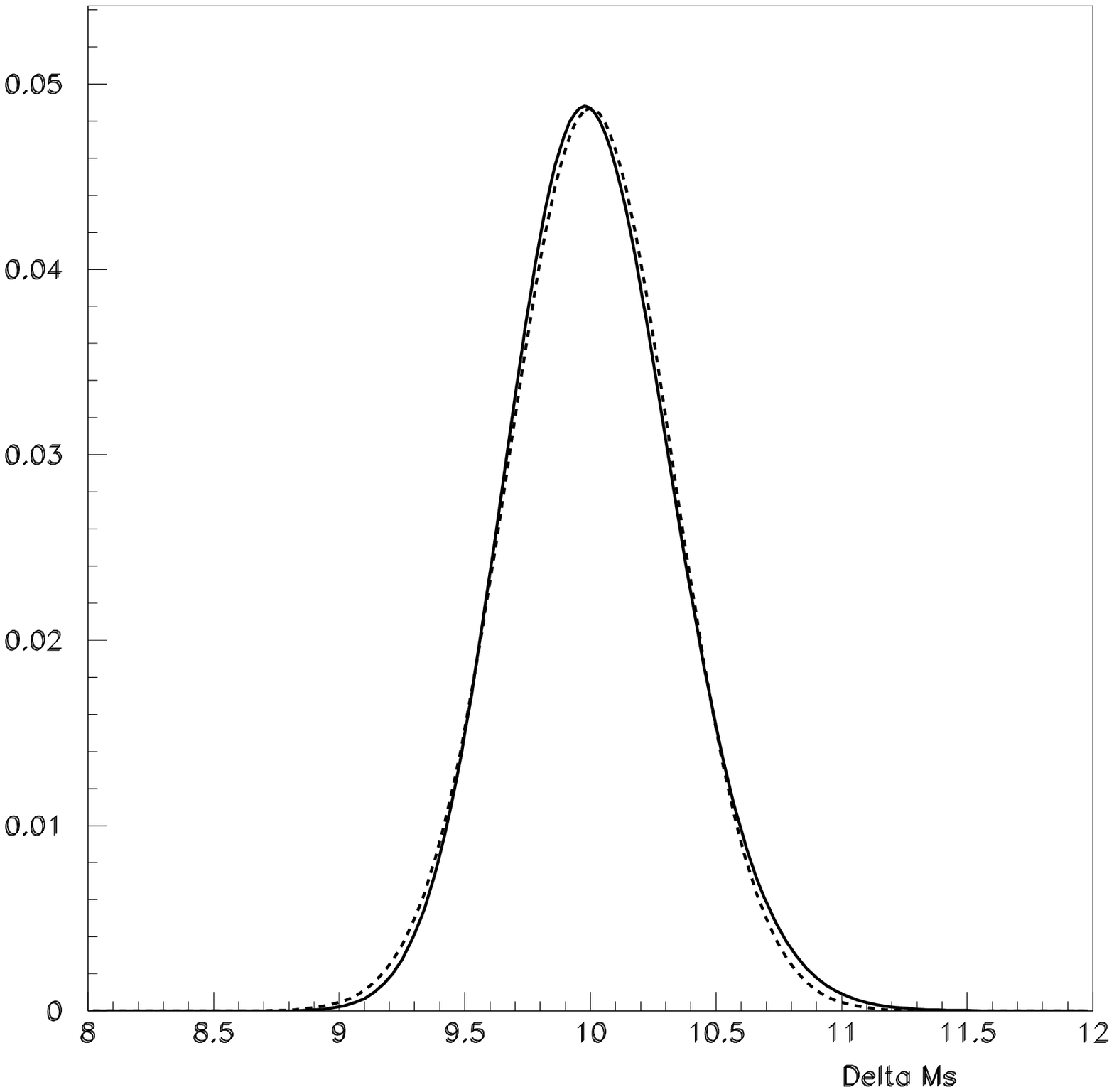}
              \epsfxsize8cm\epsffile{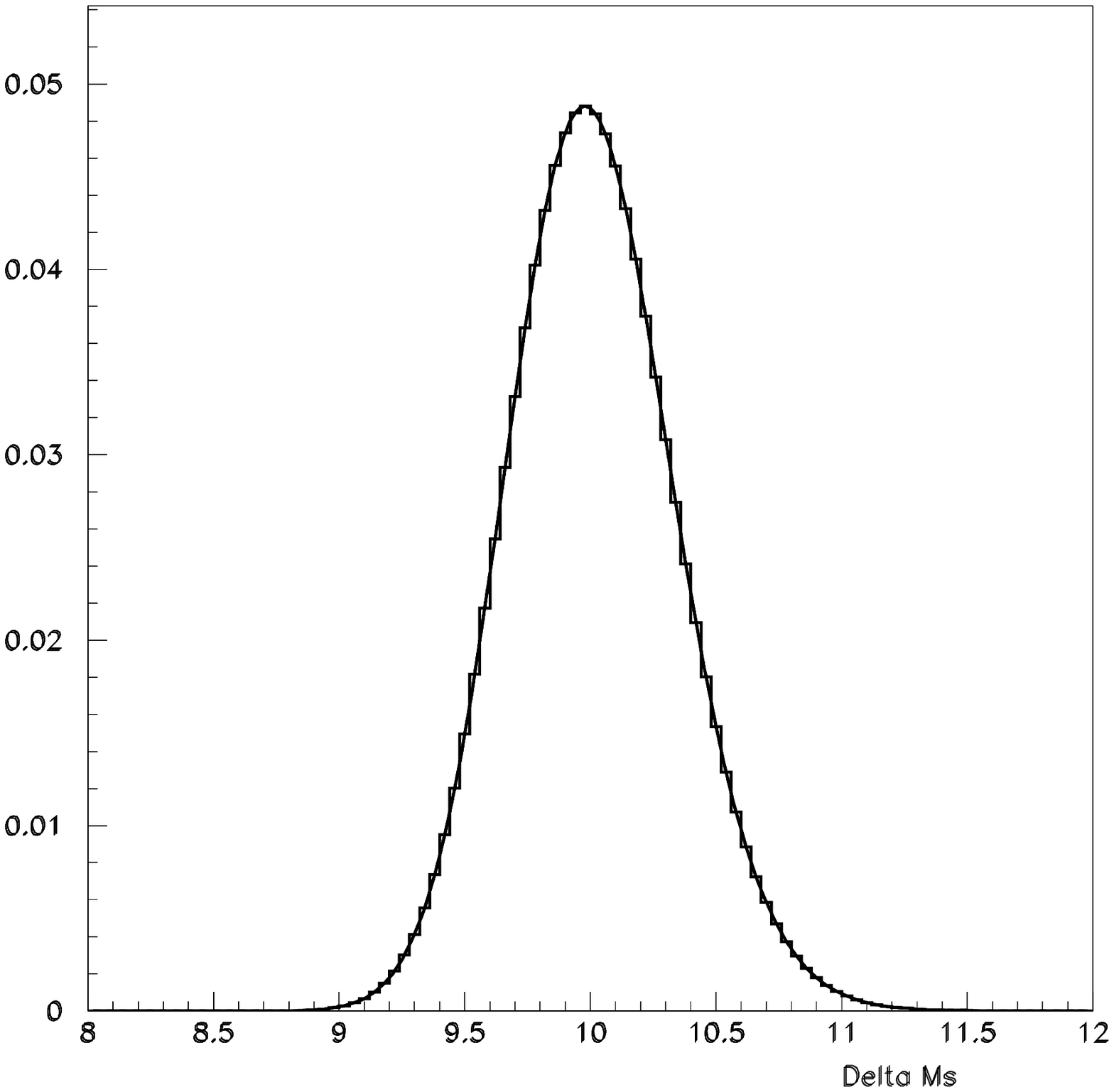}}
\caption{\label{fig:epdf10} \em\small
        \underline{Left:} next-to-leading-order $\Phinlo^{10}$ 
        \pdf\  of the $\dms$ measurement (solid line) and leading order 
        \pdf\  $\Philo^{10}$ (dotted line), for a true value $\dms=10\ps^{-1}$.
        The two PDFs are barely distinguishable. The leading order 
        approximation is excellent.
        \underline{Right:} next-to-leading-order $\Phinlo^{10}$ \pdf\  
        (solid line) and $\Phinlo^{11}$ \pdf, for $\dms=11\ps^{-1}$ 
        (dotted line). The integral of $\Phinlo^{10}$ above $\dms=11\ps^{-1}$
        is a good approximation of the CL of a true value $\dms=11\ps^{-1}$ 
        leading to a measurement $\dms\le 10\ps^{-1}$, the latter being 
        defined as the integral of $\Phix^{11}$ below $\dms=10\ps^{-1}$.}
  \centerline{\epsfxsize8cm\epsffile{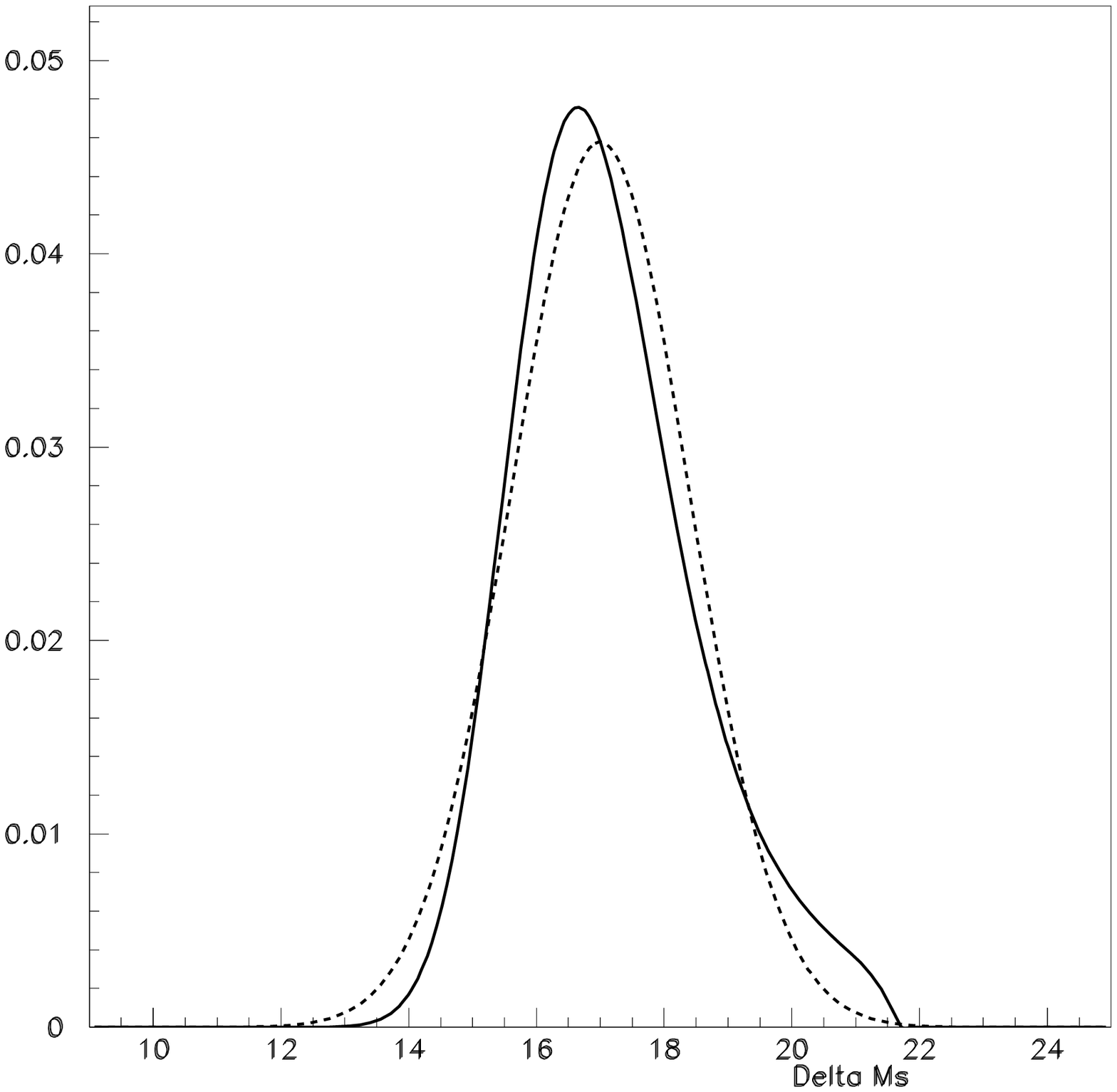}
              \epsfxsize8cm\epsffile{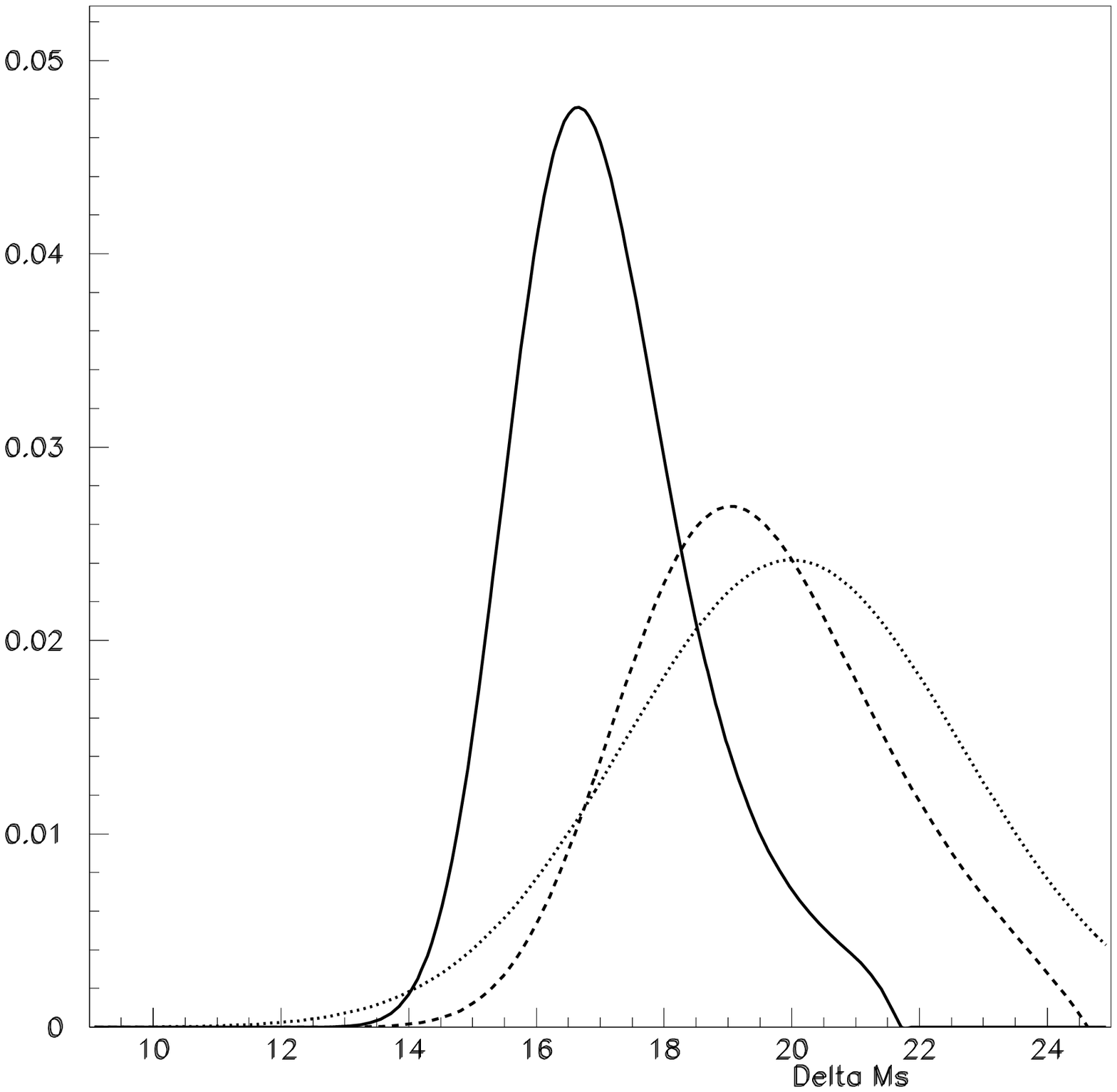}}
\caption{\label{fig:epdf17} \em\small
        \underline{Left:} next-to-leading-order $\Phinlo^{17}$ 
        \pdf\  of the $\dms$ measurement (solid line) and leading 
        order \pdf\  $\Philo^{17}$ (dotted line), for a true value 
        $\dms=17\ps^{-1}$. The two \pdfs\  differ significantly,
        but the leading order approximation remains acceptable in
        the core of the distribution. However it would underestimate 
        by a factor two about the probability to obtain a measurement 
        above $\dms=20\ps^{-1}$. 
        \underline{Right:} next-to-leading-order $\Phinlo^{17}$ 
        \pdf\  (solid line) together with $\Phinlo^{20}$ \pdf\  (dotted 
        line) and the leading order \pdf\  $\Philo^{20}$ for $\dms=20\ps^{-1}$ 
        (dashed-dotted line). The integral of $\Phinlo^{17}$ above 
        $\dms=20\ps^{-1}$ is not a good approximation of the CL of 
        a true value $\dms=20\ps^{-1}$ leading to a measurement
        $\dms=17\ps^{-1}$, the latter being defined as the integral
        of $\Phix^{20}$ below $\dms=17\ps^{-1}$. However, in the 
        present case, one cannot fully rely on the approximation 
        $\Phix^{20}\simeq\Phinlo^{20}$ owing to the large variation 
        between $\Philo^{20}$ and $\Phinlo^{20}$.}
\end{figure}
\begin{figure}[t]  
  \centerline{\epsfxsize10cm\epsffile{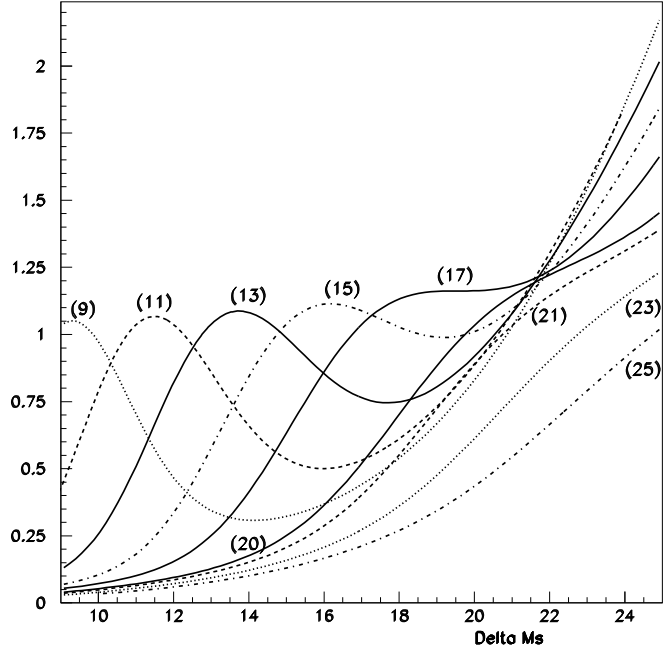}}
\caption{\label{fig:amplitudes} \em\small
        Amplitude $\Ampli$ solutions given in Eq.~(\ref{AmpExpression}) 
        (where orders up to $\K^4(\xshyp)$ are considered in the 
        expansion used here) for different true values of $\dms$. }
\end{figure}
\beqns
        N=1200~, \hspace{0.5cm}
        \frs=0.35~, \hspace{0.5cm}
        \w~\hbox{is set to zero}~,\hspace{0.5cm}
        a=0.0046~, \hspace{0.5cm}
        b=0.0090~.
\eeqns
The numerical values of the coefficients introduced previously are 
given in Table (\ref{tab:Values}) for four $\dms$ values.
\vs
Figure~\ref{fig:epdf10} shows the next-to-leading-order $\Phinlo$ \pdf\  
Eq.~(\ref{KeyNLOFormula}) of the $\dms$ measurement for a true value
of $\dms=10\ps^{-1}$ and the leading order \pdf\  $\Philo$ 
Eq.~(\ref{KeyFormula}). The two \pdfs\  being barely distinguishable, 
the leading order approximation is excellent.
\vs
Figure~\ref{fig:epdf17} gives the next-to-leading-order $\Phinlo^{17}$ 
\pdf\  of the $\dms$ measurement (solid line) and leading order \pdf\  
$\Philo^{17}$ (dotted line), for a true value $\dms=17\ps^{-1}$. The 
two \pdfs\  differ significantly, but the leading order approximation 
remains acceptable in the core of the distribution. However it would 
underestimate by a factor about two the probability to obtain a 
measurement above $\dms=20\ps^{-1}$. The right hand plot in Fig.~\ref{fig:epdf17}
shows the next-to-leading-order $\Phinlo^{17}$  \pdf\  (solid line) 
together with $\Phinlo^{20}$ \pdf\  (dotted line) and the leading 
order \pdf\  $\Philo^{20}$ for $\dms=20\ps^{-1}$ (dashed-dotted line). 
The integral of $\Phinlo^{17}$ above $\dms=20\ps^{-1}$ is not a good 
approximation of the CL of a true value $\dms=20\ps^{-1}$ leading to 
a measurement $\dms=17\ps^{-1}$, the latter being defined as the 
integral of $\Phix^{20}$ below $\dms=17\ps^{-1}$. However, in the 
present case, one cannot fully rely on the approximation 
$\Phix^{20}\simeq\Phinlo^{20}$ owing to the large variation 
between $\Philo^{20}$ and $\Phinlo^{20}$.
\vs
Figure~\ref{fig:amplitudes} shows the solutions amplitude $\Ampli$  
given in Eq.~(\ref{AmpExpression}) for different true values of $\dms$. 

\subsection{Next-to-Leading Order Likelihood}
\label{NLOLik}

We consider a likelihood
\beq
        \Lik = \sum_{i=1}^{N} \ln\P~,
\eeq
where the \pdf\ $P$ depends on the parameter $\xs$. We denote
\bei

\item   $\xs$ the true value of the parameter,

\item   $\P_0$ the \pdf\  when evaluated with $\xs$.

\eei
In the vicinity of $\xs$, 
the Taylor expansion to the fourth order is
\beq
        \Lik(\xshyp)    =
                \azero-\aone\chi+\atwo\chi^2-\athree\chi^3+\afour\chi^4~,
\eeq
with
\beqns
        \azero  =       \sum\ln(\P_0)~, \hspace{0.5cm}
        \aone   =       \phantom{\frac{1}{2}}
                        \sum\left[{\dot\P_0\over\P_0}\right]\Sigma ~,
                                \hspace{0.5cm}
        \atwo   =       \frac{1}{2}\sum
                        \left[{\ddot\P_0\over\P_0}-{\dot\P_0^2\over\P_0^2}
                        \right]\Sigma^2~, \hspace{2.0cm}\\
        \athree =       {1\over 6}\sum
                        \left[{\dot{\ddot{\P}}_0\over\P_0}
                                -3{\dot\P_0\ddot\P_0\over\P_0^2}
                                +2{\dot\P_0^3\over\P_0^3}
                        \right]\Sigma^3~, \hspace{0.5cm}
        \afour  =       {1\over 24}\sum
                        \left[
                                {\ddot{\ddot{\P}}_0\over\P_0}
                                -3{\ddot\P_0^2\over\P_0^2}
                                -4{\dot\P_0\dot{\ddot{\P}}_0\over\P_0^2}
                                +12{\dot\P_0^2\ddot\P_0\over\P_0^3}
                                -6{\dot\P_0^4\over\P_0^4}
                        \right]\Sigma^4~.
\eeqns
For a given experiment, the values of the $a_i$ coefficients are 
correlated random numbers. On the average, their values are obtained 
by replacing the sum by $N\int\P_0$. Using Eq.~(\ref{Normalization}), 
one gets
\beqns
        \aone           =       0~, \hspace{0.5cm}
        \aatwo          = -\frac{1}{2}~, \hspace{0.5cm}
        \aathree        = {1\over 6}{-3B+2C\over\sqrt{N}A^{3\over 2}}~,\hspace{0.5cm}
        \aafour         = {1\over 24}{-3D-6E+12\F-4G\over N^{3\over 2}A^2}~,
\eeqns
where the last term involves the new set of integrals
\beqns
        D(\xs)  =
                \int\limits_{-\infty}^{+\infty} 
                \left({\Pdotdot_-^2\over\P_-}
                     +{\Pdotdot_+^2\over\P_+}\right)\dtmes~,\hspace{1.0cm}
        E(\xs)  =
                \int\limits_{-\infty}^{+\infty} 
                \left({\Pdot_-^4\over\P_-^3}
                     +{\Pdot_+^4\over\P_+^3}\right)\dtmes~,\hspace{1.0cm}\\
        F(\xs)  =
                \int\limits_{-\infty}^{+\infty} 
                \left({\Pdot_-^2\Pdotdot_-\over\P_-^2}
                     +{\Pdot_+^2\Pdotdot_+\over\P_+^2}\right)\dtmes~,
                \hspace{1.0cm}
        G(\xs)  =
                \int\limits_{-\infty}^{+\infty} 
                \left({\Pdot_-\dot{\ddot{\P}}_-\over\P_-}
                +     {\Pdot_+\dot{\ddot{\P}}_+\over\P_+}
                \right)\dtmes~.
\eeqns
In effect, the maximum of $\Liknlo$ is reached for
\beqn
        0       &=&-\aone+2\atwo\chi-3\athree\chi^2\\
        \xsmes  &\simeq&\xs-
                \left({\aone\over 2\atwo}-{3\aone^2\athree\over 8\atwo^3}
                \right)~.
\eeqn

 \section{Combining Inconsistent Measurements}
\label{sec:theCombiner}

When several measurements $\xmes(i)\pm\sigmes(i)$, with $i=1,.., N$, 
of the same physical observable $X$ are available, the question arises 
on how to combine them into a single measurement $\xmesbar$. 
Combining the measurements can serve two purposes: merely,
it can be to provide a summary carrying the overall information 
within a conveniently easily-quoted global measurement, or,
more ambitiously, it can be to provide a means to incorporate the 
set of measurements into a more involved analysis, like a global 
CKM fit, where the physical observable $X$ enters as one input 
among others. 
\vs
{\em We note that the averaging method introduced below is not yet 
     applied in the present CKM analysis. We reserve its use for 
     forthcoming occasions.}
\vs
The weighted mean (WM) method defined by
\beqn
\label{eq_weightedmean}
\label{weightedmean}
        \xmesbar        &=&     \sigmesbar^2\sum_{i=1}^N 
                                \sigmes^{-2}(i)\,\xmes(i)~, \\
\label{weightedmeanerror}
        \sigmesbar^{-2} &=&     \sum_{i=1}^N \sigmes^{-2}(i)~,
\eeqn
is the optimal scheme to merge the individual measurements. However
it assumes that the measurements are consistent the ones with the others.
It leads to an easily-quoted global measurement $X=\xmesbar\pm\sigmesbar$.
Furthermore, because the underlying hypothesis is clear (the set of 
measurements is taken to be consistent) the WM method is statistically 
well-defined and its result is easy to use: the true value of the 
physical observable being assumed to be $\xtrue$, the probability for 
this value to yield for the $\chi^2$
\beq
\label{Chi2X}
        \chi^2(\xtrue) 
        =
        \left({\xmesbar-\xtrue\over\sigmesbar}\right)^{\!\!2},
\eeq
a value larger than the observed one, is given by:
\beq
\label{ProbChi2X}
        \Prob(\xtrue) =\ProbCERN(\chi^2,\,1) ~.
\eeq
A measure of the consistency of the set of measurements is provided 
by the $\chi^2$
\beq
\label{eq_chi2}
        \chi^2(\xmesbar)
        =
        \sum_{i=1}^N\left({\xmes(i)-\xmesbar\over\sigmes(i)}\right)^{\!\!2},
\eeq
and, more conveniently, by its associated confidence level
\beq
\label{eq_prob}
        \Prob(\xmesbar)= \ProbCERN(\chi^2(\xmesbar),\,N-1)~.
\eeq
If the value of $\chi^2(\xmesbar)$ is too large, one may suspect 
that some of the measurements in the set are flawed. If one insists 
on a democratic treatment of the $\xmes(i)$, \ie, if one refuses 
to remove the suspected ones, the PDG-recommended scheme~\cite{PDG},
termed the rescaled weighted mean below (RWM), consists of rescaling 
the error $\sigmesbar$ of Eq.~(\ref{weightedmeanerror}) by the scale 
factor
\beq
\label{eq_pdgRescaled}
        S = \sqrt{\chi^2(\xmesbar)/(N-1)}~,
\eeq
if the latter exceeds unity
\beq
\label{SigmaRescaled}
        \Sigmesbar=\sigmesbar\ S~ .
\eeq
The RWM method is simple and convenient, however, it suffers from two 
important drawbacks: 
\begin{enumerate}
\item   {\em Psychostatistics}

        On the average, 
        the quoted error is necessarily enlarged 
        with respect to the one of the un-rescaled weighted mean (WM), 
        even for consistent data sets.
        Tampering with Eq.~(\ref{weightedmeanerror}) implies a departure 
        from well-defined statistics to enter the realm of ill-defined
        (psycho)statistics, where working hypotheses are no longer fully
        explicited.
\item   {\em Schizostatistics}

        The average $\xmesbar$ may lie outside the range of values
        covered by the measurements.
        This is because the democratic treatment does not allow to detect 
        measurements which obviously stick out from the set.
        Such a measurement, 
        termed an outlier in the following,
        pulls the average toward its value, 
        albeit it may remain inconsistent with the resulting weighted mean,
        even though the error of the latter is rescaled.
\end{enumerate}

In this paper we advocate the use of a method, termed the {\em \Combiner},
which is an extension to the weighted mean method\footnote
{
        Another approach for the combination of inconsistent 
        measurements has been developed in Ref.~\cite{sceptical}.
}. 
Although the \Combiner\  does not provide an escape from the 
first drawback, it is shown below to be more satisfactory with 
respect to the second drawback. The discussion of the above drawbacks 
is further expanded below.

\subsection{Psychostatistics}
\label{PsychostatisticsSection}

There is no way out of the first drawback: this is because accepting 
the {\it possibility} of having in the set of measurements some that
are biased in an unspecified way implies a loss of information,
which, furthermore, is ill-defined. When the standard deviation is 
rescaled following the RWM scheme, the (re)definition of 
Eq.~(\ref{Chi2X}) is to be taken, at best, as a test statistics.
It is no longer a pure $\chi^2$ term and Eq.~(\ref{ProbChi2X}) 
does not hold. Moreover, the use of this test statistics is ill-defined.
Its distribution cannot be determined, since the underlying hypothesis 
is now unspecified (the set of measurements is taken to be inconsistent,
but this is not a precisely defined hypothesis).
\vs
However the aim of the RWM method being to be conservative, 
the price to pay is to accept the use of ill-defined statistics
and to deal with $\chi^2(\xtrue)$ as if it were a pure $\chi^2$.
Stated differently, applying Eq.~(\ref{ProbChi2X}) yields 
over-conservative confidence levels, which, after all, is precisely 
what one is looking for.

\subsection{Schizostatistics}
\label{SchizostatisticsSection}

The second drawback is worth being spelled out explicitly. If one is 
dealing with two measurements which are sufficiently apart for the 
rescaling of Eq.(\ref{SigmaRescaled}) to be enforced, namely, if 
\beq
\label{TriggerRWM}
        \DeltaxMesTwo\equiv(\xmes(2)-\xmes(1))^2 
        > \sigmes^2(1)+\sigmes^2(2)~ ,
\eeq
leading to a rescaled uncertainty (\cf\  Eq.~(\ref{SigmaRescaled}))
\beq
\label{RescaledTwo}
        \Sigmesbar 
        = 
        \frac{\sigmes(1)\sigmes(2)|\DeltaxMes|}{\sigmes^2(1)
        +\sigmes^2(2)}~,
\eeq
then the RWM method is prone to contradict itself.
\vs
On the one hand, using $\Sigmesbar$ in place of $\sigmesbar$ in 
Eq.(\ref{Chi2X}), yields
\beqn
\label{Chi2Mes1}
        \chi^2(\xtrue=\xmes(1)) &=&
                \left({\sigmes(1)\over\sigmes(2)}\right)^{\!\!2},\\
\label{Chi2Mes2}
        \chi^2(\xtrue=\xmes(2)) &=&
                \left({\sigmes(2)\over\sigmes(1)}\right)^{\!\!2},
\eeqn
independently on how far apart the two measurements are, provided 
Eq.~(\ref{TriggerRWM}) is fullfilled. Hence, if the two measurements 
have widely different uncertainties, the measurement with the largest 
uncertainty (\eg, the second one: $\sigmes(2)\gg\sigmes(1)$) is viewed 
as corresponding to a true value of the physical observable $\xtrue=\xmes(2)$
which is utterly ruled out by the data ($\chi^2(\xtrue=\xmes(2))\gg 1$),
even though the weighted mean error is rescaled.
\vs
On the other hand, the weighted mean is pulled away from both 
measurements in proportion of $\DeltaxMes$. As a result,
from the view point of both measurements $i=1,2$, the hypothesis 
that the true value of the physical observable is $\xtrue=\xmesbar$
leads to
\beq
\label{Chi2FromMesi}
        \chi^2(\xtrue=\xmesbar)(i)
        =
        \left({\xmes(i)-\xmesbar\over\sigmes(i)}\right)^2
        =
        \DeltaxMesTwo{{\sigmes}^2(i)\over({\sigmes}^2(1)+{\sigmes}^2(2))^2}~,
\eeq
and hence is liable to be ruled out as well, if $\DeltaxMes$ is large 
enough. In particular, if $\sigmes(2)\gg\sigmes(1)$
\beqn
\label{Chi2FromMes2}
        \chi^2(\xtrue   &=&
                \xmesbar)(1)\simeq S^2
                \left({\sigmes(1)\over\sigmes(2)}\right)^2\ll 1 ~, \\
        \chi^2(\xtrue   &=&
                \xmesbar)(2)\simeq S^2~ .
\eeqn
Therefore, if the second measurement has a much larger uncertainty 
than the first measurement, the conjunction of 
Eqs.~(\ref{Chi2Mes2}-\ref{Chi2FromMes2}) implies that when the 
rescaling is significant, the RWM result and the second measurement 
are mutually incompatible: this contradicts the use of the second 
measurement to define the weighted mean, especially considering its 
impact on $\Sigmesbar$ as displayed by Eq.~(\ref{RescaledTwo}).
\vs
For twin measurements with identical $\sigmes$, the RWM method is not 
self-contradictory: whereas Eq.~(\ref{Chi2FromMesi}) indicates that 
from the view point of both measurements the RWM value may be 
unacceptable, Eqs.~(\ref{Chi2Mes1}-\ref{Chi2Mes2}) guarantee that 
the rescaled uncertainty yields an acceptable $\chi^2$ for both.
\vs
If only two measurements enter into play, not much can be done to 
circumvent this second drawback, since there is no objective way to 
identify the flawed one. However, if more than two measurements are 
available, one may rely on the consistency of a subset of them to 
identify the possible outliers.

%
%
\subsection{The \Combiner}
\label{TheCombinerSection}

The \Combiner~method is explicitly build as an extension to the 
weighted mean: by construction, it tends to reproduce 
Eqs.(\ref{weightedmean}-\ref{weightedmeanerror}) in the case of 
a consistent set of measurements. The (psycho)statistical point 
of view which is taken here is that some of the measurements to
be averaged might be incorrect: if such measurements occur, they 
should be removed from the set. 

%
%
\subsubsection{Principle}
\label{PrincipleSection}

The removal of incorrect measurements relies on the clustering of 
the other measurements around a common mean. Rather than removing 
abruptly a measurement if it meets some criteria, the \Combiner\ 
does not cut but considers all possible hypotheses about the 
correctness of the measurements. A configuration being defined 
as a subset of measurements that are assumed to be consistent
the ones with others, the \Combiner\  weighs all possible 
configurations to build an overall likelihood. To reproduce 
the WM result, the \Combiner~favors the configurations involving 
the largest number of measurements, provided they have good 
probabilities.

%
%
\subsubsection{Notations}
\label{NotationsSection}

We denote:
\begin{itemize}

\item   $c$, an ordered list of $N$ bits.
        It is referred to as a {\em configuration},
        indicating which measurements are considered.
        For example, for $N=3$, the configuration $c\equiv 101$ 
        means that the two measurements $i=1$ and $i=3$ are 
        to be merged, while disregarding the measurement $i=2$.
        The void configuration being of no interest,
        the total number of configurations considered 
        amounts to $2^N-1$.

\item   $n_c$, the number of bits set to one. It is referred 
        to as the {\em multiplicity} of the configuration.

\item   $\cfull=11\dots 1$, the configuration where all measurements 
        are considered ($n_{\cfull}=N$).

\item   $\chi^2_c$, the $\chi^2$ obtained from the weighted mean 
        (\cf, Eqs.~(\ref{eq_weightedmean}-\ref{eq_chi2})) of the $n_c$ 
        measurements to be considered in the configuration $c$.

\item   $\Prob_c={\rm Prob}(\chi^2_c,\,n_c-1)$, 
        the corresponding configuration {\em probability}
        (\cf, Eq.~(\ref{eq_chi2})).

\item   $\overline\Prob_c=1-\Prob_c$

\item   $\xmesbar(c)$ and $\sigmesbar(c)$, the results of the 
        weighted mean for the configuration $c$.

\item   $G_c\equiv G_c(\xtrue)$, the Gaussian likelihood (to be 
        interpreted as a PDF when used in a Bayesian approach) 
        with mean value $\xmesbar(c)$ and standard deviation 
        $\sigmesbar(c)$. 

\item   $\w_c$, a weight characterizing the configuration $c$. 
        The sum of these weights over the $2^N-1$ configurations 
        is normalized to unity, \ie, $\sum_c\w_c=1$.

\item   $c^\prime>c$ denotes two configurations such that all the 
        bits set at one in $c^\prime$ are also set at one in $c$,
        and there is at least one bit set at one in $c^\prime$ which is not
        set at one in $c$ (\eg, $1111> 1101$, but $1101\not >1011$). 
        The configuration $c^\prime$, embedding $c$, is said to be 
        {\em larger} than $c$.

\item   Products of probabilities over void configurations  
        are set to one, \eg, 
        $\prod_{c^\prime}^{c^\prime>c}\Prob_{c^\prime}\equiv1$,
        if no $c^\prime$ exist for which $c^\prime>c$ holds
        (\ie, $c$ is the largest configuration: $c=11\dots1$).

\end{itemize}

\subsubsection{Definition}
\label{TheCombinerSubsubsection}

With these notations, the WM method consists of using only the 
configuration $\cfull$. In a likelihood analysis relying on those 
measurements, this treatment is equivalent to adding to the overall 
log-likelihood the term
\beq
\label{PseudoChi2}
        \chi^2(\xtrue) = -2\ln G(\xtrue)~,
\eeq
where the global likelihood $G$ is the fully combined Gaussian 
$G=G_{\cfull}$. The RWM method consists of using the same 
configuration, but, if $S>1$, it modifies $G_{\cfull}$ by 
rescaling its standard deviation. As pointed-out in the Introduction 
(\cf\   Section \ref{PsychostatisticsSection}) even though in that 
case $-2\ln G_{\cfull}$ defines a pseudo-$\chi^2$, it should be used 
as a pure $\chi^2$.
\vs
Instead of using a single Gaussian, the \Combiner\  uses for the global 
likelihood a compound function:
\beq 
\label{eq_combiner}
        G(\xtrue)=\sum_c \w_c G_c(\xtrue)~,
\eeq
which enters into the computation of the pseudo-$\chi^2$ of 
Eq.~(\ref{PseudoChi2}) here also to-be-used as a pure $\chi^2$.
What remains to be done is to define appropriately the weights
$\w_c$. Obviously there is an infinite number of choices.
Since the goal is to protect the analysis from biased 
measurements and over-optimistic $\sigmesbar$ values, which 
is admittedly a vague goal, one must rely on educated 
guesswork to pick-up a particular definition of the weights.

\subsubsection*{The C-\Combiner}

For the C-\Combiner, the weights are defined as
\beq
\label{eq_Ccombiner} 
\w_c^{({\rm C})}=a\ \Prob_c\prod_{c^\prime}^{c^\prime>c}
        \overline\Prob_{c^\prime}~,
\eeq
where the constant $a$ ensures the proper normalization of the weights. 
The first term is a measure of the validity of the combination, 
while the second term suppresses it, if any configuration larger 
than $c$ receives a high probability. If a configuration $c$ is such 
that all larger configurations are unlikely, it does not get suppressed 
by the above expression. For example, if an outlier measurement is 
utterly incompatible with all the others, it is kept by the C-\Combiner\
with a weight equal to $a$. It is termed the {\em Cool-\Combiner} 
because of that. However, if the number of consistent measurements 
grows, the outlier weight, $a$, decreases accordingly.

\subsubsection*{The T-\Combiner}

For the T-\Combiner, the weights are defined as
\beq 
\label{eq_Tcombiner}
\w_c=
        {\Prob_c\over\sum_{c^\prime}^{n_c=n_{c^\prime}}
        \Prob_{c^\prime}}\,
        \left(1-\prod_{c^\prime}^{n_{c^\prime}= n_c}
        \overline\Prob_{c^\prime}\right)
        \prod_{c^\prime}^{n_{c^\prime}> n_c}
        \overline\Prob_{c^\prime}~.
\eeq
The first term is a measure of the relative validity of the configuration 
$c$, with respect to configurations of the same multiplicity $n_c$. 
The second term weighs the validity of the configurations of 
multiplicity $n_c$, taken as a whole. The third term suppresses 
configurations of multiplicity $n_c$ if any higher multiplicity 
configuration receives a large probability. Therefore, 
whether or not a configuration $c^\prime$ is larger than $c$, 
if $n_{c^\prime}>n_c$, it is sufficient for $c^\prime$ to receive 
a large probability to suppress $c$. For example, if an outlier 
measurement is utterly incompatible with all the others, it is 
suppressed by the \Combiner, if some of the others are mutually 
compatible. It is termed the {\em Tough-\Combiner} because of that.
For the configuration $c=\cfull$, the third term receives no 
contribution and is defined to be equal to one. The above expression 
ensures the normalization of the weights to unity.

\subsection{Illustrations}
\label{IllustrationsSection}

\subsubsection{Twin Measurements}
\label{TwinMeasurementsSection}

To illustrate how the \Combiner\  works, we first consider the evolution 
of the likelihood $G$ as a function of the discrepancy between twin 
measurements $\xmes(1)$ and $\xmes(2)$, with 
$\sigmes(1)=\sigmes(2)=\sigma_0=1$. 
\begin{figure}[t]
  \epsfxsize12cm
  \centerline{\epsffile{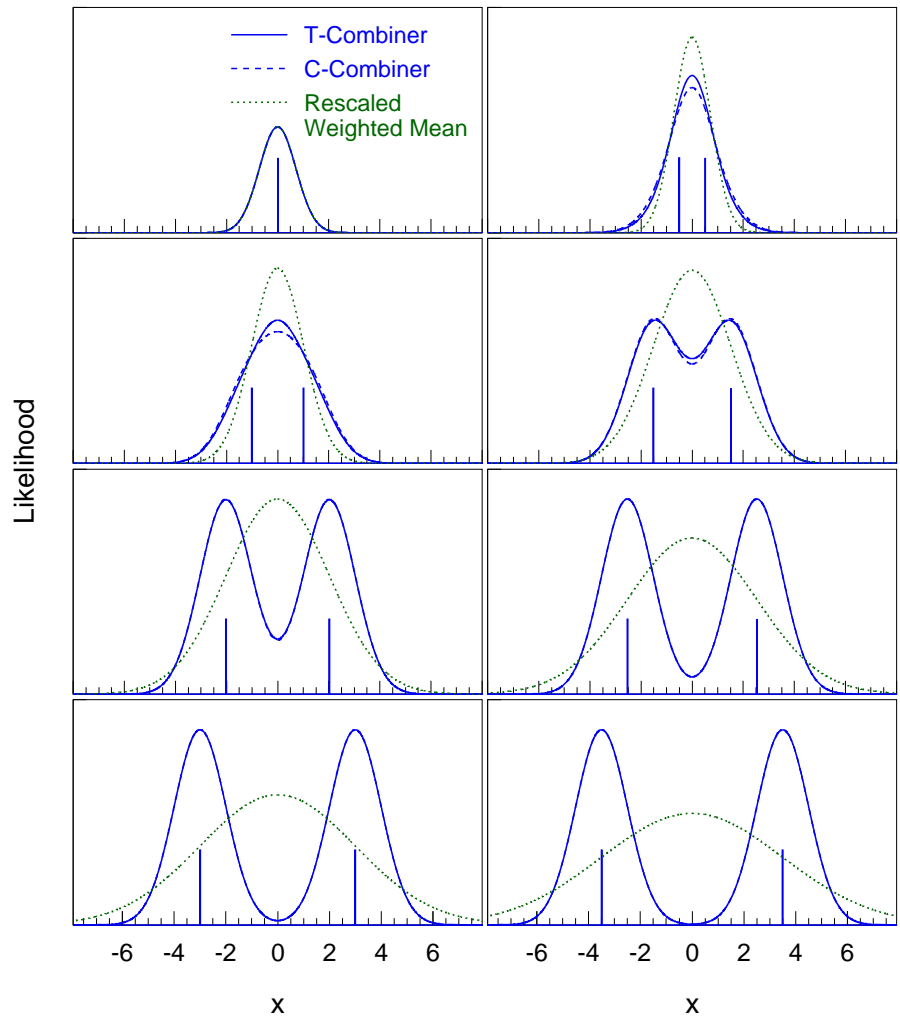}}
  \vspace{-0.5cm}
  \caption[.]{\label{fig_distance}\em
        Likelihoods $G$ for the T-\Combiner\ (solid), the C-\Combiner\
        (dashed), and for the RWM method (dotted).
        The vertical lines indicate the locations of the two individual 
        measurements which develop a mutual disagreement 
        rising from the upper left figure to the lower right figure.
        The dotted line corresponds to the likelihood of the 
        C-\Combiner.}
\end{figure}
Figure~\ref{fig_distance} shows the combined likelihood for the two
\Combiner\  approaches and for the RWM method. The measurements, 
indicated by the vertical lines, develop a mutual disagreement which 
increases from the upper left figure to the lower right figure. One 
observes that the \Combiner\  likelihoods become broader with increasing 
discrepancy and eventually splits up into two Gaussian-like likelihoods,
where the T- and C-\Combiner\ behave similarly. In the limit of extreme 
incompatibility, the two curves become genuine Gaussians with central 
values $\xmes(1)$ and $\xmes(2)$, respectively, and width $\sigma_0$. 
On the contrary, the RWM Gaussian stays on the central value while 
the error increases to always keep the two measurements within the 
(so-called) $68\%$ probability limit. The unsatisfactory behavior 
of the RWM method becomes obvious at large inconsistencies. One 
observes on the four lower curves in Fig.~\ref{fig_distance} that, 
whereas the \Combiner\  treats the center value $x=0$ as being unlikely 
--- it is favored by the weighted mean, although there exist no 
supporting measurement, while the rescaled uncertainty gives rise 
to unduly broad tails which do not show up when using the \Combiner.
\vs
This first example already exhibits advantages of using the \Combiner. 
However it does not allow to demonstrate fully the superiority of 
the method because it uses twin measurements. As discussed previously 
(\cf\  Section~\ref{SchizostatisticsSection}), the behavior of the 
RWM method is even less satisfactory if the two measurements have 
very different $\sigmes$. In addition, since only two measurements 
are available, the \Combiner\  cannot use the clustering of correct 
measurements to suppress the flawed one(s). As shown in the next 
section, the behavior of the \Combiner\  is markedly different for a 
set of more than two measurements among which a subset is consistent.

\subsubsection{Information Loss}
\label{InformationLossSection}

\begin{figure}[t]
  \epsfxsize14cm
  \centerline{\epsffile{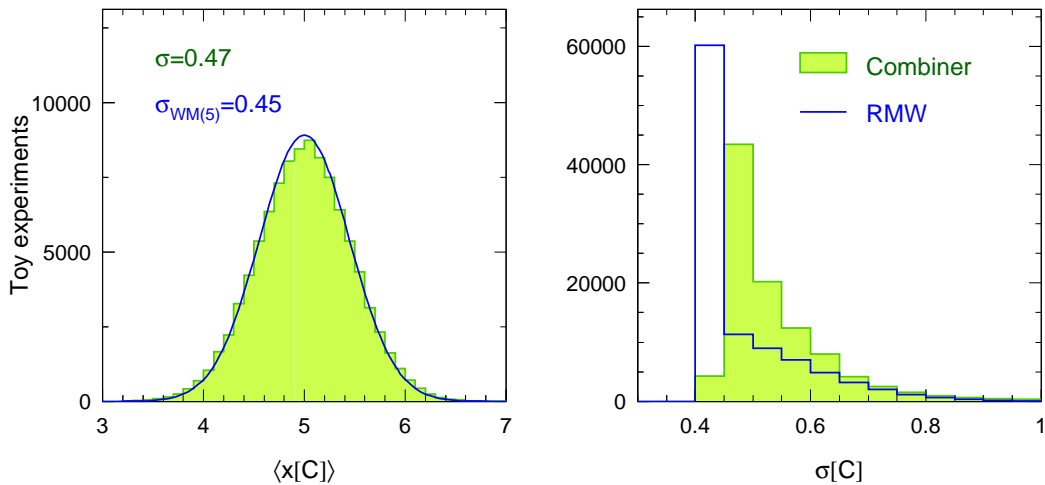}}
  \vspace{-0.5cm}
  \caption[.]{\label{combinerfig1}\it
        The left plot shows the toy Monte Carlo distribution of the mean 
        value $\meanT$ as provided by the T-\Combiner\  (shaded histogram), 
        for a set of five consistent measurements which are Gaussian 
        distributed around $\xmesbar=5$,
        each with a standard deviation $\sigma(i)=1$. A Gaussian fit 
        to the distribution results in $\sigma=0.47$. Also shown is
        the weighted mean distribution (solid line) characterized by the 
        standard deviation $\sigma_{WM:5}=5^{-1/2}\simeq0.45$.      
        The right hand plot shows the distribution of the RMS, $\rmsC$,
        of the  T-\Combiner\  (shaded histogram) compared to the RMS,
        $\rmsRWM$, of the rescaled weighted mean.}       
\end{figure}
It was mentioned before that enlarged errors (broader likelihoods) are 
an unavoidable side-effect when taking into consideration the possibility 
of biased measurements. This entails a loss of information when the set 
is consistent. To quantify this loss of information, we perform a toy Monte 
Carlo simulation of a set of $N=5$ measurements, each distributed following 
the same Gaussian, $\xmesbar(1-5)=\langle x_0\rangle=5$ and 
$\sigmes(1-5)=\sigma_0=1$. We use the RWM method and the T-\Combiner\  
to obtain the distributions of the mean value and the distributions of the 
root mean square (RMS) of the likelihood $G$. 
\vs
The results are given in
Fig.~\ref{combinerfig1}. The left plot shows the distribution of the mean 
value $\meanT$ as provided by the T-\Combiner. The distribution of $\meanT$ 
is very close to be a Gaussian of width $\sigma=0.47$. This is to be 
compared with the (optimal) WM Gaussian distribution of width 
$\sigma_{WM:5}=5^{-1/2}\simeq0.45$. This width is also obtained for the
rescaled weighted mean, as the center value of the likelihood is not 
affected by the enlargement of the width. Although statistical outliers 
are suppressed in the \Combiner, giving rise to a narrower effective 
width, the increase of the width due to the statistical occurrence of 
seeming inconsistencies superseeds the narrowing suppression effect. 
The right figure shows the distribution of the RMS, $\rmsC$, of the 
T-\Combiner, to be compared to the distribution of the RMS, $\rmsRWM$, 
of the rescaled weighted mean. The increase of the errors is stronger 
for the \Combiner.

\subsubsection{Inconsistent Set}
\label{InconsistentSetSection}

The second example uses the following set of $N=5$ measurements: 
$\xmes(1-5)=3.7,\;4.2,\;5.0,\;5.5,\;0.0$, all with identical errors 
$\sigmes(1-5)=1$. While the first four data points are mutually 
compatible with a Gaussian distribution ($\chi^2/\Ndof=1.9/3$), the 
last measurement is an outlier leading to a large overall 
$\chi^2/\Ndof=18.9/4$, translated into a scale factor of $S=2.2$. 
The C- and the T-\Combiner\  yield for the likelihood~$G$ (quoting 
only the leading terms):
\begin{figure}[t]
  \epsfxsize12cm
  \centerline{\epsffile{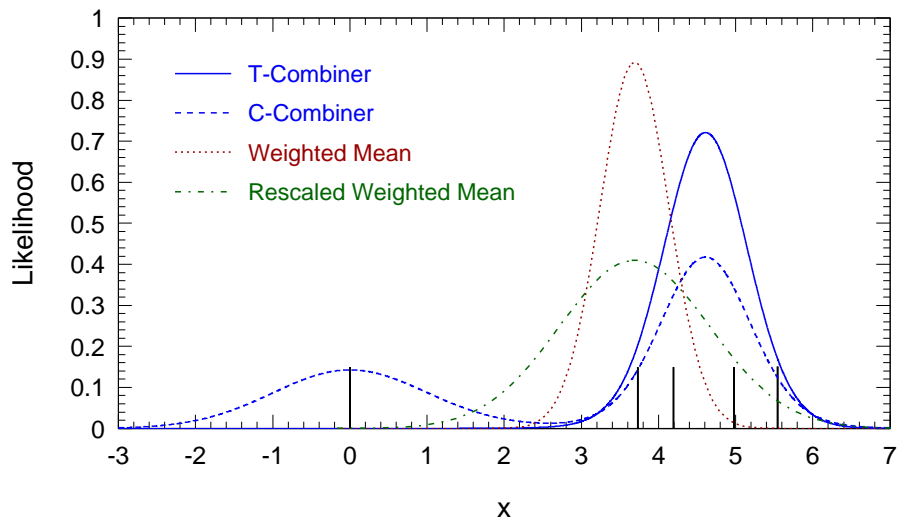}}
  \vspace{-0.5cm}
  \caption[.]{\label{fig:test_tough_cool}\it
        Likelihoods $G$ obtained by the four approaches: 
        T-\Combiner, C-\Combiner, weighted mean and the 
        rescaled weighted mean. The vertical lines indicate 
        the five individual measurements of which one is 
        inconsistent.}
\end{figure}
\begin{figure}[p]
  \epsfxsize10.5cm
  \centerline{\epsffile{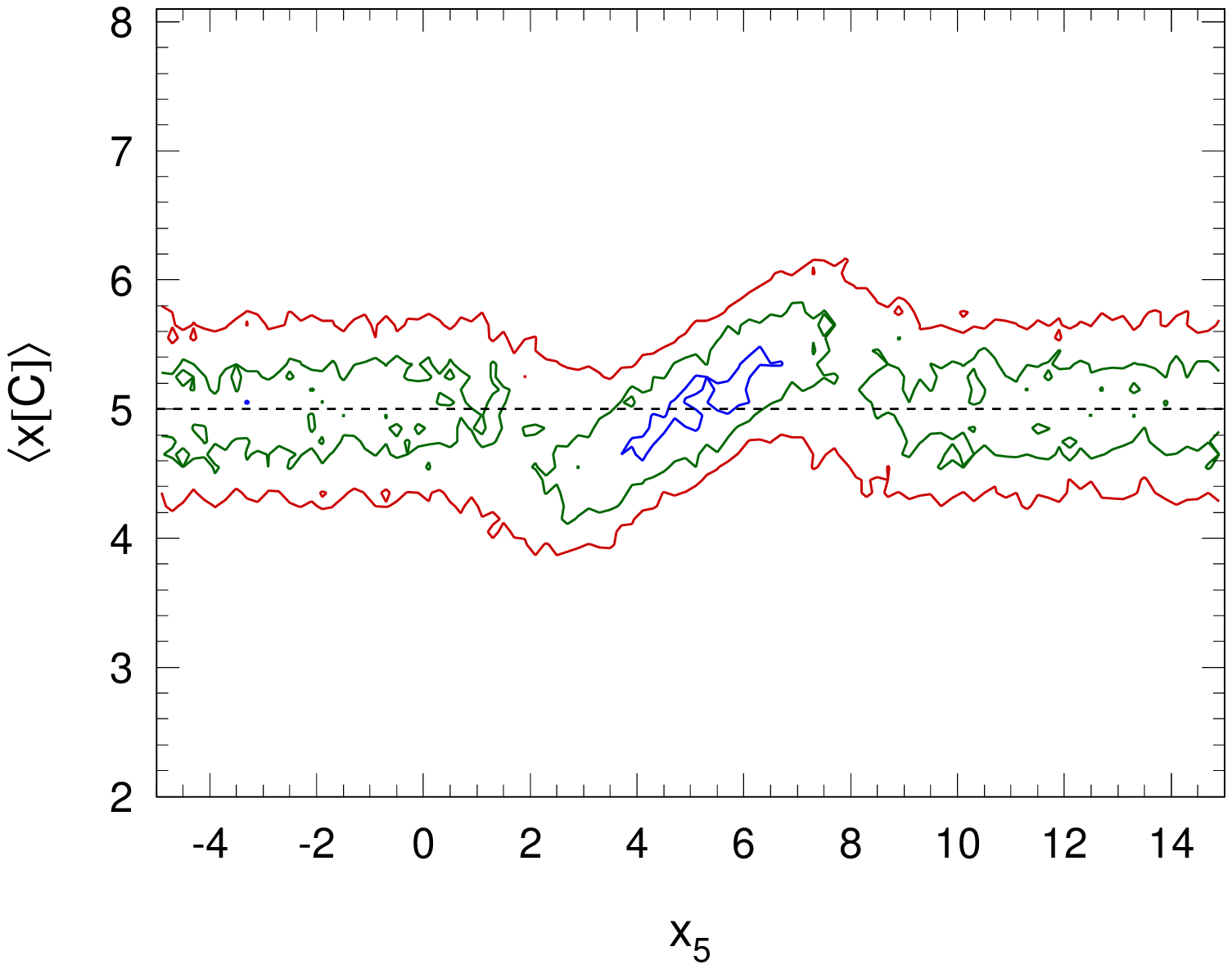}}
  \vspace{-0.5cm}
  \caption[.]{\label{fig:combinerfig4}\it
        Distribution of the mean value $\meanT$ as provided by the T-\Combiner\
        (on the vertical axis) for a set of inconsistent measurements,
        four of which are Gaussian distributed around $\xmesbar=5$,
        each with a standard deviation $\sigma(i)=1$,
        while the fifth measurement is uniformly distributed between
        $x_5=-5$ and $x_5=15$ (on the horizontal axis).
        The latter is effectively removed from the set by the T-\Combiner
        when it departs from $\xmesbar$ by about $1.5\sigma(i)$.}       
\vspace{1cm}
  \epsfxsize14cm
  \centerline{\epsffile{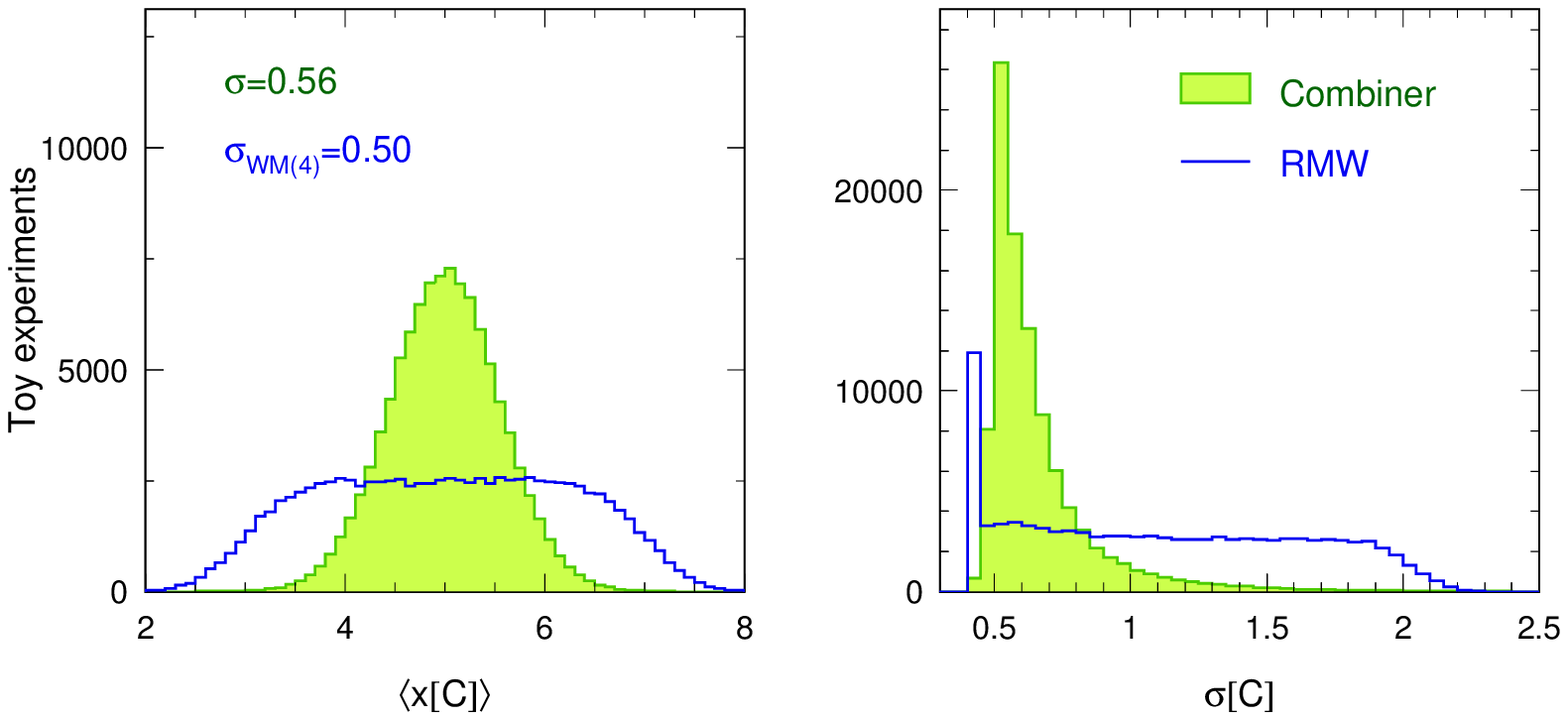}}
  \vspace{-0.5cm}
  \caption[.]{\label{combinerfig2}\it
        The left plot shows the toy Monte Carlo distribution of the mean 
        value $\meanT$ as provided by the T-\Combiner\  (shaded histogram), 
        for a set of inconsistent measurements four of which are Gaussian 
        distributed around $\xmesbar=5$, each with a standard deviation 
        $\sigma(i)=1$, while the fifth measurement is uniformly distributed 
        between $x_5=-5$ and $x_5=15$. A Gaussian fit to the distribution 
        results in $\sigma=0.56$, which is to be compared with the optimal 
        WM result $\sigma_{WM:4}=0.5$, obtained when the fifth measurement 
        is removed from the set. Also shown is the weighted mean 
        distribution (solid line).
        The right hand plot shows the distribution of the RMS, $\rmsC$,
        of the  T-\Combiner\  (shaded histogram) compared to the RMS,
        $\rmsRWM$, of the rescaled weighted mean.}       
\end{figure}
\begin{center}
\begin{tabular}{llcccccc}\hline
&&&&&& \\[-0.2cm]
 &  $c$                 & 00001 & 11110 & 01110 & 10110 & 11010 & 11100 \\
\rs{C-\Combiner:}&
  $\w_c^{({\rm C})}$    & 0.358 & 0.213 & 0.102 & 0.097 & 0.064 & 0.063 \\[0.15cm]
        \hline 
&&&&&& \\[-0.2cm]
 & $c$                  & 11110 & 01110 & 10110 & 11010 & 11100 & -     \\ 
\rs{T-\Combiner:}&
  $\w_c^{({\rm T})}$    & 0.580 & 0.125 & 0.118 & 0.079 & 0.077 & -     \\[0.15cm]
        \hline 
\end{tabular}
\end{center}
Quoted are only the configurations with $\w_c>0.05$. One observes that 
configurations where the incompatible measurements are mixed,
\ie, bit five and another one are set to one, have negligible 
weights for both the C and the T-Combiner. The C-Combiner
allows a sizable single weight for outlier (measurements five)
which, while it is suppressed by the T-Combiner.
\vs
The \Combiner\  likelihoods are shown in Fig.~\ref{fig:test_tough_cool} 
together with the WM and the RWM likelihoods. Whereas one may be satisfied
by either the C-\Combiner\  or the T-\Combiner, one observes that the 
WM and the RWM methods contradict themselves: the outlier pulls the 
mean value significantly but, even in the RWM method, the outlier and 
the mean value remain incompatible. Furthermore, in the example 
considered here, the measurement the farthest apart from the outlier, 
although correct, is also incompatible with the mean value. 
\vs
To study the recovery capability of the \Combiner\  we plot in   
Fig.~\ref{fig:combinerfig4} the distribution of the mean value $\meanT$ 
as provided by the T-\Combiner\  (on the vertical axis) for the set of 
inconsistent measurements, four of which are Gaussian distributed around 
$\xmesbar=5$, each with a standard deviation $\sigma(i)=1$, while the 
fifth measurement is uniformly distributed between $x_5=-5$ 
and $x_5=15$ (on the horizontal axis). The latter is effectively 
removed from the measurement set by the T-\Combiner\  when it departs 
from $\xmesbar$ by about $1.5\sigma(i)$. We show that, for the T-\Combiner, the distribution
of the mean value and the RMS for the toy experiments in Fig.~\ref{combinerfig2} 
  is very close to be a Gaussian of width $\sigma=0.47$
(left hand plot). This is to be compared with the optimal WM Gaussian 
distribution of width $\sigma_{WM:4}=0.5$ (discarding the inconsistent
measurement). The rescaled weighted mean exhibits a significantly 
larger scattering than the \Combiner. The right figure shows the 
distribution of the RMS, $\rmsC$, of the 
T-\Combiner, to be compared to the distribution of the RMS, $\rmsRWM$, 
of the rescaled weighted mean. The increase of the errors (\ie, the 
loss of information) is stronger for the rescaled weighted mean.

 }

%

\clearpage
 \begin{thebibliography}{999}
\markboth{\textsc{References}}
         {\textsc{References}}

\bibitem{cabibbo}       N.~Cabibbo,
                        Phys. Rev. Lett. {\bf 10}, 351 (1963)

\bibitem{kmmatrix}      M.~Kobayashi and T.~Maskawa,
                        Prog. Theor. Phys. {\bf 49}, 652 (1973)

\bibitem{wolfenstein}   L.~Wolfenstein,
                        Phys. Rev. Lett. {\bf 51}, 1945 (1983)

\bibitem{buras}         A.J.~Buras, M.E.~Lautenbacher and G.~Ostermaier,
                        Phys. Rev. {\bf D50}, 3433 (1994)

\bibitem{bigietal}      I.Y.~Bigi, V.A.~Khoze, N.G.~Uraltsev
                        and A.I.~Sanda,
                        in \textit{``CP Violation''},
                        C. Jarlskog Ed., World Scientific, Singapore
                        (1988)

\bibitem{CKMfitter}     A.~H\"ocker, H.~Lacker, S.~Laplace and F.~Le Diberder,
                        \EPJ\ {\bf C21}, 225 (2001)

\bibitem{CKMweb}        Partially updated plots and results are
                        available at
                        the CKMfitter web site: \\
                        http://ckmfitter.in2p3.fr
                        and its mirror
                        http://www.slac.stanford.edu/xorg/ckmfitter

\bibitem{Achille1}      M.~Ciuchini, G.~D'Agostini, E.~Franco, V.~Lubicz,
                        G.~Martinelli, F.~Parodi, P.~Roudeau and A.~Stocchi,
                        J. High Energy Phys. {\bf 0107}, 013 (2001)

\bibitem{CernCkmWS}     M.~Battaglia \ea, CERN-2003-002,
                        Cern Yellow Report,
                        Based on the Workshop on CKM Unitarity Triangle
                        (CERN 2002-2003), Geneva, Switzerland, 13-16 Feb 2002,
                        hep-ph/0304132

\bibitem{otherCkm}      A.J.~Buras, F.~Parodi and A.~Stocchi,
                        J. High Energy Phys., \textbf{0301}, 029 (2003);
                        A.~Ali,
                        Lectures given at the International
                        Meeting on Fundamental Physics, Soto de Cangas, Spain,
                        23-28 Feb. 2003, hep-ph/0312303;
                        G.~Eigen, G.P.~Dubois-Felsmann, D.G.~Hitlin and
                        F.C. Porter, Contributed to International Europhysics
                        Conference on High-Energy Physics (HEP 2003),
                        Aachen, Germany, 17-23 Jul 2003, hep-ex/0312062
                        (see also hep-ph/0308262);
                        K.~Schubert,
                        Talk given at 2nd Workshop on the CKM Unitarity
                        Triangle, Durham, England, 5-9 Apr 2003,
                        http://www.ippp.dur.ac.uk/\~ckm/econf/index.html, WG3;
                        see also Ref.~\cite{CKMfitter}, which cites earlier studies,
                        and Ref.~\cite{CernCkmWS}

\bibitem{chau}          L.L.~Chau and W.Y.~Keung,
                        Phys. Rev. Lett. {\bf 53}, 1802 (1984)

\bibitem{PDG}           Particle Data Group (K. Hagiwara \ea),
                        Phys. Rev. {\bf D66}, 010001 (2002 and 2003 update)

\bibitem{jarlskog}      C.~Jarlskog,
                        Phys. Rev. Lett. {\bf 55}, 1039 (1985)

\bibitem{CPV-TheBook}   G.C.~Branco, L.~Lavoura and J.P.~Silva,
                        {\em CP Violation}, The Intern. Series of
                        Monographs on Physics - 103,
                        Oxford Science Publications, Oxford, UK (1999)

\bibitem{E787}          E787 Collaboration (S.~Adler \ea),
                        Phys. Rev. Lett. {\bf 88}, 041803 (2002)

\bibitem{E949}          E949 Collaboration (A.V. Artamonov \ea),
                        hep-ex/0403036 (2004), \\
                        http://www.phy.bnl.gov/e949


\bibitem{townerhardy}   I.S.~Towner and J.C.~Hardy,
                        J. Phys. G: Nucl. Part. Phys. {\bf 29}, 197 (2003)

\bibitem{vudsaito1}     K.~Saito and A.W.~Thomas,
                        Phys. Lett. {\bf B363}, 157 (1995)

\bibitem{PERKEO}        H.~Abele \ea,
                        Phys. Rev. Lett. {\bf 88}, 211801 (2002)

\bibitem{PIBETA}        D.~Pocanic (for the PIBETA Collaboration),
                        Talk given at 2nd Workshop on the CKM Unitarity
                        Triangle, Durham, England, 5-9 Apr 2003,
                        eConf {\bf C0304052}, WG606

\bibitem{vusdhk}        J.F.~Donoghue, B.R.~Holstein and S.W.~Klimt,
                        Phys. Rev. {\bf D35}, 934 (1987)

\bibitem{vusflores}     R.~Flores-Mendieta, A.~Garcia and G.~Sanchez-Colon,
                        Phys. Rev. {\bf D54}, 6855  (1996)

\bibitem{cabibbo2}      N.~Cabibbo, E.~C.~Swallow and R. Winston,
                        APS-123-QED, hep-ph/0307214 (2003)

\bibitem{vusleutw}      H.~Leutwyler and M.~Roos,
                        Z. Phys. {\bf C25}, 91 (1984)

\bibitem{vuspaschos}    E.A.~Paschos and U.~T\"urke,
                        Phys. Rep. {\bf 178}, 145 (1989)

\bibitem{vusjaus}       W.~Jaus,
                        Phys. Rev. {\bf D44}, 2851 (1991)

\bibitem{PostSchilcher} P.~Post, K.~Schilcher, Eur. Phys. J. {\bf C25}, 427 (2002)

\bibitem{BijnensTalavera} J.~Bijnens and P.~Talavera, Nucl. Phys. {\bf B 669}, 341 (2003)

\bibitem{CiriNeufeldPichl} V.~Cirigliano, H.~Neufeld and H.~Pichl, hep-ph/0401173

\bibitem{BECIf0}        D.~Becirevic \ea, hep-ph/0403217

\bibitem{vusmarciano}   W.J.~Marciano and A.~Sirlin,
                        Phys. Rev. Lett. {\bf 56}, 22 (1986)

\bibitem{vuswilliams}   H.H.~Williams \ea,
                        Phys. Rev. Lett. {\bf 33}, 240 (1974);
                        T.~Becherrawy \ea,
                        Phys. Rev. {\bf D1}, 1452 (1970)

\bibitem{cirigliano}    V.~Cirigliano,
                        Talk given at 38th Rencontres de Moriond on
                        Electroweak Interactions and Unified Theories,
                        Les Arcs, France, 15-22 Mar 2003, hep-ph/0305154

\bibitem{E865}          BNL-E865 Collaboration (A.~Sher \ea),
                        AIP Conf. Proc. {\bf 698}, 381 (2004),
                        hep-ex/0305042

\bibitem{CkmKLOE}       B.~Sciascia (for the KLOE Collaboration),
                        Talk given at 2nd Workshop on the CKM Unitarity
                        Triangle, Durham, England, 5-9 Apr 2003,
                        eConf {\bf C0304052}, WG607

\bibitem{KLOEMoriond04} S.~Miscetti (for the KLOE Collaboration),
                        Talk given at 39th Rencontres de Moriond:
                        Electroweak Interactions and Unified Theories
                        La Thuile, Val d'Aoste, Italy,
                        Mar 21-28 2004, hep-ex/0405040

\bibitem{VusKTeV}       KTeV Collaboration, (T.~Alexopoulos \ea), hep-ex/0406001

\bibitem{CkmJamin}      E.~Gamiz, M.~Jamin, A.~Pich, J.~Prades and F.~Schwab,
                        J. High Energy Phys. {\bf 0301}, 060 (2003)

\bibitem{vcdCdhs}       CDHS Collaboration (H.~Abramowicz \ea),
                        Z. Phys. {\bf C15}, 19 (1982)

\bibitem{vcdCcfr}       CCFR Collaboration (A.O.~Bazarko \ea),
                        Z. Phys. {\bf C65}, 189 (1995)

\bibitem{vcdCharmii}    CHARM~II Collaboration (P.~Vilain \ea),
                        Eur. Phys. J. {\bf C11}, 19 (1999)

\bibitem{bolton}        T.~Bolton, KSU-HEP-97-04, hep-ex/9708014 (1997)

\bibitem{ushida}        FNAL-E531 Collaboration (N.~Ushida \ea),
                        Phys. Lett. {\bf B206}, 375 (1988)

\bibitem{kubota}        CLEO Collaboration (Y.~Kubota \ea),
                        Phys. Rev. {\bf D54}, 2994 (1996)

\bibitem{heraB}         M.~Bargiotti \ea,
                        Riv. Nuovo Cim. \textbf{23} N3, 1 (2000)

\bibitem{vcsMontanet}   L.~Montanet \ea,
                        Phys. Rev. {\bf D50}, 1173 (1994)

\bibitem{vcsAleph}      ALEPH Collaboration (R.~Barate \ea),
                        Phys. Lett. {\bf B453}, 107 (1999);
                        ALEPH Collaboration (R.~Barate \ea),
                        ALEPH 99-064 CONF-99-038;
                        ALEPH Collaboration (R.~Barate \ea),
                        Phys. Lett. {\bf B465}, 349 (1999)

\bibitem{vcsDelphi}     DELPHI Collaboration (P.~Abreu \ea),
                        Phys. Lett. {\bf B439}, 209 (1998)

\bibitem{vcsOpal2}      OPAL Collaboration (G.~Abbiendi \ea),
                        Phys. Lett. {\bf B490}, 71 (2000)

\bibitem{LEPew}         LEP Electroweak Working Group,
                        CERN-EP-2001-098, hep-ex/0112021 (2001)

\bibitem{cleoc}         K.~Benslama (for the CLEO-c Collaboration),
                        Talk given at 37th Rencontres de Moriond on
                        Electroweak Interactions and Unified Theories,
                        Les Arcs, France, 9-16 Mar 2002,
                        hep-ex/0205003

\bibitem{IsgurWise}     N.~Isgur and M.~Wise,
                        Phys. Lett. {\bf B232}, 113 (1989);
                        Phys. Lett. {\bf 237}, 527 (1990)

\bibitem{hqet}          H.~Georgi, Phys. Lett. {\bf B238}, 395 (1990)

\bibitem{luketheorem}   M.~Luke,
                        Phys. Lett. {\bf B252}, 447 (1990)

\bibitem{uraltsevfpcp03}N.~Uraltsev,
                        Talk given at Flavor Physics and \CP Violation
                        (FPCP 2003), Paris, France, 3-6 Jun 2003,
                        hep-ph/0309081

\bibitem{ShifmanSR}     M.~Shifman, N.~Uraltsev and A.~Vainshtein, Phys. Rev.
                        {\bf D51}, 2217 (1995);
                        Erratum-ibid. {\bf D52}, 3149 (1995)

\bibitem{NeubertF1}     M.~Neubert,
                        Phys. Lett. {\bf B338}, 84 (1994)

\bibitem{NeubertFalk}   A.~Falk and M.~Neubert,
                        Phys. Rev. {\bf D47}, 2965 (1993)

\bibitem{KronfeldF1}    A.~Kronfeld,
                        Phys. Rev. {\bf D62}, 014505 (2000)

\bibitem{Harada}        J.~Harada \ea,
                        Phys. Rev. {\bf D65}, 094514 (2002)

\bibitem{Hashimoto}     S.~Hashimoto \ea,
                        Phys. Rev. {\bf D66}, 014503 (2002)

\bibitem{HFAG}          The Heavy Flavor Averaging Group,
                        http://www.slac.stanford.edu/xorg/hfag/
                        (Summer 2003 and Winter 2004 averages)

\bibitem{hqt}           I.I.~Bigi, M.~Shifman and N.G.~Uraltsev,
                        Annu. Rev. Nucl. Part. Sci. {\bf 47}, 591 (1997)

\bibitem{KineticMassS}  I.I.~Bigi, M.~Shifman, N.~Uraltsev and A.~Vainshtein,
                        Phys. Rev. {\bf D56}, 4017 (1997);
                        Phys. Rev. {\bf D52},  196 (1995)

\bibitem{CLEOXsgamma}   CLEO Collaboration (S.~Chen \ea),
                        Phys. Rev. Lett. {\bf 87}, 251807 (2001)

\bibitem{GLOBALVcb}     C.W.~Bauer, Z.~Ligeti, M.~Luke, A.V.~Manohar 
                        and M.~Trott,
                        Phys. Rev. {\bf D70}, 094017 (2004)

\bibitem{Schubert}      K.~Schubert,
                        Talk given at 21st International Symposium on
                        Lepton and Photon Interactions at High Energies
                        (LP 03), Batavia, Illinois, 11-16 Aug 2003

\bibitem{BABARelemom}   \babar\  Collaboration (B.~Aubert \ea),
                        \babar-PUB-03-045, hep-ex/0403030

\bibitem{BABARhadmom}   \babar\  Collaboration (B.~Aubert \ea),
                        \babar-CONF-03-034, hep-ex/0403031

\bibitem{BABARVcbfit}   \babar\  Collaboration (B.~Aubert \ea),
                        \babar-PUB-04-007, hep-ex/0404017

\bibitem{BABARrholnu}   \babar\  Collaboration (B.~Aubert \ea),
                        Phys. Rev. Lett. {\bf 90}, 181801  (2003)

\bibitem{CLEOpirholnu}  CLEO Collaboration (S.B.~Athar \ea),
                        Phys. Rev. {\bf D68}, 072003 (2003)

\bibitem{vub_cbArgus}   ARGUS Collaboration (H.~Albrecht \ea),
                        Phys. Lett. {\bf B255}, 297 (1991)

\bibitem{vub_cbCleo1}   CLEO Collaboration (R.~Fulton \ea),
                        Phys. Rev. Lett. {\bf 64}, 16 (1990)

\bibitem{vub_cbCleo2}   CLEO Collaboration (J.~Bartelt \ea),
                        Phys. Rev. Lett. {\bf 71}, 4111 (1993)

\bibitem{modACCMM}      G.~Altarelli, N.~Cabibbo, G.~Corb\`o,
                        L.~Maiani and G.~Martinelli,
                        Nucl. Phys. {\bf B208}, 365 (1982)

\bibitem{modISGW}       N.~Isgur, D.~Scora, B.~Grinstein and M.B.~Wise,
                        Phys. Rev. {\bf D39}, 799 (1989)

\bibitem{modKS}         J.G.~K\"orner and G.A.~Schuler,
                        Z. Phys. {\bf C38}, 511 (1988);
                        Erratum-ibid. {\bf C41}, 690 (1989)

\bibitem{modWSB}        M.~Wirbel, B.~Stech and M.~Bauer,
                        Z. Phys. {\bf C29}, 637 (1985)

\bibitem{CLEOendpoint}  CLEO Collaboration (A.~Bornheim \ea),
                        Phys. Rev. Lett. {\bf 88}, 231803 (2002)

\bibitem{BABARendpoint} \babar\  Collaboration (B.~Aubert \ea),
                        BABAR-CONF-02-012,
                        Contributed to 31st International Conference
                        on High Energy Physics (ICHEP 2002), Amsterdam,
                        The Netherlands, 24-31 Jul 2002, hep-ex/0207081

\bibitem{Belleendpoint} Belle Collaboration (K.~Abe \ea),
                        BELLE-CONF-0325 (2003),
                        Contributed to International Europhysics
                        Conference on High-Energy Physics (HEP 2001),
                        Budapest, Hungary, 12-18 Jul 2001

\bibitem{SubleadingLeibovich}
                        A.K.~Leibovich, Z.~Ligeti and M.B.~Wise,
                        Phys. Lett. {\bf B539}, 242 (2002)

\bibitem{SubleadingMannel}
                        C.W.~Bauer, M.~Luke and T.~Mannel,
                        Phys. Lett. {\bf B543}, 261 (2002)

\bibitem{SubleadingNeubert}
                        M.~Neubert,
                        Phys. Lett. {\bf B543}, 269 (2002)

\bibitem{SubleadingLuke}
                        C.N.~Burrell, M.E.~Luke and A.R.~Williamson,
                        Phys.Rev. {\bf D69}, 074015 (2004)

\bibitem{DualityBigiUraltsev}
                        I.I.Y.~Bigi and N.~Uraltsev,
                        Int. J. Mod. Phys. {\bf A16}, 5201 (2001)

\bibitem{Voloshin}      M.B.~Voloshin,
                        Phys. Lett. {\bf B515}, 74 (2001)

\bibitem{DELPHImX}      DELPHI Collaboration (P.~Abreu \ea),
                        Phys. Lett. {\bf B478}, 14 (2000)

\bibitem{BABARmX}       \babar\  Collaboration (B.~Aubert \ea),
                        Phys. Rev. Lett. {\bf 92}, 071802 (2004)

\bibitem{CLEOmXq2}      CLEO Collaboration (A.~Bornheim \ea),
                        CLEO-CONF-02-08,
                        Contributed to 31st International Conference
                        on High Energy Physics (ICHEP 2002), Amsterdam,
                        The Netherlands, 24-31 Jul 2002, hep-ex/0207064

\bibitem{BellemX}       A.~Sugiyama (for the Belle Collaboration),
                        Talk given at 38th Rencontres de Moriond on
                        Electroweak Interactions and Unified Theories,
                        Les Arcs, France, 15-22 Mar 2003, hep-ex/0306020

\bibitem{ALEPHVub}      ALEPH collaboration (R.~Barate \ea),
                        \EPJ\ {\bf C6}, 555 (1999)

\bibitem{OPALVub}       OPAL collaboration, (G.~Abbiendi \ea),
                        \EPJ\ {\bf C21}, 399 (2001)

\bibitem{Gibbons}       L. Gibbons (for the CLEO Collaboration),
                        To appear in the proceedings of 9th International Conference
                        on \B Physics at Hadron Machines (Beauty 2003), Pittsburgh,
                        Pennsylvania, 14-18 Oct 2003, hep-ex/0402009

\bibitem{cata}          O.~Cat\`a and S.~Peris,
                        hep-ph/0406094

\bibitem{Trippe}        T.~Trippe, private communication (2004)

\bibitem{KTeVepsk}      KTeV Collaboration, (T.~Alexopoulos \ea), hep-ex/0406002

\bibitem{Lellouch}      L.~Lellouch,
                        Nucl. Phys. Proc. Suppl. {\bf 94}, 142 (2001)

\bibitem{BK_Nc}         J.~Bijnens and J.~Prades,
                        Nucl. Phys. \textbf{B444}, 523 (1995);
                        J. High Energy Physics \textbf{01}, 002 (2000);
                        S.~Peris and E.~de~Rafael,
                        Phys. Rev. Lett. \textbf{490}, 213 (2000).

\bibitem{bbl}           G.~Buchalla, A.J.~Buras and M.E.~Lautenbacher,
                        Rev. Mod. Phys. {\bf 68}, 1125 (1996)

\bibitem{InamiLim}      T.~Inami and C.S.~Lim,
                        Prog. Theor. Phys. {\bf 65}, 297 (1981);
                        Erratum-ibid. {\bf 65}, 1772 (1981)

\bibitem{tevatronmtop}  L.~Cerrito (for the CDF and D0 Collaboration),
                        To appear in the proceedings of 39th Rencontres de
                        Moriond on QCD and High-Energy Hadronic Interactions,
                        La Thuile, Italy, 28 Mar - 4 Apr 2004,
                        hep-ex/0405046

\bibitem{mtopmatch1}    K.~Melnikov, T.~van Ritbergen,
                        Phys. Lett. {\bf B482}, 99 (2000)

\bibitem{mtopmatch2}    N.~Gray, D.J.~Broadhurst, W.~Grafe and K.~Schilcher,
                        Z. Phys. {\bf C48}, 673 (1990)

\bibitem{mtopmatch3}    D.J.~Broadhurst, N.~Gray and K. Schilcher,
                        Z. Phys. {\bf C52}, 111 (1991)

\bibitem{Nierste}       S.~Herrlich and U.~Nierste,
                        Nucl. Phys. {\bf B419}, 292 (1994)

\bibitem{Nierste2}      U.~Nierste, Ref.~\cite{CernCkmWS}
                        and private communication (2003)

\bibitem{Becirevic2}    D.~Becirevic,
                        Talk given at 2nd Workshop on the CKM Unitarity
                        Triangle, Durham, England, 5-9 Apr 2003,
                        hep-ph/0310072

\bibitem{Booth}         M.J.~Booth,
                        Phys. Rev. {\bf D51}, 2338 (1995)

\bibitem{SharpeZhang}   S.R.~Sharpe and Y.~Zhang,
                        Phys. Rev. {\bf D53}, 5125 (1996)

\bibitem{BernardBlumSoni} C.~Bernard, T.~Blum and A.~Soni,
                        Phys. Rev. {\bf D58} 014501 (1998)

\bibitem{Yamada}        JLQCD Collaboration (N.~Yamada \ea),
                        Nucl. Phys. B Proc. Suppl. {\bf 106}, 397 (2002),
                        hep-lat/0110087

\bibitem{Kronfeld}      S.~Kronfeld and S.M.~Ryan,
                        Phys. Lett. {\bf B543}, 59 (2002)

\bibitem{Becirevic1}    D.~Becirevic \ea,
                        Phys. Lett. {\bf B563}, 150 (2003)

\bibitem{Davies}        C.~Davies,
                        Talk given at Flavor Physics and \CP Violation
                        (FPCP 2003), Paris, France, 3-6 Jun 2003,
                        hep-ph/0311040

\bibitem{Rosner}        J.L.~Rosner,
                        Nucl. Instrum. Meth. {\bf A462}, 304 (2001)

\bibitem{dmsALEPH}      ALEPH Collaboration (A.~Heister \ea),
                        Eur. Phys. J. {\bf C29}, 143 (2003)

\bibitem{dmsDELPHI}     DELPHI Collaboration (P.~Abreu \ea),
                        Eur. Phys. J. {\bf C16}, 555 (2000);
                        DELPHI Collaboration (P.~Abreu \ea),
                        Eur. Phys. J. {\bf C18}, 229 (2000);
                        DELPHI Collaboration (J.~Adballah \ea),
                        Eur. Phys. J. {\bf C28}, 155 (2003);
                        DELPHI Collaboration (J.~Adballah \ea),
                        Contribution 607 to
                        30th International Conference on High-Energy Physics
                        (ICHEP 2002), Amsterdam, The Netherlands,
                        Jul 24-31 2002

\bibitem{dmsOPAL}       OPAL Collaboration (G.~Abbiendi \ea),
                        Eur. Phys. J. {\bf C11} 587, (1999);
                        OPAL Collaboration (G.~Abbiendi \ea),
                        Eur. Phys. J. {\bf C19}, 241 (2001)

\bibitem{dmsSLD}        SLD Collaboration (K.~Abe \ea),
                        Phys. Rev. {\bf D66}, 032009 (2002);
                        SLD Collaboration (K.~Abe \ea),
                        Phys. Rev. {\bf D67}, 012006 (2003);
                        SLD Collaboration (K.~Abe \ea),
                        SLAC-PUB-8568,
                        Contributed to 30th International Conference on
                        High-Energy Physics (ICHEP 2000), Osaka, Japan,
                        27 Jul - 2 Aug 2000, hep-ex/0012043

\bibitem{dmsCDF}        CDF Collaboration (F.~Abe \ea),
                        Phys. Rev. Lett. {\bf 82}, 3576 (1999)

\bibitem{Roussarie}     H.G.~Moser and A.~Roussarie,
                        \NIM\  {\bf A384}, 491 (1997)

\bibitem{toy}           D.~Abbaneo and G.~Boix,
                        J. High Energy Phys. {\bf 08}, 4 (1999)

\bibitem{checchiaLogLik}P.~Checchia, E.~Piotto and F.~Simonetto,
                        Contributed to International Europhysics Conference
                        on High-Energy Physics (EPS-HEP 99), Tampere,
                        Finland, 15-21 Jul 1999, hep-ph/9907300

\bibitem{AliMisiak}     A.~Ali and M.~Misiak,
                        Chapter 6.2 in Ref.~\cite{CernCkmWS}

\bibitem{bellerhogam}   M.~Iwasaki (for the Belle Collaboration),
                        Talk given at 39th Rencontres de Moriond:
                        Electroweak Interactions and Unified Theories
                        La Thuile, Val d'Aoste, Italy,
                        21-28 Mar 2004

\bibitem{alirhogam}     A.~Ali, E.~Lunghi and A.Y.~Parkhomenko,
                        DESY-04-065, hep-ph/0405075 (2004)

\bibitem{BABARstb}      \babar\  Collaboration (B.~Aubert \ea),
                        \PRL\  {\bf 89}, 201802 (2002)

\bibitem{Bellestb}      Belle Collaboration (K. Abe \ea), BELLE-CONF-0344,
                        Contributed to 21st International Symposium on
                        Lepton and Photon Interactions at High Energies
                        (LP 03), Batavia, Illinois, 11-16 Aug 2003,
                        hep-ex/0308036

\bibitem{RFpeng}        R.~Fleischer,
                        Phys. Lett. \textbf{B365}, 399 (1996)

\bibitem{sonipeng}      D.~London and A.~Soni,
                        Phys. Lett. \textbf{B407}, 61(1997);
                        see also Ref.~\cite{BabarPhysBook}

\bibitem{ligetiquinn}   Y.~Grossman, Z.~Ligeti, Y.~Nir and H.~Quinn
                        Phys. Rev. {\bf D68}, 015004 (2003)

\bibitem{grossrosner}   M.~Gronau, Y.~Grossman and J.L.~Rosner,
                        Phys. Lett. {\bf B579}, 331 (2004)

\bibitem{BN}            M.~Beneke and M.~Neubert,
                        \NP~ {\bf B675} 333 (2003)

\bibitem{BABARphiks}    \babar\  Collaboration (B.~Aubert \ea),
                        \babar-PUB-04/004, hep-ex/0403026 (2004)

\bibitem{Bellephiks}    Belle Collaboration (K.~Abe \ea),
                        Phys. Rev. Lett. {\bf 91}, 261602 (2003)

\bibitem{babarallbeta}  M.~Verderi (for the \babar\  Collaboration),
                        Talk given at 39th Rencontres de Moriond:
                        Electroweak Interactions and Unified Theories
                        La Thuile, Val d'Aoste, Italy,
                        21-28 Mar 2004

\bibitem{babarf0ks}     \babar\  Collaboration (B.~Aubert \ea),
                        \babar-PUB-04-017, hep-ex/0406040 (2004)

\bibitem{babaretaprks}  \babar\  Collaboration (B.~Aubert \ea),
                        Phys. Rev. Lett. {\bf 91}, 161801 (2003)

\bibitem{grolon}        M.~Gronau and D.~London,
                        \PRL\ {\bf 65}, 3381 (1990)

\bibitem{GrQu}          Y.~Grossman and H.R.~Quinn,
                        \PRD\ {\bf D58}, 017504 (1998)

\bibitem{theseJerome}   J.~Charles,  Th\`ese de l'Universit\'e
                        Paris-Sud, April 1999, LPT-Orsay 99-31;\\
                        available (in French) at
                        http://www.tel.ccsd.cnrs.fr (ID 00002502)

\bibitem{GrLoSiSi}      M.~Gronau, D.~London, N.~Sinha and R.~Sinha,
                        Phys. Lett. {\bf B514}, 315 (2001)

\bibitem{babarrhorhoprl}\babar\ Collaboration (B.~Aubert \ea),
                        \babar-PUB-04-09, hep-ex/0404029 (2004)

\bibitem{rhorhoold}     \babar\  Collaboration (B.~Aubert \ea),
                        Phys. Rev. {\bf D69}, 031102 (2004)

\bibitem{BABARrhoplusrho0}
                        \babar\  Collaboration (B. Aubert \ea),
                        \PRL\  {\bf 91}, 171802 (2003)

\bibitem{Bellerhoplusrho0}
                        Belle Collaboration (J.~Zhang \ea),
                        \PRL\  {\bf 91}, 221801 (2003)

\bibitem{BABARrhorho}   L.~Roos (for the \babar\  Collaboration),
                        Talk given at 39th Rencontres de Moriond:
                        Electroweak Interactions and Unified Theories
                        La Thuile, Val d'Aoste, Italy,
                        21-28 Mar 2004

\bibitem{cmd2}          CMD-2 Collaboration (R.~Akhmetshin \ea)
                        Phys. Lett. {\bf B578}, 285 (2004)

\bibitem{aleph_vsf}     ALEPH Collaboration (R.~Barate \ea),
                        Z. Phys. {\bf C76}, 15 (1997)

\bibitem{ligetirhorho}  A.~Falk, Z.~Ligeti, Y.~Nir and H.~R.~Quinn,
                        Phys. Rev. {\bf D69}, 011502 (2004)

\bibitem{babarbtaunu}   \babar\  Collaboration (B. Aubert \ea),
                        \babar-CONF-03-004,
                        Contributed to 38th Rencontres de Moriond
                        on Electroweak Interactions and Unified Theories,
                        Les Arcs, France, 15-22 Mar 2003, hep-ex/0304030

\bibitem{bellebmunu}    Belle Collaboration, (K.~Abe \ea),
                        BELLE-CONF-0247,
                        Contributed to 31st International Conference
                        on High Energy Physics (ICHEP 2002), Amsterdam,
                        The Netherlands, 24-31 Jul 2002

\bibitem{nirichep}      Y.~Nir,
                        Nucl. Phys. Proc. Suppl. {\bf 117}, 111 (2003),
                        hep-ph/0208080

\bibitem{quinngrossman} Y.~Grossman and H.~R.~Quinn,
                        Phys. Rev. \textbf{D56}, 7259 (1997)

\bibitem{lassKpi}       LASS Collaboration (D.~Aston \ea),
                        Nucl. Phys. {\bf B296}, 493 (1988)

\bibitem{13r}           J.~Charles \ea,
                        Phys. Lett. {\bf B425}, 375 (1998);
                        Erratum-ibid. {\bf B433}, 441 (1998)

\bibitem{7r}            C.W.~Chiang and L.~Wolfenstein,
                        Phys. Rev. {\bf D61}, 074031 (2000)

\bibitem{15r}           A.S.~Dighe, I.~Dunietz and R.~Fleischer,
                        Eur. Phys. J. {\bf C6}, 647 (1999)

\bibitem{17r}           B.~Kayser,
                        in Proceedings of the 32nd Rencontres de
                        Moriond on Electroweak Interactions and
                        Unified Theories, Les Arcs,
                        France, edited by J. Tr\^an Thanh V\^an,
                        Ed. Fronti\`eres (1997), p.389;
                        Y.I.~Azimov,
                        JETP Lett. {\bf 50}, 447 (1989);
                        Phys. Rev. {\bf D42}, 3705 (1990)

\bibitem{18r}           H.R.~Quinn, T.~Schietinger, J.P.~Silva and A.E.~Snyder,
                        Phys. Rev. Lett. {\bf 85}, 5284 (2000)

\bibitem{19r}           Y.~Grossman and D.~Pirjol,
                        J. High Energy Phys. {\bf 006}, 029 (2000)

\bibitem{Buras03}       A.J.~Buras and M.~Jamin,
                        J. High Energy Phys. {\bf 0401}, 048 (2004)

\bibitem{PajuoMayor}    A.~J.~Buras, F.~Schwab and S.~Uhlig, TUM-HEP-547, MPP-2004-47,
                        hep-ph/0405132 (2004)

\bibitem{na31}          NA31 Collaboration (G.D.~Barr \ea),
                        Phys. Lett. {\bf B317}, 233 (1993)

\bibitem{e731}          E731 Collaboration (L.K.~Gibbons \ea),
                        Phys. Rev. {\bf D55}, 6625 (1997)

\bibitem{na48}          NA48 Collaboration (J.R.~Batley \ea),
                        Phys. Lett. {\bf B544}, 97 (2002)

\bibitem{ktev}          KTeV Collaboration (A.~Alavi-Harati \ea),
                        Phys. Rev. {\bf D67}, 012005 (2003)

\bibitem{Buras93}       A.~Buras, M.~Jamin and M.E.~Lautenbacher,
                        Nucl. Phys. {\bf B408}, 209 (1993)

\bibitem{Roma93}        M.~Ciuchini, E.~Franco, G.~Martinelli and L.~Reina,
                        Phys. Lett. {\bf B301}, 263 (1993)

\bibitem{Piketal}       E.~Pallante, A.~Pich and I.~Scimemi,
                        Nucl. Phys. {\bf B617}, 441 (2001)

\bibitem{deRafael}      T.~Hambye, S.~Peris and E.~de Rafael,
                        J. High Energy Phys. {\bf 0305}, 027 (2003)

\bibitem{Bijnens}       J.~Bijnens, E.~Gamiz and J.~Prades,
                        J. High Energy Phys. {\bf 0110}, 009 (2001)

\bibitem{SPQCDR}        D.~Becirevic (for the SPQCDR Collaboration),
                        Nucl. Phys. Proc. Suppl. {\bf 119}, 359 (2003),
                        hep-lat/0209136

\bibitem{CPPACS}        CP-PACS Collaboration (S.~Aoki \ea),
                        Nucl. Phys. Proc. Suppl.  {\bf 106}, 332 (2002)

\bibitem{RBC}           RBC Collaboration (T.~Blum \ea),
                        Phys. Rev. {\bf D68}, 114506 (2003)

\bibitem{GP}            M.~Golterman and S.~Peris,
                        Phys. Rev. \textbf{D68}, 094506 (2003)

\bibitem{GPS_B6}              E.~Gamiz, J.~Prades and I.~Scimemi,
                        J. High Energy Phys. 0310, 042 (2003)

\bibitem{BuBu}          G.~Buchalla and A.~Buras,
                        Nucl. Phys. {\bf B548}, 309 (1999)

\bibitem{kaonmarciano}  W.J.~Marciano and Z.~Parsa,
                        Phys. Rev. {\bf D53}, 1 (1996)

\bibitem{CKM}           The CKM Collaboration,
                        Proposal to the FNAL PAC, \\
                        http://www.fnal.gov/projects/ckm/documentation/public/proposal/proposal.html

\bibitem{JHF}           T.~Inagaki (for the E391 Collaboration),
                        KEK-PREPRINT-96-181,
                        Talk given at 3rd International Workshop on
                        Particle Physics Phenomenology, Taipei,
                        Taiwan, 14-17 Nov 1996

\bibitem{KOPIO}         The KOPIO Collaboration,
                        RSVP MRE Proposal, \\
                        http://pubweb.bnl.gov/people/rsvp/proposal.ps

\bibitem{babardstarpiF} \babar\  Collaboration (B.~Aubert \ea),
                        \babar-PUB-03-031, hep-ex/0309017 (2003)

\bibitem{DsKref}       Belle Collaboration (P. Krokovny \ea),
                        Phys. Rev. Lett. \textbf{89}, 231804 (2002);
                        \babar\  Collaboration (B.~Aubert \ea),
                        BABAR-CONF-02-034
                        Contributed to 31st International Conference
                         on High Energy Physics (ICHEP 2002), Amsterdam,
                          The Netherlands, 24-31 Jul 2002, hep-ex/0207053

\bibitem{babardstarpiP} \babar\  Collaboration (B.~Aubert \ea),
                        \babar-PUB-03-033, hep-ex/0310037 (2003)

\bibitem{belledstarpiF} Belle  Collaboration (K. Abe \ea),
                        BELLE-CONF-0341,
                        Contributed to 21st International Symposium
                        on Lepton and Photon Interactions at High Energies
                        (LP 03), Batavia, Illinois, 11-16 Aug 2003,
                        hep-ex/0308048

\bibitem{gronaulondon}  M.~Gronau and D.~London,
                        Phys. Lett. {\bf B253}, 483 (1991)

\bibitem{gronauwyler}   M.~Gronau and D.~Wyler,
                        Phys. Lett. {\bf B265}, 172 (1991)

\bibitem{dunietz1}      I.~Dunietz,
                        Phys. Lett. {\bf B270}, 75 (1991)

\bibitem{dunietz2}      I.~Dunietz,
                        Z. Phys. {\bf C56}, 129 (1992)

\bibitem{ADS}           D.~Atwood, I.~Dunietz and A.~Soni,
                        Phys. Rev. Lett. {\bf 78}, 3257 (1997);
                        Phys. Rev. \textbf{D63}, 036005 (2001)

\bibitem{babardk}       \babar\  Collaboration (B.~Aubert \ea),
                        \babar-PUB-04-002, hep-ex/0402024 (2004)

\bibitem{d0kidea}       A.~Giri, Y.~Grossman, A.~Soffer and J.~Zupan,
                        Phys. Rev. {\bf D68}, 054018 (2003)

\bibitem{belled0k}      Belle Collaboration (K.~Abe \ea),
                        BELLE-CONF-0343, hep-ex/0308043 (2003);
                        updated analysis presented by A.~Poluektof
                        (for the Belle Collaboration)
                        at 39th Rencontres de Moriond:
                        Electroweak Interactions and Unified Theories
                        La Thuile, Val d'Aoste, Italy,
                        Mar 21-28, 2004.

\bibitem{belled0k2}     Belle Collaboration (K.~Abe \ea),
                        BELLE-CONF-0343, hep-ex/0406067 (2004)

\bibitem{cleod0}        CLEO Collaboration (H.~Muramatsu \ea),
                        Phys. Rev. Lett. {\bf 89}, 251802 (2002);
                        Erratum-ibid. {\bf 90}, 059901 (2003);
                        CLEO Collaboration (S.~Kopp \ea),
                        Phys. Rev. {\bf D63}, 092001 (2001)

\bibitem{bib:Gif89}     J.~Ha\"{\i}ssinski,
                        21st Ecole de Gif (1989),
                        in English, and references therein

\bibitem{bib:TheseMu}   M.~Pivk,
                        Th\`ese de l'Universit\'e Paris VII, May 2003,
                        \babar-THESIS-03/012, \\
                        available  (in French) at http://tel.ccsd.cnrs.fr
                        (ID 00002991)

\bibitem{BabarPhysBook} \babar\  Collaboration (P.F~Harrison and H.~Quinn eds.),
                        ``The \babar\   Physics Book: Physics at an
                        Asymmetric \B Factory'',
                        SLAC-R-0504 (1998)

\bibitem{charles}       J.~Charles,
                        \PRD\ {\bf D59} 054007, (1999)

\bibitem{garnderpipi}   S.~Gardner,
                        Phys. Rev. {\bf D59}, 077502 (1999)

\bibitem{ourpipi}       A.~H\"ocker, H.~Lacker, M.~Pivk and L.~Roos,
                        LAL 02-103 (2002), LPNHE 2002-15; writeup available at
                        http://www.slac.stanford.edu/xorg/ckmfitter/ckm\_talks.html

\bibitem{GrRo}          M.~Gronau and J.L.~Rosner,
                        \PRD\ {\bf D65}, 013004 (2002);
                        Erratum-ibid. {\bf D65}, 079901 (2002)

\bibitem{FM2pi}         R. Fleischer and T. Mannel,
                        Phys. Lett. {\bf B397}, 269 (1997)

\bibitem{BBNS0}         M.~Beneke, G.~Buchalla, M.~Neubert and C.T.~Sachrajda,
                        \PRL\  {\bf 83}, 1914 (1999)

\bibitem{BBNS}          M.~Beneke, G.~Buchalla, M.~Neubert and C.T.~Sachrajda,
                        \NP\  {\bf B606}, 245 (2001)

\bibitem{mufrapipi}     M.~Pivk and F.R.~Le Diberder,
                        LAL 04-10, hep-ph/0406263, \EPJ\ {\bf C39}, 397 (2005)

\bibitem{BFPEW}       A.J.~Buras and R.~Fleischer,
                        Eur. Phys. J. \textbf{C11}, 93 (1999)

\bibitem{NRPew}         M.~Neubert and J.L.~Rosner,
                        \PL\  {\bf B441}, 403 (1998);
                        \PRL\  {\bf 81}, 5076 (1998)

\bibitem{silvwolf}      J.P.~Silva and L.~Wolfenstein,
                                Phys. Rev. \textbf{D49}, 1151 (1994)

\bibitem{fleischer}     R. Fleischer,
                        \PL\  {\bf B459}, 306 (1999)

\bibitem{colorTransparency}
                      J.D.~Bjorken in
                      \textit{``New Developments in High Energy
                      Physics''}, eds. E.G~Floratos and A.~Verganelakis,
                      Nucl. Phys. \textbf{B11} (Proc. Suppl.), 325
                      (1989)

\bibitem{StechNeubert}  M.~Neubert and B.~Stech,
                         Adv. Ser. Direct. High Energy Phys.
                         \textbf{15} 294 (1998)

\bibitem{KLS}           Y.Y.~Keum, H.~Li, A.I.~Sanda
                        \PL~ {\bf B504}, 6 (2001)

\bibitem{Ciuch}         M.~Ciuchini \ea,
                        \PL~ {\bf B515}, 33 (2001)

\bibitem{SCETfacto}     C.W.~Bauer, D.~Pirjol, I.Z.~Rothstein and
                        I.W.~Stewart, MIT-CTP-3469, hep-ph/0401188 (2004)

\bibitem{pipiBabar}     H.~Jawahery,
                        Talk given at 21st International Symposium on
                        Lepton and Photon Interactions at High Energies
                        (LP 03), Batavia, Illinois, 11-16 Aug 2003

\bibitem{pipiBelle}     Belle Collaboration (K.~Abe \ea),
                        BELLE-PREPRINT-2004-1, hep-ex/0401029 (2004)

\bibitem{BABARK0pi0}    T.~Browder,
                        Talk given at 21st International Symposium on
                        Lepton and Photon Interactions at High Energies
                        (LP 03), Batavia, Illinois, 11-16 Aug 2003

\bibitem{BABAR2}        \babar\ Collaboration (B.~Aubert \ea),
                        \PRL\  {\bf 91}, 021801 (2003)

\bibitem{BELLEACP}      T. Tomura (for the Belle Collaboration),
                        Proceedings of 38th Rencontres de Moriond
                        on Electroweak Interactions and Unified Theories,
                        Les Arcs, France, 15-22 Mar 2003, hep-ex/0305036

\bibitem{BABARBelleLP03}J.~Fry,
                        Talk given at 21st International Symposium on
                        Lepton and Photon Interactions at High Energies
                        (LP 03), Batavia, Illinois, 11-16 Aug 2003

\bibitem{CLEO2}         CLEO Collaboration (S.~Chen \ea),
                        \PRL{85}, 525 (2000)

\bibitem{BABAR3}        M.~Bona (for the \babar\  Collaboration),
                        Talk given at Flavor Physics and \CP Violation
                        (FPCP 2003), Paris, France, 3-6 Jun 2003

\bibitem{BABAR1}        \babar\ Collaboration (B.~Aubert \ea),
                        \PRL\  {\bf 89}, 281802 (2001).

\bibitem{BELLE1}        Belle Collaboration (Y.~Chao \ea),
                        BELLE-PREPRINT-2003-26, hep-ex/0311061 (2003)

\bibitem{CLEO1}         CLEO Collaboration (A. Bornheim \ea),
                        Phys. Rev. {\bf D68}, 052002 (2003)

\bibitem{BABAR0}        \babar\ Collaboration (B.~Aubert \ea),
                        Phys. Rev. Lett. {\bf 91}, 241801 (2003)

\bibitem{CDFKpiblessed} CDF Collaboration (C.~Paus \ea),
                        Results prepared for Summer conferences (2003),
                        http://www-cdf.fnal.gov/physics/new/bottom/030529.blessed-bhh/


\bibitem{BFRS2}          A.J.~Buras, R.~Fleischer, S.~Recksiegel and F.~Schwab,
                        Phys. Rev. Lett. {\bf 92}, 101804 (2004);
                        A.J.~Buras, R.~Fleischer, S.~Recksiegel and F.~Schwab,
                        CERN-PH-TH-2004-020, hep-ph/0402112 (2004)

\bibitem{analyses2pi}   S.~Barshay, L.M.~Sehgal and J.~van~Leusen, Phys.
                        Lett. \textbf{B591}, 97 (2004);
                        A.~Ali, E.~Lunghi and A.Ya.~Parkhomenko, DESY-04-036,
                        hep-ph/0403275 (2004);
                        M.~Gronau and J.L.~Rosner, TECHNION-PH-2004-21,
                        hep-ph/0405173 (2004);
                        G.~Buchalla and A.~Salim~Safir, LMU-24-03,
                        hep-ph/0406016 (2004)

\bibitem{ckmws03}       L. Roos,
                        Talk given at 2nd Workshop on the CKM Unitarity
                        Triangle, Durham, England, 5-9 Apr 2003,
                        eConf {\bf C0304052}, WG418

\bibitem{CGRS_PP}       C.-W.~Chiang, M.~Gronau, J.L.~Rosner and
                        D.A.~Suprun, MADPH-04-1372, hep-ph/0404073 (2004);

\bibitem{pirjolKK}      D.~Pirjol,
                        Phys. Rev. \textbf{D60}, 054020 (1999)

\bibitem{LipkinSR}      M. Gronau and J.L.~Rosner,
                        Phys. Rev. \textbf{D59}, 113002 (1999);
                        H.J.~Lipkin,
                        Phys. Lett. \textbf{B445}, 403 (1999);
                        J.~Matias,
                        Phys. Lett. \textbf{B520}, 131 (2001)

\bibitem{GRKpi03}       M. Gronau and J.L.~Rosner,
                        Phys. Lett. \textbf{B572}, 43 (2003)

\bibitem{BFRS1}       A.J.~Buras, R.~Fleischer, S.~Recksiegel and F.~Schwab,
                        Eur. Phys. J. {\bf C32}, 45 (2003)

\bibitem{KpiBarger}     V.~Barger, C.-W.~Chiang, P.~Langacker, H.S.~Lee,
                        MADPH-04-1381, hep-ph/0406126 (2004)

\bibitem{NQmethod}      Y.~Nir and H.R.~Quinn,
                        Phys. Rev. Lett. {\bf 67}, 541 (1991)

\bibitem{lavoura}       L.~Lavoura,
                        Mod. Phys. Lett. {\bf A7}, 1553 (1992)

\bibitem{london}        M.~Imbeault, A.~St-Laurent Lemerle,
                        V.~Page and D.~London,
                        Phys. Rev. Lett. {\bf 92}, 081801 (2004)

\bibitem{messGronauEtAl}
                        M.~Gronau, O.F.~Hernandez, D.~London and J.L.~Rosner,
                        Phys. Rev. \textbf{D52}, 6374 (1995)

\bibitem{GRKaPiReview}  M. Gronau and J.L.~Rosner,
                        CLNS 03/1852, TECHNION-PH-2003-41,
                        To appear in the proceedings of Workshop on the
                        Discovery Potential of an Asymmetric \B Factory at
                        $10^{36}$ Luminosity, Menlo Park, California, 8-10 May 2003,
                        hep-ph/0311280 (2003)

\bibitem{GPY}           M.~Gronau, D.~Pirjol and T.M.~Yan,
                        Phys. Rev. \textbf{D60}, 034021 (1999);
                        Erratum-ibid. \textbf{D69}, 119901 (2004)

\bibitem{rhopipaper}    \babar\   Collaboration (B.~Aubert \ea),
                        Phys. Rev. Lett. {\bf 91}, 201802 (2003);
                        updated results \babar-PLOT-0055 (2003)

\bibitem{PV_dueal}      D.~Du, J.~Sun, G.~Zhu and D.~Du,
                        Phys. Rev. {\bf D68}, 054003 (2003)
                        D.~Du, J.~Sun, D.~Yang and G.~Zhu,
                        Phys. Rev. {\bf D67}, 014023 (2003);
                        D.~Du, H.~Gong, J.~Sun, D.~Yang and G.~Zhu
                        Phys. Rev. {\bf D65}, 094025 (2002);
                        Erratum-ibid. {\bf D66}, 079904 (2002);

\bibitem{PV_orsay}      R.~Aleksan, P.-F.~Giraud, V.~Mor\'enas, O.~P\`ene
                        and A. S. Safir,
                        Phys. Rev. {\bf D67}, 094019 (2003)

\bibitem{PV_brits}      N.~de Groot, W.N.~Cottingham and I.B.~Whittingham,
                        Contributed to 38th Rencontres de Moriond on
                        QCD and Hadronic Interactions,
                        Les Arcs, France, 22-29 Mar 2003, hep-ph/0305263
                        eConf {\bf C0304052}, WG405

\bibitem{PV_pheno}      C.-W. Chiang, M.~Gronau, Z.~Luo, J.L.~Rosner and
                        D.A.~Suprun,
                        Phys. Rev. {\bf D69}, 034001 (2004)

\bibitem{lalrhopi}      A.~H\"ocker, M.~Laget, S.~Laplace and
                        J.~von~Wimmersperg-Toeller,
                        LAL 03-17 (2003), writeup available at
                        http://www.slac.stanford.edu/xorg/ckmfitter/ckm\_talks.html

\bibitem{SnyderQuinn}   H.R.~Quinn and A.E.~Snyder,
                        Phys. Rev. {\bf D48}, 2139 (1993)

\bibitem{Sophie}        S.~Versill\'e, Th\`ese de l'Universit\'e
                        Paris-Sud, April 1999; \\
                        available (in French) at
                        http://www-lpnhep.in2p3.fr/babar/public/versille/Thesis/

\bibitem{BELLErhopi}    Belle Collaboration, (K.~Abe \ea),
                        BELLE-CONF-0318,
                        Contributed to International Europhysics
                        Conference on High-Energy Physics (HEP 2001),
                        Budapest, Hungary, 12-18 Jul 2001

\bibitem{CLEOrhopi}     CLEO Collaboration (C.P.~Jessop \ea),
                        Phys. Rev. Lett. {\bf 85}, 2881 (2000)

\bibitem{rhopiBRpaper}  \babar\ Collaboration, (B.~Aubert \ea),
                        \babar-PUB-03-037, hep-ex/0311049 (2003)

\bibitem{rhopiCpaper}   \babar\ Collaboration, (B.~Aubert \ea),
                        \babar-CONF-03-14,
                        Contributed to International Europhysics
                        Conference on High-Energy Physics (HEP 2003),
                        Aachen, Germany, 17-23 Jul 2003, hep-ex/0307087

\bibitem{BELLErho+pi0}  Belle Collaboration (J.~Zhang \ea),
                        Belle-PREPRINT-2004-15, hep-ex/0406006 (2004)

\bibitem{BELLErhopi00}  Belle Collaboration (J.~Dragic \ea),
                        BELLE-PREPRINT-2004-14, hep-ex/0405068 (2004)

\bibitem{BELLEKKpi}     Belle Collaboration (K.~Abe \ea),
                        BELLE-CONF-0317,
                        Contributed to International Europhysics
                        Conference on High-Energy Physics (HEP 2003),
                        Aachen, Germany, 17-23 Jul 2003

\bibitem{CLEOKstrpi}    CLEO Collaboration (E.~Eckhart \ea),
                        Phys. Rev. Lett. {\bf 89}, 251801 (2002)

\bibitem{CLEOKstpiC}    CLEO Collaboration (D.M.~Asner \ea),
                        Phys. Rev. {\bf D53}, 1039 (1996)

\bibitem{BabarKstpiC}   \babar\  Collaboration (B.~Aubert \ea),
                        \babar-PUB-03-027, hep-ex/0308065 (2003)

\bibitem{BelleKstpiC}   Belle Collaboration (K.~Abe \ea),
                        BELLE-CONF-0338,
                        Contributed to 21st International Symposium on
                        Lepton and Photon Interactions at High Energies
                        (LP 03), Batavia, Illinois, 11-16 Aug 2003

\bibitem{aleksanRR}     R.~Aleksan, F.~Buccella, A.~Le Yaouanc, L.~Oliver,
                        O.~P\`ene and J.-C.~Raynal
                        \PL\  {\bf B356}, 96 (1995)

\bibitem{oldsnyderetal} H.J.~Lipkin, Y.~Nir, H.R.~Quinn and A.~Snyder,
                        Phys. Rev. {\bf D44}, 1454 (1991)

\bibitem{quinnsilva}    H.R.~Quinn and J.~Silva,
                        Phys. Rev. \textbf{D62}, 054002 (2000)

\bibitem{transv}        I.~Dunietz, H.R.~Quinn, A.~Snyder, W.~Toki and
                        H.J.~Lipkin,
                        Phys. Rev. {\bf D43}, 2193 (1991);
                        A.S.~Dighe, I.~Dunietz, HJ.~Lipkin and J.L.~Rosner,
                        Phys. Lett. {\bf B369}, 144 (1996)

\bibitem{kaganVV}       A.L.~Kagan, UCTP-102-04, hep-ph/0405134 (2004)

\bibitem{LEET}          J.~Charles, A.~Le~Yaouanc, L.~Oliver, O.~P\`ene
                        and J.C.~Raynal, Phys. Rev. \textbf{D60},
                        014001 (1999)

\bibitem{zito}          M.~Zito,
                        Phys. Lett. {\bf B586}, 314 (2004)

\bibitem{charmingpenguins}
                        M.~Ciuchini, E.~Franco, G.~Martinelli and L.~Silvestrini,
                        Nucl. Phys. \textbf{B501}, 271 (1997);
                        M.~Ciuchini, R~Contino, E.~Franco, G.~Martinelli
                        and L.~Silvestrini,
                        Nucl. Phys. \textbf{B512}, 3 (1998);
                        Erratum-ibid. \textbf{B531}, 656 (1998);
                        M.~Ciuchini, E.~Franco, G.~Martinelli, M.~Pierini
                        and L.~Silvestrini,
                        Phys. Lett. \textbf{B515}, 33 (2001);
                        A.~Khodjamirian, Th.~Mannel and~B. Melic,
                        Phys. Lett. \textbf{B571}, 75 (2003)

\bibitem{rescattering}  J.F.~Donoghue, E.~Golowich, A.A.~Petrov and J.M.~Soares,
                        Phys. Rev. Lett. \textbf{77}, 2178 (1996)

\bibitem{uut}           See, e.g., Ref.~\cite{GrossmanNirWorah} and
                        references therein

\bibitem{FleischerIsidoriMatias}
                        R.~Fleischer, G.~Isidori and J.~Matias,
                        J. High Energy Phys. {\bf 0305}, 053 (2003)

\bibitem{soares}         J.M.~Soares and L.~Wolfenstein,
                        Phys. Rev. \textbf{D47}, 1021 (1993)

\bibitem{GrossmanNirWorah} Y.~Grossman, Y.~Nir and M.P.~Worah,
                        Phys. Lett. {\bf B407}, 307 (1997)

\bibitem{KN}            A.L.~Kagan and M.~Neubert,
                        Phys. Lett. \textbf{B492}, 115 (2000)

\bibitem{FMpsiK}        R.~Fleischer and T.~Mannel,
                        Phys. Lett. \textbf{B506}, 311 (2001)

\bibitem{AtwoodHiller}  D.~Atwood and G.~Hiller,
                        LMU-09-03, hep-ph/0307251 (2003)

\bibitem{LaplaceLigetiNirPerez}
                        S.~Laplace, Z.~Ligeti, Y.~Nir and G.~Perez,
                        Phys. Rev. {\bf D65}, 094040 (2002)

\bibitem{SandaXing}     A.I.~Sanda, and Z-z.~Xing,
                        Phys. Rev. {\bf D56}, 6866 (1997)

\bibitem{RandallSu}     L.~Randall and S.~Su,
                        Nucl. Phys. {\bf B540}, 37 (1999)

\bibitem{CahnNir}       R.N.~Cahn and M.P.~Worah,
                        Phys. Rev. {\bf D60}, 076006 (1999)

\bibitem{BarenboimEyalNir}
                        G.~Barenboim, G.~Eyal and Y.~Nir,
                        Phys. Rev. Lett. {\bf 83}, 4486 (1999)

\bibitem{Beneke:2003az} M.~Beneke, G.~Buchalla, A.~Lenz and U.~Nierste,
                        Phys. Lett. B {\bf 576} 173 (2003)

\bibitem{Ciuchini:2003ww} M.~Ciuchini, E.~Franco, V.~Lubicz, F.~Mescia and C.~Tarantino,
                        JHEP {\bf 0308} 031 (2003)

\bibitem{OPALASL}       OPAL Collaboration (G.~Abbiendi \ea),
                        Eur. Phys. J. {\bf C12}, 609 (2000)

\bibitem{CLEOASL}       CLEO Collaboration (D.E.~Jaffe \ea),
                        Phys. Rev. Lett. {\bf 86}, 5000 (2001)

\bibitem{ALEPHASL}      ALEPH Collaboration (R.~Barate \ea),
                        Eur. Phys. J. {\bf C20}, 431 (2001)

\bibitem{BABARASL}      \babar\  Collaboration (B.~Aubert \ea),
                        Phys. Rev. Lett. {\bf 88}, 231801 (2002)

\bibitem{BABARDGAMMA}   \babar\  Collaboration (B.~Aubert \ea),
                        Phys. Rev. Lett. {\bf 92}, 181801 (2004)

\bibitem{Krueger}       G.C.~Branco, F.~Cagarrinho, F.~Kr\"uger,
                        Phys. Lett. {\bf B459} 224, (1999)

\bibitem{MFV}       M.~Ciuchini, G.~Degrassi, P.~Gambino and G.F.~Giudice,
                        Nucl. Phys. {\bf B534}, 3 (1998);
                        A.~Ali and D.~London,
                        Eur. Phys. J. {\bf C9}, 687 (1999);
                        A.~Ali and D.~London
                        Phys. Rept. {\bf320}, 79 (1999;
                   A.J.~Buras, P.~Gambino, M.~Gorbahn,
                        S.~Jager and L.~Silvestrini,
                        Phys. Lett. {\bf B500}, 161 (2001);
                      A.~J.~Buras, P.~Gambino, M.~Gorbahn,
                        S.~Jager and L.~Silvestrini,
                       Nucl. Phys. {\bf B592}, 55 (2001);
                         A.~J. Buras and R.~Buras,
                        Phys. Lett. {\bf B501}, 223 (2001);
                         A.~J.~Buras and R.~Fleischer,
                        Phys. Rev. {\bf D64}, 115010 (2001);
                        S.~Bergmann and G.~Perez,
                        Phys. Rev. {\bf D64}, 115009 (2001)

\bibitem{revueNP}       For recent references see, \eg,
                        Y.~Grossman, Int. J. Mod. Phys. \textbf{A19},
                        907 (2004);
                        M.~Neubert, CLNS-04-1878,
                        hep-ph/0405105 (2004)

\bibitem{GRphiK}        C.W.~Chiang and J.~Rosner,
                        Phys. Rev. \textbf{D68}, 014007 (2003)

\bibitem{phiKgal}       Y.~Grossman, G.~Isidori and M.P.~Worah,
                        Phys. Rev. \textbf{D58}, 057504 (1998);
                         R.~Fleischer and T.~Mannel,
                         Phys. Lett. \textbf{B511}, 240 (2001);
                         M.~Ciuchini and L.~Silvestrini,
                         Phys. Rev. Lett. \textbf{89}, 231802 (2002);
                         see also Ref.~\cite{hillerphiK}

\bibitem{GKLKSvsKL}     Y.~Grossman, A.L.~Kagan and Z.~Ligeti
                        Phys. Lett. \textbf{B538}, 327 (2002)

\bibitem{hillerphiK}    G.~Hiller,
                        Phys. Rev. \textbf{D66}, 071502 (2002)

\bibitem{PEW_phiK}      V.~Barger, C.-W.~Chiang, P.~Langacker and
                        H.-S.~Lee,
                        Phys. Lett. \textbf{B580}, 186 (2004);
                        N.G.~Deshpande and D.~K.~Ghosh, OITS-742,
                        hep-ph/0311332 (2003)

\bibitem{sceptical}     G.~D'Agostini,
                        CERN-EP-99-139, hep-ex/9910036 (1999)

\end{thebibliography}
\end{document}